\documentclass[12pt]{article}

\usepackage{amsmath, amsfonts, amssymb,latexsym,wasysym,graphicx,stmaryrd,tikz,bbold,upgreek}
\usepackage{pgfplots}
\pgfplotsset{compat=1.14}
\input epsf.sty
\usetikzlibrary{decorations.pathmorphing}
\usetikzlibrary{arrows.meta,arrows}
\usetikzlibrary{shapes,shapes.geometric}
\usetikzlibrary{decorations.markings}
\usetikzlibrary{trees}

\newtheorem{Lemma}{Lemma}[section]
\newtheorem{Proposition}[Lemma]{Proposition}

\newtheorem{Definition}[Lemma]{Definition}

\newcommand{\propagator}[2]{
\begin{tikzpicture}[scale=0.7] \label{fig:propagator}
\draw[->](0,0) -- (1.5,0); \draw(1.5,0)--(3,0);
\draw(0.3,0.4) node {$#1$};
\draw(2.7,0.4) node {$#2$};
\end{tikzpicture}}

\newcommand{\Pitilde}{
\begin{tikzpicture}[scale=1.1] \label{fig:Pitilde}
\draw(0,0) node {$\lambda$};
\draw(0.7,0.3)--(1.3,-0.3); \draw(0.7,-0.3)--(1.3,0.3);
\draw(1,0) node {$\bullet$};
\draw(0.6,0.2) node {$\alpha_1$}; \draw(1.4,0.2) node {$\alpha_3$};
\draw(0.6,-0.2) node {$\alpha_2$}; \draw(1.4,-0.2) node {$\alpha_4$};
\draw(2,0) node {$+\ \lambda^2$};

\draw(3,0) arc(180:150:1.5 and 3); \draw(3,0) arc(180:210:1.5 and 3);

\draw[rotate=90,->](0.7,-3-1) arc(30:90:0.88  and 0.44);
\draw[rotate=90](0.7-0.7,-3+0.2-1) arc(90:150:0.88  and 0.44);
\draw[rotate=90,->](-0.8,-3-1) arc(210:270:0.88 and 0.44);
\draw[rotate=90](-0.8+0.7,-3-0.2-1) arc(270:330:0.88 and 0.44);
\draw(3+1,0.7) node {$\bullet$};  \draw(3.6+1,0.6) node {$\xi_1=\xi_3$};
\draw(3+1,-0.7) node {$\bullet$};  \draw(3.6+1,-0.6) node {$\xi_2=\xi_4$};

\draw(1+2.7,1)--(1+3,0.7+0); \draw(1+3.3,1)--(1+3,0.7+0);
\draw(1+2.7,-1)--(1+3,-0.7+0); \draw(1+3.3,-1)--(1+3,-0.7+0);
\draw(1+2.6,0.9) node {$\alpha_1$}; \draw(1+3.4,0.9) node {$\alpha_3$};
\draw(1+2.6,-0.9) node {$\alpha_2$}; \draw(1+3.4,-0.9) node {$\alpha_4$};

\draw(1+4,0) node {$+$};

\draw[rotate=90](0.7,-3-2-1) arc(30:90:0.88  and 0.44);
\draw[rotate=90](0.7-0.7,-3+0.2-2-1) arc(90:150:0.88  and 0.44);
\draw[rotate=90](-0.8,-3-2-1) arc(210:270:0.88 and 0.44);
\draw[rotate=90](-0.8+0.7,-3-0.2-2-1) arc(270:330:0.88 and 0.44);
\draw(1+2+3,0.7) node {$\bullet$};  \draw(1+2+3.6,0.6) node {$\xi_1=\xi_3$};
\draw(1+2+3,-0.7) node {$\bullet$};  \draw(1+2+3.6,-0.6) node {$\xi_2=\xi_4$};
\draw(1+2+2.8,0) node {$\times$}; \draw(1+2+3.2,0) node {$\times$};

\draw(1+2+2.7,1)--(1+2+3,0.7+0); \draw(1+2+3.3,1)--(1+2+3,0.7+0);
\draw(1+2+2.7,-1)--(1+2+3,-0.7+0); \draw(1+2+3.3,-1)--(1+2+3,-0.7+0);
\draw(1+2+2.6,0.9) node {$\alpha_1$}; \draw(1+2+3.4,0.9) node {$\alpha_3$};
\draw(1+2+2.6,-0.9) node {$\alpha_2$}; \draw(1+2+3.4,-0.9) node {$\alpha_4$};

\draw(1.5+6,0) arc(0:30:1.5 and 3); \draw(1.5+6,0) arc(0:-30:1.5 and 3);

\draw(4+4,0) node {$+\ \lambda^3$};

\draw(3+5.5,0) arc(180:150:1.5 and 3); \draw(3+5.5,0) arc(180:210:1.5 and 3);

\draw(10.7,1) node {$\xi_1=\xi_3$}; \draw(9,-1) node {$\xi_2$};
\draw(11,-1) node {$\xi_4$};
\draw(10,1)--(9.3,-1); \draw(10,1)--(10.7,-1);
\draw(10,1) node {$\bullet$}; \draw(9.3,-1) node {$\bullet$};
\draw(10.7,-1) node {$\bullet$};
\draw(9.3,-1) arc(150:30:0.80 and 0.40);
\draw(9.3,-1) arc(210:330:0.80 and 0.40);

\draw(10,1)--(10.3,1.3); \draw(10,1)--(9.7,1.3);
\draw(9.3,-1)--(9,-1.3); \draw(10.7,-1)--(11,-1.3);

\draw(10+1.5,0) node {$+$};

\draw(12.3,1) node {$\xi_1=\xi_3$}; \draw(12,-1) node {$\xi_2$};
\draw(14,-1) node {$\xi_4$};
\draw(10+3,1)--(9.3+3,-1); \draw(10+3,1)--(10.7+3,-1);
\draw(10+3,1) node {$\bullet$}; \draw(9.3+3,-1) node {$\bullet$};
\draw(10.7+3,-1) node {$\bullet$};
\draw(9.3+3,-1) arc(150:30:0.80 and 0.40);
\draw(9.3+3,-1) arc(210:330:0.80 and 0.40);
\draw(10+3,-1+0.2) node {$\times$};\draw(10+3,-1-0.2) node {$\times$};

\draw(10+3,1)--(10.3+3,1.3); \draw(10+3,1)--(9.7+3,1.3);
\draw(9.3+3,-1)--(9+3,-1.3); \draw(10.7+3,-1)--(11+3,-1.3);

\draw(7+1.5+6,0) arc(0:30:1.5 and 3); \draw(7+1.5+6,0) arc(0:-30:1.5 and 3);

\end{tikzpicture}}

\newcommand{\Pitildedeux}{
\begin{tikzpicture}[scale=1.1] \label{fig:Pitildedeux}
\draw(2+4.5,0) node {$+\ \lambda^4$};

\draw(7.5,0) arc(180:150:1.5 and 3); \draw(7.5,0) arc(180:210:1.5 and 3);

\draw(1.5+2+1.5+4,1)--(1.5+2+1.7+4,1); \draw(1.5+2+1.7+4,1)--(1.5+2+3+4,1); \draw(1.5+2+3+4,1)--(1.5+2+4.3+4,1);\draw(1.5+2+4.5+4,1)--(1.5+2+4.3+4,1);
\draw(1.5+2+1.5+4,-1)--(1.5+2+1.7+4,-1); \draw(1.5+2+1.7+4,-1)--(1.5+2+3+4,-1); \draw(1.5+2+3+4,-1)--(1.5+2+4.3+4,-1);\draw(1.5+2+4.5+4,-1)--(1.5+2+4.3+4,-1);
\draw(1.5+2+2+4,1)--(1.5+2+2+4,0); \draw(1.5+2+2+4,0)--(1.5+2+2+4,-1);
\draw(1.5+2+4+4,1)--(1.5+2+4+4,0); \draw(1.5+2+4+4,0)--(1.5+2+4+4,-1);
\draw(1.5+2+2+4,1) arc(210:270:1.14  and 0.57);
\draw(1.5+2+3+4,0.7) arc(270:330:1.14  and 0.57);
\draw(1.5+2+4+4,-1) arc(30:90:1.14  and 0.57);
\draw(1.5+2+3+4,-0.7) arc(90:150:1.14  and 0.57);

\draw(1.5+2+2+4,1) node {$\bullet$}; \draw(1.5+2+4+4,1) node {$\bullet$};
\draw(1.5+2+2+4,-1) node {$\bullet$}; \draw(1.5+2+4+4,-1) node {$\bullet$};
\draw(1.5+2+2+4,1.3) node {$\xi_1$}; \draw(1.5+2+4+4,1.3) node {$\xi_3$};
\draw(1.5+2+2+4,-1.3) node {$\xi_2$}; \draw(1.5+2+4+4,-1.3) node {$\xi_4$};
\draw(-0.5+1.5+2+2+4,-0.5+1.3) node {$\alpha_1$};
\draw(-0.5+1.5+2+2+4,-1.3) node {$\alpha_2$};
\draw(2+0.5+1.5+2+2+4,-0.5+1.3) node {$\alpha_3$};
\draw(2+0.5+1.5+2+2+4,-1.3) node {$\alpha_4$};

\draw(5+1.5+6,0) node {$+\, \cdots$};

\draw(5+1.5+7,0) arc(0:30:1.5 and 3); \draw(5+1.5+7,0) arc(0:-30:1.5 and 3);

\end{tikzpicture}}

\newcommand{\RFmu}{
\begin{tikzpicture}[scale=2] \label{fig:RFmu}
\draw(-0.5,0.5) node {$-$};
\draw(2,0) -- (1.5,0); \draw(1.5,0)--(0.5,0);\draw(0.5,0)--(0,0);
\draw[->](1,0) arc(-90:90:0.5 and 0.5);
\draw(1,1) arc(90:270:0.5 and 0.5); \draw(1,0) node {$\bullet$};
\draw(1.8,0.1) node {\tiny $\bar{\psi}$}; \draw(0.2,0.1) node {\tiny $\psi$};
\draw(1.6,0.5) node {\tiny $\bar{\psi}$}; \draw(0.4,0.5) node {\tiny $\psi$};
\end{tikzpicture}}

\newcommand{\RFm}{\begin{tikzpicture}[scale=2] \label{fig:RFm}
\draw(-1,0) node {$-\frac{(m^*)^2}{p_F^*} \frac{d}{dp_{\perp}}$};
\draw[-<](0,0)--(1.25,0); \draw(1.25,0)--(2.5,0); \draw(2.3,0.1) node {\tiny $\psi$};
 \draw(0.2,0.1) node {\tiny $\bar{\psi}$};
\draw[->](2,0) arc(30:90:0.88  and 0.44); \draw(1.25,0.22) arc(90:150:0.88 and 0.44);
\draw[->](0.5,0) arc(210:270:0.88 and 0.44);
 \draw(1.25,-0.22) arc(270:330:0.88 and 0.44);
\draw(0.5,0) node {$\bullet$}; \draw(0.5,-0.2) node {\tiny $\xi$};
\draw(2,0) node {$\bullet$}; \draw(2,-0.2) node {\tiny $\xi'$};
\draw(0.1,-0.1) node {\tiny $\beta$}; \draw(0.7,0.3) node {\tiny $\alpha_1$};
\draw(0.8,0.05) node {\tiny $\alpha_2$}; \draw(0.7,-0.3) node {\tiny $\alpha_3$};
\end{tikzpicture}}

\newcommand{\RFZ}{
\begin{tikzpicture}[scale=2] \label{fig:RFZ}
\draw[-<](0,0)--(1.25,0); \draw(1.25,0)--(2.5,0); \draw(2.3,0.1) node {\tiny $\psi$};
 \draw(0.2,0.1) node {\tiny $\bar{\psi}$};
  \draw(-0.5,0) node {$-\II \frac{d}{dp^0}$};
\draw[->](2,0) arc(30:90:0.88  and 0.44);
 \draw(1.25,0.22) arc(90:150:0.88 and 0.44);

\draw[->](0.5,0) arc(210:270:0.88 and 0.44);
 \draw(1.25,-0.22) arc(270:330:0.88 and 0.44);
\draw(0.5,0) node {$\bullet$}; \draw(2,0) node {$\bullet$};
\end{tikzpicture}}

\newcommand{\RFGamma}{
\begin{tikzpicture}[scale=2] \label{fig:RFGamma}
\draw(2,0) -- (1.5,0); \draw(1.5,0)--(0.5,0);\draw(0.5,0)--(0,0);
\draw(1,0.5) circle(0.5);
\draw(1,1) node {$\times$}; \draw(1,1.15) node {\tiny $\Gamma^j$};
\draw(1.8,-0.1) node {\tiny $\bar{\psi}$}; \draw(0.2,-0.1) node {\tiny $\bar{\psi}$};
\draw(1.2,0.1) node {\tiny $\psi$}; \draw(0.8,0.1) node {\tiny $\psi$};
\draw(1,0) node {$\bullet$}; 

\end{tikzpicture}}

\newcommand{\BetheSalpeterbis}{
\begin{tikzpicture}[scale=1.5] \label{fig:BetheSalpeterbis}

\draw(1,1)--(1.2,1); \draw[<->](1.2,1)--(3,1); \draw(3,1)--(4.8,1);\draw[-<](5,1)--(4.8,1);
\draw(1,-1)--(1.2,-1); \draw[<-<](1.2,-1)--(3,-1); \draw(3,-1)--(4.8,-1);\draw[-<](5,-1)--(4.8,-1);
\draw[-<](2,1)--(2,0); \draw(2,0)--(2,-1);
\draw[->](4,1)--(4,0); \draw(4,0)--(4,-1);
\draw[<->](2.44,0.8) arc(240:300:1.14  and 0.57);
\draw(2,1) arc(210:240:1.14 and 0.57);
\draw(3,0.72) node {$\times$};  \draw(3,0.5) node {$\Gamma^*$};
\draw(3.6,0.8) arc(300:330:1.14  and 0.57);

\draw[<->](2.44,-0.8) arc(120:60:1.14  and 0.57);
\draw(2,-1) arc(150:120:1.14 and 0.57);
\draw(3,-0.72) node {$\times$}; \draw(3,-0.5) node {$\Gamma^*$};
\draw(3.6,-0.8) arc(60:30:1.14  and 0.57);

\end{tikzpicture}}

\newcommand{\BetheSalpeter}{
\begin{tikzpicture}[scale=2] \label{fig:BetheSalpeter}

\draw[-<](4.5,1)--(4.75,0.85); \draw(4.75,0.85)--(5,0.7);
 \draw[-<](4.5,-1.1)--(4.75,-0.95); \draw(4.75,-0.95)--(5,-0.8);
\draw(4.7,1.1) node {$p_1$}; \draw(5,0.9) node {$\xi_1$}; 
\draw(4.7,-1.1) node {$p_2$}; \draw(5,-1) node {$\xi_2$}; 

\draw[rotate=90,->](0.7,-5) arc(30:90:0.88  and 0.44);
\draw[rotate=90](0.7-0.7,-5+0.2) arc(90:150:0.88  and 0.44);
\draw[rotate=90,->](-0.8,-5) arc(210:270:0.88 and 0.44);
\draw[rotate=90](-0.8+0.7,-5-0.2) arc(270:330:0.88 and 0.44);
\draw[-<](5,0.7)--(6,0.35); \draw(6,0.35)--(7,0); 
 \draw[-<](5,-0.8)--(6,-0.4); \draw(6,-0.4)--(7,0);
\draw[-<](7,0)--(7.25,0.1);\draw(7.25,0.1)--(7.5,0.2); 
\draw[-<](7,0)--(7.25,-0.1);\draw(7.25,-0.1)--(7.5,-0.2);
\draw(6.9,-0.3) node {$\xi_3\!=\!\xi_4$};
\draw(7.4,0.4) node {$p_3$};
\draw(7.4,-0.4) node {$p_4$};

\end{tikzpicture}}

\newcommand{\twopointfunction}{
\begin{tikzpicture}[scale=1.1] 
\label{fig:twopointfunction}
\draw(2-4.5,-0.9) arc(30:150:0.88  and 0.44);
\draw(0.5-4.5,-0.9) arc(210:330:0.88 and 0.44);
\draw(-0.7-3.8,-0.9) node {$\Gamma(\theta')$};
\draw(0.8-2.8,-0.9) node {$\Gamma(\theta'')$};

\draw(0.5+0.3-1.5,-0.9) node {$+\sum_{k\ge 0}$};

\draw(2+2-1.5,-0.9) arc(30:150:0.88  and 0.44);
\draw(2+0.5-1.5,-0.9) arc(210:330:0.88 and 0.44);
\draw(1-0.6,-0.9) node {$\Gamma(\theta')$};
\draw(2+0.2,-0.9) node {\tiny $\sigma^{\perp}$};

\draw(2+2,-0.9) arc(30:150:0.88  and 0.44);
\draw(2+0.5,-0.9) arc(210:330:0.88 and 0.44);
\draw(2+0.8,-0.9) node {\tiny $\sigma^{\perp}$}; \draw(2+1.6,-0.9) node {\tiny $\sigma^{\perp}$};

\draw(2+2+1.5,-0.9) arc(30:150:0.88  and 0.44);
\draw(2+0.5+1.5,-0.9) arc(210:330:0.88 and 0.44);
\draw(2+0.8+1.5,-0.9) node {\tiny $\sigma^{\perp}$}; \draw(2+1.6+1.5,-0.9) node {\tiny $\sigma^{\perp}$};

\draw(2+3.9,-0.9) node {$..........$};
\draw(2+2+3.8,-0.9) arc(30:150:0.88  and 0.44);
\draw(2+0.5+3.8,-0.9) arc(210:330:0.88 and 0.44);
\draw(2+0.8+3.8,-0.9) node {\tiny $\sigma^{\perp}$}; \draw(2+1.6+3.8,-0.9) node {\tiny $\sigma^{\perp}$};

\draw(2+1.25,-0.5) node {$1$};
\draw(2+1.25+1.5,-0.5) node {$2$};
\draw(2+1.25+3.8,-0.5) node {$k$};

\draw(2+6.1,-0.9) node {\tiny $\sigma^{\perp}$};
\draw(2.3+0.7+7,-0.9) node {$\Gamma(\theta'')$};

\draw(2+2+5.3,-0.9) arc(30:150:0.88  and 0.44);
\draw(2+0.5+5.3,-0.9) arc(210:330:0.88 and 0.44);
\end{tikzpicture}}

\newcommand{\boundstate}{
\begin{tikzpicture}[scale=1.8]  \label{fig:boundstate}
\draw(-1.2,-0.9) node {$\sum_{k\ge 0}$};

\draw(0.5-0.7,-0.7-0.9) -- (0.5,-0.9);
\draw(0.5-0.7,0.7-0.9)--(0.5,-0.9);
\draw(0,-0.9) node {\tiny $\sigma(\theta+\frac{\pi}{2})$};

\draw(2,-0.9) arc(30:150:0.88  and 0.44);
\draw(0.5,-0.9) arc(210:330:0.88 and 0.44);
\draw(0.8,-0.9) node {\tiny $\sigma(\theta+\frac{\pi}{2})$}; \draw(1.6,-0.9) node {\tiny $\sigma(\theta+\frac{\pi}{2})$};

\draw(2+1.5,-0.9) arc(30:150:0.88  and 0.44);
\draw(0.5+1.5,-0.9) arc(210:330:0.88 and 0.44);
\draw(0.8+1.5,-0.9) node {\tiny $\sigma(\theta+\frac{\pi}{2})$}; \draw(1.6+1.5,-0.9) node {\tiny $\sigma(\theta+\frac{\pi}{2})$};

\draw(3.9,-0.9) node {$..........$};
\draw(2+3.8,-0.9) arc(30:150:0.88  and 0.44);
\draw(0.5+3.8,-0.9) arc(210:330:0.88 and 0.44);
\draw(0.8+3.8,-0.9) node {\tiny $\sigma(\theta+\frac{\pi}{2})$}; \draw(1.6+3.8,-0.9) node {\tiny $\sigma(\theta+\frac{\pi}{2})$};

\draw(1.25,-0.5) node {$1$};
\draw(1.25+1.5,-0.5) node {$2$};
\draw(1.25+3.8,-0.5) node {$k$};

\draw(5.8+0.7,-0.7-0.9) -- (5.8,-0.9);
\draw(5.8+0.7,0.7-0.9)-- (5.8,-0.9);
\draw(6.3,-0.9) node {\tiny $\sigma(\theta+\frac{\pi}{2})$};
\end{tikzpicture}}

\newcommand{\onepointequaltwopoint}{
\begin{tikzpicture}[scale=1.5]  \label{fig:onepointequaltwopoint}
\draw(-2,0) node {$\int d\xi'$};
\draw(0,0) circle(0.5);
\draw(-0.5,0) node {\textbullet}; \draw(0.5,0) node {$\times$};
\draw(2.5,0) node {$\simeq \ \int d\xi'\, $};
\draw(-0.5,0) -- (-0.6,0.1); \draw(-0.5,0) -- (-0.6,-0.1);

\draw(-0.5,-0.7) node {$\xi$}; 
\draw(-1,0) node {$\lambda$};

\draw(4.4,-0.7) node {$\xi$};

\draw(3.9,0) node {$\lambda$};
\draw(0+7,0) arc(30:150:1.5  and 0.75);
\draw(-2.6+7,0) arc(210:330:1.5 and 0.75);
\draw(7,0)--(7.1,0.1); \draw(7,0) -- (7.1,-0.1);
\draw(4.4,0)--(4.3,0.1); \draw(4.4,0)--(4.3,-0.1);

\draw(7,-0.7) node {$\xi'$};
\draw(-2.6+7,0) node {\textbullet};
\draw(7,0) node {\textbullet};
\draw(7.5,0) node {$\Gamma$};

\draw[<-](5.4,-0.1) arc(195:345:0.4 and 0.3);

\draw[->](5.4+0.75,0.1) arc(15:165:0.4 and 0.3);

\end{tikzpicture}}

\newcommand{\geometricseries}{
\begin{tikzpicture}[scale=0.6] \label{fig:geometricseries}
\draw(-5.8-2-0.8,-0.8) -- (0.8-5.8-2,0.8);
\draw(-5.8-2-0.8,0.8) -- (0.8-5.8-2,-0.8);
\draw(-8,-0.8) node {$\bar{\psi}_{\downarrow}$};
\draw(-7,-0.8) node {$\psi_{\downarrow}$};
\draw(-8,0.8) node {$\bar{\psi}_{\uparrow}$};
\draw(-7,0.8) node {${\psi}_{\uparrow}$};

\draw(-1-5.5,0) node{$+\lambda$};

\draw(0,0) arc(30:150:3  and 1.5);
\draw(-5.2,0) arc(210:330:3 and 1.5);
\draw (-5.2,0) -- (-6,0.8);
\draw (-5.2,0) -- (-6,-0.8);
\draw(0,0) -- (0.8,0.8);
\draw(0,0) -- (0.8,-0.8);
\draw(-5.5,-0.8) node {$\bar{\psi}_{\downarrow}$};
\draw(0.2,-0.8) node {$\psi_{\downarrow}$};
\draw(-5.5,0.8) node {$\bar{\psi}_{\uparrow}$};
\draw(0.2,0.8) node {${\psi}_{\uparrow}$};

\draw(1.4,0) node{$+\lambda^2$};

\draw(0+8,0) arc(30:150:3  and 1.5);
\draw(-5.2+8,0) arc(210:330:3 and 1.5);
\draw(0+5.3+8,0) arc(30:150:3  and 1.5);
\draw(-5.2+5.3+8,0) arc(210:330:3 and 1.5);
\draw (-5.2+8,0) -- (-6+8,0.8);
\draw (-5.2+8,0) -- (-6+8,-0.8);
\draw(0+5.3+8,0) -- (0.8+5.3+8,0.8);
\draw(0+5.3+8,0) -- (0.8+5.3+8,-0.8);
\draw(-5.5+8,-0.8) node {$\bar{\psi}_{\downarrow}$};
\draw(0.2+5.2+8,-0.8) node {$\psi_{\downarrow}$};
\draw(-5.5+8,0.8) node {$\bar{\psi}_{\uparrow}$};
\draw(0.2+5.2+8,0.8) node {${\psi}_{\uparrow}$};

\draw(15,0) node{$+\cdots$};
\end{tikzpicture}}

\newcommand{\bubble}[8]{
\begin{tikzpicture}[scale=1] \label{fig:bubble}
\draw(0,0) arc(30:150:3  and 1.5);
\draw(-5.2,0) arc(210:330:3 and 1.5);
\draw (-5.2,0) -- (-6,0.8);
\draw (-5.2,0) -- (-6,-0.8);
\draw(0,0) -- (0.8,0.8);
\draw(0,0) -- (0.8,-0.8);
\draw(-4.5,-0.8) node {$#1$};
\draw(-0.7,-0.8) node {$#2$};
\draw(-4.5,0.8) node {$#3$};
\draw(-0.7,0.8) node {$#4$};
\draw(-5.5,-0.8) node {$#5$};
\draw(0.2,-0.8) node {$#6$};
\draw(-5.5,0.8) node {$#7$};
\draw(0.2,0.8) node {$#8$};
\draw[->](-2.2,0.2) arc(15:345:0.4);
\draw(-2.6,0.1) node {$q$};
\draw[<-] (0.5,0) -- (1.5,0);
\draw(0.9,0.2) node {$q$};
\end{tikzpicture}}

\newcommand{\dimensiondeux}{
\begin{tikzpicture}[scale=1.5] \label{fig:dimensiondeux}
\draw(0,0) circle(1);
\draw(1.3,0) node {$\Sigma^*_F$};
\draw[->] (-0.5,0.866) -- (-0.025,0.043);
\draw[->] (-0.5,-0.866) -- (-0.025,-0.043);
\draw[dashed](0.025,0.043)--(0.5,0.866);
\draw[->] (0.032,0.04) -- (0.63,0.8);
\draw[dashed] (-0.35,0.92)--(-0.0175,0.046);
 \draw[->] (0.0175,-0.046) -- (0.35,-0.92);

\draw (-0.45,0.45) node {$\vec{p}_1$};
\draw (-0.45,-0.45) node {$\vec{p}_2$};
\draw (0.50,0.45) node {$\vec{p}_4$};
\draw (0.40,-0.50) node {$\vec{p}_3$};

\draw(0.34,0.68) arc(70:62:1 and 1);
\draw(-0.24,0.7) arc(110:118:1 and 1);
\draw(-0.38,1.05) node {$\theta$};

\end{tikzpicture}}

\newcommand{\oneladder}{
\begin{tikzpicture}[scale=0.8] \label{fig:oneladder}
\draw(0,1) -- (6,1);
\draw(0,-1) -- (6,-1);
\draw[decorate,decoration=snake](2,1) -- (2,-1);
\draw[decorate,decoration=snake](4,1) -- (4,-1);
\end{tikzpicture}}

\newcommand{\ladder}{
\begin{tikzpicture}[scale=0.8] \label{fig:ladder}
\draw(0,1) -- (6,1);
\draw(0,-1) -- (6,-1);
\draw[dashed] (6,1) -- (9,1);
\draw[dashed] (6,-1) -- (9,-1);
\draw(9,1) -- (11,1);
\draw(9,-1)-- (11,-1);
\draw[decorate,decoration=snake](2,1) -- (2,-1);
\draw[decorate,decoration=snake](4,1) -- (4,-1);
\draw[decorate,decoration=snake](10,1)-- (10,-1);
\end{tikzpicture}}

\newcommand{\vacuum}{
\begin{tikzpicture}[scale=0.6] \label{fig:vacuum}
\draw(0,0) arc(30:150:3  and 1.5);
\draw(-5.2,0) arc(210:330:3 and 1.5);
\draw (-6.2,0) -- (-7,0.8);
\draw (-6.2,0) -- (-7,-0.8);

\draw[decorate,decoration=snake] (-6.2,0) -- (-5.2,0);
\draw(1,0) -- (1.8,0.8);
\draw(1,0) -- (1.8,-0.8);
\draw[decorate,decoration=snake] (0,0) -- (1,0);

\draw[->](-2.2,0.2) arc(15:345:0.4);

\end{tikzpicture}}

\newcommand{\Cooperbubble}[8]{
\begin{tikzpicture}[scale=1] \label{fig:Cooperbubble}
\draw(0,0) arc(30:150:3  and 1.5);
\draw(-5.2,0) arc(210:330:3 and 1.5);
\draw (-5.2,0) -- (-6,0.8);
\draw (-5.2,0) -- (-6,-0.8);
\draw(0,0) -- (0.8,0.8);
\draw(0,0) -- (0.8,-0.8);
\draw(-4.5,-0.8) node {$#1$};
\draw(-0.7,-0.8) node {$#2$};
\draw(-4.5,0.8) node {$#3$};
\draw(-0.7,0.8) node {$#4$};
\draw(-5.5,-0.8) node {$#5$};
\draw(0.2,-0.8) node {$#6$};
\draw(-5.5,0.8) node {$#7$};
\draw(0.2,0.8) node {$#8$};
\draw[<-](-3.7,-0.2) arc(195:345:0.8 and 0.4);
\draw(-1.6,-0.3) node {$-p$};
\draw[->](-2.2,0.2) arc(15:165:0.8 and 0.4);
\draw(-1.6,0.3) node {$p-q$};
\draw[<-] (0.5,0) -- (1.5,0);
\draw(0.9,0.2) node {$q$};
\end{tikzpicture}}

\newcommand{\Medskip}{\medskip\noindent}
\newcommand{\Bigskip}{\bigskip\noindent}

\newcommand{\BEQ}{\begin{equation}}     
\newcommand{\BEA}{\begin{eqnarray}}
\newcommand{\BD}{\begin{displaymath}}
\newcommand{\EEQ}{\end{equation}}       
\newcommand{\EEA}{\end{eqnarray}}
\newcommand{\ED}{\end{displaymath}}

\newcommand{\del}{\delta}
\newcommand{\Del}{\Delta}
\newcommand{\eps}{\varepsilon}          

\newcommand{\supp}{{\mathrm{supp}}}

\newcommand{\Tr}{{\mathrm{Tr}}}

\newcommand{\Vol}{{\mathrm{Vol}}}

\newcommand{\PreSigma}{{\mathrm{Pre}}\Sigma}
\newcommand{\R}{\mathbb{R}}
\newcommand{\C}{\mathbb{C}}
\newcommand{\Z}{\mathbb{Z}}
\newcommand{\N}{\mathbb{N}}
\newcommand{\D}{\mathbb{D}}
\renewcommand{\P}{\mathbb{P}}

\newcommand{\Id}{{\mathrm{Id}}}

\def\proba{{\mathbb{P}}}

\def\T{{\mathbb{T}}}
\def\F{{\mathbb{F}}}
\def\G{{\mathbb{G}}}

\def\sgn{{\mathrm{sgn}}}

\newcommand{\eop}{\hfill $\Box$}        

\newcommand{\II}{{\rm i}}               
\renewcommand{\Re}{{\rm Re\ }}          
\newcommand{\half}{{1\over 2}}          
\renewcommand{\vec}[1]{\boldsymbol{#1}} 

\catcode`\@=11
\def\numberbysection{\@addtoreset{equation}{section}
        \def\theequation{\thesection.\arabic{equation}}}
\numberbysection

\begin{document}

\title
{\bf A mathematical derivation of  zero-temperature 2D superconductivity from 
microscopic Bardeen-Cooper-Schrieffer model}

\date {}
\vskip -2cm
\maketitle

\vskip -2cm
\begin{center}
{\bf  Jacques Magnen$^a$ and J\'er\'emie Unterberger$^b$}
\end{center}


\centerline {\small $^a$Centre de Physique Th\'eorique,\footnote{Laboratoire associ\'e au CNRS UMR 7644}
Ecole Polytechnique,} 
\centerline{91128 Palaiseau Cedex, France}
\centerline{jacques.magnen@cpht.polytechnique.fr}
\vskip 0.5 cm
\centerline {\small $^b$Institut Elie Cartan,\footnote{Laboratoire 
associ\'e au CNRS UMR 7502. J. Unterberger acknowledges the support of the ANR, via the ANR project ANR-16-CE40-0020-01.} Universit\'e de Lorraine,} 
\centerline{ B.P. 239, 
F -- 54506 Vand{\oe}uvre-l\`es-Nancy Cedex, France}
\centerline{jeremie.unterberger@univ-lorraine.fr}

\vspace{2mm}
\begin{quote}

\renewcommand{\baselinestretch}{1.0}
\footnotesize
{

Starting from H. Fr\"ohlich's
second-quantized Hamiltonian  for a $d$-dimensional electron gas  in interaction with lattice phonons
describing the quantum vibrations of a metal, we present a rigorous
mathematical
derivation of the  superconducting state, following the 
principles laid out originally in 1957 by J. Bardeen, L. Cooper and J. Schrieffer. 
As in the series of papers written on the subject in the 90es 
\cite{FMRT-fermionic}, \cite{FMRT-infinite}, \cite{FMRT-Ward}, \cite{FMRT-intrinsic}, \cite{MR-single}, \cite{DisRiv1}, \cite{DisRiv2},  of which the present paper
is a continuation, the
representation of ions as a uniform charge background allows for
a $(1+d)$-dimensional fermionic quantum-field theoretic reformulation of the model at equilibrium.  For simplicity, we restrict in this article to  $d=2$
dimensions and zero temperature, and disregard effects due to electromagnetic interactions. Under these
assumptions, we prove   transition from a Fermi liquid state to a superconducting state made up of
Cooper pairs of electrons at an energy
level $\Gamma_{\phi}\sim \hbar\omega_D e^{-\pi/m\lambda}$ equal to the mass gap, expressed in terms of the 
Debye frequency $\omega_D$, electron mass $m$ and 
coupling constant $\lambda$. The dynamical $U(1)$-symmetry breaking produces at energies lower than the energy gap $\Gamma_{\phi}$ a Goldstone boson, a
non-massive particle described by an effective $(2+1)$-dimensional non-linear sigma-model,
whose parameters and correlations are computed. The proof relies on a mixture of general concepts and tools (multi-scale cluster expansions, Ward identities),
adapted to this quantum many-body problem with its  extended
infra-red 
singularity located on the Fermi circle,   and a specific $1/N$-expansion giving the leading diagrams at intermediate energies. Ladder diagrams are proved to provide the leading
behavior in the infra-red limit, in agreement with mean-field theory predictions. 

\Medskip  Some insights about expected extensions of our method to a rigorous study of
real-world, low-temperature  superconductivity are provided.

}
\end{quote}

\vspace{4mm}
\noindent

 \medskip
 \noindent {\bf Keywords:}
BCS theory, constructive field theory, renormalization, cluster expansions, non-linear sigma-model,
Goldstone boson, low-temperature superconductivity, Cooper pairs,
Ward identities, Bethe-Salpeter kernel, Fermi liquids.

\smallskip
\noindent
{\bf Mathematics Subject Classification (2010):}  81T08, 81V70, 82D55.

\tableofcontents

\Medskip {\bf Index of notations}\  . . . . . . . . . . . . . . . . . . . . . . . . . . . . . . . . . . . .
\pageref{section:index}



\section{Introduction}



\subsection{The Bardeen-Cooper-Schrieffer model of electrons and phonons}


The Bardeen-Cooper-Schrieffer (BCS) theory has proved extremely successful at predicting the main characteristic features of  conventional, low-temperature superconductivity of metals such as tin, lead, aluminium...  see e.g. textbooks \cite{Sch}, 
\cite{DeG}, \cite{Til}, \cite{Tin}. The original objective of the Nobel prizes in their ground-breaking paper \cite{BCS}
was to account for the formation of a condensate of  {\em Cooper pairs}, bound states made of pairs of electrons which behave
like bosons, and are in particular responsible for the Meissner effect \footnote{to which is traditionally attributed the spectacular magnetic
levitation. However,  disorder-induced vortex trapping supercurrents are more directly responsible for this effect in real materials \cite{BeeLev}.}.   The explanation for this bound-state is the existence
of an effective {\em attractive force} between pairs of electrons
due to the quantum oscillations of the ionic lattice. This
remarkable effect can be easily understood starting from H. Fr\"ohlich's \cite{Fro1,Fro2}
second-quantized Hamiltonian $\cal H$, depending on a
constant $\gamma\not=0$,

\BEA && {\cal H}:=\int_V d^d \vec{x}\, \bar{\psi}_{\sigma}(\vec{x}) \Big\{
-\frac{\hbar^2 \nabla^2}{2m} - \mu \Big\} \psi_{\sigma}(\vec{x}) \nonumber\\
&& + \gamma \int_V d^d \vec{x} \, \bar{\psi}_{\sigma}(\vec{x}) \psi_{\sigma}(\vec{x}) \phi(\vec{x})  \label{eq:BCS}
\EEA
where Einstein's implicit summation convention is used for
spin states $\sigma=\{\uparrow,\downarrow\}$.
This Hamiltonian involves:

\textbullet\ an {\em electron field} $\psi_{\sigma}=(\psi_{\uparrow},
\psi_{\downarrow})$ describing a spin $1/2$, mass $m$, fermionic  particle with 
dispersion relation $\epsilon(\vec{k}):=\frac{\hbar^2 |\vec{k}|^2}{2m}-\mu$, 
where $\mu$ is a chemical potential regulating the density;

\textbullet\ a {\em phonon field}, namely, a bosonic quasi-particle describing the oscillations of the lattice,
\BEQ \phi(\vec{x}):=\sum_{\vec{k}} \Big(\frac{\hbar \omega(\vec{k})}{2V}\Big)^{1/2} \big[ c_k e^{\II \vec{k}\cdot\vec{x}} + c^{\dagger}_k 
e^{-\II \vec{k}\cdot\vec{x}}\big] \Theta(\omega_D-\omega(\vec{k}))   \label{eq:cut-off3}
\EEQ
with dispersion relation $\omega(\vec{k}):=c|\vec{k}|$, where
$\Theta$ is a smoothened Heaviside function. The superconducting material
is assumed to take up a roughly cubic volume $V\subset\R^d$ ($d\ge 1$) with
$|V|\approx L^d$, implying  a discrete sum
over momenta $\vec{k}$ rougly multiples of $2\pi/L$. The sum is
really 
a finite sum over momenta $\vec{k}$ such that $\omega(\vec{k})<\omega_D$ or equivalently $|\vec{k}|<k_D:=\frac{\omega_D}{c}$,
 where $k_D^{-1}$ is roughly equal to the mean spacing between ions.
 The frequency $\omega_D$ is called {\em Debye frequency}.
We shall eventually restrict
to space dimension $d=2$ (possible extension to $d=3$ is briefly discussed in section \ref{section:conclusion}), but let us keep $d$ arbitrary till then.
 
\Medskip In practice, only electrons with momentum $\vec{k}$ close to the Fermi circle
\BEQ \Sigma_F:=\{|\vec{k}|=k_F\}, \qquad k_F:=\hbar^{-1} \sqrt{2m\mu}  \label{eq:SigmaF} \EEQ 
defined by the vanishing of the energy $\epsilon(\cdot)$ participate in
the interactions.   We {\em assume} here that
\BEQ \hbar\omega_D\ll \mu, \label{eq:cut-off1} \EEQ
a condition satisfied in usual materials (see discussion in
\cite{FW}, \S 37),
and consider only couplings of the phonon field to electrons with momentum
$\vec{k}$ such that 
\BEQ \epsilon(\vec{k})=O(\hbar \omega_D),  \label{eq:cut-off2} \EEQ
i.e. to electrons with momenta in an annulus of radius
$\del |\vec{k}|\approx \frac{\hbar \omega_D}{\mu} k_F$ 
around the Fermi circle. The reader may
read with profit the classical book by A. Fetter and J. Walecka
\cite{FW} for: the second quantization formalism applied
to interacting many-particle systems (Chapter 1); phonons
interacting with electrons (Chapter 12), see in particular 
\S 45 where a value for $\gamma$ is derived in terms of the bulk compressibility of the ionic background seen as an elastic
medium;  fundamental properties of superconductors, and a semi-rigorous explanation of these
using the above model (\ref{eq:BCS}), see Chapter 13, following arguments due to Bardeen, Cooper
and Schrieffer and detailed computations done by L. P. Gor'kov
using finite temperature Green's functions \cite{Gor}. For a more mathematical, modern presentation of 3D
BCS theory, see the series of papers by R. Frank, C. Hainzl, E.  Hamza,
B. Schlein,   R. Seiringer and J. P. Solovej
\cite{HaiHamSeiSol07}, \cite{FraHaiSeiSol11}, \cite{HaiSei11},
\cite{HaiSch12} where the Ginzburg-Landau theory is derived
in an $\hbar$-small vicinity of the critical temperature by minimizing
in the semi-classical limit
a functional introduced by A. J. Leggett \cite{Leg1} (called BCS functional) built out of  quasi-free trial states.

\Medskip At zero temperature, the lowest energy state of the system is obtained by filling the energy levels up to $\mu$, i.e. 
summing over all $\vec{k}\in \frac{2\pi}{L}\Z$ such that
$|\vec{k}|<k_F$. The self-consistent quadratic approximation of $\cal H$ introduced by
Bogoliubov  \cite{Bog1,Bog2} involves elementary excitations known as quasi-particles, with a minimum energy $\Gamma_{\phi}$ known as the {\em energy
gap}, see e.g. \cite{AltSim}, \S 6.4.     A phenomenologically important parameter is the density
 of states  at the Fermi surface (compare to  \cite{FW}, p. 333), at temperature $T=0$,  $N(0)\equiv N(T=0)=\frac{2}{2\pi} k_F \frac{dk}{d\epsilon}\Big|_{\epsilon=\mu}=\frac{m}{\pi \hbar^2}$. Fetter
 and Walecka find the following value for the
 energy gap $\Gamma_{\phi}$ in {\em three} dimensions:
 \BEQ \Gamma_{\phi}\approx\hbar \omega_D\,  e^{-1/N(0)\gamma^2}.  \label{eq:FWgap} \EEQ
 This energy gap may also be interpreted  as  a critical temperature 
$T_c:=\Gamma_{\phi}/k_B$, where $k_B$ is the Boltzmann constant. 
In 2D, however, the critical temperature is known to be zero
 because of the celebrated Mermin-Wagner theorem, which
forbids continuous symmetry-breaking for any $T>0$.  At $T=0$,
the theory is effectively  $(1+2)$-dimensional, as we shall
presently see, so the Mermin-Wagner theorem does not apply.
   
\Bigskip
In this article, we use the well-known equivalence (shown by going over to interaction picture and using Wick's theorem) of the quantum model with a functional integral  representation in terms of Grassmann fields $\psi$, $\bar{\psi}$ living in a $(1+d)$-dimensional space. The supplementary coordinate $\tau$ plays formally the r\^ole of an {\em imaginary time}, since dynamics
are retrieved (at least formally) by letting $t:=-\II\tau$.  Space-time 
points are denoted by $\xi=(\tau,\vec{x})$. We consider here only 
vacuum expectation values at zero temperature.  Let us consider the infinite volume
limit $V\to\R^3$, so that $\vec{p}\equiv \hbar\vec{k}$
 becomes a continuous momentum variable;  and  integrate
out the phonon field (see \cite{FW}, \S 46). Then the
ground-state expectation $\Big\langle  \Big(\prod_{i=1}^n \bar{\psi}_{\sigma_i}(\vec{x}_i) \Big)
\Big(\prod_{i'=1}^{n} \psi_{\sigma_{i'}}(\vec{x}'_{i'}) \Big\rangle$
becomes the Grassmann  integral
\BEQ G_n\Big((\vec{x}_i)_{i=1,\ldots,n}, (\vec{x}'_{i'})_{i=1,\ldots,n}\Big)\equiv \frac{1}{{\cal Z}_{\lambda}} \int d\mu(\psi,\bar{\psi})\   \Big(\prod_{i=1}^n \bar{\psi}_{\sigma_i}(0,\vec{x}_i) \Big)
\Big(\prod_{i'=1}^n \psi_{\sigma_{i'}}(0,\vec{x}_{i'})\Big) e^{-\frac{1}{\hbar} {\cal V}(\psi,\bar{\psi})} 
\label{eq:0.7}
\EEQ
which may be expressed in terms of the normalized Grassmann measure 
\BEQ d\mu_{\lambda}(\psi,\bar{\psi}):=\frac{1}{{\cal Z}_{\lambda}} e^{-\frac{1}{\hbar}
{\cal V}(\psi,\bar{\psi})}
d\mu(\psi,\bar{\psi}), \label{eq:0.8} \EEQ
 where:

\textbullet\  $\psi=\psi(\xi),\bar{\psi}=\bar{\psi}(\xi)$
  are now fields living in $(1+d)$-dimensional space-time with coordinate
$\xi\equiv (\tau,\vec{x})$;

\textbullet\ $d\mu(\psi,\bar{\psi})$ is a Grassmann Gaussian
measure  with covariance kernel given by the inverse of the
quadratic form $B_0$, 
\BEQ B_0(\psi,\bar{\psi}):=
\int dp\, \bar{\psi}_{\sigma}(-p) (\II p^0-e(\vec{p}))
\psi_{\sigma}(p) \label{eq:A0} \EEQ
for fields $\psi,\bar{\psi}$ with Fourier support  satisfying
the  cut-off condition 
\BEQ e(\vec{p})\equiv \epsilon(\frac{\vec{p}}{\hbar})=\frac{|\vec{p}|^2}{2m}-\mu=O( \hbar\omega_D)
\label{cut-off3},
\EEQ
 see 
(\ref{eq:cut-off2}). A precise definition of the model will be given only at
the end of section \ref{section:gap} (see Definition \ref{def:Grassmann-measure}). It is enough to say here
that the measure involves only Fourier scales $\ge j_D$, i.e.
momenta $(p_0,\vec{p})$ such that $|p^0|=O(2^{-j_D}\mu)$, $\Big| |\vec{p}|-p_F\Big|
=O(2^{-j_D}p_F)$, where $p_F\equiv \hbar k_F$ and 
\BEQ j_D\equiv \lfloor\log_2(\mu/\hbar\omega_D)\rfloor\ge 0
\label{eq:jD}
\EEQ
 by (\ref{eq:cut-off1});

\textbullet\  and the interaction ${\cal V}(\psi,\bar{\psi})$,
formally defined in Fourier coordinates as 
\BEQ \int \prod_{i=1}^4 dp_i \ \del(p_1+p_2-p_3-p_4) 
\bar{\psi}_{\uparrow}(p_1)\psi_{\uparrow}(p_3) \ \langle p_1,p_2|U|p_3,p_4\rangle
 \, \bar{\psi}_{\downarrow}(p_2) \psi_{\downarrow}(p_4) 
 \label{eq:intro-U} 
 \EEQ
  for a general,
 spin-neutral two-body
 potential $U$ written in second-quantized form using the positive electron density
 operator $\bar{\psi}_{\sigma}\psi_{\sigma}$,  may be  chosen in this context in the form

\BEA && {\cal V}(\psi,\bar{\psi}):=\int \prod_{i=1}^4 dp_i\, \del(p_1+p_2-p_3-p_4) \bar{\psi}_{\uparrow}(p_1)\psi_{\uparrow}(p_3) \nonumber\\
&&\  \Big( -\lambda \tilde{\Theta}(\hbar\omega_D - \omega((\vec{p}_1-\vec{p}_3))/\hbar) \frac{\omega^2((\vec{p}_1-\vec{p}_3)/\hbar)}{(p_1^0-p_3^0)^2 + \omega^2((\vec{p}_1-\vec{p}_3)/\hbar)} + \frac{r_0}{r_B} \hat{v}(\vec{p}_1- 
\vec{p}_3) \Big) \bar{\psi}_{\downarrow}(p_2)\psi_{\downarrow}(p_4)  \nonumber\\
\label{eq:V0}
\EEA
where $\hat{v}$ is (the Fourier transform of) a static, 
rotation-invariant, spin-neutral two-body potential, say,  with high enough infra-red cut-off (see discussion in
section \ref{section:conclusion}), and $\frac{r_0}{r_B}\equiv \frac{r_0}{\hbar^2/me^2}$ (mean interparticle spacing, divided by the Bohr radius) is assumed to be $\ll 1$, corresponding 
to a high-density (also called: {\em degenerate}) regime of the electron gas, see \cite{FW}, \S 3.  Up to inessential issues regarding
cut-offs (with $\tilde{\Theta}\approx \Theta^2$ but not quite), (\ref{eq:BCS}) reduces exactly to this model if one sets
$\hat{v}\equiv 0$ and lets
\BEQ \lambda=\gamma^2. \label{eq:lambda} \EEQ
The well-known and essential observation is that 
the above kernel $\frac{\omega^2((\vec{p}_1-\vec{p}_3)/\hbar)}{(p_1^0-p_3^0)^2 + \omega^2((\vec{p}_1-\vec{p}_3)/\hbar)}$ is $>0$,  bounded
 above by 1, and $\approx 1$ in average,   under our cut-off conditions.
In practice,  we shall simply replace 
$\frac{\omega^2((\vec{p}_1-\vec{p}_3)/\hbar)}{(p_1^0-p_3^0)^2 + \omega^2((\vec{p}_1-\vec{p}_3)/\hbar)}$ by $1$, implying an {\em attractive
$\del$-interaction} between electrons. We still denote by
${\cal V}$ the corresponding interaction, 
\BEQ {\cal V}(\psi,\bar{\psi})=-\lambda  \int \prod_{i=1}^4 dp_i\, \del(p_1+p_2-p_3-p_4) \bar{\psi}_{\uparrow}(p_1)\psi_{\uparrow}(p_3)  \tilde{\Theta}(\hbar\omega_D - e(\vec{p}_1-\vec{p}_3))  \bar{\psi}_{\downarrow}(p_2)\psi_{\downarrow}(p_4)  
\label{eq:V}
\EEQ
and argue in  section \ref{section:conclusion} 
that a small enough, short-ranged  two-body potential $\hat{v}$ can be added to
$\cal V$ without altering the general conclusions, provided
the overall effect near the Fermi sphere remains that of an
{\em attractive} potential.

\Medskip  Let us mention the following 
result by W. Kohn and  J. M. Luttinger  \cite{KohLut,Lut}:   in $d=3$ dimensions, they proved that for essentially
{\em any} arbitrary, even {\em purely repulsive}, rotation-invariant interaction, the 
scattering amplitude for pairs of quasi-particles  of opposite momenta
had poles  on the Fermi surface, implying the possibility of creation of
a superconducting bound state in large enough {\em odd} angular momentum sectors. This is shown by analyzing the sign of
non-zero angular momentum
second-order contributions such as those of Fig. \ref{subsection:Sigma}.2, which however vanish for $d=2$. On the other hand, in $d=2$ dimensions, J. Feldman, H. Kn\"orrer, R. Sinclair and 
E. Trubowitz \cite{FKST} proved  that third-order contributions, namely, {\em triangle diagrams}

\Bigskip

{\centerline{\BetheSalpeter}}

\Medskip {\tiny \bf Fig. \ref{fig:BetheSalpeter}. Triangle diagram.}

\Bigskip
  created superconducting bound states in the  angular momentum sector $\ell=1$. Hence one
can expect that our conclusions extend to very general, not purely
attractive interactions.


\subsection{Constructive approaches for superconductivity}


\noindent Constructive methods (actually, {\em multi-scale cluster expansions}), see e.g. \cite{AbdRiv1,Erice,GalNic,GawKup,FMRS,GJb,GJa,MagUnt2,Mas,Riv,Sal,Unt-rev,FVT},  consist in implementing rigorously  Wilson's renormalization group ideas developed to study small perturbations of Gaussian models. In the case at hand,  correlation functions are computed by 
averaging w.r. to a  Grassmann Gaussian measure $d\mu(\psi , \bar{\psi })$ perturbed by a quartic interaction $e^{\lambda
\int d\tau\, d\vec{x}\, (\bar{\psi}_{\uparrow}\psi_{\uparrow})(\tau,
\vec{x}) (\bar{\psi}_{\downarrow}\psi_{\downarrow})(\tau,\vec{x})}$,
with $\lambda$ {\em small enough}. The bare covariance is infra-red singular on the set  $(p^{0}=0, |\vec p|=p_{F})$. In principle, bare
 parameters  of the model, $m,\mu ,\lambda $, become
 running coupling constants $m^j,\mu^j,\lambda^j$ through
 the renormalization procedure. In our case, $\lambda$ is
 not renormalized (see below why), and the model is
 directly rewritten in terms of its renormalized parameters
 $m^*:=\lim_{j\to +\infty} m^j,\mu^*:=\lim_{j\to +\infty}\mu^j$ and 
 \BEQ p^*_F:=\sqrt{2m^*\mu^*}. \label{eq:p*F} \EEQ 
The first step consists in splitting the covariance of the measure according to the distance to the singularity; 
namely, one rewrites $\psi,\bar{\psi}$ as sums of independent fields,  $\psi = \sum _{j}\psi ^{j}$,  $\bar\psi = \sum _{j}\bar\psi ^{j}$ ($j\ge 0$), whose covariance is supported on the set
 $|p^0|+\frac{p^*_F}{\mu^*}\Big||\vec{p}|-p^*_F\Big|\approx 2^{-j}\mu$.  Splitting each field of each vertex into its
 components, and splitting accordingly perturbative graphs
 by letting scale indices {\em grow from the top to the bottom}
  (see e.g. Fig. \ref{subsection:cluster}.1), one
 gets the following picture:   {\em high-momentum  diagrams} of lowest scale $j$ (i.e. with all covariance indices $\le j$) are {\em quasi-local} w.r. to {\em low-momentum diagrams} of highest scale $k\gg j$. For $\lambda$ small enough, this allows the sum of {\em all}
 perturbative graphs of lowest scale $j$  to be resummed
 into effective corrections to low-momentum vertices, amounting
to a scale-by-scale renormalization of parameters. Graphs of a given scale $j$ are resummed using a cluster expansion; as 
well-known, for {\em fermionic} theories, the exponential of the interaction may
be expanded to infinity, see e.g. \cite{Mas}, \cite{Sal}. The above scheme works provided
effective corrections can be shown to remain small; in particular, if the running coupling constant remains $o(1)$. 
In the present case, however,  the running coupling constant 
becomes large  around  some transition scale (logarithm of the inverse of the energy gap) called  $j_{\phi}$, implying that the perturbation around the Grassmann Gaussian measure is not  pertinent any more. This is interpreted as  the formation of a {\em bound state} made up of  {\em Cooper pairs}. Instead of merely relying  on perturbation theory, the idea is therefore to {\em resum} explicitly  a class 
of
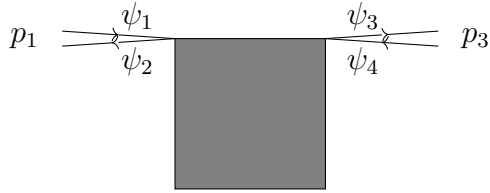
four-point diagrams, forming the {\em Bethe-Salpeter kernel},
which  contribute to this {\em non-perturbative} effect.  The sum --  mathematically,
a  kernel denoted $\Sigma_{\perp,\perp}(\tau,\vec{x};\tau',
\vec{x}')$, one of the components of a two-by-two matrix-valued kernel $\Sigma$  --, may be interpreted
as the two-point function of the bound state. The
leading behavior of this kernel may be captured
by looking simply at the geometric series of {\em ladder (bubble) diagrams}, which can be explicitly computed. These two-point
functions turn out to be the only non-massive (i.e. long-range) ones, hence they  give the main contribution to the theory at
large scale, an effective {\em bosonic} theory in the same
class as the {\em  $U(1)$ non-linear sigma model}. The underlying non-massive boson may be called {\em Goldstone boson}, by reference to the general theory of continuous
symmetry-breaking (see discussion of Regime II below), and
the kernel $\Sigma_{\perp,\perp}$  {\em Goldstone boson propagator},  the other non-vanishing component, $\Sigma_{//,//}$, being massive.  A full 
understanding of the model at all scales can be
obtained by considering
"mixed" diagrams featuring both fermionic propagators and
vertices, and some bound-state two-point functions. Contrary
to  fermionic vertices, though, $\Sigma$-kernels must not
be systematically expanded, but rather (as for most non-massive bosonic
models, see e.g. \cite{FMRS}) through some careful cluster expansion.

\Bigskip  Let us emphasize one specific aspect of the present model, which ultimately explains why  non-perturbative effects
  can be dealt with at an analytical level. The {\em small parameter} here is the expected mean value of 
the interaction, restricted to some {\em scale} $j$ and  integrated over a scaled box $\Del^j\subset\R^{1+d}$
with sides of length $\approx 2^j$, 
 $\lambda I^j(\Del^j)=O(m\lambda)$ up to spurious logarithmic corrections, see (\ref{eq:I(Del)}), 
 (\ref{eq:lambdaIjjg}), independently
of the choice of the scale $j$ defined for typical momenta as $\lfloor 
\log(\mu/|p^0|)\rfloor $ or $\lfloor \log(p_F/|\, |\vec{p}|-p_F\, |)
\rfloor$, which reflects the fact that the theory near the
Fermi sphere
is {\em just renormalizable (independently of space dimension)}.
 Note that this very fact is actually not straightforward, and
 more easily established in $d=2$ dimensions than for $d=3$
(see brief discussion in section \ref{section:conclusion}). Therefore
{\em we restrict to $d=2$ in the sequel}. 
  $n$-point functions within a
given scale will be reexpressed in terms of a geometric-like entire
series in the {\em non-dimensional parameter} 
\BEQ g:=m\lambda, \label{eq:g} \EEQ
 hence converge
{\em provided}
\BEQ g\ll 1. \label{eq:g-small} \EEQ 

\Medskip (\ref{eq:lambdaIjjg}) is based on the "sector-counting proposition", see Proposition \ref{prop:sector-counting}, which may be rephrased as follows (see \cite{FMRT-intrinsic}): {\em The theory near the Fermi surface (in this context, Fermi circle) may be
with remarkable accuracy reformulated  after a scale-dependent rescaling as a large $N$ vector model with action}  
 \BEQ \sim N \int d\xi \,  \Big(\sum_{\alpha_1=1}^N
 (\bar{\psi}_{\uparrow}^{\alpha_1}\bar{\psi}_{\downarrow}^{\alpha_1})(\xi)
 \Big)  \Big(\sum_{\alpha_2=1}^N
 (\psi_{\uparrow}^{\alpha_2}\psi_{\downarrow}^{\alpha_2})(\xi)
 \Big),   \label{eq:intro-N} \EEQ
$\xi=(\tau,\vec{x})$, where the fields $\psi^{\alpha}_{\uparrow,\downarrow}$, $\bar{\psi}^{\alpha}_{\uparrow,\downarrow}$ ($\alpha=\alpha_{1,2}$) are fermion fields restricted to a 
given momentum  angular sector indexed by $\alpha$ and rescaled in such a way that
$\sum_{\alpha=1}^N \langle  (\bar{\psi}_{\sigma}^{\alpha}\psi_{\sigma}^{\alpha})(\xi) \rangle \sim 1$.
The momentum-scale dependent number $N\equiv N_j=2^j$ increases exponentially as one
gets nearer to the Fermi circle defined by $j=+\infty$. Expanding the interaction and using Wick's theorem yields Feynman diagrams made up of vertices and fermion loops, each in a given angular sector, which may be thought of as a "color". Performing the sum over colors produces {\em for generic vertices} a factor $O(N)$ per fermion
loop, and  $O(1/N)$ per vertex. Alternatively, following a fermion loop, one can prove that {\em there is at most one sum over sectors per vertex} in a given diagram.  Then dominant diagrams
in an $1/N$-expansion are {\em chains of bubbles} (see \S \ref{subsection:bubble}), as confirmed perturbatively by a Feynman
diagram expansion; the simplicity of the theory is due to
the fact that chains of bubbles make up a geometric series which
can be resummed explicitly. The idea of the "$1/N$-expansion" is old and has
been used in many different contexts; we refer the reader to  \cite{KMR} and
\cite{Kop} respectively for a rigorous analysis of  the $N$-component Gross-Neveu and non-linear sigma model. Instead of $N$-component vectors, one sometimes also considers large
$N\times N$-matrices, either in the context of random matrices or Schr\"odinger
operators  \cite{BMR}, \cite{Poi} -- in connection to two-dimensional gravity
\cite{DiFGin}, since leading terms are then planar diagrams --
 or as an approximation to gauge theories, following a seminal paper by
 't Hooft, see \cite{tHoo} or \cite{FriSon}, chap. 7 for a review of
 two-dimensional quantum-field theory models in this limit. R. Gurau, V. Rivasseau et al. have also
applied these ideas to tensor field models, see \cite{Gur}, \cite{GurRiv}. 

\Medskip Under Hypothesis (\ref{eq:g-small}), the following scenario -- in accordance with
the original idea by Bardeen, Cooper and Schrieffer, but going beyond mean-field
regime predictions, which are valid only for an infinitesimal interaction, physically, in an ideally degenerate regime, see below (\ref{eq:cut-off2}) -- was explored in the 90'es by various theoretical physicists, including one of the authors of the
present work, see articles by M. Disertori, J. Feldman, J. Magnen,
V. Rivasseau, E. Trubowitz \cite{FMRT-fermionic}, \cite{FMRT-infinite}, \cite{FMRT-Ward}, \cite{FMRT-intrinsic}, \cite{MR-single}, \cite{DisRiv99},\cite{DisRiv1}, \cite{DisRiv2},
\cite{DisMagRiv}. Notations are as follows: energy and momentum scales are labeled by an integer index $j$ ranging
from $j_D$ to $+\infty$; typical energies $p^0$, resp. transverse momenta $|\, |\vec{p}|-p_F\, |$ of scale $j$ are
$\approx 2^{-j}\mu$, resp. $\approx 2^{-j}p_F$. The highest scale $j_D\ge 0$ is defined in
agreement with the above cut-off hypotheses (\ref{eq:cut-off1}), (\ref{eq:cut-off2}). Physical parameters, in particular,  the electron 
mass, the coupling constant and the Fermi radius (or, equivalently, the
chemical potential) are renormalized \`a la Wilson, defining scale parameters $m^j, 
\lambda^j, \mu^j$. {\em Three energy regimes were singled out:}

\begin{itemize}
\item[(i)]  {\bf (Regime I, high-energy regime)} At {\em high enough energy}, i.e. {\em for
 $j$ small enough},   no bound states can form,  and
electrons are still in their normal, Fermi liquid phase, where they behave essentially like
free fermions, see \cite{DisRiv99}. 
The scale-by-scale renormalization of the model \`a la Wilson yields the
following flow for the coupling constant,  
\BEQ \lambda^{j+1}-\lambda^j\approx  (\lambda^j)^2 \Big[{\cal A}_0^{j\to}(\Upsilon_{3,diag})-
{\cal A}_0^{(j-1)\to}(\Upsilon_{3,diag}) \Big] \label{eq:Wilson-lambda} \EEQ
where $\Upsilon_{3,diag}$ is the  amputated Cooper pair bubble diagram of \S \ref{subsection:bubble} (see {\small Fig. \ref{fig:Cooperbubble}.3}), and ${\cal A}_0^{j\to}(\Upsilon_{3,diag})$ is  the evaluation at zero external momentum
of $\Upsilon_{3,diag}$ computed for internal momenta of scale $\le j$, see \S
\ref{subsection:bubble}. The difference $({\cal A}_0^{j\to}-
{\cal A}_0^{(j-1)\to})(\Upsilon_{3,diag}) $ is approximately scale independent
(which is essentially tantamount to saying that the theory is just renormalizable) and $\approx \frac{m}{\pi}$. As long as $\lambda {\cal A}_0^{j\to}(\Upsilon_{3,diag})$ remains
$o(1)$, the solution of the flow equation is well approximated by
\BEQ (\lambda^j)^{-1}\simeq \lambda^{-1}- {\cal A}_0^{j\to}(\Upsilon_{3,diag})\EEQ
or equivalently,
\BEQ \lambda^j\simeq \frac{\lambda}{1-\lambda {\cal A}_0^{j\to}(\Upsilon_{3,diag})}.
\EEQ
Just a few scales  above a  transition scale $j_{\phi}$ defined by 
$\lambda {\cal A}_0^{(j_{\phi}-1)\to}(\Upsilon_{3,diag})<1< \lambda {\cal A}_0^{j_{\phi}\to}(\Upsilon_{3,diag})$, the renormalized coupling constant $m\lambda^j\approx 1$ becomes large. Thus the above perturbative regime analysis breaks down. Since (as shown in 
\S \ref{subsection:bubble}) ${\cal A}_0^{j\to}(\Upsilon_{3,diag})\sim \frac{m}{\pi}
(j-j_D)$, the above condition holds for
\BEQ j_{\phi}=j_D+\frac{\pi}{\lambda m} + O(1). \label{eq:jc0} \EEQ
This defines an energy level for the transition, 
\BEQ \Gamma_{\phi}\approx 2^{-j_{\phi}}\mu.  \label{eq:Gammaphijphimu} \EEQ

\Bigskip Let us add here some necessary precisions. Because one renormalizes in the vicinity of 
the Fermi circle, which is an {\em extended singularity},  it is easy to see that leading corrections to the 
vertex are "pinched" diagrams of the form

\Bigskip

{\centerline{
\begin{tikzpicture}
\draw(2,2)--(2.75,2.05); \draw[>-](2.75,2.05)--(3.5,2.1); \draw(2,2)--(2.75,1.95); \draw[<-](2.75,1.95)--(3.5,1.9);
\draw(0,2)--(-0.75,2.05); \draw[<-](-0.75,2.05)--(-1.5,2.1); \draw(0,2)--(-0.75,1.95); \draw[>-](-0.75,1.95)--(-1.5,1.9);
\draw(2.5,2.3) node {$\psi_3$}; \draw(2.5,1.7) node {$\psi_4$};
\draw(4,2) node {$p_3$}; 
 \draw(-2,2) node {$p_1$};
 \draw(-0.5,2.3) node {$\psi_1$}; \draw(-0.5,1.7) node {$\psi_2$};
\draw[fill=gray](0,0) rectangle(2,2);
\end{tikzpicture}}}

\Medskip{\bf \tiny Fig. 0.2. Bethe-Salpeter kernel.}

\Bigskip
(see Fig. \ref{subsection:Sigma}.5 for details), $\psi_i=\psi,\bar{\psi}$ $(i=1,\ldots,4)$  with external
momenta $p_1=(0,\vec{p}_1),p_3=(0,\vec{p}_3)$ near the
Fermi circle, depending only on the relative angle $\widehat{(\vec{p}_1,\vec{p}_3)}$. {\em Two-particle irreducible}
diagrams of this type {\em with Cooper pair external structure},  i.e. with $\psi_1\psi_2,\psi_3\psi_4=\bar{\psi}_{\uparrow}\bar{\psi}_{\downarrow},\psi_{\downarrow}\psi_{\uparrow}$, form the so-called {\em Bethe-Salpeter kernel}, which may be summed into a geometric series. 
 Now, the bare theory has an  ultra-local, {\em angle-independent} vertex $-\lambda(\bar{\psi}\psi)^2$, and Cooper pair bubbles lead
to angle-independent, $s$-wave corrections. However, more complicated
diagrams with extra vertices lead to  {\em angle-dependent}
corrections, that may be seen as effective $p$-wave, $d$-wave, etc.
effective vertices. To leading order these do not interfere,
because they are orthogonal Fourier modes. Hence we get
instead of (\ref{eq:Wilson-lambda})
an infinite series of flows,
\BEQ  \lambda_s^{j+1}-\lambda_s^j\approx  (\lambda_s^j)^2 \Big[{\cal A}_0^{j\to}(\Upsilon_{3,diag})-
{\cal A}_0^{(j-1)\to}(\Upsilon_{3,diag}) \Big]   \label{eq:Wilson-s} \EEQ
compare with (\ref{eq:Wilson-lambda}), and 
\BEA && \lambda_p^{j+1}-\lambda_p^j\approx (\lambda_p^j)^2 \Big[{\cal A}_0^{j\to}(\Upsilon_{3,diag})-
{\cal A}_0^{(j-1)\to}(\Upsilon_{3,diag}) \Big]  +
O((\lambda^j)^3), \nonumber\\
&&  \qquad \lambda_d^{j+1}-\lambda_d^j\approx (\lambda_d^j)^2 \Big[{\cal A}_0^{j\to}(\Upsilon_{3,diag})-
{\cal A}_0^{(j-1)\to}(\Upsilon_{3,diag}) \Big]  +
O((\lambda^j)^3), \ \  \cdots \nonumber\\  \label{eq:Wilson-p} 
\EEA

The difference between (\ref{eq:Wilson-s}) and (\ref{eq:Wilson-p}) is that
$\lambda_s^{j_D}=\lambda$ whereas $\lambda_n^{j_D}=0$, $n=p,d,\ldots$  Thus coupling constants other than $\lambda_s$
increase much more slowly, with leading term at scale $j\ll j_{\phi}$ 
bounded by the
sum $\sum_{k=j_D}^j O((\lambda_s^k)^3)\approx \frac{1}{(j_{\phi}-j)^2}$, whereas $\lambda_s^j\approx \frac{1}{j_{\phi}-j}$. This is not sufficient to conclude (see discussion in \S \ref{subsection:Sigma}, and around eq. (\ref{eq:KK-angle-dependence})), but strongly
hints at an $s$-wave superconducting behavior.

\Bigskip

\item[(ii)] {\bf (Regime II, near transition scale)} 

 In regimes II and III, that is, around and below the
gap energy, the phonon field $\phi$, initially centered around 0,  is seen by semi-perturbative arguments to choose a more favorable  
position on a circle of radius $\approx \Gamma_{\phi}$, which
{\em breaks the $U(1)$ fermionic charge symmetry}
$\left(\begin{array}{c} \psi(\xi) \\ \bar{\psi}(\xi)\end{array}\right)\mapsto 
\left(\begin{array}{c} e^{\II\alpha}\psi(\xi) \\ \bar{\psi}(\xi) e^{-\II\alpha}
\end{array}\right)$. At a semi-rigorous level,
this can be seen using a {\em Hubbard-Stratonovich transform,}
\BEQ e^{-{\cal V}}=\int d\mu(\Gamma) \, e^{-\int_V  d\xi\, \left(\begin{array}{c} \bar{\psi}_{\uparrow}\\
\psi_{\downarrow} \end{array}\right)^t (\xi)  \, {\mathbb{\Gamma}}(\xi)\,
\left(\begin{array}{c} \psi_{\uparrow} \\ \bar{\psi}_{\downarrow} \end{array}\right)(\xi)  
} ,
\label{eq:HS}
\EEQ
  ${\mathbb{\Gamma}}(\xi)\equiv 
\left(\begin{array}{cc} 0 & \Gamma^*(\xi) \\
\Gamma(\xi) & 0 \end{array}\right)$, where $d\mu(\Gamma)$ is the probability
measure on complex-valued fields $\Gamma\equiv \Gamma_1+\II\Gamma_2:V\to\C\simeq \R^2$ defined by
 \BEQ \int d\mu(\Gamma)\,  \Gamma_i(\xi)
\Gamma_{i'}(\xi') 
=\lambda \del_{i,i'} \del(\xi-\xi'), \label{eq:dmuGamma}
\EEQ corresponding to a quadratic action   $\half\lambda^{-1} \int d\xi\, |\Gamma(\xi)|^2$.
This formula is very much related to the original BCS
interaction in $\bar{\psi}_{\sigma}\psi_{\sigma}\phi$, see
(\ref{eq:BCS}), but now the random potential $\phi$ has become 
an {\em off-diagonal Hermitian matrix $\mathbb{\Gamma}$, coupling
Cooper pairs} $(\bar{\psi}_{\uparrow}, \bar{\psi}_{\downarrow})$
or $(\psi_{\uparrow}, \psi_{\downarrow})$. Fermionic charge
symmetries translate into rotations $\Gamma(\xi)\mapsto e^{2\II\alpha}\Gamma(\xi)$ of the $\Gamma$-field.   The fact that  the values of the
field $\Gamma$ concentrate statistically on a small neighborhood of a circle
is by itself
a good argument to try to split locally $\Gamma$ into the sum of a tangential component $\Gamma_{tang}$
(parallel to the circle) and a transversal (orthogonal) component $\Gamma_{transv}$; from
simpler models with the same symmetries, see e.g. \cite{LeBel}, \S 13.3.1 for
an example inspired by the Glashow-Salam-Weinberg model of electroweak
interactions, it is understood by elementary computations that the tangential,
resp. transversal
components should be massless, resp. massive. The tangential component
 is traditionally called {\em Goldstone boson} by reference to the well-known Goldstone's theorem
\cite{Gol} 
stating that {\em for every spontaneously broken continuous symmetry, a given
theory must contain a massless particle}. Here we find
as effective theory a non-linear sigma model, see e.g.
  \cite{PesSch}, Chapter 11
and \S 13.3. Applying Gaussian
integrations by parts on the measure (\ref{eq:HS}), one easily sees that $n$-point functions of the {\em Goldstone boson} $\Gamma$ are
directly related to $2n$-point functions of electron pairs in
the {\em Cooper pair channel}, $\bar{\psi}_{\uparrow}\bar{\psi}_{\downarrow}$ or 
$\psi_{\downarrow}\psi_{\uparrow}$.  Integrating
instead  w.r. to
the {\em fermions} yields now a {\em bosonic theory in terms of 
$\Gamma$}, with total (i.e., quadratic part + interaction) action
\BEQ {\cal F}(\Gamma) =\half  \lambda^{-1} \int d\xi\,  \Big\{
|\Gamma(\xi)|^2 - \log \det (\Id - C{\mathbb{\Gamma}})(\xi) \Big\}  \label{eq:A} \EEQ
up to cut-offs, where $C$ is the covariance kernel of the 
fermions. Using the identity $\log\det(\Id-A)=\Tr\log(\Id-A)=
 -\Tr\Big(A+\frac{A^2}{2}+\frac{A^3}{3}+\ldots\Big)$,  noting that odd powers do not contribute to the trace,  assuming
 $\Gamma$ to be {\em constant}, and 
 dividing by the volume, one obtains
 the so-called {\em effective potential} , ${\cal S}(\Gamma)$, for
 which an explicit formula was obtained in \cite{FMRT-fermionic},
 section 4,
 \BEQ {\cal S}(\Gamma)={\cal S}(|\Gamma|)=\half  \lambda^{-1} |\Gamma|^2 -
 (2\pi)^{-3}
 \int dp\, \log\Big(1+\frac{|\Gamma|^2}{(p^0)^2+(e(\vec{p}))^2}\Big)  \label{eq:eff-pot}
 \EEQ
with an  ultra-violet cut-off in $p$ at $|p^0|,|\vec{p}|\approx \hbar\omega_D$. (The exact correspondence is  ${\cal S}(\Gamma)\sim (\hbar\omega_D)^2 {\cal E}((\frac{\Gamma}{g\hbar\omega_D})^2)$ in the notations of \cite{FMRT-fermionic}, p. 48, with $g\equiv \sqrt{\lambda}$). By construction, ${\cal S}(\Gamma)$ is due to the
{\em ladder diagrams} obtained by resumming Cooper pair bubble diagrams, see
\S \ref{subsection:bubble}.  Searching for the minimum $\Gamma_{\phi}=$argmin$({\cal S})$ of  $\cal S$ gives an implicit
equation, known as the {\bf gap equation}, which coincides with the one found in mean-field
theory , i.e. starting from the Bogoliubov-De Gennes
Hamiltonian, see e.g. \cite{AltSim}, \S 6.4, and the
one given in our \S \ref{subsection:gap-lowest}. The solution
of the gap equation is
\BEQ |\Gamma|\sim\Gamma_{\phi}\sim \hbar\omega_D e^{-\pi/\lambda m}, \label{eq:Gamma-mean-field}
\EEQ
where $\Gamma_{\phi}$ is as in (\ref{eq:Gammaphijphimu}) and
(\ref{eq:FWgap}).  

\Bigskip  As was already well understood  from the $1/N$-asymptotic expansion in previous work on the subject,
 dominant contributions due to ladder diagrams  are actually the {\em only}
 divergent ones, thus explaining in particular the essential fact that {\em only electron pairs in the Cooper channel contribute to the superconductive phase.}
 
 \Bigskip Allowing $\Gamma=\Gamma(\xi)$ to fluctuate, one obtains an
 action for tangential fluctuations $\Gamma_{tang}(\xi)$ which is roughly
 $\approx \frac{1}{g_{\phi}}\int d\xi\,  |\nabla\Gamma_{tang}(\xi)|^2,$ in Fourier, 
 \BEQ\approx \frac{1}{g_{\phi}} \int dq\, ((q^0)^2+
 v_{\phi}^2 |\vec{q}|^2) |\hat{\Gamma}_{tang}(q)|^2, \EEQ where
 $g_{\phi}\approx \frac{\Gamma_{\phi}^2}{m}$ and 
 \BEQ v_{\phi}\approx v_F:=\frac{p_F}{m} \label{eq:vF} \EEQ
  (Fermi velocity) is a velocity, see e.g. \cite{FMRT-intrinsic}.

\Medskip Although $d=2$, the zero-temperature theory is
effectively a $(1+2)$-dimensional theory. Hence Mermin-Wagner's argument does not apply, and infinitesimal bulk or boundary
terms are in principle enough to imply a symmetry-breaking in some direction
$\theta$, by which we mean that 
\BEQ \Gamma\sim \Gamma_{\phi} e^{\II\theta} \ +\ {\mathrm{fluctuation}}. \label{eq:Gammathetafluct} 
 \EEQ

\item[(iii)] {\bf (Regime III, low-energy regime)} 
Around the transition scale, the Goldstone boson favors a position on a circle
of radius $\Gamma_{\phi}$, thus conferring the electron an effective squared mass  $\sim 
\Gamma_{\phi}>0$ which dominates the kinetic energy term. Thus electrons are "hooked
up" to a fixed scale $\simeq j_{\phi}$:  Regime III
may be thought of {\em from the point of view of unpaired fermions} as a {\em single scale} extending from $j_{\phi}$ to $+\infty$. Dominant low-momentum contributions to Feynman
diagrams come therefore exclusively from the Goldstone boson.  Now,
Ward identities associated to the (broken) $U(1)$ number symmetry prove (more or
less as
in the case of diagrams with low-momentum photons in  quantum electrodynamics)
that such diagrams vanish up to error terms coming from cut-off effects, which decrease exponentially as
one lowers the energy scale. Thus, below an energy level
\BEQ j'_{\phi}=j_{\phi}+o(\ln(1/g)), \label{eq:j'phi} \EEQ
 the effective coupling constant of diagrams is once again small enough to sum  series produced by the cluster expansion.  Contrary to
 individual fermions, the Goldstone boson (or equivalently, electron pairs in the Cooper channel) remains non-massive
in the {\em Euclidean infra-red} defined {\em not by the vicinity to
the Fermi circle}, but by the {\em vicinity to $0$}.

\end{itemize}

\Bigskip
Unfortunately, the writing of the above program stopped at the end of 
the 90'es somewhere between Regime I and Regime II  for lack of a sound
mathematical proof, due to technical difficulties coming mainly from the
necessity of integrating simultaneously the fermion fields $\psi,\bar{\psi}$ and
the Goldstone boson field $\Gamma$. 
At each scale $j$, one had to distinguish "small-field", perturbative regions, from "large-field" regions, which could
be handled only through large deviation arguments.
In theory, the idea was to   split the $\Gamma$-field into the
sum of  a "fast" perturbation field $\Gamma _{f}$ with
scale index $\le j$, and of a "slow" background field 
$\Gamma_b$ with scale index $>j$, giving a local orientation
and making it possible to define a "tangential" and a
"transversal" component. In the small field region,
 $\Gamma _{f}$ is small, and $\Gamma_b$ minimizes
 to a good approximation the effective potential computed with the contributions of the scales $\le j$.
The main difficulty was to prove that a given
region was  "small-field"  with high probability.

\Medskip    
 Also, both the  implementation of Ward identities and the use of the $1/N$-expansion    are  awkward in this representation, requiring partial integrations
 of the bosonic field in order to go back to a purely fermionic representation.
 Finally, integrating out fermions in order to deal with the infra-red behavior of the
 boson field naturally leads to functional determinants such as (\ref{eq:A}) 
 which are  delicate
to define properly and handle in combination with cluster decompositions  (see
the above cited paper \cite{Kop} by C. Kopper for a successful result in this direction for a {\em scalar} intermediate boson field).

\vskip 1cm
{\centerline{**************************************************}}
\vskip 1cm

\noindent Our strategy here is different; let us describe it briefly, highlighting differences with previous attempts.  We introduce by hand a fixed symmetry-breaking term ("adding and subtracting" it) with a given, arbitrary but fixed, orientation $\theta$
(see (\ref{eq:Gammathetafluct})), and a module computed by a fixed-point argument; this spares us the trouble of having to define a local orientation around every point. Thus the 
transversal, resp. tangential, direction is {\em parallel} $(//)$, resp. {\em perpendicular} $(\perp)$, to $\theta$. We carefully avoid introducing the  Goldstone field in the first place, producing  instead through an explicit expansion  an effective, non-local interaction kernel $\Sigma_{\perp,\perp}$ in the perpendicular $(\perp)$ direction,
called {\em Goldstone boson propagator}, which is obtained by resumming a geometric
series made up of alternating Cooper pair bubble diagrams and  Bethe-Salpeter kernels, themselves sums of two-particle irreducible diagrams as in Fig. \!0.2; the  $\Sigma_{\perp,\perp}$-kernel  may be interpreted as
the propagator of a bosonic particle which is never introduced.
 Decomposing fields
into angular sectors, we are able to prove that the two-point
function of Cooper pairs diverge only in the $s$-wave.
Then -- taking the $s$-wave projection --  we write down in terms of $\Sigma_{\perp,\perp}$ an exact, non-perturbative version of the
gap equation mentioned in the discussion of the neighborhood
of the
transition scale (see Regime II above). The actual value
of the gap $\Gamma_{\phi}$ is shown to be close to the mean-field one  (\ref{eq:Gamma-mean-field}) by a fixed-point argument. Cooper pair two-point functions are shown to be
proportional to the $\Sigma_{\perp,\perp}$-kernel, itself roughly inverse
of the conjectural action associated to tangential fluctuations
of the $\Gamma$-field (again, see  Regime II).

\Bigskip Proceeding this way, we fill this important gap in the literature, thus
hopefully laying the foundational stones
of a rigorous mathematical analysis of 2D and 3D low-temperature superconductivity
from first principles, going beyond Ginzburg-Landau theory. We also conjecture that our methods will 
help understand related models featuring a dynamical
symmetry-breaking transition in a large $N$ regime, like 3D Anderson's model in its
delocalized phase, or the integer quantum Hall effect.
\vskip 1cm


\subsection{Our results}  \label{subsection:results}


\noindent  Instead of singling out
three regimes, we proceed as follows. 

\Medskip{ \bf A. (fermionic regime, or high-momentum
theory)} We  {\em choose} a value for $j_{\phi}$ such that 
(\ref{eq:jc0}) holds,  and consider the fermionic theory directly with
{\em effective parameters} $\Gamma_{\phi}\approx \hbar\omega_D e^{-\pi/m\lambda}$,
$m^*\approx m$, $\mu^*\approx\mu$, namely, we modify the Grassmann Gaussian measure $d\mu(\psi,\bar{\psi})$ of eq. (\ref{eq:A0}) by multiplying
it with  the exponential of a quadratic weight, 
\BEQ d\mu^*_{\theta}(\psi,\bar{\psi}) \propto e^{-\int  d\xi\, 
\left(\begin{array}{c} \bar{\psi}_{\uparrow} \\ \psi_{\downarrow} \end{array}\right)^t (\xi) \, \Big\{ {\mathbb{\Gamma}}(\theta)+(\frac{1}{m}-\frac{1}{m^*}) \frac{|\vec{\nabla}|^2}{2} +(\mu^*-\mu)\Big\}\,  \left(
\begin{array}{c} \psi_{\uparrow}\\
\bar{\psi}_{\downarrow} \end{array}\right) (\xi)} \ d\mu(\psi,\bar{\psi})  \label{eq:dmu*(theta)} 
\EEQ
where $\xi=(\tau,\vec{x})$, and ${\mathbb{\Gamma}}(\theta)=\left(\begin{array}{cc} 0 & \Gamma^* \\
\Gamma & 0 \end{array}\right)\equiv \left(\begin{array}{cc} 0 & \Gamma_{\phi} e^{-\II\theta}
\\  \Gamma_{\phi} e^{\II \theta} & 0\end{array}\right)$, $\theta\in\R/2\pi\Z$ is a {\em constant} off-diagonal, Hermitian
matrix; see Definition \ref{def:Grassmann-measure} for a precise definition. Up to cut-offs, the covariance kernel
of $d\mu^*_{\theta}$ is the inverse of the quadratic form
\BEQ B(\psi,\bar{\psi})=\int dp\, \left(\begin{array}{c} \bar{\psi}_{\uparrow}(-p) \\ \psi_{\downarrow}(-p) \end{array}\right)^t  \, \Big\{ \II p^0 -e^*(p)\sigma^3-{\mathbb{\Gamma}} 
\Big\} \left(
\begin{array}{c} \psi_{\uparrow}(p)\\
\bar{\psi}_{\downarrow}(p) \end{array}\right)  
\label{eq:B},
\EEQ
compare with (\ref{eq:A0}), where $e^*(p):=\frac{|\vec{p}|^2}{2m^*}-m^*$ is the effective (or renormalized) dispersion
relation. 
 This extra term is compensated by subtracting it
to the interaction, which takes   the form of a bare, scale $j_D$ counterterm 
$\del {\cal V}_{\theta}:=\int d\xi\, \del{\cal V}_{\theta}(\xi)$ added to 
the interaction $\cal V$ defined in (\ref{eq:V}), namely,
$$\del {\cal V}_{\theta}(\xi)=-\left(\begin{array}{c} \bar{\psi}_{\uparrow} \\ \psi_{\downarrow} \end{array}\right)^t(\xi) \, \Big\{\del{\mathbb{\Gamma}}^{j_D}+ \frac{\del 
m^{j_D}}{(m^*)^2}  \frac{|\vec{\nabla}|^2}{2}+\del\mu^{j_D} \Big\}\,  \left(
\begin{array}{c} \psi_{\uparrow}\\
\bar{\psi}_{\downarrow} \end{array}\right) (\xi),$$
and
\BEQ \del {\mathbb{\Gamma}}^{j_D}:={\mathbb{\Gamma}}(\theta), \qquad
\frac{\del m^{j_D}}{(m^*)^2}:=\frac{1}{m}-\frac{1}{m^*}, \qquad \del\mu^{j_D}:=\mu^*-\mu. 
\EEQ
Because of the ultra-violet, scale $j_D$  cut-off, the resulting measure 
\BEQ d\mu_{\theta;\lambda}(\psi,\bar{\psi}):= \frac{1}{{\cal Z}^*_{\lambda}} 
e^{-({\cal V}+\del{\cal V}_{\theta})} d\mu^*_{\theta}(\psi,\bar{\psi})
\label{eq:0.29}
\EEQ
 depends on $\theta$, on $\Gamma_{\phi}$, on $m^*$ and on
 $\mu^*$. In particular -- as will be shown --, 
it induces an infinite-volume symmetry-breaking precisely in the direction $\theta$.  Then we study the renormalization flow
from scale $j_D$ to scale $j'_{\phi}=j_{\phi}+o(\ln(1/g))$. Fermions are shown to become massive
at energy scales $\simeq j_{\phi}$, making it possible to integrate out fermions
with energy scales $>j_{\phi}$ as if they were of the same scale, {\em provided} they do not form Cooper pairs with
low transfer momentum. 

\Medskip The considerable difference with the above discussion of Regime I
is that, should one start from $\Gamma=0$, then the number symmetry would 
imply that
$\Gamma$ is {\em not} renormalized: the fact that $\Gamma\not=0$ in average 
is a {\em non-perturbative} effect. Proceeding that way,  the symmetry-breaking would be read
only indirectly through the apparition of a  large,
renormalized coupling constant depending on the momentum angular sectors 
of its four legs, favoring  $\bar{\psi}_{\uparrow}\bar{\psi}_{\downarrow}, \psi_{\downarrow}\psi_{\uparrow}$ in Cooper channels. These, in turn, 
might be subtracted by changing the value of $\Gamma$. Our approach looks much
simpler.

\Medskip Let us remark at this point that another (maybe
physically
more natural)
procedure would have led to an equivalent effective, large-scale behavior. Namely, we could have chosen a lattice ultra-violet
cut-off  with lattice constant $a$  on a large volume $V\subset \R\times\R^2=$span$(\vec{e}_0,\vec{e}_1,\vec{e}_2)$, with inverse covariance kernel ${\cal A}(\psi,
\bar{\psi}):=\sum_i \Big( \half(\bar{\psi}_{\sigma}(\xi_i)\psi_{\sigma}(\xi_i+a\vec{e}_0)-\bar{\psi}_{\sigma}(\xi_i+a\vec{e}_0)\psi_{\sigma}(\xi_i)) - \frac{1}{4m} \sum_{j=1,2}
 (\bar{\psi}_{\sigma}(\xi_{i}+a\vec{e}_j)-\bar{\psi}_{\sigma}(\xi_i)) ({\psi}_{\sigma}(\xi_{i}+a\vec{e}_j)-{\psi}_{\sigma}(\xi_i)) -\mu\bar{\psi}_{\sigma}(\xi_i)\psi_{\sigma}(\xi_i)
 \Big)$. Adding
  $\sum_i \left(\begin{array}{c} \bar{\psi}_{\uparrow} \\
   \psi_{\downarrow} \end{array}\right)^t(\xi_i) {\mathbb{\Gamma}}(\theta)
    \left(\begin{array}{c}
 \psi_{\uparrow}\\ \bar{\psi}_{\downarrow} \end{array} \right)(\xi_i)$
 to ${\cal A}(\cdot)$, and subtracting the same term to
 ${\cal V}$, would have led back formally to the initial
 theory -- up to boundary terms on $\partial V$. Then
 it would have been natural to add to the interaction an
 infinitesimal bulk term, 
 ${\cal V}+\del {\cal V}_{\theta}\to {\cal V}+\del {\cal V}_{\theta}+ \sum_i  \left(\begin{array}{c} \bar{\psi}_{\uparrow} \\ \psi_{\downarrow} \end{array} \right)^t(\xi_i) 
 \, \left(\begin{array}{cc} 0 & \eps e^{-\II\theta} \\ \eps e^{\II\theta} & 0 \end{array}\right)\,  \left(\begin{array}{c}
 \psi_{\uparrow}\\ \bar{\psi}_{\downarrow} \end{array} \right)(\xi_i)$
and let
\BEQ d\mu_{\theta;\lambda}(\psi,\bar{\psi}):=
\lim_{\eps\to 0^+} \lim_{|V|\to\infty}  d\mu^{\eps,V}_{\theta;\lambda}(\psi,\bar{\psi})  \label{eq:0.30} \EEQ
where  $ d\mu_{\theta;\lambda}^{\eps,V}(\psi,\bar{\psi})$ is the normalized measure obtained by modifying ${\cal V}$ as indicated. Our (unproven but unlikely to be difficult) claim is that models (\ref{eq:0.29})
and (\ref{eq:0.30}) have the same infra-red behavior. 

\Bigskip {\bf B. (bosonic regime, or low-energy theory)}  Diagrams involving
a small number  ($\le 6$) of low-momentum Cooper pairs are a priori divergent, because
the Goldstone boson propagator is not integrable (see Theorem 2 below).  {\em Ward identities}, however, imply that these are 
 actually convergent. On the other hand, it would be a very bad
 idea to reorganize the series of perturbations in such a way
 as to produce systematically {\em all} Goldstone bosons, since
 this would lead to a local accumulation of low-momentum bosonic 
 fields which cannot be controlled perturbatively. Thus
 what is needed here is a complementary, multi-scale Cooper pair expansion procedure producing only a limited number of Goldstone bosons.
 This procedure makes it possible to give at last the large scale behavior of $n$-point
functions of the theory.

\Bigskip

**********************************************************

\Bigskip Let us now present the main results of our article.  Our starting point, namely, the symmetry-broken
measure $d\mu_{\theta;\lambda}$,  is equal to
the measure  defined in (\ref{eq:0.29}),
 with a precise choice of
cut-offs.  First, we have:

\Bigskip 
{\bf Theorem 1 (construction of the model).} {\em Assume  $g:=\lambda m>0$  is  small enough. Consider the Grassmann
measure $d\mu_{\theta;\lambda}$ of Definition \ref{def:Grassmann-measure},  re-expressed
in terms of  
modified
parameters $\Gamma_{\phi},m^*,\mu^*$.  Denote by  $\langle \ \cdot\
\rangle_{\theta;\lambda}$ the expectation of a fermionic functional with
respect to $d\mu_{\theta;\lambda}$. Let 
\BEQ j_{\phi}:=\lfloor j_D+ \frac{\pi}{\lambda m}\rfloor
\label{eq:jphi} \EEQ 
and
\BEQ j'_{\phi}:=j_{\phi}+ \lfloor \frac{1}{4}\ln(1/g) \rfloor. \EEQ
 Then, for an adequate choice of  parameters such that
\BEQ  \Gamma_{\phi}\approx  \hbar \omega_D\,  e^{-\pi/m\lambda} \approx 2^{-j_{\phi}}
\mu, \qquad 
m^*=m(1+O(g^2)), \qquad \mu^*=\mu(1+O(g)), \EEQ
the renormalization flow  is well-defined down to scale
$j'_{\phi}-1$, leading to vanishing renormalized
counterterms  $ \del \mu^{j'_{\phi}},\del m^{j'_{\phi}}$, and a
small value of 
$\del\Gamma^{j'_{\phi}}=O(g\Gamma_{\phi})$. The coefficient
of $p^0$ in the effective covariance kernel, see (\ref{eq:A}),
is $\II Z^{j'_{\phi}}$, with $Z^{j'_{\phi}}=1+O(g^2)$.
The parameters $m^*, \mu^*$, together with the {\em
pre-gap energy} $\Gamma^{(j'_{\phi}-1)\to}$ --
defined as solution of an approximate gap equation, 
see Definition \ref{def:pre-gap} -- are fixed simultaneously
by a fixed-point argument. }

\Bigskip
Theorem 1 covers exactly the fermionic theory of  {\bf A.}, and spans
the entire section \ref{section:fermion}. Because fermions
become massive in the infra-red region, the "true" effective
mass (defined in terms of decay of fermionic two-point functions, see Theorem 3) is not directly accessible (it is
given in terms of the distance to the real axis of the first pole of the Fourier-transformed two-point function). Nevertheless,
$m^*,\mu^*$ may be thought as {\em effective parameters} giving
the essentially {\em exact location of the renormalized Fermi circle}. Going slightly {\em beyond} scale $j_{\phi}$, 
downto scale $j'_{\phi}$ -- which is required for technical reasons pertaining to the {\em bosonic regime} --  can only improve the precision.  As
for the scale $j'_{\phi}$ counterterm coefficient $\del\Gamma^{j'_{\phi}}$, thanks to its supplementary $O(g)$ prefactor, it is sufficiently close to 0 to enter only as a
small, single-scale correction to the value of $\Gamma_{\phi}$, which
is chosen so as to satisfy the gap equation (see Theorem 2).  The gap energy may equivalently be
expressed in terms of the renormalized mass, $\Gamma_{\phi}\approx \hbar\omega_D \, e^{-\pi/m^*\lambda}$ since $e^{-\frac{\pi}{\lambda} |\frac{1}{m}-\frac{1}{m^*}|}=O(1)$.

\Bigskip {\em Cooper pairs} are defined as linear combinations of electron pairs
in Cooper channel, $(\bar{\psi}_{\uparrow}\bar{\psi}_{\downarrow})(\xi)$ and
$({\psi}_{\downarrow}\psi_{\uparrow})(\xi)$, which may be represented in matrix
form as $\left(\begin{array}{c}
\bar{\psi}_{\uparrow} \\ \psi_{\downarrow} \end{array}\right)^t 
{\mathbb{\Gamma}} \left(\begin{array}{c} \psi_{\uparrow} \\ \bar{\psi}_{\downarrow}\end{array}\right)$ where ${\mathbb{\Gamma}}=\left(\begin{array}{cc} 0 & \Gamma^ *\\ \Gamma & 0 \end{array}\right)$ is an arbitrary
off-diagonal Hermitian matrix. Cooper pairs in the $(\perp,\perp)$ channel, i.e. with associated matrix 
${\mathbb{\Gamma}}^{\perp}=\Gamma_{\phi} \left(\begin{array}{cc} 0 & e^{-\II(\theta+\frac{\pi}{2})}  \\ e^{\II(\theta+\frac{\pi}{2})} & 0 \end{array}\right)$, are dominant only in the low-momentum
regime (Regime II and III, or {\bf B.} in our terminology), so relevant observables are
averages of these composite fields in a large volume, or correlations at
distances $|\tau-\tau'|+\frac{m^*}{p^*_F}|\vec{x}-\vec{x}'|\gtrsim \Gamma_{\phi}^{-1}$.

\Bigskip
{\bf Theorem 2 (gap equation, Cooper pair correlations at large scale).} \label{Theorem2}
 {\em 
\begin{itemize} 
\item[(i)] The gap equation for $d\mu^*_{\theta;\lambda}$
(see Remark below Theorem 1), defined by fixed point as the value of $\Gamma_{\phi}$ for which the local part of the fermionic two-point function vanishes, see
Definition \ref{def:gap-one-fct}, has a unique solution
\BEQ \Gamma_{\phi}\approx\hbar\omega_D e^{-\pi/m\lambda}.
\label{eq:Gammaphi}
\EEQ
\item[(ii)] Fix  $\Gamma_{\phi}$ as in (i), and consider
the associated theory with measure $d\mu^*_{\theta;\lambda}$. 
Let $\xi_1\not=\cdots\not=\xi_{2n}$ ($n\ge 1$). If $\eps$ is a two-by-two matrix, we define  for $\eta>0$ (small)
\BEA && :\, \Big(\left(\begin{array}{c}
\bar{\psi}_{\uparrow} \\ \psi_{\downarrow} \end{array}\right)^t 
\eps \left(\begin{array}{c} \psi_{\uparrow} \\ \bar{\psi}_{\downarrow}\end{array}\right)\Big)(\eta^{-1}\xi_i) \, :\  = \nonumber\\
&&\qquad  \Big(\left(\begin{array}{c}
\bar{\psi}_{\uparrow} \\ \psi_{\downarrow} \end{array}\right)^t 
\eps \left(\begin{array}{c} \psi_{\uparrow} \\ \bar{\psi}_{\downarrow}\end{array}\right)\Big)(\eta^{-1}\xi_i) \, \  - \Big\langle \left(\begin{array}{c}
\bar{\psi}_{\uparrow} \\ \psi_{\downarrow} \end{array}\right)^t 
\eps \left(\begin{array}{c} \psi_{\uparrow} \\ \bar{\psi}_{\downarrow}\end{array}\right)\Big)(\eta^{-1}\xi_i) \,  \Big\rangle_{\theta;\lambda}.
\nonumber\\ 
\EEA
Then  $2n$-point functions of Cooper pairs   have the following asymptotic large-scale
behavior:
\BEA &&  \Big\langle \prod_{i=1}^{2n}\,  : \, \Big(\left(\begin{array}{c}
\bar{\psi}_{\uparrow} \\ \psi_{\downarrow} \end{array}\right)^t 
{\mathbb{\Gamma}}^{\perp} \left(\begin{array}{c} \psi_{\uparrow} \\ \bar{\psi}_{\downarrow}\end{array}\right)\Big)(\eta^{-1}\xi_i) \, :\ 
\Big\rangle_{\theta;\lambda} \sim_{\eta\overset{>}{\to} 0}  \ ( 1+o(1))\ \cdot\ \nonumber\\
&&\qquad \  \cdot\  \lambda^{-2n} \sum_{(i_1,i_2),\ldots,(i_{2n-1},i_{2n})} 
\, \prod_{k=1}^n  \big\{ \Gamma_{\phi}^2\Sigma_{\perp,\perp}\ 
(\eta^{-1}(\xi_{i_{2k-1}}-\xi_{i_{2k}})) \big\},  \label{eq:main1} \nonumber\\
\EEA
where ${\bf{\Gamma}}^{\perp}:=\Gamma_{\phi}\left(\begin{array}{c} e^{-\II(\theta+\frac{\pi}{2})} \\
e^{\II(\theta+\frac{\pi}{2})}\end{array}\right)$ is perpendicular to ${\bf{\Gamma}}^{//}:=\Gamma_{\phi}\left(\begin{array}{c} e^{-\II\theta} \\
e^{\II\theta}\end{array}\right)$;  the sum ranges over all pairings $\Big\{ \{i_1,i_2\},
\cdots,\{i_{2n-1},i_{2n}\} \Big\}$ of indices, and the 
two-by-two matrix kernel
$\Sigma$,  whose entry $ \Sigma_{\perp,\perp}=\frac{1}{\Gamma_{\phi}^2} \, ^t {\bf{\Gamma}}^{\perp}\Sigma {\bf{\Gamma}}^{\perp}$ in the $(\perp,\perp)$-channel is  interpreted as {\em propagator of the Goldstone boson},  is defined and studied in \S \ref{subsection:complementary} and \ref{subsection:fixed-point}.
 In particular, see (\ref{eq:GammaSigmaGamma}) and
(\ref{eq:Sigmaj+Sigma}), if $\xi\not=0$, 

\BEQ  \Sigma_{\perp,\perp}(\eta^{-1}\xi) \sim_{\eta\overset{>}{\to} 0} \eta   \frac{g_{\phi}/
4\pi v_{\phi}}{\sqrt{|\vec{x}|^2+v^2_{\phi}\tau^2}} \Big(1+O(\frac{\eta}{\Gamma_{\phi}(|\tau|+
|\vec{x}|/v_{\phi})})\Big), \label{eq:intro-Sigmaperp}
\EEQ
 where 
\BEQ v_{\phi}\approx v^*_F\equiv \frac{p_F^*}{m^*} \label{eq:vphi} \EEQ
is roughly equal to the (renormalized) Fermi velocity
$v^*_F$; and   
\BEQ g_{\phi}\approx
\frac{\Gamma_{\phi}^2}{m^*}.  \label{eq:gphi} \EEQ
 The kernel $\Sigma$  is {\em not integrable}, in the sense that the integral $\int_{\R^3} d\xi\  \Sigma_{\perp,\perp}(\xi-\xi')$ diverges. 
 \end{itemize}
 }

\vskip 1cm\noindent  The gap parameter $\Gamma_{\phi}$, as well as the Goldstone boson parameters $v_{\phi},g_{\phi}$, are
 obtained as limits of scale-dependent converging series, $(\Gamma^{j_+\to})_{j_+\ge j'_{\phi}}$,  $(v_{\phi}^{j_+\to})_{j_+\ge j'_{\phi}}, (g_{\phi}^{j_+\to})_{j_+\ge j'_{\phi}}$, see Definition \ref{def:scalej+param} and \S \ref{subsection:fixed-point}.

\Medskip {\em Remark 1.} The Fourier transform of $\Sigma_{\perp,\perp}$, as follows by straightforward computations from (\ref{eq:intro-Sigmaperp}), is of the form
\BEQ \Sigma_{\perp,\perp}(q)\sim_{q\to 0} \frac{g_{\phi}}{(|q|^0)^2+(v_{\phi}|\vec{q}|)^2}
\EEQ
and {\em diverges for $q=0$.} Quite remarkably, the fact that $\Sigma_{\perp,\perp}$ {\em has a p\^ole 
precisely at $q=0$}, implying a non-massive behavior, is shown by a {\em Ward identity} 
(see \S \ref{subsection:Ward}) to be {\em equivalent to the  gap equation}. We actually
use most of the time the latter characterization in terms of $\Sigma_{\perp,\perp}$ as a definition of the gap equation, a convenient choice for estimates.
 
\Medskip {\em Remark 2.} Error terms $o(1)$ in Theorem 2   (ii)
go to 0 as $g\to 0$.  Details of the proof,
see in particular (\ref{eq:GFMSB-P}) in section \ref{section:fermion} and
(\ref{eq:GBMSB-P}) in section \ref{section:boson}, give error terms
denoted by $"o(1)$" which are bounded by $O(g^{1/4})$ or equivalently  $O(2^{-(j'_{\phi}-j_{\phi})})$; a more detailed
computation of lowest-order terms would allow to replace
$O(g^{1/4})$ by $O(g)$. The leading order of $\Gamma_{\phi}$ 
in (i), on the other hand, depends both on the ultra-violet cut-off and on subleading
diagrams.

\Medskip {\em Remark 3.} The above Theorem only involves
the $(\perp,\perp)$-component of $\Sigma$. Intermediate
computations produce a two-by-two matrix 
propagator $\Sigma$, diagonal in the basis $({\bf\Gamma}^{//},{\bf\Gamma}^{\perp})$. However, the other diagonal coefficient
$\Gamma_{\phi}^2\Sigma_{//,//}(\xi-\xi')=^t {\bf\Gamma}^{//} \Sigma {\bf\Gamma}^{//}(\xi-\xi')$ is {\em massive}, 
which translates into a quasi-exponential decay at large
distances, similar to that of the fermions. This case
is studied alongside the case of isolated fermions in
Theorem 3 below.  

\Bigskip We can now tentatively recast our results in terms of  the Hubbard-Stratonovich transformed theory
(\ref{eq:HS}) involving the complex boson field $\Gamma=\Gamma(\xi)$. Namely, by
Gaussian integration by parts (see Appendix), $n$-point functions of the
$\Gamma$-field are in one-to-one correspondence with the Cooper pair $n$-point
functions of (\ref{eq:main1}), from which we can conclude
(leaving aside mathematical rigor) to the following.
The Goldstone boson $\Gamma$ behaves like a non-massive particle described by a non-linear
sigma model with coupling constant $g_{\phi}\approx  \frac{\Gamma_{\phi}^2}{m^*}$ 
having the dimension of an energy, namely, the statistical weight for a function $\Gamma$ with values on the circle 
$\{|\Gamma|=\Gamma_{\phi}\}$ 
 is $\sim e^{-\frac{1}{g_{\phi}} \int d\xi\, |\nabla \Gamma(\xi)|^2}$,
where  $\nabla:=\left(\begin{array}{c} \partial_{\tau} \\ v_{\phi} \vec{\nabla}\end{array}\right)$ . In rescaled, non-dimensional units, $(\tau,\vec{x},\Gamma)\to (\tau',\vec{x}',\Gamma'):=
(\Gamma_{\phi} \tau,p_c \vec{x},\Gamma_{\phi}^{-1}\Gamma)$, with $p_c=\frac{\Gamma_{\phi}}{v_{\phi}} $, the weight
for $\Gamma$ becomes $\sim e^{-\frac{1}{g'_{\phi}}\int d\xi' |\nabla'\Gamma'(\xi')|^2}$ with
\BEQ g'_{\phi}=\frac{g_{\phi}}{\Gamma_{\phi} v^2_{\phi}}  \approx  \frac{\Gamma_{\phi}}{\mu} \ll 1.  \label{eq:g'phi} \EEQ
 This is the signature of  a weakly coupled, super-critical $(1+2)$-dimensional   non-linear sigma model,  known (at least perturbatively) to be 
a free theory at large scales (see e.g. \cite{PesSch}, \S 13.3), from
which the qualitative behavior (\ref{eq:main1}) can be expected.  Note, however, that the interaction between Cooper pairs is mediated by
{\em massive} individual electrons.

\Bigskip
{\bf Theorem 3 (isolated electrons).}  {\em $n$-point functions
of electrons  which are not in the $(\perp,\perp)$ Cooper pair channel
  decay quasi-exponentially at large distance with
a decay rate $\approx \Gamma_{\phi}$, namely, for every $n\ge 0$, there exists $C_n>0$ such that, if $\xi\not=\xi'$,

\BEQ \Big|\, \langle \bar{\psi}_{\sigma}(\eta^{-1}\xi)\psi_{\sigma'}(\eta^{-1}\xi')
\rangle_{\theta;\lambda} \, \Big| \le C_n   (p^*_F)^2\del_{\sigma,\sigma'} \,  \Big( 1+ \Gamma_{\phi}\eta^{-1}(|\tau-\tau'|+\frac{m^*}{p_F^*}|\vec{x}-\vec{x}'|) \Big)^{-n}  \label{eq:0.39}
\EEQ
and similarly,
 \BEQ \Big|\, \langle \, : (\bar{\psi}_{\uparrow}\psi_{\uparrow} )(\eta^{-1}\xi)  :\,  \, : (\bar{\psi}_{\uparrow}\psi_{\uparrow} ) (\eta^{-1}\xi')  :\, \rangle_{\theta;\lambda} 
\, \Big| \le C_n (p^*_F)^4
\Big(1+\Gamma_{\phi}\eta^{-1}(|\tau-\tau'|+\frac{m^*}{p_F^*}|\vec{x}-\vec{x}'|)\Big)^{-n}.  \label{eq:0.40} \EEQ
and, letting ${\mathbb{\Gamma}}^{//}:=\Gamma_{\phi} \left(\begin{array}{cc} 0 & e^{-\II\theta} \\ e^{\II\theta} & 0\end{array}\right)$, 
\BEQ \Big|\,  \langle \, : (\bar{\Psi}{\mathbb{\Gamma}}^{//}\Psi)(\eta^{-1}\xi) :\, \, :   (\bar{\Psi}{\mathbb{\Gamma}}^{//}\Psi) (\eta^{-1}\xi') :\,   \rangle_{\theta;\lambda} 
\, \Big| \le C_n (p^*_F)^4
\Big(1+\Gamma_{\phi}\eta^{-1}(|\tau-\tau'|+\frac{m^*}{p_F^*}|\vec{x}-\vec{x}'|)\Big)^{-n}.  \label{eq:0.40bis} \EEQ

 }

\Bigskip {\em Remark.} We fall short of proving {\em exponential decay} for
two-point functions of electrons. This is mainly due to our choice
of cut-offs for the Gaussian covariance kernel, which
yields {\em quasi-exponentially} decaying scaled kernels. However, 
non-Gaussian cut-offs of the form $\chi^j(p/\mu)=e^{-2^{2j}\mu^{-2}[(p^0)^2+(e(\vec{p}))^2+\Gamma_{\phi}^2]}-e^{-2^{2(j+1)}\mu^{-2} [(p^0)^2+(e(\vec{p}))^2+\Gamma_{\phi}^2]}$, similar to those used in \cite{DisMagRiv},  would produce {\em exponentially} decaying
kernels, and most likely allow us to prove an exponential
decay in $e^{-\Gamma_{\phi}(1+o(1))\eta^{-1}(|\tau-\tau'|+\frac{m^*}{p_F^*}|\vec{x}-\vec{x}'|)}$, resp.  $e^{-2\Gamma_{\phi}(1+o(1))\eta^{-1}(|\tau-\tau'|+\frac{m^*}{p_F^*}|\vec{x}-\vec{x}'|)}$ for two-point, resp. four-point
functions, instead of the quasi-exponential decay factors
$\Big( 1+ \Gamma_{\phi}\eta^{-1}(|\tau-\tau'|+\frac{m^*}{p_F^*}|\vec{x}-\vec{x}'|) \Big)^{-n} $ in (\ref{eq:0.39}), 
resp. (\ref{eq:0.40}), implying that the {\em effective mass}
of electrons is $\approx \Gamma_{\phi}$.

\Medskip Let us add a few comments. Cooper pairs are the main observables at low energy. On the other
hand, isolated electrons have become {\em massive}.
 The large-scale behavior of isolated fermions and Cooper pairs 
written down in Theorem 3 points out to a {\em superconductive phase},
qualitatively different from the {\em normal phase}, which is that of the free
electron gas obtained for $\lambda=0$, or more generally of Fermi liquids. Namely, for $\lambda=0$ (i.e. in the {\em normal phase}), specializing to vanishing $\tau$-coordinates, the
one-point density kernel is proved by a standard computation (see (\ref{eq:5.18}) and
(\ref{eq:5.19})),
\BEQ \Big|\langle \bar{\psi}_{\sigma}(0,\eta^{-1}\vec{x})\psi_{\sigma'}(0,\eta^{-1}\vec{x}')\rangle_{\lambda=0} \Big| \approx_{\eta\overset{>}{\to} 0} (p_F^*)^{1/2} \, \del_{\sigma,\sigma'}\  
|\eta^{-1}(\vec{x}-\vec{x}')|^{-3/2},  \label{eq:1pt-density-kernel0}
\EEQ
 and the  density-density correlation kernel (connected two-point function)
\BEQ \Big|\langle \, :(\bar{\psi}_{\uparrow}\psi_{\uparrow})(0,\eta^{-1}\vec{x})\, :\, :\, 
(\bar{\psi}_{\uparrow}\psi_{\uparrow})(0,\eta^{-1}\vec{x}'):\, \rangle_{\lambda=0} \Big|
\approx_{\eta\to 0}  p_F^*\   |\eta^{-1}(\vec{x}-\vec{x}')|^{-3}
. \EEQ
These  correlation functions decrease polynomially at large distances, 
to be compared with the quasi-exponential decrease exhibited in Theorem 3. Two-point
functions of Cooper pairs in arbitrary channels behave similarly with a {\em cubic} inverse  distance
decrease, 
\BEQ  \Big\langle \prod_{i=1}^{2} \, :\,  \Big(\left(\begin{array}{c}
\bar{\psi}_{\uparrow} \\ \psi_{\downarrow} \end{array}\right)^t 
{\mathbb{\Gamma}}(\theta_i) \left(\begin{array}{c} \psi_{\uparrow} \\ \bar{\psi}_{\downarrow}\end{array}\right)\Big)(\eta^{-1}\xi_i)\, :\,
\Big\rangle_{\lambda=0}   =O_{\eta\to 0}\Big(  p_F^*\   |\eta^{-1}(\vec{x}-\vec{x}')|^{-3}
\Big),
\EEQ
where ${\mathbb{\Gamma}}(\theta_i):=\Gamma_{\phi}\left(\begin{array}{cc} 0 & e^{-\II\theta_i} \\ e^{\II\theta_i} 
& 0 \end{array}\right)$, $\theta_i\in\R/2\pi\Z,i=1,2$, 
as opposed to the {\em linear} inverse distance decrease shown in  Theorem 2 for the
superconductive phase.

\Bigskip {\bf Theorem 4 (phase transition).}
{\em Let $\theta,\theta_1\in\R/2\pi\Z$ and $\xi_1\in\R\times\R^2$. Then
\BEA \langle (\bar{\Psi}{\mathbb{\Gamma}}(\theta_1)\Psi)(\xi_1)\rangle_{\theta;\lambda} =\frac{c}{\lambda} \Tr({\mathbb{\Gamma}}(\theta_1){\mathbb{\Gamma}}(\theta))
 \EEA
with $c=1+o(1)$.
}

\Medskip This is in agreement with the mean-field prediction for the average of Cooper pairs in a superconducting ground-state oriented in direction $\theta$, namely, e.g.
\BEQ  \langle \bar{\psi}_{\uparrow}\bar{\psi}_{\downarrow}+ \psi_{\downarrow}  \psi_{\uparrow}\rangle_{\theta=0;\lambda}\sim\frac{\Gamma_{\phi}}{\lambda} \Tr((\sigma^1)^2)=
2\frac{\Gamma_{\phi}}{\lambda}, \qquad \langle \bar{\psi}_{\uparrow}\bar{\psi}_{\downarrow} - \psi_{\downarrow} \psi_{\uparrow}\rangle_{\theta=0;\lambda}=0 \EEQ
if $\theta=0$. Thus an infinitesimally small "Cooper magnetic field" (i.e.
parameter conjugate to Cooper pair field, not to be confused with a true 
magnetic field pairing to individual electrons) in a given direction $\theta$ is enough
to orient Cooper pairs in that direction.

\Bigskip   In the following theorem, we consider the integral $X_{\Omega}$ of a Cooper pair
field over a space-time domain $\Omega$ "comparable"
to a scale $j_{\phi}$ box. Polynomial and exponential moments of $X_{\Omega}$, i.e.
 expectations $\langle P(X_{\Omega})\rangle_{\theta;\lambda}$, $\langle e^{cX_{\Omega}}\rangle_{\theta;\lambda}$  may be computed as the integral w.r. to
 a probability measure denoted by $\proba_{\theta;\lambda}$, e.g. $\langle e^{cX_{\Omega}}\rangle_{\theta;\lambda}=\int d\proba_{\theta;\lambda}(x) \, e^{cx}$. Then upper bounds for quantities $\proba_{\theta;\lambda}[X_{\Omega}< \eta]\equiv 
\int_{-\infty}^{\eta} d\proba_{\theta;\lambda}(x)$,  $\proba_{\theta;\lambda}[X_{\Omega}> \eta]\equiv 
\int_{\eta}^{+\infty} d\proba_{\theta;\lambda}(x)$, may be interpreted as  large deviation estimates.

\Bigskip 
{\bf Theorem 5 (local transverse behavior of Goldstone bosons).}\\
{\em Let $\Omega\subset\R\times\R^2$ be a domain such that $\{\xi=(\tau,\vec{x})\ |\ \Gamma_{\phi}\Big(|\tau|+\frac{m^*}{p^*_F}|\vec{x}|
\Big)\le c\}\subset\Omega\subset \{\xi=(\tau,\vec{x})\ |\ \Gamma_{\phi} \Big(|\tau|+\frac{m^*}{p^*_F}|\vec{x}|
\Big)\le C\}
$, where $0<c<C$ are constants, and  
 \BEQ X_{\Omega}:=\Re \int_{\Omega} d\xi\, (\bar{\Psi}\sigma(\theta)
 \Psi)(\xi), \EEQ
   where $\sigma(\theta):=\cos(\theta)\sigma^1 +
   \sin(\theta)\sigma^2=\left(\begin{array}{cc} 0 & e^{-\II\theta} \\
   e^{\II\theta} & 0 \end{array}\right)$ is a linear combination of off-diagonal 
   Pauli matrices, see (\ref{eq:Pauli}). Let 
\BEQ \bar{X}_{\Omega}:=\int d\mu_{\theta;\lambda} 
   \, X_{\Omega}(\theta)=|\Omega|\,   \langle (\bar{\Psi}\sigma(\theta)\Psi)(0)\rangle_{\theta;\lambda}  =2 |\Omega|\, \frac{\Gamma_{\phi}}{\lambda}(1+o(1)).
   \EEQ 
      Then, for $\eta>0$ small enough,
 \BEQ \proba_{\theta;\lambda}[|X_{\Omega}-\bar{X}_{\Omega}|>\eta
 \frac{\Gamma_{\phi}}{\lambda}\, \Vol(\Omega)]
 \lesssim e^{-g_{\perp} \eta^2} \label{eq:transverse-behavior}
\EEQ
 where 
 \BEQ g_{\perp}\approx \frac{\mu}{\Gamma_{\phi}}\approx 2^{j_{\phi}}.
 \EEQ
 }

\Bigskip This is in general agreement with mean-field estimates for the curvature of the effective potential ${\cal S}$ (see
(\ref{eq:eff-pot})) near its
minimum,  
 \BEQ {\cal S}(|\Gamma|)\sim \frac{m}{4\pi} 
\Big\{ -(\hbar\omega_D)^2 e^{-2\pi/m\lambda} + 2 (|\Gamma|-\Gamma_{\phi})^2  \Big\} \label{eq:curvature}
\EEQ
 in a neighborhood of $\Gamma_{\phi}$, 
see  \cite{FMRT-fermionic}. Namely, the field $\Gamma(\xi)$ may
be considered as roughly constant over $\Omega$. Through the 
correspondence between the Cooper pair field and its conjugate
field $\Gamma$, one may assimilate the event 
$\big\{|X_{\Omega}-\bar{X}_{\Omega}|>\eta
 \frac{\Gamma_{\phi}}{\lambda}\, \Vol(\Omega)|\big\}$ with 
 the event $\big\{ \big|\, |\Gamma(\xi)|-\Gamma_{\phi}\big|>\eta \Gamma_{\phi},
 \ \xi\in\Omega\big\}$, which has probability $\lesssim 
 e^{-\frac{m}{2\pi} \Vol(\Omega)\eta^2\Gamma_{\phi}^2} \approx 
 e^{-g_{\perp}\eta^2}$.  For larger domains $\Omega$,
the large deviation rate $-\log\Big( \proba \Big[ \int_{\Omega}d\xi\,  \big|\, |\Gamma(\xi)|-\Gamma_{\phi}\big| > \eta\Gamma_{\phi} \Vol(\Omega) \Big] \Big)$ is expected from
(\ref{eq:curvature}) to increase
linearly in $\Vol(\Omega)$.


\subsection{Outline and notations}


 The paper is organized as follows. 

\Medskip Section 1 is a long,
introductory section, where we gradually introduce notations, recall
 standard facts about the BCS model, reformulate the initial model
 as a multi-scale model, and evaluate bubble diagrams. Main
 points are  {\em Definition \ref{def:Grassmann-measure}}, where our model
 is precisely defined, and \S  \ref{subsection:Sigma}, where a
two-by-two matrix-valued kernel $\Sigma$ --  whose $(\perp,\perp)$-component is interpreted as {\em propagator of the Goldstone boson} -- and a  {\em gap equation} are 
 introduced in perturbative terms ({\em Definition \ref{def:gap}}).  Section \S \ref{subsection:Sigma} goes 
 well beyond mean-field theory, but remains perturbative; hence the
 way in which the gap energy $\Gamma_{\phi}$ and the  $\Sigma$-kernel are defined differs in details from the non-perturbative definition of section \ref{section:boson}. The precise construction in section \ref{section:boson} is
 heavily inspired from, but does not rely upon, the one in 
\S \ref{subsection:Sigma}, so details in \S \ref{subsection:Sigma} are not really important, and the whole subsection -- apart from the {\em displacement procedure}, which is
accurately described there -- may be skipped from a mathematical point of view. Yet we hope that \S \ref{subsection:Sigma} will help the reader find his way through the much more technical section \ref{section:boson}.
 
\Medskip Section 2  is dedicated to the {\em high-momentum} or {\em  fermionic theory} (part {\bf A.} of our scheme).
It relies mainly on {\em expansions}. Formally, these
apply to the {\em dressed} model, see {\em Definition \ref{def:dressing}}, which interpolates between the
coupled model of section 1 and a fully-decoupled model.  
The horizontal (cluster) and momentum-decoupling (vertical) expansions 
implement the decoupling between degrees of freedom required to
renormalize the diverging contributions, and show
convergence of the series obtained by formally expanding $e^{-{\cal V}}=\sum_{n\ge 0} (-1)^n \frac{{\cal V}^n}{n!}$. A
simple bubble resummation procedure provides the "$1/N$-argument" required to control the series in the neighborhood of the transition scale. 

\Medskip Section 3 is concerned with the {\em low-momentum} or {\em bosonic theory} (part {\bf B.}) As in QED,  Ward identities are
proved, which show that -- despite the fact that Goldstone
bosons is non-massive -- all diagrams are infra-red convergent.
By a careful complementary multi-scale expansion of Cooper pairs, $n$-point functions are rewritten in terms of a sum over multi-scale trees with vertices connected by $\Sigma_{\perp,\perp}$-kernels.
The end of the section is dedicated to the  proof of
the Theorems stated in \S \ref{subsection:results}.

\Medskip Perspectives are presented in section 4.

\Medskip Finally, some technical lemmas are collected in section 5.

\vskip 1cm\noindent
{\bf Important notations.} The reader will find many more notations in the body of the article. We hope that this selection will help the
reader find his way through the article and  prevent any confusion. A more comprehensive index of notations with reference pages is given at the end of the text.

\Medskip Following the physicists' convention, the star ($^*$)
denotes {\em complex conjugation}. Alternatively, {\em real} parameters with a star $(^*)$ denote {\em effective parameters};
Gaussian measures and covariance kernels with a star are computed with
the effective parameters.

\Medskip The   gap parameter $\Gamma_{\phi}$ is a positive
real number.

\Medskip -- Space-time points are typically denoted by 
$\xi=(\tau,\vec{x})$, $\tau\in\R$, $\vec{x}\in\R^2$.

\Medskip -- {\em Fermions} are denoted either by $\psi=(\psi_{\sigma})_{\sigma=\uparrow,\downarrow}=\left(\begin{array}{c} \psi_{\uparrow} \\
\psi_{\downarrow} \end{array}\right)$, $\bar{\psi}=(\bar{\psi}_{\sigma})_{\sigma=\uparrow,\downarrow}=\left(\begin{array}{cc} 
\bar{\psi}_{\uparrow} & \bar{\psi}_{\downarrow}\end{array}\right)$ or 
$\Psi=(\Psi_{\sigma})_{\sigma=\uparrow,\downarrow}=\left(\begin{array}{c} 
\psi_{\uparrow} \\ \bar{\psi}_{\downarrow} \end{array}\right)$, $\bar{\Psi}=(\bar{\Psi}_{\sigma})_{\sigma=\uparrow,\downarrow}= \left(\begin{array}{cc} 
\bar{\psi}_{\uparrow} & {\psi}_{\downarrow}\end{array}\right)$. The latter are {\em Nambu spinors}, which mix $\psi$ and
$\bar{\psi}$-components. Thus {\em Cooper pairings} are (depending on the
context) written as $\psi_{\downarrow}\psi_{\uparrow}, \bar{\psi}_{\uparrow}\bar{\psi}_{\downarrow}$ or $\bar{\Psi}_{\downarrow}\Psi_{\uparrow},\bar{\Psi}_{\uparrow}
\Psi_{\downarrow} $, or (in matrix notation) $\bar{\Psi} \left(\begin{array}{cc} 0 & 0\\ 1 & 0 \end{array}\right)\Psi$,
 $\bar{\Psi}\left(\begin{array}{cc} 0 & 1\\ 0 & 0 \end{array}\right)
\Psi$,  combining into $\bar{\Psi}\sigma^{1,2}\Psi$ (in terms of
Pauli matrices, see (\ref{eq:Pauli})), or more generally, using an
off-diagonal hermitian matrix,
$\bar{\Psi}{\mathbb{\Gamma}}\Psi$, where ${\mathbb{\Gamma}}:=\left(
\begin{array}{cc} 0 & \Gamma^* \\ \Gamma & 0\end{array}\right)$, with
$\Gamma\in\C$. If $\Gamma=\Gamma_{\phi} e^{\II\theta}$, resp. $\Gamma_{\phi}e^{\II(\theta+\frac{\pi}{2})}$, then we note $\Gamma\equiv \Gamma^{//}=
\Gamma(\theta)$, ${\mathbb{\Gamma}}\equiv  {\mathbb{\Gamma}}^{//}={\mathbb{\Gamma}}(\theta)$, resp. $\Gamma^{\perp}$, ${\mathbb{\Gamma}}^{\perp}$. We  also consider the complex-valued vector
$\vec{\Gamma}=\left(\begin{array}{c} \Gamma^* \\ \Gamma\end{array}\right)$, in particular,
${\bf\Gamma}^{//}$ and ${\bf\Gamma}^{\perp}$. Seeing $\Gamma$ either as a complex number, a vector or a matrix will turn
out to be useful. In particular, $\half \Tr ({\mathbb{\Gamma}}{\mathbb{\Gamma}}')=\half(\vec{\Gamma},\vec{\Gamma}')=\Re(\Gamma(\Gamma')^*)$ is
the scalar product of $\mathbb{\Gamma}$ with $\mathbb{\Gamma}'$, or indifferently $\vec{\Gamma}$ with $\vec{\Gamma}'$, or  $\Gamma$ with $\Gamma'$. All quantities are
equivariant w.r. to rotations by an angle $\theta'$, acting as follows,
\BEQ {\mathbb{\Gamma}}(\theta)\mapsto \left(\begin{array}{cc} e^{-\II\theta'} & \\ & e^{\II\theta'}
\end{array}\right) {\mathbb{\Gamma}}  \left(\begin{array}{cc} e^{\II\theta'} & \\ & e^{-\II\theta'}
\end{array}\right)={\mathbb{\Gamma}}(\theta+2\theta'), \ \ 
{\bf\Gamma}(\theta)\mapsto  \left(\begin{array}{cc} e^{-\II\theta'} & \\ & e^{\II\theta'}
\end{array}\right) {\bf\Gamma}(\theta)={\bf\Gamma}(\theta+\theta')  \label{eq:equivariant}
\EEQ
and  $
\Gamma(\theta)\mapsto \Gamma(\theta) e^{\II\theta'}=\Gamma(\theta+\theta').
$

\Medskip -- Part {\bf A.} is dedicated in a large part to a {\em multi-scale analysis}, in which power-counting is directly dependent on the
{\em vicinity to the Fermi circle}. Thus the {\em norm} on momenta,
\BEQ |p|:=\sqrt{(p^0)^2 + (\frac{p^*_F}{m^*}p_{\perp})^2} \EEQ
with $p_{\perp}:=|\vec{p}|-p^*_F$, measures the distance to
the (extended to $(1+2)$-dimensional description) Fermi circle defined by $p^0=0,|\vec{p}|=p^*_F$. Dualizing, we let 
\BEQ |\xi|:=\sqrt{|\tau|^2 + (\frac{m^*}{p^*_F} |\vec{x}|)^2}. \EEQ

\Medskip -- On the other hand, the analysis in part {\bf B.} focuses
on {\em Cooper pairs} with {\em small} total momentum $\vec{q}$, entering
into   fermionic diagrams of {\bf A.} as a small {\em transfer
momentum}. Therefore, the {\em Euclidean norm} adapted to this context,
\BEQ |q|_+:=\sqrt{(q^0)^2 +(v_{\phi} |\vec{q}|)^2}   \label{eq:qplus} \EEQ
vanishes when $q=0$. Dualizing, we let
\BEQ  |\xi|_+:=\sqrt{|\tau|^2 + (|\vec{x}|/v_{\phi} )^2}.
\label{eq:xi+} 
\EEQ  
 Note that $|p|$ and $|p|_+$ are as different
as can be. In order to further avoid confusions, we systematically denote {\em  electron momenta} by the letter $p$ and  {\em Cooper pair momenta}
by the letter $q$. On the other hand, $|\xi|\approx |\xi|_+$
may be used interchangeably most of the time.

\Medskip -- A fermion momentum $p$ has {\em scale} $j\in\{j_D,\ldots,
j'_{\phi}\}$ if 
$|p|\approx 2^{-j}\mu$. A Cooper pair momentum $q$ has
{\em Euclidean scale} $j_+$ $(j_+\ge j'_{\phi})$ if $|q|_+\approx 2^{-j_+}\mu$. 
Momentum cut-offs $\chi^j$, resp. $\chi_+^{j_+}$ select momenta
of scale $j$, resp. Euclidean scale $j_+$, namely 
$\chi^j(p)$, resp. $\chi^{j_+}(q)$ vanish except if 
$|p|\approx 2^{-j}\mu$, resp. $|q|_+\approx 2^{-j_+}\mu$. 
The sum $\chi^{j\to}:=\sum_{k\le j} \chi^k$, resp. $\chi^{\to j}:=
\sum_{k\ge j} \chi^k$ are infra-red, resp. ultra-violet cut-offs;
similarly for $\chi_+^{j_+\to}:=\sum_{k_+\le j_+}\chi_+^{k_+}$, 
resp. $\chi_+^{\to j_+}:=\sum_{k_+\ge j_+} \chi_+^{k_+}$. Infra-red,
resp. ultra-violet cut-off fermion fields are similarly denoted by
$\psi^{j\to},\bar{\psi}^{j\to},\Psi^{j\to},\bar{\Psi}^{j\to}$, resp.
$\psi^{\to j},(\bar{\psi})^{\to j},\Psi^{\to j},(\bar{\Psi})^{\to j}$.

\Medskip -- The amplitude of an amputated Feynman diagram
$\Upsilon$ is denoted by ${\cal A}(\Upsilon)$. Letting $\xi_{int,i}, i=1,\ldots,N_{int}$, resp.
$\xi_{ext,i}, i=1,\ldots,N_{ext}$ be the location  of the internal, resp.  external vertices
of $\Upsilon$, ${\cal A}(\Upsilon)$ is equal to the integral
$\big(\prod_{i=2}^{N_{ext}} \int d\xi_{ext,i})
\big(\prod_{i=1}^{N_{int}} \int  d\xi_{int,i})$ of the product
of the internal vertices by the internal propagators. Because
of translation invariance, the location $\xi_{ext,1}$ of  one of the external vertices has been kept fixed. Restricting to propagators of momentum scales $\le j$, one obtains instead
a scale $j$ infra-red cut-off amplitude denoted by ${\cal A}^{j\to}(\Upsilon)$.

\Medskip -- A {\em Cooper pair with momentum $q$} (or equivalently, a pair of electrons in
Cooper channel) is an integrated composite field 
$\int d\xi\, e^{-\II (q,\xi)} (\psi_{\downarrow}\psi_{\uparrow})
(\xi)$ or $\int d\xi\, e^{-\II (q,\xi)}
(\bar{\psi}_{\uparrow}\bar{\psi}_{\downarrow})(\xi)$ with total
momentum $q$ such that $|q|_+ \lesssim \Gamma_{\phi}$. Such fields are 
the main subject of part {\bf B.}. However, they also
contribute in an unessential way to the analysis of part {\bf A.}, since
they can enter even high-energy diagrams as external legs of fermion loops, though with a very small relative volume. By abuse of
language, we shall sometimes also speak of "Cooper pairs" even when 
the total momentum $q$ is such that  $|q|_+\gg \Gamma_{\phi}$; 
such pairs are simply unphysical quantities entering as virtual
particles into diagrams.

\Medskip -- Sign conventions regarding Grassmann integrals
are recalled in Appendix, \S \ref{subsection:Grassmann}.

\Medskip -- Finally, $N_0:=8$ is a  constant. Bounds for multi-scale polymers with $<N_0$, resp.
$\ge N_0$ external legs proceed differently due to
the complicated angular dependence of external legs.

\Bigskip {\bf \em Acknowledgements.} 
One of us (J. Magnen)  would like to express his warmest and most sincere thanks to  J. Feldman, V. Rivasseau and E. Trubowitz  for 
extended discussions on  the topics of this article and related
subjects. 


\section{The gap equation}  \label{section:gap}


This long section starts with a series of reformulations of the
initial model defined by the Grassmann measure
\BEQ d\mu_{\lambda}(\psi,\bar{\psi})=\frac{1}{{\cal Z}_{\lambda}} 
e^{-{\cal V}(\psi,\bar{\psi})} d\mu(\psi,\bar{\psi}); \EEQ
recall that $d\mu(\psi,\bar{\psi})$ is  Grassmann Gaussian
measure cut-off at some scale $j_D$, with covariance kernel  given (up to cut-off issues) by the operator
$\Big(\II p^0-e(\vec{p})\Big)^{-1}$, see (\ref{eq:A0}), and 
\BEQ {\cal V}(\psi,\bar{\psi}):=-\lambda \int \prod_{i=1}^4 dp_i\, 
\del(p_1+p_2-p_3-p_4) \bar{\psi}_{\uparrow}(p_1)\psi_{\uparrow}(p_3) \bar{\psi}_{\downarrow}(p_2)
\psi_{\downarrow}(p_4)  \label{eq:VV}
\EEQ
see (\ref{eq:V}),  again up to cut-off issues. The issue is
to motivate and define precisely the symmetry-broken Grassmann measure $d\mu_{\theta;\lambda}$ (see Theorems in the Introduction) defining the model.

\Medskip Let us describe informally the various stages:

\Medskip\textbullet\ first (see \S \ref{subsection:Nambu}), the $U(1)$-number (charge) symmetry $(\psi,\bar{\psi})\to 
(e^{\II \theta}\psi,\bar{\psi}e^{-\II\theta})$ is {\em broken}
at energies comparable to the energy gap (i.e. in Regime II), yielding  a non-zero value in 
the ground state for
products of {\em spin-neutral} annihilation operators, for instance 
\BEQ \langle (\bar{\psi}_{\uparrow} \bar{\psi}_{\downarrow}+\psi_{\downarrow}
\psi_{\uparrow})(\xi)\rangle_{\theta=0;\lambda} \approx 2\frac{\Gamma_{\phi}}{\lambda}. \label{eq:Gammamlambda}
\EEQ
This is a direct consequence of the fact that {\em Cooper pairs}, i.e. 
{\em neutral} pairs of electrons $
\bar{\psi}_{\uparrow}\bar{\psi}_{\downarrow},\  \psi_{\downarrow}\psi_{\uparrow}$ form bound states. On the other hand, non spin-neutral bound states do not form. This
motivates the introduction of {\em Nambu fields} $(\Psi,
\bar{\Psi})$. The 
well-known 
Bogolioubov-De Gennes-Gor'kov mean-field, quadratic Hamiltonian  \cite{Bog1,Bog2} (which
we do not discuss here) is  diagonalized in a
rotated basis expressed in terms of $\bar{\Psi}$, and the
mean-field ground-state is similarly obtained from the 
normal state ground-state by applying products of $\bar{\Psi}$'s, see e.g. 
\cite{AltSim}, \S 6.4.

\Medskip\textbullet\ second (see \S \ref{subsection:broken}), the Cooper pair contribution to effective action is written in the Nambu basis in terms of
an off-diagonal matrix $\int d\xi\, \bar{\Psi}(\xi)  \left( 
\begin{array}{cc} 0 & \Gamma^* \\ \Gamma & 0 \end{array}
\right) \Psi(\xi)$, where the {\em energy gap} $\Gamma_{\phi}:=|\Gamma|$
(coinciding with the above ground state expectation value)  satisfies a consistency 
equation called {\em gap equation} (discussed later on). Pauli matrices help encode the
separation of  $\Gamma$ into its real/imaginary parts. This leads to a modified Grassmann
Gaussian measure $d\mu^*_{\theta}(
\Psi,\bar{\Psi})$ with covariance kernel $C^*_{\theta}(p)$ in the form of
a rational function, whose denominator  
is bounded near the Fermi circle. Its
 inverse Fourier transform therefore decreases exponentially at large distances like  $O(e^{-c\Gamma_{\phi}|\xi|_+)}$,
with $|\xi|_+:=\sqrt{|\tau|^2+(\frac{|\vec{x}|}{v_{\phi}})^2}$. Thus the energy gap  
also plays the r\^ole of an inverse correlation length for the fermions.

\Medskip\textbullet\ further (see \S \ref{subsection:multi-scale}), the energy gap $\Gamma_{\phi}$ (as follows from the analysis in \cite{FMRT-Ward}) may be defined (neglecting
small corrections due to scales $\ge j'_{\phi}$)
as {\em minus the inverse of the off-diagonal interacting Green function} evaluated
at $(p^0,\vec{p})$ such that $p^0=0$ and $\vec{p}$ is located
on the Fermi sphere. However,
the  {\em radius} $p_F$ {\em of the Fermi sphere} -- defined as the singular
locus of the theory -- must be {\em renormalized}.  Since renormalization
is implemented \`a la Wilson as a flow of the parameters 
$m,\mu,\Gamma_{\phi}$ of the theory, we actually rewrite the model
in terms of its effective (i.e. far infra-red) parameters
$m^*:= m^{j'_{\phi}},\mu^*:= \mu^{j'_{\phi}}$ and $\Gamma_{\phi}$. This implies introducing quadratic counterterms proportional to $m^*,\mu^*,
\Gamma_{\phi}$ at
bare scale $j_D$. If $m^*,\mu^*$ have been chosen
correctly, then  counterterms $\del m^{j'_{\phi}}$, $\del \mu^{j'_{\phi}}$ for parameters $m,\mu$  vanish. Now, the flow is computed using a {\em 
multi-scale decomposition} of the fields, 
\BEQ \Psi:=\sum_{j\ge j_D} \Psi^j, \qquad \bar{\Psi}:=\sum_{j\ge j_D} \bar{\Psi}^j  \label{eq:multi-scale-dec}
\EEQ
Scale $j$ counter-terms are then  determined by
computing two-point functions of the scale $j$ 
infra-red cut-off theory  with external momenta
located on the renormalized Fermi circle. The above arguments
explain {\em why}
the precise form of the UV cut-off (\ref{eq:cut-off1}) (in
itself largely irrelevant) is not fixed from the beginning,
but rather as part of (\ref{eq:multi-scale-dec}).

\vskip 1cm

{\centerline{************************************************************
}}

\vskip 1cm

\noindent The model used as basis for discussion in the
present article is that introduced in \S \ref{subsection:multi-scale}, though (for pedagogical reasons) some of its features
are specified only in \S \ref{subsection:angular}.   As for the rest of the section, it is dedicated to the
study of the {\em gap equation}.

\Bigskip Starting with perturbation theory, we  discuss in \S \ref{subsection:bubble}
and \S \ref{subsection:gap-lowest}
{\em ladder diagrams} (see \cite{FW}, \S 11), which we call
here for obvious diagrammatic reasons {\em bubble diagrams}. 
The evaluation of these diagrams with transfer momentum close
to zero yields the main contribution to the energy gap $\Gamma_{\phi}$.
As a concession to the mean-field approximation scheme, we
resum ladder diagrams  (see \S \ref{subsection:pre-boson}) in an
intermediate step, and deduce the approximate value for 
$\Gamma_{\phi}$ known in textbooks. In our rigorous approach, the sum
of ladder diagrams plays a quintessential r\^ole  -- as we shall see,  ladder diagrams in their {\em Cooper
pairing channel} are alone responsible for the superconducting
transition, other diagrams yielding only finite corrections -- but does not
give the exact value of $\Gamma_{\phi}$.

\Bigskip  The last subsection 
(\S \ref{subsection:Sigma}) -- which lies half-way betweeen
mean-field theory and rigorous arguments -- shows  how
to re-sum  {\em perturbatively} all diagrams forming the Goldstone boson
propagator $\Sigma_{\perp,\perp}$  (the task is done beyond perturbation theory in \S \ref{subsection:complementary}). The series is formed by alternating bubble
diagrams with two-point irreducible diagrams forming the
{\em Bethe-Salpeter kernel}.  Definition \ref{def:gap} defines $\Gamma_{\phi}$ 
as the value of the parameter $\Gamma$ such that  $\Sigma_{\perp,\perp}$ diverges when $q=0$, a requirement that -- as mentioned in a Remark in the introduction -- is equivalent to
the exact gap equation.  This means that
Goldstone bosons, as expected, are not massive. The diagram series naturally forms a two-by-two, diagonal matrix kernel $\Sigma=\left(\begin{array}{cc} \Sigma_{//,//} & 0 \\ 0 &
\Sigma_{\perp,\perp}\end{array}\right)$, whose $(\perp,\perp)$ component is shown to
be {\em massive}, contrary to $\Sigma_{\perp,\perp}$.


\subsection{Green functions and Nambu formalism}  \label{subsection:Nambu}


The main quantity controlling the electron pairing is
the $2\times 2$ Green function,
\BEQ G(\tau,x;\tau',x'):=\left(\begin{array}{cc} 
\langle\psi_{\uparrow}(\tau,x) \bar{\psi}_{\uparrow}(\tau',x')\rangle &  \langle\psi_{\uparrow}(\tau,x) \psi_{\downarrow}(\tau',x')\rangle
\\ \langle\bar{\psi}_{\downarrow}(\tau,x) \bar{\psi}_{\uparrow}(\tau',x')\rangle
& \langle  \bar{\psi}_{\downarrow}(\tau,x) \psi_{\downarrow}(\tau',x') \rangle
\end{array} \right)   \label{eq:G}
\EEQ

In terms of the so-called {\em Nambu fields}, 
\BEQ \Psi(\cdot):=\left(\begin{array}{c} \psi_{\uparrow} \\
\bar{\psi}_{\downarrow} \end{array} \right), \qquad
\bar{\Psi}(\cdot):=\Big( \bar{\psi}_{\uparrow} \ \ 
\psi_{\downarrow} \Big)
\EEQ
the above matrix is simply the tensor quantity
$\langle \Psi(\tau,\vec{x})\underset{,}{\otimes} \bar{\Psi}(\tau',\vec{x}')
\rangle_c$. The particle number symmetry 
\BEQ (\psi,\bar{\psi})(\xi)\longrightarrow (e^{\II\theta}\psi,
\bar{\psi}e^{-\II\theta}) \EEQ
is broken by Cooper pairing. However, the  remaining symmetries, namely, {\em $SU(2)$ spin symmetry}
\BEQ (\psi,\bar{\psi})(\xi)\longrightarrow (g\psi,
\bar{\psi} g^{-1})(\xi), \qquad g\in SU(2), \EEQ
 invariance under  {\em $CT$ involution}
\BEQ (\psi,\bar{\psi}) \longrightarrow (\II \bar{\psi}, \II \psi) \EEQ
and spatial rotational invariance
 imply
(see \cite{FMRT-Ward}, Lemma II.1)
that $G=\left(\begin{array}{cc} G_{1,1} & G_{1,2} \\ G_{2,1} & 
G_{2,2} \end{array}\right)$ is rotationally invariant, and   (in Fourier coordinates)
\BEQ G_{1,1}(p^0,\vec{p})=\Big(G_{1,1}(-p^0,\vec{p})\Big)^*,
\ G_{2,2}(p^0,\vec{p})=-G_{1,1}(-p^0,\vec{p}) \label{eq:1.9} \EEQ
\BEQ G_{1,2}(p^0,\vec{p})=\Big(G_{2,1}(-p^0,\vec{p})\Big)^*. 
\label{eq:1.10} \EEQ 
In particular, {\em $G(p^0=0,\vec{p})$ is a Hermitian matrix
of the form $\left(\begin{array}{cc} a & b\\ b^* & -a \end{array}\right)$
with $a\in\R$.} 

\Bigskip In this language, the quadratic part of the action, see (\ref{eq:A0}),  is (up to the ultra-violet cut-off)
\BEQ {\cal A}_0(\Psi,\bar{\Psi})=\int dp \, 
\bar{\Psi}(-p) \Big(\II p^0 \mathbb{1}
-e(\vec{p})\sigma^3\Big) \Psi(p)
\EEQ
So  the  Grassmann Gaussian measure $d\mu(\Psi,\bar{\Psi})$ obtained by letting $\lambda=0$
is characterized (again, up to UV cut-off) by
\BEQ C_0:=G(\lambda=0)= \frac{1}{\II p^0 \mathbb{1}
-e(\vec{p})\sigma^3}= \left(\begin{array}{cc} 
[\II p^0-e(\vec{p})]^{-1} & 0 \\ 0 & [\II p^0+e(\vec{p})]^{-1}
\end{array} \right).
\EEQ

\medskip
Then the interaction ${\cal V}(\psi,\bar{\psi})$, see (\ref{eq:VV}),  rewrites
as ${\cal V}(\Psi,\bar{\Psi})\equiv \int_V d\xi\, {\cal V}(\Psi,\bar{\Psi};\xi)$,
with
\BEQ {\cal V}(\Psi,\bar{\Psi};\xi):=-\lambda (\bar{\Psi}\sigma^i\Psi)^2(\xi) \EEQ
where $i=1,2$ or $3$ indifferently (since the square depends only on the
determinant of the matrix inserted between $\bar{\Psi}$ and $\Psi$). Choosing
$\sigma^1, \sigma^2$ or more generally $\sigma(\theta):=\cos(\theta)\,  \sigma^1+\sin(\theta)\, \sigma^2$ emphasizes the r\^ole of Cooper pairings in a given  direction $\theta$.


\subsection{Symmetry-broken Grassman Gaussian measure}
\label{subsection:broken}


The BCS phase transition and the above symmetry considerations  suggest that  we: 

\begin{itemize}
\item[(i)] substitute to the bare dispersion relation $e(\vec{p})$ the {\bf effective dispersion relation},
\BEQ e^*(\vec{p})=\frac{|\vec{p}|^2}{2m^*}-\mu^*
\label{eq:e*(p)} \EEQ
in terms of the {\bf effective parameters} $\mu^*,m^*$ where:

  \BEQ \boxed{\del\mu:=\mu^*-\mu, \qquad |\del\mu|=O(g)\mu\ll \mu }
  \EEQ
   is the {\bf Fermi sphere radius renormalization}; and

 \BEQ \boxed{\del m:=\frac{m}{m^*}(m^*-m), \qquad |\del m|=O(g^2) m
\ll m}
 \EEQ 
 is the {\bf mass renormalization} defined
in such a way that $\frac{\del m}{m^2}=
\frac{1}{m}-\frac{1}{m^*}$; 
 
\item[(ii)] and   rewrite the functional
integral in terms of the unknown renormalized parameters
$\mu^*,m^*$ and unknown gap $\Gamma_{\phi}\not=0$ as

\BEA
&& d\mu_{\theta;\lambda}(\Psi,\bar{\Psi})= \frac{1}{{\cal Z}_{\lambda}^*}   e^{-{\cal V}(\Psi,\bar{\Psi})} \nonumber \\
&& \qquad  \exp\Big(+\int dp \, \bar{\Psi}(-p) \Big(
\Gamma_1(\theta)\sigma^1+\Gamma_2(\theta)\sigma^2+\frac{\del m}{m^2} \frac{|\vec{p}|^2}{2} +\del \mu\Big) \Psi(p) \Big)
 \ d\mu^*_{\theta}(\Psi,\bar{\Psi}), 
  \label{eq:measure}
\nonumber\\ \EEA 
where:

\textbullet\  
\BEQ \sigma^1=\left(\begin{array}{cc} 0 & 1 \\ 1 & 0 \end{array}
\right), \qquad \sigma^2=\left(\begin{array}{cc} 0 &  -\II \\ \II & 0
\end{array}\right), \qquad \sigma^3=\left(\begin{array}{cc} 1 & 0
\\ 0 & -1 \end{array}\right) \label{eq:Pauli} 
\EEQ
are the usual Pauli matrices, and $\left(\begin{array}{c}
\Gamma_1(\theta) \\ \Gamma_2(\theta) \end{array}\right)= 
\left(\begin{array}{cc} \cos \theta&\ -\sin\theta \\ 
\sin\theta & \cos\theta \end{array} \right) \left(\begin{array}{c} \Gamma_{\phi} \\ 0 \end{array}\right)=\Gamma_{\phi} \left(\begin{array}{c} 
\cos\theta\\ \sin\theta\end{array}\right)$ is a vector in $\R^2$ of
norm $\Gamma_{\phi}$ pointing in the direction $\theta$;

\textbullet\
{\em $d\mu^*_{\theta}(\Psi,\bar{\Psi})$ is the free Grassmann
measure characterized by its covariance kernel,}
\BEQ \boxed{C^*_{\theta}:= \frac{1}{\II p^0 \mathbb{1}
-e^*(\vec{p})\sigma^3-{\mathbb{\Gamma}}(\theta)}= -\frac{\II p^0 \mathbb{1} +e^*(\vec{p})
\sigma^3+{\mathbb{\Gamma}}(\theta)}{(p^0)^2+(e^*_{|\Gamma|}(\vec{p}))^2}, } \label{eq:C*}
\EEQ
and
\BEQ e^*_{|\Gamma|}(\vec{p}):=\sgn(e^*(\vec{p}))\, \sqrt{e^*(\vec{p})^2+|\Gamma|^2}. \label{eq:e*Gamma}
\EEQ
In  (\ref{eq:C*}), ${\mathbb{\Gamma}}(\theta):=\Gamma_1(\theta)\sigma^1 +
\Gamma_2(\theta)\sigma^2=\left(\begin{array}{cc} 0 & \bar{\Gamma}(\theta) \\ \Gamma(\theta) & 0\end{array}\right)$ is  an off-diagonal Hermitian matrix, 
in conformity with (\ref{eq:1.10}). 

\end{itemize}

Let us emphasize that the  measure $\exp\Big(+\int dp \, \bar{\Psi}(-p) \Big(
\Gamma_1(\theta)\sigma^1+\Gamma_2(\theta)\sigma^2+\frac{\del m}{m^2} \frac{|\vec{p}|^2}{2} +\del \mu\Big) \Psi(p) \Big) d\mu^*_{\theta}(\Psi,\bar{\Psi})$ is
by construction {\em independent} of  $\Gamma_{\phi}$ and $\theta$, and equal to the
original measure $d\mu_{\lambda}(\psi,\bar{\psi})$ of (\ref{eq:0.8}), {\em up to error terms due to the ultra-violet cut-off}.


\subsection{Multi-scale analysis}  \label{subsection:multi-scale}


Let $p_F^*:=\sqrt{2m^* \mu^*}$ be the radius of the effective
Fermi circle 
\BEQ \Sigma_F^*:=\{|\vec{p}|=p_F^*\}, \label{eq:SigmaF*}
\EEQ defined by $e^*(\vec{p})=0$ for $|\vec{p}|=p^*_F$.
Near the Fermi circle, 
 one has
\BEQ e^*(\vec{p})\sim \frac{de^*}{dp}\Big|_{p_F^*}  p_{\perp}= 
\frac{p_F^*}{m^*}  p_{\perp}  \label{eq:de/dp} \EEQ
where $ p_{\perp}:=|\vec{p}|-p_F^*$ is a signed measure of  the distance of $\vec{p}$ to the Fermi circle. 
Let $\vec{p}_D$ be some momentum such that $e^*(\vec{p}_D)\equiv \hbar \omega_D$, see Assumption (\ref{eq:cut-off2}). Given the form of the 
denominator of $C^*_{\theta}$, see (\ref{eq:C*}), it is natural to
impose the UV cut-off conditions $\max(|p^0|,e^*(\vec{p}))\lesssim
e^*(\vec{p}_D)=\hbar \omega_D$;  namely, it can
 easily be seen that the region $|p^0|\gg \hbar\omega_D\gtrsim \frac{p^*_F}{m^*}
 |p_{\perp}|$  plays practically no r\^ole in the theory.   Assumption (\ref{eq:cut-off1}) $\hbar \omega_D\ll \mu$ implies 
\BEQ  p_{D,\perp}\sim \frac{d|\vec{p}|}{de^*}\Big|_{|\vec{p}|=p_F^*} \hbar\omega_D=
\frac{m^*}{p_F^*} \hbar \omega_D \ll  \frac{2m^*}{(p_F^*)^2}\mu^* \ p_F^*=p_F^*. 
\label{eq:Domega} 
\EEQ 

\Medskip A generic point in $V\subset\R\times\R^2$ will in general be denoted by
$\xi=(\tau,\vec{x})$, or $\xi'=(\tau',\vec{x}')$, $\xi_k=(\tau_k,\vec{x}_k)$, etc. By convention $(p,\xi)=p^0\tau-\vec{p}\cdot\vec{\xi}$ is the Minkowski scalar product, while 
\BEQ  \boxed{|p|:=\sqrt{(p^0)^2+
(\frac{p_F^*}{m^*} p_{\perp})^2}, \qquad |\xi|:=\sqrt{|\tau|^2+(\frac{m^*}{p_F^*}|\vec{x}|)^2}}
\EEQ
is a homogeneous norm chosen in such a way that $|p|\sim 
e^*(\vec{p})$ if $p=(0,\vec{p})$ is close to the Fermi circle, see
(\ref{eq:de/dp}).

\Medskip {\bf Remark.} Note that $|p|$ has nothing to do with the
Euclidean norm $|p|_+\approx \sqrt{(p^0)^2+(\frac{p_F^*}{m^*}|\vec{p}|)^2}$
used in the low-momentum regime (see section \ref{section:boson}).

\Medskip Eq. (\ref{eq:Domega}) yields some upper momentum scale $j_D\gg 0$ for $p^0$ and $p_{\perp}$,
\BEQ |p^0|\lesssim 2^{-j_D} \mu, \qquad  |p_{\perp}|\lesssim 2^{-j_D} p_F, \qquad j_D:=\log_2(\mu/\hbar\omega_D).
\EEQ
Finally, we note that $e^*_{|\Gamma|}(\vec{p})\sim e^*(\vec{p})\sim \frac{p^*_F}{m^*} p_{\perp}$ for $|\Gamma|\ll  \frac{p^*_F}{m^*} |p_{\perp}|\lesssim  2^{-j_D} \mu$.
Thus
\BEQ j_{\phi}:=\lfloor \log_2(\mu/|\Gamma|) \rfloor = \ln(1/g) + O(1)
\label{eq:jc}
 \EEQ 
fixes an {\em energy scale associated to the phase transition}. On
the other hand, if $|p|\ll |\Gamma|$, then $e^*_{|\Gamma|}(\vec{p})\approx |\Gamma|$ is essentially constant; thus, when $\lambda=0$ (i.e. for the free theory),
all momenta $p$ such that $|p|\ll |\Gamma|$ may be treated as being of the same scale. 
It turns out {\em not} to be the case for the interacting theory. In order to make the
bridge between the high- and the low-momentum regimes, we fix a {\em lower} energy scale
 $j'_{\phi}\ge j_{\phi}$, 
\BEQ j'_{\phi}:=j_{\phi}+O(\ln(1/g)).
\EEQ
 In this section, we only discuss momenta $p$ such that 
\BEQ 2^{-j'_{\phi}}\mu\lesssim |p|\lesssim 2^{-j_D}\mu. \label{eq:j'c} \EEQ

\Bigskip We now define a multi-scale version of our model according to the above principles. Fix some smooth function $\chi:\R_+\to [0,1]$ supported on
$[\half,\frac{3}{2}]$ such that 
$\sum_{j\in\Z} \chi^j\equiv 1$ on $\R_+$, where 
$\chi^j(\cdot):=\chi(2^j\cdot)$.  {\em Assume} $\chi^j(\cdot)\equiv \sum_{\alpha} \chi^{j,\alpha}(\cdot)$ is further decomposed as a finite sum of
smooth functions such that $\supp(\chi^{j,\alpha})\subset\supp(\chi^j)$; the
$\chi^{j,\alpha}$ will be specified only  in \S \ref{subsection:angular}
(with $\alpha$ indexing momentum angular sectors), since it does not matter
at this stage.  Let
\BEQ \boxed{\chi^{j\to}:=\sum_{k=j_D}^j \chi^k, \qquad \chi^{\to j}:=\sum_{k=j}^{+\infty} \chi^k, \qquad j_D\le j\le j'_{\phi}} \EEQ
so that $\chi^{j\to}(|p|/\mu)\approx {\bf 1}_{|p|\gtrsim 2^{-j}\mu}$ is a scale $j$
infra-red cut-off (coupled with a scale $j_D$ UV cut-off), resp.
$\chi^{\to j}(p)\approx {\bf 1}_{|p|\lesssim 2^{-j}\mu}$ is a scale $j$ 
UV cut-off.   Define  as in \cite{FMRT-infinite} the $2\times 2$ matrix-valued kernel
\BEQ \boxed{ C^j_{\theta}(\xi,\xi'):=
\int \frac{d^3 p}{(2\pi)^3} \frac{e^{\II(p,\xi-\xi')}}{\II p^0-e^*(\vec{p})\sigma_3 -{\mathbb{\Gamma}}(\theta)} \chi^j(|p|/\mu), \qquad j_D\le j<j'_{\phi}}   \label{eq:Cj} \EEQ
\BEQ \boxed{C^{j'_{\phi}}_{\theta}(\xi,\xi'):=\int \frac{d^3 p}{(2\pi)^3} \frac{e^{\II(p,\xi-\xi')}}{\II p^0-e^*(\vec{p})\sigma_3 -{\mathbb{\Gamma}}(\theta)} \chi^{\to j'_{\phi}}(|p|/\mu)} \label{eq:Cj'c}
 \EEQ
and similarly (assuming $\chi^{\to j'_{\phi}}(\cdot)=\sum_{\alpha} \chi^{\to j'_{\phi},\alpha}(\cdot)$) 
\BEQ \boxed{ C^{j,\alpha}_{\theta}(\xi,\xi'):=
\int \frac{d^3 p}{(2\pi)^3} \frac{e^{\II(p,\xi-\xi')}}{\II p^0-e^*(\vec{p})\sigma_3 -{\mathbb{\Gamma}}(\theta)} \chi^{j,\alpha}(p/\mu), \qquad j_D\le j<j'_{\phi}}   \label{eq:Cjalpha} \EEQ
\BEQ \boxed{C^{j'_{\phi},\alpha}_{\theta}(\xi,\xi'):=\int \frac{d^3 p}{(2\pi)^3} \frac{e^{\II(p,\xi-\xi')}}{\II p^0-e^*(\vec{p})\sigma_3 -{\mathbb{\Gamma}}(\theta)} \chi^{\to j'_{\phi},\alpha}(p/\mu)}, \label{eq:Cj'calpha}
 \EEQ
and let $(\Psi^{j,\alpha})_{j_D\le j\le j'_{\phi}}, (\bar{\Psi}^{j,\alpha})_{j_D\le j\le j'_{\phi}}$ be independent free spin $1/2$ fermion fields, either on $\R^3$ or on
a finite volume
$ V$, with covariance kernel
\BEQ 
\int d\mu^*_{\theta}(\Psi^{j,\alpha},\bar{\Psi}^j_{\alpha}) \  {\Psi}^{j,\alpha}(\xi) \bar{\Psi}^{j',\alpha'}(\xi')
\equiv \del_{j,j'}\del_{\alpha,\alpha'}  C^{j,\alpha}_{\theta}(\xi,\xi').
\EEQ 
If $V$ is a finite volume, then $C^{j,\alpha}_{\theta}(\xi,\xi')$ is simply
extended outside of $V\times V$ by $0$.
Let 
\BEQ \Psi^j:=\sum_{\alpha} \Psi^{j,\alpha},\qquad \bar{\Psi}^j:=\sum_{\alpha} \bar{\Psi}^{j,\alpha}.
\EEQ
Then 
\BEQ \Psi(\cdot):=\sum_{j=j_D}^{j'_{\phi}} \Psi^j(\cdot),
\bar{\Psi}(\cdot):=\sum_{j=j_D}^{j'_{\phi}} \bar{\Psi}^j(\cdot)
 \label{eq:Psi=sumPsij} \EEQ
has covariance kernel 
\BEQ C^*_{\theta}(\xi,\xi')=\sum_{j=j_D}^{j'_{\phi}} \sum_{\alpha} C^{j,\alpha}_{\theta}(\xi,\xi')=\sum_{j=j_D}^{j'_{\phi}} C^j_{\theta}(\xi,\xi').  \label{eq:C0} \EEQ
Its Fourier transform $C^*_{\theta}(p)$ is equal to $\frac{1}{\II p^0-e^*(\vec{p})\sigma_3-{\mathbb{\Gamma}}(\theta)}$ for $|p|\ll \hbar\omega_D$.

\Medskip Summarizing:
for $j_D\le j<j'_{\phi}$, $C^j_{\theta}$ is given by an integral over momenta $p$ such that
$|p|\approx \max\Big(|p^0|,|e^*(\vec{p})|\Big)\approx 2^{-j}\mu$, or equivalently: $\max(|\vec{p}|,\frac{m^*}{p_F^*} |p^0|)\approx
2^{-j}p_F^*$.

\begin{Definition}[Grassmann measure]  \label{def:Grassmann-measure}
\begin{itemize}
\item[(i)] 
Let 
 \BEQ d\mu_{\theta;\lambda}(\Psi,\bar{\Psi})=
 \frac{1}{{\cal Z}^*_{\lambda}}    e^{- {\cal L}_{\theta}(\Psi,\bar{\Psi})}  \ d\mu^*_{\theta}(\Psi,\bar{\Psi})  
\EEQ  
where $d\mu^*_{\theta}(\Psi,\bar{\Psi})=d\mu^*_{\theta}((\Psi^j)_{j_D\le j\le j'_{\phi}},
(\bar{\Psi}^j)_{j_D\le j\le j'_{\phi}})$ is the Grassmann Gaussian measure associated
 to the fields $(\Psi^j)_{j_D\le j\le j'_{\phi}}, (\bar{\Psi}^j)_{j_D\le j\le j'_{\phi}}$ as in 
 (\ref{eq:Cj}), (\ref{eq:Psi=sumPsij}), and 
\BEQ {\cal L}_{\theta}(\cdot)\equiv \int d\xi\, {\cal L}_{\theta}(\cdot;\xi), \EEQ
\BEQ \boxed{ {\cal L}_{\theta}(\Psi,\bar{\Psi};\xi):=\lambda (\bar{\Psi}\Psi)^2(\xi) + \sum_{j=j_D}^{j'_{\phi}} \sum_{\alpha} \bar{\Psi}^{j,\alpha}(\xi) \Big(
{\mathbb{\Gamma}}(\theta)- \Big(\frac{\del m}{m^2} \frac{|\vec{\nabla}|^2}{2}+\del\mu\Big) \sigma^3 \Big)\Psi^{j,\alpha}(\xi)  }
\label{eq:Grassmann-measure} 
\EEQ

\item[(ii)] (same with scale $j$ infra-red cut-off)
For $j_D\le j\le  j'_{\phi}$, we let
\BEQ   d\mu_{\theta;\lambda}(\Psi^{j\to},\bar{\Psi}^{j\to})=
 \frac{1}{{\cal Z}_{\lambda}^{j\to}}   e^{- {\cal L}_{\theta}(\Psi^{j\to},\bar{\Psi}^{j\to})}  \ d\mu^*_{\theta}(\Psi^{j\to},\bar{\Psi}^{j\to})  
\EEQ  
where $d\mu^*_{\theta}(\Psi^{j\to},\bar{\Psi}^{j\to})=d\mu^*_{\theta}((\Psi^k)_{j_D\le k\le j},
(\bar{\Psi}^k)_{j_D\le k\le j})$ is the Grassmann Gaussian measure associated
 to the fields $(\Psi^k)_{j_D\le k\le j},
  (\bar{\Psi}^k)_{j_D\le k\le j}$ as in 
 (\ref{eq:Cj}), (\ref{eq:Psi=sumPsij}), and 
\BEQ {\cal L}_{\theta}^{\to j}(\cdot)\equiv \int d\xi\, {\cal L}_{\theta}^{\to j}(\cdot;\xi), \EEQ
\BEQ  \boxed{{\cal L}_{\theta}^{\to j}(\Psi^{j\to},\bar{\Psi}^{j\to};\xi):=\lambda (\bar{\Psi}^{j\to}\Psi^{j\to})^2(\xi) + \sum_{k=j_D}^j \sum_{\alpha} \bar{\Psi}^{k,\alpha}(\xi) \Big(
{\mathbb{\Gamma}}(\theta)-\Big(\frac{\del m}{m^2} \frac{|\vec{\nabla}|^2}{2}+\del\mu\Big)\sigma^3 \Big)\Psi^{k,\alpha}(\xi)  }
\EEQ

\end{itemize}
\end{Definition}
The whole point in introducing the $\Psi^{j,\alpha}$'s at this very early stage 
is that  the quadratic part ${\cal L}_{\theta}$ in the above Definition is 
diagonal in that basis.

\Medskip
Up to change of normalization constant ${\cal Z}^*_{\lambda}$, 
$d\mu_{\lambda}(\Psi^{j\to},\bar{\Psi}^{j\to})$ is simply obtained from
$d\mu(\Psi,\bar{\Psi})$ by letting $\Psi^k,\bar{\Psi}^k\equiv 0$
for $k>j$. In practice, this  $j$-dependent measure will result from the  momentum-decoupling expansion and
renormalization by replacing $\Psi,\bar{\Psi}$ with
$\Psi(t^j;\cdot):=\sum_{k=j_D}^j \Psi^k(\cdot) + t^j 
\sum_{k>j} \Psi^k(\cdot)$, $\bar{\Psi}(t^j;\cdot):=\sum_{k=j_D}^j \bar{\Psi}^k(\cdot) + t^j 
\sum_{k>j} \bar{\Psi}^k(\cdot)$ and letting $t^j=0$.
It is the right place to mention that these infra-red cut-off
theories enjoy the same invariance properties as the
original theory, see \S \ref{subsection:Nambu}, which is
crucial for the renormalization step (see \S \ref{subsection:cluster}).


\subsection{Bubble diagrams and fermion four-point function in Coooper channel}       \label{subsection:bubble}


We primarily consider in this subsection the simplest Feynman diagrams
of the theory -- the {\em bubble diagrams}.  Feynman rules
can be found in Chapter 46 of the book by Fetter and Walecka \cite{FW}. The amplitude of an
amputated bubble diagram $\Upsilon$, computed using the free Grassmann measure $C^*_{\theta}$, is denoted in whole generality
 by ${\cal A}_q(\Upsilon)$, where $q$ is the external momentum flowing into
 the diagram, or ${\cal A}_q(\Gamma,\Upsilon_3)$ if one 
 wishes to emphasize the dependence on the parameter $\Gamma$. Replacing $C^*_{\theta}$ with $(C^*_{\theta})^{j\to}:=\sum_{k=j_D}^j 
 C^k_{\theta}$, one obtains the scale $j$ infra-red cut-off evaluation
 ${\cal A}_q^{j\to}(\Upsilon)$.

\Medskip  The symmetry-broken theory with gap parameter $\Gamma\in\C$ involves

\textbullet\ {\bf diagonal} or {\bf normal propagators}, 
 $\propagator{\bar{\psi}_{\sigma_1}}{\psi_{\sigma_2}}$
evaluated as $-\del_{\sigma_1,\sigma_2} \frac{\II p^0+e^*(\vec{p})}{(p^0)^2+(e^*_{|\Gamma|}(\vec{p}))^2}$.  Graphically, we orient such propagators using arrows from $\bar{\psi}$ to $\psi$, following the convention of 
e.g. quantum electrodynamics, in order to follow fermion loops;

\textbullet\ {\bf  off-diagonal} or {\bf symmetry-broken propagators},

\Bigskip

{\centerline{\begin{tikzpicture}[scale=0.7] 
\draw(0,0) -- (3,0);
\draw(0.3,0.4) node {$\psi_{\sigma_1}$};
\draw(2.7,0.4) node {$\psi_{\sigma_2}$};
\draw(1.3,0.2)--(1.7,-0.2);\draw(1.3,-0.2)--(1.7,0.2);
\draw(3,0) node {$,$};
\begin{scope}[shift={(6,0)}]
\draw(0,0) -- (3,0);
\draw(0.3,0.4) node {$\bar{\psi}_{\sigma_1}$};
\draw(2.7,0.4) node {$\bar{\psi}_{\sigma_2}$};
\draw(1.3,0.2)--(1.7,-0.2);\draw(1.3,-0.2)--(1.7,0.2);
\end{scope}
\end{tikzpicture}}}

\medskip evaluated as $-\del_{\sigma_1,\sigma_2} \frac{\Gamma^*(\theta)}{(p^0)^2+(e^*_{|\Gamma|}(\vec{p}))^2}$, resp. $-\del_{\sigma_1,\sigma_2} \frac{\Gamma(\theta)}{(p^0)^2+(e^*_{|\Gamma|}(\vec{p}))^2}$. Such propagators are non-oriented.
Because we introduced them by hand in the Gaussian measure,
they exist even in Regime I, i.e. above the gap energy,
but they are very {\em small} for scales $j\ll j_{\phi}$ due to the coefficient
$|\Gamma|=\Gamma_{\phi}=O(2^{-j_{\phi}})$ in the numerator.
 
The names {\em diagonal} and {\em off-diagonal} refer to the location of these
quantities inside the two-by-two free Green function $G\Big|_{\lambda=0}$, see
(\ref{eq:G}).

{\bf \textbullet\  $\Gamma$-counterterms} \begin{tikzpicture}    \draw(-1,0) node {$-$}; \draw(0,0)--(-0.3,0.3); \draw(0,0)--(-0.3,-0.3); \draw[fill=black]
(-0.1,-0.1) rectangle(0.1,0.1); \draw(0.6,0) node {$\Gamma$};
\end{tikzpicture}  coming from the term $\int dp \bar{\Psi}(-p)
{\mathbb{\Gamma}}(\theta) \Psi(p)$ in (\ref{eq:measure}).

\Medskip In subsequent computations, we rewrite diagonal propagators as follows,
\BEQ \boxed{ \frac{\II p^0+e^*(\vec{p})}{(p^0)^2+(e^*_{|\Gamma|}(\vec{p}))^2}=
\frac{1}{-\II p^0 + e^*_{|\Gamma|}(\vec{p})} -\frac{\del e^*(\vec{p})}{(p^0)^2+
(e^*_{|\Gamma|}(\vec{p}))^2}}   \label{eq:dele} \EEQ
with 
\BEQ \del e^*(\vec{p}):=e^*_{|\Gamma|}(\vec{p})-e^*(\vec{p})\sim  \begin{cases}\half 
\frac{p_F^*}{m^*} (\frac{\Gamma_{\phi}}{p_{\perp}})^2 p_{\perp} \qquad 
(|p_{\perp}|\gg \Gamma_{\phi}) \\ \frac{\Gamma_{\phi}}{|p_{\perp}|}\, p_{\perp} \qquad 
(|p_{\perp}|\lesssim \Gamma_{\phi}) .  \end{cases} \label{eq:approx-dele} \EEQ
 The integration measure $dp=dp^0 d\vec{p}$ for angle-independent quantities is of the two-dimensional form
$2\pi |\vec{p}|\, dp^0\, dp_{\perp}\sim 2\pi \, p_F \, dp^0\, dp_{\perp}$. Thus
the theory is {\em effectively $1+1$-dimensional} in the neighborhood of the Fermi sphere;
this remarkable fact actually holds true for any dimension $d\ge 2$.

\Bigskip {\bf Vacuum polarization bubbles.}
In an intermediate boson picture, where density fields
$\bar{\psi}_{\downarrow}\psi_{\downarrow}$, $\bar{\psi}_{\uparrow}\psi_{\uparrow}$ interact via a delta-potential
represented by a wavy line, such bubbles may be
represented as 

\bigskip

{\centerline{$\vacuum$}}

\medskip
\noindent yielding a renormalization of the intermediate boson.

\Medskip Bubble diagrams are evaluated without taking
into account the usual $(-1)$-factor per loop, characteristic
of fermionic theories. One obtains {\em three possibilities:}

\Bigskip \textbullet\  The {\em normal  $(\psi,\bar{\psi})$-bubble diagram} $\Upsilon_1$

\medskip
$\bubble{\bar{\psi}_{\downarrow}}{\psi_{\downarrow}}{\psi_{\downarrow}}{\bar{\psi}_{\downarrow}}
{\bar{\psi}_{\uparrow}}{\psi_{\uparrow}}{\psi_{\uparrow}}
{\bar{\psi}_{\uparrow}}$

{\tiny\bf  Fig. \ref{fig:bubble}.1. Normal bubble diagram $\Upsilon_1$.}

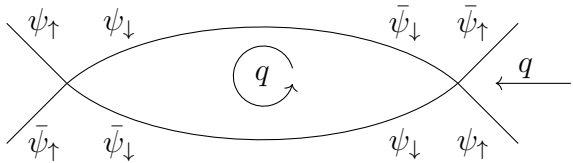
\Bigskip is evaluated up to error terms (see (\ref{eq:dele}) and below) as
\BEQ {\cal A}_q^{j\to}(\Upsilon_1)=-\int \frac{ \chi^{j\to}(|p+q|/\mu) \chi^{j\to} (|p|/\mu)\, dp}{\Big(\II (p^0+q^0) +e^*_{|\Gamma|}(
\vec{p}+\vec{q})\Big)\Big(\II p^0 +e^*_{|\Gamma|}(
\vec{p})\Big)}
\EEQ

A naive power-counting argument in terms of the integration measure, also valid
for the Cooper-pair bubble $\Upsilon_{3,diag}$ below,  yields
\BEQ |{\cal A}_0^{j\to}(\Upsilon_1)|\lesssim \int dp^0\, d\vec{p}
\, \frac{(\chi^{j\to}(p))^2}{|p|^2}
\sim 2\pi\, p_F\, \int_{2^{-j}\mu<|p|<2^{-j_D}\mu} \frac{dp^0\, dp_{\perp}}{|p|^2}\sim m^* \int_{2^{-j}\mu}^{2^{-j_D}\mu} \frac{d|p|}{|p|}\approx \sum_{j=j_D}^{j_{\phi}} m^*,  \label{eq:naive-log}
\EEQ
 a logarithmically divergent integral.
However, $ {\cal A}_0^{j\to}(\Upsilon_1)$ is UV convergent, as can be seen from a contour deformation into
the upper/lower half-plane
of the integral in $p^0$.  Namely, 
$\int \frac{dp^0}{(\II p^0+e^*_{|\Gamma|}(\vec{p}))^2}=\II 
\int dp^0 \, \frac{d}{dp^0} \Big(\frac{1}{\II p^0+e^*_{|\Gamma|}(\vec{p}))}\Big)=0$, so (by integration by parts in $p^0$ and parity
in $p^0$)

\BEQ {\cal A}_0^{j\to}(\Upsilon_1)=\II \int \frac{ \frac{d}{dp^0}( (\chi^{j\to} (|p|/\mu))^2)\, dp}{\II p^0 +e^*_{|\Gamma|}(
\vec{p}) }.
\EEQ
Now, $\frac{d}{dp^0} (\chi^{j\to} (|p|/\mu))$ vanishes except for momenta $p$ of 
scales $j_D+O(1)$ or $j+O(1)$, which thus suppresses the diverging sum
over scales.
 Generalizing, we let $q$ be a small transfer momentum,  $|q|_+\ll 2^{-j}\mu$, and subtract ${\cal A}_0^{j\to}(\Upsilon_1)$, yielding
 \BEQ {\cal A}_q^{j\to}(\Upsilon_1)= \int dp\,  \frac{ \II q^0 + (e^*_{|\Gamma|}(\vec{p}+\vec{q})-e^*_{|\Gamma|}(\vec{p})) }{\Big(\II (p^0+q^0) +e^*_{|\Gamma|}(
\vec{p}+\vec{q})\Big)\Big(\II p^0 +e^*_{|\Gamma|}(
\vec{p})\Big)^2} \ \, \chi^{j\to} (|p|/\mu)\,  \chi^{j\to}(|p+q|/\mu))
\EEQ
with $q^0, e^*_{|\Gamma|}(\vec{p}+\vec{q})-e^*_{|\Gamma|}(\vec{p}) 
=O(\frac{|q|_+}{|p|}) |p|$.
Restricting the integration domain to
momenta $p$ of scale $k$  and  moving the  contour in $p^0$ to a distance $\approx 2^{-k} \mu$
from the real axis yields   
${\cal A}_q^{j\to}(\Upsilon_1)=\sum_{k=j_D}^j {\cal A}^k_q(\Upsilon_1)$, with ${\cal A}^k_q(\Upsilon_1)=m^*\, O(\frac{|q|_+}{2^{-k}\mu})$, summing
to $O(1)$.

\smallskip\noindent The same bound may also be obtained  by noting 
that $\II (p^0+q^0)-e^*_{|\Gamma|}(\vec{p}+\vec{q})\simeq \II (p^0+q ^0) -
\frac{p_F^*}{m^*} (p_{\perp}+q_{\perp})$, $q_{\perp}:=\frac{\vec{q}\cdot 
\vec{p}}{|\vec{p}|}$,  and 
\BEA &&  (\II q^0-\frac{p^*_F}{m^*} q_{\perp})\ \frac{1}{\Big(\II (p^0+q^0) -
\frac{p_F^*}{m^*} (p_{\perp}+q_{\perp}) \Big) \Big(\II p^0 -\frac{p_F^*}{m^*} p_{\perp})\Big)}
\nonumber\\
&&\qquad =  \frac{1}{\II p^0-\frac{p_F^*}{m^*} p_{\perp}} \, -\,  \frac{1}{\II (p^0+q^0) 
-\frac{p_F^*}{m^*} (p_{\perp}+q_{\perp}) }
\EEA
which vanishes by translation invariance upon integration in
$p$ up to cut-off issues; we skip details since we do not need
a second argument, but refer the reader to  Ward identities in
quantum electrodynamics which can be proved in a similar way,
see e.g. \cite{PesSch}. 

\Medskip {\em Error terms} involve the subtacted terms, $\frac{\del e^*(\vec{p})}{(p^0)^2+(e^*_{|\Gamma|}(\vec{p}))^2}$ or $\frac{\del e^*(\vec{p}+\vec{q})}{(p^0+q^0)^2+(e^*_{|\Gamma|}(\vec{p}+\vec{q}))^2}$ or both.  However, the  small factor $\frac{\Gamma_{\phi}}{|p_{\perp}|}$ appearing in (\ref{eq:approx-dele}) 
transforms the logarithmically divergent estimates (\ref{eq:naive-log})
into a convergent integral; the same argument holds for $\Upsilon_3$ below.

\Bigskip
\textbullet\ The {\em off diagonal} or {\em symmetry-broken $(\psi,\bar{\psi})$-bubble diagrams} $\Upsilon_{3,off},\Upsilon'_{1}$

\Bigskip

{\centerline{\begin{tikzpicture}[scale=1] 
\draw(0,0) arc(30:150:3  and 1.5);
\draw(-5.2,0) arc(210:330:3 and 1.5);
\draw(-2.6-0.2,0.75+0.2)--(-2.6+0.2,0.75-0.2); \draw(-2.6-0.2,0.75-0.2)--(-2.6+0.2,0.75+0.2);
\draw(-2.6-0.2,-0.75+0.2)--(-2.6+0.2,-0.75-0.2);\draw(-2.6-0.2,-0.75-0.2)--(-2.6+0.2,-0.75+0.2);
\draw (-5.2,0) -- (-6,0.8);
\draw (-5.2,0) -- (-6,-0.8);
\draw(0,0) -- (0.8,0.8);
\draw(0,0) -- (0.8,-0.8);
\draw(-4.5,-0.8) node {$\psi_{\downarrow}$};
\draw(-0.7,-0.8) node {$\psi_{\uparrow}$};
\draw(-4.5,0.8) node {$\psi_{\uparrow}$};
\draw(-0.7,0.8) node {$\psi_{\downarrow}$};
\draw(-5.5,-0.8) node {$\bar{\psi}_{\downarrow}$};
\draw(0.2,-0.8) node {$\bar{\psi}_{\downarrow}$};
\draw(-5.5,0.8) node {$\bar{\psi}_{\uparrow}$};
\draw(0.2,0.8) node {$\bar{\psi}_{\uparrow}$};
\draw[<-] (0.5,0) -- (1.5,0);
\draw(0.9,0.2) node {$q$};
\begin{scope}[shift={(8,0)}]
\draw(0,0) arc(30:150:3  and 1.5);
\draw(-5.2,0) arc(210:330:3 and 1.5);
\draw(-2.6-0.2,0.75+0.2)--(-2.6+0.2,0.75-0.2); \draw(-2.6-0.2,0.75-0.2)--(-2.6+0.2,0.75+0.2);
\draw(-2.6-0.2,-0.75+0.2)--(-2.6+0.2,-0.75-0.2);\draw(-2.6-0.2,-0.75-0.2)--(-2.6+0.2,-0.75+0.2);
\draw (-5.2,0) -- (-6,0.8);
\draw (-5.2,0) -- (-6,-0.8);
\draw(0,0) -- (0.8,0.8);
\draw(0,0) -- (0.8,-0.8);
\draw(-4.5,-0.8) node {$\bar{\psi}_{\uparrow}$};
\draw(-0.7,-0.8) node {$\bar{\psi}_{\downarrow}$};
\draw(-4.5,0.8) node {$\psi_{\downarrow}$};
\draw(-0.7,0.8) node {$\psi_{\uparrow}$};
\draw(-5.5,-0.8) node {$\bar{\psi}_{\downarrow}$};
\draw(0.2,-0.8) node {$\bar{\psi}_{\uparrow}$};
\draw(-5.5,0.8) node {${\psi}_{\uparrow}$};
\draw(0.2,0.8) node {${\psi}_{\downarrow}$};
\draw[<-] (0.5,0) -- (1.5,0);
\draw(0.9,0.2) node {$q$};
\end{scope}
\end{tikzpicture}}}

\Medskip
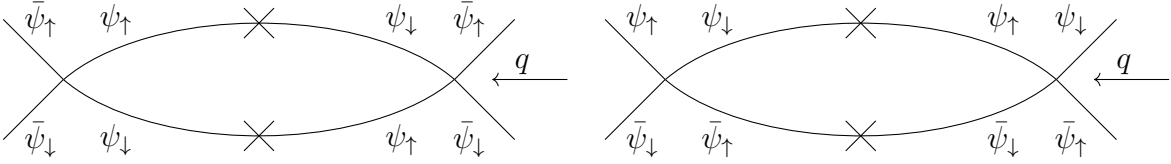
{\tiny \bf Fig. \ref{subsection:bubble}.2. Symmetry broken bubble diagrams $\Upsilon_{3,off}$, resp. $\Upsilon'_{1}$ (from left to right).}

\Bigskip are evaluated at zero external momentum as 
\BEQ {\cal A}_0^{j\to}(\Gamma,\Upsilon_{3,off})=-\bar{\Gamma}^2 \int 
\frac{\chi^{j\to}(|p|/\mu)  dp}{\Big(p_0^2+(e^*_{|\Gamma|}(\vec{p}))^2\Big)^2}, \qquad 
{\cal A}_0^{j\to}(\Upsilon'_{1})=|\Gamma|^2 \int 
\frac{\chi^{j\to}(|p|/\mu)  dp}{\Big(p_0^2+(e^*_{|\Gamma|}(\vec{p}))^2\Big)^2}
\label{eq:A0Upsilon3off}
\EEQ 
which are equal by simple scaling to a constant times $(\frac{\Gamma}{\mu})^2$,
resp. $(\frac{|\Gamma|}{\mu})^2$ times  $2^{2j} m^* (1+O((2^{-(j-j_D)})^2))$. Letting
$|\Gamma|\equiv \Gamma_{\phi}\approx 2^{-j_{\phi}}\mu$ and  summing
over scales $j_D\le j\le j_{\phi}$ (in this computation
 one may simply set $j'_{\phi}=j_{\phi}$), one gets 
\BEQ O(m^*)\sum_{j\le j_{\phi}} 2^{-2(j_{\phi}-j)}
=O(m^*) \label{eq:eval-Upsilon2},
\EEQ a finite contribution.

\Medskip The diagram $\Upsilon_{3,off}$ has a {\em Cooper pair external structure}; it
connects a Cooper pair $\bar{\psi}_{\uparrow}\bar{\psi}_{\downarrow}$ to another one. Later on (see \S \ref{subsection:Sigma}), following the Nambu convention, see
(\ref{eq:G}),  we shall denote with a lower index
"diag" diagrams connecting a Cooper pair $\bar{\psi}_{\uparrow}\bar{\psi}_{\downarrow}$ to the adjoint pair ${\psi}_{\downarrow}{\psi}_{\uparrow}$ or
conversely, and
"off" diagrams connecting two Cooper pairs of the same kind. This accounts for
the notation $\Upsilon_{3,off}$. On the other hand, $\Upsilon'_{1}$ is
a companion diagram to $\Upsilon_1$, connecting two conventional pairs
$\bar{\psi}\psi$.
\Medskip Rewriting the product of two vertices
\BEQ \big\{ (\bar{\psi}_{\uparrow}\psi_{\uparrow})(\xi)
(\bar{\psi}_{\downarrow}\psi_{\downarrow})(\xi)\big\} 
\  \big\{ (\bar{\psi}_{\uparrow}\psi_{\uparrow})(\xi')
(\bar{\psi}_{\downarrow}\psi_{\downarrow})(\xi')\big\} 
\label{eq:two-vertex}
\EEQ
in the form $(\bar{\psi}_{\uparrow}\bar{\psi}_{\downarrow})(\xi)\ \big\{ (\psi_{\uparrow}(\xi)\psi_{\downarrow}(\xi'))\, \cdot\, (\psi_{\downarrow}(\xi)\psi_{\uparrow}(\xi')) \big\}\ (\bar{\psi}_{\uparrow}\bar{\psi}_{\downarrow})(\xi')$,
one sees that ${\cal A}^{j\to}(\Gamma,\Upsilon_{3,off})$ is
a contribution to the off-diagonal channel $(\bar{\psi}_{\uparrow}\bar{\psi}_{\downarrow})\otimes (\bar{\psi}_{\uparrow}\bar{\psi}_{\downarrow})$, equal to $e^{-2\II\theta}$
times a quantity which is {\em negative}, equal to 
$-|{\cal A}_0^{j\to}(\Gamma,\Upsilon_{3,off}))|$, when $q=0$.

\Bigskip
\textbullet\ The main diagram in  BCS theory is the {\em Cooper pair bubble $\Upsilon_{3,diag}$},  \label{Cooper-pair-bubble}

\medskip

$\Cooperbubble{\psi_{\downarrow}}{\bar{\psi}_{\downarrow}}{\psi_{\uparrow}}{\bar{\psi}_{\uparrow}}{\bar{\psi}_{\downarrow}}{\psi_{\downarrow}}{\bar{\psi}_{\uparrow}}{\psi_{\uparrow}}$

\Medskip {\tiny \bf Fig. \ref{fig:Cooperbubble}.3. Cooper pair bubble.}

\Bigskip
{\em Remark. Ladder diagrams.}
In the intermediate boson picture as above, such bubbles take the form

\medskip
{\centerline{$\oneladder$}}
\Bigskip
where wiggling lines \begin{tikzpicture} 
\draw[decorate,decoration=snake] (0,0)--(1,0); 
\draw(-0.3,0) node {$\xi$}; \draw(1.3,0) node {$\xi'$};
\end{tikzpicture} represent the $\del$-interaction $\del(\xi-\xi')$,
and may be concatenated in the form of ladders of arbitrary length,

\medskip
{\centerline{$\ladder$}}

\Medskip {\bf \tiny Fig. \ref{fig:ladder}.4. Ladder diagram.}

\Bigskip
conventionally called {\em ladder diagrams} in connection to
the Bethe-Salpeter kernel, see \cite{FW}, or equivalently in
the following dual form,

\bigskip

{\centerline{
\begin{tikzpicture}[scale=1.5]  
\begin{scope}[shift={(-1,0)}]
\draw[decorate,decoration=snake](-0.5,-0.9)--(0.5,-0.9);
\draw(2,-0.9) arc(30:150:0.88  and 0.44);
\draw(0.5,-0.9) arc(210:330:0.88 and 0.44);
\end{scope}
\draw[decorate,decoration=snake](1,-0.9)--(2,-0.9);
\draw(2+1.5,-0.9) arc(30:150:0.88  and 0.44);
\draw(0.5+1.5,-0.9) arc(210:330:0.88 and 0.44);
\draw[dashed](3.5,-0.9)--(4.3,-0.9);
\draw[dashed](3.9,-0.9) ;
\draw(2+3.8,-0.9) arc(30:150:0.88  and 0.44);
\draw(0.5+3.8,-0.9) arc(210:330:0.88 and 0.44);
\draw[decorate,decoration=snake](2+3.8,-0.9)--(3+3.8,-0.9);
\end{tikzpicture}
}}

\Medskip {\bf\tiny Fig. \thesubsection.5. Bubble chain.}

\Bigskip The above Cooper pair bubble diagram is evaluated up to convergent
error terms as
\BEQ {\cal A}_q^{j\to} (\Gamma,\Upsilon_{3,diag})=\frac{1}{(2\pi)^3} \int dp\,  \frac{ \chi^{j\to} (|p|/\mu) \chi^{j\to}(|p+q|/\mu) }{\Big(\II (p^0+q_0) -e^*_{|\Gamma|}(
\vec{p}+\vec{q})\Big)\Big(-\II p^0 -e^*_{|\Gamma|}(
\vec{p})\Big)}
\EEQ

\Medskip It is UV divergent: at zero external momenta,
\BEA  {\cal A}_0(\Gamma,\Upsilon_{3,diag}) &=& \frac{1}{(2\pi)^3} \int dp \, |C^*_{\theta}(p)|^2 \nonumber\\
&=& \frac{1}{(2\pi)^3} \int \frac{dp}{| \II p^0 -e^*_{|\Gamma|}(
\vec{p})|^2}  + O(m^*) \label{eq:approx-formula-A0-3} \nonumber\\
&=& \frac{1}{(2\pi)^2} \int \frac{d\vec{p}}{|e^*_{|\Gamma|}(\vec{p})|} + O(m^*)  \label{eq:A0-3-vecp}
\nonumber\\
&= & \frac{1}{2\pi} p^*_F \int_{-(2m^*/p^*_F)\hbar\omega_D}^{(2m^*/p^*_F)\hbar\omega_D} \frac{dp_{\perp}}{\sqrt{(\frac{p^*_F}{2m^*})^2 p_{\perp}^2 +|\Gamma|^2}}  \Big(1+\frac{m^*}{p_F^*} O(\frac{p_{\perp}}{p_F^*})\Big)\nonumber\\
&= & \frac{m^*}{\pi}  \Big( \sinh^{-1} \frac{\hbar \omega_D}{|\Gamma|} +O(1)\Big) \nonumber\\ 
&=&  \frac{m^*}{\pi}
\Big(  \log (\hbar\omega_D/|\Gamma|)+O(1)\Big). \label{eq:sinh-1}
\EEA

\Medskip Rewriting (\ref{eq:two-vertex}) in the form $(\bar{\psi}_{\uparrow}\bar{\psi}_{\downarrow})(\xi)\ \big\{ (\psi_{\uparrow}(\xi)\bar{\psi}_{\uparrow}(\xi'))\, \cdot\, (\psi_{\downarrow}(\xi)\bar{\psi}_{\downarrow}(\xi')) \big\}\ (\psi_{\downarrow}\psi_{\uparrow})(\xi')$, one sees
that ${\cal A}_q(\Gamma,\Upsilon_{3,diag})$  is a contribution
to the diagonal channel $(\bar{\psi}_{\uparrow}\bar{\psi}_{\downarrow})\otimes (\psi_{\downarrow}\psi_{\uparrow})$,
which is {\em positive} when $q=0$.

\Medskip 
This quantity diverges logarithmically  in the neighbourhood of the Fermi circle when $\Gamma=0$.  The estimates
(\ref{eq:sinh-1}) also holds for ${\cal A}_0^{j\to}(\Gamma,\Upsilon_{3,diag})$ 
when $|\Gamma|\approx 2^{-j_{\phi}}\mu$ and $j \ge j_{\phi}+O(1)$. On the other hand, if 
$2^{-j}\mu\gg |\Gamma|$, then the above integral has an infra-red cut-off
of scale $j$, so that 
\BEQ {\cal A}_0^{j\to}(\Gamma,\Upsilon_{3,diag})\sim \frac{m^*}{\pi}(j-j_D) \label{eq:A0j3} 
\EEQ
is essentially independent of $\Gamma$.

\Medskip Both the transfer momentum $q$ and the infra-red
cut-off at fermionic scale $j$ play the r\^ole of an infra-red
cut-off; the resulting effective IR cut-off fermionic scale
is $k\simeq \min\Big(j,\lfloor \log(\mu/|q|_+)\rfloor\Big)$.
In particular,
\BEQ {\cal A}_q^{j\to}(\Gamma,\Upsilon_{3,diag})\sim
 {\cal A}_0^{j\to}(\Gamma,\Upsilon_{3,diag}) \sim
 \frac{m^*}{\pi}(j-j_D) \label{eq:Aqj3} 
\EEQ
if the transfer momentum is small, i.e. $|q|_+\lesssim
2^{-j}\mu$. 

\Medskip On the other hand, as proved in Appendix, see \S \ref{subsection:anisotropic},  if $ |q|_+\gtrsim 2^{-j} \mu$, 
\BEQ |{\cal A}_q^{j\to}(\Gamma,\Upsilon_{3,diag})- {\cal A}_0^{k\to}(\Gamma,\Upsilon_{3,diag})
|\lesssim  m^*, \qquad k:=\lfloor \log(\mu/|q|_+)\rfloor
\label{eq:Aq3}
\EEQ 
so ${\cal A}^{j\to}_q(\Gamma,\Upsilon_{3,diag})$ is roughly equal to the evaluation
of  a bubble with zero transfer momentum and
infra-red cut-off $\approx |q|_+$; the reason is that (say, for $q^0=0$),
the integrand in $\vec{p}$ (a generalization of (\ref{eq:A0-3-vecp}))  vanishes when $\vec{p}$ and $\vec{p}+\vec{q}$ are
not on the same side of the Fermi sphere. 

\Medskip It is a remarkable fact
{\em per se} that, due to symmetry-breaking,  the above integral (\ref{eq:sinh-1}) has an effective cut-off,
making the diagram finite.


\subsection{The gap equation at lowest order} \label{subsection:gap-lowest}


We consider here only the Cooper pair bubble diagram $\Upsilon_{3,diag}$ and 
interpret the gap equation as a self-consistent identity ensuring the vanishing
of the one-point function in the direction $\theta$.

\Medskip We first {\em remark} that, as an immediate
consequence of (\ref{eq:C*}), 
\BEQ \lambda \int dp\, (C^*_{\theta})_{12}(p) \simeq \lambda \Gamma \int dp\, |C^*_{\theta}(p)|^2 =\lambda\Gamma {\cal A}_0(\Upsilon_{3,diag})
 \EEQ
with error terms coming from $\del e^*(\vec{p})$, see (\ref{eq:dele}), 
graphically,

\Medskip
{\centerline{ $\onepointequaltwopoint$}}

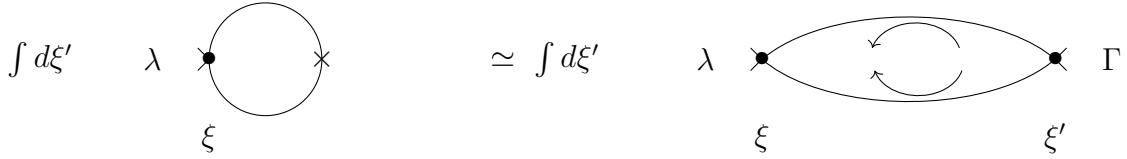
\Medskip {\tiny \bf Fig. \ref{fig:onepointequaltwopoint}.1. Off-diagonal covariance is  equal to Cooper pair bubble diagram.}

\Bigskip The diagram on the left-hand side may be interpreted as  {\em one} of the two {\em   main contributions
to the one-point function of the Goldstone boson}, the other coming from the counterterm insertion
$-\Gamma (\psi_{\downarrow}\psi_{\uparrow})(\xi)$ at $\xi$, \qquad
\begin{tikzpicture}    \draw(-1,0) node {$-$}; \draw(0,0)--(-0.3,0.3); \draw(0,0)--(-0.3,-0.3); \draw[fill=black]
(-0.1,-0.1) rectangle(0.1,0.1); \draw(0.6,0) node {$\Gamma$};
\end{tikzpicture}
\qquad . Then these two main contributions cancel provided
\BEQ \boxed{\lambda {\cal A}_0(\Gamma,\Upsilon_{3,diag})\simeq 1} \label{eq:pre-Gap} \EEQ
or equivalently, taking (\ref{eq:A0j3}) into account,
\BEQ \boxed{j_{\phi}-j_D=\frac{\pi}{g}+O(1).} \label{eq:pre-Gap-bis} \EEQ
This is the crudest approximation to the {\bf gap equation}. Letting 
$\lambda {\cal A}_0(\Gamma,\Upsilon_{3,diag})\equiv 1$ implies that the bubble series,

\Bigskip
{\centerline{$\geometricseries$}}

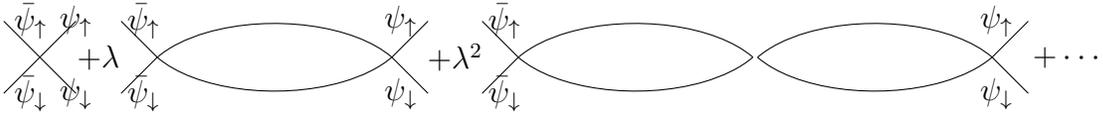
\Medskip {\bf \tiny Fig. \ref{fig:geometricseries}.2. Cooper pair bubble
geometric series.}

\Bigskip which is the leading order approximation of the Goldstone boson propagator, diverges at zero transfer momentum. These two properties -- vanishing of the one-point function
in the direction $\theta$, singularity of the  Goldstone boson propagator --
remain simultaneously true beyond mean-field theory and even non-perturbatively, as we shall see later on, see \S \ref{subsection:Sigma} and section \ref{section:boson}. 
 Whatever
the exact value of $\Gamma$, and the cut-off scale
$j'_{\phi}\ge j_{\phi}$, one has as soon as
$|\Gamma|\approx \hbar\omega_D e^{-\pi/g}$
\BEQ \lambda {\cal A}_0^{(j'_{\phi}-1)\to}(\Gamma,\Upsilon_{3,diag})=1+O(g). \label{eq:approx-Gamma} \EEQ

\Medskip The infra-red behavior of chains of bubbles, and later on, of Goldstone
boson propagators, are best understood after a change of basis,
choosing for external structures the linear combinations
$(\bar{\Psi}{\mathbb{\Gamma}}^{//,\perp}\Psi)\otimes 
(\bar{\Psi}{\mathbb{\Gamma}}^{//,\perp}\Psi)$ instead of 
the diagonal channels $(\bar{\psi}_{\uparrow}\bar{\psi}_{\downarrow})\otimes (\psi_{\downarrow}\psi_{\uparrow}),(\psi_{\downarrow}\psi_{\uparrow})\otimes (\bar{\psi}_{\uparrow}\bar{\psi}_{\downarrow})$ and the off-diagonal channels 
 $(\bar{\psi}_{\uparrow}\bar{\psi}_{\downarrow})\otimes (\bar{\psi}_{\uparrow}\bar{\psi}_{\downarrow}), (\psi_{\downarrow}\psi_{\uparrow})\otimes (\psi_{\downarrow}\psi_{\uparrow})$, where $\Gamma^{//}:={\mathbb{\Gamma}}(\theta), \Gamma^{\perp}:={\mathbb{\Gamma}}(\theta+\frac{\pi}{2})$. 
 (Explicit change-of-basis formulae are provided in \S \ref{subsection:heuristic} in the case $\theta=0$, see 
 in particular (\ref{eq:Sigma-//-//}) and below).
 Letting ${\cal A}_0^{j\to}(\Gamma,\Upsilon_{3,diag})=:
 a_{diag}^{j\to} {\mathbb{1}}$, ${\cal A}_0^{j\to}(\Gamma,\Upsilon_{3,off})=:-a_{off}^{j\to} \sigma(\theta)$, one
 obtains in the $(//,\perp)$-basis a diagonal matrix
 ${\cal A}_0^{j\to}$=diag$(({\cal A}_0)^{j\to}_{//,//} , ({\cal A}^{j\to}_0)_{\perp,\perp})$, with  
\BEQ ({\cal A}^{j\to}_0)_{//,//} =2(a_{diag}^{j\to}-a_{off}^{j\to}),
\qquad ({\cal A}^{j\to}_0)_{\perp,\perp} =2(a_{diag}^{j\to}+a_{off}^{j\to})  \label{eq:AaaAaa}
\EEQ
and $a_{diag}^{j\to}\approx \frac{m^*}{g}>0$, $a_{off}^{j\to}
\approx  m^*>0$.

\Bigskip It is useful at this point to introduce the 
following

\begin{Definition}[pre-gap equation] \label{def:pre-gap}
Let $\Gamma=\Gamma^{(j'_{\phi}-1)\to}$ be the solution
of the {\em pre-gap equation},
\BEQ \lambda ({\cal A}^{(j'_{\phi}-1)\to}_0)_{\perp,\perp}(\Gamma)=1. \label{eq:pre-gap-scal}
\EEQ
\end{Definition}

Because $a_{diag}^{j\to}\gg a_{off}^{j\to}>0$, $\Gamma$ may also be understood as the solution
 with largest
module of the eigenvalue equation,
\BEQ \Big(-{\mathbb{1}}+ \lambda {\cal A}_0^{(j'_{\phi}-1)\to}(\Gamma,\Upsilon_3)\Big)
\left(\begin{array}{c} \Gamma^{\perp} \\ (\Gamma^{\perp})^* \end{array}\right)=0.
\label{eq:pre-gap} 
\EEQ

\Medskip The matrix ${\mathbb{1}}- \lambda {\cal A}_0^{(j'_{\phi}-1)\to}(\Gamma,\Upsilon_3)$ has eigenvectors
$\left(\begin{array}{c} \Gamma^{//} \\ (\Gamma^{//})^* \end{array}\right), \left(\begin{array}{c} \Gamma^{\perp} \\ (\Gamma^{\perp})^* \end{array}\right)$, with
respective eigenvalues $4\lambda a_{off}^{(j'_{\phi}-1)\to}\approx g>0$ and $0$. As emphasized in \S \ref{subsection:pre-boson}, this
means that {\em the geometric series of Cooper pair bubble
diagrams is {\bf massive} in the parallel direction ($//$), i.e.
along ${\bf\Gamma}(\theta)$, whereas it is {\bf non-massive}
in the perpendicular direction ($\perp$), i.e.
along ${\bf\Gamma}(\theta+\frac{\pi}{2})$}, corresponding
to the longitudinal direction along the circle $|\Gamma|=$
Cst, in line with Goldstone's insight.   Compare with Definition \ref{def:gap}. As shown in \S \ref{subsection:Sigma} for
a corrected version of the gap equation including the
Bethe-Salpeter kernel $\bar{\Pi}_0$, this equation has a solution
$\approx \hbar\omega_D e^{-\pi/g}$, as expected. Computations
in \ref{subsection:Sigma} reduce to the results of \S
\ref{subsection:pre-boson} if one sets $\bar{\Pi}_0\equiv \lambda\Id$.   The pre-gap equation is corrected in the bosonic regime, 
see Definition \ref{def:gap-j},
leading to a sequence $\Gamma^{j_+\to}$, $j_+=j'_{\phi},j'_{\phi}+1,\cdots$ converging when $j_+\to\infty$ to the
correct value of $\Gamma_{\phi}$.


\subsection{The pre-Goldstone boson propagator}  \label{subsection:pre-boson}


\Bigskip The pre-Goldstone boson propagator  is directly related to the second derivatives
of ${\cal A}_q^{j\to}(\Gamma,\Upsilon_{3,diag})$ and ${\cal A}_q^{j\to}(\Gamma,\Upsilon_{3,off})$  w.r. to $q$ around $q=0$. A general argument, relying
on the fact that
\BEQ {\cal A}_0^{j\to}(\Gamma,\Upsilon_{3,diag})= \int d\xi'\, |C^{j\to}_{diag}(\xi-\xi')|^2
\ge \int d\xi'\, e^{-\II (q,\xi-\xi')}\, |C_{diag}^{j\to}(\xi-\xi')|^2= {\cal A}_q^{j\to} (\Gamma,\Upsilon_{3,diag})  \label{eq:positivity-argument}
\EEQ
and
\BEQ -{\cal A}_0^{j\to}(\Gamma,\Upsilon_{3,off})= \int d\xi'\, |C^{j\to}_{off}(\xi-\xi')|^2
\ge \int d\xi'\, e^{-\II (q,\xi-\xi')}\, |C_{off}^{j\to}(\xi-\xi')|^2= -{\cal A}_q^{j\to} (\Gamma,\Upsilon_{3,off})  \label{eq:positivity-argument-off}
\EEQ
implies that $\nabla^2 {\cal A}_q^{j\to}(\Gamma,\Upsilon_{3,diag})\Big|_{q=0}\le 0$, resp. $\nabla^2 (-{\cal A}_q^{j\to}(\Gamma,\Upsilon_{3,off}))\Big|_{q=0}\le 0$ . By
symmetry, $\partial_{q^0} {\cal A}_q(\Gamma,\Upsilon_{3,\eps})\Big|_{q=0}=0$, $\vec{\nabla}{\cal A}_q(\Gamma,\Upsilon_{3,\eps})\Big|_{q=0}=0$,  $\partial_{q^0} \vec{\nabla} {\cal A}_q(\Gamma,\Upsilon_{3,\eps})\Big|_{q=0}=0$ and
$\vec{\nabla}^2 {\cal A}_q(\Gamma,\Upsilon_{3,\eps})\Big|_{q=0}$ is a scalar matrix, where $\eps$=diag,off.

\begin{Lemma} \label{lemma:bubble}

\BEQ  \boxed{\ \ \partial_{q^0}^2 {\cal A}_q(
\Gamma^{(j'_{\phi}-1)\to},\Upsilon_{3,diag})\Big|_{q=0}=:-\frac{1}{g^0_{\phi,diag}}, \qquad\partial_{q^0}^2 {\cal A}_q(\Gamma^{(j'_{\phi}-1)\to},\Upsilon_{3,off})\Big|_{q=0}=\frac{1}{g^0_{\phi,off}} \ \ }
\EEQ

\BEQ \boxed{\ \  \vec{\nabla}^2{\cal A}_q(\Gamma^{(j'_{\phi}-1)\to},\Upsilon_{3,diag})\Big|_{q=0}=:
 -\frac{(v^0_{\phi,diag})^2}{g^0_{\phi,diag}} \, \Id, \ \  \vec{\nabla}^2{\cal A}_q(\Gamma^{(j'_{\phi}-1)\to},\Upsilon_{3,off})\Big|_{q=0}=:
 \frac{(v^0_{\phi,off})^2}{g^0_{\phi,off}} \, \Id\ \ }
\EEQ

where 
\BEQ \boxed{\ g^0_{\phi,diag},\ g^0_{\phi,off}\approx \frac{\Gamma_{\phi}^2}{m^*}\ } \EEQ
have the dimension of an energy, and
\BEQ \boxed{\ v^0_{\phi,diag},v^0_{\phi,off} \approx \frac{p_F^*}{m^*}\ } \EEQ
are  velocities.

\end{Lemma}

\noindent The proof is given in Appendix.

\begin{Definition}[Goldstone pre-boson coupling constant and velocity]
\label{def:gphivphi}

Let 
\BEQ g^0_{\phi}:=\frac{g^0_{\phi,diag}g^0_{\phi,off}}{g^0_{\phi,diag}+g^0_{\phi,off}},
\qquad v^0_{\phi}:=\sqrt{ \frac{g^0_{\phi,off}(v^0_{\phi,diag})^2+g^0_{\phi,diag}(v^0_{\phi,off})^2}{g^0_{\phi,diag}+g^0_{\phi,off}}}.
\EEQ 

\end{Definition}
The Goldstone pre-boson coupling constant $g^0_{\phi}$ and velocity $v^0_{\phi}$ are
obtained by inverting $F(q):=\Big({\cal A}_q(\Gamma^{(j'_{\phi}-1)\to},\Upsilon_{3,diag})-{\cal A}_0(\Gamma^{(j'_{\phi}-1)\to},\Upsilon_{3,diag})\Big)
- \Big({\cal A}_q(\Gamma^{(j'_{\phi}-1)\to},\Upsilon_{3,off})-{\cal A}_0(\Gamma^{(j'_{\phi}-1)\to},\Upsilon_{3,diag})\Big)$ for $q\to 0$, namely,
\BEQ  \frac{1}{F(q)}\sim_{q\to 0} \Big(\sum_{\eps={\mathrm{diag}},{\mathrm{off}}} \frac{1}{g^0_{\phi,\eps}} (q^0)^2+
\frac{(v^0_{\phi,\eps})^2}{g^0_{\phi,\eps}} |\vec{q}|^2\Big)^{-1}= \frac{g^0_{\phi}}{(|q|^0_+)^2}
\EEQ
where
\BEQ |q|^0_+:=\sqrt{(q^0)^2+(v^0_{\phi}|\vec{q}|)^2}
 \EEQ
similarly to (\ref{eq:qplus}).

\Bigskip The meaning of Definition \ref{def:gphivphi} is explained in the
next subsection.

\Medskip Summing the  geometric series of Cooper pair bubble diagrams, one gets a
$2\times 2$ matrix kernel, called {\bf pre-Goldstone boson propagator} or simply {\bf pre-kernel},

\BEQ \boxed{{\mathrm{Pre}}\Sigma(q):= \lambda \sum_{n=0}^{+\infty} \Big\{ \lambda
{\cal A}^{(j'_{\phi}-1)\to}_q(\Gamma^{(j'_{\phi}-1)\to},\Upsilon_3) \Big\}^n = \frac{\lambda}{{\mathbb{1}}-\lambda {\cal A}_q^{(j'_{\phi}-1)\to}(\Gamma^{(j'_{\phi}-1)\to},\Upsilon_3)}}
\label{eq:PreSigma}
\EEQ
with (choosing a basis)

\medskip

\BEQ
\begin{tikzpicture} \label{eq:sandwich-kernel-A}
\draw(-2.5,0) node {$ {\cal A}^{(j'_{\phi}-1)\to}_q(\Gamma^{(j'_{\phi}-1)\to},\Upsilon_3):=$};
\draw(0.5,0.3) node {$ \bar{\psi}_{\uparrow} \bar{\psi}_{\downarrow}$}; 
\draw(0.5,-0.3) node {$  \psi_{\downarrow} \psi_{\uparrow} $};
\begin{scope}[shift={(2,0)}]
\draw(4.5,0) node {$ \left(\begin{array}{cc}  {\cal A}_q^{(j'_{\phi}-1)\to}(\Gamma^{(j'_{\phi}-1)\to},\Upsilon_{3,diag}) &  \overline{{\cal A}_q^{(j'_{\phi}-1)\to}(\Gamma^{(j'_{\phi}-1)\to},\Upsilon_{3,off})} \\  {\cal A}^{(j'_{\phi}-1)\to}_q(\Gamma^{(j'_{\phi}-1)\to},\Upsilon_{3,off}) &  {\cal A}^{(j'_{\phi}-1)\to}_q(\Gamma^{(j'_{\phi}-1)\to},\Upsilon_{3,diag})
\end{array}\right)$};
\end{scope}
\draw(4.5,1.3) node  {$ \psi_{\downarrow} \psi_{\uparrow}$};
\draw(8.5,1.3) node  {$ \bar{\psi}_{\uparrow} \bar{\psi}_{\downarrow}$}; 
\end{tikzpicture}
\EEQ

\Medskip
This kernel is in sandwich between adjoint fermion pairs in Cooper pairings, $\bar{\psi}_{\uparrow}\bar{\psi}_{\downarrow}$ or $\psi_{\downarrow}\psi_{\uparrow}$. Note that the amplitude
${\cal A}_q^{(j'_{\phi}-1)\to}(\Gamma^{(j'_{\phi}-1)\to},\Upsilon_3)$ is computed
{\em not} for the exact value $\Gamma_{\phi}$ of $\Gamma$, but
for $\Gamma=\Gamma^{(j'_{\phi}-1)\to},$ in such a way that
$\PreSigma(q)$ {\em diverges} when $q\to 0$.

\Medskip As seen in the previous subsection, the above kernel is divergent in the infra-red limit, i.e. when $q\to 0$, in
the $(\perp,\perp)$-channel. Let us now consider  $q\not=0$ but small. The behavior of Pre$\Sigma_{\perp,\perp}(q)$ is obtained (see (\ref{eq:AaaAaa}))
as the inverse of the quadratic form $-\half \sum_{i,j=0}^3  \partial_{q_i}
\partial_{q_j}\Big\{ ({\cal A}_q(\Upsilon_{3,diag})+{\cal A}_q(\Upsilon_{3,off})
\Big\}\Big|_{q=0} q^i q^j$. Lemma \ref{lemma:bubble} yields:
 \BEQ \boxed{\PreSigma_{\perp,\perp}(q)\sim_{q\to 0}  \half
\frac{g^0_{\phi}}{(|q|_+^0)^2} (1+O(g))}
\label{eq:PreSigma-asymptotics-perp}
\EEQ

 with the notations of Definition \ref{def:gphivphi}. On the
 other hand, 
\BEQ \PreSigma_{//,//}(q)\sim_{q\to 0} \frac{\lambda}{4\lambda a^{(j'_{\phi}-1)\to}_{off}}\approx \frac{1}{m^*}.
\label{eq:PreSigma-asymptotics-//}
\EEQ

\Medskip
As we shall see in the next subsection, the kernel component
$\PreSigma_{\perp,\perp}$  is a good approximation of
the (scalar) {\em Goldstone boson propagator} $\Sigma_{\perp,\perp}$ for large transfer momenta $|q|_+\gg \Gamma_{\phi}$. For this reason, and also because it is explicitly produced through the bubble resummations in the fermionic regime, we may consider it as the
covariance  of fictitious bosonic particles which we call {\em pre-Goldstone bosons}, whence the name ${\mathrm{Pre}}\Sigma$; in Feynman diagrams it is 
represented as a dashed wavy line, 
\begin{tikzpicture}[scale=1] \draw[dashed,decorate,decoration=snake](0,0)--(2,0);
\end{tikzpicture}
\qquad . This kernel is dressed by more complicated
diagrams, which play an important r\^ole only in the bosonic regime, but 
do {\em} not modify the general picture described in the previous paragraphs; perturbatively
(as shown in (\ref{eq:1.110})),  
\BEQ \Sigma_{\perp,\perp}(q)\sim_{q\to 0} \half \frac{g_{\phi}}{|q|_+^2} \label{eq:1.71} 
\EEQ
where $g_{\phi}=g^0_{\phi}(1+O(g))$, $v_{\phi}=v^0_{\phi}(1+O(g))$, see 
(\ref{eq:gphivphi}), and
$|q|_+:=\sqrt{(q^0)^2+(v_{\phi}|\vec{q}|)^2}
 $, see  (\ref{eq:qplus}). 
 The rigorous definition of $\Sigma$ in section \ref{section:boson} -- which relies instead on a somewhat
 technical multi-scale construction --  will be slightly different, but
 (\ref{eq:1.71}) will be shown to hold for some modified
 values of $g_{\phi},v_{\phi}$ and  up to error terms, for
 which we only prove a crude bound in $O(g^{1/4})$   (though
 one should be able to prove that they are
 $O(g)$ by looking more precisely at lowest-order terms with
 a small number of vertices).

\Bigskip  Let us now describe the (non-rigorous) connection to
the effective non-linear sigma model theory described in the
Introduction (see Regime II).  Let $\Gamma(q):=\int d\xi
\, e^{-\II (q,\xi)}   (\bar{\Psi}{\mathbb{\Gamma}}^{\perp} \Psi)(q)$. Bubbles in the channel $(\perp,\perp)$  may be resummed into the following counterterm, 
\BEQ  \ \Gamma(-q)  {\mathrm{Pre}}\Sigma_{\perp,\perp}(q)\   \Gamma(q) \sim_{q\to 0}  \Gamma(-q) \ 
\frac{g^0_{\phi}}{  (q^0)^2+ (v^0_{\phi})^2 
|\vec{q}|^2} \  \Gamma(q) \label{eq:1.51}
\EEQ
implying an effective action, see (\ref{eq:HS}),  of the form
\BEA &&
 \exp\Big(- \int dq\,   \Gamma(-q)\ \frac{   q_0^2+ (v^0_{\phi})^2 
|\vec{q}|^2}{g^0_{\phi}} \Gamma(q) \Big) \nonumber\\
&&\qquad \propto \int {\cal D}\Gamma\, 
\exp -\Big( \frac{1}{g^0_{\phi}} \int d\xi\,  |\nabla\Gamma(\xi)|^2  \Big),
\EEA
where by definition
\BEQ \nabla\Gamma(\xi):=\left(\begin{array}{c} \partial_{\tau} \Gamma(\xi) \\ v_{\phi}^0\vec{\nabla}\Gamma(\xi) \end{array}\right),\qquad |\nabla\Gamma(\xi)|^2:= |\partial_{\tau} \Gamma(\xi)|^2+
(v^0_{\phi})^2  |\vec{\nabla}\Gamma(\xi)|^2. \label{eq:nabla-gamma}
\EEQ
Removing the upper $0$ indices, we get a similar resummation of the 
Goldstone {\em boson}. 
 So $g_{\phi}$ plays the {\em r\^ole of a coupling constant for the non-linear
sigma model} describing the infra-red behavior of the Goldstone boson.

\Bigskip Let us investigate corrections to the leading-order behavior of $\PreSigma_{\perp,\perp}(q)$ when $|q|_+\to 0$ (as we shall see
in \S \ref{section:boson}, the same estimates hold for $\Sigma_{\perp,\perp}(q)$, up to the replacement of $g_{\phi}^0,v_{\phi}^0$ by their counterparts $g_{\phi},v_{\phi}$). 
Corrections  involve fourth-order derivatives of 
${\cal A}_q(\Upsilon_3)$, 
\BEQ \PreSigma_{\perp,\perp}(q)=\frac{g_{\phi}}{q_0^2+v_{\phi}^2 |\vec{q}|^2 +
O(|q|_+^4/\Gamma^2_{\phi})} = \frac{g_{\phi}}{q_0^2+v_{\phi}^2 |\vec{q}|^2} + O(|q|_+^2/\Gamma_{\phi}^2)  \label{eq:1.59} \EEQ
Let 
\BEQ |\xi|_+:=\sqrt{\tau^2+(|\vec{x}|/v_{\phi})^2} \EEQ
(see Notations in the introduction), $|\xi_+|\approx|\xi|$. 
An inverse Fourier transform yields
\BEQ \PreSigma_{\perp,\perp}(\xi):= (2\pi)^{-3} \int dq\, e^{\II (q,\xi)}
\PreSigma(q) \sim_{|\xi|_+\to\infty} \frac{1}{4\pi} \frac{g_{\phi}/v_{\phi}}{
\sqrt{|\vec{x}|^2+
v^2_{\phi}\tau^2}} = \frac{1}{4\pi} \frac{g_{\phi}/v_{\phi}^2}{|\xi|_+}.
\EEQ
Error terms can be bounded, yielding
\BEQ \PreSigma_{\perp,\perp}(\xi)\sim_{|\xi|_+\to\infty} \frac{1}{4\pi}  \frac{g_{\phi}/v_{\phi}^2}{|\xi|_+} (1+O(\frac{1}{\Gamma_{\phi}|\xi|_+})). 
\label{eq:Sigma-error-terms} 
\EEQ
as proved in the Appendix.

\Bigskip Finally, let us bound the infra-red cut-off matrix  kernel
kernel $\PreSigma^{j\to}(\xi)$ with $j_{\phi}-j\gg 1$. For that, we need further estimates on the Fourier transformed kernel
$\PreSigma^{j\to}(q)$. {\em Consider first the case when 
$|q|_+\gtrsim 2^{-j}\mu$.} 
We deduce from the computations of \S \ref{subsection:bubble} that 
$|\PreSigma^{j\to}(q)|\lesssim |\frac{\lambda}{1-\lambda {\cal A}^{j\to}(q)}| \approx \frac{1}{m^*} \frac{1}{\log(|q|_+/\Gamma_{\phi})}$.  
 Then, as proved in the Appendix, if $|\kappa|\ge 1$, 
\BEQ |\nabla_q^{\kappa} \PreSigma(q) |\lesssim \frac{1}{m^*|q|_+^{|\kappa|}} \frac{1}{\log^2 (|q|_+/\Gamma_{\phi})}
\label{eq:bound-nabla-kappa-Sigma}
\EEQ
where $\nabla_q^{\kappa}:=(\nabla_{q^0})^{\kappa^0} (\frac{\mu}{p_F^*}\nabla_{q^1})^{\kappa^1} 
(\frac{\mu}{p_F^*}\nabla_{q^2})^{\kappa^2}$. {\em The
other case, namely when $|q|_+\lesssim 2^{-j}\mu$}, is
easier. As already observed in \S \ref{subsection:bubble}
${\cal A}_q^{j\to}(\Upsilon_{3,diag})\sim {\cal A}_0^{j\to}(\Upsilon_{3,diag})$, and similarly, 
\BEQ |\nabla^{\kappa}{\cal A}_q^{j\to}(\Upsilon_{3,diag})|\lesssim m^* \int_{2^{-j}\mu}^{2^{-j_D}\mu} \frac{\rho \, d\rho}{\rho^{\kappa+2}} \approx
\frac{m^*}{(2^{-j}\mu)^{\kappa}}  \qquad (|\kappa|\ge 1),
\EEQ
so that $1-\lambda{\cal A}_q^{j\to}(\Upsilon_{3,diag})\approx
m^*\lambda(j_{\phi}-j)$, and finally, by an elementary computation,
\BEQ |\nabla_q^{\kappa} \PreSigma(q) |\lesssim \frac{1}{m^*(2^{-j}\mu)^{|\kappa|}} \frac{1}{(j_{\phi}-j)^2}.
\EEQ 
 Introduce a reduced 
partition of unity for transfer momenta, fix $\chi_+\equiv \chi$, and let $\chi_+^{j_+}(\cdot):=\chi_+(2^{j_+}\cdot)$ $\qquad (j_+<j)$, $\chi_+^{j}(\cdot):=\sum_{j_+=-\infty}^{j}
\chi_+(2^{j_+}\cdot)$. Then 
\BEA && m^* |\PreSigma^{j\to}(\xi)|=m^* \Big|\sum_{k=j_D}^j \int dq\,  \chi_+^j(q) \ e^{\II \langle q,\xi\rangle} \Sigma^{j\to}(q)
\Big| \nonumber\\
&&\qquad \lesssim \Big\{\sum_{j_+=j_D}^j \frac{1}{(j_{\phi}-j_+)^2}\, \Vol(\supp(\chi_+^{j_+}))\,   \Big(1+
2^{-j_+}|\xi|\Big)^{-n}\Big) \Big\} \nonumber\\
&& + \frac{1}{j_{\phi}-j_D}
\Vol(\supp(\chi_+^{j_D}))\, \Big(1+
2^{-j_D}|\xi|\Big)^{-n} +  \frac{1}{j_{\phi}-j}
\, \Vol(\supp(\chi_+^{j})) \Big(1+
2^{-j}|\xi|\Big)^{-n} \nonumber\\
\EEA
for every $n\ge 1$, as shown by successive integrations by parts.
Whence, in particular,
\BEA  m^*\int d\xi\, |\PreSigma^{j\to}(\xi)| &\lesssim& \Big\{\sum_{k=j_D}^{j}
 \frac{1}{(j_{\phi}-k)^2} \Big\} + \frac{1}{j_{\phi}-j_D} +
 \frac{1}{j_{\phi}-j} \nonumber\\
 &\lesssim &\frac{1}{j_{\phi}-j}, \qquad j_{\phi}-j\gg 1.
\label{eq:bound-integral-sigma-j-arrow} \EEA

When $j\ll j_{\phi}$, a good approximation for $\PreSigma^{j\to}(\xi)$
is simply its leading order, which is the $(\bar{\Psi}\Psi)^2$-vertex $\lambda\del(\xi)$.
The bound (\ref{eq:bound-integral-sigma-j-arrow}) is coherent
with the value of  the integrated kernel $\int d\xi\, \lambda\del(\xi)=\lambda$, since $\frac{1}{j_{\phi}}\approx g\approx
\lambda m^*$.

\Bigskip  On the other hand, the kernel $\PreSigma_{//,//}^{(j'_{\phi}-1)\to}(\xi)$, which is the Fourier transform of the massive
kernel 
\BEQ \PreSigma_{//,//}^{(j'_{\phi}-1)\to}(q)\approx \frac{1}{m^*}\approx \frac{1}{m^*} \frac{\Gamma_{\phi}^2}{|q|_+^2
+\Gamma_{\phi}^2} \qquad (|q|_+\lesssim \Gamma_{\phi}), 
\EEQ
 has a quasi-exponential decay at distances $\gg 1/\Gamma_{\phi}$,
\BEQ |\PreSigma_{//,//}^{(j'_{\phi}-1)\to}(\xi)|\le C_n \frac{1}{m^*} \Vol(\supp(\chi_+^{j_{\phi}})) \, (1+2^{-j_{\phi}}|\xi|)^{-n}
\approx C_n \frac{\Gamma_{\phi}^3}{\mu}\, (1+2^{-j_{\phi}}|\xi|)^{-n}, \qquad |\xi|\gtrsim \Gamma_{\phi}^{-1}
\label{eq:decay-PreSigma//}
\EEQ
see (\ref{eq:vol-box}), as can be shown in a standard way by repeated integrations by parts.


\subsection{Energy gap and Goldstone boson}  \label{subsection:Sigma}


\Medskip The above  Cooper pair bubble diagrams $\Upsilon_{3,\eps}$, $\eps=$diag,off,  may be composed an arbitrary number of 
times, yielding  $\lambda$ times the geometric series
of bubbles, see Fig. \ref{fig:geometricseries}.1
 with transfer momentum $q$, evaluated as $\frac{\lambda}{1-\lambda
{\cal A}_q(\Upsilon_3)}$. A quick but not rigorous computation, taking into account
only these diagrams, yields an approximate value for $\Gamma_{\phi}$, based 
on the requirement that {\em the denominator of the series of bubbles
vanishes for $q=0$},   see {\bf Gap equation} in Definition \ref{def:gap} below, 
namely (considering only the $\Upsilon_{3,diag}$-diagram, and summarizing briefly the
findings of the two previous subsections)
\BEQ \lambda {\cal A}_0(\Upsilon_{3,diag})\sim 1 \EEQ 
from which
\BEQ \Gamma_{\phi}\approx \hbar\omega_D e^{-\pi/g} 
\label{eq:epig} \EEQ
and
\BEQ  \frac{\lambda}{1-\lambda
{\cal A}_q(\Upsilon_{3,diag})} \approx \frac{g_{\phi}}{|q|_+^2}. \label{eq:approx-Sigma}\EEQ

\Bigskip
 As explained in the Introduction
to this section, what we really need in section \ref{section:boson} is
the  {\em effective contribution $\Sigma_{\perp,\perp}$ to the Cooper pair bound state (Goldstone boson) propagator of  fermionic four-point functions}. We need some preliminary
explanations before we give a closed formula for this quantity, but let us
mention already at this stage that the approximation (\ref{eq:epig}) gives
a correct order of magnitude for $\Gamma_{\phi}$ -- as shall be checked self-consistently --, and that the value obtained for the bubble series (\ref{eq:approx-Sigma}) for small $q$ is a good approximation for $\Sigma(q)$ --
except that the latter is a two-by-two matrix.

\Medskip First,
a Feynman diagram with Cooper pair external structure

\BEA && (\bar{\psi}_{\uparrow}\bar{\psi}_{\downarrow})\otimes (\psi_{\downarrow}\psi_{\uparrow})= \Big(\bar{\psi}_{\uparrow}(\xi_1)\bar{\psi}_{\downarrow}(\xi_2)\Big) \Big(\psi_{\downarrow}(\xi_3)
\psi_{\uparrow}(\xi_4)\Big),\nonumber\\
&& \qquad 
(\psi_{\downarrow}\psi_{\uparrow})\otimes(\bar{\psi}_{\uparrow}\bar{\psi}_{\downarrow})=\Big(\psi_{\downarrow}(\xi_1)\psi_{\uparrow}(\xi_2)\Big)\Big(\bar{\psi}_{\uparrow}(\xi_3)
\bar{\psi}_{\downarrow}(\xi_4)\Big) \label{eq:non-mixing-case}
\EEA
({\em non-mixing case}), or 
\BEA && (\bar{\psi}_{\uparrow}\bar{\psi}_{\downarrow})\otimes(\bar{\psi}_{\uparrow}\bar{\psi}_{\downarrow})=\Big(\bar{\psi}_{\uparrow}(\xi_1)\bar{\psi}_{\downarrow}(\xi_2)\Big)\Big(\bar{\psi}_{\uparrow}(\xi_3)
\bar{\psi}_{\downarrow}(\xi_4)\Big),\nonumber\\
&&\qquad
(\psi_{\downarrow}\psi_{\uparrow})\otimes(\psi_{\downarrow}\psi_{\uparrow})= \Big(\psi_{\downarrow}(\xi_1)\psi_{\uparrow}(\xi_2)\Big)\Big(\psi_{\downarrow}(\xi_3)
\psi_{\uparrow}(\xi_4)\Big)  \label{eq:mixing-case}
\EEA
({\em mixing case}), is called {\em two-particle irreducible} if {\em the
compound made up of vertices $\{\xi_1,\xi_2\}$ is connected to the
compound made up of vertices $\{\xi_3,\xi_4\}$}. Note that 
two-particle reducible diagrams may be generated from the partition
function by a double Legendre transform, see e.g. \cite{Ram}, eq.
(10.132). We could, but have chosen not to, use this nice trick. As detailed in \S \ref{subsection:angular},  external legs possibly have indices --   scale $j$ and momentum angular
indices $\alpha$ for {\em sectors} ${\cal S}^{j,\alpha}$ with $j< j'_{\phi}$, 
scale $j'_{\phi}$, momentum angular indices $\alpha$ and
an extra integer index $k$ for {\em micro-sectors} ${\cal S}^{j,\alpha,k}$ --
, numbered as the corresponding vertex locations, e.g.
$(\bar{\psi}^{j_1,\alpha_1}_{\uparrow} \bar{\psi}^{j_2,\alpha_2}_{\uparrow})\otimes (\psi^{j_3,\alpha_3}_{\downarrow} \psi^{j_4,\alpha_4}_{\uparrow})=\Big((\bar{\psi}^{j_1,\alpha_1}_{\uparrow}(\xi_1) \bar{\psi}^{j_2,\alpha_2}_{\uparrow}
(\xi_2)\Big) \Big( (\psi^{j_3,\alpha_3}_{\downarrow}(\xi_3) \psi^{j_4,\alpha_4}_{\uparrow}(\xi_4)\Big)$. As explained later
on in \S \ref{subsection:cluster}, only $\tilde{\Pi}$-kernels
with external scales $j_1=j_2=j_3=j_4=j'_{\phi}$  need be considered; their external momenta are very close to, but not on, the Fermi
circle. Triples $(j'_{\phi},\alpha,k)$ index regions ${\cal S}^{j'_{\phi},\alpha,k}$ of the
momentum space of dimensions scaling like $2^{-j_+}\times 2^{-j_+}\times 2^{-j_+}$, so that, denoting by $p^{j'_{\phi}\alpha,k}$ the center of ${\cal S}^{j'_{\phi},\alpha,k}$, 
$|p-p^{j'_{\phi},\alpha,k}|_+\lesssim 2^{-j_+}\mu$ for $p$
lying inside ${\cal S}^{j'_{\phi},\alpha,k}$. None of this
is really required to read the present subsection, which remains at a descriptive level based on finite number of
graphs produced by perturbation.

\Medskip
{\em We let $\tilde{\Pi}(\xi_1,\xi_2;\xi_3,\xi_4)$ be the sum of all two-particle
irreducible, four-point Feynman diagrams with fixed external structure}; it may be represented as a two-by-two matrix, with {\em non-mixing coefficients} on the diagonal, and {\em mixing coefficients} off-diagonal, 

\BEQ
\begin{tikzpicture} 
\draw(-0.5,0) node {$ \tilde{\Pi}:=$};
\draw(1,0.3) node {$ \bar{\psi}_{\uparrow} \bar{\psi}_{\downarrow}$}; 
\draw(1,-0.3) node {$  \psi_{\downarrow} \psi_{\uparrow} $};
\draw(4.5,0) node {$ \left(\begin{array}{cc} \Pi_{\bar{\psi}_{\uparrow}\bar{\psi}_{\downarrow},\psi_{\downarrow} \psi_{\uparrow}}  \ &  \Pi_{\bar{\psi}_{\uparrow}\bar{\psi}_{\downarrow},\bar{\psi}_{\uparrow}\bar{\psi}_{\downarrow}} \\ 
\Pi_{\psi_{\downarrow} \psi_{\uparrow}, \psi_{\downarrow},\psi_{\uparrow}} &  \Pi_{\psi_{\downarrow} \psi_{\uparrow},\psi_{\downarrow} \psi_{\uparrow}}
\end{array}\right)$};
\draw(6,1) node  {$ \bar{\psi}_{\uparrow} \bar{\psi}_{\downarrow}$};
\draw(3.5,1) node  {$ \psi_{\downarrow} \psi_{\uparrow}$}; 
\draw(7.5,0) node {$\equiv$};
\draw(10,0) node {$\left(\begin{array}{cc} \tilde{\Pi}_{diag} & (\tilde{\Pi}_{off})^{\dagger} \\ \tilde{\Pi}_{off} & \tilde{\Pi}_{diag} \end{array}\right),$};  
\end{tikzpicture}  \label{eq:Pitilde}
\EEQ

compare with (\ref{eq:sandwich-kernel-A}). Lowest-order terms are 

\bigskip
{\centerline{$\Pitilde$}}

$\Pitildedeux$

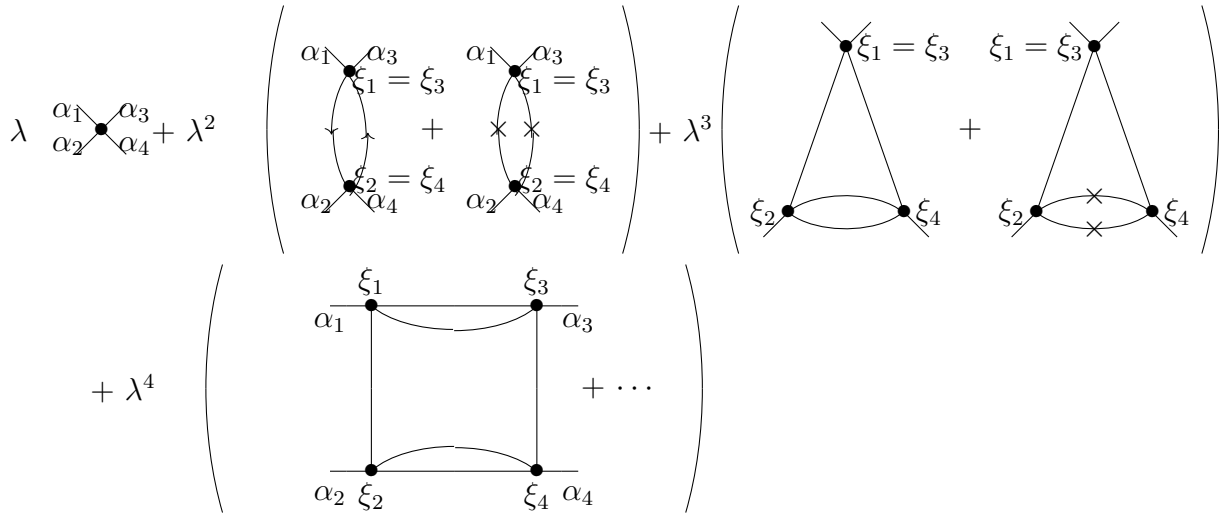
{\bf \tiny Fig. \ref{fig:Pitilde}.1. Lowest-order terms of the Bethe-Salpeter
kernel.}

\Bigskip Note that the $O(\lambda^3)$ triangle diagram, {\small Fig. \ref{fig:BetheSalpeter}}, is a two-particle {\em reducible} diagram \begin{tikzpicture} 
 \draw(2,0) node{obtained
by composing};
 \draw[rotate=90,->](0.7,-5) arc(30:90:0.88  and 0.44);
\draw[rotate=90](0.7-0.7,-5+0.2) arc(90:150:0.88  and 0.44);
\draw[rotate=90,->](-0.8,-5) arc(210:270:0.88 and 0.44);
\draw[rotate=90](-0.8+0.7,-5-0.2) arc(270:330:0.88 and 0.44);
\draw(7,0) node {and};
\draw(8.7,-0.3)--(9.3,0.3); \draw(8.7,0.3)--(9.3,-0.3);
\draw(9,0) node {$\bullet$};
 \end{tikzpicture} 
 On the other hand, the $O(\lambda^3)$ diagrams on  the first line of Fig.  \ref{fig:Pitilde}.1 (obtained by tilting the triangle diagram)
 are two-particle irreducible --  but note that by cutting the two long
 slanted lines, vertices $\{\xi_1,\xi_3\}$ are cut off from the
 other two, $\{\xi_2,\xi_4\}$.

\Medskip The first diagram is the vertex, simply evaluated as the coupling constant $\lambda$. The second and third diagrams look superficially like bubble diagrams
tilted to the side, but the impression is wrong because the momentum  circulating
inside the "tilded bubble" diagrams $\tilde{\Upsilon}_1,\tilde{\Upsilon}'_1,\tilde{\Upsilon}_{3,off}$, named with a tilde by reference to the original
bubble diagrams of the previous subsection,

\bigskip

\begin{tikzpicture}[scale=1.8]  \label{fig:tiltedbubbles}
\draw(5,0.62)--(5.25,0.87); \draw(5,0.62)--(4.75,0.87);
\draw[->](4.5,1.12)--(4.75,0.87); \draw(4.75,1.1) node {$p_1$};
\draw[>-](5.25,0.87)--(5.5,1.12); \draw(5.25,1.1) node {$p_3$};
\draw(4.65,0.74) node {$\psi$}; \draw(5.35,0.74) node {$\bar{\psi}$};
\draw(4.75,-0.1-0.74) node {$\psi$}; \draw(5.25,-0.1-0.74) node {$\bar{\psi}$};
\draw(5,0.62) node {$\bullet$};\draw(5,-0.72) node {$\bullet$};
\draw(5,-0.72)--(5.25,-0.97); \draw[-<](5,-0.72)--(4.75,-0.97);
\draw(4.2,-0.1-1) node {$q\!-\!p_1$}; \draw(5.9,-0.1-1) node {$-\!q-\!p_3$};
\draw[>-](5.25,-0.97)--(5.5,-1.22); \draw(4.75,-0.97)--(4.5,-1.22);
 \draw[rotate=90,->](0.7,-5) arc(30:90:0.88  and 0.44);
\draw[rotate=90](0.7-0.7,-5+0.2) arc(90:150:0.88  and 0.44);
\draw[rotate=90,->](-0.8,-5) arc(210:270:0.88 and 0.44);
\draw[rotate=90](-0.8+0.7,-5-0.2) arc(270:330:0.88 and 0.44);
\draw(4.1,-0.05) node {$p_1\!-\!p_3+r$}; \draw(5.4,-0.05) node {$r$};

\draw(3+5,0.62)--(3+5.25,0.87); \draw(3+5,0.62)--(3+4.75,0.87);
\draw[->](3+4.5,1.12)--(3+4.75,0.87); \draw(3+4.75,1.1) node {$p_1$};
\draw[>-](3+5.25,0.87)--(3+5.5,1.12); \draw(3+5.25,1.1) node {$p_3$};
\draw(3+4.65,0.74) node {$\psi$}; \draw(3+5.35,0.74) node {$\bar{\psi}$};
\draw(3+4.75,-0.1-0.74) node {$\psi$}; \draw(3+5.25,-0.1-0.74) node {$\bar{\psi}$};
\draw(3+5,0.62) node {$\bullet$};\draw(3+5,-0.72) node {$\bullet$};
\draw(3+5,-0.72)--(3+5.25,-0.97); \draw[-<](3+5,-0.72)--(3+4.75,-0.97);
\draw(3+4.2,-0.1-1) node {$q\!-\!p_1$}; \draw(3+5.9,-0.1-1) node {$-\!q-\!p_3$};
\draw[>-](3+5.25,-0.97)--(3+5.5,-1.22); \draw(3+4.75,-0.97)--(3+4.5,-1.22);
 \draw[rotate=90](0.7,-3-5) arc(30:90:0.88  and 0.44);
\draw[rotate=90](0.7-0.7,-3-5+0.2) arc(90:150:0.88  and 0.44);
\draw[rotate=90](-0.8,-3-5) arc(210:270:0.88 and 0.44);
\draw(3+5-0.22,-0.05) node {$\times$}; \draw(3+5+0.22,-0.05) node {$\times$};
\draw[rotate=90](-0.8+0.7,-3-5-0.2) arc(270:330:0.88 and 0.44);
\draw(3+4.1,-0.05) node {$p_1\!-\!p_3+r$}; \draw(3+5.4,-0.05) node {$r$};

\draw(6+5,0.62)--(6+5.25,0.87); \draw(6+5,0.62)--(6+4.75,0.87);
\draw[-<](6+4.5,1.12)--(6+4.75,0.87); \draw(6+4.75,1.1) node {$p_1$};
\draw[>-](6+5.25,0.87)--(6+5.5,1.12); \draw(6+5.25,1.1) node {$p_3$};
\draw(6+4.65,0.74) node {$\bar{\psi}$}; \draw(6+5.35,0.74) node {$\bar{\psi}$};
\draw(6+4.75,-0.1-0.74) node {$\bar{\psi}$}; \draw(6+5.25,-0.1-0.74) node {$\bar{\psi}$};
\draw(6+5,0.62) node {$\bullet$};\draw(6+5,-0.72) node {$\bullet$};
\draw(6+5,-0.72)--(6+5.25,-0.97); \draw[->](6+5,-0.72)--(6+4.75,-0.97);
\draw(6+4.2,-0.1-1) node {$q\!-\!p_1$}; \draw(6+5.9,-0.1-1) node {$-\!q-\!p_3$};
\draw[>-](6+5.25,-0.97)--(6+5.5,-1.22); \draw(6+4.75,-0.97)--(6+4.5,-1.22);
 \draw[rotate=90](0.7,-6-5) arc(30:90:0.88  and 0.44);
\draw[rotate=90](0.7-0.7,-6-5+0.2) arc(90:150:0.88  and 0.44);
\draw[rotate=90](-0.8,-6-5) arc(210:270:0.88 and 0.44);
\draw[rotate=90](-0.8+0.7,-6-5-0.2) arc(270:330:0.88 and 0.44);
\draw(6+4.1,-0.05) node {$p_1\!-\!p_3+r$}; \draw(6+5.4,-0.05) node {$r$};
\draw(6+5-0.22,-0.05) node {$\times$}; \draw(6+5+0.22,-0.05) node {$\times$};
\end{tikzpicture}

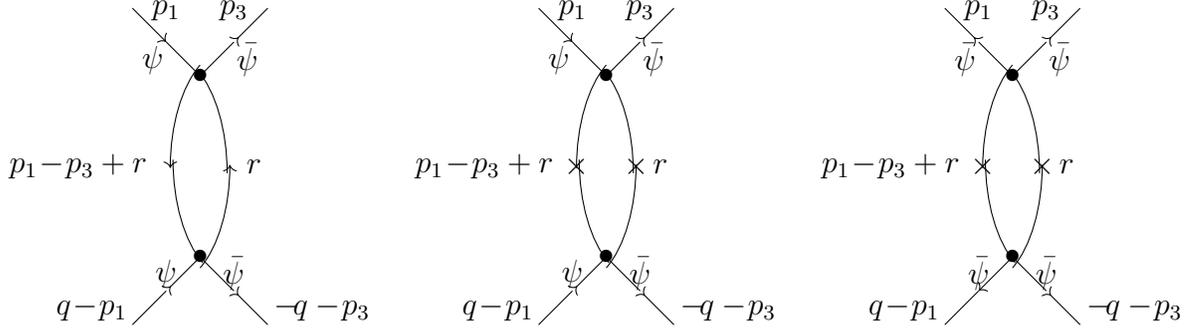
\Medskip {\tiny \bf Fig.  \ref{fig:tiltedbubbles}.2. "Tilted bubble" diagrams
$\tilde{\Upsilon}_1,\tilde{\Upsilon}'_1,\tilde{\Upsilon}_{3,off}$ (from left
to right).}

\Bigskip
isn't $q$ as in the $\Upsilon_1$-diagram, see {\small  Fig. \ref{fig:bubble}}.1,
but $p_1-p_3$, therefore, generically a {\em large} momentum. Note incidentally
that the third diagram of {\small Fig. \ref{fig:Pitilde}.1} decomposes as
the sum of two diagrams, $\tilde{\Upsilon}'_1$ and $\tilde{\Upsilon}_{3,off}$.

\Bigskip The first three diagrams, namely, the vertex, $\tilde{\Upsilon}_1$ and
$\tilde{\Upsilon}'_1$,  are of non-mixing type. The fourth one, $\tilde{\Upsilon}_{3,off}$,  is
mixing, proportional (when considered together with the conjugate diagram as
entries of a matrix) to the off-diagonal matrix $\left(\begin{array}{cc}
 0 & \Gamma^2 \\ (\Gamma^*)^2 & 0 \end{array}\right)$.  
 
 \Medskip Finally, the last ones (the
 terms in $O(\lambda^3)$ and $O(\lambda^4)$ in {\small Fig. \ref{fig:Pitilde}.1}),  can be either of non-mixing or mixing  type,
 see the figure below for the mixing version of the $O(\lambda^{4})$-diagram.

\medskip
{\centerline{$\BetheSalpeterbis$}}
\Medskip {\tiny\bf  Fig. \thesubsection.3. Mixing diagram $\tilde{\Upsilon}_{4,off}$.}

\Medskip Consider a general (mixing or non-mixing) diagram with an arbitrary number of
$\Gamma$- or $\Gamma^*$-insertions. Cutting the symmetry-broken propagators in their middle, thus leaving only neutral vertices, i.e. four-leg vertices
$(\bar{\psi}_{\uparrow}\bar{\psi}_{\downarrow})(\psi_{\downarrow}\psi_{\uparrow})$ with the same number of entering and exiting lines, one finds: 
$\sharp$\  entering lines=$\sharp$
exiting lines, whence (considering the whole diagram): $\# \, \Gamma-\#\, 
\Gamma^*=2$, resp. $0$, resp. $-2$ depending on whether the external structure is of the type
$(\psi_{\downarrow}\psi_{\uparrow})\otimes (\psi_{\downarrow}\psi_{\uparrow})$, 
resp. 
$ (\psi_{\downarrow}\psi_{\uparrow}) \otimes  (\bar{\psi}_{\uparrow}\bar{\psi}_{\downarrow})$ or $(\bar{\psi}_{\uparrow}\bar{\psi}_{\downarrow})\otimes (\psi_{\downarrow}\psi_{\uparrow})$, resp. $(\bar{\psi}_{\uparrow}\bar{\psi}_{\downarrow}) \otimes (\bar{\psi}_{\uparrow}\bar{\psi}_{\downarrow})$. 
If $\Gamma=|\Gamma_{\phi}|\, e^{\II\theta}$, then the corresponding two-by-two
symmetrized matrix $\half\Big\{\tilde{\Pi}(\xi_1,\xi_2;\xi_3,\xi_4) + \tilde{\Pi}(\xi_2,\xi_1;\xi_4,\xi_3)\Big\}$   is therefore   of the form
 $\left(\begin{array}{cc} c_{1,1} & c_{1,2} e^{-2\II
\theta} \\ c_{1,2} e^{2\II\theta} & c_{2,2} \end{array}\right)$, where
$c_{1,1},c_{1,2},c_{2,2}$ are {\em real-valued} functions.

\Bigskip {\em Define now $\tilde{\Sigma}(\xi_1,\xi_3)$ to  be $\frac{1}{\lambda}$ times the sum of all {\em connected} but {\em not necessarily two-particle irreducible} Feynman diagrams with
external stucture as in (\ref{eq:non-mixing-case},\ref{eq:mixing-case}), but
with $\xi_1=\xi_2$, $\xi_3=\xi_4$}:

\begin{tikzpicture}[scale=1]
\draw(-2.5,0) node {$\tilde{\Sigma}(\xi_1,\xi_3)=\del(\xi_1-\xi_3) + \lambda\Big\{$};

\draw(0.2,0) arc(180:150:0.6 and 1.2);
\draw(0.2,0) arc(180:210:0.6 and 1.2);
\draw(0.5,0) node {$\bullet$}; \draw(2,0) node {$\bullet$};
\draw(2,0) arc(30:150:0.88  and 0.44);
\draw(0.5,0) arc(210:330:0.88 and 0.44);
\draw(0.5,0.4) node {$\xi_1$};
\draw(2,0.4) node {$\xi_3$};

\draw(2.25,0) node {$+$}; \draw(4.6,0) node {$+$};

\begin{scope}[shift={(2,0)}]
\draw(0.5,0) node {$\bullet$}; \draw(2,0) node {$\bullet$};
\draw(2,0) arc(30:150:0.88  and 0.44);
\draw(0.5,0) arc(210:330:0.88 and 0.44);
\draw(0.5,0.4) node {$\xi_1$};
\draw(2,0.4) node {$\xi_3$};
\draw(2.25,0) arc(0:30:0.6 and 1.2);
\draw(2.25,0) arc(0:-30:0.6 and 1.2);
\draw(1.25,0.2) node {$\times$};
\draw(1.25,-0.2) node {$\times$};
\end{scope}

\draw(0.5+5,0) node {$\bullet$}; 
\draw(1.2+5,0.45) node {$\bullet$};\draw(1.2+5,-0.45) node {$\bullet$};
\draw(0.5+5,0) arc(150:90:0.88  and 0.88);
\draw(0.5+5,0) arc(210:270:0.88 and 0.88);
\draw(0.2+5,0) node {$\xi_1$};
\draw(1+5,1) node {$\xi'_1$}; \draw(1+5,-1) node {$\xi'_2$};

\draw(7.7,0) node{$\tilde{\Pi}(\xi'_1,\xi'_2;\xi'_3,\xi'_4)$};


\draw(2+8,0) node {$\bullet$}; 
\draw(1.3+8,0.45) node {$\bullet$};\draw(1.3+8,-0.45) node {$\bullet$};
\draw(2+8,0) arc(30:90:0.88  and 0.88);
\draw(2+8,0) arc(-30:-90:0.88 and 0.88);
\draw(1+8,1) node {$\xi'_3$}; \draw(1+8,-1) node {$\xi'_4$};
 \draw(2.4+8,0) node {$\xi_3$};

\end{tikzpicture}


{\center{\begin{tikzpicture}[scale=1]  \label{fig:formal-expansion-Bethe-Salpeter}
\draw(4,0) node {$+\ $};

\draw(0.5+5,0) node {$\bullet$}; 
\draw(1.2+5,0.45) node {$\bullet$};\draw(1.2+5,-0.45) node {$\bullet$};
\draw(0.5+5,0) arc(150:90:0.88  and 0.88);
\draw(0.5+5,0) arc(210:270:0.88 and 0.88);
\draw(0+5,0) node {$\xi_1$};
\draw(1+5,1) node {$\xi'_1$}; \draw(1+5,-1) node {$\xi'_2$};

\draw(7.7,0) node{$\tilde{\Pi}(\xi'_1,\xi'_2;\xi'_3,\xi'_4)$};

\draw(9.3,0) arc(180:170:5 and 10); \draw(9.3,0) arc(180:190:5 and 10);

\draw(9.8,0.6) node {$\bullet$}; \draw(1.5+9.8,0.6) node {$\bullet$};
\draw(9.8,-0.6) node {$\bullet$}; \draw(1.5+9.8,-0.6) node {$\bullet$};
\draw(0+9.8,0.6)--(1.5+9.8,0.6); \draw(0+9.8,-0.6)--(1.5+9.8,-0.6);
\draw(0+9.8,1) node {$\xi'_3$}; \draw(1.5+9.8,1) node {$\xi''_1$};
\draw(0+9.8,-1) node {$\xi'_4$}; \draw(1.5+9.8,-1) node {$\xi''_2$};

\draw(11.55,0) node {$+$};

\draw(2+9.8,0.6) node {$\bullet$}; \draw(3.5+9.8,0.6) node {$\bullet$};
\draw(2+9.8,-0.6) node {$\bullet$}; \draw(3.5+9.8,-0.6) node {$\bullet$};
\draw(2+9.8,0.6)--(3.5+9.8,0.6); \draw(2+9.8,-0.6)--(3.5+9.8,-0.6);
\draw(2+9.8,1) node {$\xi'_3$}; \draw(3.5+9.8,1) node {$\xi''_1$};
\draw(2+9.8,-1) node {$\xi'_4$}; \draw(3.5+9.8,-1) node {$\xi''_2$};
\draw(2.75+9.8,0.6) node {$\times$}; \draw(2.75+9.8,-0.6) node {$\times$};

\draw(13.8,0) arc(0:10:5 and 10); \draw(13.8,0) arc(0:-10:5 and 10);
\draw(15.5,0) node{$\tilde{\Pi}(\xi''_1,\xi''_2;\xi''_3,\xi''_4)$};


\draw(2+16,0) node {$\bullet$}; 
\draw(1.3+16,0.45) node {$\bullet$};\draw(1.3+16,-0.45) node {$\bullet$};
\draw(2+16,0) arc(30:90:0.88  and 0.88);
\draw(2+16,0) arc(-30:-90:0.88 and 0.88);
\draw(1+16,1) node {$\xi''_3$}; \draw(1+16,-1) node {$\xi''_4$};
 \draw(2.4+16,0) node {$\xi_3$};

\end{tikzpicture} }}

\Bigskip 
\qquad $+ \cdots \qquad \Big\}$

\Bigskip {\tiny \bf Fig. \thesubsection.4. Formal expansion of the two-particle
propagator in terms of two-particle irreducible diagrams.}

\Bigskip and let $\tilde{\Sigma}(q)$ be its Fourier transform, with transfer 
momentum $q$ entering the chain of diagrams at the left
end $\xi_1$, and exiting at the right end $\xi_3$. Note that connecting pairs of propagators preserve momenta, so that
corresponding pairs of external momenta for neighboring pairs of $\tilde{\Pi}$-kernels are equal, e.g. 
$p'_3=p''_1,p'_4=p''_2$.

\Bigskip Let us now introduce the following {\em projection procedure}, a simplified procedure giving the order of magnitude of the main contributions to the Bethe-Salpeter kernel near
the infra-red singularity, i.e. for a transfer momentum 
$q\to 0$ (of which a corrected version, called
{\em averaging procedure}, is presented in \S \ref{subsection:Ward} {\bf D.} (ii)).

\begin{Definition}
[Fermi surface projection of 2 P.I. diagrams]   \label{def:averaging}
For any two-particle irreducible, four-point diagram $\tilde{\Upsilon}$ with Cooper pair external structure, we denote by $\tilde{\cal A}_q(\tilde{\Upsilon})$  the {\em averaged} evaluation of
$\tilde{\Upsilon}$ on the Fermi circle, namely,
\BEQ \tilde{\cal A}_q(\tilde{\Upsilon}):=\frac{1}{(\Vol(\Sigma^*_F))^2}
\int_{\Sigma^*_F\times\Sigma^*_F} dp_1\, dp_3\, 
{\cal A}(p_1,-p_1+q;p_3,-p_3-q).
\EEQ
\end{Definition}

\Medskip
\begin{Lemma} \label{lem:lowest-Pi}
\BEQ |\tilde{\cal A}_q(\tilde{\Upsilon}_1)|\lesssim m^* \EEQ
\BEQ
 |{\cal A}_q(\tilde{\Upsilon}'_1)|,
|\tilde{\cal A}_q(\tilde{\Upsilon}_{3,off})| \lesssim m^*/N_{j_{\phi}}
\EEQ
\BEQ |\tilde{\cal A}_q(\tilde{\Upsilon}_{4,diag})|\lesssim m^*, |\tilde{\cal A}_q(\tilde{\Upsilon}_{4,off})|\lesssim m^*/N_{j_{\phi}}
\EEQ
\BEQ
|\frac{d}{d\Gamma_{\phi}} \tilde{\cal A}_q(\tilde{\Upsilon}_1)|, 
|\frac{d}{d\Gamma_{\phi}} \tilde{\cal A}_q(\tilde{\Upsilon}'_1)|,
|\frac{d}{d\Gamma_{\phi}} \tilde{\cal A}_q(\tilde{\Upsilon}_{3,off})|
\lesssim \frac{m^*}{\mu}
\EEQ
where $N_{j_{\phi}}:=2^{-j_{\phi}}$, see \S \ref{subsection:angular}.
\end{Lemma}

\Medskip{\bf Proof.} The bound on $\tilde{\cal A}_q(\tilde{\Upsilon}_1)$ derives by
simple averaging
from the bound in $O(m^*)$ showed for ${\cal A}_{p_1-p_3}(\Upsilon_1)$. 
Consider now the $r$-momentum scale $j$ contribution to $\tilde{\cal A}_q(\tilde{\Upsilon}'_1)$ or $\tilde{\cal A}_q(\tilde{\Upsilon}_{3,off})$: one finds $\Gamma^2$ or
$|\Gamma|^2$ times
\BEA &&  I^j:=\frac{1}{\Vol(\Sigma^*_F))^2} \int_{\Sigma^*_F\times\Sigma^*_F} dp_1\, dp_3\,  \int dr\, \chi^j(r) \nonumber\\
&&\qquad  \frac{1}{(p_1^0-p_3^0+r^0)^2+(e^*_{|\Gamma|}(\vec{p}_1-\vec{p}_3+
\vec{r}))^2}  
\frac{1}{(r^0)^2+(e^*_{|\Gamma|}(\vec{r}))^2}
\EEA
Briefly said, both propagators have a scale $j_{\phi}$ infra-red cut-off; 
main contribution comes a priori from the region 
\BEQ |r|,|p_1-p_3+r|\approx \Gamma, \label{eq:constraint-on-p1-p3} \EEQ
 in which the product of the two propagators by  the integration volume $\int dr \ \chi^j(r)\approx m^* \Gamma^2$  yields $O(\frac{m^*}{|\Gamma|^2})$. Taking into account the
prefactor $\Gamma^2$ or $|\Gamma^2|$, the missing prefactor $N_{j_{\phi}}=2^{-j_{\phi}}$ comes
from the constraint (\ref{eq:constraint-on-p1-p3}) on $p_3$. For a
more precise computation, one notes that the sector $\alpha\in\Z/2^j\Z$ of $\vec{r}$ is fixed by the sectors $\alpha_1,\alpha_3\in\Z/2^j\Z$ of $\vec{p}_1,\vec{p}_3$. Hence
\BEA  \sum_{j\le j_{\phi}} I^j &\lesssim& \sum_{j\le j_{\phi}} \Big(2^{-2j} \sum_{\alpha_1,\alpha_3\in\Z/2^j\Z}\Big)\ (2^{-3j}\mu^2 m^*) \ 
(2^{-j}\mu)^{-4} \nonumber\\
&\lesssim & \sum_{j\le j_{\phi}} 2^j \mu^{-2}m^*\lesssim 2^{j_{\phi}} \mu^{-2}m^*\approx m^*/(N_{j_{\phi}}|\Gamma|^2).
\EEA

\Medskip The bounds for $\tilde{\cal A}_q(\tilde{\Upsilon}_{4,\eps})$, 
$\eps$=diag,off,  follow from similar bounds for
the quantities ${\cal A}_{\tilde{\Upsilon}_{\eps}}(p_1,-p_q;p_3,-p_3-q)$, themselves
consequences of the general bounds of section \ref{section:fermion}, in particular
of the subsection (\S \ref{subsection:complementary}) on the complementary
$1/N$-expansion. As in the case of $\tilde{\Upsilon}'_1$ or $\tilde{\Upsilon}_{3,off}$, in the case of scale $j$ inner momenta, one gets compared to the
naive power-counting estimates in $O(m^*)$ two supplementary small prefactors, 
(i) one in $O(1/N_j)$, $N_j:=2^{-j}$; (ii) and one in $O(2^{-(j_{\phi}-j)})$ for
each symmetry-broken propagator since $\frac{|\Gamma|}{(p^0)^2+(e^*_{|\Gamma|}(\vec{p})} \approx \frac{|\Gamma|}{|p|} \, \cdot\, \frac{1}{|p|}$. Multi-scale diagrams
with inner momenta of scales varying between $j_{min}$ and $j_{max}$, $j_D\le 
j_{min}<j_{max}\le j_{\phi}$ enjoy a supplementary "spring factor" $O(2^{-(j_{max}-j_{min})})$, allowing a multi-scale generalization of the previous argument.

\Medskip Finally, the bound for $\frac{d}{d\Gamma_{\phi}} \tilde{\cal A}_q(\tilde{\Upsilon}_1)$, say, may be obtained along the same line by noting that
$\Gamma_{\phi} |\frac{d}{d\Gamma_{\phi}}  C^*_{diag}(p)|\approx \Gamma_{\phi}\Big|\frac{d}{d\Gamma_{\phi}} 
\Big(\frac{1}{-\II p^0+e^*_{|\Gamma|}(\vec{p})} \Big) \Big| \lesssim \frac{\Gamma_{\phi}}{|p|^2+\Gamma^2}$ is bounded by a symmetry-broken propagator.
Thus, using similar arguments as for ${\cal A}_q(\tilde{\Upsilon}'_1)$, one
obtains a bound in $O(\frac{1}{\Gamma_{\phi}} \ \cdot\ (m^*/N_{j_{\phi}})) \approx
\frac{m^*}{\mu}$. The bound for $\frac{d}{d\Gamma} \tilde{\cal A}_q(\tilde{\Upsilon}'_1)$ and $\frac{d}{d\Gamma} \tilde{\cal A}_q(\tilde{\Upsilon}_{3,off})$ follows more directly from the inequality
$|\frac{d}{d\Gamma} C^*_{diag}(p)|\lesssim \frac{C^*_{diag}(p)}{|\Gamma|}$ by
noting that $N_{j_{\phi}}\Gamma\approx\mu$.

 \hfill \eop

\Bigskip  In order to get the leading behavior of $\tilde{\Sigma}$ in the infra-red limit $(q\to 0)$,  we first want to {\bf displace the lower external
legs}, namely, those on  
 the second line, $\xi'_4,\xi''_2,\xi''_4,\xi''_2,\cdots$ of
Fig. \thesubsection.4, to the
corresponding locations on the first line $\xi'_3,\xi''_1,
\xi''_3,\xi'''_1,\cdots$, so that intermediate ladders form
{\em bubbles} inserted between $\tilde{\Pi}$-kernels, see
Fig. \thesubsection.5 below. To this end, we need to specify
both the scales $j'_i,j''_i,\cdots$ and the angular sector
indices $\alpha'_i,\alpha''_i,\cdots$, $i=2,4$, of the
{\em lower} momenta entering and exiting the $\tilde{\Pi}$-kernels,
see  (\ref{eq:Cjalpha}); the Bethe-Salpeter kernel is
then obtained by summing over  indices of all lower momenta.
Next, we rewrite {\em lower} external propagators of, say, $\tilde{\Pi}(\xi'_1,\xi'_2;\xi'_3,\xi'_4)$ 
\BEQ C^{j'_2,\alpha'_2}(\xi_4-\xi'_2) C^{j'_4,\alpha'_4}(\xi'_4-\xi''_2) \label{eq:debut-displacement}
\EEQ
as convolutional squares,
\BEQ \Big(\int dy'_2\, \sqrt{C^{j'_2,\alpha'_2}}(\xi_4-y'_2) 
\sqrt{C^{j'_2,\alpha'_2}}(y'_2-\xi'_2)\Big)
\Big(\int dy'_4\, \sqrt{C^{j'_4,\alpha'_4}}(\xi'_4-y'_4) 
\sqrt{C^{j'_4,\alpha'_4}}(y'_4-\xi''_2)\Big).
\EEQ
Then we displace lower external legs $\xi'_2,\xi'_4$ to
the corresponding upper locations $\xi'_1,\xi'_3$, 
while taking into account the oscillations.  The simplest {\bf displacement procedure},
independent -- contrary to the one used for renormalization in 
\S \ref{subsection:dressing}, see e.g. (\ref{eq:R1}) -- of a choice of angular sector momenta $p^{k,\alpha}$,  
is done in Fourier coordinates. Namely, fix the momenta
$p'_1,p'_2=-p'_1+q,p'_3,p'_4=-p_3-q$
of  the upper external propagators
 $C^{j'_1,\alpha'_1}(p'_1), C^{j'_3,\alpha'_3}(p'_3)$, and 
  of the lower external propagators,
 \BEQ \int dy'_4\,  e^{-\II(p'_2,y'_4)} \sqrt{C^{j'_2,\alpha'_2}}(y'_4-\xi'_2) , 
 \int dy'_2\,  e^{\II(p'_4,y'_2)}  \sqrt{C^{j'_4,\alpha'_4}}(\xi'_4-y'_2). \label{eq:lower-ext-prop} 
 \EEQ 
   We may restrict to $p'_i\in {\cal S}^{j_i,\alpha_i}$, $i=1,\ldots,4$  (otherwise $C^{j'_i,\alpha'_i}(p'_i)=0$ 
 by construction, see Definition \ref{def:angular}).  Then
 we change integration variables in (\ref{eq:lower-ext-prop}), $y'_4\to y'_4+(\xi'_1-\xi'_2)$, 
 $y'_2\to y'_2+(\xi'_3-\xi'_4)$, so that $\sqrt{C^{j'_2,\alpha'_2}}(y'_4-\xi'_2)\longrightarrow  \sqrt{C^{j'_2,\alpha'_2}}(y'_4-\xi'_1)$, $\sqrt{C^{j'_4,\alpha'_4}}(\xi'_4-y'_2) \longrightarrow  \sqrt{C^{j'_4,\alpha'_4}}(\xi'_3-y'_2)$
 are now attached to the $\tilde{\Pi}$-kernel at the locations of the {\em upper} vertices $\xi'_1,\xi'_3$.  Next, we {\em replace} the resulting extra oscillations $e^{\II(p'_2,\xi'_2-\xi'_1)}$, 
 $e^{-\II(p'_4,\xi'_4-\xi'_3)}$ by $1$, and rewrite 
 $\int dy'_4 e^{-\II(p'_2,y'_4)} \sqrt{C^{j'_2,\alpha'_2}}(y'_4-\xi'_1)$, resp. 
 $\int dy'_2\, e^{\II(p'_4,y'_2)} \sqrt{C^{j'_4,\alpha'_4}}(\xi'_3-y'_2)$, as $e^{\II(p'_2,\xi'_1)} \sqrt{C^{j'_2,\alpha'_2}}(p'_2)$, resp. $e^{-\II(p'_4,\xi'_2)} \sqrt{C^{j'_4,\alpha'_4}}(p'_4)$.
 The remaining oscillations $e^{\II(p'_2,\xi'_1)},e^{-\II(p'_4,\xi'_2)}$ may now be attached to the partially Fourier transformed kernel
$$  \tilde{\Pi}^{disp}_q(p'_1,p'_3;\xi'_1,\xi'_3):=
e^{\II(-p'_1+q,\xi'_1)} \Pi(\xi'_1,-p'_1+q;\xi'_3,-p'_3-q)
e^{\II(p'_3+q,\xi'_3)}.$$
By momentum conservation, the above kernel depends only on
$\xi'_1-\xi'_3$, so  its integral
\BEQ \boxed{\tilde{\Pi}^{disp}_q(p'_1,p'_3):=\int d\xi'_3\, 
e^{\II(-p'_1+q,\xi'_1)} \Pi(\xi'_1,-p'_1+q;\xi'_3,-p'_3-q)
e^{\II(p'_3+q,\xi'_3)}}
\label{eq:Pidisp}
\EEQ
is independent of $\xi'_1$. We call it the {\bf displaced
kernel}.  {\em Error terms} due to extra oscillations $e^{\II(p'_2,\xi'_2-\xi'_1)}$, 
 $e^{-\II(p'_4,\xi'_4-\xi'_3)}$ 
involve quantities 
\BEQ \approx \del\xi'\, \cdot\, 
\nabla_{\xi'}(e^{-\II(p',\xi')} \sqrt{C^{j',\alpha'}}(\xi'-y')) 
\sim \del\xi'\, \cdot\,  (\nabla_{\xi'}-\II p^{\alpha'}) \sqrt{C^{j',\alpha'}}(\xi'-y') \label{eq:CCerror}
\EEQ
 or conjugate, with $\del\xi'=\xi'_1-\xi'_2$,
 $y'=y'_2$, $\xi'=\xi'_2$, resp.    $\del\xi'=\xi'_3-\xi'_4$,
  $y'=y'_4$, $\xi'=\xi'_4$. The derivative $\nabla_{\xi'}-\II p^{\alpha'}$ 
 acting on a sector covariance $C^{j,\alpha'}$ or a micro-sector
 covariance $C^{j,\alpha',k}$ 
 generates (see (\ref{eq:I-II}), (\ref{eq:I-IIbis}) or
 (\ref{eq:I-IIter})) a small prefactor  $2^{-j_{ext}}$ proportional to the diameter of the (micro-)sector,
 where $j_{ext}=j$ or (in the case of a micro-sector) $2^{-j_+}$,
 while the  displacement distance $\del\xi'$, associated with
 the decay factors of the two-particle irreducible diagram,
 produces a factor $\lesssim 2^{j_{int}}$, where $j_{int}$ is the lowest
 internal scale of the diagram,  all together   an {\em extra
small prefactor} 
\BEQ O(2^{-(j_{ext}-j_{int})})  \label{eq:jextjint} \EEQ
playing the r\^ole of a spring factor in section \ref{section:boson}, see in particular \ref{subsection:complementary} {\bf D.}

\Medskip Note that the $\sqrt{C}$-kernels are now attached to the
 {\em upper} external vertices of $\tilde{\Pi}$. Contracting them with their counterparts yields pairs of $C$-kernels
 connecting upper vertices,

 \Bigskip 
 
{\centerline{ 
\begin{tikzpicture}
\draw(0,2.3) node {$\xi'_1$}; \draw(2,2.3) node {$\xi'_3$};
\draw(3,2.3) node {$C^{\alpha'_3}$}; \draw(3,1.5) node {$C^{\alpha'_4}$}; \draw(4,2.3) node {$\xi''_1$};
\draw(6,2.3) node {$\xi''_3$};
\draw(0,-0.3) node {$\xi'_2$}; \draw(2,-0.3) node {$\xi'_4$};
\draw(4,-0.3) node {$\xi''_2$}; \draw(6,-0.3) node {$\xi''_4$};
\draw[dashed](0,2) arc(-60:-90:2 and 2);
\draw[dashed](-1,2)--(0,2);\draw(-0,2)--(6,2);
\draw[dashed](6,2)--(7,2);
\draw[fill=gray](0,0) rectangle(2,2);
\draw[fill=gray](4+0,0) rectangle(4+2,2);
\draw(2,2) arc(-120:-60:2 and 2);
\draw[dashed](6,2) arc(-120:-90:2 and 2);
\end{tikzpicture}}}

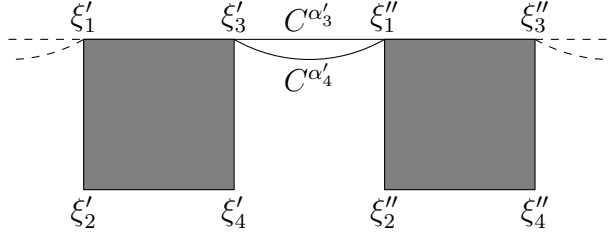
\Medskip {\bf \tiny Fig. \thesubsection.5. Chains
contributing to $\tilde{\Sigma}^{disp}$. Gray squares stand
for $\tilde{\Pi}^{disp}$-kernels.}

\Bigskip 
This procedure leads to a kernel 
 $\tilde{\Sigma}^{disp}$
 made up of kernels $\tilde{\Pi}_q^{disp}(p_1,p_3)\equiv 
\tilde{\Pi}^{disp}(p_1,-p_1+q;p_3,-p_3-q)$,  alternating with pairs of
propagators
$C^*_{\theta}(p'_3)C^*_{\theta}(q-p'_3),\  C^*_{\theta}(p''_3)C^*_{\theta}(q-p''_3),\cdots$, which may be arrayed into  the  Hermitian matrix\\ ${\cal A}_q(p_3):=\left(\begin{array}{cc} {\cal A}_q(\Upsilon_{3,diag};p_3) & {\cal A}_q(\Upsilon_{3,off};p_3) \\ {\cal A}_q^*(\Upsilon_{3,off};p_3) &
{\cal A}_q(\Upsilon_{3,diag};p_3) \end{array}\right)$, $p_3=p'_3,p''_3,\cdots$ Note
that $p_3$-momenta  {\em cannot} be  integrated over in  ${\cal A}_q(p_3)$  since
the $\tilde{\Pi}^{disp}$-kernels also depend on them, {\em 
except} for the leading-order contribution to $\tilde{\Sigma}^{disp}$, which is proportional to the  Hermitian bubble matrix ${\cal A}_q(\Upsilon_3)$ of eq. (\ref{eq:sandwich-kernel-A}). 

\begin{Definition}    {\bf ($\Sigma$-kernel, preliminary version)} 
\label{def:Sigmadisp}
 Let
\BEA && \tilde{\Sigma}^{disp}(q):=\lambda {\mathbb{1}}+ \lambda^2 \Big\{ {\cal A}_q(\Upsilon_3)+ \int dp'_1 \int dp'_3\ 
 {\cal A}_q(p'_1) \tilde{\Pi}^{disp}_q(p'_1,p'_3) {\cal A}_q(p'_3)
 \nonumber\\
 &&\qquad
+ \int dp'_1 \int dp'_3 \int dp''_1 \int dp''_3\ \del(p''_1-p'_3)\  {\cal A}_q(p'_1) \Big(\tilde{\Pi}^{disp}_q(p'_1,p'_3)
 {\cal A}_q(p''_1)\Big)  \ \Big( \tilde{\Pi}^{disp}_q(p''_1,p''_3)
 {\cal A}_q(p''_3)\Big) \nonumber\\
&&
+\cdots+ \    \Big\{\prod_{i=1}^n \int dp_{i,1} \int dp_{i,3}\Big\} \ \Big( \prod_{i=1}^{n-1} \del(p_{i,3}-p_{i+3,1})\Big)\ 
 {\cal A}_q(p_{1,1})  \Big\{ \prod_{i=1}^n  \Big(\tilde{\Pi}_q^{disp}(p_{i,1},p_{i,3}){\cal A}_q(p_{i,3}) \Big)\Big\} \nonumber\\
 && + \cdots
 \nonumber\\   \label{eq:Sigmadispkernel}
\EEA
\end{Definition}

\Bigskip As stated below (\ref{eq:CCerror}),  the above displacement procedure neglects error
terms smaller by a prefactor $O(2^{-(j_{ext}-j_{int})})$ that goes to $0$ with $q$. This requires however
 angular (micro-)sectors
whose thickness in all directions is comparable to $q$; this is explained in details in section \ref{section:boson}.

\Bigskip {\bf The Fermi and $s$-wave projections.} The above description of the
$\tilde{\Sigma}^{disp}$ makes it plain  that the two-particle irreducible
contributions $\tilde{\Pi}_{disp}$ are not  factorized from the bubbles, making the
above expressions intractable and not ready for use in the gap equation. In order
to obtain a factorization, while keeping the leading-order contribution when $q\to 0$, we take
two further steps.

\begin{itemize}
\item[(i)] {\bf (Fermi projection)}  We {\em replace external
momenta $p'_1,p'_3,p''_1,p''_3,\cdots$ by their orthogonal
projection $p'_{1,F}=(0,\vec{p}'_{1,F}), p'_{3,F}=(0,\vec{p}'_{3,F}), p''_{1,F}=(0,\vec{p}''_{1,F}), p''_{3,F}=(0,\vec{p}''_{3,F}),\cdots$  onto the Fermi circle $\Sigma^*_F$.} This
defines the {\em Fermi projected kernel} $\tilde{\Pi}^{proj}$,
\BEQ \tilde{\Pi}^{proj}_q(p_1,p_3):=\tilde{\Pi}^{disp}_q(p_{1,F},p_{3,F}).
\label{eq:PitildePi}
\EEQ

As mentioned before, the correct procedure, described in section \ref{section:boson},
is rather an averaging procedure in the neighborhood of the Fermi circle, but it does not matter
 at this stage.

\item[(ii)]  {\bf ($s$-wave projection)} 
 {\em If $q=0$,}  momenta
$(\vec{p}_3,\vec{p}_4)=(\vec{p}_3,-\vec{p}_3)$ are obtained from $(\vec{p}_1,\vec{p}_2)=(\vec{p}_1,-\vec{p}_1)$ by a rotation of 
angle $\theta$; the kernel $\tilde{\Pi}^{proj}_0(p_1,p_3)$ depends only
on $|\vec{p}_1|,|\vec{p}_3|$ and $\theta$, not separately on
the angular directions of $\vec{p}_1$ and $\vec{p}_3$. (For $q\not=0$ but close to $0$, this is only 
approximately true.) Let $\theta'_1,\theta''_1=\theta'_3,\theta'''_1=\theta''_3,\ldots$ be the angles of $\vec{p}'_1,
\vec{p}''_1=\vec{p}'_3,\vec{p}'''_1=\vec{p}''_3,\ldots$ w.r. to a fixed direction
$\vec{e}_1$, and $\theta:=\theta'_1-\theta''_1=\theta'_1-\theta'_3; \theta':=\theta''_1-\theta'''_1=\theta''_1-\theta''_3,\ldots$
(The definition makes sense also for $q\not=0$.) {\em Assume
first that $q=0$.}
 Bubble diagrams have by construction $\theta=0$. Hence -- taking a Fourier transform -- the
convolution of kernels turns into a geometric series which may be resummed for each individual Fourier mode $k\in\Z$, 
different Fourier modes being prevented from interacting by
orthogonality. From the arguments
in the Introduction, it may be conjectured that only 
the mode $k=0$ (or {\em $s$-wave}) diverges in the infra-red
limit, which is itself a consequence that the interaction
vertex itself is in the $s$-wave. Thus the kernel
$\tilde{\Pi}^{proj}_0(p_{1,F},p_{3,F})$ reduces in the end to the averaged 
$2\times 2$ matrix $\bar{\Pi}_0=\frac{1}{(\Vol(\Sigma^*_F))^2} \int_{\Sigma^*_F\times\Sigma^*_F} dp_1\, dp_3\, 
\Pi_0(p_1,p_3)$. {\em When $q\not=0$}, we introduce similarly the averaged $2\times 2$ matrix 
\BEQ \bar{\Pi}(q)\equiv\bar{\Pi}_q:=\frac{1}{(\Vol(\Sigma^*_F))^2} \int_{\Sigma^*_F\times\Sigma^*_F} dp_1\, dp_3\, 
\tilde{\Pi}^{proj}_q(p_1,p_3). \label{eq:Pibar(q)} 
\EEQ
\end{itemize}
 
\Medskip Now, bubbles may be factorized, yielding
 
 \begin{Definition}    {\bf ($\Sigma$-kernel,  or Goldstone boson propagator)} \label{def:Sigma}
Let
\BEQ  \Sigma(q):=\lambda {\mathbb{1}}+ \lambda^2 \Big\{ {\cal A}(q)+ {\cal A}(q)\bar{\Pi}(q){\cal A}(q)+ {\cal A}(q)\bar{\Pi}(q){\cal A}(q)\bar{\Pi}(q){\cal A}(q)+\cdots\}.
\label{eq:Sigma-kernel}\EEQ
\end{Definition}

where $ {\cal A}(q)={\cal A}_q(\Upsilon_3)$ is 
the Hermitian  bubble matrix of (\ref{eq:sandwich-kernel-A}),
which we represent for simplicity as
a bubble 
\begin{tikzpicture}[scale=1]
\draw(2,0) arc(30:150:0.88  and 0.44);
\draw(0.5,0) arc(210:330:0.88 and 0.44);
\end{tikzpicture}
:   graphically (compare to Fig. \thesubsection.5)
 
\medskip
\begin{tikzpicture}[scale=1]
\draw(-1.5,0) node {${\Sigma}(q) \, =\, \lambda{\mathbb{1}}  \, +\, \lambda^2\ \Big\{$};
\draw(2,0) arc(30:150:0.88  and 0.44);
\draw(0.5,0) arc(210:330:0.88 and 0.44);
\draw(0.5,0) node {\textbullet};
\draw(2,0) node {\textbullet};

\draw(2.5,0) node {$+$};

\draw(2+3,0) arc(30:150:0.88  and 0.44);
\draw(0.5+3,0) arc(210:330:0.88 and 0.44);
\draw(0.5+3,0) node {\textbullet};
\draw(2+3,0) node {\textbullet};

\draw(5.5,0) node {$\bar{\Pi}$};

\draw(6+2,0) arc(30:150:0.88  and 0.44);
\draw(6+0.5,0) arc(210:330:0.88 and 0.44);
\draw(6+0.5,0) node {\textbullet};
\draw(6+2,0) node {\textbullet};

\draw(10,0) node {$+\ \cdots \ \Big\}(q)$};
\end{tikzpicture}

\Bigskip Let us consider specifically the case of
{\em zero transfer momentum} $(q=0)$. Then,  replacing 
$\tilde{\Pi}^{disp}_q(p_1,p_3)$  by
 $\tilde{\Pi}^{proj}_q(p_1,p_3)$ in (\ref{eq:Sigmadispkernel}), the $s$-wave projection is exact, and defines
\BEQ \Sigma(0)=\lambda {\mathbb{1}}+\lambda^2 \Big\{
{\cal A}_0+ {\cal A}_0 \bar{\Pi}_0 {\cal A}_0+\cdots\Big\}=
\lambda \Id+\frac{\lambda^2 {\cal A}_0 }{
\Id-\bar{\Pi}_0{\cal A}_0 },
\EEQ
where ${\cal A}_0={\cal A}_0(\Gamma,\Upsilon_3)$ and 
\BEQ \bar{\Pi}_0=\bar{\Pi}_0(\Gamma):=\frac{1}{(\Vol(\Sigma^*_F))^2}\int \int_{\Sigma^*_F\times\Sigma^*_F} dp_1\, dp_3\,  
\tilde{\Pi}^{disp}_0(p_1,p_3).
\EEQ

\Medskip 
First-order terms are
\Bigskip

{\center{\begin{tikzpicture}[scale=0.8]
\draw(-1,0) node {$\lambda{\mathbb{1}} \, +\, \lambda^2$};

\draw(0.5,0) node {$\bullet$}; \draw(1.5,0) node {$\bullet$};
\draw(1.5,0) arc(30:150:0.60  and 0.44);
\draw(0.5,0) arc(210:330:0.60 and 0.44);

\draw(2.5,0) node {$+\, \lambda^3\, $};

\draw(0+3.5,0) node {$\bullet$}; 
\draw(1+3.5,0) arc(30:150:0.60  and 0.44);
\draw(0+3.5,0) arc(210:330:0.60 and 0.44);
\draw(0+4.5,0) node {$\bullet$}; \draw(1+4.5,0) node {$\bullet$};
\draw(1+4.5,0) arc(30:150:0.60  and 0.44);
\draw(0+4.5,0) arc(210:330:0.60 and 0.44);

\draw(7,0) node {$+\, \lambda^4 \Big($};

\draw(-1.5+0.5+9,0) node {$\bullet$}; \draw(-1.5+1.5+9,0) node {$\bullet$};
\draw(-1.5+1.5+9,0) arc(30:150:0.6  and 0.44);
\draw(-1.5+0.5+9,0) arc(210:330:0.6 and 0.44);
\draw(-1.5+10.5,0) node {$\bullet$}; \draw(-1.5+2+10.5,0) node {$\bullet$};
\draw(-2+1.5+10.5,0) arc(30:150:0.6  and 0.44);
\draw(-2+0.5+10.5,0) arc(210:330:0.6 and 0.44);
\draw(-2.5+0.5+12,0) node {$\bullet$}; \draw(-2.5+1.5+12,0) node {$\bullet$};
\draw(-2.5+1.5+12,0) arc(30:150:0.6  and 0.44);
\draw(-2.5+0.5+12,0) arc(210:330:0.6 and 0.44);

\draw(-3+15,0) node {$+\, $};

\draw(-3.5+0.5+16,0) node {$\bullet$}; \draw(-3.5+2+16,0) node {$\bullet$};
\draw(-3.5+2+16,0) arc(30:150:0.88  and 0.66);
\draw(-3.5+0.5+16,0) arc(210:330:0.88 and 0.66);
\draw[rotate=90](0.3,3.5-1.25-16) arc(30:150:0.33  and 0.33);
\draw[rotate=90](0.3,3.5-1.25-16) arc(330:210:0.33  and 0.33);
\draw(-3.5+1.25+16,0.25) node {$\bullet$}; \draw(-3.5+1.25+16,-0.25) node {$\bullet$};

\draw(15,0) node {$\Big)$};

\end{tikzpicture}}}

\begin{tikzpicture}[scale=0.8]
\draw(6,0) node {$ +\, \lambda^5$}; \draw(6.75,0) node {$\Big($};
\draw(-1.5-0.5+9,0) node {$\bullet$};
\draw(-1.5+0.5+9,0) node {$\bullet$}; \draw(-1.5+1.5+9,0) node {$\bullet$};
\draw(-1.5+0.5+9,0) arc(30:150:0.6  and 0.44);
\draw(-1.5-0.5+9,0) arc(210:330:0.6 and 0.44);
\draw(-1.5+1.5+9,0) arc(30:150:0.6  and 0.44);
\draw(-1.5+0.5+9,0) arc(210:330:0.6 and 0.44);
\draw(-1.5+10.5,0) node {$\bullet$}; \draw(-1.5+2+10.5,0) node {$\bullet$};
\draw(-2+1.5+10.5,0) arc(30:150:0.6  and 0.44);
\draw(-2+0.5+10.5,0) arc(210:330:0.6 and 0.44);
\draw(-2.5+0.5+12,0) node {$\bullet$}; \draw(-2.5+1.5+12,0) node {$\bullet$};
\draw(-2.5+1.5+12,0) arc(30:150:0.6  and 0.44);
\draw(-2.5+0.5+12,0) arc(210:330:0.6 and 0.44);

\draw(-3+15,0) node {$+\, $};

\draw(-3.5+0.5+16,0) node {$\bullet$}; \draw(-3.5+2+16,0) node {$\bullet$};
\draw(-3.5+2+16,0) arc(30:150:0.88  and 0.66);
\draw(-3.5+0.5+16,0) arc(210:330:0.88 and 0.66);
\draw[rotate=90](0.3,3.5-1.25-16) arc(30:150:0.33  and 0.33);
\draw[rotate=90](0.3,3.5-1.25-16) arc(330:210:0.33  and 0.33);
\draw(-3.5+1.25+16,0.25) node {$\bullet$}; \draw(-3.5+1.25+16,-0.25) node {$\bullet$}; 

 \draw(-3.5+3+16,0) node {$\bullet$};
\draw(-3.5+3+16,0) arc(30:150:0.6  and 0.44);
\draw(-3.5+2+16,0) arc(210:330:0.6 and 0.44);

\draw(16.5,0) node {$+\, $};

 \draw(17.5,0) node {$\bullet$};
\draw(18.5,0) arc(30:150:0.6  and 0.44);
\draw(17.5,0) arc(210:330:0.6 and 0.44);
\draw(0.5+18,0) node {$\bullet$}; \draw(2+18,0) node {$\bullet$};
\draw(2+18,0) arc(30:150:0.88  and 0.66);
\draw(0.5+18,0) arc(210:330:0.88 and 0.66);
\draw[rotate=90](0.3,-1.25-18) arc(30:150:0.33  and 0.33);
\draw[rotate=90](0.3,-1.25-18) arc(330:210:0.33  and 0.33);
\draw(1.25+18,0.375) node {$\bullet$}; \draw(1.25+18,-0.375) node {$\bullet$}; 

\draw(21,0) node {$+$};

\draw(22,0) node {$\bullet$};
\draw(22,0) arc(150:90:1  and 1);
\draw(22,0) arc(210:270:0.6 and 1);
\draw(22.9,0.5) node {$\bullet$};
\draw(22.65,-0.5) node {$\bullet$};
\draw(22.9,0.5)--(22.65,-0.5); \draw(22.9,0.5)--(23.15,-0.5);
\draw(23.15,-0.5) node {$\bullet$};
\draw(22.9,0.5) arc(90:30:1 and 1);
\draw(23.15,-0.5) arc(270:330:0.65 and 1);
\draw(23.75,0) node {$\bullet$};
\draw(22.65,-0.5) arc(150:30:0.3 and 0.22);
\draw(22.65,-0.5) arc(210:330:0.3 and 0.22);

\draw(24.5,0) node {$\Big)$};
\end{tikzpicture}


\Medskip where propagators are either diagonal or 
off-diagonal.

\Bigskip
  The  one-point counterterm coefficient $\Gamma\in \C$  is chosen to be {\em the solution with largest  module} of the 

\begin{Definition}[gap equation] \label{def:gap}
\BEQ  \boxed{\Big(-{\mathbb{1}}+\bar{\Pi}_0(\Gamma){\cal A}_0(\Gamma)\Big) 
\left(\begin{array}{c} \Gamma^{\perp}\\
(\Gamma^{\perp})^* \end{array}\right) =0,}  \label{eq:gap-0} \EEQ
\end{Definition}

\noindent i.e. $ \left(\begin{array}{c} \Gamma^{\perp}\\
(\Gamma^{\perp})^* \end{array}\right)$ is a null eigenvector of the matrix $-{\mathbb{1}}+\bar{\Pi}_0{\cal A}_0$. 
Recall $\Gamma^{\perp}:=\Gamma(\theta+\frac{\pi}{2})$.  Graphically,

\bigskip

{\centerline{
\begin{tikzpicture}[scale=0.5]
\draw[color=blue](-1,3)--(19.5,3); \draw[color=blue](-1,-3.5)--(19.5,-3.5); \draw[color=blue](-1,3)--(-1,-3.5); 
\draw[color=blue](19.5,3)--(19.5,-3.5);
\draw(0,0) node {$-$};  
\draw(2,0)--(7,0);
\draw[fill=black](3.9,1.9) rectangle(4.1,2.1); \draw(4.6,2) node{$\Gamma$};
\draw[dashed](4,2)--(4,-2);
\draw[decorate,decoration=snake,->](2,-2)--(3,-2);
\draw[decorate,decoration=snake](3,-2)--(4,-2);
 \draw(3,-1.2) node {$\Sigma_{\perp,\perp}$};
 \draw(3,-2.7) node {\tiny $q=0$};
\draw(8.5,0) node {$+$};
\draw(10,0)--(16.5,0);
\draw[dashed](12,2)--(12,-2);
\draw[decorate,decoration=snake,->](10,-2)--(11,-2); 
\draw[decorate,decoration=snake](11,-2)--(12,-2); 
\draw(11,-2.7) node {\tiny $q=0$};\draw(11,-1.2) node {$\Sigma_{\perp,\perp}$}; \draw(12,-2) node {\textbullet};
\draw(12,2) node {\textbullet}; \draw(13.5,2) node { $\bar{\Pi}_{\perp,\perp}$}; 
\draw(14.5,2) node {\textbullet};
\draw(16.5,2) arc(30:150:1.2 and 1); \draw(14.5,2) arc(210:330:1.2 and 1);
 \draw(16,2) node {\tiny ${\mathbb{\Gamma}}^{\perp}$};
 \draw(16.5,2) node {\textbullet};
\draw(18,0) node {$=0$};
\end{tikzpicture}
}}

\Medskip {\bf {\tiny Fig. \ref{subsection:Sigma}.6. Gap equation.}}

\Bigskip
By the arguments below {\small Fig. \thesubsection.3}, the matrix $\bar{\Pi}_0 {\cal A}_0$
 is of the form
 $\left(\begin{array}{cc} a & -b\omega \\ -b\omega^{-1} & a\end{array}\right)$ 
  with $\omega:=e^{-2\II\theta}$ if $\Gamma=|\Gamma|\, e^{\II\theta}$. Hence the
matrix $\bar{\Pi}_0{\cal A}_0$ can be made {\em real} by the conjugation 
\BEQ  \bar{\Pi}_0  {\cal A}_0
\mapsto \left(\begin{array}{cc} e^{\II\theta} & \\ & e^{\II\theta}\end{array}\right) 
\bar{\Pi}_0{\cal A}_0
 \left(\begin{array}{cc} e^{\II\theta} & \\ & e^{-\II\theta}\end{array}\right),  \label{eq:conjugation}
\EEQ
 a condition which fixes $\theta$ mod $\pi/2$. Then  eigenvectors $
\left(\begin{array}{c} |\Gamma|\, e^{\II(\theta+\frac{\pi}{2})} \\ |\Gamma| e^{-\II(\theta+\frac{\pi}{2})} \end{array}\right)$, resp.  $
\left(\begin{array}{c} |\Gamma|\, e^{\II\theta} \\ 
|\Gamma|e^{-\II\theta} \end{array}\right)
$ are eigenvalues for two different values, $\Gamma_{\phi}:=\Gamma_{\phi}^+$ and $\Gamma_{\phi}^{-}$, of $|\Gamma|$
 obtained by solving the equations $a-1=\mp b $, or equivalently (conjugating
 as in (\ref{eq:conjugation}) to reduce to the case $\theta=0$)

\BEQ f_{\pm}(g,\Gamma):=\frac{1}{g} \Big\{ \Big(\bar{\Pi}_0 {\cal A}_0-{\mathbb{1}}\Big)_{diag} \mp\Big(\bar{\Pi}_0{\cal A}_0\Big)_{off} \Big\}\equiv 0 
\EEQ
for $\Gamma>0$, yielding {\em two} solutions, $\Gamma_{\phi}^{\pm}$.
 Using previous computations, see (\ref{eq:sinh-1}), (\ref{eq:eval-Upsilon2}), one obtains leading-order
terms as functions of the non-dimensional variables $g,\tilde{\Gamma}:=\frac{\Gamma}{\hbar\omega_D}$:
\BEA {\cal A}_0(\Gamma)&=& {\cal A}_0(\Gamma;\Upsilon_3) - {\cal A}_0(\Gamma;
\Upsilon_{3,off}) \sigma(\theta) \nonumber\\
&=& \frac{m^*}{\pi}\Big( -\log \tilde{\Gamma} + c_0+ O(\tilde{\Gamma})\Big)\, {\mathbb{1}} \, + m b_2(\tilde{\Gamma}) \,
\sigma(\theta)  \label{eq:1.95}
\EEA
with 
\BEQ b_2(\tilde{\Gamma}):=-\frac{1}{m} {\cal A}_0(\Gamma;\Upsilon_{3,off})=b_2+
O(\tilde{\Gamma})=O(1) \qquad  (b_2>0);
\label{eq:b2}
\EEQ

\BEQ \bar{\Pi}_{diag}(0)=\frac{g}{m}(1+a_2(\tilde{\Gamma})g+O(g^2))  \, {\mathbb{1}}, 
\qquad 
\bar{\Pi}_{off}(0)=\frac{g^2}{m} O(\tilde{\Gamma}), 
 \EEQ
see Lemma \ref{lem:lowest-Pi}, 
with $a_2(\tilde{\Gamma}):=\tilde{\cal A}_0(\Gamma;\tilde{\Upsilon}_1)=a_2+
O(\tilde{\Gamma})$. The coefficients $c_0,b_2,a_2$ that we introduced -- and the
coefficients that we introduce below -- are $O(1)$. Finally,
 $m^*=m(1+m_1(\tilde{\Gamma})g^2+O(g^3))$. The dependence on $\tilde{\Gamma}$ of $a_2,b_2$ and $m_1$
is secondary since $\tilde{\Gamma}=O(e^{-\pi/g})$ vanishes to all orders at $g=0$,
implying in particular that $\Pi_0$ is essentially {\em diagonal}. The change of variable $\tilde{\Gamma}\mapsto \gamma:=\tilde{\Gamma} e^{\pi/g}$ allows one to rewrite the function $({\cal A}_0)_{diag}$ as
\BEQ ({\cal A}_0)_{diag}(\gamma)=\frac{m^*}{\pi}\Big(\frac{\pi}{g}-\log\gamma+ c_0+O(\tilde{\Gamma})\Big)\, {\mathbb{1}}  \label{eq:A0-gamma},
\EEQ
 whence
\BEA  \Big(\bar{\Pi}_0{\cal A}_0-{\mathbb{1}}\Big)_{diag} &=& \Big(\frac{m^*}{m} \Big\{ (1-\frac{g}{\pi}\log\gamma +\frac{g}{\pi}c_0+O(\tilde{\Gamma}))(1+a_2 g+O(g^2))\Big\}-1 \Big) \, {\mathbb{1}} \nonumber\\
&=& \Big(g(-\frac{1}{\pi}\log\gamma +m_1+a_2+\frac{c_0}{\pi})+ O(g^2)\Big)\ {\mathbb{1}}
\EEA
\BEQ -\Big(\bar{\Pi}_0 {\cal A}_0\Big)_{off}=\Big(b_2 g+O(g^2)\Big)\sigma(\theta)\sim b_2 g\, \sigma(\theta) \EEQ
with $b_2 >0$  provided $g$ is small enough.  Thus, by the implicit function theorem,
\BEQ \boxed{\Gamma_{\phi}^{\pm}\sim_{g\to 0} \hbar\omega_D e^{-\pi/g} e^{\pi(m_1+a_2+\frac{c_0}{\pi})+O(g)\pm \del}}
\EEQ
with 
\BEQ \del:=\pi b_2+O(g).\EEQ
In particular,
\BEQ \boxed{({\cal A}_0)_{diag}=\frac{1}{\lambda}+O(m^*)}. \label{eq:A0=1/lambda}\EEQ
We note for further use that, {\em choosing $|\Gamma|=\Gamma_{\phi}^+$}, one
gets modulo the conjugation (\ref{eq:conjugation}), or for $\theta=0$,
\BEQ {\mathbb{1}}-\bar{\Pi}_0{\cal A}_0=\left(\begin{array}{cc}
gb & gb \\ gb & gb\end{array}\right) \label{eq:1-APi}
\EEQ
 with $b\sim \pi b_2$, implying a null
eigenvector proportional to $\left(\begin{array}{c} -\II \\ \II \end{array}\right)$. 

\Bigskip Thus the net effect of diagrams other than the Cooper pair bubble $\Upsilon_3$ is to change by a finite amount the pre-factor in front of $\Gamma_{\phi}$ and
produce a finite multiplicative splitting $e^{\pm\del}$ between two solutions.

\Bigskip

\Bigskip

\Bigskip {\bf Estimates for $\Sigma(q)$ near $q=0$.} We first need a Taylor
expansion of ${\mathbb{1}}-\Pi(q){\cal A}(q)$ near $q=0$. Lemma
\ref{lemma:bubble} and power-counting estimates for $\Pi(q)$ imply that
\BEQ 
\Big({\mathbb{1}}-\Pi_{diag}(q){\cal A}_q(\Gamma_{\phi}^+;\Upsilon_{3,i}) \Big)=\Big({\mathbb{1}}-\Pi_{diag}(0){\cal A}_0(\Gamma_{\phi}^+;\Upsilon_{3,i})  \Big)+ \lambda Q_{i}(q)+ O(|q|_+^4),
\EEQ
$i$=diag,off, 
 where $\left(\begin{array}{c} Q_{diag} \\ Q_{off} \end{array}\right)$
  is a couple of positive-definite quadratic forms,  $Q_{diag}(q)=
  \frac{1}{g_{\phi,diag}} (q^0)^2+ \frac{v^2_{\phi,i}}{g_{\phi,diag}}
   |\vec{q}|^2$, $Q_{off}(q)=-\big(
  \frac{1}{g_{\phi,off}} (q^0)^2+ \frac{v^2_{\phi,off}}{g_{\phi,off}}
   |\vec{q}|^2\big)$, see
 Lemma \ref{lemma:bubble}, with 
 \BEQ g_{\phi,i}=g^0_{\phi,i}(1+O(g)), v_{\phi,i}=v^0_{\phi,i}(1+O(g))
 \label{eq:gphivphi}.
 \EEQ
  Remaining terms involving $\Pi_{off}(q)$ have an extra
 $1/N_{j_{\phi}}=2^{-j_{\phi}}$ prefactor.  Using 
(\ref{eq:1-APi}), one gets for $\theta=0$
\BEQ  \det({\mathbb{1}}-\Pi(q){\cal A}(q)) =: \det M(q) \sim 2b \frac{g^2}{m}  Q(q) \EEQ
where
\BEQ  M(q):=\Big( \begin{array}{cc}  g\Big(b+\frac{1}{m} Q_{diag}(q)+ O(|q|_+^4))\Big) & g\Big(b+ \frac{1}{m} Q_{off}(q) +O(|q|_+^4) \Big) \\
g\Big(b+ \frac{1}{m} Q_{off}(q) + O(|q|_+^4) \Big) & 
g\Big(b+
\frac{1}{m} Q_{diag}(q) + O(|q|_+^4) \Big)  \end{array}
\Big)  \label{eq:M}
\EEQ
and $Q(q):=Q_{diag}(q)-Q_{off}(q)=\frac{|q|_+^2}{g_{\phi}}$ by Definition
\ref{def:gphivphi}. At this point it is natural to
shift to the $(//,\perp)$-basis (compare with
\S \ref{subsection:gap-lowest}), in which $M(q)$=diag$(M_{//,//}(q),M_{\perp,\perp}(q))$, with 
\BEQ M_{//,//}(q)=\sum_{i,i'=1,2} M_{i,i'}(q)\sim_{|q|_+\to 0} 4gb, \EEQ
\BEQ
\qquad M_{\perp,\perp}(q)=\big(\sum_{i=i'} - \sum_{i\not=i'}
\big) M_{i,i'}(q)\sim_{|q|_+\to 0} 2\frac{g}{m}Q(q)=2\frac{g}{m}
\frac{|q|_+^2}{g_{\phi}}.
\EEQ 
Multiplying by $ \lambda^2 ({\cal A}_0)_{//,//}, 
\lambda^2 ({\cal A}_0)_{\perp,\perp}\sim \lambda$ yields
(see (\ref{eq:Sigma-kernel}), (\ref{eq:1.95}), (\ref{eq:A0=1/lambda}))
\BEQ \Sigma_{//,//}(q)\approx \frac{1}{m^*} \EEQ
and
\BEQ \boxed{\Sigma_{\perp,\perp}(q)\sim\half \frac{g_{\phi}}{|q|_+^2} . }
\label{eq:1.110}
\EEQ


\Bigskip {\em Connection to pre-kernel.}
Far enough from $q=0$, it is possible to expand further the denominator: letting
$R(q):=\Pi(q)-\lambda=O(\lambda^2 m)$, and representing by
\begin{tikzpicture}[scale=1] \draw[dashed,decorate,decoration=snake](0,0)--(2,0);
\end{tikzpicture}
the pre-kernel $\frac{\lambda}{1-\lambda{\cal A}(q)}$ as in the previous subsection, we can write down
the geometric series

\Bigskip
 \begin{tikzpicture}[scale=1]
\draw(-0.3,0) node {$\frac{\lambda}{1-\Pi(q){\cal A}(q) } \ =\ 
$};
 \draw[dashed,decorate,decoration=snake](2,0)--(4,0);
 \draw(5,0) node{$+\,$};
  \draw[dashed,decorate,decoration=snake](5.5,0)--(7.5,0);
 \draw(9,0) node {$ \frac{1}{\lambda} {\cal A}(q) R(q)$};
 \draw[dashed,decorate,decoration=snake](10.5,0)--(12.5,0);
 \draw(13.5,0) node{$+\ \cdots$};
\end{tikzpicture}

However, the  above expansion is faulty for $|q|_+\lesssim\Gamma_{\phi}$ (i.e. in our Regime {\bf B.}), since there (replacing \begin{tikzpicture}[scale=1] \draw[dashed,decorate,decoration=snake](0,0)--(2,0);
\end{tikzpicture} by the kernel $\Sigma_{\perp,\perp}$, of which it is assumed to be
an approximation)

{\centerline{
$\Sigma_{\perp,\perp}(q)\ \cdot\ \frac{1}{\lambda} {\cal A}(q) R(q)\approx 
  \frac{g_{\phi}}{|q|_+^2}\, \cdot \, m
\approx (\frac{\Gamma_{\phi}}{|q|_+})^2\gg 1.$}}


\section{Fermionic theory}  \label{section:fermion}


The section is organized as follows. Recall $j'_{\phi}=j_{\phi}+O(\ln(1/g))$. We
concentrate exclusively on momenta with scales $\le j'_{\phi}$. 

\Medskip First (\S \ref{subsection:angular}), we refine the multi-scale analysis of \S \ref{subsection:multi-scale} by  decomposing scale propagators according to {\em angular
sectors} (see Definitions \ref{def:angular} and \ref{def:Cjm}).

\Medskip Then (\S \ref{subsection:dressing}), we introduce a {\em dressed action} ${\cal L}(\vec{t})$, see Definition \ref{def:dressing}, which
is the basis of all subsequent computations. This  action
is dressed by scale dependent $\vec{t}$-parameters which enact the
scale decoupling necessary to separate local parts of two-point functions
and renormalize. We take the opportunity to give qualitative statements about
the renormalization flow of parameters.

\Medskip With ${\cal L}(\vec{t})$ at hand, it is possible to present the 
general cluster expansion scheme (\S \ref{subsection:cluster}),
involving  scale-by-scale {\em horizontal expansions}, 
a {\em bubble resummation}, a vertical
{\em momentum-decoupling expansion}, the {\em separation of local parts} of
divergent diagrams, and a {\em Mayer expansion}. On the way, we
present a preliminary, single-scale version of the bounds for
polymers.

\Medskip Finally, {\em general bounds for  multi-scale fermionic
polymers}, based on the preliminary bounds proved in
\S \ref{subsection:cluster}, and confirming at the same time
the  predictions of \S \ref{subsection:dressing} about the renormalization flow,  are proved in \S \ref{subsection:bounds}.


\subsection{Angular decompositions}  \label{subsection:angular}


We continue here the analysis of \S \ref{subsection:multi-scale}
by  decomposing the $j$-th momentum shell into $2^{j}$ 
spatial sectors of aperture  angle $\approx 2\pi/N_j$, where 
\BEQ N_j:=2^{j}.  \label{eq:def-Nj} \EEQ 
Technically, one uses a smooth partition of unity
\BEQ 1=\sum_{\alpha\in \Z/2^{j}\Z}  \eta^{j,\alpha}(\frac{\vec{p}}{|\vec{p}|} p^*_F) \EEQ
where the support of  $\eta^{j,\alpha}$ intersects the Fermi circle
$\Sigma_F^*$ roughly on the circular arc\\
 $\{(p_F^* \cos\theta,p_F^*\sin\theta)\ |\ 2\pi \alpha 2^{-j}<\theta<2\pi (\alpha+1) 2^{-j}\}$, in such a way that the distance dist($\supp(\eta^{j,\alpha})\cap \Sigma_F^*,\supp(\eta^{j,\alpha'})\cap\Sigma_F^*$)
$\approx 2^{-j} |\alpha-\alpha'|$, with $|\alpha-\alpha'|:=\min_{k\in\Z} |\alpha-\alpha'+2k\pi|$. For definiteness, we 
choose $\eta^{j,\alpha}$ (defined below, see Definition
\ref{def:Cjm})
in such a way that $\supp(\chi^{j,\alpha})\subset {\cal S}^{j,\alpha}$, where 

\begin{Definition}[angular sectors]  \label{def:angular}
\begin{itemize}
\item[(i)] \BEQ {\cal S}^{j}:=\Big\{p=(p^0,\vec{p})\ |\  \half \, 2^{-j}\mu <\max\Big(|p^0|,|e(\vec{p})|\Big)<\frac{3}{2} \, 2^{-j}\mu \Big\},  \qquad j<j'_{\phi}
\EEQ
\BEQ {\cal S}^{j'_{\phi}}:=
\Big\{p=(p^0,\vec{p})\ |\  \max\Big(|p^0|,|e(\vec{p})|\Big)<\frac{3}{2} \, 2^{-j'_{\phi}}\mu \Big\};
\EEQ
\BEQ {\cal S}^{j,\alpha}:=\Big\{p=(p^0,\vec{p})\in {\cal S}^j\ |\ 
2\pi(\alpha-1) 2^{-j} <\theta(\vec{p})<2\pi (\alpha+1) 2^{-j} \Big\},
\qquad \alpha\in\Z/2^{j}\Z,
\EEQ
where $(|\vec{p}|,\theta(\vec{p}))\in\R_+\times \R/2\pi\Z$ are the polar  coordinates of $\vec{p}$,
 namely, $\frac{\vec{p}}{|\vec{p}|}=\left(\begin{array}{c} \cos \theta(\vec{p}) \\ \sin \theta(\vec{p})
\end{array}\right)$.
\item[(ii)] $\vec{{\cal S}}^j:=\{\vec{p}\in\R^2\ |\ \exists p^0\in\R,
(p^0,\vec{p})\in{\cal S}^j\}$ and $\vec{{\cal S}}^{j,\alpha}:=\{\vec{p}\in\R^2\ |\ \exists p^0\in\R,
(p^0,\vec{p})\in{\cal S}^{j,\alpha}\}$ are the {\em spatial projections} of
${\cal S}^j,{\cal S}^{j,\alpha}$.  
\end{itemize}
\end{Definition}

\Medskip The number  $N_j$ of angular sectors is chosen in such a way 
that sectors are essentially isotropic: they are small
deformations of 3-dimensional cubes of side
scaling like $O(2^{-j})$. Note that the
momentum scale and angular sector of a given spatial momentum $\vec{p}$ is
defined up to $O(1)$, namely, if $\vec{p}\in\R^3$, $2^{-j'_{\phi}}\mu \lesssim\frac{p^*_F}{m^*}
|p_{\perp}| \lesssim 2^{-j_D}\mu$, then $(\vec{p}\in\vec{{\cal S}}^{j,\alpha}\cap 
\vec{{\cal S}}^{j',\alpha'}) \Rightarrow (|j-j'|=O(1),|\alpha-\alpha'|=O(1)).$

\Medskip {\bf Remark.} As discussed in (\cite{FMRT-infinite}, Lemma 3) in 
details, see also Proposition \ref{prop:sector-counting} below, this choice
of isotropic sectors is not optimal, and produces spurious (but not really
disturbing) $\log N_j\approx j$ prefactors when resumming over angular
sectors. However, the choice of curvelet-like anisotropic sectors
 (see e.g. \cite{CanDem,CanDon,Unt-cur}) with angular width $\approx 2^{-j/2}$ instead
 of $2^{-j}$ -- taking into account the {\em curvature} of the Fermi circle --
 would have led to a tedious wavelet-like phase-space analysis which we
 want to spare to the reader. 

\begin{Definition}  \label{def:Cjm}
Let
\BEQ \chi^{j,\alpha}(p/\mu):=\chi^j(|p|/\mu)\  \eta^{j,\alpha}(\frac{\vec{p}}{|\vec{p}|} p^*_F), \qquad j_D\le j\le j'_{\phi}-1, \ \alpha\in\Z/2^j\Z,
\EEQ
\BEQ \chi^{\to j'_{\phi},\alpha}(p/\mu):=\chi^{\to j'_{\phi}}(|p|/\mu)\  \eta^{j'_{\phi},\alpha}(\frac{\vec{p}}{|\vec{p}|} p^*_F), \qquad \alpha\in\Z/2^{j'_{\phi}}\Z.
\EEQ
Let us rewrite (\ref{eq:Cjalpha}),(\ref{eq:Cj'calpha}) for the sake of the reader,

\BEQ \boxed{C^{j,\alpha}_{\theta}(\xi,\xi'):=
\int \frac{d^3 p}{(2\pi)^3} \frac{e^{\II(p,\xi-\xi')}}{\II p^0-e^*(\vec{p}) \sigma^3-{\mathbb{\Gamma}}(\theta)} \chi^{j,\alpha}(\frac{p}{\mu}) 
, \qquad j<j'_{\phi}}
\label{eq:Cjm}
\EEQ
\BEQ \boxed{C^{j'_{\phi},\alpha}_{\theta}(\xi,\xi'):=
\int \frac{d^3 p}{(2\pi)^3} \frac{e^{\II(p,\xi-\xi')}}{\II p^0-e^*(\vec{p}) \sigma^3-{\mathbb{\Gamma}}(\theta)} \chi^{\to j'_{\phi},\alpha}(\frac{p}{\mu}) }
\label{eq:Cj'cm}
\EEQ
\end{Definition}

\Medskip For every $(j,\alpha)$, one chooses  a vector $\vec{p}^{j,\alpha}\in \Sigma_F$ lying
inside the support of $\eta^{j,\alpha}$, and
lets $p^{j,\alpha}=(0,\vec{p}^{j,\alpha})$. As explained in \S \ref{subsection:multi-scale},
see discussion in the paragraph between (\ref{eq:jc}) and (\ref{eq:j'c}), all
fermions of scale $j_{\phi}\le j\le j'_{\phi}$ may actually be considered as (non-optimal) decompositions
of scale $j_{\phi}$ fermions.  

\Medskip {\bf Remark.} In
section \ref{section:boson}, we shall need to further decompose scale $j'_{\phi}$ sectors  ${\cal S}^{j'_{\phi},\alpha}$, $\alpha\in \Z/2^{j'_{\phi}}\Z$,  into smaller angular sectors
of angle aperture $\approx 2\pi/2^{j_+}$, with $j_+\ge j'_{\phi}$;  we write
\BEQ \chi^{\to j'_{\phi},\alpha}(p/\mu):=\chi^{\to j'_{\phi}}(|p|/\mu)\  \eta^{j_+,\alpha}(\frac{\vec{p}}{|\vec{p}|} p^*_F), \qquad \alpha\in\Z/2^{j_+}\Z, \qquad j_+\ge j'_{\phi} \label{eq:further-angular}
\EEQ 
and  define accordingly angular sectors ${\cal S}^{j'_{\phi},\alpha}$ $(\alpha\in\Z/2^{j_+}\Z)$ roughly equal to the support of the cut-off functions $\chi^{\to j'_{\phi},\alpha}(\cdot/\mu)$, and covariance kernels $C^{j'_{\phi},\alpha}$, 
$\alpha\in\Z/2^{j_+}\Z$ as in (\ref{eq:Cj'cm}). Note that such angular sectors are
strongly {\em anisotropic} if $j_+\gg j'_{\phi}$. If necessary, 
they may be further "chopped" into isotropic "microsectors"
${\cal S}^{j'_{\phi},\alpha,k}$, $k=1,2,\ldots,2^{j_+-j'_{\phi}}$ with size $\approx 2^{-j_+}\mu$ in the direction
transverse to the Fermi circle. We then let $p^{j'_{\phi},\alpha,k}=(0,\vec{p}^{j'_{\phi},\alpha,k})$ be some momentum lying inside ${\cal S}^{j'_{\phi},\alpha,k}$, and define covariance kernels $C^{j'_{\phi},\alpha,k}$ with
characteristic functions $\chi^{\to j'_{\phi},\alpha,k}(\frac{|p|}{\mu})\eta^{j_+,\alpha}(\frac{\vec{p}}{|\vec{p}|} p^*_F)$ instead of $\chi^{\to j'_{\phi}}(\frac{|p|}{\mu})\eta^{j'_{\phi},\alpha}(\frac{\vec{p}}{|\vec{p}|} p^*_F)$.
\Medskip

\begin{Proposition}
\label{prop:C}

For every $n>0$, there exists a constant $C_n$
such that, for every scale $j_D\le j\le j'_{\phi}$,  sector $\alpha\in \Z/2^{j}\Z$ and multi-index $\kappa\ge 0$ , 
\begin{itemize}
\item[(i)] if $j\le j_{\phi}$,
\BEQ |(\nabla_{\xi}-\II p^{j,\alpha})^{\kappa} \, C^{j,\alpha}_{\theta}(\xi,\bar{\xi})|\le  C_n\,  2^{-2j} (2^{-j}\mu)^{\kappa_0} (2^{-j}p^*_F)^{\kappa_1+\kappa_2} 
\Big( 1+ 2^{-j}\mu |\xi-\bar{\xi}| \Big)^{-n} (p^*_F)^2;
\label{eq:I-II}
\EEQ

\item[(ii)] if $j> j_{\phi}$,
\BEQ |(\nabla_{\xi}-\II p^{j,\alpha})^{\kappa} \, C^{j,\alpha}_{\theta}(\xi,\bar{\xi})|\le  C_n\,  2^{-2j} 2^{-(j-j_{\phi})}  (2^{-j}\mu)^{\kappa_0} (2^{-j}p^*_F)^{\kappa_1+\kappa_2} 
\Big( 1+ 2^{-j}\mu |\xi-\bar{\xi}| \Big)^{-n} (p^*_F)^2 .
\label{eq:I-IIbis}
\EEQ
\end{itemize}
\end{Proposition}

\Medskip Eq. (\ref{eq:I-II}) is proved in (\cite{FMRT-infinite}, Lemma 4). If $j\le j_{\phi}$ and $\kappa=0$, then the factor
$2^{-2j}$ is equal to the scaling factor $2^j$ of  $\frac{1}{\II p^0-e^*_{|\Gamma|}(\vec{p})\sigma^3-{\mathbb{\Gamma}}(\theta)}$, times the
scaling of the volume of the sector, $2^{-j}\times 2^{-j}\times 2^{-j}=2^{-3j}$.   The supplementary prefactor $(2^{-j}\mu)^{\kappa_0} (2^{-j}p^*_F)^{\kappa_1+\kappa_2} $ comes from the Fourier multiplier $\II(p-p^{j,\alpha})$ 
associated to the operator $\nabla_{\xi}-\II p^{j,\alpha}$, 
since $|p-p^{j,\alpha}|$ is bounded over the support of
$\chi^{j,\alpha}$  by the diameter of the sector, which is
proportional to $2^{-j}$.  Assume
then that $j>j_{\phi}$, so that  $e^*_{|\Gamma|}(\vec{p})\approx \Gamma_{\phi}$,
$\Big|\frac{1}{\II p^0-e^*_{|\Gamma|}(\vec{p}) \sigma^3-{\mathbb{\Gamma}}(\theta)} \Big| 
\lesssim  \Gamma_{\phi}^{-1} \approx (2^{-j_{\phi}}\mu)^{-1}$. Hence $|\tilde{C}^{j,\alpha}_{\theta}(\xi,\bar{\xi})|\lesssim  (2^{-j_{\phi}}\mu)^{-1} \Vol({\cal S}^{j,\alpha})  \approx 
 (2^{-j_{\phi}}\mu)^{-1} (2^{-j}\mu)(2^{-j}p^*_F)^2   $. The summable spring factor $j\mapsto 2^{-(j-j_{\phi})}$ shows
clearly that this further decomposition would be useless for the free theory. The decay rate
is only $2^{-j}\mu$ instead of the expected fermion mass $2^{-j_{\phi}}\mu$ because of
the scale $j$ cut-offs, which are thus clearly not optimal.

\medskip\noindent  In other words:  the average magnitude of $\psi^{j,\alpha}(\xi)$ -- defined as
the {\em square-root} of $C^{j,\alpha}_{\theta}(\xi,\bar{\xi})$
for nearby $\xi,\bar{\xi}$  -- is
\BEQ \boxed{\psi^{j,\alpha}(\xi) \sim  
 2^{-j}p^*_F \qquad  (j \le j_{\phi}), \qquad  2^{-(j-j_{\phi})/2} 2^{-j}p^*_F \qquad  (j > j_{\phi})} \label{eq:scaling-psijalpha}
 \EEQ

A similar computation -- with a supplementary $O(2^j)$ relative
factor coming from integration in Fourier coordinates --
holds for the sector-symmetric covariance kernel $C^{j}_{\theta}(\xi,\bar{\xi})$,
yielding
\BEQ \psi^{j}(\xi) \sim  
 2^{-j/2}p^*_F \qquad  (j \le j_{\phi}), \qquad  2^{-(j-j_{\phi})/2} 2^{-j/2}p^*_F \qquad  (j > j_{\phi}) \label{eq:scaling-psij}
 \EEQ

\Medskip{\bf Remark.}  The above Proposition generalizes
to microsector covariances (see previous Remark),
\BEQ |(\nabla_{\xi}-\II p^{j'_{\phi},\alpha,k})^{\kappa} \, C^{j'_{\phi},\alpha,k}_{\theta}(\xi,\bar{\xi})|\le  C_n\,  2^{-2j_+} 2^{-(j_+-j_{\phi})}   (2^{-j_+}\mu)^{\kappa_0} (2^{-j_+}p^*_F)^{\kappa_1+\kappa_2} 
\Big( 1+ 2^{-j_+}\mu |\xi-\bar{\xi}| \Big)^{-n} (p^*_F)^2  
\label{eq:I-IIter}
\EEQ
if $\alpha\in\Z/2^{j_+}\Z$ ($j_+\ge j'_{\phi}$) and 
$k=1,2,\ldots,2^{j_+-j'_{\phi}}.$ Namely, compared to 
(\ref{eq:I-IIbis}),   the volume of the microsector scales now
like $2^{-j_+}\times 2^{-j_+}\times 2^{-j_+}=2^{-3j_+}$, and
its size along any direction like $2^{-j_+}$.


\Bigskip{\bf Boxes.} For every $j=j_D,\ldots,j'_{\phi}$, we decompose 
$\R^3$ into a disjoint union of {\em scale $j$ boxes} defined as follows.

\begin{Definition}[boxes]  \label{def:boxes}
A {\em scale $j$ box} is a "cube" $\Del:=(k_0 2^j\mu^{-1},(k_0+1)2^j\mu^{-1}))\times (k_1 2^j (p^*_F)^{-1},
(k_1+1)2^j (p^*_F)^{-1}) \times (k_2 2^j (p^*_F)^{-1},(k_2+1)2^j (p^*_F)^{-1})$, $k_0,k_1,k_2\in\Z$. We denote
by $\D^j$ the set of scale $j$ boxes, so that (up to a subset of measure zero)
$\D^j$ defines a partition of $\R^3$. Let finally $\D:=\uplus_{j=j_D,\ldots,
j'_{\phi}} \D^j$.
\end{Definition}

In order not to spoil completely momentum preservation at vertices, we
further  introduce for each $j=j_D,\ldots,j'_{\phi}$ a smoothened partition of unity, 
$1=\sum_{\Del\in\D^j} \chi_{\Del}(\xi)$, where each $\chi_{\Del}$ is a smooth,
compactly supported function and :

\Medskip\textbullet\ (scaling property) $\chi_{2^k\Del^{j_D}}(2^k \xi)=\chi_{\Del^{j_D}}(\xi)$
if $\Del^{j_D}\in\D^{j_D}$, $k\ge 0$;

\Medskip\textbullet\ (translation invariance) if $\vec{v}= 2^j \left( 
\begin{array}{c} i_0 \mu^{-1} \\ i_1   (p^*_F)^{-1} \\ i_2 (p^*_F)^{-1} \end{array}
\right)$ is a scaled
integer translation, then  $\chi_{\Del+\vec{v}}(\xi)  =\chi_{\Del}(\xi-\vec{v})$;

\Medskip\textbullet\ (support) $\chi_{\Del}(\xi)\equiv 0$ except if $\xi$
belongs to $\Del$ or its (direct of diagonal) neighbors;

\Medskip\textbullet\ $\chi_{\Del}\equiv 1$ in some neighbourhood of the
center $\xi_{\Del}$ of $\Del$.

\Medskip Then (by the scaling and translation invariance properties), for
every multi-index $\kappa=(\kappa_0,\vec{\kappa})$,
\BEQ |\nabla^{\kappa}\chi_{\Del}(\xi)|\le  C_{|\kappa|} (2^{-j}\mu)^{\kappa_0}
(2^{-j}p^*_F)^{\kappa_1+\kappa_2}, \qquad \Del\in\D^j.  \label{eq:2.14}
\EEQ
The volume of a scale $j$ box $\Del^j$ is roughly inverse to that of a sector,
\BEQ \Vol(\Del^j)\approx \Vol(\supp(\chi^{j,\alpha}))^{-1}\approx  (2^j\mu^{-1})(2^j (p_F^*)^{-1})^2
\approx  \frac{1}{m^*\mu^2} (2^{j})^3  \label{eq:vol-box}.
\EEQ
The bound (\ref{eq:2.14}) implies a bound for the Fourier transform with a quasi-exponential decay,
\BEQ |\hat{\chi}_{\Del^j}(q)|\le C_n \Vol(\Del^j) \, \Big(
1+|q|_+/2^{-j}\mu\Big)^n
\label{eq:exp-decay-chi-Del}
\EEQ
for every $n\ge 0$.

\Bigskip 
{\bf Sector counting.}
A characteristic feature of dimension 2 is that ({\em generically}, as we shall see)
four spatial momenta $\vec{p}_1,\ldots,\vec{p}_4$ in the
immediate neighborhood of the Fermi circle and such that
$\vec{p}_1+\vec{p}_2=\vec{p}_3+\vec{p}_4=0$ are essentially
two-by-two equal, namely, 
$\vec{p}_1\simeq \vec{p}_3, \vec{p}_2\simeq \vec{p}_4$ or $\vec{p}_1\simeq 
\vec{p}_4,\vec{p}_2\simeq\vec{p}_3$, by which we mean (letting $\alpha_i+O(1)$ be
the angular sector of $\vec{p}_i$, $i=1,\ldots,4$) that, to leading order,
\BEQ \Big(\alpha_1=\alpha_3+O(1), \alpha_2=\alpha_4+O(1)\Big) \qquad {\mathrm{or}} \qquad 
\Big(\alpha_1=\alpha_4+O(1), \alpha_2=\alpha_3+O(1)\Big)
\EEQ
  as 
illustrated by the following picture

\bigskip

{\centerline{$\dimensiondeux$}}

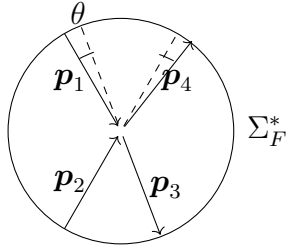
\medskip {\bf \tiny Fig. \ref{fig:dimensiondeux}.1. Close to the Fermi
circle, there are only two
free sectors per vertex in dimension 2.}

\Bigskip Thus, splitting each of the four fields of a vertex
$\int d\xi\ (\bar{\psi}_{\uparrow}\bar{\psi}_{\downarrow})(\xi) 
(\psi_{\downarrow}\psi_{\uparrow})(\xi)$ into its $(j,\alpha)$-components
leads for "quasi-scale-diagonal" terms in a near-Cooper pairing,
\BEQ \int d\xi \, \int d\xi'\, (\bar{\psi}^{j_1,\alpha_1}_{\uparrow}\bar{\psi}^{j_3,\alpha_3}_{\downarrow})(\xi) \ \del(\xi-\xi')\ 
(\psi_{\downarrow}^{j_2,\alpha_2}\psi_{\uparrow}^{j_4,\alpha_4})(\xi')
\label{eq:two-half-vertices} 
\EEQ with
$j_1\simeq j_2\simeq j_3\simeq j_4$, to essentially only {\em two sums} over
angular sectors. More precisely:

\begin{Proposition} (see \cite{FMRT-infinite}, Lemma 3).  \label{prop:sector-counting} 
\begin{itemize}
\item[(i)]
Let ${\cal S}^{j_i,\alpha_i}$, $i=1,\ldots,4$ be the spatial projection of  four  angular sectors of scales $j_1,j_2,j_3,j_4=j+O(1)$
 and $A\ge 1$. 
Then the set 
\BEA && \Omega_4:=\{(\vec{p}_1,\ldots,\vec{p}_4)\in \vec{\Sigma}^{j_1,\alpha_1}\times\ldots\times \vec{\Sigma}^{j_4,\alpha_4} \ |\ |\vec{p}_1-\vec{p}_3|<|\vec{p}_1+\vec{p}_2|,\,  \nonumber\\
&&\qquad\qquad\qquad |\vec{p}_1+\vec{p}_2-\vec{p}_3-\vec{p}_4|\le A2^{-j} p^*_F\}
\EEA
is empty unless 
there exists $k\in\{\lfloor \frac{j}{2}\rfloor,\ldots,j\}$ such that
\BEQ |\alpha_1-\alpha_{3}|,|\alpha_2-\alpha_{4}|\lesssim A2^{j-k} \lesssim   |\alpha_1+\alpha_{2}|\le CA 2^{k}. \EEQ 
\item[(ii)] (generalization)  Let ${\cal S}^{j_i,\alpha_i}$, $i=1,\ldots,2n$\ 
$(2n\ge 6)$  be the spatial projection of  $2n$  angular sectors of scales $j_1,j_2,\ldots,j_{2n}=j+O(1)$. Then the set $\Omega_{2n}:=\{(\vec{p}_1,\ldots,\vec{p}_{2n})\in \vec{\Sigma}^{j_1,\alpha_1}\times\ldots\times \vec{\Sigma}^{j_{2n},\alpha_{2n}} \ |\ |\sum_{i=1}^n \vec{p}_i-\sum_{i=n+1}^{2n} \vec{p}_i|\le A2^{-j} p^*_F\}$
has cardinal $O(A^2 2^{(2n-2)j})$. 
\end{itemize}
\end{Proposition}

{\em Sketch of proof for (i) (case of a  vertex).}
The condition  $|\vec{p}_1-\vec{p}_3|<|\vec{p}_1+\vec{p}_2|$
ensures that the aperture angle $\theta:=\widehat{(\vec{p}_1,\vec{p}_3)}$ is smaller than the relative pair angle
$\alpha:=\widehat{(\vec{p}_1,-\vec{p_2})}$. Since 
$|\vec{p}_2-\vec{p}_4|\simeq |\vec{p}_1-\vec{p}_3|\approx p^*_F |\sin(\frac{\theta}{2})|$,
we deduce that $|\alpha_1-\alpha_3|\approx |\alpha_2-\alpha_4|\lesssim |\alpha_1+\alpha_2|\approx |\alpha_1+\alpha_4|$. In other words, $\alpha_3$ is the sector
closest to $\alpha_1$, and the pair $(\vec{p}_2,\vec{p}_4)$
is essentially obtained from $(\vec{p}_1,\vec{p}_3)$ through
a rotation of angle $\alpha$. (Permuting $\alpha_3$ with $\alpha_4$
or $-\alpha_2$ yields two equivalent possibilities). 
Using the approximate equality,  $\vec{p}_1+\vec{p}_2\simeq \vec{p}_3+\vec{p}_4$, and projecting $\vec{p}_2-\vec{p}_4$
onto the axis perpendicular to $\vec{p}_1-\vec{p}_3$  yields
$|2\sin(\alpha)\sin(\frac{\theta}{2})|\lesssim A2^{-j}$, whence $|\alpha| \ |\theta|\lesssim A2^{-j}.$ \hfill \eop

\Bigskip 
The most important case is $2n=4$, for the above Proposition gives then the
sector counting factor of a vertex: accepting an error $O(A2^{-j}p^*_F)$ on
the momentum preservation condition, one has  $O(j2^{2j}A^2)$ possibilities for the choices of the sector
indices $\alpha_1,\ldots,\alpha_4$; namely, for fixed
$k=j+\lfloor \log_2|\alpha_1-\alpha_3|^{-1} \rfloor$,  there are $\approx 2^{j-k}$ choices for $\alpha_3$,  $\lesssim A2^{k}$ 
possibilities for $\alpha_2$, and finally, $\lesssim A$ possibilities for $\alpha_4$ given $\alpha_2,\alpha_3$. Summing over $k\in \{\lfloor \frac{j}{2}\rfloor,\ldots,j\}$
yields the result. We call {\bf generic} a vertex for which the {\em relative pair
angle} $\alpha\approx 1$ is {\em large}, which implies that the {\em aperture angle}
$\theta$
is small, of order $O(2^{-j})$; diagrams with only generic vertices have the following
very important property: {\em vertices may be cut into two parts in such a way that the
transfer momentum between the half-vertices is very small}. In (\ref{eq:two-half-vertices}),
half-vertices are $\bar{\psi}_{\uparrow}^{j_1,\alpha_1}\bar{\psi}_{\downarrow}^{j_3,\alpha_3}$ and $\psi_{\downarrow}^{j_2,\alpha_2}\psi^{j_4,\alpha_4}_{\uparrow}$. Sectors $\alpha_1,\alpha_3$ are very close. Splitting all vertices in this way yields {\em fermion loops
on which all sectors are very close}, so that there is essentially only {\em one sum over
sectors per loop}. This gives birth to the {\em "1/N" factors}. This approach is pursued
in a systematic way in \S \ref{subsubsection:single-scale-bounds}. In this ideal case, the
$1/N$-factor is $1/N_j=2^{-j}$. Taking into consideration the opposite {\bf degenerate} case when $k$
is closer to $j/2$ than to $j$, the small factor $2^{-k}$ is really (at worst) $2^{-j/2}$.

\Medskip   Alternatively, assuming $\alpha_1$ to be fixed,
this leaves only $O(2^j A^2)$ possibilities. Later on, we shall see that
(at least  perturbatively) sectors are in average shared by two
vertices, so fixing one of the sectors is the correct way of counting; thus, for a 
vertex $(2n=4)$, there is -- up to the logarithmic correction  -- only {\em one} sum over sectors. Also, as
mentioned in the Remark just above Definition \ref{def:Cjm}, choosing {\em anisotropic sectors} with angular width $\approx 2^{-j/2}$ instead, the
logarithmic correction disappears, so there is {\em exactly one sum over sectors.} 

\Medskip
An immediate corollary is the following. Let
\BEQ  I^j(\Del^j):= \int d\xi\, 
\chi_{\Del^j}(\xi) \sum_{\alpha_2,\alpha_3,\alpha_4}
\Big(\bar{\psi}^{j,\alpha_1}_{\uparrow}(\xi)\bar{\psi}^{j,\alpha_2}_{\downarrow}(\xi)\Big)
\Big(\psi^{j,\alpha_3}_{\downarrow}(\xi)\psi^{j,\alpha_4}_{\uparrow}(\xi)\Big),  \label{eq:I(Del)}
\EEQ 
$\Del^j\in\D^j$, with $\alpha_1$ fixed, and $\chi_{\Del}$ as above.   Then, 
taking a partial Fourier transform ${\cal F}$ w.r. to spatial coordinates,
and considering a family of smooth, scaled Fourier cut-offs $\chi^k_+(\vec{q})$
such that $\sum_k \chi^k_+\equiv 1$ and $\chi^k_+(\vec{q})\equiv 0$ except if
$\sqrt{(q^1)^2+(q^2)^2}\approx 2^{-k} p_F^*$, so that (in particular) $|{\cal F}^{-1}(\chi_+^k)(\vec{x})|\le C_n 
2^{-2k}(p^*_F)^2 \, (1+2^{-k}p^*_F |\vec{x}|)^{-n}$ for all $n\ge 0$ (compare with the bounds of Proposition \ref{prop:C}),  we get
\BEA  && I^j(\Del^j) =  \int d\tau \, \sum_{k} \int d\vec{q}\, \hat{\chi}_{\Del^j}(\tau,\vec{q}) \chi_+^k(\vec{q}) \sum_{(\alpha_2,\alpha_3,\alpha_4)\in {\cal M}^k_{\alpha_1}} {\cal F}\Big(\bar{\psi}^{j,\alpha_1}_{\uparrow} 
\bar{\psi}^{j,\alpha_2}_{\downarrow} \psi^{j,\alpha_3}_{\downarrow}\psi^{j,\alpha_4}_{\uparrow}\Big)(\tau,\vec{q}) \nonumber\\
&& = \sum_{k} \int d\tau\,  \int d\vec{x} \, (\chi_{\Del^j}(\tau,\cdot)\ast{\cal F}^{-1}({\chi}_+^k))(\vec{x})
\sum_{(\alpha_2,\alpha_3,\alpha_4)\in {\cal M}^k_{\alpha_1}}\Big(\bar{\psi}^{j,\alpha_1}_{\uparrow}(\xi)\bar{\psi}^{j,\alpha_2}_{\downarrow}(\vec{x})\Big)
\Big(\psi^{j,\alpha_3}_{\downarrow}(\vec{x})\psi^{j,\alpha_4}_{\uparrow}(\vec{x})\Big).
\nonumber\\   \label{eq:2.13}
\EEA 
where ${\cal M}_{\alpha_1}^k:=\{(\alpha_2,\alpha_3,\alpha_4) \ |\ \exists\,  (\vec{p}_1,\ldots,\vec{p}_4)\in \vec{\Sigma}^{j,\alpha_1}\times\ldots\vec{\Sigma}^{j,\alpha_4} , 
 \vec{p}_1+
\vec{p}_2-\vec{p}_3-
\vec{p}_4\in\supp(\chi_+^k)\}$ has cardinal $O(j2^j (2^{j-k})^2)$ for $k\le j$; in practice, we restrict the sum in (\ref{eq:2.13}) to $k\le j$
and replace without further mention $\chi_+^j$ by $\chi_+^{\to j}=\sum_{k\ge j} \chi^k_+$. Then 
\BEA &&  \int d\tau \int d\vec{x}\,  \Big|(\chi_{\Del^j}(\tau,\cdot)\ast{\cal F}^{-1}(\chi_+^k))(\vec{x})\Big| \lesssim \int d\tau \int d\vec{x}\, 
\Big|(\vec{\nabla}^3 \chi_{\Del^j}(\tau,\cdot) \ast {\cal F}^{-1}(\vec{p
}\mapsto |\vec{p}|^{-3} \chi_+^k(\vec{p}))(\vec{x})\Big| \nonumber\\
&&\qquad  \lesssim 
(2^{-(j-k)})^3 \,   \Vol(\supp(\chi_{\Del^j}))  \ \cdot \ \int  d\vec{x}\, |{\cal F}^{-1}(\chi_+^k)(\vec{x})| \nonumber\\
&&\qquad \lesssim  (2^{-(j-k)})^3 \ \cdot\  \frac{1}{m^*\mu^2} (2^j)^3.
\EEA
Taking into account the $O(2^{-j}p_F^*)=O(2^{-j} \sqrt{m^*\mu})$ scaling of each
field,  one obtains finally $I^j(\Del^j)=O(jm^*)$.

\Medskip In other words:
{\em  
 The {\bf vertex}, i.e. the quartic term in the interaction, integrated 
over a scale $j$ box $\Del^j$, fixing one of the sectors, 
\BEQ  \lambda I^j(\Del^j)=O(j\lambda m^*)=O(jg)
\label{eq:lambdaIjjg} \EEQ
is small}  -- up to a logarithmic correction  --, as follows from our Assumption on the coupling constant $g$.  
 
\Medskip We shall see in \S \ref{subsection:bounds} 
how to get rid of this logarithmic correction in {\em diagram}
bounds.  Note that there is no logarithmic correction as soon as  $2n\ge 6$.


\subsection{Dressed action}   \label{subsection:dressing}


 The principle of multi-scale expansions is to rewrite the bare action 
${\cal L}_{\theta}(\Psi,\bar{\Psi}) \equiv \int_V d\xi\, {\cal L}_{\theta}(\Psi,\bar{\Psi};\xi)$ of Definition \ref{def:Grassmann-measure}
as a sum over $j\ge j_D$ of effective  actions ${\cal L}_{\theta}^{\to j}(\Psi,\bar{\Psi})
  \equiv \int_V d\xi\, {\cal L}_{\theta}^{\to j}(\Psi,\bar{\Psi};\xi)$ with
  highest scale $j$, and  renormalized, scale-dependent coefficients $\lambda^j, \Gamma^j,
  \mu^j,  m^j$. 

\Medskip {\em Field dressing.}  Recall $\D=\sum_{j=j_D}^{j'_{\phi}} \D^j$. 
Given a function $\vec{t}:\D\to[0,1]$, we let
$(T^{\to j}\Psi)^j(\xi)=\Psi^j(\xi)$,
\BEQ (T^{\to k}\Psi)^j(\xi)=t^k_{\xi}t^{k+1}_{\xi}\cdots t^{j+1}_{\xi} \Psi^j(\xi), \qquad k<j  \EEQ
with 
\BEQ t^k_{\xi}:=\sum_{\Del\in\D^k} \chi_{\Del}(\xi) t^k_{\Del} \label{eq:txiDel}
\EEQ
and
\BEQ (T\Psi)^{\to k}:=\sum_{j\ge k} (T^{\to k}\Psi)^j. \EEQ
The sum in (\ref{eq:txiDel}) contains only
 at most  $6=O(1)$ terms by the support condition of $(\chi_{\Del})_{\Del}$.
This definition implies
\BEQ  \frac{\partial}{\partial t^j_{\Del}} (T\Psi)^{\to k}(\xi)=
 t^k_{\xi}\cdots t^{j+1}_{\xi}  \chi_{\Del}(\xi) (T\Psi)^{\to j}(\xi), \qquad k<j.
\EEQ
The above expression vanishes except in some neighborhood of $\Del$. The adjoint field $\bar{\Psi}$ is dressed similarly.

\Medskip Let 
\BEQ \boxed{\lambda^{j_D}:=\lambda, \qquad \del\Gamma^{j_D}:=\Gamma_{\phi},
\qquad 
\del \mu^{j_D}:=\del \mu=\mu^*-\mu, \qquad \del m^{j_D}:=\del m=
\frac{m^*}{m}(m^*-m) }.  \label{eq:bare-parameters}
\EEQ

\begin{Definition}[dressed action] \label{def:dressing}

Let
 \BEQ d\mu_{\theta;\lambda}(\vec{t};\Psi,\bar{\Psi})=
  \frac{1}{{\cal Z}^*_{\lambda}} 
  e^{- {\cal L}_{\theta}(\vec{t};\Psi,\bar{\Psi})}  \,
   d\mu^*_{\theta}(\Psi,\bar{\Psi}),
\EEQ

where  
\BEQ {\cal L}_{\theta}(\vec{t})\equiv
\int_V d\xi\, {\cal L}_{\theta}(\vec{t};\xi), \EEQ
\BEQ {\cal  L}_{\theta}(\vec{t};\xi)\equiv \sum_{j=j_D}^{j'_{\phi}}   {\cal  L}^{\to j}(\vec{t};\xi) \EEQ

\BEA && {\cal L}_{\theta}^{\to j_D}(\vec{t};\xi):=
\lambda^{j_D} ((T\bar{\Psi})^{\to j_D}(T\Psi)^{\to j_D})^2(\xi) \nonumber \\
&& \qquad  +
(T\bar{\Psi})^{\to j_D}(\xi) \Big(
\del{\mathbb{\Gamma}}^{j_D}(\theta)- \Big(\frac{\del m^{j_D}}{(m^*)^2} \frac{|\vec{\nabla}|^2}{2}+\del\mu^{j_D}\Big)\sigma^3\Big) (T\Psi)^{\to j_D}(\xi); 
\nonumber\\
\EEA

and, for $j=j_D+1,\ldots,j'_{\phi}$,

\BEA
 && {\cal L}_{\theta}^{\to j}(\vec{t};\xi):=(1-(t^{j-1}_{\xi})^4)\  \lambda^{j_D}
 ((T\bar{\Psi})^{\to j}(T\Psi)^{\to j})^2(\xi) \nonumber\\
&&\qquad + (1-(t^{j-1}_{\xi})^2) (T\bar{\Psi})^{\to j}(\xi) \Big(
 \del Z^j \partial_{\tau}- \Big( \frac{\del m^j}{(m^*)^2} \frac{|\vec{\nabla}|^2}{2}+ \del\mu^j \Big)
\sigma^3 +
 \del {\mathbb{\Gamma}}^j(\theta)\Big) (T\Psi)^{\to j}(\xi)  \nonumber\\
 \label{eq:Lthetatoj}
\EEA

where $\vec{\nabla}=(\partial_{x_1},\partial_{x_2})$, 
$|\vec{\nabla}|^2:=\partial^2_{x_1}+\partial^2_{x_2}$.

\end{Definition}

\Medskip Compare with Definition \ref{def:Grassmann-measure}. The principle of the dressing is the following:

\Medskip \textbullet\  If $\vec{t}\equiv \vec{1}$, then $d\mu_{\theta;\lambda}(\vec{t})
\equiv d\mu_{\theta;\lambda}$, i.e. one retrieves the bare theory (see (\ref{eq:check-bare-parameters}) below);

\Medskip\textbullet\  If $t_{\Del}=0$ for all $\Del\in\D^j$, then 
$d\mu_{\theta;\lambda}(\vec{t};\Psi,\bar{\Psi}) \propto 
e^{-{\cal L}_{\theta}(\vec{t}^{(j-1)\to};\Psi^{j\to},\bar{\Psi}^{j\to})} 
d\mu^*_{\theta}(\Psi^{j\to},\bar{\Psi}^{j\to}) \ \otimes\ 
e^{-{\cal L}_{\theta}(\vec{t}^{\to (j+1)};\Psi^{\to(j+1)},\bar{\Psi}^{\to(j+1)})} 
d\mu^*_{\theta}(\Psi^{\to(j+1)},\bar{\Psi}^{\to(j+1)})$ is the tensor product
of two {\em independent} measures, a measure for the  {\em high-momentum fields}
$\Psi^{j\to},\bar{\Psi}^{j\to}$, and a measure for the {\em low-momentum fields}
$\Psi^{\to(j+1)},\bar{\Psi}^{\to(j+1)}$. On the other hand,
derivatives $\frac{d}{dt_{\Del}}$, $\Del\in\D^j$ acting
on ${\cal L}_{\theta}(\vec{t};\Psi,\bar{\Psi})$ produce
 by construction
either  (i) {\em mixed} terms involving both high-momentum fields
$(\overset{(-)}{\Psi})^{j\to}$ and low-momentum fields $(\overset{(-)}{\Psi})^{\to(j+1)} $; or (ii) {\em subtracted terms}
$(\lambda^j-\lambda^{j+1}) ((T\bar{\Psi})^{\to(j+1)}(T\Psi)^{\to(j+1)})^2$ or $ (T\bar{\Psi})^{\to(j+1)} \Big\{ (\del Z^j-\del Z^{j+1})\partial_{\tau} - \Big( (\del m^j-\del m^{j+1})
\frac{|\vec{\nabla}|^2}{2m^2}  + (\del \mu^j-\del \mu^{j+1})\Big)\sigma^3 + (\del{\mathbb{\Gamma}}^j(\theta)-\del {\mathbb{\Gamma}}^{j+1}(\theta)) \Big\} (T\Psi)^{\to (j+1)}$.
Subtracted terms (ii) associated with parameter renormalization 
are there precisely to compensate mixed terms of type (i).

\Bigskip
Looking at the theory at scale $j$, and combining the covariance $C^j_{\theta}$
with the quadratic term in ${\cal L}_{\theta}^{\to j}$,  we obtain an effective
fermion covariance
\BEA && \Big( \Big\{\II p^0-\Big(\frac{|\vec{p}|^2}{2m^*}-\mu^*\Big)\sigma^3 -{\mathbb{\Gamma}}(\theta) \Big\}
+\Big\{\II \del Z^j p^0-\Big(\frac{\del m^j}{2(m^*)^2} |\vec{p}|^2+\del\mu^j
\Big)\sigma^3 +\del{\mathbb{\Gamma}}^j(\theta)\Big\}\Big)^{-1} \nonumber\\
&&\qquad \equiv (\II Z^j p^0-\Big( \frac{|\vec{p}|^2}{2 m^j} -\mu^j \Big)\sigma^3-{\mathbb{\Gamma}}^j(\theta))^{-1} 
\EEA
where
\BEQ \Gamma^j=\Gamma-\del\Gamma^j, \EEQ
\BEQ  m^j=\frac{m^*}{1+\frac{\del m^j}{m^*}} =m^*-\del m^j + O(\frac{(\del m^j)^2}{m^*}), \EEQ
\BEQ  \qquad Z^j=1+
\del Z^j, \qquad \mu^j=\mu^* -\del \mu^j.
\EEQ
In particular, {\em when $j=j^D$, one retrieves the bare parameters,}
\BEQ \Gamma^{j_D}=\Gamma,\qquad m^{j_D}=m,
\qquad \mu^{j_D}=\mu. \label{eq:check-bare-parameters} 
\EEQ

\Bigskip  {\em We determine $\mu^*=\mu^*(\Gamma)$,
 $m^*=m^*(\Gamma)$,  and simultaneously $\Gamma\equiv \Gamma^{(j'_{\phi}-1)\to}$,    by 
a fixed point argument in such a way that }
\BEQ \boxed{\del\mu^{j'_{\phi}}=0,\  \del m^{j'_{\phi}}= 0.} 
\EEQ 
and the pre-gap equation (see Definition \ref{def:pre-gap}) holds.
As emphasized in the Introduction, this procedure implies that
{\em the effective radius of the Fermi circle is $p_F^*:=\sqrt{2m^*\mu^*}$.}

\Bigskip As we shall see (see \S \ref{subsubsection:single-scale-bounds}), the order of divergence $\omega(\Upsilon)$ of
 a lowest scale $j$ amputated diagram $\Upsilon$ with $N_{ext}$ external fermion legs is $\half(4-N_{ext})$ {\em above} symmetry-breaking
 momentum scale $j_{\phi}$, which means that ${\cal A}^{j\to}(\Upsilon)$ scales at most like $O(2^{\frac{j}{2}(4-N_{ext})})$.
 Thus diagrams with $\le 4$ external legs must be
 renormalized.
 This power-counting estimate is not necessarily true for 
the amplitude of a polymer, which suffers  from non-overlapping constraints; we shall therefore be led to renormalize diagrams
with up to $6 $ external legs, whose local parts may be
freed from non-overlapping constraints by the Mayerization
procedure described in \S \ref{subsubsection:Mayer}.    In this theory, one renormalizes with external
 legs on the effective Fermi circle $\Sigma^*_F$ where the IR singularity lies.  By symmetry,
 non-vanishing $N_{ext}$-point functions must have $\half N_{ext}$ external
 $\Psi$-fields and $\half N_{ext}$ external $\bar{\Psi}$-fields, with $N_{ext}$
 {\em even}.  So we need only discuss  two-, four- and six-point functions. {\em Counterterms} are associated to  two-point functions only; they require a precise analysis, which
 will lead to the renormalization flow.

\begin{enumerate}
\item {\bf (two-point functions)} The two external legs -- 
$(\psi,\psi)$, $(\psi,\bar{\psi})$ or $(\bar{\psi},\bar{\psi})$ --
are in the vicinity of the Fermi sphere.  Let $\vec{i}$ be a
unit vector, and $\vec{i}^{\perp}$ its image through the rotation of angle $\pi/2$. Expanding with respect
to an external momentum $(p^0,\vec{p})=(p^0, (p^*_F+p_{\perp})\vec{i}+
p_{//}\vec{i}^{\perp}
)$ in space direction $\vec{i}$,  with $|p_{\perp}|, |p_{//}|\ll 2^{-j}p_F^*$, 
$|p^0|\ll 2^{-j}\mu$, and using $p_{\perp}\sim \frac{1}{2p_F^*} 
(|\vec{p}|^2-(p_F^*)^2)  $, one gets 

\BEA &&  
{\cal A}^{j\to}_{p}(\Upsilon)=
{\cal A}^{j\to}_{(0,p_F^* \vec{i})} (\Upsilon)+ p^0\frac{d}{dp^0} {\cal A}^{j\to}_p(\Upsilon)\Big|_{p=(0,p^*_F \vec{i})}+  p_{//}\frac{d}{dp_{//}} {\cal A}^{j\to}_p(\Upsilon)\Big|_{p=(0,p^*_F \vec{i})} \nonumber\\
&&\qquad  +  \frac{1}{2p_F^*} 
(|\vec{p}|^2-(p_F^*)^2) 
\frac{d}{dp_{\perp}} {\cal A}^{j\to}_p(\Upsilon)\Big|_{p=(0,p^*_F \vec{i})}   +O\big(|p|^2\frac{d^2}{dp^2} {\cal A}^{j\to}_p(\Upsilon) \big) \nonumber\\
\EEA
and similarly (summing over all diagrams with internal legs
of momentum scales $\le j$), letting $G^{j\to}(p)$ be as in
\S \ref{subsection:Nambu} the 
Fourier transform of the two-point function $\langle ^t \bar{\Psi}^{j\to}(\cdot)  \Psi^{j\to}(\cdot) \rangle_{\theta;\lambda}$,
\BEA && G^{j\to}(p)=G^{j\to}(0,p^*_F \vec{i})+p^0 \frac{d}{dp^0} G^{j\to}(p)\Big|_{p=(0,p^*_F \vec{i})}  + p_{//} \frac{d}{dp_{//}} G^{j\to}(p)\Big|_{p=(0,p^*_F \vec{i})} \nonumber\\
&&\qquad  +
\frac{1}{2p_F^*} 
(|\vec{p}|^2-(p_F^*)^2)  \frac{d}{dp_{\perp}} G^{j\to}(p)\Big|_{p=(0,p^*_F \vec{i})} + O\big(
|p|^2 \frac{d^2}{dp^2} G^{j\to}(p) \big). \nonumber\\
\label{eq:local-part}
 \EEA

The difference $ \del G^j:=G^{j\to}-G^{(j-1)\to}$ selects the
contribution of diagrams of lowest internal scale precisely
equal to $j$. 
Second-order derivatives need not be renormalized because the effective
order of divergence is 
$\half(4-2)-2<0$. By rotation invariance, $\frac{d}{dp_{//}}\del G^{j}(p)\Big|_{p=(0,p^*_F \vec{i})}=0$.  

\Medskip Let us  discuss  to begin with {\em diagonal components} $\del G^j_{i,i}$,
$i=1,2$. First,  $\del G^{j}_{i,i}(0,p^*_F \vec{i})$ and $\frac{d}{dp_{\perp}}\del G^{j}_{i,i}(p)\Big|_{p=(0,p_F^* \vec{i})}$ yield 
spin-independent counterterms $\del\mu^{j+1}-\del\mu^j$ for the chemical
potential and 
$\del m^{j+1}-\del m^j$ for the mass.   Finally, $\frac{d}{dp^0} \del G^j_{i,i}(p)\Big|_{p=(0,p_F^*\vec{i})}$ yields a $Z$-counterterm
 $\del Z^{j+1}-\del Z^j$.

\Medskip The {\em off-diagonal part} $\del G^j_{off}=\left(
\begin{array}{cc} 0 & \del G^j_{1,2} \\ \del G^j_{2,1} & 0
\end{array}\right)$ of $\del G^j$ involves an  {\em odd} number of  off-diagonal propagators with matrix coefficient 
${\mathbb{\Gamma}}$ or $\del{\mathbb{\Gamma}}^k$, $k<j$, 
 whence 
 (as
shown by induction on $j$) one extra prefactor
$O(2^{-j_{\phi}})=O(2^{-j} 2^{-(j_{\phi}-j)})$, thus lowering the effective degree
of divergence of diagrams by $1$. Consequently, we need only Taylor expand
to order $0$, 
\BEQ G_{off}^{j\to}(p)=G_{off}^{j\to}(0,p^*_F\vec{i})+
O(|p|\frac{d}{dp}G_{off}^{j\to}(p)). \label{eq:local-part0}
\EEQ

\Medskip
It follows crucially from the fact that the scale $j$ infra-red cut-off theory, see Definition \ref{def:Grassmann-measure} (ii),
retains the symmetries of the original theory, so  that  (by (\ref{eq:1.9}) and (\ref{eq:1.10}))
\BEQ \del G_{1,1}^{j\to}(0,p_F^*\vec{i})=\Big( \del G_{1,1}^{j\to}(0,
p_F^*\vec{i}) \Big)^* = -\del G_{2,2}^{j\to}(0,p_F^*\vec{i});
\EEQ

\BEQ \frac{d}{dp^0} \del G_{1,1}^{j\to}\Big|_{p=(0,p_F^*\vec{i})}
= \frac{d}{dp^0} \del G_{2,2}^{j\to}\Big|_{p=(0,p_F^*\vec{i})} =-\Big( 
\frac{d}{dp^0} \del G_{1,1}^{j\to}\Big|_{p=(0,p_F^*\vec{i})} \Big)^*
\EEQ

\BEQ \frac{d}{dp_{\perp}} \del G_{1,1}^{j\to}\Big|_{p=(0,p_F^*\vec{i})}
= -\frac{d}{dp_{\perp}} \del G_{2,2}^{j\to}\Big|_{p=(0,p_F^*\vec{i})}
=\Big(
\frac{d}{dp_{\perp}} \del G_{1,1}^{j\to}\Big|_{p=(0,p_F^*\vec{i})} \Big)^*
\EEQ
\BEQ \del G^{j\to}_{1,2}(0,p_F^*\vec{i})=\Big(\del G^{j\to}_{2,1}(0,p_F^*\vec{i})\Big)^* \EEQ

\Medskip
Hence: $\del\mu^j_1=-\del\mu^j_2=:\del\mu^j, \del m^j_1=-\del m^j_2=:\del m^j$, $\del Z^j_1=\del Z^j_2=:\del Z^j$ and $c^j$ are all {\em real}
parameters, while $\del G^j_{off}(0,p_F^*\vec{i})$ is a {\em Hermitian} matrix
of the form $\del{\mathbb{\Gamma}}^{j+1}-\del{\mathbb{\Gamma}}^j\equiv \left(\begin{array}{cc} 0 & \bar{\del\Gamma}^{j+1}-\bar{\del\Gamma^j} \\ \del\Gamma^{j+1}-\del\Gamma^j & 0 \end{array}\right)$, with $\del\Gamma^{j+1}-\del\Gamma^j\in\C$.

\Medskip {\em Remark.} Note that diagrams contributing to $\del{\mathbb{\Gamma}}$-counterterms are not necessarily in the
$\theta$-direction ($//$) because of mixing diagrams. However, the leading-order contribution is a tadpole, see Fig. \thesubsection.4 (iv), which {\em is} in the $\theta$-direction. Furthermore, a  Ward identity (see \S \ref{subsection:Ward}) ensures that the 
local part of the fermionic two-point function vanishes in the direction perpendicular ($\perp$) to $\theta$.

Summarizing: 

\Medskip 
\begin{Definition}[counterterms]  \label{def:counterterms}
Let $\vec{i}\in\R^2$ be a unit vector. The following definitions
do not depend on the orientation of $\vec{i}$:
\begin{itemize}
\item[(i)] (chemical potential counterterm) let
\BEQ  \boxed{\del\mu^{j+1}-\del\mu^j:= -\del G_{1,1}^j(0,p_F^*\vec{i}) +
\frac{p^*_F}{2} \frac{d}{dp_{\perp}}\Big|_{p_{\perp}=0} \del G_{1,1}^{j}(0,p^*_F \vec{i})  .} \EEQ
\item[(ii)] (mass counterterm)
\BEQ \boxed{\del m^{j+1}-\del m^j:=-\frac{m^2}{p^*_F} \frac{d}{dp_{\perp}}\Big|_{p_{\perp}=0}
 \del G_{1,1}^j(0,p_F^* \vec{i}).}
 \EEQ 
\item[(iii)] (energy gap counterterm) let
\BEQ \boxed{\del\Gamma^{j+1}-\del\Gamma^j:=\del G_{2,1}^j(0,p_F^*\vec{i}).} \EEQ
\item[(iv)] ($Z$ counterterm) let 
\BEQ \boxed{\del Z^{j+1}-\del Z^j:=-\II\frac{d}{dp^0}\Big|_{p^0=0} \del G_{1,1}^j(p^0,p_F^*\vec{i}).} 
\EEQ
\end{itemize}
\end{Definition}

\Bigskip Subtracting counterterms is equivalent to subtracting
{\em local parts} of diagrams, which is done as follows. Let
\BEQ {\cal A}^{j\to}(\Upsilon;\xi,\xi'):=\int dp\, e^{\II (p,\xi-\xi')} {\cal A}^{j\to}_p(\Upsilon) 
\EEQ 
be the amplitude of $\Upsilon$ in direct space.
The external leg at $\xi'$ is paired to a field $\Psi^{k,\alpha}(\xi'')$ or $\bar{\Psi}^{k,\alpha}(\xi'')$, with $k\ge j$, giving rise to an external covariance kernel $C^{k,\alpha}(\xi',\xi'')$. The direct space renormalization algorithm proceeds by applying Taylor's formula to the above kernel, so
as to displace its attachment point from $\xi'$ to $\xi$. 
For diagrams contributing to $G^j_{diag}$, we need to expand
to order 1 so as to lower the order of divergence by two.
For diagrams contributing to $G^j_{off}$, a simple subtraction
is enough: 

\begin{itemize} 
\item[(i)] {\em  (off-diagonal diagrams)}
Setting apart only the leading-order term, we get 
\BEA && \int d\xi'\, {\cal A}^{j\to}(\Upsilon;\xi,\xi')\,  C^{k,\alpha}(\xi',\xi'') =
 \int d\xi'\, {\cal A}^{j\to}(\Upsilon;\xi,\xi')
 e^{\II \,  (p^{k,\alpha},\xi'-\xi)} \, 
C^{k,\alpha}(\xi,\xi'') \nonumber\\
&&+ \int_0^1 dt\, \int d\xi'\, {\cal A}^{j\to}(\Upsilon;\xi,\xi')
\ \partial_t\Big( e^{\II (1-t) \,  (p^{k,\alpha},\xi'-\xi)}
C^{k,\alpha}((1-t)\xi+t\xi',\xi'')\Big) \nonumber\\
&&\equiv {\cal A}^{j\to}_{p^{k,\alpha}}(\Upsilon) C^{k,\alpha}(\xi,\xi'') + 
{\cal R}\bar{\cal A}^{j\to}(\Upsilon;\xi,\xi''),
\label{eq:R1}
\EEA
a sum of a {\em local part} as in (\ref{eq:local-part0}), and
of a renormalized amplitude ${\cal R}\bar{\cal A}^{j\to}(\Upsilon;\xi,\xi'')$ involving the scalar product
$
(\xi'-\xi,(\nabla-\II p^{k,\alpha})C^{k,\alpha}(\cdot,\xi''))
$  is
manifest. Since the lowest internal scale of $\Upsilon$ is $j$,
 distances $|\xi'-\xi|$ between vertices contribute 
 a scaling factor $\lesssim 2^j$ to the expected power-counting,
 as can be shown using part of the decay factors in the
 covariance (\ref{eq:I-II}); this is a standard argument, see
 e.g. \cite{MagUnt2}. On the other hand, see again (\ref{eq:I-II}),
 the gradient operator $\nabla-\II p^{k,\alpha}$ produces
 a small factor $O(2^{-k})$. All together, we have obtained a small spring factor $O(2^{-(k-j)})$, having the effect
of lowering the degree of divergence of the diagram from $0$ to $-1$.

\item[(ii)]
{\em (diagonal diagrams)}
For such diagrams, we need to  set apart the two leading-order terms,

\BEA && \int d\xi'\, {\cal A}^{j\to}(\Upsilon;\xi,\xi') C^{k,\alpha}(\xi',\xi'') = \int d\xi'\, {\cal A}^{j\to}(\Upsilon;\xi,\xi')
 e^{\II \,  (p^{k,\alpha},\xi'-\xi)}
C^{k,\alpha}(\xi,\xi'') \nonumber\\
&&+ \int d\xi'\, {\cal A}^{j\to}(\Upsilon;\xi,\xi') \Big(
\xi'-\xi, \partial_{\xi}\Big( e^{\II \,  (p^{k,\alpha},\xi'-\xi)}
C^{k,\alpha}(\xi,\xi'')\Big)\Big) \nonumber\\
&&+ \int_0^1 dt\, \int d\xi'\, {\cal A}^{j\to}(\Upsilon;\xi,\xi')\ \partial_t^2\Big( e^{\II (1-t) \,  (p^{k,\alpha},\xi'-\xi)}
C^{k,\alpha}((1-t)\xi+t\xi',\xi'')\Big) \nonumber\\
&&\equiv \Big( {\cal A}^{j\to}_{p^{k,\alpha}}(\Upsilon)
 + \Big(p,
\frac{d}{dp} {\cal A}^{j\to}_p(\Upsilon)\big|_{p=p^{k,\alpha}} \big) \Big) \,   C^{k,\alpha}(\xi,\xi'') + 
{\cal R}^2\bar{\cal A}^{j\to}(\Upsilon;\xi,\xi'')
 \nonumber\\   \label{eq:R2}
 \EEA
where ${\cal R}^2(\cdot)$ subtracts the local part of the diagram to order 1.
Compared to  (i), the renormalized amplitude involves now
the square of the scalar product $
(\xi'-\xi,(\nabla-\II p^{k,\alpha})C^{k,\alpha}(\cdot,\xi''))
$, 
 hence a spring factor $O(2^{-2(k-j)})$, having the effect
of lowering the degree of divergence of the diagram from $1$ to $1-2=-1$.

\end{itemize}

\item {\bf ($N_{ext}$-point functions with $4\le N_{ext}<N_0$)}  
A simple subtraction (as for off-diagonal  two-point diagrams) is required.
Fixing one of the external vertices $\xi_{ext,1}$, we rewrite
$\Big(\prod_{i=2}^{N_{ext}}\int  d\xi_{ext,i}\Big) {\cal A}^{j\to}(\Upsilon;\vec{\xi}_{ext}) \ \prod_{i=2}^{N_{ext}} 
C^{k_i,\alpha_i}(\xi_{ext,i},\xi''_i)$ as a local part,\\
$ \Big\{ \Big(\prod_{i=2}^{N_{ext}} \int d\xi_{ext,i}\, 
e^{\II (p^{k_i,\alpha_i},\xi_{ext,i}-\xi_{ext,1})} \Big) {\cal A}^{j\to}(\Upsilon;\vec{\xi}_{ext}) \Big\} \ \prod_{i=2}^{N_{ext}} 
C^{k_i,\alpha_i}(\xi_{ext,1},\xi''_i)$, plus an error term which
may be written in integral form as in (i) above.

\Medskip Four-point diagrams are in principle logarithmically divergent. However, as we shall prove in \S \ref{subsection:complementary},
the {\em only} logarithmically divergent contributions come from
Goldstone boson insertions, which are explicitly resummed
into the $\Sigma$-kernel. So {\em no counterterms need to
be introduced for four-point functions}: the coupling
constant $\lambda=\lambda^{j_D}$ keeps its bare value
throughout.

\end{enumerate}

\Bigskip Leading-order diagrams in $g$ or $g^2$  are obtained as follows,

{\em (i)}

 {\centerline{$\RFmu$}}

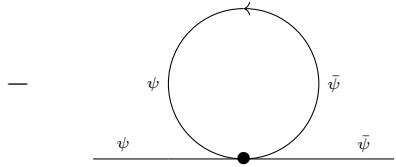
\Medskip{\bf \tiny Fig. \ref{subsection:dressing}.1. Leading contribution to chemical potential renormalization
$\del\mu^{j+1}-\del \mu^j$.}

\Bigskip {\em (ii)}

{\centerline{$\RFm$}}

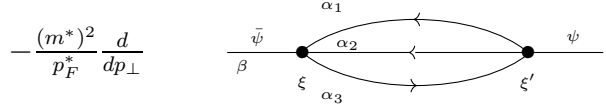
\Medskip{\bf \tiny Fig. \ref{subsection:dressing}.2. Leading contribution to mass renormalization $\del m^{j+1}-\del m^j$.}

\Bigskip {\em (iii)}

{\centerline{$\RFZ$}}

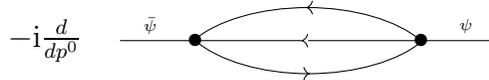
\Medskip{\bf \tiny Fig. \ref{subsection:dressing}.3. Leading contribution to $Z$ renormalization $\del Z^{j+1}-\del Z^j$.}

\Bigskip {\em (iv)}

{\centerline{$\RFGamma$}}

\Medskip{\bf \tiny Fig. \ref{subsection:dressing}.4. Leading contribution to gap renormalization $\del\Gamma^{j+1}-\del\Gamma^j$.}

\Bigskip We denote by $I_{(i)},\, I_{(ii)},\, I_{(iii)}$ and $I_{(iv)}$ their respective values.  We can now restate
Theorem 1 of the Introduction.

\Bigskip
\begin{Lemma} (see Theorem 1) Assume $\Gamma_{\phi}\approx \hbar \omega_D e^{-\pi/m\lambda}\approx 2^{-j_{\phi}}\mu$ . Then:
\begin{enumerate}
\item
For $j=j_D,\ldots,j_{\phi}-1$, the renormalization flow is to leading order

\BEQ \del \mu^{j+1}-\del\mu^j=-(\mu^{j+1}-\mu^j) \approx g 2^{-2j} \mu \label{eq:RFmu} \EEQ

\BEQ \del m^{j+1}-\del m^j\approx -(m^{j+1}-m^j)\approx g^2  2^{-j} m^*  \label{eq:RFm}\EEQ

\BEQ \del Z^{j+1}-\del Z^j=Z^{j+1}-Z^j \approx g^2 2^{-j}  \label{eq:RFZ} \EEQ

\BEQ \del \Gamma^{j+1}-\del \Gamma^j= -(\Gamma^{j+1}-\Gamma^j) \approx  -\frac{g}{\pi}(1+O(g))\Gamma
 \label{eq:RFGamma}
 \EEQ

For $j=j_{\phi},\ldots,j'_{\phi}-1$, one has instead
\BEQ |\del m^{j+1}-\del m^j| \lesssim 2^{-(j-j_{\phi})}g^2  2^{-j} m^*, \qquad |\del \mu^{j+1}-\del\mu^j|\lesssim 2^{-(j-j_{\phi})} g 2^{-2j} \mu \EEQ
\BEQ |\del Z^{j+1}-\del Z^j|\lesssim 2^{-(j-j_{\phi})} 
g^2 2^{-j}, \qquad |\del \Gamma^{j+1}-\del \Gamma^j|\lesssim
2^{-(j-j_{\phi})} g\Gamma_{\phi}.
\EEQ
In particular (summing over scales $j=j_D,\ldots,j'_{\phi}-1$)
 \BEQ Z^{j'_{\phi}}-1\approx g^2. \EEQ
 
 \item 
Furthermore, for an adequate choice of 
 $m^*, \mu^*$, $\Gamma=\Gamma^{(j'_{\phi}-1)\to}$  such that 
\BEQ m^*-m\approx -g^2 m, \qquad \mu^*-\mu\approx g\mu, \qquad |\Gamma|\approx \hbar\omega_D e^{-\pi/g}
\label{eq:mmmumu}
\EEQ
the  scale $j'_{\phi}$ mass and chemical potential 
 counterterms vanish, more precisely,
 \BEQ \del m^{j'_{\phi}}=0, \qquad \del \mu^{j'_{\phi}}=0,
 \EEQ
and the pre-gap equation, see Definition \ref{def:pre-gap}, holds.
 Also,  \BEQ  \boxed{\del 
 \Gamma^{j'_{\phi}}=O(g\Gamma_{\phi}).}  \label{eq:delGammajphi}
 \EEQ
\end{enumerate}
\end{Lemma}

{\bf Proof.}

\begin{enumerate}
\item  We shall be content here with computing the values
of $I_{(i)},I_{(ii)},I_{(iii)},I_{(iv)}$; the arguments of
\S \ref{subsection:multi-scale} show that more complicated
diagrams give contributions which are smaller by a factor
$O(g)$.   The evaluation of the diagrams of  Fig. 
\thesubsection.1,\thesubsection.2,\thesubsection.3,\thesubsection.4 (see
\cite{FMRT-infinite} for details) is  based on the decomposition
(\ref{eq:dele}) of diagonal propagators, implying\\ $C^*_{\theta}\approx 
\left(\begin{array}{cc}  \frac{1}{-\II p^0+e^*(\vec{p})} & \\ & \frac{1}{\II p^0+
e^*(\vec{p})}
\end{array}\right) 
\approx
\left(\begin{array}{cc}  \frac{1}{-\II p^0+p_{\perp}} & \\ & \frac{1}{\II p^0+p_{\perp}}
\end{array}\right)$ for momenta of scale $j\ll j_{\phi}$. Recall that the integration
measure
$dp=dp^0 \, d\vec{p}$ averaged over the angular coordinates (i.e. summed over
sectors) is $\approx 2\pi p_F\, dp^0\, dp_{\perp}$.  We do
the computations for $j\le j_{\phi}-1$; for larger $j$, one
need just remark that 
the covariance kernel (\ref{eq:I-IIbis}) has a supplementary 
$2^{-(j-j_{\phi})}$ prefactor.

\begin{itemize}
\item[(i)] (chemical potential)  Evaluate the amputated
diagram $I_{(i)}:=\int \frac{dp}{\II p^0+e^*(\vec{p})}$ over momenta  $\frac{|p^0|}{\mu},\frac{|\vec{p}|}{p_F}\approx 
2^{-j}$. By symmetry, the result is real. Using $e^*(\vec{p})\sim \frac{1}{2m^*}
p_{\perp}(p_{\perp}+2p^*_F)$, one gets $I_{(i)}\approx \frac{1}{2m^*}
\int dp^0 \, dp_{\perp}\ \frac{p_{\perp}^2}{(p^0)^2+(\frac{p^*_F}{m^*})^2 p_{\perp}^2} \approx 2^{-2j} \mu.$ In particular, $I_{(i)}>0$.

\item[(ii)] (mass) As in \S \ref{subsection:anisotropic}, the best way to get a correct order of magnitude of $I_{(ii)}$ is to choose
{\em anisotropic} sectors $\alpha_1,\alpha_2,\alpha_3$. If $|\xi-\xi'|\gg
2^{j/2} \mu^{-1}$, then the main contribution to the integrand comes from sectors
$\alpha_i$, $i=1,2,3$ roughly parallel to $\xi-\xi'$. Thus: the sum over sectors
produces a factor $O(1)$; the product $(C^*_{\theta}(\xi,\xi'))^3$, a factor $O((2^{-3j/2})^3)$; 
the derivative $\partial_{p_{\perp}}$, a factor $O(2^{j})$; the integral over $\xi'$, a factor
$O(2^{5j/2})$, i.e. the volume of an anisotropic box. Whence  all together, a contribution $O(2^{-j})$.  The
integral over $|\xi-\xi'|\lesssim 2^{j/2}\mu^{-1}$  yields a contribution
smaller by a factor $O(2^{-3j/2})$, since: each propagator $C^*_{\theta}(\xi,\xi')$ is a $O(2^{-2j})$ instead of 
$O(2^{-3j/2})$, see (\ref{eq:5.15}); the integration volume is $O(2^{2j})$ instead of 
$O(2^{5j/2})$, while (for $\beta$ fixed) the sum over sectors $\alpha_i$, $i=1,2,3$,
reduces essentially in fact to a single sum,
yields a supplementary factor $O(2^{j/2})$ only (see Proposition \ref{prop:sector-counting}, with logarithmic factor removed for anisotropic sectors, following the
Remark just above Definition \ref{def:Cjm}).

\item[(iii)] ($Z$-coefficient) Similar to (ii).

\item[(iv)] ($\Gamma$-counterterm renormalization)
Main scale $j$  diagram is a tadpole involving the symmetry-broken propagator,
 
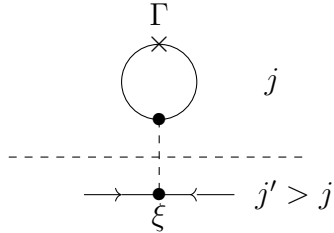
{\centerline{\begin{tikzpicture} \draw(0,0.5) 
circle(0.5); \draw(0,1) node {$\times$}; \draw(0,1.4) node {$\Gamma$}; \draw(0,0) node {$\bullet$}; \draw[dashed](0,0)--(0,-1); \draw[dashed](-2,-0.5)--(2,-0.5); \draw(0,-1) node {$\bullet$}; \draw(0,-1.3) node {$\xi$};
\draw[->](-1,-1)--(-0.5,-1); 
\draw[-<](-0.5,-1)--(0.5,-1); \draw(0.5,-1)--(1,-1);
\draw(1.5,0.5) node {$j$}; \draw(1.8,-1) node {$j'>j$};\end{tikzpicture}}}
As already noted in \S \ref{subsection:gap-lowest}, see Fig. \ref{subsection:gap-lowest}.1, this
diagram involves to leading order the same integral as the Cooper pair bubble diagram $\Upsilon_{3,diag}$, 
and is evaluated (see (\ref{eq:sinh-1}))  as $\sim -\frac{g}{\pi}\Gamma$.

\end{itemize}

\item (fixed point) 
Summing over scales $j=j_D,\ldots,j'_{\phi}-1$, one gets
\BEQ \del m^{j'_{\phi}}=\del m^{j_D}+\sum_{j=j_D}^{j'_{\phi}-1} (\del m^{j+1}-\del m^j)=\frac{m^*}{m} \Big\{ (m^*-m) + c_1
g^2 mf_1(m^*,\mu^*,\Gamma_{\phi};g)\Big\} \label{eq:del1}
\EEQ

\BEQ \del \mu^{j'_{\phi}}=\del \mu^{j_D}+\sum_{j=j_D}^{j'_{\phi}-1} (\del \mu^{j+1}-\del \mu^j)=  (\mu^*-\mu) + c_2
g \mu f_2(m^*,\mu^*,\Gamma_{\phi};g) \label{eq:del2}
\EEQ

\BEQ \del\Gamma^{j'_{\phi}}=\del\Gamma^{j_D}+ \sum_{j=j_D}^{j'_{\phi}-1} (\del \Gamma^{j+1}-\del\Gamma^j)=
\Gamma \Big\{1 -\frac{g}{\pi} (j_{\phi}-j_D) f_3(m^*,\mu^*,\Gamma_{\phi};g)  \Big\} \label{eq:del3}
\EEQ
(see proof of 1.(iv) for the last equality),
where $c_1,c_2>0$ are non-dimensional constants, $f_1,f_2,f_3=1+O(g)$, and  $\frac{\partial f_i}{\partial g}=O(1)$, $\frac{\partial f_i}{\partial m^*}=O(1/m)$,   $\frac{\partial f_i}{\partial \mu^*}, 
\frac{\partial f_i}{\partial \Gamma_{\phi}}=O(1/\mu)$ $(i=1,2,3)$ for any
$(m^*,\mu^*)$ such that $m^*-m=o(m),\mu^*-\mu=o(\mu)$.
Eq. (\ref{eq:del3}), combined with (\ref{eq:pre-Gap-bis}), imply
\BEQ  \del\Gamma^{j'_{\phi}}=O(g\Gamma_{\phi})
\EEQ
  for such values of $m^*,\mu^*$. 
Then the  implicit function theorem yields for $g$ small enough a unique
solution $(m^*=m^*(g),\mu^*=\mu^*(g),\Gamma=\Gamma^{(j'_{\phi}-1)\to})$ to the system of equations $(\del m^{j'_{\phi}}=0,\del \mu^{j'_{\phi}}=0)$
complemented with the pre-gap equation (\ref{eq:pre-gap-scal}).
Furthermore, they satisfy (\ref{eq:mmmumu}) and 
(\ref{eq:delGammajphi}).

\end{enumerate}


\subsection{Cluster expansions} \label{subsection:cluster}


The art of cluster expansions for a  general renormalizable theory consists  in  resumming separately the leading contributions of diverging $n$-point functions, and absorbing them into 
a {\em renormalization} of the measure.  By "renormalization of the measure", we intend a change of normalization by a constant prefactor (which disappears when one evaluates connected
correlations), and a shift of its parameters, here $\mu,m$ and 
$\Gamma$.    This is done in an organized way by evaluating $n$-point functions in terms of a {\em sum over clusters} or {\em polymers}, which themselves
may be expanded perturbatively into an infinite series of diagrams involving only fields
located inside the image of the polymer. Clusters
are arrays of scaled  {\em boxes} connected  by a tree of links -- the linking consisting of propagators
$C^j(\xi,\xi')$ between two boxes of $\D^j$. Polymers are synonymous for
clusters, or multi-scale versions of them, involving both {\em horizontal links}
connecting boxes of the same scale, and {\em vertical links} capturing the interaction
between scales through multi-scale vertices. Here is a
 graphical representation of a multi-scale vertex $(\bar{\Psi}\Psi)^2$,
with the four half-propagators connecting it to other vertices;
the scale of the vertex is by definition {\em the scale of
the half-propagator with highest momentum}, here $j$.

\bigskip
\begin{tikzpicture}[scale=0.7]
\draw[color=blue](2,-1) node {$\Del^j_{\xi}$};
\draw[color=blue](2,-3) node {$\Del^k_{\xi}$};
\draw[color=blue](2,-5) node {$\Del^{k'}_{\xi}$};
\draw(13,-1) node {$j$};
\draw(13,-3) node {$k$}; \draw(13,-5) node {$k'$};
\draw(0,0)--(15,0); \draw(0,-2)--(15,-2);
\draw(0,-4)--(15,-4); \draw(0,-6)--(15,-6);
\draw(4,-1) node {\textbullet};
\draw(4,-0.5) node {$\xi$}; \draw(4,-1)--(3,-1);
\draw[dashed](4,-1)--(4,-5); \draw(4,-3)--(5,-3);
\draw(4,-5)--(4.6,-5.6); \draw(4,-5)--(3.4,-5.6);
\draw[color=blue](3.8,-1.2) rectangle(4.2,-0.8);
\draw[color=blue](3.6,-2-1.4) rectangle(4.4,-2-0.6);
\draw[color=blue](3.2,-5-0.8) rectangle(4.8,-5+0.8);

\begin{scope}[shift={(4,0)}]
\draw(4,-0.5) node {$\xi$};
\draw(4,-1) node {\textbullet}; \draw(5,-1)--(3,-1);
\draw[dashed](4,-1)--(4,-5); \draw(4,-3)--(5,-3);
\draw(4,-5)--(5,-5); 
\draw[color=blue](3.8,-1.2) rectangle(4.2,-0.8);
\draw[color=blue](3.6,-2-1.4) rectangle(4.4,-2-0.6);
\draw[color=blue](3.2,-5-0.8) rectangle(4.8,-5+0.8);
\end{scope}
\end{tikzpicture}

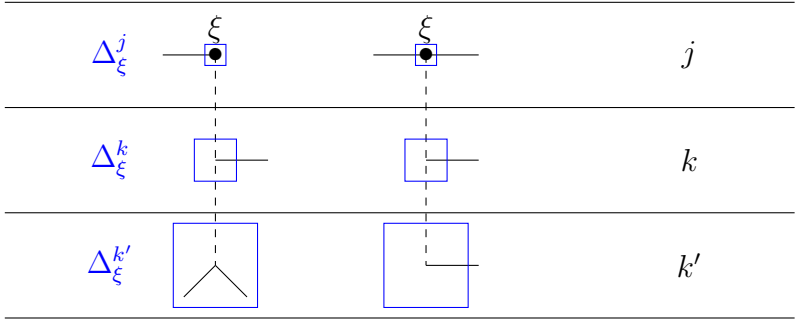
\Medskip {\tiny \bf Fig. \thesubsection.1. Two multi-scale
vertices of scale $j$, from left to right:\\ $\Psi^j(\xi)\bar{\Psi}^{k}(\xi) (\Psi^{k'}\bar{\Psi}^{k'})(\xi)$, and
$(\Psi^j\bar{\Psi}^j)(\xi)(\Psi^{k}\bar{\Psi}^{k'})(\xi)$. By assumption $j<k\le k'$.}

\Bigskip
  {\em Horizontal cluster expansions}
(one per scale) produce horizontal links. {\em Vertical cluster expansions} produce
vertical links, and single out in particular divergent polymers, which are polymers with a small (here $\le 4$) number of external legs. The renormalization step resums the
leading term -- called {\em local part} -- of divergent polymers into a scale-dependent
redefinition of the parameters of the interaction. As a minor
correction, a further cluster-like expansion (called {\em Mayer expansion}) is required to get rid of the non-overlapping conditions between boxes belonging to two different polymers, so as to regain translation invariance for divergent polymers. So far for fundamentals.

\Medskip
We start from scale $j=j_D$ and apply inductively the following {\bf sequence of fermionic
expansions}, down to a scale $j'_{\phi}=j_{\phi}+o(\ln(1/g))$,

\Medskip
(Horizontal cluster expansion of scale $j_D$) 
$\longrightarrow$ (Resummation of (lowest) scale $j_D$ chains of
bubbles) $\longrightarrow$
(Momentum-decoupling expansion of scale $j_D$) $\longrightarrow$
(Displacement of external legs and Mayer expansion of scale $j_D$ )  $\longrightarrow\cdots$

\Medskip
$\longrightarrow$ (Horizontal cluster expansion of scale $j$)
$\longrightarrow$ (Resummation of  lowest scale $(j+1)$ chains of
bubbles) $\longrightarrow$
(Momentum-decoupling expansion of scale $j$) $\longrightarrow$
(Displacement of external legs and Mayer expansion of scale $j$ )       $\longrightarrow \cdots$

\Medskip $\longrightarrow$
(Horizontal cluster expansion of scale $(j'_{\phi}-1)$) 
$\longrightarrow$ (Resummation of  lowest scale $(j'_{\phi}-1)$ chains of
bubbles)  $\longrightarrow$
(Momentum-decoupling expansion of scale $(j'_{\phi}-1)$) $\longrightarrow$
(Displacement of external legs and Mayer expansion of scale $(j'_{\phi}-1)$ )

\Medskip Once this program has been completed, we shall
be led to apply in the next section (see \S \ref{section:boson}) 
a sequence of  {\bf complementary horizontal} or
{\bf bosonic expansions} of {\em bosonic} scales $j_+\ge j'_{\phi}$,

\Medskip
(Horizontal and Complementary horizontal expansions of  scale $j'_{\phi}$) $\longrightarrow $ (Complementary horizontal expansion of bosonic scale $j'_{\phi}+1$)   $\longrightarrow\cdots $ (Complementary horizontal expansion of bosonic scale $j_+$) $\longrightarrow\cdots $

\Medskip with scales ranging from $j'_{\phi}$ to $+\infty$. Scale $j'_{\phi}$ is very particular, since
purely fermionic connectivity issues (see introduction to this subsection) mingle with
the necessity of factorizing Goldstone boson propagators to show infra-red summability of the
expansion. Therefore the scale $j'_{\phi}$ horizontal cluster expansion is postponed to
section \ref{section:boson}.

\Bigskip Apart from the intertwining sequence of bubble
resummations,  the sequence of {\em fermionic expansions} is fairly
standard, see e.g.  \cite{Unt-rev} for a
review. Through the horizontal and momentum-decoupling  expansions, $n$-point functions are rewritten as sums
of products of {\em polymer}-dependent expressions of
the type $F(\P)$. A polymer $\P$ is by definition a 
union of boxes in $\D$ with the following properties:

(i) two boxes $\Del,\Del'\in\P\cap\D^j$ may or may not be
connected by a {\em horizontal  link} produced by the {\em  scale $j$ horizontal
cluster expansion}. The set of all boxes in $\P\cap\D^j$,
connected by horizontal links, is a {\em cluster forest}, i.e. a disjoint
union of a finite number of {\em cluster trees};

(ii) two boxes $\Del\in \P\cap\D^j,\Del'\in \P\cap \D^{j+1}$
such that $\Del\subset \Del'$, i.e. $\Del'$ lies just
below $\Del$, may or may not be connected by a {\em vertical
link} produced by the {\em scale momentum-decoupling} (or
{\em vertical}) {\em expansion}. The set of all boxes 
in $\P$, connected by {\em cluster links}, i.e. {\em horizontal and vertical links}, is connected.

The set of all such polymers with lowest scale $j_{max}$ will
be denoted by ${\cal P}^{j_{max}\to}$.  \label{Pjto} Furthermore, if
$\P\in{\cal P}^{j_{max}\to}$, then

(iii) ({\em external structure of a polymer}) there exists a subset ${\bf \Del}^{j_{max}}={\bf\Del}^{j_{max}}(\P)
\subset\P\cap\D^{j_{max}}$ of scale $j_{max}$ boxes vertically connected
to the boxes in $\D^{j_{max}+1}$ lying just below them. 

 The function $F(\P)=F(\P; \Psi^{\to j_{max}}\Big|_{{\bf \Del}^{j_{max}}},
\bar{\Psi}^{\to j_{max}}\Big|_{{\bf \Del}^{j_{max}}}) $ is
obtained by Gaussian integrations by parts, which produce
 a number of {\em vertices}, each lying in one of the
 boxes contained in $\P$; say, $n(\Del)$ vertices in
 $\Del\in\P$. If $\Del,\Del'\in\P\cap\D^k$
 $(k\le j_{max})$ are connected, then $F(\P)$ contains
 a propagator $C^{k,\alpha}(\xi,\xi')$, with $\xi\in\Del,\xi'\in\Del'$. Furthermore, $F(\P)$ contains at least one
 external field $\Psi^{\to j_{max}}$ or $\bar{\Psi}^{\to j_{max}}$ per
 external box $\Del\in{\bf\Del}^{j_{max}}$, namely, 
 $\Big|\frac{\del F(\P; \Psi^{\to j_{max}}|_{{\bf \Del}^{j_{max}}},
\bar{\Psi}^{\to j_{max}}|_{{\bf \Del}^{j_{max}}})}{\del \Psi^{\to j_{max}}(\xi)} \Big| + \Big|\frac{\del F(\P; \Psi^{\to j_{max}}|_{{\bf \Del}^{j_{max}}},
\bar{\Psi}^{\to j_{max}}|_{{\bf \Del}^{j_{max}}})}{\del \bar{\Psi}^{\to j_{max}}(\xi)} \Big|\, \not=0$ if $\xi\in\Del$. The {\em number
of external fields} of $\P$ is denoted by $N_{ext}(\P)$.
\label{eq:Next}

\Bigskip In principle, the sum over all polymers can be bounded using {\bf perturbative arguments} as follows. First we {\em fix} the location of
one of the lowest-scale boxes, say, $\Del_0\in\D^{j_{max}}$, in order to
avoid introducing diverging volume factors proportional to $|V|$, and let 
${\cal P}^{j_{max}\to}_{\Del_0}$ be the set of polymers $\P$ in
${\cal P}^{j_{max}\to}$ such that $\Del_0\in\P$.  In practice, 
$\Del_0$ may be seen as the approximate location of one of
the external legs of the $n$-point function.  Then the sum $ \sum_{\P\in{\cal P}^{j_{max}\to}_{\Del_0}} F(\P)$ is controlled by
two types of factors,

-- {\em local factors}, $O(\lambda^{\kappa})$ $(\kappa>0)$ for
each vertex, in particular, for each box;

-- {\em geometric factors} due to the fast-decaying propagators $C^{k,\alpha}(\xi,\xi')$.

\noindent Geometric factors make it possible to sum over all possible
geometries and vertex number assignations $(n(\Del))_{\Del}$ for a given number of vertices, say, $n=n(\P)$. Local factors
ensure that the sum over $n$ is bounded by a geometric
series of the type $\sum_n (\lambda^{\kappa})^n$.

\Bigskip  Things are however not so simple. We have overlooked in the above quick overview  extra {\em volume factors} due to the necessity, for $\Del'\in\P\cap\D^{k}$ fixed, of summing over all $\Del\in\P\cap\D^{k-1}$ lying above
$\Del'$; this necessity comes from the fact that only
the locations of the {\em external legs} of the polymer  (or, more or 
less equivalently, of $\Del_0\in \D^{j_{max}}$)  are fixed.   
If there is a vertex in $\Del$, then this can be remedied
by simply replacing $\lambda^{\kappa}$ by a constant times $\lambda^{\kappa}$. However, it may well be that successive
vertical links $\Del\subset\Del^{j+1}\subset\cdots\subset
\Del^{k-1}\subset\Del'$ connect $\Del\in\D^j$ to $\Del'$
through boxes $\Del^{j+1},\ldots,\Del^{k-1}$ {\em void}
of any vertex. The sum over $\Del^j\in\D^j$ such that
$\Del^j\subset\Del'$ becomes an arbitrarily large {\em volume
factor} $2^{3(k-j)}$ as the
scale difference $k-j$ increases, eventually swallowing up
the small factor $\lambda^{\kappa}$. However, multi-scale diagrams
extending over several scales can be shown by simple power-counting arguments to be smaller than
expected
by some similar, exponential factor $\le 2^{-c(k-j)}$, where
$c$ depends on the number $N_{ext}$  of external legs. This is typically a situation which calls for {\em renormalization}; if there
are  enough links between the part of $\P$ lying
above $\Del$ and $\Del'$, then the vertex power-counting
beats the volume factor because $c>3$; in this theory $6$ links are a priori enough, but for technical reasons (due to the
complexity of the sum over angular sectors of external legs,
see \S \ref{subsection:bounds} below), we shall follow
the same procedure whenever $N_{ext}< N_0:=8$. When $N_{ext}< N_0$,  the
renormalization procedure described in \S \ref{subsection:dressing} should be applied. Subtracting
the local part of a polymer is equivalent to {\em displacing
all its external legs} to the location of one of them. 
Error terms (polymer) - (local part) have the same
relative power-counting as if they had $\ge N_0$ external legs,
whence they {\em are indeed} considered as if they did have
$\ge N_0$ external legs.
On the other hand, local parts are {\em resummed} by defining scale-dependent
constants $\mu^j,m^j,\Gamma^j$, at the price of
 some easy partial resummation of the series of polymers, called
 {\em Mayer expansion}, which takes into account the
 non-overlapping constraints between boxes, and whose
 outcome is a sum over {\em Mayer polymers}. At this
 stage, it is enough to know that the Mayerization step
 involves a "multicolor" version of the theory, with one
 color per polymer. Then previously developed links connect
 boxes of the same color, while Mayer polymers are multicolor
 polymers connected by {\em overlap links} between different
 colored versions of the same two boxes. Painting all boxes
 gray, Mayer polymers are then polymers in the usual sense,
 with some extra overlap links connecting a box to itself.
Summing over Mayer polymers in ${\cal P}_0^{j_D}=\{\P\in 
{\cal P}^{j_D\to} \ |\ N_{ext}(\P)=0\}$, one obtains e.g. for the theory
restricted to scale $j_D$
\BEQ {\cal Z}^{j_D}_{\lambda,V}=\sum_{N=0}^{\infty} \frac{1}{N!}
\sum_{\P_1,\ldots,\P_N\in {\cal P}_0^{j_D}} \prod_{n=1}^N
F(\P_n) \approx_{|V|\to\infty} e^{|V|f^{j_{D}}(\lambda)} \EEQ
in the sense that $\frac{1}{|V|} \log {\cal Z}_{\lambda,V}^{j_D}\to_{|V|\to\infty} f^{j_{D}}(\lambda)$, 
where (fixing some box $\Del_0\in\D^{j_{D}}$ in an arbitrary way) $|V|$ is the volume of $V$, defined as the number of
scale $j_D$ boxes lyind inside $V$, and  $f^{j_{D}}(\lambda)=\sum_{\P\in {\cal P}_0^{j_D}\ |\ \Del_0\in \P} F(\P)$ is interpreted as the {\em scale $j_{D}$ free energy density}. Going further down
the scales, the general task is to prove that ${\cal Z}^{j\to}_{\lambda,V}\approx_{|V|\to\infty} \prod_{k=j}^{j_D} e^{|V|f^k(\lambda)}$, where $f^k(\lambda)$ -- similarly interpreted as  a scale $j$ free
energy density -- is obtained by summing over Mayer polymers
in ${\cal P}^{k\to}$.
 
 \Bigskip In general, the {\em coupling constant} $\lambda$
 should also be made scale-dependent, following the above
 local part resummation pattern. However, the specificity
 of this model (as explained in the Introduction) is that
 the "local part" of the main diagrams contributing to the renormalization of
 $\lambda$, making up the Bethe-Salpeter kernel, must be resummed {\bf nonperturbatively} as a series. This strongly
 model-dependent, nonperturbative step is performed  separately, by expanding successively Cooper pairs with transfer momentum
 $q$ of  scales $j_+=j'_{\phi},j'_{\phi}+1,j'_{\phi}+2,\ldots$
 downto the lowest infra-red scales. 
 It will be described in  the  section about the bosonic regime (section \ref{section:boson}). 

\Medskip Downto scale $j'_{\phi}-1$, however, namely, in the present section, much simpler
arguments suffice. Namely, one may {\em systematically expand
chains of bubbles}. Looking more specifically e.g. at the
first resummation step of the sequence of fermionic expansions,
 chains of bubbles with scale $j_D$ fermionic propagators
 are first considered as contracted external legs of scale $j_D$ fermionic
 diagrams, and  resummed systematically before expanding
 external legs of scale $\ge j_D+1$; in some sense, this procedure
 may be seen as a expansion in itself,  of  scale
 $"j_D+\half"$ intermediate between $j_D$ and $j_D+1$. Similarly, the resummation of lowest scale $j$ chains of 
 bubbles may be considered as an expansion step of scale $"j+\half"$.

\Medskip The reader will have to wait until \S \ref{subsection:bounds} to have a comprehensive image
of how polymers of lowest scale $\le j'_{\phi}-1$ are bounded. It is however
not straightforward to understand how all arguments fit
together. Therefore, we shall first present   a single-scale bound for the outcome 
of the horizontal cluster expansion in \S
\ref{subsubsection:hor-cluster}. The bounds of \S \ref{subsection:bounds}
cover the whole fermionic regime down to scale $j'_{\phi}-1$ and
take into account  bubble resummations.


\subsubsection{Horizontal cluster expansion}\label{subsubsection:hor-cluster}


Let $\vec{s}$ be a family of functions $\vec{s}:=(s^{j})_{j=j_D,\ldots,j'_{\phi}}$,
with
\BEQ s^{j}: \{\Del,\Del'\}\to s^j_{\Del,\Del'}\in[0,1] \qquad \Del,\Del'\in\D^j,
\Del\not=\Del' \EEQ
extended trivially to the diagonal by letting $s^j_{\Del,\Del}\equiv 1$
$(\Del\in\D^j)$. (Later on, we shall use slightly
generalized functions, defined on pairs of {\em clusters},
where clusters are unions of boxes of the same scale). As we shall see, the set of coefficients $(s^j_{\Del,\Del'})$ for $j$ fixed, called {\em scale $j$ weakening coefficients}, 
defines an interpolation between the initial covariance $C^{j,\alpha}(s^j=1)\equiv C^{j,\alpha}$ and the box-diagonal covariances $C^{j,\alpha}(s^j=0)$ for which fields lying in different boxes have been made independent, namely,
$C^{j,\alpha}(s^j=0;\xi,\xi')\equiv 0$ if $\xi\in\Del,\xi'\in\Del'$, $\Del\not=\Del'\in\D^j$.

\Medskip {\em Index notations:} it is often useful or simply convenient to
use different indices for external $\Psi$- and $\bar{\Psi}$-fields; thus we shall often
index the $\Psi$-fields and their coordinates $\xi$ by an index $i$ ranging in some abstract set $\subset\{1,\ldots,n\}\subset\N$, and the $\bar{\Psi}$-fields 
and their coordinates $\bar{\xi}$ by an index $\bar{i}$ ranging in some
abstract set $\subset\{\bar{1},\ldots,\bar{n}\}\subset\bar{\N}$, where $\bar{\N}$ is a  copy of $\N$. On the other hand, on other occasions it will be more convenient to use "neutral" notations which do not distinguish between 
$\Psi$'s and $\bar{\Psi}$'s; for instance, if $I\equiv J\uplus \bar{J}$, $J\subset\{1,\ldots,n\}$, $\bar{J}\subset
\{\bar{1},\ldots,\bar{n}\}$, then $\vec{\xi}_I:=(\xi_{ext,i})_{i\in I}$, $\vec{\Psi}_I:=({\Psi}_i(\xi_{ext,i}))_{i\in J},(\bar{\Psi}_{\bar{i}}(\bar{\xi}_{ext,\bar{i}})_{\bar{i}\in\bar{J}})$.

\Bigskip {\bf Horizontal cluster expansion of scale $j_D$.} 
 In order to compute the connected $n$-point function
\BEA &&  G_{n,\bar{n}}({\vec{\xi}}_{ext},\bar{\vec{\xi}}_{ext}) \equiv G_{\{1,\ldots,n\}\uplus
\{\bar{1},\ldots,\bar{n}\}}((\xi_{ext,i})_{i=1,\ldots,n},(\bar{\xi}_{ext,\bar{i}})_{\bar{i}=\bar{1},\ldots,\bar{n}})
\nonumber\\
&& :=\Big\langle
 \Big(\prod_{i=1}^n {\Psi}_{\sigma_i}(\xi_{ext,i}) \Big)
\Big(\prod_{i'=1}^n \bar{\Psi}_{\sigma_{\bar{i}}}(\bar{\xi}_{ext,\bar{i}}) \Big)\Big
\rangle_{\theta;\lambda}^c \nonumber\\
&&\qquad \equiv   \int d\mu_{\theta;\lambda}(\Psi,\bar{\Psi}) \ \Big( \prod_{i=1}^n {\Psi}_{\sigma_i}(\xi_{ext,i}) \Big) \Big( \prod_{\bar{i}=1}^{\bar{n}}\bar{\Psi}_{\sigma_{\bar{i}}}(\bar{\xi}_{ext,\bar{i}}) \Big)  \nonumber\\
&& \qquad -\sum_{k\ge 2}\sum_{I_1\uplus\ldots\uplus I_k=\{1,\ldots,n\}\uplus
\{\bar{1},\ldots,\bar{n}\}} \prod_{i=1}^k G_{I_i}(\vec{\xi}_{I_i})
\EEA
one applies a  Brydges-Kennedy-type expansion, see  Appendix, \S
\ref{subsection:integration-by-parts}, 
  to the numerator 
\BEQ F_{n,\bar{n}}(s^{j_D};\vec{\xi}_{ext},\bar{\vec{\xi}}_{ext}):= \int d\mu^*_{\theta} 
(\vec{s};\Psi,\bar{\Psi})\  e^{-{\cal L}_{\theta}(\Psi,\bar{\Psi})} \Big( \prod_{i=1}^n {\Psi}_{\sigma_i}(\xi_{ext,i}) \Big)\Big(\prod_{\bar{i}=\bar{1}}^{\bar{n}} \bar{\Psi}_{\sigma_{\bar{i}}}(\bar{\xi}_{ext,\bar{i}}) \Big), \EEQ
where $d\mu^*_{\theta}({s}^{j_D};\cdot)$ is the free Grassmann
measure on $V\times V$ with weakened two-by-two matrix covariance kernel
\BEQ C^*_{\theta}(s^{j_D})\equiv C_{\theta}^{j_D,\alpha}(s^{j_D})+  C_{\theta}^{\to(j+1)}, \EEQ
\BEQ  C^{j_D,\alpha}_{\theta}(s^{j_D};\xi,\xi'):=\sum_{\Del,\Del'\in\D^{j_D}} \chi_{\Del}(\xi)\chi_{\Del'}(\xi')\,  s^{j_D}_{\Del,\Del'} \, 
C^{j_D,\alpha}_{\theta}(\xi,\xi'), 
\EEQ
where $s^{j_D}:L(\D^{j_D})\equiv \{ \{\Del,\Del'\}\in\D^{j_D}\times\D^{j_D}\ |\ \Del\not=\Del'\}\to[0,1]$ are sets of  {\em weakening coefficients} which (as their name indicates) weaken the covariance kernels $C^{j_D,\alpha}_{\theta}(\xi,\xi')$ when $\xi,\xi'$ lie in different boxes. The function $s^{j_D}$ 
is implicitly trivially extended to the diagonal by letting
$s^{j_D}_{\Del,\Del}\equiv 1,\ \Del\in\D^{j_D}$.  The unnormalized $(n,\bar{n})$-point  functions $F_{n,\bar{n}}(s^{j_D};\vec{\xi},\vec{\bar{\xi}})$ interpolate between
the original $(n,\bar{n})$-point function $F_{n,\bar{n}}(\vec{1};\vec{\xi},\vec{\bar{\xi}})\propto G_{n,\bar{n}}(\vec{\xi},\vec{\bar{\xi}})$ and the unnormalized
$(n,\bar{n})$-point function $F_{n,\bar{n}}(\vec{0};\vec{\xi},\vec{\bar{\xi}})$ computed in a totally decoupled theory. 
The BKAR formula (\S \ref{subsection:integration-by-parts},
Proposition \ref{prop:BKAR}) yields 
\BEA && F_{n,\bar{n}}(\vec{\xi}_{ext},\vec{\bar{\xi}}_{ext})=
\sum_{I^{j_D}\subset I} \sum_{\F^{j_D}\in{\cal F}^{j_D}}
\left[ \prod_{\ell\in L(\F^{j_D})} \int_0^1 dw^{j_D}_{\ell} \int_{\Del_{\ell}} d\xi_{\ell} \int_{\Del'_{\ell}}d\xi'_{\ell}\,  \right] \nonumber\\
&& \  \ \int d\mu^*_{\theta}(s^{j_D}(\vec{w}^{j_D}))(\Psi,\bar{\Psi})\,  {\mathrm{Hor}}^{j_D} \Big(
\prod_{i\in I^{j_D}}\Psi^{j_D}(\xi_{ext,i}) \ \cdot\ e^{-{\cal L}_{\theta}(\Psi,\bar{\Psi})}\Big) \ \cdot\ \Big(\prod_{i\not\in I^{j_D}}
\Psi^{\to(j_D+1)}(\xi_{ext,i})\Big) \nonumber\\
\label{eq:jDBKAR}
\EEA

\Bigskip 
{\bf Induction on $j$.} {\em Let} ${\cal P}^{j\to}_{N_{ext}}$, resp. ${\cal P}^{j\to}_{\ge N_{ext}}$  {\em be the set of Mayer polymers with lowest scale $j$ and 
$N_{ext}$, resp. $\ge N_{ext}$ external fields.} We assume by induction on $j$ that $F_{n,\bar{n}}(\vec{\xi}_{ext},\vec{\bar{\xi}}_{ext})$ has been
rewritten as 
\BEQ  \prod_{k=j_D}^{j-1} e^{|V| f^{k}(\theta;\lambda)} \   \sum_{I\subset\{1,\ldots,n\}\uplus\{\bar{1},\ldots,\bar{n}\}} F_I^{(j-1)\to}(\vec{\xi}_{ext},\vec{\bar{\xi}}_{ext}),
\EEQ
 where
\BEQ  F_I^{(j-1)\to}(\vec{\xi}_{ext},\vec{\bar{\xi}}_{ext})=\int d\mu^*_{\theta} 
(\Psi^{\to j},\bar{\Psi}^{\to j})\  e^{-{\cal L}_{\theta}(\Psi^{\to j},\bar{\Psi}^{\to j})} \
 \Big(\prod_{i\not\in I} \Psi^{\to j}_{\sigma_i}(\xi_i)\Big) \,  G_I^{(j-1)\to}(\vec{\xi}_{I}),
\label{eq:induction-j}
\EEQ
and $G_I^{(j-1)\to}(\cdot)$ is expressed in terms of a sum over Mayer polymers
$\P_1,\ldots,\P_N$
extending down to scale $j-1$,
\BEQ G_I^{(j-1)\to}(\vec{\xi}_{I})=
 \sum_{n=1}^{\infty} \frac{1}{n!} \sum_{
 \P_1,\ldots,\P_n\in {\cal P}^{(j-1)\to}_{\ge 4}}
 \sum_{I_1,\ldots,I_n} \prod_{i=1}^n F^{j-1}(\P_i;\Psi^{\to j},\bar{\Psi}^{\to j}; \vec{\xi}_{I_i}).  
 \EEQ
The expression involves a sum over all possible assignments
of the external field indices to the various polymers, so
that $I=I_1\uplus\cdots\uplus I_n$; the function
$F^{j-1}(\P_i;\Psi^{\to j},\bar{\Psi}^{\to j}; \vec{\xi}_{I_i})$ is obtained by contracting in a certain way the external fields
$\Psi^{(j-1)\to}(\xi_p),p\in I_i$ with the interaction or
between themselves. The function
$F^{j-1}(\P_i;\Psi^{\to j},\bar{\Psi}^{\to j}; \vec{\xi}_{I_i})$  depends
(polynomially) on the fields $\Psi^{\to j},\bar{\Psi}^{\to j}$
through the external structure of $\P_i$.  By assumption only Mayer polymers  with $N_{ext}\ge 4$
external fields are considered in the sum; the reason is
that

(i) {\em vacuum polymers} (i.e. polymers with $N_{ext}=0$)
have been resummed into the scale free energies;

(ii) {\em two-point polymers}  (i.e. polymers with $N_{ext}=2$) have been absorbed into the renormalization of 
the parameters $\mu,m,\Gamma$;

\Medskip We shall {\em not} decouple boxes $\Del,\Del'\in\D^j$ vertically
connected to two boxes of scale $j-1$ belonging to the
same polymer $\P\in{\cal P}^{(j-1)\to}$, since they are
{\bf connected  from above}. Identifying all such sets of boxes
defines a set of objects ${\cal O}^j$, which are
unions of boxes in $\D^j$. Thus it is enough to introduce
a weakening function $s^j:L({\cal O}^j)\to[0,1]$, with
$L({\cal O}^j)\equiv \{\{o,o'\}\in {\cal O}^j\times {\cal O}^j \ |\ o\not=o'\}$. Replace in (\ref{eq:induction-j}) the integral
$\int d\mu^*_{\theta}(\Psi^{\to j},\bar{\Psi}^{\to j})
\  e^{-{\cal L}_{\theta}(\Psi^{\to j},\bar{\Psi}^{\to j})} $
with 
\BEQ \int d\mu^*_{\theta}(s^j)((T\Psi)^{\to j},(T\bar{\Psi})^{\to j}) \ 
\  e^{-{\cal L}_{\theta}((T\Psi)^{\to j},(T\bar{\Psi})^{\to j})} \label{eq:induction-j-bis}. \EEQ

\Bigskip
{\bf Horizontal cluster expansion of scale $j\le j'_{\phi}-1$.} Start from the expression
(\ref{eq:induction-j-bis}), and proceed as we did
at scale $j_{D}$: 

\BEA && F_{I}^{(j-1)\to}(\vec{\xi}_{ext},\vec{\bar{\xi}}_{ext})=
\sum_{I^{j}\subset I} \sum_{\F^{j}\in{\cal F}^{j}}
\left[ \prod_{\ell\in L(\F^{j})} \int_0^1 dw^{j}_{\ell} \int_{\Del_{\ell}} d\xi_{\ell} \int_{\Del'_{\ell}}d\xi'_{\ell}\,  \right]
\ \cdot\ \nonumber\\
&& \qquad \cdot\  \int d\mu^*_{\theta}(s^{j}(\vec{w}^{j}))(\Psi^{\to j},\bar{\Psi}^{\to j})\,  \tilde{G}_{I^j}^{(j-1)\to}(\vec{\xi}_I),
\nonumber\\
&&\qquad\qquad  \tilde{G}_{I^j}^{(j-1)\to}(\vec{\xi}_I):= {\mathrm{Hor}}^{j} \Big(
\prod_{i\in I^{j}}\Psi^{j}(\xi_{ext,i}) \ \cdot\ e^{-{\cal L}_{\theta}(\Psi^{\to j},\bar{\Psi}^{\to j})}\Big) \ \cdot\ \Big(\prod_{i\not\in I^{j}}
\Psi^{\to(j+1)}(\xi_i)\Big) \nonumber\\ \label{eq:2.75}
\EEA

\Bigskip 
The above scale $j$ horizontal cluster expansion has
produced {\em step $j$ polymers} $\P_1,\ldots,\P_n$ in ${\cal P}^{j\to}$, described as
follows:

-- $\P_i\cap\D^{(j-1)\to}$ is a union of the polymers
in ${\cal P}^{(j-1)\to}$ obtained at the end of the
scale $(j-1)$ perturbative expansions; 

-- $\P_i\cap\D^j, i=1,\ldots,n$ is a union of  clusters
belonging to ${\cal O}^j$.

\noindent More precisely: two step $(j-1)$ polymers $\P_1,\P_2\in{\cal P}^{(j-1)\to}$ 
are connected after step $j$ if and only if there exist
two horizontally connected boxes $\Del'_1,\Del'_2\in\D^j$ (i.e.
boxes belonging to the same step $j$ polymer) and two
boxes $\Del_1\in\P_1\cap\D^{j-1},\Del_2\in\P_2\cap\D^{j-1}$ 
such that $\Del_1$, resp. $\Del_2$ is vertically connected
to $\Del'_1$, resp. $\Del'_2$; in other words, if
$\P_1,\P_2$ are {\bf connected  from below}.  

\Medskip Polymers $\P_1,\ldots,\P_n$ with $N_{ext}<N_0:=8$ external legs  have been
"mayerized" at scales $\le j-1$, but not at scale $j$, which means that they are {\bf non-$j$-overlapping}: $\P_i\cap\D^{(j-1)\to}$, $i=1,\ldots,n$ are free to overlap, but $(\P_i\cap\D^j) \cap (\P_{i'}\cap\D^j)=\emptyset$ if $i\not=i'$.

\Medskip  The final outcome of the set of  scale $j$ expansions for a given scale $j$  can be found
in \S \ref{subsubsection:vert-cluster}, see (\ref{eq:HV-expansion}).


\subsubsection{Single-scale bounds}  \label{subsubsection:single-scale-bounds}


 In this  preliminary version of our final bounds,
we present a bound for a single-scale {\em amputated} diagram. It is correct, but
not fully conclusive, because it fails to take into account bubbles in an effective way, which will be done only in \S \ref{subsubsection:bubble-sum}; as we shall see, the core principles are left unchanged by bubble resummations, but power-counting estimates are modified in an
essential way.  
Fix $j_D\le j\le j'_{\phi}-1$, and assume we want to bound
some connected, amputated $N_{ext}$-point function 
$F^j(\vec{\xi}_{I}),$ $
\vec{\xi}_{I}\equiv (\xi_{ext,1},\ldots,\xi_{ext,\frac{N_{ext}}{2}},
\bar{\xi}_{ext,1},\ldots,\bar{\xi}_{ext,\frac{N_{ext}}{2}}),$
 
\BEA && F^j(\vec{\xi}_{I})= \Big(\prod_{i=1}^{N_{ext}/2}
\frac{\del}{\del\Phi^j(\xi_i)}\Big)\Big|_{\Phi^j=0} \,
\Big(\prod_{i=1}^{N_{ext}/2}
\frac{\del}{\del\bar{\Phi}^j(\xi_i)}\Big)\Big|_{\bar{\Phi}^j=0}  \nonumber\\
&&\qquad \log\Big(  \int d\mu^*_{\theta}(\Psi^{j}+\Phi^j,\bar{\Psi}^{j}+\bar{\Phi}^j) \Big(
e^{-{\cal L}_{\theta}(\Psi^{j}+\Phi^j,\bar{\Psi}^{j}+
\bar{\Phi}^j)} \Big) 
\Big),  \label{eq:dphidphilog}
\EEA
a sum of connected diagrams.
  In this subsection, we
do not want to distinguish between $\xi$'s and $\bar{\xi}$'s,
and shift consequently to  a neutral notation, $\vec{\xi}_{ext}:=(\xi_{ext,1},\ldots,\xi_{ext,N_{ext}})$. Scale $j$ cluster expansions do not produce directly $F^j(\vec{\xi}_I)$, but  
a sum $\sum_{\P} {\cal A}(\P;\vec{\xi}_{ext})$ over scale $j$ polymers $\P$ with $N_{ext}$ external fields located
in $\xi_{ext,i}\in\Del_{ext,i}$, $\Del_{ext,1},\ldots,\Del_{ext,N_{ext}}\in\D^j$. Polymers are in principle evaluated by expanding as a sum over Feynman diagrams; however, diagrams differing by merely {\em exchanging} vertices belonging
to the same box should be evaluated together: their sum is
much smaller than expected due to
 sign
compensations (see arguments in  \S \ref{subsection:local-bounds-det} and later explanations). In coherence with the  logic of
multi-scale cluster expansions, we are really interested in
 power-counting estimates for the integrated quantities
 \BEQ {\cal A}(\P;\xi_{ext,1}):=\Big(\prod_{i=2}^{N_{ext}} m^*\mu^2\int_{\Del_{ext,i}} d\xi_{ext,i}\Big) {\cal A}(\P;\vec{\xi}_{ext}).  \label{eq:A(P)}
 \EEQ 
The dimensional factors $m^*\mu^2\approx \frac{1}{\Vol(\Del^0)}$, $\Del^0\in\D^0$ are chosen in such a way that 
$m^*\mu^2 \int_{\Del_{ext,i}} d\xi_{ext,i}\approx 2^{3j}$ is
a non-dimensional scaling factor, so that ${\cal A}(\P;\xi_{ext,1})$ has the same dimensionality as that of an amputated $N_{ext}$-point function.
The fixed vertex location $\xi_{ext,1}$ will be integrated 
later on at
some lower scale (equal to the scale of one of the low-momentum external
fields attached at $\xi_{ext,1}$). 

\Bigskip {\bf Single-scale bounds for a given diagram with
fixed sectors.} 
Let us first  bound the contribution to $ {\cal A}(\P;\xi_{ext,1})$ of  a {\em single diagram} ${\cal A}^{j,\vec{\alpha}}(\Upsilon;\xi_1)$, with  angular sectors of all internal
lines fixed. Because of the sign issue mentioned above  (easily solved by summing over
permutations of vertices located in the same box), but mainly because the present power-counting principle overlooks the main problem -- which consists in summing over the different choices of angular sectors --, this paragraph can only be 
of pedagogical value, therefore it can be skipped. Here are two examples of polymers with $N_{ext}=4$ external legs. Black lines
are cluster links. We actually selected contributions
to the polymers  containing only one vertex
per box (or cluster),

\Bigskip

\begin{tikzpicture}[scale=0.4]

\begin{scope}[shift={(-14,0)}]
\draw(-1,1+4)--(1,1+4); \draw(1,1+4)--(1,-1+4); \draw(1,-1+4)--(-1,-1+4);
\draw(-1,-1+4)--(-1,1+4);

\draw(0,4)--(4,4); \draw(0,4) node {\textbullet};
\draw(4,4)node {\textbullet};
\draw[color=green](-2.5,4) node {$\Psi^2_{ext}$}; 
\draw[color=green](6.5,4) node {$\Psi^2_{ext}$};
\draw(0,6) node {$\Del_1$}; \draw(4,6) node {$\Del_2$};
\draw[color=blue,dashed](0,4)--(1.5,5.5);
\draw[color=blue,dashed](4,4)--(2.5,5.5);

\draw(-1+4,1+4)--(1+4,1+4); \draw(1+4,1+4)--(1+4,-1+4); \draw(1+4,-1+4)--(-1+4,-1+4);
\draw(-1+4,-1+4)--(-1+4,1+4);
\end{scope}

\draw(2,8) node {$\Del_1$}; \draw(0,-2) node {$\Del_2$};
\draw(8,-2) node {$\Del_3$}; \draw(8,6) node {$\Del_4$};
\draw(0,0) node {\textbullet};
\draw(4,0) node {\textbullet};
\draw(8,0) node {\textbullet};
\draw(-1,1)--(1,1); \draw(1,1)--(1,-1); \draw(1,-1)--(-1,-1);
\draw(-1,-1)--(-1,1);
\draw(-1+4,1)--(1+4,1); \draw(1+4,1)--(1+4,-1); \draw(1+4,-1)--(-1+4,-1);
\draw(-1+4,-1)--(-1+4,1);
\draw(-1+8,1)--(1+8,1); \draw(1+8,1)--(1+8,-1); \draw(1+8,-1)--(-1+8,-1);
\draw(-1+8,-1)--(-1+8,1);
\draw(0,0)--(8,0);

\draw(4,4) node {\textbullet};
\draw(8,4) node {\textbullet};
\draw(4,8) node {\textbullet};
\draw(-1+4,1+4)--(1+4,1+4); \draw(1+4,1+4)--(1+4,-1+4); \draw(1+4,-1+4)--(-1+4,-1+4);
\draw(-1+4,-1+4)--(-1+4,1+4);
\draw(-1+8,1+4)--(1+8,1+4); \draw(1+8,1+4)--(1+8,-1+4); \draw(1+8,-1+4)--(-1+8,-1+4);
\draw(-1+8,-1+4)--(-1+8,1+4);
\draw(4,4)--(8,4); \draw(4,0)--(4,8);

\draw(4,8) node {\textbullet};
\draw(-1+4,1+8)--(1+4,1+8); \draw(1+4,1+8)--(1+4,-1+8); \draw(1+4,-1+8)--(-1+4,-1+8);
\draw(-1+4,-1+8)--(-1+4,1+8);

 \draw[color=green](-2.5,0) node {$\Psi_{ext}$};
\draw[color=green](10.5,0) node {$\Psi_{ext}$};
\draw[color=green](10.5,4) node {$\Psi_{ext}$};
\draw[color=green](4,10) node {$\Psi_{ext}$};
\draw[color=blue,dashed](0,0)--(1.5,3);
\draw[color=blue,dashed](0,0)--(1.5,1.5);
\draw[color=blue,dashed](4,0)--(5.5,0.5);
\draw[color=blue,dashed](8,0)--(6.5,0.5);
\draw[color=blue,dashed](8,0)--(8,1.5);
\draw[color=blue,dashed](8,4)--(8,2.5);
\draw[color=blue,dashed](8,4)--(6.5,5.5);
\draw[color=blue,dashed](4,8)--(5.5,6.5);
\draw[color=blue,dashed](4,8)--(2.5,5);
\draw[color=blue,dashed](4,4)--(2.5,2.5);
\end{tikzpicture} 

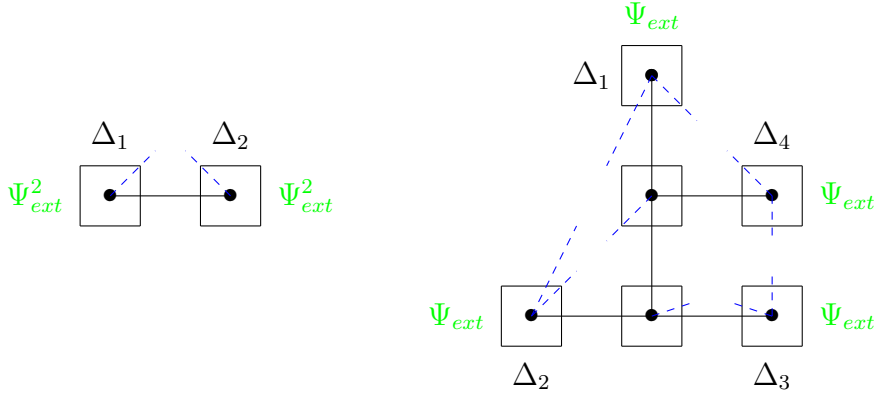
\Medskip {\bf\tiny Fig. \ref{subsection:cluster}.2. Examples of cluster trees for
a diagram with four external legs.}

\Bigskip with $\xi_{ext,i}\in\Del_i$, $\Del_i\in{\cal O}^j$. Locations of external fields are indicated by the
symbol $\Psi_{ext}$ (in green). Dashed lines (in blue) 
indicate unpaired fields. In general there may be more
than one vertex per box, hence many more unpaired lines
than cluster links. Pairing unpaired fields using Wick's theorem, we obtain Feynman diagrams, e.g.

\bigskip

\begin{tikzpicture}[scale=0.4]

\begin{scope}[shift={(-14,0)}]
\draw(-1,1+4)--(1,1+4); \draw(1,1+4)--(1,-1+4); \draw(1,-1+4)--(-1,-1+4);
\draw(-1,-1+4)--(-1,1+4);

\draw(0,4)--(4,4); \draw(0,4) node {\textbullet};
\draw(4,4)node {\textbullet};
\draw[color=green](-2.5,4) node {$\Psi^2_{ext}$}; 
\draw[color=green](6.5,4) node {$\Psi^2_{ext}$};
\draw(0,6) node {$\Del_1$}; \draw(4,6) node {$\Del_2$};
\draw[color=blue](4,4) arc(30:150:2.3 and 1);

\draw(-1+4,1+4)--(1+4,1+4); \draw(1+4,1+4)--(1+4,-1+4); \draw(1+4,-1+4)--(-1+4,-1+4);
\draw(-1+4,-1+4)--(-1+4,1+4);
\end{scope}

\draw(2,8) node {$\Del_1$}; \draw(0,-2) node {$\Del_2$};
\draw(8,-2) node {$\Del_3$}; \draw(8,6) node {$\Del_4$};
\draw(0,0) node {\textbullet};
\draw(4,0) node {\textbullet};
\draw(8,0) node {\textbullet};
\draw(-1,1)--(1,1); \draw(1,1)--(1,-1); \draw(1,-1)--(-1,-1);
\draw(-1,-1)--(-1,1);
\draw(-1+4,1)--(1+4,1); \draw(1+4,1)--(1+4,-1); \draw(1+4,-1)--(-1+4,-1);
\draw(-1+4,-1)--(-1+4,1);
\draw(-1+8,1)--(1+8,1); \draw(1+8,1)--(1+8,-1); \draw(1+8,-1)--(-1+8,-1);
\draw(-1+8,-1)--(-1+8,1);
\draw(0,0)--(8,0);

\draw(4,4) node {\textbullet};
\draw(8,4) node {\textbullet};
\draw(4,8) node {\textbullet};
\draw(-1+4,1+4)--(1+4,1+4); \draw(1+4,1+4)--(1+4,-1+4); \draw(1+4,-1+4)--(-1+4,-1+4);
\draw(-1+4,-1+4)--(-1+4,1+4);
\draw(-1+8,1+4)--(1+8,1+4); \draw(1+8,1+4)--(1+8,-1+4); \draw(1+8,-1+4)--(-1+8,-1+4);
\draw(-1+8,-1+4)--(-1+8,1+4);
\draw(4,4)--(8,4); \draw(4,0)--(4,8);

\draw(4,8) node {\textbullet};
\draw(-1+4,1+8)--(1+4,1+8); \draw(1+4,1+8)--(1+4,-1+8); \draw(1+4,-1+8)--(-1+4,-1+8);
\draw(-1+4,-1+8)--(-1+4,1+8);

 \draw[color=green](-2.5,0) node {$\Psi_{ext}$};
\draw[color=green](10.5,0) node {$\Psi_{ext}$};
\draw[color=green](10.5,4) node {$\Psi_{ext}$};
\draw[color=green](4,10) node {$\Psi_{ext}$};
\draw[color=blue](0,0)--(4,4);
\draw[color=blue](0,0)--(4,8);
\draw[color=blue](8,0) arc(60:120:4.3 and 3);
\draw[color=blue](8,0)--(8,4);
\draw[color=blue](8,4)--(4,8);
\draw[color=blue](4,4)--(0,0);
\end{tikzpicture}

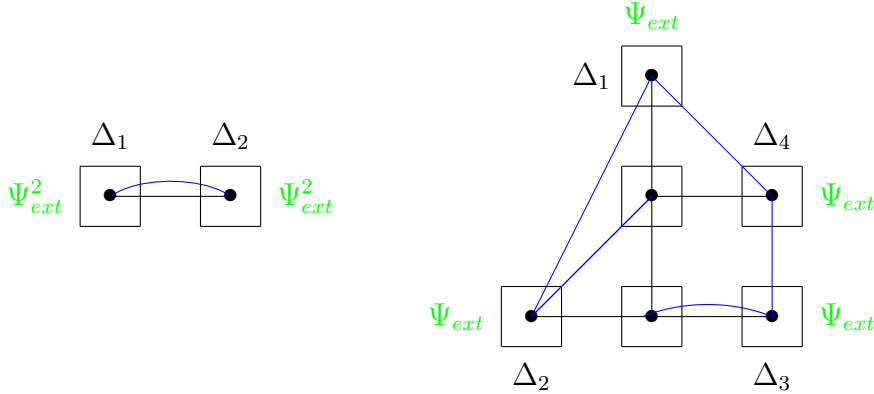
\Medskip {\tiny \bf Fig. \ref{subsection:cluster}.3. For the bubble diagram (left), 
$N_{ext}=4$ (in green), $n=2$, $I_{tree}=1$ (in black), $L=1$
(in blue), $I=I_{tree}+L=2$. Similarly, for the other diagram, $N_{ext}=4$, $n=6$, $I_{tree}=5$, $L=5$, $I=I_{tree}+L=10$.}

\Bigskip 
General topological properties of a Feynman graph $\Upsilon$ with $n$ four-valent vertices,
$L$ loops, $I$ internal lines ($2I$ internal fields) and $N_{ext}$ external legs imply that 
\BEQ L-I+n=1, \qquad N_{ext}=4n-2I. \EEQ
These relations give a quick power-counting argument for
the evaluation of an amputated diagram {\em with fixed angular
sectors}.  Choose some spanning tree $\mathfrak{t}$ (which may be e.g.  the cluster tree in
the above examples, but in general, $\mathfrak{t}$ has much
more vertices and edges than $\T$; in fact one may
take $\mathfrak{t}\supset\T$). 
Internal lines are split into $I_{tree}=n-1$ internal lines on the spanning
tree (in black) -- in this very particular example, cluster links -- and 
$L=I_{tree}-\half(N_{ext}-4)$ lines generating the loops
(in blue). All together, $I=I_{tree}+L=2I_{tree}-
\half(N_{ext}-4)$. 

\Medskip Now, fixing $\xi_{ext,1}$ and integrating the other
$\xi_{ext,i}$'s in their respectives boxes, one obtains
an overall scaling factor
\BEQ (2^{3j})^{n-1} (2^{-2j})^I=(2^{3j})^{I_{tree}}
(2^{-2j})^{2I_{tree}-
\half(N_{ext}-4)}=(2^{-j})^{L} (2^{-j})^{\half(4-N_{ext})} ,
\label{eq:overall-scaling-factor}
 \EEQ
which takes into account the integration of vertices over their respectives boxes, and the pre-scaling  factor
$O(2^{-2j})$ of each propagator $C^{j,\alpha}_{\theta}$, see
Proposition \ref{prop:C}.  

\Medskip Counting one sum over angular sector assignments per
loop yields a {\em total scaling factor $(2^{-j})^{\half(4-N_{ext})}$ , characteristic of a just renormalizable theory with four-valent vertices.} However, our aim is to prove that diagrams are
{\em smaller} than that by a factor of the type $O(N_j^{-cn})$
for some constant $c>0$, with $N_j:=2^j$, providing the basis
for the $1/N$-expansion.

\Bigskip {\bf Single-scale bounds for a given polymer.}
We shall actually use a different power-counting argument, not based on general topological properties of a Feynman graph, but on the concept of {\em total weight per vertex} (see
below). Previous considerations -- see Proposition \ref{prop:sector-counting} and the bound just below for the
integrated vertex $I^j(\Del^j)$ -- imply that the leading
behavior is obtained by assuming that sectors $\alpha_1,\alpha_2,\alpha_3,\alpha_4$ on a given vertex satisfy
up to some permutation
\BEQ |\alpha_1-\alpha_3|,|\alpha_2-\alpha_4|\approx 2^{j-k}\lesssim  |\alpha_1+\alpha_2|\lesssim 2^{k} \label{eq:2.80} \EEQ
for some $k\in\{\lfloor \frac{j}{2}\rfloor,\cdots,j\}$.
We then {\em decide} that $\alpha_1,\alpha_3$ are on the
same fermion loop, and similarly for $\alpha_2,\alpha_4$. 
Generically,  $k=j$, $|\alpha_1+\alpha_2|\approx 2^j$, $|\alpha_1-\alpha_3|\approx 1$ so it makes sense (at least graphically) to assume that the angular sector along a fermion loop
is constant. 
Following loops, one obtains e.g. 

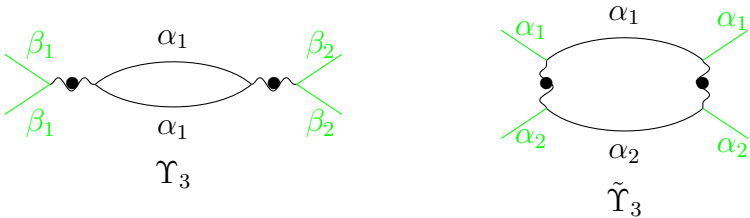
\bigskip

\begin{tikzpicture}[scale=0.4]

\draw[decorate,decoration=snake](-6.7,0)--(-5.2,0); 
\draw[decorate,decoration=snake](0,0)--(1.5,0);
\draw(0,0) arc(30:150:3  and 1.5);
\draw(-5.2,0) arc(210:330:3 and 1.5);
\draw[color=green](-8.2,1)--(-6.7,0);  \draw[color=green](-8.2,-1)--(-6.7,0);
\draw[color=green](1.5,0)--(3,1);\draw[color=green](1.5,0)--(3,-1);
\draw(-5.95,0) node {\textbullet};
\draw(0.75,0) node {\textbullet};
\draw(-2.6,1.5) node {$\alpha_1$};
\draw(-2.6,-1.5) node {$\alpha_1$};
\draw[color=green](-7,1.3) node {$\beta_1$};
\draw[color=green](-7,-1.3) node {$\beta_1$};
\draw[color=green](2.3,1.3) node {$\beta_2$};
\draw[color=green](2.3,-1.3) node {$\beta_2$};
\draw(-2.6,-3) node {$\Upsilon_3$};

\begin{scope}[shift={(15,0)}]
\draw[decorate,decoration=snake](-5.2,-0.8)--(-5.2,0.8); 
\draw[decorate,decoration=snake](0,0.8)--(0,-0.8);
\draw(0,0.8) arc(30:150:3  and 1.5);
\draw(-5.2,-0.8) arc(210:330:3 and 1.5);
\draw[color=green](-8.2+1.5,1+0.8)--(-6.7+1.5,0.8);  \draw[color=green](-8.2+1.5,-1-0.8)--(-6.7+1.5,-0.8);
\draw[color=green](1.5-1.5,0+0.8)--(3-1.5,1+0.8);\draw[color=green](1.5-1.5,0-0.8)--(3-1.5,-1-0.8);
\draw(-5.95+0.75,0) node {\textbullet};
\draw(0,0) node {\textbullet};
\draw(-2.6,1.5+0.8) node {$\alpha_1$};
\draw(-2.6,-1.5-0.8) node {$\alpha_2$};
\draw[color=green](-5.7,1+0.8) node {$\alpha_1$};
\draw[color=green](-5.7,-1-0.8) node {$\alpha_2$};
\draw[color=green](1,1.3+0.8) node {$\alpha_1$};
\draw[color=green](1,-1.3-0.8) node {$\alpha_2$};

\draw(-2.6,-3-0.8) node {$\tilde{\Upsilon}_3$};
\end{scope}

\end{tikzpicture}

\bigskip 

for the bubble diagram, and

\bigskip

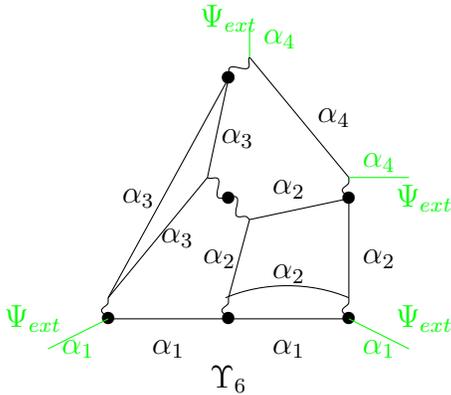
\begin{tikzpicture}[scale=0.4]

\draw(0,0) node {\textbullet};
\draw(4,0) node {\textbullet};
\draw(8,0) node {\textbullet};

\draw(0,0)--(8,0);
\draw[decorate,decoration=snake](3.3,4.7)--(4.7,3.3);

\draw(4,4) node {\textbullet};
\draw(8,4) node {\textbullet};
\draw(4,8) node {\textbullet};

\draw(4.7,3.3)--(8,4); \draw(3.3,4.7)--(4,8);

\draw(4,8) node {\textbullet};

\draw[decorate,decoration=snake](0,0)--(0,0.7);
\draw[decorate,decoration=snake](4,0)--(4,0.7);
\draw[decorate,decoration=snake](8,0)--(8,0.7);
\draw[decorate,decoration=snake](4,8)--(4.7,8.7);
\draw[decorate,decoration=snake](8,4)--(8,4.7);

\draw(4,0.7)--(4.7,3.3);

 \draw[color=green](-2.5,0) node {$\Psi_{ext}$};
\draw[color=green](10.5,0) node {$\Psi_{ext}$};
\draw[color=green](10.5,4) node {$\Psi_{ext}$};
\draw[color=green](4,10) node {$\Psi_{ext}$};
\draw(0,0.7)--(3.3,4.7);
\draw(0,0.7)--(4,8);
\draw(8,0.7) arc(60:120:4.1 and 3);
\draw(8,0.7)--(8,4);
\draw(8,4.7)--(4.7,8.7);

\draw(1,4) node {$\alpha_3$}; \draw(2.3,2.7) node {$\alpha_3$};
\draw(4.3,6) node {$\alpha_3$};
\draw(7.5,6.7) node {$\alpha_4$}; \draw[color=green](9,5.2) node {$\alpha_4$}; \draw[color=green](5.7,9.3) node {$\alpha_4$};
\draw(2,-1) node {$\alpha_1$}; \draw(6,-1) node {$\alpha_1$};
\draw(3.7,2) node {$\alpha_2$}; \draw(6,4.3) node {$\alpha_2$};
\draw(9,2) node {$\alpha_2$}; \draw(6,1.5) node {$\alpha_2$};
\draw[color=green](-1,-1) node {$\alpha_1$}; \draw[color=green](9,-1) node {$\alpha_1$};
\draw[color=green](4.7,8.7)--(4.7,9.7);
\draw[color=green](0,0)--(-2,-1);
\draw[color=green](8,0)--(10,-1);
\draw[color=green](8,4.7)--(10,4.7);

\draw(4,-2) node {$\Upsilon_6$};

\end{tikzpicture} 

\Medskip {\bf\tiny Fig. \ref{subsection:cluster}.4. Loop decompositions. Loop lengths
are $L(\alpha_1)=2,L(\beta_1)=L(\beta_2)=0$ ($\Upsilon_3$), $L(\alpha_1)=L(\alpha_2)=1$ ($\tilde{\Upsilon}_3$), $L(\alpha_4)=1$, $L(\alpha_1)=2$, $L(\alpha_3)=3$, $L(\alpha_2)=4$.}

\Bigskip for the other diagram. The diagram $\Upsilon_3$ is
the usual Cooper bubble diagram. Define the {\em length} $L$ of a loop to be the
number of internal propagators belonging to it, see Figures above.
Considering the generic case, we see that there are e.g.  only
$L'=4<L=5$ independent sector assignments for $\Upsilon_6$.   In $\tilde{\Upsilon}_3$,
the external sectors $\beta_1,\beta_2$ keep {\em outside} the
diagram; in all other cases, there  are {\em broken loops},
 finishing with two external legs, for instance, the loop
 with sector $\alpha_1$ in $\Upsilon_6$. 
 
 \Medskip  Note that, by splitting vertices into half vertices, one has rewritten the original
 bubble diagram into the sum of two contributions: a {\em bubble}, $\Upsilon_3$, and another
 diagram, $\tilde{\Upsilon}_3$, which is of a different nature. {\em Bubbles}, in this sense,
 are characterized that they make up a {\em loop of length 2}. {\em Loops of length 1}, such
  as $\alpha_1,\alpha_2$ in $\tilde{\Upsilon}_3$ or $\alpha_4$ in $\Upsilon_6$, are called
  {\bf incomplete bubbles} because external propagators may contract at some lower 
  scale to form a full, multi-scale bubble, implying the possibility of obtaining 
  chains of multi-scale bubbles of arbitrary lengths,
  
  \bigskip

{\centerline{  
 \begin{tikzpicture}[scale=0.4]
\draw[color=blue](-7.5,-2)--(18,-2); \draw(16.5,0) node{$j$};
\draw[decorate,decoration=snake](-6.7,0)--(-5.2,0); 
\draw(0,0) arc(30:150:3  and 1.5);
\draw[dotted](-5.2,0)--(-5.2,-3);\draw[dotted](0,0)--(0,-3);\draw[dashed](-5.2,-3) arc(210:330:3 and 1.5);
\draw(-5.95,0) node {\textbullet};
\draw(-2.6,1.5) node {$\alpha_1$};
\begin{scope}[shift={(6.7,0)}]
\draw[decorate,decoration=snake](-6.7,0)--(-5.2,0); 
\draw(0,0) arc(30:150:3  and 1.5);
\draw[dotted](-5.2,0)--(-5.2,-5);\draw[dotted](0,0)--(0,-5);\draw[dashed](-5.2,-5) arc(210:330:3 and 1.5);
\draw(-5.95,0) node {\textbullet};
\draw(0.75,0) node {\textbullet};
\draw(-2.6,1.5) node {$\alpha_2$};
\end{scope}
\begin{scope}[shift={(6.7+6.7,0)}]
\draw[decorate,decoration=snake](-6.7,0)--(-5.2,0); 
\draw(0,0) arc(30:150:3  and 1.5);
\draw[dotted](-5.2,0)--(-5.2,-4);\draw[dotted](0,0)--(0,-4);\draw[dashed](-5.2,-4) arc(210:330:3 and 1.5);
\draw(-5.95,0) node {\textbullet};
\draw(0.75,0) node {\textbullet};
\draw(-2.6,1.5) node {$\alpha_3$};
\draw[decorate,decoration=snake](0,0)--(1.5,0);
\draw(0.75,0) node {\textbullet};
\end{scope}
\end{tikzpicture}
}}

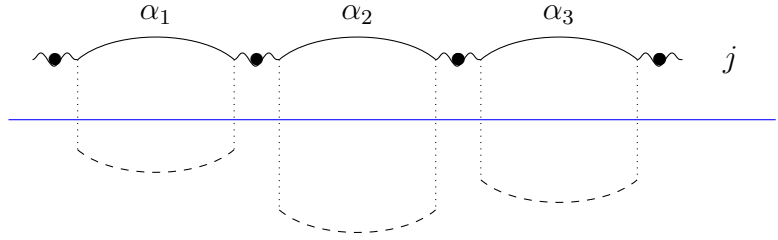
\Medskip {\bf \tiny Fig. \thesubsection.5.  A chain of 3 incomplete upper scale $j$  bubbles, possibly completed
at lower scales into a chain of multi-scale bubbles.}

\Bigskip Let us start with the argument.   First of all, 
single scale polymers must be prepared as in \cite{IagMag},
 Proposition 5.3, by displacing all fields lying in a scale 
 $j$ box or sub-box to the same location; see Appendix, \S \ref{subsection:local-bounds-det} for a
 description of the procedure. Because of the fermionic
 nature of the  fields $(\Psi^{j,\alpha})_{\alpha},(\bar{\Psi}^{j,\alpha})_{\alpha}$, there remains only contributions with
 at most {\em one}  field $\Psi^{j,\alpha}$ or $\bar{\Psi}^{j,\alpha}$ per box, and  terms featuring gradient fields
 $(\nabla-\II  p^{\alpha})^{\kappa} \Psi^{j,\alpha}$, 
  $(\nabla+\II  p^{\alpha})^{\kappa} \bar{\Psi}^{j,\alpha}$ integrated over sub-boxes. The net outcome is a sum
 over  Feynman diagrams with supplementary gradients, inducing
 an overall scaling factor bounded by 
 \BEQ C_r^n\prod_{\Del}\prod_{\alpha} 
 \frac{1}{(n_{\Del,\alpha}!)^r (\bar{n}_{\Del,\alpha}!)^r}
 \label{eq:Iago-factor}
 \EEQ  
 for every $r\ge 0$, where $n_{\Del,\alpha}$, resp. $\bar{n}_{\Del,\alpha}$  is the number of $\Psi^{j,\alpha}$, resp. 
 $\bar{\Psi}^{j,\alpha}$-fields in the scale $j$ box $\Del$. 
 
 \Medskip Choose now  a set of (not necessarily distinct) boxes 
${\bf\Del}\equiv\{\Del_1,\ldots,\Del_n\}\subset\D^j$ such that $\Del_{ext,i}\subset{\bf\Del}$, and   some
{\bf loop structure ${\bf\gamma}\equiv\{\gamma_1,\ldots,
\gamma_{L'}\}$}, namely, an equivalence set of Feynman diagrams 
with fixed topological structure  -- allowing the permutation
of vertices located in the same box -- and sector assignment
 compatible with the above set of loops, i.e.
satisfying (\ref{eq:2.80}). We shall compute  to begin with an upper
bound to the power-counting of the sum of all Feynman diagrams
in the equivalence set $({\bf\Del},{\mathbf{\gamma}})$, {\em including
the integration of $\xi_1$ in its box,  the power-counting
of  external legs, and the sum over external sectors}.  The main task consists in summing over angular sector attributions
of each vertex. In order not to lose the benefit of
(\ref{eq:Iago-factor}), we  {\em fix} a set of integers
$n_{\Del,i},\bar{n}_{\Del,i}\ge 0$, $i=1,\ldots,2^j$ and restrict to diagrams such that each list $\{n_{\Del,\alpha},\alpha\in 
\Z/2^j\Z\}$ is some {\em permutation} of the list $\{n_{\Del,i}, i=1,\ldots,2^j\}$.  The power-counting proceeds
as follows:

(i) First select a "loop spanning tree", i.e. a tree 
${\mathfrak{t}}'$ connecting the loops. This may be done inductively by
choosing $\gamma_1$ for a root; then connecting $\gamma_1$  
 successively to all loops $\gamma_{i_1},\ldots,\gamma_{i_{n_1}}$
$(1\not=i_1<\ldots<i_{n_1})$ such that there is a vertex connecting
two successive moments along $\gamma_1$ to two successive 
moments along $\gamma_{i_k}$, $k=1,\ldots,n_1$; then
connecting $\gamma_{i_1}$ successively to all loops 
$\gamma_{i'_1},\ldots,\gamma_{i'_{n_{i_1}}}$ {\em other than
those previously chosen} such that there is a vertex connecting
two successive moments along $\gamma_{i_1}$ to two successive 
moments along $\gamma_{i'_k}$, $k=1,\ldots,n_{i_1}$; and 
so on, until all loops have been exhausted. For instance,
choosing $\gamma_1$ to be the loop with sector $\alpha_1$ in
$\Upsilon_6$, the loop spanning tree associated to $\Upsilon_6$ 
is \qquad  \begin{tikzpicture}[scale=0.3]  \draw(0,0) node {\textbullet};  \draw(0,-1) node {$1$}; 
  \draw(0,5) node {$4$};\draw(1.5,3) node {$3$};
 \draw(0,2) node {\textbullet}; \draw(0,4) node {\textbullet};
\draw(1.5,2) node {\textbullet}; \draw(0,0)--(0,4);
 \draw(0,0)--(1.5,2); \draw(-1,2) node {$2$};
 \end{tikzpicture}

(ii) Reorder loops by using depth-first search algorithm
on ${\mathfrak{t}}'$. In the case of $\Upsilon_6$, the
algorithm visits successively vertices $1,2,4$ and $3$, so
loops are reordered by permuting indices $3$ and $4$.

(iii)
  Order the vertices $v_{i,p}$, $p=1,\ldots,n_i$
along $\gamma_i$ in such a way that vertices connecting
$\gamma_i$ to its {\em descendant loops} on the loop
spanning tree   (e.g. $\gamma_1$ to $\gamma_2,
\gamma_4$ for ${\mathfrak{t}}'$, with its loop reindexing defined
in (ii)) are $v_{i,1},\ldots,v_{i,n'_i}$. Then choose some
propagator along $\gamma_i$, denote by $\alpha_{i,1}$ its 
sector, go around the loop in some arbitrary direction starting
from $\alpha_{i,1}$, and index the successive sectors by
$\alpha_{i,1},\ldots,\alpha_{i,n_i}$. 
 Write $k_v=k$ if $v$ is a
vertex along $\gamma_i$ connecting the momentum in the sector $\alpha_{i,p}$ to the momentum in the sector $\alpha_{i,p+1}$, and 
$ \lfloor \log_2\Big(1+|\alpha_{i,p}-\alpha_{i,p+1}| \Big) \rfloor=j-k$, $k\in \{\lfloor\frac{j}{2}\rfloor,\ldots,j\}$.
This means the following: {\em there are $O(2^{j-k})$ possibilities
for $\alpha_{i,p+1}$ for $\alpha_{i,p}$ fixed.}

(iv)  Let $\del j:=\begin{cases} 0 \qquad (j\le j_{\phi}) \\ j-j_{\phi}
\qquad (j>j_{\phi})\end{cases}$.  The power-counting associated to a vertex $v$
 with its 4  legs (see (\ref{eq:scaling-psijalpha}))
is  $O(\lambda\times \frac{1}{m^*\mu^2} 2^{3j}\times (2^{-j}2^{-\del j/2} p^*_F)^4)=O(g2^{-j}
2^{-2\del j})$ (hence roughly a factor $O(2^{-j})$ per vertex or per loop, compare with (\ref{eq:overall-scaling-factor})), {\em times} $2^{j-k_v}$, 
all together $O(g\times 2^{-k_v})$, that is, a {\em $(\frac{1}{N})$-type}
factor, generically  but not always $O(N_j)=O(2^{-j})$. 

(v) {\em (overall rotation factor $O(2^j)$)} Consider one of the connecting vertices $v=v_{i,1},
\ldots,v_{i,n'_i}$  along $\gamma_i$, see  (iii). Denote by
$(\alpha_1,\alpha_3)=(\alpha_{i,p},\alpha_{i,p+1})$, 
$(\alpha_2,\alpha_4)=(\alpha_{i',p'},\alpha_{i',p'+1})$
$(i<i')$ its two halves, one on $\gamma_i$, the other on
$\gamma_{i'}$. Summing over $\alpha_2$ for $\alpha_1$ fixed
(see (\ref{eq:2.80})) yields a factor $O(2^{k_v})$ compensating
the factor $O(2^{-k_v})$ in (iv). Now {\em local rotation invariance}  has almost been  fixed: choose some sector along $\gamma_1$  {\em ($2^j$ possibilities)}, then 
the factors
$O(2^{k_v})$ associated to connecting vertices fix one
sector on every other loop.

(vi)  ({\em total weight associated to a connecting vertex})
Let $v=v_{i,p}$, $p\le n'_i$ be a connecting vertex. Given
(iv), (v),  there remains only a small
factor $O(g)$, and a sum over all possible values of 
$k_v$ in $\{\lfloor \frac{j}{2}\rfloor,\ldots,j\}$, all together
a factor $O(g j 2^{-2\del j})=O(1)$ if $j\le j_{\phi}\approx 
g^{-1}$. If $j>j_{\phi}$ then $O(gj2^{-2\del j})$ is small, but this is actually
very deceptive because of bubble resummations (see \S \ref{subsubsection:bubble-sum}). The power-counting for connecting vertices {\em associated
to  bubbles} is better than that and computed in (viii) below.

(vii) ({\em total weight associated to a non-connecting vertex})
Let $v=v_{i,n}$, $n> n'_i$ be a non-connecting vertex.
Given (iv), its total weight is $\le \sum_{k_v} g
2^{-k_v} 2^{-2\del j}\lesssim g2^{-j/2} 2^{-2\del j}$, featuring
a $O(\frac{1}{\sqrt{N_j}})$-prefactor.

(viii) {\em (subcase of (vi): bubbles)} A loop $\gamma_i$ is called a {\bf bubble}  when
$n_i=2$, $n'_i=1$. As explained above, it can be {\em complete} or {\em incomplete}, depending on whether
$L(\gamma_i)=2$ or $1$. Then the outgoing transfer momentum is
equal to the ingoing transfer momentum, so that in (vi)
no sum over $k_v$ is required. Hence the total weight associated to the connecting vertex $v_{i,1}$ is $O(g2^{-2\del j})$ 
instead of $O(1)$. Call {\bf strongly connecting vertex} 
a connecting vertex which is {\em not} originated from a bubble
as just described.
Note that a vertex which is {\em not} a strongly connecting
vertex (see cases (vii) and (viii)) has a {\em small} 
{\bf weight} $\lesssim g2^{-2\del j}$ attached to it. On the other hand,
strongly connecting vertices have a weight $O(1)$.

\Medskip {\em Assume for simplicity that $j\le j_{\phi}$.} 
Let $n'\le n$ be the number of strongly connecting vertices,
and $n'':=n-n'$.	The total weight obtained by multiplying all the
above factors is $O((g2^{-2\del j})^{n''} 2^j)$. Let $B$ be the number of bubbles (in the sense defined in (viii)).  Using the equality
 $n'+B+1=L'$ (number of loops) -- which is modified into
 $n'+B=L'$ in the case of diagrams containing only bubbles,
 e.g. $\Upsilon_3,\tilde{\Upsilon}_3$ -- and the bound $L'-B\le \frac{2}{3} (n-B)$ (expressing the fact that
each loop $\gamma_i$  different from a bubble has at least three half-vertices attached
to it), one obtains
\BEQ (g2^{-2\del j})^{n''}=(g2^{-2\del j})^{n-n'}\le (g2^{-2\del j})^{\frac{n}{3}+\frac{2}{3} B+1} \le (g2^{-2\del j})^{\frac{n}{3}+1},  \label{eq:ggg0}
\EEQ
or $(g2^{-2\del j})^{n/3}$ in the case of diagrams containing only bubbles.
For instance, $n'=0, n=n''=2$ and $L'=B=1$, resp. $2$ for $\Upsilon_3$, resp.
$\tilde{\Upsilon}_3$; and $L'=4,n=6$ and (depending on how
the bubble tree is generated) $n'=n''=3,B=0$ or $n'=2,n''=4,B=1$ for $\Upsilon_6$. 

\Medskip The lowest possible value of $n''$ for small diagrams
with $N_{ext}\ge 4$ {\em which are not only made up of bubbles}
is $n''=3$. Namely, such diagrams have at least {\em two}
loops $\alpha_1,\alpha_2$ and {\em two external vertices}. 
One (connecting) vertex is needed to connect $\alpha_1$ and $\alpha_2$,
while another, non-connecting one is needed to assure that
these do not make a chain of two bubbles. At least two other non-connecting vertices are needed to connect $\alpha_1,\alpha_2$ either to
other loops or to external legs. Therefore (disregarding 
diagrams containing only bubbles), one may replace the
exponent $\frac{n}{3}+1$ by $\max(3,\frac{n}{3}+1))=2+\max(1, \frac{n}{3}-1)$, so that, improving on (\ref{eq:ggg0}),
\BEQ (g2^{-2\del j})^{n''}\le (g2^{-2\del j})^{2+\max(1, \frac{n}{3}-1)},  \label{eq:ggg}
\EEQ
Individual pre-factors $g2^{-2\del j}$, which can be interpreted as the weight of {\em bare vertices}, compare
with (\ref{eq:I(Del)}),  are modified in multi-scale bounds because  effective vertices  are scale-dependent, but the
exponent $2+\max(1, \frac{n}{3}-1)$ remains the same, and is essential for polymer bounds.

\Medskip
 One also gets a product of "$1/N$"-factors, $O(2^{-j/2})$
per {\em non-connecting vertex},  see (vii), which however is simply $1$ 
 for diagrams consisting
only of bubbles (chains of bubbles); this feature will however be exploited only later on, once
chains of bubbles will have been resummed, see \S \ref{subsubsection:bubble-sum}. 
 {\em  The overall scaling factor
must however be corrected:} first, the location of $\xi_1$ is
fixed, leading to a correcting factor $\frac{1}{\Vol(\Del_{ext,1})}=O(m^* \mu^2 2^{-3j})$.
Second, the above weight mistakingly includes the power-counting of external fields $\Psi^j_{ext}$ together with
their sectors, see (\ref{eq:scaling-psij}), which leads to a correcting factor 
$O((2^{\frac{j}{2}}(p^*_F)^{-1})^{N_{ext}})$. Finally, assuming
the underlying cluster tree is
$\T$, the above may be multiplied by the decay factor
$C_p^n \prod_{\ell\in L(\T)}\Big(1+2^{-j}
\frac{ |\xi_{\ell}-\xi'_{\ell}|}{\mu} \Big)^{-p}$ for every $p\ge 0$.

\Medskip {\em All together, we have proved the following bound for
a single-scale diagram $\Upsilon$ not made up only of bubbles, with underlying cluster 
tree $\T$ and angular sector distribution $\{n_{\Del,i}\}\uplus \{\bar{n}_{\Del,i}\}$:}
\BEA && |{\cal A}(\Upsilon;\xi_1)|\le   \Big\{C_{p,r}^n\prod_{\Del}\prod_{i=1}^{2^j} 
 \frac{1}{(n_{\Del,i}!)^r (\bar{n}_{\Del,i}!)^r} 
  \prod_{\ell\in L(\T)}\Big(1+2^{-j}
\frac{ |\xi_{\Del_{\ell}}-\xi_{\Del'_{\ell}}|}{\mu} \Big)^{-p} \Big\}\nonumber\\
&& \qquad\qquad (g2^{-2\del j})^{2+\max(1,\frac{n}{3}-1)}\   (2^{-j})^{\half(4-N_{ext})} 
(p^*_F)^{-N_{ext}}, \label{eq:Aamputated}
\EEA
where $\xi_{\Del_{\ell}},\xi_{\Del'_{\ell}}$ are the centers
of the boxes $\Del_{\ell},\Del'_{\ell}$. {\em Diagrams made up
only of bubbles feature a small factor $(g2^{-2\del j})^{n/3}$ instead.}

\Bigskip {\bf Sum over trees.} This is a standard argument
in statistical mechanics and constructive field theory; 
see e.g. \cite{MagUnt2}, Corollary 5.3. Recall
that there is no sum over permutations of  vertices. So
all trees may be generated by the following algorithm. (All boxes in the argument are in $\D^j$). Fix
a box $\Del_1$. Sum at  step 1 over all possible 
boxes $\Del'_1\in\D^j$ (including $\Del_1$), and add a  link between $\Del_1$ and $\Del'_1$,
i.e. a pairing $\langle \Psi^{j,\alpha}(\xi_1)\bar{\Psi}^{j,\alpha}(\xi'_1)\rangle$ or $\langle \bar{\Psi}^{j,\alpha}(\xi_1)\Psi^{j,\alpha}(\xi'_1)\rangle$
between fields located at $\xi_1\in\Del_1,\xi'_1\in\Del'_1$.
The corresponding multiplicative spatial decrease factor  is $C:=\sum_{\Del'_1}\Big(1+d^j(\Del_1,\Del'_1)\Big)^{-p}$, where 
\BEQ d^j(\Del,\Del'):=\sup_{\xi\in\Del,\xi'\in\Del'}  2^{-j}|\xi-\xi'| \EEQ
is a scaled distance between two scale $j$ boxes. Ordering
boxes by their distance  to $\Del_1$, one obtains
\BEQ C\lesssim \sum_{i\ge 1} i^{2-p}<\infty \EEQ
provided $p\ge 4$. Continue by picking a second link ) between $\Del_1$ and a box $\Del''_1$, and so on, until all pairings between fields in $\Del_1$ and fields either in $\Del_1$ or in any other box
have been exhausted. There are local factorials involved,
since fields located inside $\Del_1$ may be permuted, but
inverse local factorials  $\frac{1}{(n_{\Del_{1},i}!)^r (\bar{n}_{\Del_1,i}!)^r}$ beat them for $r>1$; a precise argument may
be found in \cite{MagUnt2}, Corollary (5.3) 2., where inverse local factorials
are deduced from the quasi-exponential decay of propagators, 
and not from the fermionic nature of the theory (see discussion
in \S \ref{subsection:bounds}).   Then, at step 2, one looks
for all possible pairings between a field located in $\Del_2$ defined as $\Del'_i$, where $i:=\min\{i'\ge 1\ |\ \Del'_i\not=\Del_1\}$,  
and a field located in $\Del'_2\not=\Del_1$, and so on. Once
all vertices of the tree have been explored, the procedure
stops.  In the end, one finds for the sum over all scale $j$  polymers $\P$
with  fixed number $n$ of vertices:

\BEQ   \sum_{\P\ |\ |\P|=n} |{\cal A}(\P;\xi_{ext,1})|\lesssim
(Cg2^{-2\del j})^{2+\max(1,\frac{n}{3}-1)}  (2^{-j})^{\half(4-N_{ext})} (p^*_F)^{-N_{ext}},
\EEQ
which is the general term of a converging series for
$\lambda$ small enough. Hence, finally,

\BEQ \Big| \sum_{\P} {\cal A}(\P;\xi_{ext,1}) \Big|\le \sum_n \sum_{\P\ |\ |\P|=n} |{\cal A}(\P;\xi_1)|\lesssim (2^{-j})^{\half(4-N_{ext})} (p^*_F)^{-N_{ext}}.
\EEQ


\subsubsection{Bubble resummations} \label{subsubsection:bubble-sum}


Let $j_+\le j'_{\phi}-1$. We shall now detail the resummation of chains of bubbles
with lowest  fermionic scale $j_+$, also called "$j_++\half$"-scale
expansion. The notation $"j_+$" for a {\em fermionic} scale seemingly contradicts
the principles laid out in the "Important notations" section
of the Introduction.  However, as emphasized in \S \ref{subsection:bubble}, see (\ref{eq:Aqj3}),
the kernel $\PreSigma^{j_+\to}(q)$ has an effective infra-red
cut-off for $|q|_+\le 2^{-j_+}\mu$. Therefore it makes sense
in this context
to identify the lowest {\em fermionic} scale with
a {\em bosonic} cut-off scale.

\Medskip This step proceeds "by inspection", namely, it
requires no supplementary expansion, rather an explicit
resummation by hand of {\em structures} found in the sum
of  perturbative
diagrams.

\Medskip The structures we want to single out are chains
of bubbles made up of pairs of  propagators with lowest
scale $k$, see Fig. 
\ref{subsection:bubble}.5. The lowest-order term, $\lambda\Id$,  connecting the  two halves of any vertex,

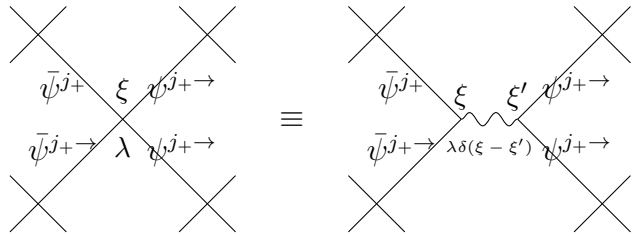
\Bigskip

{\centerline{
\begin{tikzpicture}[scale=0.75]
\draw(-2,2)--(2,-2); \draw(-2,-2)--(2,2);
\draw(1,2)--(2,1); \draw(1,-2)--(2,-1);
\draw(-1,2)--(-2,1); \draw(-1,-2)--(-2,-1);
\draw(-1.05,0.55) node {$\bar{\psi}^{j_+}$};
\draw(1.05,0.55) node {$\psi^{{j_+}\to}$};
\draw(-1.05,-0.55) node {$\bar{\psi}^{{j_+}\to}$};
\draw(1.05,-0.55) node {$\psi^{{j_+}\to}$};
\draw(0,0.5) node{\small $\xi$}; \draw(0,-0.5) node{\small $\lambda$};
\draw(3,0) node {$\equiv$};
\begin{scope}[shift={(6,0)}]
\draw(-2,2)--(0,0); \draw(1,0)--(3,-2);
 \draw(-2,-2)--(0,0); \draw(1,0)--(3,2);
\draw(2,2)--(3,1); \draw(2,-2)--(3,-1);
\draw(-1,2)--(-2,1); \draw(-1,-2)--(-2,-1);
\draw[decorate,decoration=snake](0,0)--(1,0);
\draw(-1.05,0.55) node {$\bar{\psi}^{j_+}$};
\draw(2.05,0.55) node {$\psi^{{j_+}\to}$};
\draw(-1.05,-0.55) node {$\bar{\psi}^{{j_+}\to}$};
\draw(2.05,-0.55) node {$\psi^{{j_+}\to}$};
\draw(0,0.4) node{\small $\xi$}; \draw(1,0.4) node{\small $\xi'$};
\draw(0.5,-0.5) node {\tiny  $\lambda\del(\xi-\xi')$};
\end{scope}
\end{tikzpicture}}}

\Medskip {\bf \tiny Fig. \thesubsection.6. Two equivalent ways
a representing a local vertex.}

\Bigskip is complemented by a sum of terms involving $\ge 1$ bubbles
forming a ladder diagram,

\Bigskip

{\centerline{
\begin{tikzpicture}[scale=0.75]
\draw(-2,2)--(0,0); 
 \draw(-2,-2)--(0,0); 
\draw(-1,2)--(-2,1); \draw(-1,-2)--(-2,-1);
\draw(0,0.4) node{$\xi$};
\draw[decorate,decoration=snake](0,0)--(1,0);
\draw(-1.3,0.55) node {$\bar{\psi}^{j_+\to}$};
\draw(0.3+9.5+2.05,0.55) node {$\psi^{j_+\to}$};
\draw(-1.3,-0.55) node {$\bar{\psi}^{j_+\to}$};
\draw(0.3+9.5+2.05,-0.55) node {$\psi^{j_+\to}$};
\draw(1,0) arc(150:30:2 and 1);
\draw(1,0) arc(210:330:2 and 1);
\draw(5,0) node {$\cdots$}; \draw(5.5,0) node {$\cdots$};
\draw(6,0) arc(150:30:2 and 1);
\draw(6,0) arc(210:330:2 and 1);
\draw[decorate,decoration=snake](9.5,0)--(10.5,0);
\draw(9.5+1,0)--(9.5+3,-2);
\draw(9.5+1,0)--(9.5+3,2);
\draw(9.5+2,-2)--(9.5+3,-1);
\draw(9.5+2,2)--(9.5+3,1); 
\draw(1.5,0.65) node {$\psi^{j_+\to}$};
\draw(1.5,-0.65) node {$\psi^{j_+\to}$};
\draw(2.5+1.5,0.65) node {$\bar{\psi}^{j_+\to}$};
\draw(2.5+1.5,-0.65) node {$\bar{\psi}^{j_+\to}$};
\draw(6.5,0.65) node {$\psi^{j_+\to}$};
\draw(6.5,-0.65) node {$\psi^{j_+\to}$};
\draw(9,0.65) node {$\bar{\psi}^{j_+\to}$};
\draw(9,-0.65) node {$\bar{\psi}^{j_+\to}$};
\draw(10.5,0.4) node {$\xi'$};
\end{tikzpicture}}}

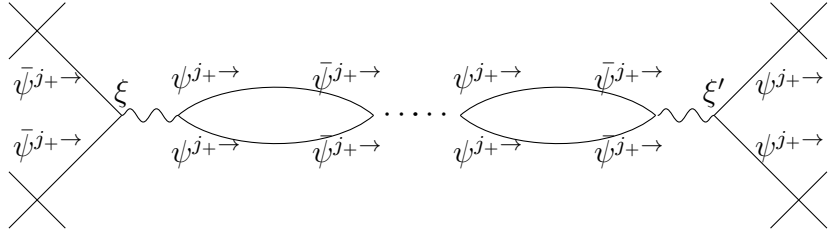
\Medskip {\bf \tiny Fig. \thesubsection.7. Bubble resummation.
At least one of the fields $\psi^{j_+\to}$ or $\bar{\psi}^{j_+\to}$ is assumed to be of scale $j_+$.}

\Bigskip Resumming  the above series yields -- up to the discrepancy between the current value
$\Gamma^{(j'_{\phi}-1)\to}$ and the asymptotic infra-red value
$\Gamma_{\phi}$ of the energy gap --  the kernel
Pre$\Sigma^{{j_+}\to}(\xi-\xi')$-Pre$\Sigma^{({j_+}-1)\to}(\xi-\xi')$. 
The error term due to the difference $\Gamma_{\phi}-\Gamma^{(j'_{\phi}-1)\to}$ will be
bounded only in section \ref{section:boson}, following the arguments of 
\S \ref{subsection:fixed-point} {\bf C.}, 2. and 3.  Now internal 
vertices of the chain may in turn be replaced by chains of
bubbles with lowest momentum of scale $\le {j_+}-1$, yielding
all together Pre$\Sigma^{{j_+}\to}(\xi,\xi')$.

\Medskip This has the following consequence; see   the preliminary bounds of \S \ref{subsubsection:single-scale-bounds} for notations. All  loops $\gamma_i$ have at least $3$ vertices along them
($n_i\ge 3$), so case (viii) is absent. On the other hand,
half-vertices are now possibly connected by Pre$\Sigma^{j_+\to}$-kernels with $j_+\ge j$, which is materialized by the substitution of {\bf delocalized vertices} to the original,  local vertices,

\Bigskip

{\centerline{\begin{tikzpicture}[scale=0.8]
\draw(-2,1)--(-0.5,0); \draw(-2,-1)--(-0.5,0);
\draw[decorate,decoration=snake](-0.5,0)--(0.5,0);
\draw(0.5,0)--(2,1); \draw(0.5,0)--(2,-1);
 \draw(-0.5,0.7) node {$\xi$}; \draw(0.5,0.7) node {$\xi'$};
 \draw(0,-0.7) node {\small $\lambda\del(\xi-\xi')$};
\draw(4,0)--(7,0); \draw(7,0)--(6.5,0.3); \draw(7,0)--(6.5,-0.3);
\draw(9,1)--(10.5,0); \draw(9,-1)--(10.5,0);
  \draw[dashed,decorate,decoration=snake](10.5,0)--(13,0);
 \draw(11.75,-0.7) node {\small Pre$\Sigma^{j_+\to}(\xi-\xi')$}; 
 \draw(10.5,0.7) node {$\xi$}; \draw(13,0.7) node {$\xi'$};
\draw(13,0)--(14.5,1); \draw(13,0)--(14.5,-1);
\end{tikzpicture} }}

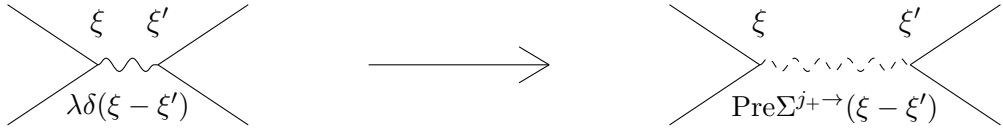
\Medskip {\bf \tiny Fig. \thesubsection.8. Local vertices 
(on the left), substituted by delocalized vertices (on the right).}

\Bigskip

Let $\Del^{j_+}$ be a scale ${j_+}$ box, and $\Vol(\Del^{j_+})\approx  \frac{1}{m^*\mu^2}2^{3{j_+}}$ its volume. {\em Assume first
that ${j_+}< j_{\phi}$. }
Then the integrated vertex $m^*\int d\xi'\, |\PreSigma^{{j_+}\to}(\xi-\xi')| $
is $O(\frac{1}{j_{\phi}-{j_+}})$, see (\ref{eq:bound-integral-sigma-j-arrow}), a bound substituting itself to the 
bound obtained by considering only the lowest-order term in $O(\lambda)$ in the Goldstone boson propagator, $m^*\int d\xi'\, \lambda\del(\xi-\xi')=O(g)$. For the highest
scales, say ${j_+}<j_{\phi}-c(j_{\phi}-j_D)$, $0<c<1$, the integrated
vertex is $O(g)$ since $j_{\phi}-j_D\approx g^{-1}$. However,
for larger scale indices ${j_+}$, that is, close to the transition
scale, the bound is only $O(1)$.  {\em For $j_{\phi}\le {j_+}\le j'_{\phi}-1$}, the bounds of \S \ref{subsection:pre-boson} 
yield instead a large bound $O(2^{2({j_+}-j_{\phi})})$. Recapitulating, one may define an equivalent {\bf scale ${j_+}$ coupling constant} $g^{j_+}$ replacing the bare constant $g$:

\begin{Definition}[scale ${j_+}$ coupling constant $g^{j_+}$]
\label{eq:scale-j-coupling}
Let $g^{j_+}:=m^*\int d\xi'\, |\PreSigma^{{j_+}\to}(\xi-\xi')|$
$({j_+}\le j'_{\phi}-1)$.
\end{Definition}
Previous computations show that
\BEQ g^{j_+} \lesssim
\begin{cases} \frac{1}{j_{\phi}-{j_+}} \qquad ({j_+}<j_{\phi}) \\
2^{2({j_+}-j_{\phi})} \qquad (j_{\phi}\le {j_+}\le j'_{\phi}-1)
\end{cases} \EEQ
Thus the {\bf weight} $g^{j_+} 2^{-2\del j}$  of a 
 vertex other than strongly connecting {\em including its four
half-propagators} (called {\bf effective vertex} in the
literature), compare with \S \ref{subsubsection:single-scale-bounds} (viii), is $\lesssim {\cal W}^{j,j_+}$ ($j_D\le j\le j_+\le j'_{\phi}-1$), where one has set

\begin{Definition}[scale $(j,j_+)$ effective vertex]
\label{def:scale-j-effective}
Let
\BEQ {\cal W}^{j,j_+}:= \begin{cases} \frac{1}{j_{\phi}-j_+} 
\qquad (j\le j_+<j_{\phi}) \\
2^{2(j_+-j)} \qquad (j_{\phi}\le j\le j_+\le j'_{\phi}-1)
\\ 
2^{2(j_+-j_{\phi})} \qquad (j\le j_{\phi}\le j_+\le j'_{\phi}-1)
\end{cases} \EEQ
and 
\BEQ {\cal W}^j:=\max_{j\le j_+\le j'_{\phi}-1} {\cal W}^{j,j_+}= \begin{cases} 2^{2(j'_{\phi}-j)} \qquad (j_{\phi}\le j\le j'_{\phi})
\\
2^{2(j'_{\phi}-j_{\phi})} \qquad (j\le j_{\phi})
\end{cases} \EEQ
\end{Definition}
 {\em Strongly connecting vertices}, see
 \S \ref{subsubsection:single-scale-bounds} (vi) and (viii),  have
 an extra logarithmic prefactor $O(j)$.
 
\Medskip 
{\em Above the transition scale}, i.e. if $j,j_+\le j_{\phi}$, ${\cal W}^{j,j_+}$  {\em grows logarithmically} in terms of $j_+$ from the UV cut-off scale $j_D$ to
the transition scale $j_{\phi}$. {\em Below transition scale},
i.e. if $j_{\phi}\le j_+\le j'_{\phi}-1$, ${\cal W}^{j,j_+}$
is the {\em inverse of a spring factor} between scales $\max(j_{\phi},j)$ and $j'_{\phi}$. This shows clearly why
other arguments (namely, Ward identities, developed in
section \ref{section:boson}) are required to bound polymers
with $j_+\gg j_{\phi}$. However, keeping
$j'_{\phi}-j_{\phi}=o(\ln(1/g))$, in line with (\ref{eq:j'phi}),
ensures that large factors ${\cal W}^{j,j_+}$, $j\ll j_+\le j'_{\phi}$, bounded by ${\cal W}^j\le 2^{2(j'_{\phi}-j_{\phi})}$,  may be absorbed by putting aside some power
of $g$, say, 
\BEQ {\cal W}^{j,j_+}g^{1/2}=O(1) \label{eq:Wgeps}
\EEQ
for all $j\le j_+\le j'_{\phi}-1$ provided $j'_{\phi}-j_{\phi}\le\frac{1}{4}\ln(1/g)$.
{\em For definiteness,  we  choose}

\begin{Definition}[choice of $j'_{\phi}$] \label{def:j'phi}
\BEQ j'_{\phi}:=j_{\phi}+\lfloor\frac{1}{4}\ln(1/g)\rfloor. \label{eq:1/4} \EEQ
\end{Definition}

Hence the total weight of ${\cal W}^{j,j_+}$ for a diagram
with $n$ vertices is $\lesssim g^{-n/2}$.  For single-scale
diagrams, or multi-scale diagrams with no external  legs
in Cooper pair configuration, later to form a pre-Goldstone
boson, only {\em small} weights ${\cal W}^{j,j}=\begin{cases}
\frac{1}{j_{\phi}-j} \qquad (j\le j_{\phi}) \\
1 \qquad (j_{\phi}\le j\le j'_{\phi}-1) \end{cases}$
show up. The problem of potentially large ${\cal W}^{j,j_+}$ weights is therefore raised up again only in \S \ref{subsection:bounds}. 

\Medskip Summarizing, and referring to the details of the
computations in \S \ref{subsubsection:single-scale-bounds},  the small factor $g^{2+\max(1,\frac{n}{3}-1)}$ in (\ref{eq:Aamputated}) undergoes the following
transformations:

\Medskip (vi)$\longrightarrow$(vi)' {\em (total weight associated to a connecting vertex)} $O(gj2^{-2\del j})\longrightarrow
O({\cal W}^{j,j}j)$, featuring now a (probably spurious but
not very disturbing) logarithmic factor $j$;

\Medskip (vii)$\longrightarrow$(vii)' {\em (total weight associated to a non-connecting vertex)} 
\BEQ O(g2^{-j/2} 2^{-2\del j})\longrightarrow O({\cal W}^{j,j} 2^{-j/2})\lesssim O({\cal W}^{j,j} 2^{-j/2})=O(\frac{{\cal W}^{j,j}}{\sqrt{N_j}}), \label{eq:viivii'}
\EEQ
 featuring a $"1/N$"-type factor $N_j=2^j$. 

\Medskip Most importantly, we note here that -- {\em after} 
bubble chains have been resummed -- {\em connecting} or {\em strongly connecting} vertices are now the same. Hence the number of factors
${\cal W}^{j,j}j$ in (vi) is $n'=n-n''\le 2n/3\le 2\Big(2+\max(1,\frac{n}{3}-1\Big)$ (since $n''=n-n'\ge 2+\max(1,\frac{n}{3}-1)\ge n/3$).  Combining these
with the $n''$ factors $O(\frac{{\cal W}^{j,j}}{\sqrt{N_j}})$ coming
from (vii), we finally get instead of 
(\ref{eq:Aamputated})

\BEA && |{\cal A}(\Upsilon;\xi_{ext,1})|\le   \Big\{(C_{p,r} 
{\cal W}^{j,j})^n\prod_{\Del}\prod_{i=1}^{2^j} 
 \frac{1}{(n_{\Del,i}!)^r (\bar{n}_{\Del,i}!)^r} 
  \prod_{\ell\in L(\T)}\Big(1+2^{-j}
\frac{ |\xi_{\Del_{\ell}}-\xi_{\Del'_{\ell}}|}{\mu} \Big)^{-p} \Big\}\nonumber\\
&& \qquad\qquad  (j^2 2^{-j/2})^{2+\max(1,\frac{n}{3}-1)}\   (2^{-j})^{\half(4-N_{ext})} 
(p^*_F)^{-N_{ext}}.   \label{eq:single-scale-bound-t}
\EEA

\Medskip By the same argument as in \S \ref{subsubsection:single-scale-bounds}, one may sum over diagrams. 
This may be done in two stages; first one sums over all
diagrams spanning a given polymer $\P^j$ of scale $j$, yielding
\BEA && |{\cal A}(\P^j;\xi_{ext,1})|\lesssim 
\Big\{ (C_p{\cal W}^{j,j})^{n(\P^j)} \prod_{\ell\in L(\T)}\Big(1+2^{-j}
\frac{ |\xi_{\Del_{\ell}}-\xi_{\Del'_{\ell}}|}{\mu} \Big)^{-p}
\Big\}
\nonumber\\
&&\qquad\qquad (j^2 2^{-j/2})^{2+\max(1,\frac{n(\P^j)}{3}-1)} (2^{-j})^{\half(4-N_{ext})} (p^*_F)^{-N_{ext}}
   \label{eq:single-scale-bound-P}
\EEA
where $\T$ is a cluster tree spanning $\P^j$, and  $n(\P^j)$ is the number of boxes of $\P^j$. Then the sum
over polymers may be bounded using  space-time decay,  so that
\BEQ
\sum_{\P^j\in{\cal P}^j_{N_{ext}}\ |\ n(\P^j)=n}  |{\cal A}(\P^j;\xi_{ext,1})|\lesssim  (C_p {\cal W}^{j,j})^n (j^2 2^{-j/2})^{2+\max(1,\frac{n}{3}-1)} (2^{-j})^{\half(4-N_{ext})} 
(p^*_F)^{-N_{ext}} \label{eq:single-scale-bound-n} 
\EEQ
and finally, summing over all polymers of scale $j$ with one
external leg located at $\xi_{ext,1}$

\BEQ |\sum_{\P\in {\cal P}^j_{N_{ext}}} {\cal A}(\P;\xi_{ext,1})|\lesssim
(j^2 2^{-j/2})^{N_{ext}/9} (j^2 2^{-j})^{\half(4-N_{ext})}
(p^*_F)^{-N_{ext}}
\label{eq:single-scale-bound}
\EEQ 
since (considering the worst case, for which external 
momenta are grouped 3 by 3, see Figure below) $n(\P^j)\ge \frac{1}{3} N_{ext}(\P^j)=\frac{1}{3} N_{ext}$.

\bigskip

{\centerline{
\begin{tikzpicture}
\draw[fill=gray](0,0) rectangle(2,2);
\draw(0,1)--(-0.5,1.5); \draw(0,1)--(-0.5,1); \draw(0,1)--(-0.5,0.5);
\draw(2,1)--(2.5,1.5); \draw(2,1)--(2.5,1); \draw(2,1)--(2.5,0.5);
\end{tikzpicture}
}}

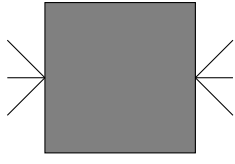
\Medskip {\bf \tiny Fig. \thesubsection.9. One example of external momentum configuration for a polymer with $N_{ext}=6$. The minimal number of vertices is 2.}


\subsubsection{Momentum-decoupling expansion and displacement of external legs}
\label{subsubsection:vert-cluster}



One must now {\em test} how step $j$ polymers $\P_1,\ldots,\P_n$ are connected from below to boxes in $\D^{j+1}$. 
To this end, we apply to each polymer $\P$ in ${\cal P}^{j\to}$ the following expansion operator

\BEQ {\mathrm{Vert}}^j(\P)=  \prod_{\Del\in\P\cap \D^j} \left( \sum_{\mu_{\Del}=0}^{N_0-1} \frac{1}{\mu_{\Del}!} \partial^{\mu_{\Del}}_{t_{\Del}}\big|_{t_{\Del}=0} +  \int_0^1  dt_{\Del}
\frac{(1-t_{\Del})^{N_0-1}}{
(N_0-1)!} \partial_{t_{\Del}}^{N_0} \right).  \label{eq:Vert} \EEQ

Expanding the product yields a sum of terms with 
$\mu_{\Del}$, $\Del\in\P\cap\D^j$, ranging in $\{0,\ldots,N_0\}$, $N_0$ coding for the  integral remainder term.
Let $N_{ext}(\P):=\sum_{\Del\in\P\cap\D^j} \mu_{\Del}$, and denote by ${\bf\Del}_{ext}(\P)\equiv \{\Del_1,\Del_2,\ldots\}$ the
boxes in $\P\cap\D^j$ containing the external fields.
If $N_{ext}(\P)< N_0$, then all $\mu_{\Del},\Del\in\P\cap\D^j$ are
$< N_0$, so $N_{ext}(\P)$ is equal to the number of
external fields. If $N_{ext}(\P)\ge N_0$ then
(i) either all $\mu_{\Del}$ are $< N_0$ and the same conclusion holds; or (ii) some $\mu(\Del)$ is equal to $N_0$, which
means that $t_{\Del}>0$: when this happens, we decide
somewhat arbitrarily that the polymer has exactly $N_0$ vertical
links connecting $\Del$ to the box below it in $\D^{j+1}$.
As we shall see, polymers with $N_{ext}< N_0$ will undergo
a further treatment. 

\Medskip The final outcome of the horizontal and momentum-decoupling expansion may be written as

\BEQ
F_{I}(\vec{\xi}_{ext},\vec{\bar{\xi}}_{ext})=\sum_{\F^j\subset{\cal F}^j}\, \prod_i {\mathrm{Vert}}^j(\P_i)\Big( \int d\vec{w}
\, \int d\mu^*_{\theta}(s^j(\vec{w}); \Psi^{\to j},\bar{\Psi}^{\to j}) \ \tilde{G}_I^{(j-1)\to}(\vec{\xi}_I)
\Big),
\label{eq:HV-expansion}
\EEQ
where:

-- $\F^j$ ranges over the set ${\cal F}^j$ of scale $j$ cluster forests;

-- $\P_1,\ldots,\P_n\in{\cal P}^{j\to} $ are the  non-$j-$overlapping polymers introduced at the end of
the last subsection;

--  and $\tilde{G}_I^{(j-1)\to}(\cdot)$ is as in (\ref{eq:2.75}).

\Medskip Instead of summing over cluster forests,
it is equivalent, but more convenient, to sum over non-$j$-overlapping polymers, so that
\BEQ \sum_{\F}\, \prod_i {\mathrm{Vert}}^j(\P_i) (\cdots) \ \longrightarrow \sum_n \sum_{\P_1,\ldots,
\P_n \ {\mathrm{non-}}j{\mathrm{-overlapping}}} 
 \, \prod_i \sum_{\T_i}{\mathrm{Vert}}^j(\P_i) (\cdots)
\EEQ
in (\ref{eq:HV-expansion}), where $\T_i$,
$i=1,\ldots,n$ 
are cluster trees compatible with the choice of
non-$j$-overlapping polymers $\P_1,\ldots,\P_n\in{\cal P}^{j\to}$, namely: $\T_i,i=1,\ldots,n$ are the connected components of $\F$, and $\T_i=\P_i\cap\D^j$. It is in this form that we shall be using (\ref{eq:HV-expansion}). Thus: 
\BEQ F_I(\vec{\xi}_{ext},\vec{\bar{\xi}}_{ext})=\sum_n \sum_{\P_1,\ldots,
\P_n \ {\mathrm{non-}}j{\mathrm{-overlapping}}}
\prod_{i=1}^n F^{HV}(\P_i).
\EEQ

\Bigskip {\bf Displacement of external legs.}  Let
us discuss the smallest values for $N_{ext}(\P)$:

\Medskip (i) if $N_{ext}(\P)=0$, then $\P$ has no external field, so
$\P$ is a vacuum polymer;

\Medskip (ii) if $N_{ext}(\P)=2$, then $\P$ is a two-point polymer,
with corresponding contribution 
\BEA  &&\int d\xi_1\,\chi_{\Del_1}(\xi_1)
\int d\xi_2 \, \chi_{\Del_2}(\xi_2)\,  {\cal A}^j(\P;\xi_1,\xi_2)\  \Psi^{\to(j+1)}(\xi_1)
\bar{\Psi}^{\to(j+1)}(\xi_2) \nonumber\\
&&=F^j_{local}(\P;\Psi^{\to (j+1)},\bar{\Psi}^{\to (j+1)}) +
\del F^j(\del \P;\Psi^{\to (j+1)},\bar{\Psi}^{\to (j+1)})
\EEA

where
\BEA && F^j_{local}(\P;\Psi^{\to (j+1)},\bar{\Psi}^{\to (j+1)}):=\nonumber\\
&&\ \ \sum_{k_1,k_2\ge j+1}\sum_{\alpha_1\in \Z/2^{k_1}\Z,\alpha_2\in \Z/2^{k_2}\Z}
\int d\xi_1\,\chi_{\Del_1}(\xi_1)
\int d\xi_2 \, \chi_{\Del_2}(\xi_2)\,
 {\cal A}^j(\P;\Psi^{\to (j+1)},\bar{\Psi}^{\to (j+1)};\xi_1,\xi_2) \ \cdot\ 
 \nonumber\\
&& \qquad\qquad \cdot\  \half\Big[ \Psi^{k_1,\alpha_1}(\xi_1) \Big(\bar{\Psi}^{k_2,\alpha_2}(\xi_1)+ \Big(\xi_2-\xi_1,\partial_{\xi_1} 
(e^{\II ( p^{k_2,\alpha_2},\xi_2-\xi_1)} 
\Psi^{k_2,\alpha_2}(\xi_1)\Big)\Big) \nonumber\\
&&\qquad+
\bar{\Psi}^{k_2,\alpha_2}(\xi_2)\Big(\Psi^{k_1,\alpha_1}(\xi_2)+ \Big(\xi_1-\xi_2,\partial_{\xi_2} 
(e^{\II ( p^{k_1,\alpha_1},\xi_1-\xi_2)} 
\Psi^{k_1,\alpha_1}(\xi_2)\Big)\Big) \Big] ,  \label{eq:2.93}
\EEA
see (\ref{eq:R2}),
is the local part of the polymer, obtained by {\em displacing} symmetrically {\em one of the external fields
to the location of the other}. The error term $\del F^j(\cdot)$, once its 
external legs are
contracted (at some later stage, namely, at scales $k_1$ and
$k_2$), may be rewritten as a  Taylor remainder as in
(\ref{eq:R2}), and comes with an extra spring factor $O(2^{-2 (\min(k_1,k_2)-j)})$. Since two-point functions are
{\em linearly divergent}, subtracting local part to order 1
 makes the remainder convergent. Referring to the 
 logarithmically divergent, four-point bubble diagram, and to
 the power-counting of a contracted pair $C^j(\cdot,\cdot)=\langle \Psi^j(\cdot)
 \Psi^j(\cdot)\rangle=O(2^{-j})$, $\P$ may be thought of
as having $6$ external legs; however, for pure book-keeping 
reasons, we shall consider it as a polymer with $N_0=8$
external legs -- which simply means that it is convergent
enough for our purposes. For future use we indeed
replace $\P$ with a polymer $\del \P$ which is identical to
$\P$, except that  it has $N_0$ external legs in the box
where the external fields have been displaced to,  $\Del_1$,
resp. $\Del_2$, and
{\em none} in the other box, $\Del_2$, resp. $\Del_1$. 

\Medskip (iii) if $4\le N_{ext}(\P)<N_0=8$, then 
we proceed as in (\ref{eq:2.93}), Taylor expanding to order 2
symmetrically
in the neighborhood of each external vertex location
$\xi_1,\ldots,\xi_{N_{ext}}$. All such diagrams -- recall Cooper pair bubble diagrams have already been resummed by hand  -- are already
convergent {\em before} displacing external legs. However
(see \S \ref{subsection:bounds}), the above operation will
help us sum over external angular sectors in the multi-scale
bounds. Local contributions $F^j_{local}(\P;\Psi^{\to(j+1)},\bar{\Psi}^{\to(j+1)})$ are {\em translation
invariant} polymers with $N_{ext}(\P)$ external legs; the error term $\del F^j_{local}(\del \P;\Psi^{\to(j+1)},\bar{\Psi}^{\to(j+1)})$, on the other hand, is   considered as as a polymer 
$\del \P$ with $N_0$ external legs.


\subsubsection{Mayer expansion}
\label{subsubsection:Mayer}


We shall now   apply
the {\em restricted cluster expansion}, see Proposition \ref{prop:BK2}, to
the result of our expansion.  Cluster expansions have allowed us to rewrite Green functions as sums over
non-$j$-overlapping polymers denoted by $\P_1,\ldots,\P_n$.  The {\em objects} are now
 polymers $\P $ in    ${\cal O}=\{\P_1,\ldots,\P_n\}\subset{\cal P}^{j\to}$ ; a link
$\ell\in L({\cal O})$ is a pair of polymers $\{\P_i,\P_{i'}\}$, $i\not=i'$.
Objects of type 2 are  polymers with $\ge N_0$ external legs, whose non-overlap conditions we shall not remove.
Then objects of type $1$  are   polymers with $<N_0$  external legs. Due to the displacement of external legs operated in
 \S \ref{subsubsection:vert-cluster},
all external legs are located in the {\em same} scale $j$ box.

\medskip\noindent Implicit in the outcome of the cluster expansions is  the {\bf non-overlapping condition},
 \BEA {\mathrm{NonOverlap}}(\P_1,\ldots,\P_n) &:=& \prod_{(\P_i,\P_{i'}) } {\bf 1}_{\P_i,\P_{i'}\ {\mathrm{non}}-j-{\mathrm{overlapping}}} \nonumber\\
 &=&
\prod_{(\P_i,\P_{i'}) } \prod_{\Del\in\vec{\Del}^j(\P_i),\Del'\in
\vec{\Del}^j(\P_{i'})}
 \left( 1 + \left(
 {\bf 1}_{\Del\not=\Del'}-1 \right)\right) \nonumber\\ \EEA
 stating that a scale $j$ box $\Del$ belonging to  $\P_i$ and
 a scale $j$ box $\Del'$ belonging to $\P_{i'}$ are necessarily
 distinct.
Similarly to what we did during the horizontal cluster expansion, we
choose some polymer  with $<N_0$ external legs, say $\P_1$, and 
weaken the non-overlap condition between $\P_1$ and all the other polymers $\P_{i'},i'\not=1$ by introducing a parameter $S_1$,

 \BEA && {\mathrm{NonOverlap}}(\P_1,\ldots,\P_n) (S_1) 
 \nonumber\\
 &&\qquad =
 \Big( \prod_{\{\P_i,\P_{i'}\}_{i,i'\not=1} } \prod_{\Del\in{\bf\Del}^j(\P_i),\Del'\in
{\bf\Del}^j(\P_{i'})} {\bf 1}_{\Del\not=\Del'} \Big) \Big( 
\prod_{(\Del,\Del')\in \vec{\Del}_{ext}(\P_1)\times \vec{\Del}_{ext}(\P_{i'}) }  {\bf 1}_{\Del\not=\Del'} \Big) \ 
 \cdot\nonumber\\
 &&\qquad 
\prod_{ (\Del,\Del')\in\vec{\Del}^j(\P_1)\times \vec{\Del}^j(\P_{i'})\setminus \vec{\Del}_{ext}(\P_1)\times \vec{\Del}_{ext}(\P_{i'}) } \left( 1 +  S_1 \left( {\bf 1}_{\Del\not=\Del'}-1 \right)\right),  \nonumber\\ \label{eq:Mayer1}\\ \EEA
where $\vec{\Del}_{ext}(\P)\subset\vec{\Del}^j(\P)$ is the subset of boxes $\Del$ with external legs - i.e. that have been differentiated
with respect to $t^j_{\Del}$ -, 
 and Taylor expand in $S_1$ to order $1$; each factor 
 \BEQ {\bf 1}_{\Del\not=\Del'}-1 =-{\bf 1}_{\Del=\Del'} \EEQ
  produced by
 differentiation is a {\bf Mayer link} between $\P_1$ and
 some $\P_{i'}, i'\not=1$, or more precisely, some  box $\Del\in\vec{\Del}(\P_1)$ and some
 box $\Del'\in\vec{\Del}(\P_{i'})$, implying an explicit overlap between $\P_1$ and
 $\P_{i'}$, and adding a link to the forest $\F^j$.
 Iterating the procedure and applying Proposition \ref{prop:BK2}
 to the weakened non-overlap condition
 \BEA &&{\mathrm{NonOverlap}}(\P_1,\ldots,\P_n) (\vec{S}):
=  \prod_{\{\P_i,\P_{i'}\} } \prod_{\Del\in{\bf\Del}_{ext}(\P_i),\Del'\in
{\bf\Del}_{ext}(\P_{i'})} {\bf 1}_{\Del\not=\Del'} \ \cdot\nonumber\\
&&
 \qquad 
\prod_{(\Del,\Del')\in\vec{\Del}^j(\P_i)\times\vec{\Del}^j(\P_{i'})\setminus
{\bf\Del}_{ext}(\P_i)\times
{\bf\Del}_{ext}(\P_{i'})} \left( 1 +  S_{i ,i'} \left( {\bf 1}_{\Del\not=\Del'}-1 \right)\right),  \nonumber\\ \label{eq:Mayer2}
\EEA
 
 The outcome is a sum 
 \BEA &&
 \sum_{\G\in {\cal F}_{res}({\cal O})} \Big( \prod_{\ell \in 
 L(\G)} \int_0^1 dW_{\ell}\Big)\ \   {\mathrm{Mayer}}(\vec{S}(\vec{W})), \nonumber\\
 &&\qquad  {\mathrm{Mayer}}(\vec{S}(\vec{W})):= \Big[ \Big(\prod_{\ell\in L(\G)} \frac{\partial}{\partial S_{\ell}} \Big) {\mathrm{ NonOverlap}}(\P_1,\ldots,\P_n)\Big] (\vec{S}(\vec{W})) \nonumber\\  \label{eq:Mayer3}
 \EEA
 Links $\ell=\ell_{\P_i,\P_{i'}}\in L(\G)$ are obtained as links between {\em polymers}, however the corresponding differentiation 
$\frac{\partial}{\partial S_{\ell}}$ is immediately rewritten as
a sum over pairs of boxes $(\Del,\Del')\in\vec{\Del}^j(\P_i)\times 
\vec{\Del}^j(\P_{i'})$. Thus we see Mayer links as {\em links between boxes}.  As such they add up to the set of links $L(\F^j)$ produced
by the horizontal cluster expansion, producing a forest $\bar{\F}^j$
with same vertices as $\F^j$ but larger set of links $L(\bar{\F}^j)\equiv 
L(\F^j)\uplus L_{{\mathrm{Mayer}}}$, where $L_{{\mathrm{Mayer}}}$
(in bijection with $L(\G)$) is the set of Mayer links. Since
a forest is characterized by its set of links, we  rewrite in practice 
(\ref{eq:Mayer3}) as
\BEA && \sum_{L_{{\mathrm{Mayer}}}} \Big( \prod_{\ell \in 
 L_{{\mathrm{Mayer}}}} \int_0^1 dW_{\ell}\Big)\ \   {\mathrm{Mayer}}(\vec{S}(\vec{W})), \nonumber\\
 &&\qquad  {\mathrm{Mayer}}(\vec{S}(\vec{W})):= \Big[ \Big(\prod_{\ell\in L_{{\mathrm{Mayer}}}} \frac{\partial}{\partial S_{\ell}} \Big) {\mathrm{ NonOverlap}}(\P_1,\ldots,\P_n)\Big] (\vec{S}(\vec{W})) \nonumber\\  \label{eq:Mayer4}.
 \EEA

\medskip\noindent The number of external legs of a set of polymers connected by Mayer links is the
sum of the number of external legs of each of the polymers. In particular, 
\begin{itemize}
\item[(i)] new vacuum polymers without any non-overlap conditions have been produced; they are resummed into
the {\em scale $j$ free energy function $f^{j}(\lambda)$};
\item[(ii)] two-point polymers have been {\em dressed by a
cloud} of vacuum polymers; they can now be resummed into
a {\em renormalization of the two-point function};
\item[(iii)] links from polymers of type 1 to polymers of type 2 produce new polymers with $\ge N_0$ external legs, whose overlap conditions have not been removed.
\end{itemize}
Other possibilities include e.g. polymers with $4,6,8,\cdots$ 
external legs without any non-overlap condition, produced
by Mayer-linking $2,3,4,\cdots$ two-point polymers.

\medskip\noindent   Let us now give some necessary precisions. Since the Mayer expansion is really applied to the non-overlap function NonOverlap and {\em not} to
the outcome of the expansion,  one must
still extend the outcome of the expansion  to the case when the $\P_i$, $i=1,\ldots,n$ have some overlap. The natural way to do this is to assume that
 the fields $(\Psi^{j,\alpha}\big|_{\P_i}, \bar{\Psi}^{j,\alpha}\big|_{\P_i})_{i=1,\ldots,n}$ remain independent even when they overlap. This may be understood in the following way. Choose a different color for each polymer $\P_i=\P_1,\ldots,\P_n$, and paint with that color
 {\em all} boxes $\Del\in \P_i\cap\D^j$. If $\Del\in{\bf \Del}_{ext}(\P_i)$, then its
 external vertical links to $\D^{j+1}$ are left in black. The
 previous discussion implies that boxes  with different colors may superpose; on the other hand, {\em external inclusion links} may {\em not}, so that
 {\em low-momentum fields} $\Psi^{\to(j+1)},\bar{\Psi}^{\to(j+1)}$
do not superpose and may be left in black.

\medskip\noindent Hence one must see $\Psi,\bar{\Psi}$ as living on a two-dimensional set, $\D^{j}\times\{ {\mathrm{colors}}\}$, so that copies of $\Psi,\bar{\Psi}$ with different colors are independent of each other. This defines  new, {\em extended} fields $\Psi,\bar{\Psi} :\R_\times\R^2\times \{{\mathrm{colors}}\}\to\R$, and {\em Mayer-extended polymers}.
 By abuse of notation, we shall skip the tilde in the sequel, and always implicitly extend the fields and the measures  by taking into account colors.

\Bigskip As a general principle (see \cite{Bry} for a 
single-scale version, or \cite{MagUnt2} for the general, multi-scale version), sums over Mayer-extended polymers are
bounded exactly as sums over {\em non}-Mayer-extended polymers, $\sum_{n} \sum_{\P\in {\cal P}^{j\to}\ |\ |\P|=n} (C_p)^n g^{n/3}$, see (\ref{eq:single-scale-bound-P}), (\ref{eq:GFMSB-P}), with the constant $C_p$ of non-Mayer-extended
polymers replaced by $eC_p$.




\subsection{Multi-scale bounds and fermionic fixed-point} \label{subsection:bounds}


 Our starting point is the single-scale bound

\BEA && |{\cal A}(\P^j;\xi_{ext,1})|\lesssim 
\Big\{ (C_p{\cal W}^{j,j})^{n(\P^j)} \prod_{\ell\in L(\T)}\Big(1+2^{-j}
\frac{ |\xi_{\Del_{\ell}}-\xi_{\Del'_{\ell}}|}{\mu} \Big)^{-p}
\Big\}
\nonumber\\
&&\qquad\qquad (j^2 2^{-j/2})^{2+\max(1,\frac{n(\P^j)}{3}-1)} (2^{-j})^{\half(4-N_{ext})} (p^*_F)^{-N_{ext}}  \nonumber
\EEA
see (\ref{eq:single-scale-bound-P}), complemented by 
multi-scale bounds
for the effective vertex, 
(\ref{eq:single-scale-bound-n}) and (\ref{eq:single-scale-bound}), and Definition \ref{def:scale-j-effective}, which imply  a total supplementary  weight
$\le (2^{2(j'_{\phi}-j_{\phi})})^{n(\P^j)}$ in a multi-scale
setting. Recall ${\cal W}^{j,j}=\begin{cases}
\frac{1}{j_{\phi}-j} \qquad (j\le j_{\phi}) \\
1 \qquad (j_{\phi}\le j\le j'_{\phi}-1) \end{cases}$.  We shall now prove
by scale induction the following {\em multi-scale bound}.
Fix $N_0:=8$ and consider a polymer $\P\in {\cal P}^{k\to}$
with external legs $\Psi^{k_{ext,1},\alpha_{ext,1}},\ldots$, 
$\Psi^{k_{ext,N_{ext}},\alpha_{ext,N_{ext}}}$, $k\le  k_{ext,1}, \ldots,  
k_{ext,N_{ext}}$. Let $k_{ext}:=\min(k_{ext,1},\ldots,k_{ext,N_{ext}})$ be
the highest external momentum scale, considered as the
external scale of $\P$. The procedure considered
in \S \ref{subsubsection:vert-cluster} has produced three types
of polymers: 
\begin{itemize}
\item[(i)] local parts of polymers with $4\le N_{ext}<N_{0}$ external legs;
\item[(ii)]  polymers with $\ge N_0$ external
legs;  and
\item[(ii')] error terms $\del F^k(\del\P; \Psi^{\to(k+1)},\bar{\Psi}^{\to(k+1)})$ associated with
polymers $\del \P$ with $2\le N_{ext}<N_0$ external legs, which have a spring factor $2^{-2(k_{ext}-k)}$, and have been considered as polymers with $N_0$ external legs in the above
book-keeping.
\end{itemize}
 As we shall see, the  power-counting associated
to  external leg angular sectors is more favorable  for
translation-invariant polymers (case (i)), for which Proposition \ref{prop:sector-counting} holds. On the other hand, polymer contributions of type (ii), (ii')
enjoy (though for different reasons) a spring factor $2^{-2(k_{ext}-k)}$, making it possible to control them
in a similar way.

\Medskip Note that, for every $c>0$,
\BEQ  \frac{1}{j_{\phi}-j}2^{-cj} \lesssim 
\frac{1}{j_{\phi}}\approx g,\qquad j2^{-cj}\lesssim 1
\EEQ
and (by (\ref{eq:1/4})) $2^{2(j'_{\phi}-j_{\phi})}=O(g^{-1/2})$,  hence  
\BEQ  ({\cal W}^{j,j} 2^{2(j'_{\phi}-j_{\phi})})^{n(\P^j)} (j^2 2^{-j/2})^{2+\max(1,\frac{n(\P^j)}{3}-1)}\lesssim g^{n(\P^j)/2}
2^{j(-3/2+o(1))}. \label{eq:NextN0+} \EEQ
For multi-scale polymers of lowest internal scale $k$ and
highest external scale $j$, the factor $2^{-3k/2}$ in
(\ref{eq:NextN0+}) is multiplied by the spring factor
$2^{-(j-k)\frac{N_{ext}-4}{2}}$. For $N_{ext}=6$, the sum
\BEQ \sum_{k<j} 2^{-(j-k)\frac{N_{ext}-4}{2}} 2^{-3k/2}
\label{eq:k3k2}
\EEQ is
only $O(2^{-j})$, thus one expects the decay exponent
$3j/2$ to be lowered to $j$ at most. With a decay $O(2^{-k})$ however, the sum (\ref{eq:k3k2}) yields $O(2^{-j})$ but with
a logarithmic correction. Hence it is safer to say that a lower
decay exponent holds, say, $(\half-o(1))j$ only, compare 
with the exponent in (\ref{eq:NextN0+}).

\Medskip Our claims are therefore the following.  (\ref{eq:single-scale-bound-P}), 
(\ref{eq:single-scale-bound-n}), (\ref{eq:single-scale-bound})
and (\ref{eq:NextN0+}) ensure that they hold for a 
single-scale polymer. Let $N_0:=8$ and
\BEQ (N_{ext}-N_0)_+:=\begin{cases} 1 \qquad (N_{ext}\ge N_0) \\ 0
\qquad (N_{ext}<N_0) \end{cases}.
\EEQ 

\Medskip {\bf  General fermionic multi-scale bound.}
{\em Let $\P^{j\to}\in{\cal P}^{j\to}$ not made up only of bubbles, then 
\BEQ \boxed{ |{\cal A}(\P^{j\to};\xi_{ext,1})|\lesssim g^{n(\P^{j\to})/2}  2^{j(-\half+o(1))} 2^{-j(N_{ext}-N_0)_+}  (2^{j})^{\half(N_{ext}-4)}
(p^*_F)^{-N_{ext}} }
\label{eq:GFMSB-P}
\EEQ

Furthermore, if $\P^{j\to}$ ranges in the set of  polymers
in ${\cal P}^{j\to}$ with $n\ge 1$ vertices, 
\BEQ  \boxed{ \sum_{\P^{j\to}\in{\cal P}^{j\to}\ |\ n(\P^{j\to})=n} |{\cal A}(\P^{j\to};\xi_{ext,1})|\lesssim g^{n/2}  2^{-j/2} (2^{j})^{\half(N_{ext}-4)}(p^*_F)^{-N_{ext}} }
\label{eq:GFMSB-n}
\EEQ}
from which, finally, summing over all polymers in $\P^{j\to}$
 with one external vertex located in $\xi_{ext,1}$,
 \BEQ  \sum_{\P^{j\to}\in{\cal P}^{j\to}} |{\cal A}(\P^{j\to};\xi_{ext,1})|\lesssim g^{N_{ext}/6}  2^{-j/2} (2^{j})^{\half(N_{ext}-4)}(p^*_F)^{-N_{ext}},
\label{eq:GFMSB}
\EEQ
compare with the discussion above Fig. \ref{subsection:cluster}.9. 
The single remaining factor $2^{-j/2}$, coming  originally from the $1/N$-expansion, is sufficient to perform the sum
over scales, as we shall see.

\Medskip We now consider $p$ polymers $\P_1\in{\cal P}^{k_1\to},
\ldots,\P_p\in{\cal P}^{k_p\to}$  $(k_1,\ldots,k_p<j)$ with
$n_1,\ldots,n_p$ vertices, 
forming at scale $j$ a polymer $\P^{j\to}$ through pairings of (some of)
their external legs between themselves or with  scale $j$ propagators expanded at scale $j$.

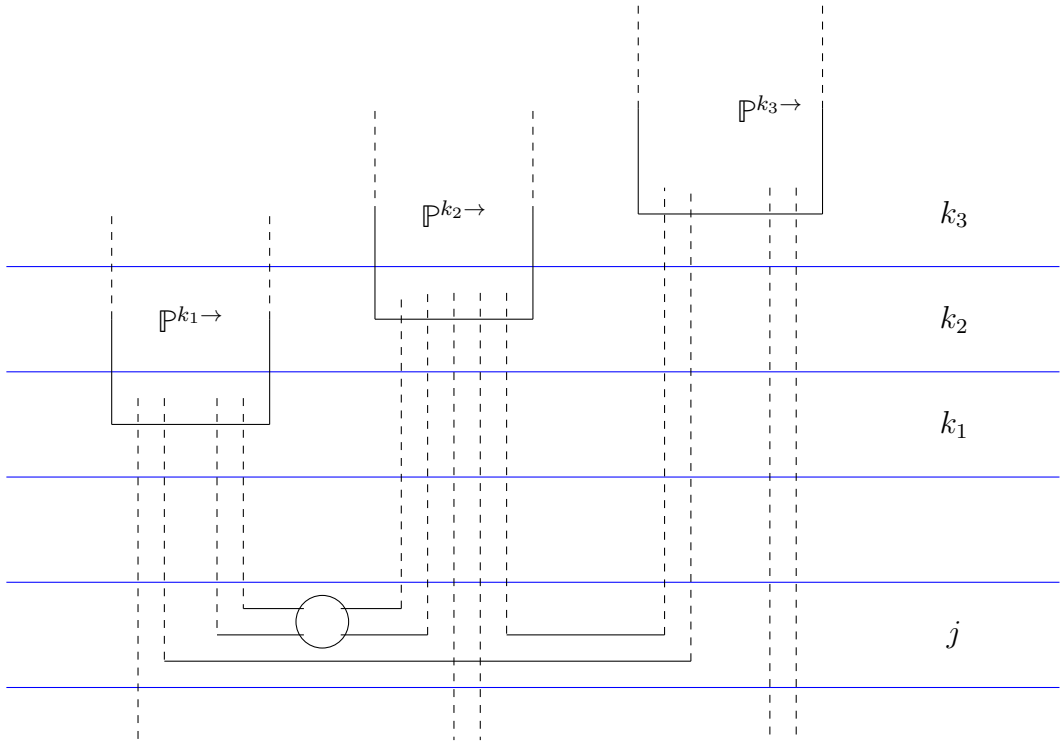
\Bigskip

{\centerline{
\Bigskip
\begin{tikzpicture}[scale=0.7]
\draw[color=blue](0,0)--(20,0);
\draw[color=blue](0,2)--(20,2);
\draw[color=blue](0,4)--(20,4);
\draw[color=blue](0,-2)--(20,-2);
\draw[color=blue](0,-4)--(20,-4);
\draw(2,1)--(5,1); \draw(2,1)--(2,3); \draw[dashed](2,3)--(2,5); \draw(5,1)--(5,3); \draw[dashed](5,3)--(5,5);
 \draw[dashed](2.5,1.5)--(2.5,-5);
\draw[dashed](3,1.5)--(3,-3.5); \draw[dashed](4,1.5)--(4,-3);
\draw[dashed](4.5,1.5)--(4.5,-2.5);
 \draw(4.5,-2.5)--(5.65,-2.5); \draw(6.35,-2.5)--(7.5,-2.5);
  \draw(4,-3)--(5.65,-3); \draw(6.35,-3)--(8,-3);
   \draw(6,-2.75) circle(0.5);
\draw[dashed](8,-3)--(8,3.5); \draw[dashed](7.5,-2.5)--(7.5,3.5); 
\draw[dashed](8.5,3.5)--(8.5,-5); \draw[dashed](9,3.5)--(9,-5);
\draw[dashed](9.5,3.5)--(9.5,-3); \draw(9.5,-3)--(12.5,-3);
\draw[dashed](12.5,-3)--(12.5,5.5); 
\draw(3,-3.5)--(13,-3.5); \draw[dashed](13,-3.5)--(13,5.5);
\draw[dashed](14.5,5.5)--(14.5,-5); \draw[dashed](15,5.5)--(15,-5);
\draw(18,-3) node {$j$};
\draw(18,1) node {$k_1$}; \draw(18,3) node {$k_2$};
\draw(18,5) node {$k_3$};
\draw(3.5,3) node {$\P^{k_1\to}$}; \draw(8.5,5) node {$\P^{k_2\to}$}; \draw(14.5,7) node {$\P^{k_3\to}$};
\begin{scope}[shift={(5,2)}]
\draw(2,1)--(5,1); \draw(2,1)--(2,3); \draw[dashed](2,3)--(2,5); \draw(5,1)--(5,3); \draw[dashed](5,3)--(5,5);
\end{scope}
\begin{scope}[shift={(10,4)}]
\draw(2,1)--(5.5,1); \draw(2,1)--(2,3); \draw[dashed](2,3)--(2,5); \draw(5.5,1)--(5.5,3); \draw[dashed](5.5,3)--(5.5,5);
\end{scope}
\end{tikzpicture}
}}

\Medskip {\bf \tiny Fig. \thesubsection.1. Multi-scale
polymer $\P^{j\to}$ with lowest scale $j$ and $N_{ext}=5$, obtained by contracting $\P_i\in {\cal P}^{k_i\to}$, $i=1,2,3$,  with
$N_{ext,1}=4$, resp.  $N_{ext,2}=5$,  $N_{ext,3}=4$ external legs. The blob represents a polymer expanded at scale $j$ and connecting the 
scale $j$ external legs of $\P^{k_1\to}$ and $\P^{k_2\to}$.}

\Bigskip
  Assuming (\ref{eq:GFMSB-P})
holds for the connected components $\P_1,\ldots,\P_p$ of 
$\P^{j\to}\cap\D^{(j-1)\to}$, we want to show that it also
holds for $\P^{j\to}$.  For that, we  must take into account several factors
related to the scale $j$ integrations.  The following discussion is relative to a polymer $\P\in {\cal P}^{k\to}$ with external scale $k_{ext}$, which is one of the 
$\P_i$'s. By assumption $k< k_{ext}\le j$. 
For simplicity, we remove all dimensional constants ($\mu,p_F^*,m^*\cdots$) from our
estimates.

\Medskip{\bf Volume factors.}
External legs of $\P$ which are contracted at scale $j$ are
integrated in a box $\Del^j$ of scale $j$. This implies an
absolute volume factor $O(2^{3j})$. 

\Medskip {\bf Field scaling factors.} 
Each external field $\Psi^{j,\alpha_i}$, $j=k_{ext,i}$ of $\P$ which is contracted at scale $j$ comes with a supplementary
absolute  prefactor $2^{-j}$.

\Medskip Taking into account volume and field scaling factors yields the general bound
\BEA  |{\cal A}(\P;\xi_{ext,1})| &\lesssim&
2^{jN_{ext}} 2^{-3j}
\prod_{i=1}^p  \Big\{ |{\cal A}(\P_i;\cdot)|\ 
2^{-jN_{ext,i}} 2^{3j}\Big\} \ \times {\cal N}_{sec} \nonumber\\
&\lesssim & 2^{jN_{ext}} 2^{-3j}
\prod_{i=1}^p  \Big\{
 g^{n_i/2} \,  2^{-k_i/2} \, 2^{-k_i(N_{ext,i}-N_0)_+} \,  2^{\frac{k_i}{2}(N_{ext,i}-4)} \, 
2^{-jN_{ext,i}}\,  2^{3j}\Big\} \ \times {\cal N}_{sec} \nonumber\\
&=& 2^{j(N_{ext}-3)}
\prod_{i=1}^p  \Big\{ g^{n_{i}/2}\,  2^{-(j-k_i)\frac{N_{ext,i}-4}{2}} \, 2^{-k_i/2}\, 2^{-k_i(N_{ext,i}-N_0)_+} \, 
2^{-j\frac{N_{ext,i}-2}{2}}\Big\} \ \times {\cal N}_{sec},\nonumber\\  \label{eq:2.106}
\EEA
where ${\cal N}_{sec}$ is a bound on the sum over all possible
 sector assignments of external legs of polymers contracted
 at scale $j$. The $2^{jN_{ext}}2^{-3j}$ prefactor comes
 from an over-counting: the number of scale $j$ contracted
 external fields is not $\sum_i N_{ext,i}$, but $\Big\{\sum_i N_{ext,i}\Big\}-N_{ext}$, which changes the overall field scaling factor;
 and the overall volume factor $(2^{3j})^n$ is corrected by
 a factor $2^{-3j}$, since amputated diagrams are evaluated
 by definition with one  vertex location fixed. {\em Polymers with 4 external
 legs may be considered as four-valent vertex insertions in a scale $j$ diagram.}  From the diagrammatic
 analysis of \S \ref{subsubsection:single-scale-bounds} and
 \S \ref{subsubsection:bubble-sum},  it follows in particular that {\em a compound polymer obtained by contracting only polymers with 4 external legs} has at least one extra non-connecting vertex, which produces
 an extra factor $\lesssim  2^{-j}$, see (\ref{eq:NextN0+}), w.r. to the bound
 (\ref{eq:2.106}).

\Medskip {\bf Sector assignment factors.} 
We  rewrite the overall sector assignment factor
${\cal N}_{sec}$ as  
 ${\cal N}'_{sec} \times  {\cal N}''_{sec}$,
where ${\cal N}'_{sec}:=\prod_{i=1}^p {\cal N}'_{sec,i}$, and
\BEQ {\cal N}'_{sec,i}=(2^{j/2})^{N_{ext,i}-2} \, \overline{{\cal N}}_{sec,i}, \EEQ
and
\BEQ  {\cal N}''_{sec}:=(2^{j/2})^{-N_{ext}+2} \, \overline{\cal N}''_{sec}.\EEQ

Factors ${\cal N}'_{sec,i}$, ${\cal N}''_{sec}$ combine
nicely with the factors between brackets $\Big\{\ \cdot\ \Big\}$ in (\ref{eq:2.106}): 
 $2^{j(N_{ext}-3)}\times {\cal N}''_{sec}=(2^{j})^{\half(N_{ext}-4)},$ while $\Big\{ \prod_{i=1}^p
 2^{-j\frac{N_{ext,i}-2}{2}}\Big\}\times {\cal N}'_{sec}=\prod_{i=1}^p 
 \overline{\cal N}_{sec,i}$.
Thus
\BEQ |{\cal A}(\P;\xi_{ext,1})|\lesssim 
\prod_{i=1}^p \Big\{ g^{n_i/2} 2^{-(j-k_i)\frac{N_{ext,i}-4}{2}} 2^{-k_i/2} \,  2^{-k_i(N_{ext,i}-N_0)_+}\, \overline{\cal N}_{sec,i}\Big\} \ \times\  (2^{j})^{\half(N_{ext}-4)}.
\EEQ
The value of the {\em correcting factors} $\overline{{\cal N}}_{sec,i}$  and $\overline{\cal N}''_{sec}$ will be presently computed. The induction  is
successful provided we manage to prove
that 
\BEQ \prod_{i=1}^p  \Big\{\sum_{k_i<j} 2^{-(j-k_i)\frac{N_{ext,i}-4}{2}}\, 
2^{-k_i/2}\,  2^{-k_i(N_{ext,i}-N_0)_+}\, \overline{\cal N}_{sec,i}\Big\} \overset{?}{\lesssim} 2^{-j/2}  2^{-j(N_{ext}-N_0)_+} 
\label{eq:2.110}
\EEQ
if $\max_{i=1,\ldots,p} N_{ext,i}>4$, or
\BEQ \prod_{i=1}^p  \Big\{ \sum_{k_i<j} 2^{-(j-k_i)\frac{N_{ext,i}-4}{2}}\, 
2^{-k_i/2} \,  2^{-k_i(N_{ext,i}-N_0)_+}\,  \overline{\cal N}_{sec,i}\Big\} \overset{?}{\lesssim} 2^{-j(N_{ext}-N_0)_+}
\label{eq:2.110bis}
\EEQ
  in the specific case when $N_{ext,i}=4$ for all $i=1,\ldots,p$. Sums $\sum_{k_i<j}$ take care of all possible choices for
  scales $k_1,\ldots,k_p$.

\Medskip{\em Scale $j$ momentum conservation for
local parts of polymers with $6\le N_{ext}(<N_0)$.}
Local parts may be seen as sums of Fourier diagrams with
external momenta $p_{ext,1},\ldots,p_{ext,N_{ext}}$, all
coming from the same external vertex.  {\em Consider a local part whose external momenta are all contracted at scale $j$.}
Let $\alpha_1,\ldots,\alpha_{N_{ext,i}}\in\Z/2^j\Z$ be their
angular sectors. Since (by momentum
conservation) $\sum_{i=1}^{N_{ext}} p_{ext,i}=0$, from which
$\Big|\sum_{i=1}^{N_{ext}} p^{j,\alpha_i} \Big|\lesssim 
2^{-j}p^*_F$,  and sectors of contracted legs $\langle 
\Psi^{j,\alpha}(\xi_{ext})\bar{\Psi}^{j,\alpha}(\xi'_{ext})\rangle\equiv C^{j,\alpha}_{\theta}(\xi_{ext},\xi'_{ext})$  are  {\bf shared} between the two contracted fields, Proposition \ref{prop:sector-counting} gives  an external sector assignment $\lesssim 
{\cal N}'_{sec,i}:=(2^{j})^{\frac{N_{ext,i}-2}{2}}$.  Thus one
sets 
$\overline{\cal N}_{sec,i}=1$ for such polymers. Then
 \BEQ \sum_{k_i<j} 2^{-(j-k_i) \frac{N_{ext,i}-4}{2}} 2^{-k_i/2}\le \sum_{k_i<j} 2^{-(j-k_i)} 2^{-k_i/2} \le 2^{-j/2}\sum_{k_i<j} 2^{-(j-k_i)/2}=O(2^{-j/2}).
\EEQ

\Medskip{\em Polymers with $N_{ext}\ge N_0$ external legs.}
For polymers of type (ii) or (ii'), momentum conservation  does not hold in general, so that (disregarding the sharing issue) the number of
{\em scale $k_i$} external sector assignments is $O(2^{k_i N_{ext,i}})$. The factor $2^{-k_i(N_{ext,i}-N_0)_+}=2^{-k_i}$ in the
term between brackets $\big\{ \ \cdot\ \big\}$ in
(\ref{eq:2.110}) can however be put to good use, enhancing
the factor $O(2^{k_i N_{ext,i}})$ to
$O(2^{k_i(N_{ext,i}-2)})$, with the same exponent
$N_{ext,i}-2$ as for local parts of polymers.

\Medskip   Next, each possible scale $k_i$ sector assignment for each of the $N_{ext,i}$ external legs must be redivided into $O(2^{j-k_i})$
scale $j$ subsectors.  {\em Assume once
again that all external momenta are contracted at scale $j$.} 
 Compared to the external sector
assignment factor for local parts (see previous paragraph), the
outcome is that {\em  polymers of type (ii)}   have  an extra (shared) relative factor $O(2^{j-k_i})$. For polymers of type (ii'), this factor is compensated by the spring factor $2^{-2(j-k_i)}$, so one may set $\overline{\cal N}_{sec,i}=1$ as for polymers of type (i). So let us concentrate on the case of polymers of type (ii), for which
$\overline{\cal N}_{sec,i}=2^{j-k_i}$ is precisely this
extra relative factor.
The factor $2^{j-k_i}$ is then compensated by the factor
$2^{-(j-k_i)\frac{N_{ext,i}-4}{2}}2^{-3k_i/2}$ 
found in the terms
between brackets $\big\{ \ \cdot\ \big\}$ in (\ref{eq:2.106})
or (\ref{eq:2.110}):
namely,
\BEQ 2^{-(j-k_i) \frac{N_{ext,i}-4}{2}} 2^{-3k_i/2}  \ \cdot\ 
2^{j-k_i} = 2^{-3k_i/2} (2^{-\half(j-k_i)})^{N_{ext,i}-6} \le 2^{-\half(j-k_i)} \, 2^{-j} 
\EEQ
if $N_{ext,i}\ge N_0$. The relative factor $2^{-\half(j-k_i)}$ makes it possible to sum over scales $k_i<j$.  The extra remaining factor,
$2^{-j}$ per polymer with $N_{ext}\ge N_0$ external legs, may be used to produce the required overall $2^{-j/2}$ factor
in (\ref{eq:GFMSB-P}), except, of course, if {\em all} polymers are local parts. 

\Medskip {\em Scale $j$ external legs.} The above arguments
double-count the external sector assignment for legs which
are {\em not} contracted at scale $j$, and therefore remain external legs of the polymer $\P^{j\to}$. Removing this
double-counting produces an extra sector assignment factor
$\le (2^{j/2})^{-N_{ext}}= 2^{-j}\, {\cal N}''_{sec} $ when scale $j$ momentum conservation does not hold globally, i.e. when 
there is at least one polymer with $\ge N_0$ external legs.
On the other hand, if {\em all} polymers $\P_i$ are local
parts, then momentum conservation for each of them implies
momentum conservation for $\P$, hence the extra sector
assignment factor is $\le (2^{j/2})^{-N_{ext}+2}={\cal N}''_{sec}$. Thus $\overline{\cal N}''_{sec}=2^{-j}$ in
 the first case, $1$ in the second case.

\Medskip {\em Case when all polymers are local parts.} 
In that case, all correcting factors $\overline{\cal N}_{sec,i}$ are equal to 1, so (\ref{eq:2.110bis}) holds when
all $N_{ext,i}=4$. Otherwise, assuming e.g. that
$N_{ext,1}\ge 6$, one has $2^{-(j-k_1)\frac{N_{ext,1}-4}{2}} 2^{-k_1/2}\le 2^{-\half(j-k_i)}\, 2^{-j/2}$, whence (\ref{eq:2.110}) holds.

\Bigskip {\bf Combinatorial factors.}
Compared to {\em single-scale} trees, {\em multi-scale} trees involve supplementary combinatorial factors. This is a well-known problem in constructive field theory, which has been
e.g. exposed in details and some generality for {\em bosonic
theories} in \cite{MagUnt2}. The fact that the present model
is {\em fermionic}, hence obeys Pauli's principle, leads to substantial simplifications.

\Medskip
{\em Accumulation of low-momentum fields} (see \cite{MagUnt2},
Remark below Definition 2.9). Let $\Del^k\in\D^k$, $j_D\le k\le j'_{\phi}-1$,  and $\Del^j_1,\ldots,\Del^j_n$, $n=2^{3(k-j)}$,
be the boxes of given scale $j\le k$  included in $\Del^k$. 
The scale $j$  $t$-derivatives   may have produced   up to $O(1)$ low-momentum fields $\Psi^{\to (j-1)}$ or $\bar{\Psi}^{\to(j-1)}$  in each box $\Del^j_i$.
Decomposing $\Psi^{\to(j-1)},\bar{\Psi}^{\to(j-1)}$ into scales, this means that
there are for
each sector $\alpha\in\Z/2^k\Z$ up to $N=O(n)$ fermion fields $\Psi^{k,\alpha},\bar{\Psi}^{k,\alpha}$ located
in a single box $\Del^k$. The inverse local factorial factor $\prod_{\alpha}  \frac{1}{(n_{\Del,\alpha}!)^r (\bar{n}_{\Del,\alpha}!)^r}$ of (\ref{eq:single-scale-bound}) makes it
possible to sum over all pairing possibilities.


\section{Low-energy theory for Cooper pairs}  \label{section:boson}


\Bigskip The picture emerging after the expansions of section
\ref{section:fermion} is that of {\em fermionic polymers}
connected by chains of bubbles evaluated as Pre$\Sigma$-kernels. The infra-red cut-off at scale $j'_{\phi}-1$ keeps
away infra-red singularities, and the $1/N$ expansions ensures
that the series of perturbations converges. This is  still
clearly one step behind the conjectured,  
divergent infra-red behavior in $O(1/|q|^2)$
of the Goldstone boson propagator (or $\Sigma_{\perp,\perp}-$ kernel), found in principle by means
of the gap equation as in \S \ref{subsection:Sigma}.  

\Bigskip We need in the first place a proper definition of
the Goldstone boson propagator that goes beyond perturbation theory. 
Instead of developing blindly  an infinite sum of individual  four-point diagrams,  and selecting
those which are two-particle irreducible by checking all possible ways of cutting it into two pieces, we implement a decision rule relying only on the number of links explicitly
constructed by hand between pairs of  polymers seen as
small, almost pointwise vertex insertions. This is the object
of the {\bf complementary cluster expansions} of
\S \ref{subsection:complementary}. Those are
horizontal cluster expansions,  between scale $j'_{\phi}$ fermions only,  obtained by
expanding  links {\em to third order}  between boxes of
pairs of polymers. The idea is that we exhaust explicitly in this way {\em all possible} configurations of the type

\bigskip

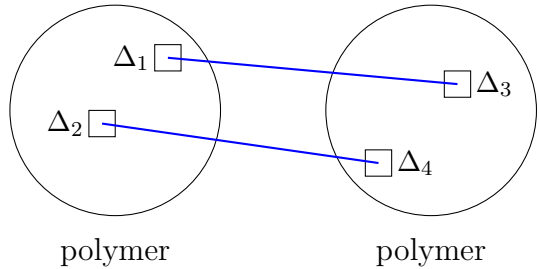
{\centerline{
\begin{tikzpicture}[scale=0.7]
\draw(0,0) circle(2);
\draw(0.75,0.75) rectangle(1.25,1.25); \draw(0.3,1) node {\small $\Del_1$};
\draw(-0.5,-0.5) rectangle(0,0); \draw(-0.95,-0.25) node
{\small $\Del_2$};
\begin{scope}[shift={(2,0)}]
\draw(4,0) circle(2);
\draw(4+0.25,0.25) rectangle (4+0.75,0.75);
\draw(4+0.5+0.7,0.5) node {\small $\Del_3$};
\draw(4-1.25,-1.25) rectangle (4-0.75,-0.75);
\draw(4-1+0.7,-1) node {\small $\Del_4$};
\end{scope}
\draw[thick,color=blue](1,1)--(6+0.5,0.5);
\draw[thick,color=blue](-0.25,-0.25)--(6-1,-1);
\draw(0,-2.7) node {polymer}; \draw(6,-2.7) node {polymer}; 
\end{tikzpicture}}}

\Bigskip where two polymers are connected by exactly two
links; displacing one of the blue fermion links so as to form a bubble as in
Fig. \ref{subsection:Sigma}.5, one obtains eventually  a rung connecting Bethe-Salpeter kernels. Such configurations contribute to the infra-red divergence of the Goldstone boson propagator, therefore the
two polymers are not {\em sufficiently} connected to be able
to sum over the location of one of them w.r. to that of the
other. On the other hand, {\em three} links suffice to
ensure {\bf strong connectivity} between these. 
 Building
upon this concept of strong connectivity, we develop multi-scale trees whose vertices are polymers, and edges --
 three-fold links. Because polymers with $\ge 6$ external
  legs are not diverging, emphasis is laid on four-point polymers.
Those are considered to have a bosonic scale $j_+\ge j'_{\phi}$
given by the scale of their transfer momentum $q$, which
may be determined to some precision by refining the sector
decomposition of the fermions.   

\Bigskip The outcome of these complementary expansions is 
a expression of the  kernel $\Pi(q)$ for $|q|\approx 2^{-j_+}\mu$  in terms of a sum over
 polymer configurations, 
 $n$-point functions are then rewritten in terms 
 of sums over simplified trees called {\bf bosonic trees},  whose vertices are still polymers, but edges are now
 {\bf single-scale Goldstone boson propagators} (not yet
 Goldstone boson propagators, see below) of
 given scales, see Fig. \ref{subsection:complementary}.14.

\Medskip However, even at this stage, we are not able to prove that
$n$-point functions are infra-red summable, because the theory is  a priori plagued with infra-red divergences coming from Goldstone boson propagators
with low transfer momentum,  connecting arbitrarily far-away fermionic polymers. This problem is  solved in \S
\ref{subsection:Ward} using {\em Ward identities} similar
to those used to solve the problem of "soft photons"
(infra-red photons) in
QED. First of all, multi-scale {\em two-point functions} of Cooper
pairs,  inductively extracted from bosonic trees (see
Fig. \ref{subsection:complementary}.14), have
no associated Ward identity, hence must be resummed by hand.
This is a simple geometric series, who sum  gives at last
the  {\bf Goldstone boson propagator} we were looking for. As a
consequence of these partial resummations, previous bosonic trees are replaced by very similar bosonic trees with the
same type of vertices, but whose
edges are Goldstone boson propagators, and that furthermore have
no internal two-point functions. 
We may now solve the gap equation by an easy fixed-point
argument,  and verify the predictions
 of \S \ref{subsection:Sigma} concerning the infra-red 
 behavior of the geometric series $\Sigma$.

\Bigskip In the large-scale limit, i.e. at space-time scales much
larger than $1/\Gamma_{\phi}$,  polymers of bosonic trees appear
in our multi-scale picture as tiny islands giving subleading
contributions to $n$-point functions. This makes plausible
the general structure of
the formula of Theorem 2,
\BEA && \Big\langle \prod_{i=1}^{2n}\ :\, (\bar{\Psi} {\mathbb{\Gamma}}^{\perp}\Psi)(\eta^{-1}\xi_i) \,:\ 
\Big\rangle_{\theta;\lambda} \sim_{\eta\overset{>}{\to} 0}  \nonumber\\
&&\qquad \lambda^{-2n} \sum_{(i_1,i_2),\ldots,(i_{2n-1},i_{2n})} 
\, \prod_{k=1}^n \ \Big( \, ^t {\bf{\Gamma}}^{\perp}
\Sigma{\bf{\Gamma}}^{\perp}
\Big)(\eta^{-1}(\xi_{i_{2k-1}}-\xi_{i_{2k}})).
\nonumber\\ \label{eq:3.2}
\EEA
 As apparent in (\ref{eq:3.2}), the effective theory is roughly
that of a {\em free bosonic particle} -- the {\em 
Goldstone boson} -- with covariance kernel $\Sigma$.  
This Theorem is proved in \S \ref{subsection:proof-Theorem2}. 

\vskip 1cm

{\centerline{************************************}}

\vskip 1cm

\noindent 
Let us now turn to another question. The  effective potential approach
described in the Introduction (Regime II) suggested that
the  {\em field} $\mathbb{\Gamma}(\xi)$ conjugate to
Cooper pairs should behave like a non-linear sigma-model valued
in the immediate neighborhood of the circle of radius $\Gamma_{\phi}$. Thus, it is 
in principle  possible to distinguish {\em transverse} (i.e.
orthogonal to the circle) fluctuations -- "slow" degrees
of freedom --  from {\em  tangential}
fluctuations -- "fast" degrees of freedom. 

\Medskip Here, by comparison, a preferential direction
$\theta$ has been fixed by the symmetry-breaking term. Thus
all there remains to be done is to check that with high
probability, the Cooper-pair field is to a good approximation aligned with this
direction, and has a modulus fixed by the gap equation. 
 This is the content of  Theorems 4 and 5
in the Introduction, proved resp. in \S \ref{subsection:phase-transition} and \S \ref{subsection:transverse-mass}.

\Bigskip The introductory, heuristic subsection \S \ref{subsection:heuristic}, recapitulates  the 
main findings of mean-field theory concerning the 
infra-red behavior of the theory. Classical results presented there without proof  are only half-quantitative since the Goldstone boson propagator has been replaced
by its leading order obtained by resumming Cooper pair bubbles, but notations introduced there are important for the
subsection on Ward identities.


\subsection{A heuristic introduction} \label{subsection:heuristic}


We present in this subsection a heuristic derivation of the
large-scale behavior of the theory in the Cooper channel, based 
on elementary considerations, and the fundamental assumption that
asymptotics may be obtained by considering only chains of Cooper pair bubble diagrams. Notations introduced here are fundamental for \S \ref{subsection:Ward} where Ward identities will be proved, justifying
the present heuristics beyond its pedagogical value. 

\Medskip Let us first write down the various quadratic fields built out of 
one $\bar{\Psi}$ and one $\Psi$:

\Medskip\textbullet\  {\em Cooper pairs}:
\BEA &&  \bar{\Psi}{\mathbb{\Gamma}}(\theta)\Psi=
\Gamma_{\phi} \bar{\Psi} \left(\begin{array}{cc} 0 & e^{-\II\theta} \\ e^{\II\theta} 
& 0 \end{array}\right) \Psi=\Gamma_{\phi}\Big(e^{-\II\theta} \bar{\psi}_{\uparrow} \bar{\psi}_{\downarrow} + e^{\II\theta} \psi_{\downarrow}\psi_{\uparrow} \Big) 
; \EEA
dividing by $\Gamma_{\phi}$, we get a different possible normalization, 
$\bar{\Psi}\sigma(\theta)\Psi=\bar{\Psi} \left(\begin{array}{cc} 0 & e^{-\II\theta} \\ e^{\II\theta} 
& 0 \end{array}\right) \Psi$, where
\BEQ \sigma(\theta)\equiv \frac{1}{\Gamma_{\phi}}{\mathbb{\Gamma}}(\theta)=\cos\theta \ \sigma^1+
\sin\theta\  \sigma^2.\EEQ

\Medskip\textbullet\  {\em normal pairs}:
\BEQ \bar{\Psi}\sigma^3 \Psi=\bar{\psi}_{\uparrow} \psi_{\uparrow} + 
\bar{\psi}_{\downarrow} \psi_{\downarrow}; 
\EEQ

\Medskip\textbullet\  {\em identity pairings}:
\BEQ \bar{\Psi} {\mathbb{1}} \Psi=\bar{\psi}_{\uparrow} \psi_{\uparrow} -
\bar{\psi}_{\downarrow} \psi_{\downarrow}.\EEQ

\Bigskip Let us now translate  expressions (\ref{eq:PreSigma-asymptotics-perp},\ref{eq:PreSigma-asymptotics-//})
 for $\PreSigma(q)$ in a  new base: denoting by $//$ the direction
 parallel to $\theta$, and $\perp$ the direction rotated by
 an angle $\pi/2$, so that 
 \BEQ \Gamma^{//}\equiv \Gamma^{//}(\theta)= \Gamma(\theta),
 \qquad  \Gamma^{\perp}\equiv \Gamma^{\perp}(\theta)=\Gamma(\theta+\frac{\pi}{2}), \label{eq:Gamma//Gammaperp}
 \EEQ
  we consider the
 Pre$\Sigma$-kernel in the $(//,\perp)$ basis:
\BEQ \left(\begin{array}{cc} \PreSigma_{//,//} & \PreSigma_{//,\perp} \\ \PreSigma_{\perp,\//} & \PreSigma_{\perp,\perp} \end{array}\right).
\EEQ
When $\theta=0$, one gets $\bar{\Psi}\Gamma^{//}\Psi=
\Gamma_{\phi}(\bar{\psi}_{\uparrow}\bar{\psi}_{\downarrow}+
\psi_{\downarrow}\psi_{\uparrow})$, $\bar{\Psi}\Gamma^{\perp}\Psi=-\II
\Gamma_{\phi}(\bar{\psi}_{\uparrow}\bar{\psi}_{\downarrow}-
\psi_{\downarrow}\psi_{\uparrow})$ and
\BEA  \PreSigma_{//,//}&=&\PreSigma_{(\bar{\psi}_{\uparrow}\bar{\psi}_{\downarrow}+\psi_{\downarrow}\psi_{\uparrow})\otimes (\bar{\psi}_{\uparrow}\bar{\psi}_{\downarrow}+\psi_{\downarrow}\psi_{\uparrow})} \nonumber\\
&:=& \PreSigma_{\bar{\psi}_{\uparrow}\bar{\psi}_{\downarrow},\bar{\psi}_{\uparrow}\bar{\psi}_{\downarrow}}+
\PreSigma_{\bar{\psi}_{\uparrow}\bar{\psi}_{\downarrow},{\psi}_{\downarrow}{\psi}_{\uparrow}} +
\PreSigma_{\psi_{\downarrow}\psi_{\uparrow},\bar{\psi}_{\uparrow}\bar{\psi}_{\downarrow}} +
\PreSigma_{\psi_{\downarrow}\psi_{\uparrow},\psi_{\downarrow}\psi_{\uparrow}}  \nonumber\\  \label{eq:Sigma-//-//}
\EEA
and similarly
\BEQ \PreSigma_{//,\perp}=-\II\PreSigma_{(\bar{\psi}_{\uparrow}\bar{\psi}_{\downarrow}+\psi_{\downarrow}\psi_{\uparrow})\otimes (\bar{\psi}_{\uparrow}\bar{\psi}_{\downarrow}-\psi_{\downarrow}\psi_{\uparrow})}, \qquad 
\PreSigma_{\perp,//}=-\II\PreSigma_{(\bar{\psi}_{\uparrow}\bar{\psi}_{\downarrow}-\psi_{\downarrow}\psi_{\uparrow})\otimes (\bar{\psi}_{\uparrow}\bar{\psi}_{\downarrow}+\psi_{\downarrow}\psi_{\uparrow})}
 \EEQ

\BEQ \PreSigma_{\perp,\perp}=- \PreSigma_{(\bar{\psi}_{\uparrow}\bar{\psi}_{\downarrow}-\psi_{\downarrow}\psi_{\uparrow})\otimes (\bar{\psi}_{\uparrow}\bar{\psi}_{\downarrow}-\psi_{\downarrow}\psi_{\uparrow})}.
\EEQ

\Medskip Using rotation equivariance, it is easy to prove that
 \BEQ ^t {\bf\Gamma}^{//} \PreSigma {\bf\Gamma}^{//}=\Gamma_{\phi}^2\PreSigma_{//,//}, 
 ^t {\bf\Gamma}^{//} \PreSigma {\bf\Gamma}^{\perp}=\Gamma_{\phi}^2\PreSigma_{//,\perp},
 \nonumber
 \EEQ
 \BEQ
 ^t {\bf\Gamma}^{\perp} \PreSigma {\bf\Gamma}^{//}=\Gamma_{\phi}^2\PreSigma_{\perp,//},
 ^t {\bf\Gamma}^{\perp} \PreSigma {\bf\Gamma}^{\perp}=\Gamma_{\phi}^2\PreSigma_{\perp,\perp}.
 \label{eq:GammaSigmaGamma}
 \EEQ 
Hermitian symmetry implies that $\PreSigma$ is diagonal
in the $(//,\perp)$-basis. Now, (\ref{eq:PreSigma-asymptotics-//},\ref{eq:PreSigma-asymptotics-perp})
implies that
\BEQ  \PreSigma_{\perp,\perp}(q)\sim \frac{g^0_{\phi}}{(|q|_+^0)^2}, \qquad \PreSigma_{//,//}(q)\approx \frac{1}{m^*}.\EEQ

\Bigskip The kernel $\PreSigma_{//,//}(q)\approx \frac{1}{m^*}
\approx \frac{1}{m^*} \frac{\Gamma_{\phi}^2}{|q|_+^2+\Gamma_{\phi}^2}\approx\frac{g_{\phi}^0}{|q|_+^2+\Gamma_{\phi}^2}$ has a characteristic massive decay for 
$|q|_+\lesssim \Gamma_{\phi}$, with mass $\approx \Gamma_{\phi}$. Hence the  main term in the  large-scale behavior of  $N$-point functions of Cooper pairs is obtained by considering
only bound states  $\bar{\Psi}{\mathbb{\Gamma}}^{\perp}\Psi$
propagating through space-time through chains of  Cooper space 
bubble diagrams

\medskip

{\centerline{$\boundstate$}}

\noindent {\bf\tiny \ref{fig:boundstate}.1. Bound state propagation.}

\Bigskip
which is simply the Pre$\Sigma$-kernel
\begin{tikzpicture}
\draw[dotted,decorate,decoration=snake](0,0)--(2,0);
 \end{tikzpicture}
in the perpendicular direction.

\Medskip If $\theta',\theta''\not=\theta$, the leading-order term of the
function\\  $\Big\langle \ :\, (\bar{\Psi}{\mathbb{\Gamma}}(\theta')\Psi)(\xi)\, :  \ \   :\,
(\bar{\Psi}{\mathbb{\Gamma}}(\theta'')\Psi)(\xi')\, :  \ \Big\rangle_{\theta;\lambda}$ is  (letting $\sigma^{\perp}:=\sigma(\theta+\frac{\pi}{2})$)

\medskip
{\centerline{$\twopointfunction$}
\medskip
\noindent {\bf\tiny Fig. \ref{fig:twopointfunction}.2. Two-point function.}

\Bigskip evaluated as the inverse Fourier transform of   

\BEA && q\mapsto \Tr({\mathbb{\Gamma}}(\theta') \sigma(\theta) {\mathbb{\Gamma}}(\theta''))  ({\cal A}_q)_{//,//} 
+  \Tr({\mathbb{\Gamma}}(\theta') \sigma(\theta+\frac{\pi}{2}) {\mathbb{\Gamma}}(\theta''))  ({\cal A}_q)_{\perp,\perp}
\nonumber\\
&&\qquad 
+ \Tr({\mathbb{\Gamma}}(\theta')\sigma^{\perp})\Tr(\sigma^{\perp}{\mathbb{\Gamma}}(\theta''))
 ({\cal A}_q)_{\perp,\perp} \PreSigma(q)  ( {\cal A}_q)_{\perp,\perp}
 \nonumber\\
 &&\qquad \sim \Big\{\Tr({\mathbb{\Gamma}}(\theta'){\mathbb{\Gamma}}(\theta''))-\Tr({\mathbb{\Gamma}}(\theta')\sigma^{\perp}) -\Tr({\mathbb{\Gamma}}(\theta')\sigma^{\perp})
 \Tr(\sigma^{\perp}{\mathbb{\Gamma}}(\theta'')) \Big\}  ({\cal A}_q)_{\perp,\perp} \nonumber\\
 &&\qquad\qquad
+ \Tr({\mathbb{\Gamma}}(\theta')\sigma^{\perp}) \Tr(\sigma^{\perp}{\mathbb{\Gamma}}(\theta'')) ({\cal A}_q)_{\perp,\perp}
 \Big(1+\PreSigma(q)  ({\cal A}_q)_{\perp,\perp} \Big)  \nonumber\\
 &&=\Big\{\Tr({\mathbb{\Gamma}}(\theta'){\mathbb{\Gamma}}(\theta''))-\Tr({\mathbb{\Gamma}}(\theta')\sigma^{\perp})
 \Tr(\sigma^{\perp}{\mathbb{\Gamma}}(\theta'')) \Big\}
 ({\cal A}_q)_{\perp,\perp} \nonumber\\
 &&\qquad\qquad + \frac{1}{\lambda}
 \Tr({\mathbb{\Gamma}}(\theta')\sigma^{\perp})
 \Tr(\sigma^{\perp}{\mathbb{\Gamma}}(\theta'')) ({\cal A}_q)_{\perp,\perp} \PreSigma(q).
\nonumber\\
 \EEA
where one has neglected the difference $({\cal A}_q)_{//,//}-({\cal A}_q)_{\perp,\perp}$ in the term without $\PreSigma$ kernel.
Since $({\cal A}_q)_{//,//},({\cal A}_q)_{\perp,\perp}\sim_{q\to 0} \frac{1}{\lambda}$ are not singular in the infra-red
limit, this
gives to leading order in $g$ if $\theta',\theta''\not=\theta$
\BEQ  \boxed{\Big\langle  : (\bar{\Psi}{\mathbb{\Gamma}}(\theta')\Psi)(\xi) : \  \ :
(\bar{\Psi}{\mathbb{\Gamma}}(\theta'')\Psi)(\xi')   :  \Big\rangle_{\theta;\lambda} \approx_{|\xi-\xi'|_+\to \infty}  \Tr({\mathbb{\Gamma}}(\theta')\sigma^{\perp}) \ \Tr(\sigma^{\perp} {\mathbb{\Gamma}}(\theta''))\frac{1}{\lambda^2} \PreSigma(\xi-\xi'). } \label{eq:expected} 
\EEQ

Graphically,  

{\centerline{
\begin{tikzpicture} 
\draw(-5,0) node {$\Big\langle \ :\, (\bar{\Psi}{\mathbb{\Gamma}}(\theta')\Psi)(\xi) \,: \ \ :\,  
(\bar{\Psi}{\mathbb{\Gamma}}(\theta'')\Psi)(\xi')\, :  \ \Big\rangle_{\theta;\lambda} \approx$};
\draw[dotted,decorate,decoration=snake](0,0)--(2,0);
\draw(0,0) node {\textbullet}; \draw(2,0) node {\textbullet}; \draw(-0.5,0) node {$\xi$}; \draw(2.5,0) node {$\xi'$}; \draw(0,0.5) node {$\frac{\Gamma}{\lambda}$};
\draw(2,0.5) node {$\frac{\Gamma}{\lambda}$};
\end{tikzpicture}
}}

Generalizing, one obtains a semi-perturbative version of
Theorem 2:  the leading term of $\Big\langle \prod_{i=1}^{2n}  \, :\, (\bar{\Psi}{\mathbb{\Gamma}}^{\perp}\Psi)(\xi_i) \, : \ 
 \ \Big\rangle_{\theta;\lambda}$ for widely separated 
 points $\xi_1,\ldots,\xi_{2n}$ is graphically (taking e.g.
 $n=3$)
 
 \Bigskip
 
 {\centerline{
 \begin{tikzpicture}
 \draw(0,0.5) node {$\frac{\Gamma}{\lambda}$}; 
  \draw(2,0.5) node {$\frac{\Gamma}{\lambda}$}; 
   \draw(-1,-1.5) node {$\frac{\Gamma}{\lambda}$}; 
    \draw(3,-1.5) node {$\frac{\Gamma}{\lambda}$}; 
     \draw(0,-3.5) node {$\frac{\Gamma}{\lambda}$}; 
      \draw(2,-3.5) node {$\frac{\Gamma}{\lambda}$}; 
 \draw(0,0) node {\textbullet}; \draw(2,0) node {\textbullet};
 \draw(-1,-2) node {\textbullet}; \draw(3,-2) node {\textbullet};
 \draw(0,-4) node {\textbullet}; \draw(2,-4) node {\textbullet};
 \draw(-0.5,0) node {$\xi_1$}; \draw(2.5,0) node {$\xi_4$};
 \draw(-1.5,-2) node {$\xi_2$}; \draw(3.5,-2) node {$\xi_5$};
 \draw(-0.5,-4) node {$\xi_3$}; \draw(2.5,-4) node {$\xi_6$};
 \draw[dotted,decorate,decoration=snake] (0,0)--(2,0);
 \draw[dotted,decorate,decoration=snake] (-1,-2)--(3,-2);
 \draw[dotted,decorate,decoration=snake] (0,-4)--(2,-4);
 \draw(8,-2) node {+ perm.};
 \end{tikzpicture}
 }}
 
\Medskip {\bf \tiny Fig. \thesubsection.2. $6$-point function
of Cooper pairs.}

\Bigskip where "perm." indicates the sum over all possible pairings. 

\Medskip Let us now consider the one-point function in 
the parallel direction $(//)$. Main term is

\Bigskip

{\centerline{\begin{tikzpicture}
 \draw(-1.5,0) node {$\langle (\bar{\Psi}\sigma(\theta)\Psi)(\xi)
 \rangle_{\theta;\lambda}=$};
 \draw(1.5,0) node {$\Tr((\sigma(\theta))^2)$};
\draw(3,0) node {\textbullet}; 
\draw(3,0) arc(150:30:1.2 and 0.6); 
\draw(3,0) arc(210:330:1.2 and 0.6);
\draw(6.3,0) node {$\times\, \Gamma \qquad \approx 2 \frac{\Gamma_{\phi}}{\lambda}$};
\end{tikzpicture}}}
where the coefficient 2 comes from the trace $\Tr((\sigma(\theta))^2)$.
The gap equation, see Fig. \ref{subsection:Sigma}.6,
implies that the correction due to  the Goldstone boson

\begin{tikzpicture}[scale=1]
\draw(2+0.5-1.5,-0.9) node {\textbullet};
\draw(2+2-1.5,-0.9) arc(30:150:0.88  and 0.44);
\draw(2+0.5-1.5,-0.9) arc(210:330:0.88 and 0.44);
\draw(1-0.6,-0.9) node {$\xi$};
\draw(2+0.2,-0.9) node {\tiny $\sigma(\theta)$};

\draw(2+2,-0.9) arc(30:150:0.88  and 0.44);
\draw(2+0.5,-0.9) arc(210:330:0.88 and 0.44);
\draw(2+0.8,-0.9) node {\tiny $\sigma(\theta)$}; \draw(2+1.6,-0.9) node {\tiny $\sigma(\theta)$};

\draw(2+2+1.5,-0.9) arc(30:150:0.88  and 0.44);
\draw(2+0.5+1.5,-0.9) arc(210:330:0.88 and 0.44);
\draw(2+0.8+1.5,-0.9) node {\tiny $\sigma(\theta)$}; \draw(2+1.6+1.5,-0.9) node {\tiny $\sigma(\theta)$};

\draw(2+3.9,-0.9) node {$..........$};
\draw(2+2+3.8,-0.9) arc(30:150:0.88  and 0.44);
\draw(2+0.5+3.8,-0.9) arc(210:330:0.88 and 0.44);
\draw(2+0.8+3.8,-0.9) node {\tiny $\sigma(\theta)$}; \draw(2+1.6+3.8,-0.9) node {\tiny $\sigma(\theta)$};

\draw(2+6.3,-0.9) node {\tiny $\sigma(\theta)$};

\draw(2+2+5.3,-0.9) arc(30:150:0.88  and 0.44);
\draw(2+0.5+5.3,-0.9) arc(210:330:0.88 and 0.44);

\draw(9.5,-0.9) arc(180:150:1.5 and 1.5);
\draw(9.5,-0.9) arc(180:210:1.5 and 1.5);
\draw(10,-0.9) node {$-$};
\draw[fill=black](10.4,-1) rectangle(10.6,-0.8); \draw(10.6,-0.5) node{$\Gamma$};
\draw(11.5,-0.9) node {$+$};
\draw(6+2+2+3.8,-0.9) arc(30:150:0.88  and 0.44);
\draw(6+2+0.5+3.8,-0.9) arc(210:330:0.88 and 0.44);
\draw(6+2+0.5+3.8,-0.9) node {\textbullet}; \draw(6+2+2+3.8,-0.9) node {\textbullet}; 
\draw(13.5,-0.9) node {\small ${\mathbb{\Gamma}}^{\perp}$};

\draw(15,-0.9) arc(0:30:1.5 and 1.5);
\draw(15,-0.9) arc(0:-30:1.5 and 1.5);

\end{tikzpicture}}

\Bigskip vanishes identically.

\Bigskip Let us finally mention another identity,
\BEQ \boxed{\langle (\bar{\Psi}{\mathbb{1}}\Psi)(\xi)\rangle_{\theta;\lambda}=0,} \label{eq:W1}
\EEQ
obvious by the symmetry exchanging $\uparrow$- and $\downarrow$-states.



\subsection{Complementary expansion}
\label{subsection:complementary}


After these heuristics and general power-counting considerations,  let us now come back to where we left the story. At the end of section \ref{section:fermion}, the situation was as
follows: $n$-point functions
have been rewritten as a sum over polymers $\P_1,\ldots,\P_p\in {\cal P}^{\to(j'_{\phi}-1)}$
with lowest scale $\le j'_{\phi}-1$. Because two-point functions have been renormalized
at scale $j'_{\phi}-1$, and vacuum polymers have been divided out using Mayer's expansion,
there remain only polymers with $4$ external legs, and polymers with $\ge 6$ (not fully produced) external legs.  Since there is only one scale left, all external legs are
of scale $j'_{\phi}$. The measure has retained an exponential weight $e^{-\sum_{\Del}
\int d\xi\, \chi_{\Del}(\xi) {\cal L}_{\theta}^{j'_{\phi}}(\xi)}$, where $\Del\in\D^{j'_{\phi}}$ ranges over
the set of scale $j'_{\phi}$ boxes -- thereafter called {\bf isolated boxes} -- that are not connected by a vertical link to boxes of
higher scale $\le j'_{\phi}-1$, and  
\BEQ {\cal L}_{\theta}^{j'_{\phi}}(\xi):={\cal L}_{\theta}^{\to j'_{\phi}}(\vec{t}=0;
\xi)=\lambda (\bar{\Psi}^{j'_{\phi}}\Psi^{j'_{\phi}})^2(\xi)+ \bar{\Psi}^{j'_{\phi}}(\xi) \, \big(\del Z^{j'_{\phi}} \partial_{\tau}+ \del{\mathbb{\Gamma}}^{j'_{\phi}}
\big)\, 
\Psi^{j'_{\phi}}(\xi), \label{eq:Ljphi}
\EEQ
\noindent compare with (\ref{eq:Lthetatoj}) and (\ref{eq:delGammajphi}), since by hypothesis
 $\del\mu^{j'_{\phi}}=0,\del m^{j'_{\phi}}=0$.  Recall that $\del{\mathbb{\Gamma}}^{j'_{\phi}}=O(g\Gamma_{\phi})$ has an extra $O(g)$ pre-factor. As apparent from \S
\ref{subsection:Sigma}, a particular r\^ole will be played by what are called
{\bf polymers of type 2} in the following

\begin{Definition}[polymers of type 1 and type 2]
\label{def:type1type2}
Let $\P\in{\cal P}^{(j'_{\phi}-1)\to}$ be a four-point polymer
with scale $j'_{\phi}$ external legs in angular sectors 
$\alpha_1,\ldots,\alpha_4\in\Z/2^{j'_{\phi}}\Z$. Then  $\P$ is said to be of {\bf type 2}
if all the following conditions are satisfied:

\begin{itemize}
\item[(i)] $N_{ext}=4$;
\item[(ii)] $|\alpha_1+\alpha_2|,|\alpha_3+\alpha_4|\le 1$. 
\end{itemize}
Its  external structure is 
either non-mixing, $(\bar{\psi}^{j'_{\phi},\alpha_1}_{\uparrow}
\bar{\psi}^{j'_{\phi},\alpha_2}_{\downarrow})\otimes(\psi^{j'_{\phi},\alpha_3}_{\downarrow}\psi^{j'_{\phi},\alpha_4}_{\uparrow})$, or mixing,
$(\bar{\psi}^{j'_{\phi},\alpha_1}_{\uparrow}
\bar{\psi}^{j'_{\phi},\alpha_2}_{\downarrow})\otimes(\bar{\psi}^{j'_{\phi},\alpha_3}_{\uparrow}\bar{\psi}^{j'_{\phi},\alpha_4}_{\downarrow})$ or 
$({\psi}^{j'_{\phi},\alpha_1}_{\downarrow}
{\psi}^{j'_{\phi},\alpha_2}_{\uparrow})\otimes(\psi^{j'_{\phi},\alpha_3}_{\downarrow}\psi^{j'_{\phi},\alpha_4}_{\uparrow})$.
Furthermore, if the polymer is mixing, we require that
\begin{itemize}
\item[(iii)] $|\alpha_1+\alpha_4|,|\alpha_2+\alpha_3|>1$.
\end{itemize}

All other polymers $\P$ in ${\cal P}^{(j'_{\phi}-1)\to}$,
and also isolated scale $j'_{\phi}$ boxes, are said to be  of {\bf type 1}. 
\end{Definition}

{\em Polymers of type 2 are exactly those which make part of the Bethe-Salpeter kernel, 
accounting for the large-scale behavior of the theory.}
Roughly speaking: $\P$ is a polymer of type 2 if it is a
four-point polymer with transfer momentum $q=p_1+p_2=p_3+p_4$
such that $|q|\lesssim 2^{-j'_{\phi}}\mu$.  Note that 
conditions (ii), (iii) have some degree of arbitrariness.
For instance, because 
angular sectors do not have sharp cut-offs, one may
have $|\alpha_1+\alpha_2|\le 1$ but $|\alpha_3+\alpha_4|>1$,
 or vice-versa. Also, in the case of a mixing diagram, say with external
 structure $(\bar{\psi}^{j'_{\phi},\alpha_1}_{\uparrow}
\bar{\psi}^{j'_{\phi},\alpha_2}_{\downarrow})\otimes(\bar{\psi}^{j'_{\phi},\alpha_3}_{\uparrow}\bar{\psi}^{j'_{\phi},\alpha_4}_{\downarrow})$, one may exchange $\alpha_2$ and $\alpha_4$. 
Condition (iii) ensures that our criterion is unambiguous:
 there can be only {\em one} numbering of external legs
 such that $\P$ is of type 2. As we shall see later on,
 such subtleties are irrelevant in the infra-red limit.

\Bigskip 
{\bf A. General introduction.} 
The general philosophy in this subsection is to perform a
partial, carefully devised separation of {\em fermionic} degrees
of freedom from  {\em bosonic} degrees of freedom. 
In theory, this means the following. Imagine the interaction
$e^{-{\cal L}_{\theta}}$ has been wholly expanded into series, after which 
Wick's formula has turned an $n$-point function into an (infinite) sum of diagrams. Then sequences 
\bigskip

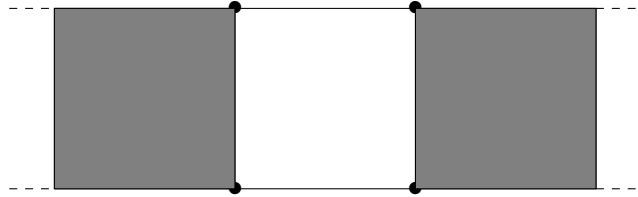
{\centerline{\begin{tikzpicture}[scale=1.2]
\draw[dashed](-0.5,1)--(0,1);\draw[dashed](-0.5,-1)--(0,-1);
\draw(0,1)--(2,1); \draw(0,-1)--(2,-1);
\draw(0,1)--(0,-1); \draw(2,1)--(2,-1);
\draw(0,0.75)--(2,0.75); \draw(0,0.5)--(2,0.5); \draw(0,0.25)--(2,0.25); \draw(0,0)--(2,0); 
\draw(0,-0.75)--(2,-0.75); \draw(0,-0.5)--(2,-0.5); \draw(0,-0.25)--(2,-0.25); 
\begin{scope}[shift={(4,0)}]
\draw(0,1)--(2,1); \draw(0,-1)--(2,-1);
\draw(0,1)--(0,-1); \draw(2,1)--(2,-1);
\draw(0,0.75)--(2,0.75); \draw(0,0.5)--(2,0.5); \draw(0,0.25)--(2,0.25); \draw(0,0)--(2,0); 
\draw(0,-0.75)--(2,-0.75); \draw(0,-0.5)--(2,-0.5); \draw(0,-0.25)--(2,-0.25);
\end{scope}
 \draw(2,1) node {$\bullet$};\draw(2,-1) node {$\bullet$};
  \draw(4,1) node {$\bullet$};\draw(4,-1) node {$\bullet$};
\draw(2,1)--(4,1);
\draw(2,-1)--(4,-1);
\draw[dashed](6,1)--(6.5,1);\draw[dashed](6,-1)--(6.5,-1);
\draw[fill=gray](0,-1) rectangle(2,1);
\draw[fill=gray](4,-1) rectangle(6,1);
\end{tikzpicture}}}

\Bigskip
see Fig. \ref{fig:formal-expansion-Bethe-Salpeter}.4, of 
two-particle irreducible,
four-point diagrams with Cooper pair external structure connected  by two fermionic lines, should be resummed into a Goldstone boson propagator, plus
error terms due to displacements and projections, as in
\S \ref{subsection:Sigma}. Then diagrams would be bound
as in \S \ref{subsection:cluster} by following fermionic
loops.

\Bigskip Things are however not so simple, because of
the well-known problem of accumulation of low-momentum
fields characteristic of non-massive bosonic theories
 (see e.g. \cite{MagUnt2}). Let us give a sketchy
 description of this problem, assuming for simplicity that
 $j'_{\phi}=j_{\phi}$.  
Suppose  that one has produced one Cooper pair with
transfer momentum $|q|\approx 2^{-j_+}\mu$ ($j_+>j_{\phi}$) 
per scale $j_{\phi}$ box $\Del$ included in a large 
scale $j_+$ box $\Del^{j_+}$; in total, $n= 2^{3(j_+-j_{\phi})}$ fields Cooper pairs. Then the (supposedly asymptotically Gaussian) contribution of all possible $\Sigma$-pairings of Cooper pairs through Wick's formula would lead to a combinatorial
factor $\approx \Gamma(n/2)$, in other words, $O( (2^{j_+-j_{\phi}})^{3/2})$ per Cooper pair. This is more than can be
compensated by the weight due to the products of the 
$\Sigma$-kernels, which is (leaving out dimensional constants) roughly $(\Sigma(|\xi-\xi'|\approx 2^{-j_+}\mu))^{n/2}
\approx (g_{\phi}/|\xi-\xi'|_+)^{n/2}\approx 2^{-j_{\phi}n/2} (2^{-\half(j_+-j_{\phi})})^{n}$. Taking further
into account small factors due to Ward identities, see
\S \ref{subsection:Ward}, would contribute in the best case
$O(2^{-(j_+-j_{\phi})})$ per Cooper pair: a global $O(1)$ 
counting, not sufficient though to sum over all scales
$j_+>j_{\phi}$. 

\Bigskip Strategies often developed to deal with this
problem in equilibrium statistical physics, relying mainly on large-deviation estimates (see
\cite{MagUnt2}), are difficult to implement here because
the Goldstone boson has not been introduced in the first place.
Instead we choose to expand {\em some} and not {\em all} Goldstone boson propagators, in such a way as to form a multi-scale
forest, where scales refer to the bosonic scales of transfer
momenta. As briefly mentioned  in the introduction to \S \ref{subsection:cluster}, we 
develop links between {\em objects}, which are  polymers
$\P_1,\ldots,\P_{n}$ of lowest
fermionic scale $\le j'_{\phi}$, {\em with external legs
of scale $j'_{\phi}$}. 
Fix some bosonic scale $j_+\ge j'_{\phi}$. At that scale, we must still
deal with potentially diverging four-point functions with
transfer momentum of scale $j_+$.   In order to read easily
the transfer momentum of a Cooper pair $\bar{\psi}^{j'_{\phi}}_{\uparrow} \bar{\psi}^{j'_{\phi}}_{\downarrow}$ or $\psi^{j'_{\phi}}_{\downarrow}
\psi^{j'_{\phi}}_{\uparrow}$, we decompose $\psi^{j'_{\phi}},\bar{\psi}^{j'_{\phi}}$ into
$2^{j_+}$ angular sectors ${\cal S}^{\alpha}\equiv {\cal S}^{j'_{\phi},\alpha}$, following (\ref{eq:further-angular}), $\psi^{j'_{\phi}}\equiv \sum_{\alpha\in \Z/2^{j_+}\Z}
\psi^{j'_{\phi},\alpha}$, $\bar{\psi}^{j'_{\phi}}\equiv \sum_{\alpha\in \Z/2^{j_+}\Z}
\bar{\psi}^{j'_{\phi},\alpha}$. Then the total momentum of
a pair $\bar{\psi}^{j'_{\phi},\alpha_1}_{\uparrow}
\bar{\psi}^{j'_{\phi},\alpha_2}_{\downarrow}$ or 
$\psi^{j'_{\phi},\alpha_1}_{\downarrow}\psi^{j'_{\phi},\alpha_2}_{\uparrow}$ is $\lesssim 2^{-j_+}\mu$ if and only if $|\alpha_1+\alpha_2|\lesssim 1$. 
Apart if $j_+=j'_{\phi}$ (in which case we also want to
deal with the remaining interaction ${\cal L}_{\theta}^{j'_{\phi}}$),  we mainly want to understand how polymers of type 2
are connected at scale $j_+$. Compared to the fermionic
cluster expansion of 
\S \ref{subsection:cluster}, things are more involved,
because expansion rules are strongly dependent on the
scales of  transfer momenta. Furthermore, we shall also need to
resum (by inspection) two-particle irreducible four-point  diagrams with
Cooper pair external structure, so as to produce the
Goldstone boson propagator $\Sigma$.

\Medskip Extending Definition \ref{def:type1type2}, we may say that a four-point 
polymer with external structure 
\BEQ (\bar{\psi}^{j'_{\phi},\alpha_1}_{\uparrow} \bar{\psi}^{j'_{\phi},\alpha_2}_{\downarrow})\otimes(\psi^{j'_{\phi},\alpha_3}_{\downarrow}
\psi^{j'_{\phi},\alpha_4}_{\uparrow}), \ (\bar{\psi}^{j'_{\phi},\alpha_1}_{\uparrow} \bar{\psi}^{j'_{\phi},\alpha_2}_{\downarrow})\otimes(\bar{\psi}^{j'_{\phi},\alpha_3}_{\uparrow}
\bar{\psi}^{j'_{\phi},\alpha_4}_{\downarrow})\ {\mathrm{or}}\  ({\psi}^{j'_{\phi},\alpha_1}_{\downarrow} \psi^{j'_{\phi},\alpha_2}_{\uparrow})\otimes(\psi^{j'_{\phi},\alpha_3}_{\downarrow}
\psi^{j'_{\phi},\alpha_4}_{\uparrow})
\label{eq:psipsipsipsi}
\EEQ
 ($\alpha_1,\ldots,\alpha_4\in\Z/2^{j_+}\Z$) 
is a  {\em scale $j_+$ polymer of type 2} provided $|\alpha_1+\alpha_2|,
|\alpha_3+\alpha_4|\le 1$, and (if the polymer is mixing)
$|\alpha_1+\alpha_4|,|\alpha_2+\alpha_3|> 1$, so that, in particular, the transfer momentum has scale $\ge j_+$ . However, as we shall see, this notion is
essentially redundant as soon as $j_+>j'_{\phi}$, because four-point polymers not satisfying these conditions have already
been taken care of at some higher scale $<j_+$.

\Medskip {\em As a general principle,  polymers will be of two types:}

-- {\bf scale $j_+$ four-point polymers} ($j_+\ge j'_{\phi}$),
generically  called {\bf low transfer momentum four-point polymers},  whose  transfer momentum is $|q|\approx 2^{-j_+}\mu$; 

-- and  {\bf "scale-neutral" polymers} -- that is, all
other types of polymers --, whose external leg
transfer momenta we do not need to determine, because they 
cannot by themselves produce divergences in the infra-red
limit.

\Medskip Particularizing, we have {\bf irreducible polymers},

\bigskip

\begin{tikzpicture}[scale=0.6]
\draw[color=blue](0,0)--(25,0);
\draw[color=blue](0,-0.4)--(25,-0.4);
\draw(22,2) node{scale-neutral};

\draw(2,2) circle(1);  \draw(2,2) node{$4$};
\draw(1.5,1.134)--(1,-1); \draw(1.8,1.05)--(1.65,-1);
\draw(2.2,1.05)--(2.35,-1);
\draw(2.5,1.134)--(3,-1);

\begin{scope}[shift={(5,0)}]
\draw(2,2) circle(1);  \draw(2,2) node{$\ge 6$};
\draw(1.25,1.25)--(0.6,-1);
\draw(1.5,1.134)--(1,-1); \draw(1.8,1.05)--(1.65,-1);
\draw(2.2,1.05)--(2.35,-1);
\draw(2.5,1.134)--(3,-1);
\draw(2.75,1.25)--(3.4,-1);
\draw[dashed](2,1)--(2,-1);
\end{scope}

\begin{scope}[shift={(10,0)}]
\draw(0.5,-1)--(1,2); \draw(1.5,-1)--(1,2);
\draw[decorate,decoration=snake,dashed](1,2)--(3,2);
\draw(2.5,-1)--(3,2);\draw(3.5,-1)--(3,2);
\end{scope}

\begin{scope}[shift={(15,0)}]
\draw(2,2) circle(1); \draw(2,2) node{$2$};
\draw(1,2)--(0,2); \draw(4,2)--(3,2);
\end{scope}

\draw(22,-1.15) node {$j'_{\phi}$};
\draw[color=blue](0,-2)--(25,-2);
\draw(22,-3) node {$\vdots$};
\draw[color=blue](0,-4)--(25,-4);
\draw(22,-6) node {$j_+$};

\draw(2,-6) circle(1);  \draw[dotted](2,-5)--(2,-7);
\draw(3,-6)--(3.5,-5.5); \draw(3,-6)--(3.5,-6.5);
\draw(1,-6)--(0.5,-5.5);\draw(1,-6)--(0.5,-6.5);


\end{tikzpicture}

\Medskip {\bf \tiny Fig. \thesubsection.1. Irreducible
polymers.}

\Bigskip Polymers are represented by {\em blobs}.

\Medskip Figures at the center of {\em scale-neutral polymers} indicate the number of external legs. Polymers with 2 scale $j'_{\phi}$ external legs contribute to an inessential renormalization of the two-point function (whereas two-point polymers with higher momentum
external legs have already been resummed in section \ref{section:fermion}). Scale-neutral four-point polymers are
four-point polymers with a large transfer momentum $\gtrsim 2^{-j'_{\phi}}\mu$.  Six-point polymers are convergent, whence
one does not need to investigate the scale of  transfer momenta of external
leg pairs.  The same holds for lonely Goldstone pre-boson propagators coming from "half-integer" fermionic scales, which
have not become part of a fermionic polymer:
they may couple to scale $j_+$ four-point polymers because
their transfer momentum may be arbitrarily small; however, due
to their scale $j'_{\phi}-1$ infra-red cut-off, such propagators have quasi-exponential decay at distances larger than 
$\Gamma_{\phi}^{-1}$, hence they
may be considered at lower energies as irreducible.

\Medskip The dotted line at the center of
a {\em scale $j_+$ four-point polymer} separates external legs
with index $1,2$ from external legs with index $3,4$, in such a way
that the transfer momentum flowing from either side is
$\approx 2^{-j_+}\mu$. 

\Medskip  Assembling irreducible polymers 
yields irreducible and  {\bf reducible polymers}. There are
no scale-neutral reducible polymers; contracting
scale-neutral polymers simply gives a new scale-neutral
polymer, whose number of external legs is at least
equal to the sum of external legs of each constituent
polymer, minus twice the number of pairings ("at least"
because pairings of not fully expanded polymers may yield
new vertices instead of contracting previously developed
external legs),

\Bigskip

\begin{tikzpicture}[scale=0.55]
\draw[color=blue](0,0)--(25,0);
\draw[color=blue](0,-0.4)--(25,-0.4);

\draw(2,2) circle(1);  \draw(2,2) node{$4$};
\draw(1.5,1.134)--(1,-1); \draw(1.8,1.05)--(1.65,-1);
\draw(2.2,1.05)--(2.35,-1);
\draw(2.5,1.134)--(3,-1);

\draw(4.5,2) node {$\times$}; \draw(9.5,2) node {$\times$};

\begin{scope}[shift={(5,0)}]
\draw(2,2) circle(1);  \draw(2,2) node{$\ge 6$};
\draw(1.25,1.25)--(0.6,-1);
\draw(1.5,1.134)--(1,-1); \draw(1.8,1.05)--(1.65,-1);
\draw(2.2,1.05)--(2.35,-1);
\draw(2.5,1.134)--(3,-1);
\draw(2.75,1.25)--(3.4,-1);
\draw[dashed](2,1)--(2,-1);
\end{scope}

\begin{scope}[shift={(10,0)}]
\draw(2,2) circle(1); \draw(2,2) node{$2$};
\draw(1,2)--(0,2); \draw(4,2)--(3,2);
\end{scope}

\draw(17,2) node {$\longrightarrow$};

\begin{scope}[shift={(20,2)}]
\draw[dashed](-0.5,-0.866)--(-0.5,-3);
\draw[dashed](0.5,-0.866)--(0.5,-3);
\draw(0,0) circle(1); \draw(1.7,0) node {$\ge 6$};
\draw(0.5,-0.866) arc(-120:-240:1 and 1);
\draw[dashed](0.866,0.5) arc(-30:-150:1 and 1);
\draw(0,1)--(0,2);
\draw(0,3) circle(1); \draw(-1.5,3) node {$4$};
\draw(-0.707,3-0.707) arc(225:315:1 and 1);
\draw(0.866,3.5) arc(-30:-150:1 and 1);
\draw(1,3)--(2,3); 
\draw(3,3) circle(1); \draw(3-0.707,3-0.707)--(0.707,0.707);
\draw(3,3) node {$2$};
\end{scope}
\end{tikzpicture}

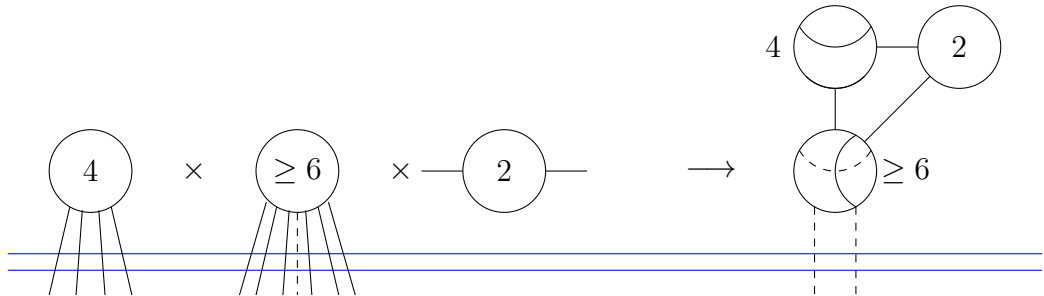
\Medskip {\bf \tiny Fig. \thesubsection.2. Assembly
rules of scale-neutral polymers: an example.}

\Bigskip
 Scale $j_+$ reducible
polymers
are made up of scale $j_+$ four-point polymers and possibly
scale-neutral polymers, interspersed
with bubbles: pure scale $j_+$ polymers generalizing
chains of bubbles,

\bigskip

{\centerline{\begin{tikzpicture}[scale=0.7]
\begin{scope}[shift={(-5,0)}]
\draw(2,-6) circle(1);  \draw[dotted](2,-5)--(2,-7);
\draw(3,-6)--(3.5,-5.5); \draw(3,-6)--(3.5,-6.5);
\draw(1,-6)--(0.5,-5.5);\draw(1,-6)--(0.5,-6.5);
\draw(4.5,-6) node {$+$};
\end{scope}
\draw(2,-6) circle(1);  \draw[dotted](2,-5)--(2,-7);
\draw(1,-6)--(0.5,-5.5);\draw(1,-6)--(0.5,-6.5);
\draw(2.866,-5.5)--(4.134,-5.5); \draw(2.866,-6.5)--(4.134,-6.5); \draw(5,-6) circle(1);\draw[dotted](5,-5)--(5,-7);
\draw(6,-6)--(6.5,-5.5); \draw(6,-6)--(6.5,-6.5);
\draw(7.5,-6) node {$+\cdots$};
\end{tikzpicture}}}

\Medskip {\bf\tiny Fig. \thesubsection.3. Generalized
chains of bubbles.}

\Bigskip and chains including one or more contribution
from scale-neutral polymers with $\ge 4$ external legs,

\begin{tikzpicture}[scale=0.65]

\draw(6.3,0) circle(1);
\draw[color=blue](-0,-4)--(15,-4); \draw[color=blue](0,-3.7)--(15,-3.7); \draw(13,0) node {scale-neutral};
\draw(13,-6) node {$j_+$};
\draw(2.866,-5.5)--(4.134,-5.5); \draw(2.866,-6.5)--(4.134,-6.5); \draw(5,-6) circle(1);\draw[dotted](5,-5)--(5,-7);
\draw(5.5,-5.134)--(5.5,-0.7);
\draw(5.866,-5.5)--(5.866,-0.92);
\draw(7.1,-5.134)--(7.1,-0.7);
\draw(2*6.3-5.866,-5.5)--(2*6.3-5.866,-0.92);
\draw(2*6.3-2.866,-5.5)--(2*6.3-4.134,-5.5); \draw(2*6.3-2.866,-6.5)--(2*6.3-4.134,-6.5); \draw(2*6.3-5,-6) circle(1);\draw[dotted](2*6.3-5,-5)--(2*6.3-5,-7);
\draw[dashed](6.3,-1) arc(135:90:1.5 and 3);
\draw[dashed](1,-6)--(2,-6); \draw[dashed](10,-6)--(11,-6);
\end{tikzpicture}

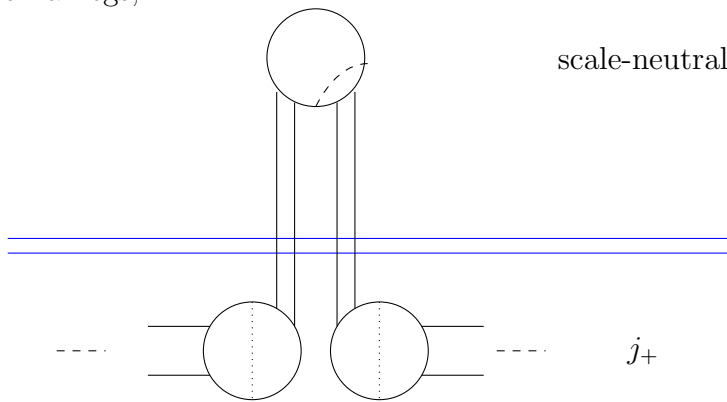
\Medskip {\bf\tiny Fig. \thesubsection.4. Insertion of
scale-neutral polymers into
chains of bubbles.}
 
\Bigskip

\Medskip Double lines are Cooper bubbles with dressed
propagators obtained by resumming scale-neutral two-point
polymers. Fig. \thesubsection.3 may be seen as the leading
term of Fig. \thesubsection.4,  the scale-neutral
polymer being just a four-point vertex with a $\lambda$ coefficient. After a Mayer expansion and an $s$-wave projection (see  {\bf D.} below for the detailed procedure),   all these chains are resummed to form 
so-called {\bf single-scale Goldstone boson propagators}  with transfer momentum 
$|q|\approx 2^{-j_+}\mu$, denoted by a dotted wiggling line
\begin{tikzpicture} \draw[dotted,decorate,decoration=snake](0,0)--(2,0); \end{tikzpicture} 
(not to be confused with the previous Pre-$\Sigma$ kernel {\em dashed} wiggling lines). A final resummation procedure produces at last our
{\em Goldstone boson propagator} $\Sigma$, which contain internal
single-scale Goldstone boson propagators with  transfer
momentum with higher scales $\le j_+$.

\Medskip There is a slight inaccuracy in Fig. \thesubsection.1
and Fig. \thesubsection.2, in that the scales $j_{+,1},j_{+,2}$ of the
low transfer momentum polymers need not be exactly the same;
because  momentum cut-offs are not sharp, one has instead  $|j_{+,1}-j_{+,2}|\lesssim 1$. Note also that a Mayer
resummation is necessary to have  momentum conservation.

\Medskip
If the inserted scale-neutral polymer has exactly 4 external legs, then this configuration forms a chain. If, however,
it has $\ge 6$ external legs, then its supplementary legs
may also contract to an arbitrary number $n\ge 3$ of low transfer momentum four-point polymers according to the following pattern (where one has chosen $n=3$),

\Bigskip

\begin{tikzpicture}[scale=0.7]
\draw(6.3,0) circle(1);
\draw[color=blue](0,-4)--(15,-4); \draw[color=blue](0,-3.7)--(15,-3.7); \draw(13,0) node {scale-neutral};
\draw(13,-6) node {$j_{+,1}$};
\draw(2.866,-5.5)--(4.134,-5.5); \draw(2.866,-6.5)--(4.134,-6.5); \draw(5,-6) circle(1);\draw[dotted](5,-5)--(5,-7);
\draw(5.5,-5.134)--(5.5,-0.7);
\draw(5.866,-5.5)--(5.866,-0.92);

\draw[color=blue](0,-9)--(15,-9);
\draw(7.1,-5-5.134)--(7.1,-0.7);
\draw(2*6.3-5.866,-5-5.5)--(2*6.3-5.866,-0.92);
\draw(2*6.3-2.866,-5-5.5)--(2*6.3-4.134,-5-5.5); \draw(2*6.3-2.866,-5-6.5)--(2*6.3-4.134,-5-6.5); \draw(2*6.3-5,-5-6) circle(1);\draw[dotted](2*6.3-5,-5-5)--(2*6.3-5,-5-7);
\draw(6.1,-0.9)--(6.1,-15);
\draw(6.5,-0.9)--(6.5,-15.2);
\draw(13,-11) node{$j_{+,2}$};
\draw[color=blue](0,-14)--(15,-14);

\begin{scope}[shift={(0.8,-10)}]
\draw(2.866,-5.5)--(4.134,-5.5); \draw(2.866,-6.5)--(4.134,-6.5); \draw(5,-6) circle(1);\draw[dotted](5,-5)--(5,-7);
\draw(12.2,-6) node{$j_{+,3}$};
\end{scope}
\end{tikzpicture}

\Medskip {\bf\tiny Fig. \thesubsection.5 (same as Fig.
\thesubsection.6).}

\Bigskip Scale-neutral polymers can also appear at either
end of a chain.

\Medskip Since external legs of a scale-neutral polymer $\P$ with
$\ge 6$ external legs -- which is virtually connected
to an infinite number of polymers, since the scale $j'_{\phi}$
interaction ${\cal L}^{j'_{\phi}}$ has not been expanded
inside $\P$ --  can be paired  with an arbitrary number of  low transfer momentum four-point polymers with arbitrary scales, it
is only at the very end of the expansion -- i.e., once
{\em all} low transfer momentum four-point polymers have
been connected -- that the general connectivity structure
of $\P$ emerges. Then the scale-neutral polymer of Fig. \thesubsection.5 will be  interpreted
as a vertex of the multi-scale bosonic cluster tree,

\Bigskip

\begin{tikzpicture}[scale=0.7]
\draw(6.3,0) circle(1);
\draw[dashed](6.3+0.707,-0.707)--(6.3+0.707,-6);
\draw[dashed](6.3,-1)--(6.3,-11);
\draw[dashed](6.3-0.707,-0.707)--(6.3-0.707,-16);
\draw[dotted,decorate,decoration=snake](6.3+0.707,-6)--(9.3+0.707,-6);
\draw[dotted,decorate,decoration=snake](6.3,-11)--(9.3,-11);
\draw[dotted,decorate,decoration=snake](6.3-0.707,-16)--(3.3-0.707,-16);
\draw(13,-6) node {$j_{+,1}$};
\draw[color=blue](0,-14)--(15,-14);
\draw[color=blue](0,-9)--(15,-9);
\draw[color=blue](0,-4)--(15,-4);
\draw(13,-11) node{$j_{+,2}$};
\draw(0.8+12.2,-10-6) node{$j_{+,3}$};
\end{tikzpicture}

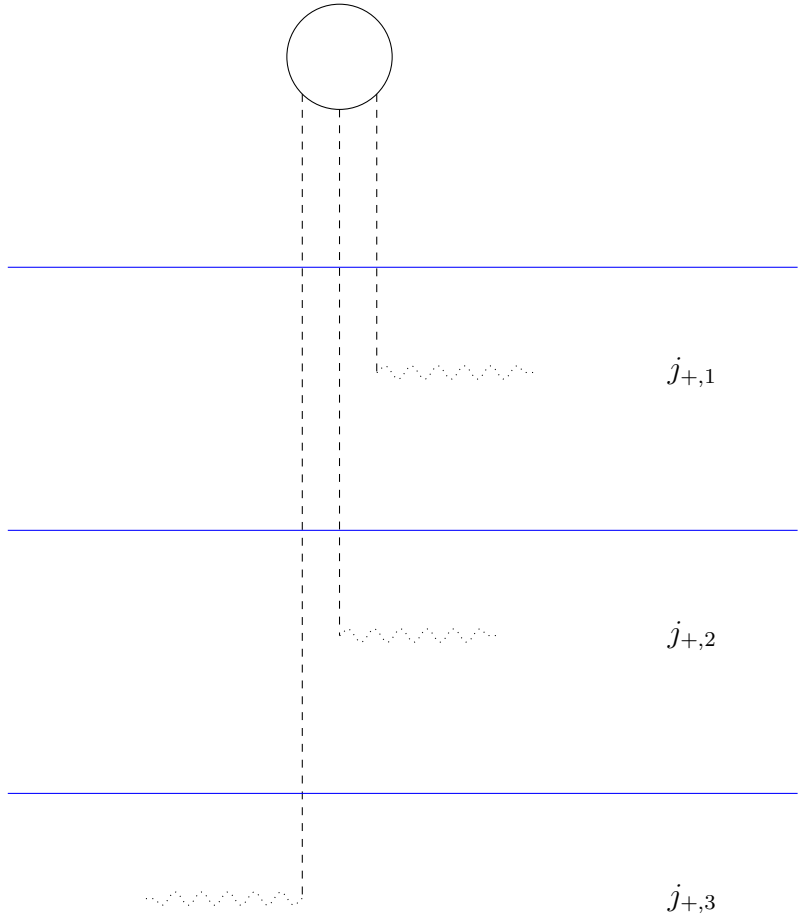
\Medskip   {\bf\tiny Fig. \thesubsection.6. A vertex
of the multi-scale bosonic cluster tree (same as 
Fig. \thesubsection.5).}

\Bigskip   Finally (see next Figure) a low transfer momentum four-point polymer
may contract one  of its two Cooper pairs to 
two {\em different} (scale neutral or low transfer
momentum) polymers (left part of the figure), in which case the three
polymers may be considered as  a single, compound scale-neutral polymer (blue rectangle on the right part of the figure). 

\bigskip

\begin{tikzpicture}[scale=0.55]
\draw[color=blue](-2.5,0.2)--(24.5,0.2);
\draw[color=blue](-2.5,-0.2)--(24.5,-0.2);
\draw(4,0) circle(1); \draw(8,0) circle(1);
\draw[->](4,1)--(4,1.3); \draw[->](4,-1)--(4,-1.3);
\draw[->](8,1)--(8,1.3); \draw[->](8,-1)--(8,-1.3);
\draw(0,-2) circle(1); \draw[dotted](0,-1)--(0,-3);
\draw(0.707,-2+0.707)--(4-0.707,-0.707);
\draw(1,-2)--(8,-1);
\draw(-0.707,-2+0.707)--(-1.4,-2+1.4);
\draw(-0.707,-2-0.707)--(-1.4,-2-1.4);

\draw[->] (10,-1.5) arc(200:340:2 and 1);
\draw(19,4) circle(1); \draw(21,2) circle(1);
\draw[->](16+0.707,2+0.707)--(17.5,3); \draw(17.5,3)--(19-0.707,4-0.707); \draw(17,3.7) node {\small $p$};
\draw[->](17,2)--(18.5,2); \draw(18.5,2)--(20,2); \draw(18.5,1.5) node {\small $-p+q$};
\draw[color=blue](14.5,0.5) rectangle(22.5,5.5);
\draw(16,2) circle(1); \draw[dotted](16,1)--(16,3);
\draw[dashed](16,1)--(16,-2);
\draw[dashed](16-0.707,2-0.707)--(16-0.707,-1.5);
\draw(16,-2)--(15,-2); \draw(16-0.707,-1.5)--(15-0.707,-1.5);
\draw[->](16-0.707/2,-1.5)--(16-0.707/2,-0.5);  \draw(16,-1) node {$q$};
\draw(23.5,-2) node {$j_+$};
\end{tikzpicture}

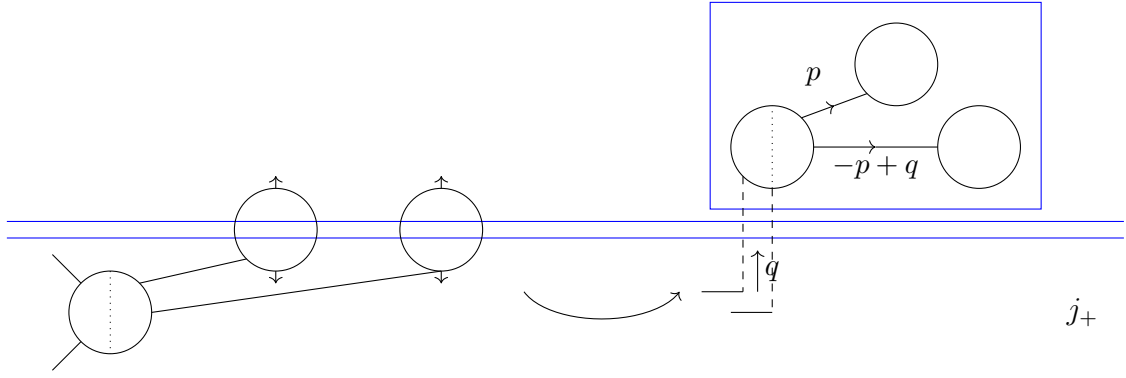
\Medskip {\bf \tiny Fig. \thesubsection.7. One scale-neutral
Cooper pair contraction. Blobs with ascending/descending
arrows $\uparrow,\downarrow$ are located above ($\uparrow$),
resp. below ($\downarrow$) the double line, and represent
scale-neutral, resp. low transfer momentum four-point polymers.}

\Bigskip The remaining Cooper pair may  contract
to a single scale $j_+$ four-point polymer, yielding the
first bricks of a  scale $j_+$ Goldstone boson attached
to a scale-neutral polymer; otherwise it does not show up
any more as a Cooper pair, and becomes part
of a scale-neutral polymer.

\vskip 2cm
Let us  start with a precise
typology of irreducible diagrams, depending on the scale:

\Medskip -- {\em at scale $j'_{\phi}$, there are polymers in ${\cal P}^{(j'_{\phi}-1)\to}$  with $\ge 6$ external legs,
four-point polymers of type 1, and four-point polymers of type 2.}  Four-point polymers of
type 1 are not divergent in the infra-red limit Four-point polymers of type 2 are further split (see below) into four-point polymers of type 2 with transfer 
momentum of scale $j'_{\phi}$ -- whose connections are closely examined through the
scale $j'_{\phi}$ cluster expansion -- and four-point polymers of type 2 with {\em low transfer 
momentum} of scale $>j'_{\phi}$. The latter ones are left for further investigation
at scale $j'_{\phi}+1$. As for four-point polymers of type 2 with transfer 
momentum of scale $j'_{\phi}$, they are  assembled together in a line into a single-scale Goldstone boson propagtor
with transfer 
momentum of scale $j'_{\phi}$. The two ends of each scale $j'_{\phi}$ Goldstone boson propagator are either
connected to one (or several) of the external legs $\xi_{ext,1},\ldots,\xi_{ext,N_{ext}}$; or
connected to  
diagrams with $\ge 6$
external legs, whose links with four-point polymers of type 2 with low transfer momentum  have not been clarified at this stage. As mentioned earlier, scale $j'_{\phi}$ Goldstone boson
propagators may also be connected to Goldstone boson propagators of lower scale $>j'_{\phi}$ because momentum cut-offs are not sharp;

\Medskip -- {\em the final diagrammatical outcome of the  scale $j'_{\phi}$ cluster expansion is:}

(i)  a new set
of polymers in ${\cal P}^{j'_{\phi}\to}$ with $\ge 6$ external legs;

(ii)  the left-over set of
four-point polymers of type 2 with low transfer momentum;

(iii) newly developed vacuum polymers (which must be divided out by a Mayer
expansion), and  {\bf two-point  polymers}, which contribute to a small, finite
{\em dressing of fermion propagators} (therefore requiring no mass or chemical
potential renormalization), resulting in the replacement
of $C^{j'_{\phi}}_{\theta}$ by the {\bf dressed propagator}
$C_{dressed}$;

\Medskip -- {\em at scale $j'_{\phi}+1$}, as  discussed in the
two previous paragraphs, all scale $j'_{\phi}$ four-point polymers of type 2 with
transfer momentum $|q|\approx 2^{-j'_{\phi}}\mu$ have been resummed and integrated
into larger polymers. Therefore there remain only four-point polymers with
transfer momentum scale $\ge j'_{\phi}+1$: all four-point polymers (save for inessential cut-off
effects) are already of type 2 for the scale $j'_{\phi}+1$. After performing (as at scale $j'_{\phi}$) a splitting according
to whether the transfer momentum is of scale $j'_{\phi}+1$ or higher, and leaving the
latter case for further investigation at scale $j'_{\phi}+2$, we are left with an assembly
composed of {\em scale $(j'_{\phi}+1)$ four-point polymers, and scale-neutral 
two-point polymers and polymers with $\ge 6$ external legs.}

 \Medskip  {\bf By convention,
scale-neutral polymers (i.e. two-point polymers, polymers with $\ge 6$ external legs, and
and scale $j'_{\phi}$ four-point polymers of type 1) are drawn on multi-scale figures
in a separate upper line} (above the double blue line).

\Medskip -- the above  description of the final diagrammatical outcome of the scale $j'_{\phi}$ cluster expansion, and of the scale $(j'_{\phi}+1)$-cluster expansion,  extends to lower scales  without modification.

\Bigskip Let us first describe diagrammatically the various expansions that shall be 
needed. We add some short comments and refer to {\bf B.} and {\bf C.} below for
detailed explanations.

\Bigskip {\em Preliminary second-order expansion in isolated boxes (step $\sharp 0$) at scale $j'_{\phi}$.}

\bigskip

{\centerline{
\begin{tikzpicture}[scale=0.6]
\draw[fill=gray](-2.5,-1) rectangle(-0.5,1); \draw(1,0) node {$=\ 1 \ + $}; 
\draw(3,-1) rectangle(5,1); \draw(6,0) node {$+$}; 
\draw(7,-1) rectangle(9,1);
\draw(3.2,0.8)--(3.8,0.2); \draw(3.8,0.8)--(3.2,0.2); \draw(3.5,0.5) node {\textbullet};
\draw(4+3.2,0.8)--(4+3.8,0.2); \draw(4+3.8,0.8)--(4+3.2,0.2); \draw(4+3.5,0.5) node {\textbullet};
\draw(4+4.8,-0.8)--(4+4.2,-0.2); \draw(4+4.2,-0.8)--(4+4.8,-0.2); \draw(4+4.5,-0.5) node {\textbullet};
\draw(8.5,0.5) node {$\cdots$}; \draw(7.5,-0.5) node {$\cdots$};
\begin{scope}[shift={(0,-0.3)}]
\draw(1,-3) node {$=\ 1 \ + $}; 
\draw(4,-3) circle(1); \draw(6,-3) node {$+$}; 
\draw(8,-3) circle(1); \draw(8,-3) node {$\ge 6$};
\draw(3.5,-3+0.5) node {\textbullet}; \draw(3.5,-1.7) node {\tiny $\xi_1=\xi_2=\xi_3=\xi_4$};
 \draw(3.5,-2.5)--(2.5,-2.5); \draw(3.5,-2.5)--(5.5,-2.5);
 \draw(3.5,-2.5)--(2.5,-3.5); \draw(3.5,-2.5)--(5.5,-3.5); 
\draw(9,-3)--(9.5,-3); \draw(8.5,-3+0.866)--(8+1.5*0.5,-3+1.5*0.866);
 \draw(7.5,-3+0.866)--(8-1.5*0.5,-3+1.5*0.866);
 \draw(7,-3)--(6.5,-3);
 \draw(8.5,-3-0.866)--(8+1.5*0.5,-3-1.5*0.866);
 \draw(7.5,-3-0.866)--(8-1.5*0.5,-3-1.5*0.866);
 \draw[dashed](8,-4)--(8,-4.5);
\end{scope}
 \end{tikzpicture}}}

\medskip {\tiny \bf Fig. \thesubsection.8. Preliminary second-order expansion in
isolated boxes.}

\Bigskip Expanding the interaction  ${\cal L}^{j'_{\phi}}_{\theta}$ to order 2 in each isolated box $\Del\in\D^{j'_{\phi}}$ yields $1$, plus
a term with one vertex -- interpreted as a scale $j'_{\phi}$ polymer with
$4$ external legs, on equal footing with four-point polymers in ${\cal P}^{(j'_{\phi}-1)\to}$  --, plus a term with $\ge 2$ vertices -- interpreted as a scale 
$j'_{\phi}$ polymer with $\ge 8\ge 6$ external legs --. This is done only once -- as a
preliminary step to scale $j'_{\phi}$ cluster expansion.

\Bigskip {\em Scale $j_+$ splitting phase for four-point polymers of type $2$.}

\bigskip

\begin{tikzpicture}[scale=0.7]
\draw[dotted](0,1)--(0,-1);
\draw(0,0) circle(1);
\draw(-0.5,0.5) node {\textbullet}; 
\draw(0.5,0.5) node {\textbullet}; 
\draw(0.5,-0.5) node {\textbullet}; 
\draw(-0.5,-0.5) node {\textbullet}; 
\draw(-1.5,0.5)--(-0.5,0.5); \draw(0.5,0.5)--(1.5,0.5);
\draw(-1.5,-0.5)--(-0.5,-0.5); \draw(0.5,-0.5)--(1.5,-0.5);
\draw(-0.5,1) node {$\xi_1$};
\draw(0.5,1) node {$\xi_3$};
\draw(-0.5,-1) node {$\xi_2$};
\draw(0.5,-1) node {$\xi_4$};
\draw(2.5,0) node {$\simeq$};
\begin{scope}[shift={(5,0)}]
\draw(0,0) circle(1);
\draw(-0.5,0.5) node {\textbullet}; 
\draw(0.5,0.5) node {\textbullet}; 
\draw(-1.5,0.5)--(-0.5,0.5); \draw(0.5,0.5)--(1.5,0.5);
\draw(-1.5,-0.5)--(-0.5,0.5); \draw(0.5,0.5)--(1.5,-0.5);
\draw(-0.5,1) node {$\xi_1$};
\draw(0.5,1) node {$\xi_3$};
\draw(2.5,0) node {$=$};
\end{scope}
\begin{scope}[shift={(10.5,0)}]
\draw(0,0) circle(1);
\draw(-0.5,0.5) node {\textbullet}; 
\draw(0.5,0.5) node {\textbullet}; 
\draw[decorate,decoration=coil](-0.5,0.5)--(-1,0);
\draw[decorate,decoration=coil](0.5,0.5)--(1,0);
\draw(-0.5-1.5,-0.5+0.5)--(-0.5-0.5,-0.5+0.5); \draw(0.5+0.5,-0.5+0.5)--(0.5+1.5,-0.5+0.5);
\draw(-0.5-1.5,-0.5-0.5)--(-0.5-0.5,-0.5+0.5); \draw(0.5+0.5,-0.5+0.5)--(0.5+1.5,-0.5-0.5);
\draw(-0.5,1) node {$\xi_1$};
\draw(0.5,1) node {$\xi_3$};
\draw(3,0) node {$+$};
\draw[<-] (-0.6,0.2)--(0,-0.5); \draw[<-] (0.6,0.2)--(0,-0.5); 
\draw(0,-0.5)--(0,-1.3); \draw(0,-1.6) node{$\chi_+^{j_+}$};
\end{scope}
\begin{scope}[shift={(17,0)}]
\draw(0,0) circle(1);
\draw(-0.5,0.5) node {\textbullet}; 
\draw(0.5,0.5) node {\textbullet}; 
\draw[decorate,decoration=coil](-0.5,0.5)--(-2,-1);
\draw[decorate,decoration=coil](0.5,0.5)--(2,-1);
\draw(-1.5-1.5,-1.5+0.5)--(-1.5-0.5,-1.5+0.5); \draw(1.5+0.5,-1.5+0.5)--(1.5+1.5,-1.5+0.5);
\draw(-1.5-1.5,-1.5-0.5)--(-1.5-0.5,-1.5+0.5); \draw(1.5+0.5,-1.5+0.5)--(1.5+1.5,-1.5-0.5);
\draw(-0.5,1) node {$\xi_1$};
\draw(0.5,1) node {$\xi_3$};
\draw[<-] (-0.6,0.2)--(0,-0.5); \draw[<-] (0.6,0.2)--(0,-0.5); 
\draw(0,-0.5)--(0,-1.3); \draw(0,-1.6) node{$\chi_+^{\to(j_+ +1)}$};
\end{scope}
\end{tikzpicture} 

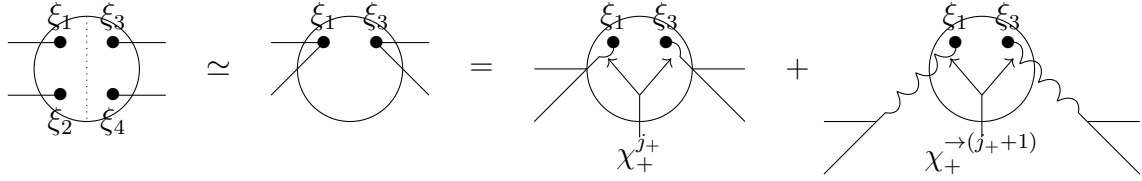
\Medskip {\bf \tiny Fig. \thesubsection.9. Scale $j_+$ splitting phase for
four-point polymers.}

\Bigskip Consider a four-point polymer of type $2$ with external legs located at $\xi_1,\xi_2,\xi_3,
\xi_4$ {\em which have not been contracted at scales $< j_+$}. Displace external legs as in \S \ref{subsection:Sigma},  $\xi_2\longrightarrow\xi_1$ and
$\xi_4\longrightarrow\xi_3$. Then insert characteristic functions $\chi_+^{j_+}$ at $\xi_1,\xi_3$ singling
out the case when the transfer momentum is of scale $j_+$. Since the polymer is
of type $2$, the remainder (last
diagram on the right) involves only {\em low transfer momenta} of scale $>j_+$. Low
transfer momentum four-point diagrams
are left for future investigation at scale $j_+ +1$.

\Bigskip {\em A single-scale diagram with trivial strongly connected
components.} Specifically at scale $j_+=j'_{\phi}$,  we develop links between  scale $j'_{\phi}$ isolated boxes and fermionic polymers in ${\cal P}^{(j'_{\phi}-1)\to}$, except four-point polymers with low transfer momentum,
$|q|\ll 2^{-j'_{\phi}}\mu$. The algorithm is as follows. Choose some arbitrary ordering
of the set of polymers (four-point polymers with low transfer momentum excluded), $\P_1,\ldots,\P_p$. Start from $\P_1$ and try to develop $\ge 3$ links between $\P_1$ and
$\P_2$. Say that $\P_1$ and $\P_2$ are {\em strongly connected} if there 
exist $\ge 3$ such links. In the  example below -- where all polymers are  four-point polymers with scale
$j_{+}$ transfer momenta, 
represented by numbered blobs -- no pair of polymers is strongly connected. The general algorithm
is  simple: expand all links between $\P_1$ and $\P_2$, $\P_3,\P_4,\P_5,\ldots$ (first figure). Then
expand all links between $\P_2$ and $\P_3,\P_4,\P_5,\ldots$
 (second figure), and so on, until one
has obtained a new polymer which is isolated at scale $j'_{\phi}$.  
 The figures below present  the successive stages of the expansion. Arrows point successively from $\P_i, i=1,2,3,4$, uncovering links between $\P_i$   and 
 new polymers $\P_{i+1},\ldots,\P_5$.

\bigskip

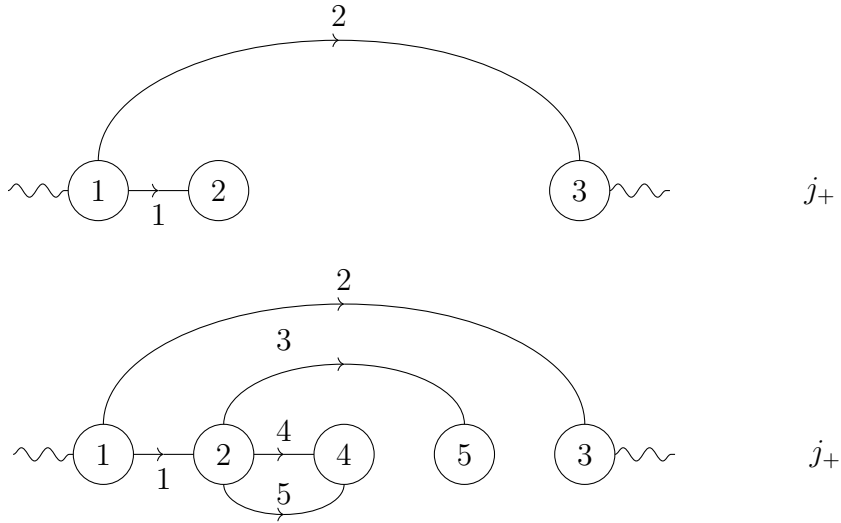
{\centerline{\begin{tikzpicture}[scale=0.4]
\draw[decorate,decoration=snake](-2,0)--(0,0);
\draw(1,0) circle(1); \draw(1,0) node {$1$}; \draw[->](2,0)--(3,0); \draw(3,0)--(4,0); \draw(5,0) circle(1); \draw(5,0)
node {$2$};
\draw(17,0) circle(1); \draw(17,0) node {$3$};
\draw[decorate,decoration=snake](18,0)--(20,0);
\draw(3,-0.8) node {$1$};
\draw[->](1,1) arc(180:90:8 and 4); \draw(9,5) arc(90:0:8 and 4); \draw(9,5.8) node {$2$};
\draw(25,0) node {$j_+$};
\end{tikzpicture}}} 

\bigskip

{\centerline{
\begin{tikzpicture}[scale=0.4]
\draw[decorate,decoration=snake](-2,0)--(0,0);
\draw(1,0) circle(1); \draw(1,0) node {$1$}; \draw[->](2,0)--(3,0); \draw(3,0)--(4,0); \draw(5,0) circle(1); \draw(5,0)
node {$2$};
\draw(17,0) circle(1); \draw(17,0) node {$3$};
\draw[decorate,decoration=snake](18,0)--(20,0);
\draw(3,-0.8) node {$1$};
\draw[->](1,1) arc(180:90:8 and 4); \draw(9,5) arc(90:0:8 and 4); \draw(9,5.8) node {$2$};
\draw(9,0) circle(1); \draw(9,0) node {$4$};
\draw(13,0) circle(1); \draw(13,0) node {$5$};
\draw[->](5,1) arc(180:90:4 and 2); \draw(9,3) arc(90:0:4 and 2); \draw(7,3.8) node {$3$};
\draw[->](6,0)--(7,0); \draw(7,0)--(8,0);  \draw(7,0.8)
node {$4$};
\draw[->](5,-1) arc(-180:-90:2 and 1); \draw(7,-2) arc(-90:0:2 and 1); \draw(7,-1.3) node {$5$};
\draw(25,0) node {$j_+$};
\end{tikzpicture}}} 

\bigskip

{\centerline{\begin{tikzpicture}[scale=0.4]
\draw[decorate,decoration=snake](-2,0)--(0,0);
\draw(1,0) circle(1); \draw(1,0) node {$1$}; \draw[->](2,0)--(3,0); \draw(3,0)--(4,0); \draw(5,0) circle(1); \draw(5,0)
node {$2$};
\draw(17,0) circle(1); \draw(17,0) node {$3$};
\draw[decorate,decoration=snake](18,0)--(20,0);
\draw(3,-0.8) node {$1$};
\draw[->](1,1) arc(180:90:8 and 4); \draw(9,5) arc(90:0:8 and 4); \draw(9,5.8) node {$2$};
\draw(9,0) circle(1); \draw(9,0) node {$4$};
\draw(13,0) circle(1); \draw(13,0) node {$5$};
\draw[->](5,1) arc(180:90:4 and 2); \draw(9,3) arc(90:0:4 and 2); \draw(7,3.8) node {$3$};
\draw[->](6,0)--(7,0); \draw(7,0)--(8,0);  \draw(7,0.8)
node {$4$};
\draw[->](5,-1) arc(-180:-90:2 and 1); \draw(7,-2) arc(-90:0:2 and 1); \draw(7,-1.3) node {$5$};
\draw[->](16,0)--(15,0); \draw(15,0)--(14,0); \draw(15,-0.8) 
node {$6$};
\draw(25,0) node {$j_+$};
\end{tikzpicture} }}

\bigskip

{\centerline{\begin{tikzpicture}[scale=0.4]
\draw[decorate,decoration=snake](-2,0)--(0,0);
\draw(1,0) circle(1); \draw(1,0) node {$1$}; \draw[->](2,0)--(3,0); \draw(3,0)--(4,0); \draw(5,0) circle(1); \draw(5,0)
node {$2$};
\draw(17,0) circle(1); \draw(17,0) node {$3$};
\draw[decorate,decoration=snake](18,0)--(20,0);
\draw(3,-0.8) node {$1$};
\draw[->](1,1) arc(180:90:8 and 4); \draw(9,5) arc(90:0:8 and 4); \draw(9,5.8) node {$2$};
\draw(9,0) circle(1); \draw(9,0) node {$4$};
\draw(13,0) circle(1); \draw(13,0) node {$5$};
\draw[->](5,1) arc(180:90:4 and 2); \draw(9,3) arc(90:0:4 and 2); \draw(7,3.8) node {$3$};
\draw[->](6,0)--(7,0); \draw(7,0)--(8,0);  \draw(7,0.8)
node {$4$};
\draw[->](5,-1) arc(-180:-90:2 and 1); \draw(7,-2) arc(-90:0:2 and 1); \draw(7,-1.3) node {$5$};
\draw[-<](13,-1) arc(0:-90:2 and 1); \draw(11,-2) arc(-90:-180:2 and 1); 
\draw(11,-3) node {$8$};
\draw[->](16,0)--(15,0); \draw(15,0)--(14,0); \draw(15,-0.8) 
node {$6$};
\draw[->](10,0)--(11,0); \draw(11,0)--(12,0); \draw(11,0.8)
node {$7$};
\draw(25,0) node {$j_+$};
\end{tikzpicture} }}

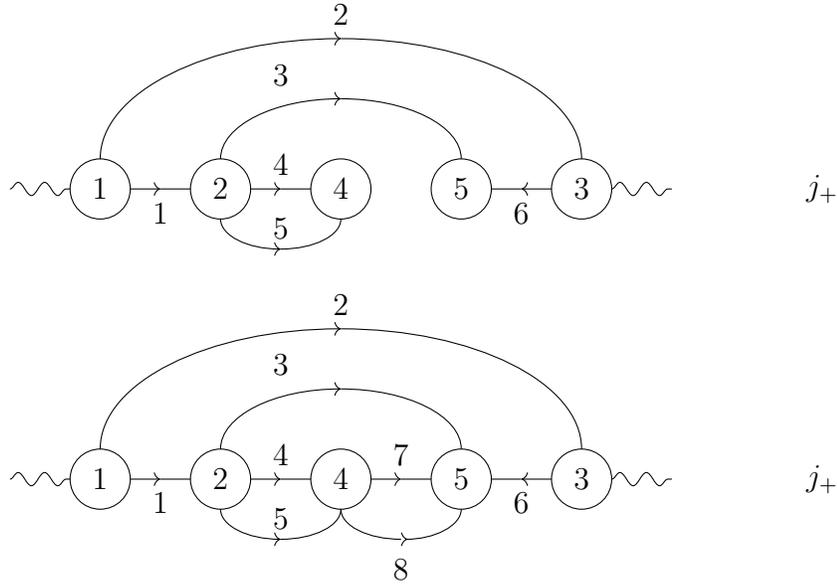
\Bigskip {\bf\tiny Fig. \thesubsection.10. Dressed propagator, dressed bubble. Wiggling lines at the two ends stand for low transfer momentum Cooper pairs.}

\Bigskip This diagram is actually a "dressed" bubble, involving one bare propagator
(the one with index $2$), and a dressed propagator from blob number $1$ to blob
number $3$, involving blobs $2,4$ and $5$.

\Bigskip The above algorithm also produces non-trivial
strongly connected clusters of polymers, discussed in the
next two examples.

\Bigskip {\em A single-scale diagram with a single, non-trivial
strongly connected component.} 

\bigskip

{\centerline{\begin{tikzpicture}[scale=0.5]
\draw[decorate,decoration=snake](-3,0)--(-1,0);
\draw(0,0) circle(1); \draw(0,0) node {$1$};
\draw(4,0) circle(1); \draw(4,0) node {$2$};
\draw(0.37,0.8)--(3.63,0.8);
\draw(0.65,0.4)--(3.35,0.4);
\draw[dashed](1,0)--(3,0);
\draw(0.65,-0.4)--(3.35,-0.4);
\draw(12,0) circle(1); \draw(12,0) node {$3$};
\draw(0,-1) arc(180:360:6 and 3);
\draw(4.5,1) arc(180:0:3.5 and 1.5);
\draw(3.5,1) arc(180:0:4.5 and 2.5);
\draw[color=blue](-2,-2) rectangle(6,2);
\draw[dashed](6,0)--(11,0);
\draw[decorate,decoration=snake](13,0)--(15,0);
\draw[color=blue](-2.5,-4.5) rectangle(14,4);
\draw(20,0) node {$j_+$};
\end{tikzpicture}}}

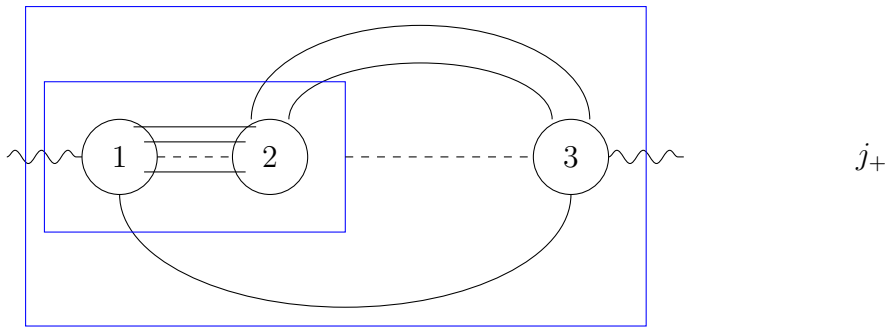
\Medskip {\bf \tiny Fig. \thesubsection.11. One contribution to the $\Pi$-kernel.}

\Bigskip Blob $1$ has $\ge 3$ links to blob $2$. Hence blobs 1 and 2 are assembled into
a single cluster $\{12\}$. Then cluster $\{12\}$ has $\ge 3$ links to blob 3, thereby forming
a new cluster $\{123\}$.  This four-point irreducible polymer contributes to the $\Pi$-kernel.

\Bigskip {\em Other contributions to the  dressed propagators.} This dressed bubble diagram involves strongly connected
two-point polymers. 

\bigskip

\begin{tikzpicture}[scale=0.5]
\draw[decorate,decoration=snake](-3,0)--(-1,0);
\draw(0,0) circle(1); \draw(0,0) node {$1$}; 
\draw(12,0) circle(1); \draw(12,0) node {$4$};
\draw(0,-1) arc(180:360:6 and 2);
\draw(0,1) arc(180:130:4 and 2);
\draw(2,2.3) circle(0.7); \draw(2,2.3) node{$2$};
\draw(2+0.7*0.37,2.3+0.7*0.8)--(2+0.7*3.63,2.3+0.7*0.8);
\draw(2+0.7*0.65,2.3+0.7*0.4)--(2+0.7*3.35,2.3+0.7*0.4);
\draw[dashed](2+0.7*1,2.3+0.7*0)--(2+0.7*3,2.3+0.7*0);
\draw(2+0.7*0.65,2.3-0.7*0.4)--(2+0.7*3.35,2.3-0.7*0.4);
\draw(2+0.7*4,2.3+0.7*0) circle(0.7);  \draw(2+0.7*4,2.3+0.7*0) node {$3$};
\draw(2.7+0.7*4,2.3)--(3.7+0.7*4,2.3);
\draw(4.4+0.7*4,2.3) circle(0.7); \draw(4.4+0.7*4,2.3) node {$7$};
\draw(5.1+0.7*4,2.3)--(6.3+0.7*4,2.3);
\draw(7+0.7*4,2.3) circle(0.7);\draw(7+0.7*4,2.3) node {$5$};

\draw(0.7*0.866+4.4+0.7*4,0.7*0.5+2.3)--(0.5+0.7*0.866+4.4+0.7*4,0.5+0.7*0.5+2.3);
\draw(0.7*0.5+4.4+0.7*4,0.7*0.866+2.3)--(0.5+0.7*0.5+4.4+0.7*4,0.5+0.7*0.866+2.3);
\draw(5.8+0.7*4,3.7) circle(0.7); \draw(5.8+0.7*4,3.7) node {$6$};
\draw(0.7*0.866+5.8+0.7*4,-0.7*0.5+3.7)--
(0.4+0.7*0.866+5.8+0.7*4,-0.4-0.7*0.5+3.7);
\draw(0.7*0.7+5.8+0.7*4,-0.7*0.7+3.7)--
(0.4+0.7*0.7+5.8+0.7*4,-0.4-0.7*0.7+3.7);
\draw(0.7*0.5+5.8+0.7*4,-0.7*0.866+3.7)--
(0.4+0.7*0.5+5.8+0.7*4,-0.4-0.7*0.866+3.7);

\draw(10.5,2.3) arc(50:0:3 and 2);
\draw[color=blue](1,1.3) rectangle(6,3.2);
\draw[color=blue](6.3,1.3) rectangle(11.5,4.7);
\draw[decorate,decoration=snake](13,0)--(15,0);
\draw(20,0) node {$j_+$};
\end{tikzpicture}

\Medskip {\bf\tiny  Fig. \thesubsection.12. Dressed propagators.}

\Bigskip {\em At scales $j_+>j'_{\phi}$}, we keep the idea of a third-order cluster
expansion, but explore only links starting from  four-point polymers  with low transfer momentum of scale $j_+$,
and connecting them to arbitrary polymers, {\em except} four-point polymers with transfer momenta of higher scale $>j_+$.

\bigskip
\noindent{\bf B. Scale $j'_{\phi}$ complementary expansion.}

\Bigskip {\bf The scale $j'_{\phi}$ horizontal cluster
expansion.}

\Medskip We resume from expression (\ref{eq:HV-expansion})
for $F_I({\bf \xi}_{ext},\bar{\bf \xi}_{ext})$. 
 
\Medskip {\bf Step $\# 0$. Preliminary second-order expansion in isolated
boxes} (see Fig. \thesubsection.8).  Let $\Del\in\D^{j'_{\phi}}$. Expand 
$ e^{-\int  d\xi \, \chi_{\Del}(\xi) {\cal L}_{\theta}^{j'_{\phi}}(\xi)}$, see (\ref{eq:Ljphi}),
into 
\BEA && 1- \int  d\xi \, \chi_{\Del}(\xi) {\cal L}_{\theta}^{j'_{\phi}}(\xi) + 
\Big(\int  d\xi \, \chi_{\Del}(\xi) {\cal L}_{\theta}^{j'_{\phi}}(\xi)\Big)^2  \int_0^1
dt\, (1-t) e^{-t\int  d\xi \, \chi_{\Del}(\xi) {\cal L}_{\theta}^{j'_{\phi}}(\xi)}\nonumber\\
&&= 1-\lambda\int d\xi\, \chi_{\Del}(\xi)  (\bar{\Psi}^{j'_{\phi}}\Psi^{j'_{\phi}})^2(\xi) + \cdots
\EEA
The first non-trivial term is seen as a four-point polymer. Neglected terms $(\cdots)$ involve at least 8 fields in $\Del$, hence are considered as polymers with $\ge 6$ external legs.

\Bigskip {\bf Step $\# 1$. Splitting phase for four-point polymers of type $2$} (see Fig. \thesubsection.9).

\Medskip Let $\P$ be a four-point polymer of type $2$, with external structure  $(\bar{\Psi}_1\Psi_2)\otimes (\bar{\Psi}_3\Psi_4)$ as in (\ref{eq:psipsipsipsi}), with external
sectors $\alpha_1,\ldots,
\alpha_4\in \Z/2^{j'_{\phi}}\Z$.
By definition, $|\alpha_1+\alpha_2|,|\alpha_3+\alpha_4|\le 1$.
The argument below can be repeated without any modification
at scale $j_+$ with $\alpha_1,\ldots,\alpha_4\in\Z/2^{j_+}\Z$, so we state the general version. 
Consider first the {\em left part} of $\P$. Take the Fourier transform,
\BEQ F(\xi_1,\xi_2):=
\bar{\psi}^{j'_{\phi},\alpha_1}_{\uparrow,1} (\xi_1)
 \bar{\psi}_{\downarrow}^{j'_{\phi},\alpha_2}(\xi_2) =
(2\pi)^{-6} \int d{p}_{1} \int dp_2\,  
 e^{\II\sum_{i=1}^2 (p_i,\xi_i)}  \, 
\bar{\psi}^{j'_{\phi},\alpha_1}_{\uparrow}(p_1)  \bar{\psi}^{j'_{\phi},\alpha_2}_{\downarrow}(p_2)
 \EEQ
 or conjugate.
Insert inside the last expression the kernel
$1=\chi^{j_+\to}_+(p_1+p_2)+(1-\chi^{j_+\to}_+)(p_1+p_2)$.
Because $|\alpha_1+\alpha_{2}|=O(1)$, there are two cases: (i) either the
transfer momentum
$q:=p_1+p_{2}$ has bosonic scale $\simeq j_+$, and   $\chi^{j_+\to}_+(p_1+p_{2})>0$; or (ii) $|q|_+\ll 2^{-j_+}\mu$, and   $\chi_+^{j_+\to}(p_1+p_2)=0$. Hence
 \BEQ (\chi^{j_+\to}_+ F)(\xi_1,\xi_2) =
\int d\zeta 
 \chi_+^{j_+\to}(\zeta) F(\zeta+\xi_1,\zeta+\xi_2)
 \label{eq:zetaxi1xi2}
\EEQ
with a convolution kernel $\chi_+^{j_+\to}(\cdot)$ essentially
supported on $\{|\zeta|_+\lesssim 2^{j_+}\mu\}$. 

\Medskip As a result, $\P$ has been rewritten as the sum of two terms interpreted
as follows: 

(i) a polymer with {\bf upper transfer momentum scale} $j_+$ (term with $\chi^{j_+\to}_+$ 
in factor);

(ii) a polymer with {\bf lower transfer momentum scale} $>j_+$ (term with $1-\chi^{j_+\to}_+$ 
in factor).

\Bigskip
\noindent{\bf Step $\#2$. External leg displacement for four-point
polymers of type 2 with upper transfer momentum scale} (see 
Fig. \thesubsection.9).

\Medskip
Consider four-point polymers obtained in (i) just above. Displace the fields $\Psi_2$, resp. $\Psi_4$ on the lower line to the location
 of the fields $\bar{\Psi}_{1}$, resp.  $\bar{\Psi}_{3}$ on the upper line, see Fig.  \ref{subsection:Sigma}.5 and eq.
(\ref{eq:Pidisp}), 
   namely, replace $\Psi^{j'_{\phi},\alpha_2}(\xi_{2})$ by
 $e^{\II p^{\alpha_2}\cdot(\xi_{2}-\xi_{1})} \Psi^{j'_{\phi},\alpha_2}(\xi_{1})$ and similarly for $\Psi(\xi_{4})$.

\Bigskip

{\centerline{*********************}}

\Bigskip {\bf Step $\#3$. Scale $j'_{\phi}$ expansion.} 

\Medskip {\em One step of expansion.} Fermionic expansions
have produced scale $j'_{\phi}$ clusters, as described in
\S \ref{subsubsection:hor-cluster}. Consider two scale $j_{\phi}$ clusters, $\Del_1\in \P_1$, and $\Del\equiv \Del_2\in\P_2$, where $\P_1,\P_2$ are arbitrary polymers, {\em 
except} four-point polymers with low transfer momentum scale $>j'_{\phi}$.   Introduce an {\em interpolation coefficient} 
 $s^{\alpha}_{\Del_1,\Del_2}$
 between $\Del_1$ and  $\Del_2$, namely, multiply 
 $ \chi_{\Del_1}(\xi)\ \chi_{\Del_2}(\xi') C_{\theta}^{j'_{\phi},\alpha}(\xi,\xi')$ by $s^{\alpha}_{\Del_1,\Del_2}$
 ($\alpha\in\Z/2^{j'_{\phi}}\Z$); and differentiate $F_{I}(\vec{\xi}_{ext},\bar{\vec{\xi}}_{ext})$
 in (\ref{eq:HV-expansion}) w.r. to $s^{\alpha}_{\Del_1,\Del_2}$. The connecting operator associated
 to the  cluster link $\ell=(\Del_1,\Del_2)$ as in Proposition
 \ref{prop:BKAR}, 
\BEQ D_{\ell}^{\alpha}:=
\int _{\Del_1} d\xi_{\ell}\, \int_{\Del_2} d\xi'_{\ell}\  C^{j'_{\phi},\alpha}(\xi_{\ell},\xi'_{\ell}) \frac{\del^2}{\del \Psi^{j'_{\phi},\alpha}(\xi_{\ell}) \del \bar{\Psi}^{j'_{\phi},\alpha}(\xi'_{\ell})},
\EEQ
either (i) transforms a pair of dangling fields into a 
 cluster link, or produces (ii) one or (iii) two
extra vertices, one in $\Del_1$, one in $\Del_2$. The procedure may be iterated
an arbitrary number of terms by differentiating $F_{I}(\vec{\xi}_{ext},\bar{\vec{\xi}}_{ext})$ w.r. to
$s^{\alpha_1}_{\Del_1,\Del_2},s^{\alpha_2}_{\Del_1,\Del_2},\ldots$ We shall actually do so
at most $3$ times per pair of boxes (see below).

\Medskip We say that
 {\em $\Del_1$ and $\Del_2$ are  strongly connected at scale $j'_{\phi}$} if {\em there
 exist at least 3 cluster or complementary cluster links 
 $C^{j'_{\phi},\alpha}(\xi_1,\xi_2), C^{j'_{\phi},\alpha'}(\xi'_1,\xi'_2), 
 C^{j'_{\phi},\alpha''}(\xi''_1,\xi''_2)$ between 
 $\Del_1$ and $\Del_2$.} 

\Medskip We now introduce a linking number $LN(\Del_1,\Del_2)$,
which is initially zero, and is incremented as links are added. If there
is {\em one} link between $\Del_1$ and $\Del_2$, we set $LN(\Del_1,\Del_2):=1$.  If there are {\em two}
links, set $LN(\cdot,\cdot)=2$. 

\Medskip We repeat the above procedure until LN($\Del_1,\Del_2$)=$3$, so
that $\Del_1$ and $\Del_2$ are strongly connected.  By choosing
one of the terms in the cluster expansion in which an  interpolation
coefficient has been set to $0$, one also gets {\em factorized
expressions} of the form $G(\Del_1,\Del_2)$, times a quantity
averaged w.r. to a measure $ d\mu^*_{\theta}(s;\cdot)$, with
$s_{\Del_1,\Del_2}=0$, for which fields in $\Del_1$ and $\Del_2$ are independent one from the other. The quantity $G(\Del_1,\Del_2)$ is either 1 (in which case
$\Del_1$ and $\Del_2$ are totally factorized), a single
propagator, or a product of two propagators. In the latter case,  we  consider that we have
{\em one external Cooper pair propagator} connecting $\Del_1$ to $\Del_2$.  
 
\Medskip After completing the cluster expansion between $\Del_1$ and $\Del_2$, we
define a new set of objects as follows. First of all, we are
not really interested in the linking number between
two boxes belonging to two given polymers $\P_1,\P_2$, but
rather in the {\em total number of links connecting $\P_1$ and
$\P_2$} (see Figure in the introduction to section 
\ref{section:boson}), which is defined as
\BEQ LN(\P_1,\P_2):=\sum_{\Del_1\in\P_1}\sum_{\Del_2\in\P_2}
LN(\Del_1,\Del_2). \label{eq:LN}
\EEQ
{\em Then we stop the cluster expansion between $\P_1$ and $\P_2$
as soon as $LN(\P_1,\P_2)\ge 3$.} This mutualized stopping rule is
more economical, sufficient for our purposes and consistent
with the general organization of the expansion in terms of
polymers.

\Medskip Let us come back to the outcome of the first
step of expansion. As in \S \ref{subsubsection:hor-cluster},
we describe (super)-clusters, called {\em objects}, produced by connecting clusters.
 Scale $j'_{\phi}$ clusters
$\not=\Del_1,\Del_2$ are objects. If LN($\Del_1$,$\Del_2)\le 2$, then
clusters $\Del_1,\Del_2$ are also objects; otherwise a new object $o_1:=\{\Del_1,\Del_2\}$ has
been formed. The expansion goes on by considering links between the object
containing $\Del_1$ and clusters $\Del\not=\Del_1,\Del_2$. Explore in this way all
clusters $\Del\not=\Del_1$. In the end, one has an object $o_1$ containing $\Del_1$, and a
set of clusters which are not strongly connected to $o_1$. Then one starts from a new box
$\Del_2\not\subset o_1$ and explores in the same manner its links to clusters $\not=\Del_2,o_1$,
and so on. 
Here is one example.

\bigskip

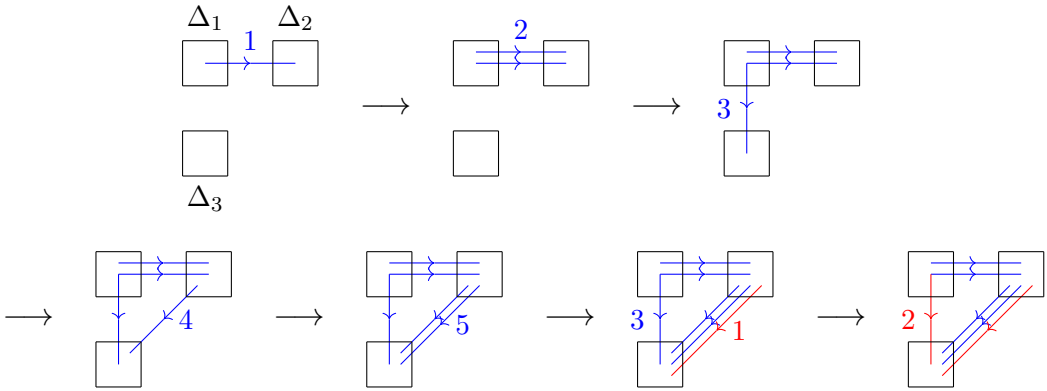
{\centerline{
\begin{tikzpicture}[scale=0.3]
\draw(0,2) node {\small $\Del_1$};
\draw(4,2) node {\small $\Del_2$}; 
\draw(0,-6) node {\small $\Del_3$};
\draw[->,color=blue](0,0)--(2,0); \draw[color=blue](2,0)--(4,0); \draw[color=blue](2,1)
node {\small 1};
\draw(-1,1)--(1,1); \draw(1,1)--(1,-1); \draw(1,-1)--(-1,-1);
\draw(-1,-1)--(-1,1);
\draw(-1+4,1)--(1+4,1); \draw(1+4,1)--(1+4,-1); \draw(1+4,-1)--(-1+4,-1);
\draw(-1+4,-1)--(-1+4,1);
\draw(-1,1-4)--(1,1-4); \draw(1,1-4)--(1,-1-4); \draw(1,-1-4)--(-1,-1-4);
\draw(-1,-1-4)--(-1,1-4);
\begin{scope}[shift={(12,0)}]
\draw(-4,-2) node {$\longrightarrow$};
\draw[->,color=blue](0,0)--(2,0); \draw[color=blue](2,0)--(4,0);
\draw[->,color=blue](0,0.5)--(2,0.5); \draw[color=blue](2,0.5)--(4,0.5);
\draw[color=blue](2,1.5) node {\small 2};
\draw(-1,1)--(1,1); \draw(1,1)--(1,-1); \draw(1,-1)--(-1,-1);
\draw(-1,-1)--(-1,1);
\draw(-1+4,1)--(1+4,1); \draw(1+4,1)--(1+4,-1); \draw(1+4,-1)--(-1+4,-1);
\draw(-1+4,-1)--(-1+4,1);
\draw(-1,1-4)--(1,1-4); \draw(1,1-4)--(1,-1-4); \draw(1,-1-4)--(-1,-1-4);
\draw(-1,-1-4)--(-1,1-4);
\end{scope}
\begin{scope}[shift={(24,0)}]
\draw(-4,-2) node {$\longrightarrow$};
\draw[->,color=blue](0,0)--(2,0); \draw[color=blue](2,0)--(4,0);
\draw[->,color=blue](0,0.5)--(2,0.5); \draw[color=blue](2,0.5)--(4,0.5);
\draw[->,color=blue](0,0)--(0,-2); \draw[color=blue](0,-2)--(0,-4);
\draw[color=blue](-1,-2) node {\small 3};
\draw(-1,1)--(1,1); \draw(1,1)--(1,-1); \draw(1,-1)--(-1,-1);
\draw(-1,-1)--(-1,1);
\draw(-1+4,1)--(1+4,1); \draw(1+4,1)--(1+4,-1); \draw(1+4,-1)--(-1+4,-1);
\draw(-1+4,-1)--(-1+4,1);
\draw(-1,1-4)--(1,1-4); \draw(1,1-4)--(1,-1-4); \draw(1,-1-4)--(-1,-1-4);
\draw(-1,-1-4)--(-1,1-4);
\end{scope}
\end{tikzpicture}}}

\bigskip

{\centerline{
\begin{tikzpicture}[scale=0.3]
\begin{scope}
\draw(-4,-2) node {$\longrightarrow$};
\draw[->,color=blue](0,0)--(2,0); \draw[color=blue](2,0)--(4,0);
\draw[->,color=blue](0,0.5)--(2,0.5); \draw[color=blue](2,0.5)--(4,0.5);
\draw[->,color=blue](0,0)--(0,-2); \draw[color=blue](0,-2)--(0,-4);
\draw[->,color=blue](3.5,-0.5)--(2,-2); \draw[color=blue](2,-2)--(0.5,-3.5);
\draw[color=blue](3,-2) node {\small 4};
 \draw(-1,1)--(1,1); \draw(1,1)--(1,-1); \draw(1,-1)--(-1,-1);
\draw(-1,-1)--(-1,1);
\draw(-1+4,1)--(1+4,1); \draw(1+4,1)--(1+4,-1); \draw(1+4,-1)--(-1+4,-1);
\draw(-1+4,-1)--(-1+4,1);
\draw(-1,1-4)--(1,1-4); \draw(1,1-4)--(1,-1-4); \draw(1,-1-4)--(-1,-1-4);
\draw(-1,-1-4)--(-1,1-4);
\end{scope}
\begin{scope}[shift={(12,0)}]
\draw(-4,-2) node {$\longrightarrow$};
\draw[->,color=blue](0,0)--(2,0); \draw[color=blue](2,0)--(4,0);
\draw[->,color=blue](0,0.5)--(2,0.5); \draw[color=blue](2,0.5)--(4,0.5);
\draw[->,color=blue](0,0)--(0,-2); \draw[color=blue](0,-2)--(0,-4);
\draw[->,color=blue](3.5,-0.5)--(2,-2); \draw[color=blue](2,-2)--(0.5,-3.5);
\draw[->,color=blue](4,-0.5)--(4-1.75,-4+1.75); \draw[color=blue](4-1.75,-4+1.75)--(0.5,-4);
\draw[color=blue](3.25,-2.25) node {\small 5};
\draw(-1,1)--(1,1); \draw(1,1)--(1,-1); \draw(1,-1)--(-1,-1);
\draw(-1,-1)--(-1,1);
\draw(-1+4,1)--(1+4,1); \draw(1+4,1)--(1+4,-1); \draw(1+4,-1)--(-1+4,-1);
\draw(-1+4,-1)--(-1+4,1);
\draw(-1,1-4)--(1,1-4); \draw(1,1-4)--(1,-1-4); \draw(1,-1-4)--(-1,-1-4);
\draw(-1,-1-4)--(-1,1-4);
\end{scope}
\begin{scope}[shift={(24,0)}]
\draw(-4,-2) node {$\longrightarrow$};
\draw[->,color=blue](0,0)--(2,0); \draw[color=blue](2,0)--(4,0);
\draw[->,color=blue](0,0.5)--(2,0.5); \draw[color=blue](2,0.5)--(4,0.5);
\draw[->,color=blue](0,0)--(0,-2); \draw[color=blue](0,-2)--(0,-4);\draw[color=blue](-1,-2)
node {\small 3};
\draw[->,color=blue](3.5,-0.5)--(2,-2); \draw[color=blue](2,-2)--(0.5,-3.5);
\draw[->,color=blue](4,-0.5)--(4-1.75,-4+1.75); \draw[color=blue](4-1.75,-4+1.75)--(0.5,-4);
\draw[->,color=red](4.5,-0.5)--(2.5,-2.5); \draw[color=red](2.5,-2.5)--(0.5,-4.5);
\draw[color=red](3.5,-2.5) node {\small 1};
\draw(-1,1)--(1,1); \draw(1,1)--(1,-1); \draw(1,-1)--(-1,-1);
\draw(-1,-1)--(-1,1);
\draw(-1+4,1)--(1+4,1); \draw(1+4,1)--(1+4,-1); \draw(1+4,-1)--(-1+4,-1);
\draw(-1+4,-1)--(-1+4,1);
\draw(-1,1-4)--(1,1-4); \draw(1,1-4)--(1,-1-4); \draw(1,-1-4)--(-1,-1-4);
\draw(-1,-1-4)--(-1,1-4);
\end{scope}
\begin{scope}[shift={(36,0)}]
\draw(-4,-2) node {$\longrightarrow$};
\draw[->,color=blue](0,0)--(2,0); \draw[color=blue](2,0)--(4,0);
\draw[->,color=blue](0,0.5)--(2,0.5); \draw[color=blue](2,0.5)--(4,0.5);
\draw[->,color=red](0,0)--(0,-2); \draw[color=red](0,-2)--(0,-4); \draw[color=red](-1,-2)
node {\small 2};
\draw[->,color=blue](3.5,-0.5)--(2,-2); \draw[color=blue](2,-2)--(0.5,-3.5);
\draw[->,color=blue](4,-0.5)--(4-1.75,-4+1.75); \draw[color=blue](4-1.75,-4+1.75)--(0.5,-4);
\draw[->,color=red](4.5,-0.5)--(2.5,-2.5); \draw[color=red](2.5,-2.5)--(0.5,-4.5);
\draw(-1,1)--(1,1); \draw(1,1)--(1,-1); \draw(1,-1)--(-1,-1);
\draw(-1,-1)--(-1,1);
\draw(-1+4,1)--(1+4,1); \draw(1+4,1)--(1+4,-1); \draw(1+4,-1)--(-1+4,-1);
\draw(-1+4,-1)--(-1+4,1);
\draw(-1,1-4)--(1,1-4); \draw(1,1-4)--(1,-1-4); \draw(1,-1-4)--(-1,-1-4);
\draw(-1,-1-4)--(-1,1-4);
\end{scope}
\end{tikzpicture}}}

\Medskip {\bf\tiny Fig. \thesubsection.13. Third order cluster graph.}

\Bigskip  Blue lines are cluster links. Because we expand to third order, we obtain
in general a graph, not a tree. However, it is easy to extract a tree out of this graph (in
red on the figure).  Decide that every third link between two objects is a tree link (red
link number 1). It may happen that, upon introducing a third link (here between 
$\Del_2$ and $\Del_3$), and forming the corresponding cluster (here $o_{1}=\{\Del_2,\Del_3\}$),
previously non strongly-connected objects become strongly connected, here $\Del_1$ and 
$o_1$. One then adds a tree link, here between $\Del_1$ and $o_1$; for the sake of
clarity, it is convenient to represent it as a link between $\Del_1$ and one of the
boxes of $o_1$, here (but it is an arbitrary choice) $\Del_3$. Thus a new cluster
$o_2=\{\Del_1,\Del_2,\Del_3\}$ is formed, and so on. These rules generate a tree.

\Bigskip {\bf C.  Complementary expansions of scales $j_+\ge j'_{\phi}+1$.} The series of steps Step $\#1 \longrightarrow$ Step $\#2\longrightarrow$ Step $\#3$ may be repeated at
scale $j_+=j'_{\phi}+1,j'_{\phi}+2,\cdots$ 
until all Cooper pairs of all scales have been connected. 
Note that the preliminary second-order expansion in
isolated boxes   (Step $\#0$) is performed only once, at
the beginning of the scale $j'_{\phi}$ complementary expansion.
As discussed above, Step $\#1$ (splitting phase for low
transfer momentum for four-point polymers) is applied to
Cooper pairs belonging to four-point polymers of type 2 (in
the sense of Definition \ref{def:type1type2}) which already have $1-\chi^{(j_+-1)\to}(p_1+p_2)$ in factor, therefore there is
no need to add a further type specification at scales $j_+>j'_{\phi}$.  Step $\#3$ is
somewhat simplified w.r. to the case $j_+=j'_{\phi}$, since (as explained above) there
is no need to test pairings between scale-neutral polymers any more; so the exploration
process starts exclusively from scale $j_+$ four-point polymers, 
whose sectors $\alpha$ range not in $\Z/2^{j'_{\phi}}\Z$, but
in $\Z/2^{j_+}\Z$.  Generally speaking, the scale $j_+$ complementary
expansion inserts scale $j_+$ Cooper pairs and the four-point polymers they belong to inside a multi-scale tree (see Fig.
\thesubsection.6).  Also, newly developed Goldstone
boson propagators of scales $\ge j'_{\phi}$ produce
inductively scale $j'_{\phi},j'_{\phi}+1,\ldots$ contributions to the  two-point function, e.g.

\bigskip

{\centerline{
\begin{tikzpicture}[scale=0.85]
\draw[color=blue](0,0)--(10,0);
\draw[color=blue](0,-1)--(10,-1);
\draw[color=blue](0,-6)--(10,-6);
\begin{scope}[shift={(-1,0)}]
\draw[dashed](2,-3)--(3,-3);
\draw(4,-3) circle(1); \draw(5,-3)--(6,-3);\draw(7,-3) circle(1); \draw[dashed](8,-3)--(9,-3);
\draw[dashed](4,-4)--(4,-7); \draw[dashed](7,-4)--(7,-7);
\draw[decorate,decoration=snake](4,-7)--(7,-7);
\end{scope}
\draw(9,-3) node {$j'_{\phi}$}; \draw(9,-7) node {$j_+$};
\end{tikzpicture}
}}

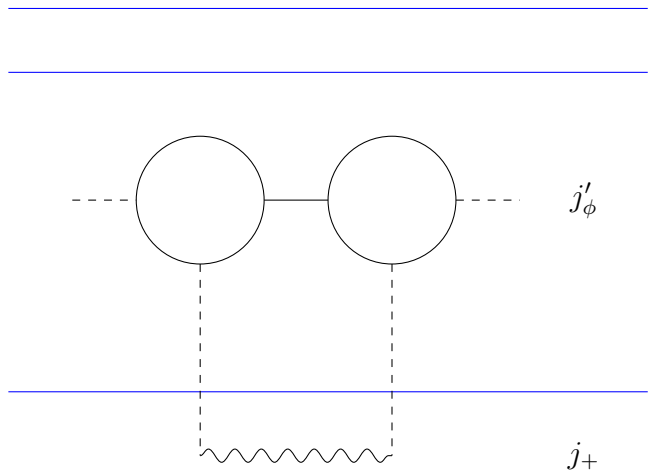
\Medskip {\tiny\bf Fig. \thesubsection.14. Scale $j_+$ dressing of the fermion propagator
by internal Goldstone boson propagators.}

\Bigskip yielding in the end of the story the dressed fermion propagator $C_{dressed}$.

\Bigskip {\bf D. Resummation of Goldstone boson propagators.}
 Once all Cooper pairs of all scales have been connected,
the remains at each scale $j_+\ge j'_{\phi}$ a sum of
structures as in Fig. \thesubsection.3 or  \thesubsection.4.
 However, it isn't 
yet in the form of a geometric series; rather of a non-factorized expression of the form 
\BEA && \tilde{\Sigma}_{disp}(q):=\lambda {\mathbb{1}}+ \lambda^2 \Big\{ {\cal A}_q(\Upsilon_3)+ \int dp'_1 \int dp'_3\ 
 {\cal A}_q(p'_1) \tilde{\Pi}_q^{disp}(p'_1,p'_3) {\cal A}_q(p'_3)
 \nonumber\\
&&
+\cdots+ \    \Big\{\prod_{i=1}^n \int dp_{i,1} \int dp_{i,3}\Big\} \ \Big( \prod_{i=1}^{n-1} \del(p_{i,3}-p_{i+3,1})\Big)\ 
 {\cal A}_q(p_{1,1})  \Big\{ \prod_{i=1}^n  \Big(\tilde{\Pi}^{disp}_q(p_{i,1},p_{i,3}){\cal A}_q(p_{i,3}) \Big)\Big\} \nonumber\\
 && + \cdots
\EEA
as in Definition \ref{def:Sigmadisp} (and \S \ref{subsection:Sigma} for notations), {\em except} that
${\cal A}_q(p_{i,3})$ is now computed as a product of two
{\em dressed} propagators, $C_{dressed}(p_{i,3})
C_{dressed}(-p_{i,3}+q)$. 
The displacement procedure of \S \ref{subsection:Sigma}, see eq. (\ref{eq:Pidisp}),  is defined in terms of 
isotropic "micro-sectors" ${\cal S}^{\alpha,k}\equiv{\cal S}^{j'_{\phi},\alpha,k}$ obtained by chopping
the angular sectors ${\cal S}^{\alpha}:={\cal S}^{j'_{\phi},\alpha}$ as in the Remark
following Definition \ref{def:Cjm}.  Error terms have by
(\ref{eq:jextjint}) a supplementary $O(2^{-(j_+ - j'_{\phi})})$ small pre-factor corresponding to the difference of scales
between the external legs, and the internal fermions, which are of scale $\le j'_{\phi}$. One must first
\begin{itemize}
\item[(i)] resum over the locations of all intermediate
scale-neutral polymers $\P_1,\ldots,\P_n$, implying in particular the possibility of overlaps between polymers.
This is a standard problem, solved  by means of a  Mayer
expansion as in \S \ref{subsubsection:Mayer}.
\end{itemize}

\Medskip Once this is settled, one must still take into account
the dependence of $\tilde{\Pi}^{disp}_q(p_{i,1},p_{i,3})$ in
$p_{i,1}$ and $p_{i,3}$. 
In order to extract factorized contributions, we mimic the
 procedure sketched in a perturbative context in \S \ref{subsection:Sigma} under the name of {\em Fermi and s-wave
 projections}. As suggested in Definition \ref{def:averaging}, 
 it is rather an {\bf averaging procedure}, a double one in fact. 
  we must get
rid of the dependence on the norms $|p_{i,1}|,|p_{i,3}|$, and
(if $q=0$) on the angles $\theta_{i,1}-\theta_{i,3}:=\widehat{\vec{p}_1,\vec{p}_3}$. (If $q\not=0$ then $\tilde{\Pi}^{disp}_q(p_{i,1},p_{i,3})$ depends
separately on $\theta_{i,1}$ and $\theta_{i,3}$).  Let ${\cal A}_{{\cal S}^{\alpha}}(p):=\chi^{j'_{\phi},\alpha}(p)C_{dressed}(p)C_{dressed}(q-p)$.

\begin{itemize}
\item[(ii)] {\em (norm averaging procedure)} The kernel $\tilde{\Pi}^{disp}_q(p_{i,1},p_{i,3})$ has a support
${\cal S}^{\alpha_{i,1}}\times {\cal S}^{\alpha_{i,3}}$ centered on $p^{\alpha_{i,1}}$, resp.
$p^{\alpha_{i,3}}$. 
We absorb the intermediate ${\cal A}_{{\cal S}^{\alpha}}$-functions by rewriting products
\BEA && \int_{{\cal S}^{\alpha_1}} dp_{1}\cdots \int_{{\cal S}^{\alpha_n}} dp_n\, {\cal A}_{{\cal S}^{\alpha_{1}}}(p_{1})
\tilde{\Pi}^{disp}(p_{1},p_{2}) {\cal A}_{{\cal S}^{\alpha_{2}}}(p_{2}) \nonumber\\
&&\qquad\qquad
\tilde{\Pi}^{disp}(p_2,p_3) {\cal A}_{{\cal S}^{\alpha_{3}}}(p_{3}) \cdots
\tilde{\Pi}^{disp}(p_{n-1},p_n) {\cal A}_{{\cal S}^{\alpha_{n}}}(p_{n})
\label{eq:ppppp}
\EEA
 as
\BEA && \Big\{ \prod_{i=1}^n\int dp_i\, \Big\}\ 
({\cal A}_{{\cal S}^{\alpha_{1\ }}}(p_{1}))^{1/2} K^{\alpha_1,\alpha_2}(p_1,p_2) \nonumber\\
&&\qquad
K^{\alpha_2,\alpha_3}(p_2,p_3)\cdots K^{\alpha_{n-1},\alpha_n}(p_{n-1},p_n)  ({\cal A}_{{\cal S}^{\alpha_{n}}}(p_{n}))^{1/2},
\EEA
 where
\BEQ K^{\alpha_{i,1},\alpha_{i,3}}(p_{i,1},p_{i,3}):= ({\cal A}_{{\cal S}^{\alpha_{i,1}}}(p_{i,1}))^{1/2} \tilde{\Pi}^{disp}(p_{i,1},p_{i,3}) ({\cal A}_{{\cal S}^{\alpha_{i,3}}}(p_{i,3}))^{1/2}.
\EEQ

 Define $P^{\alpha}$, $\alpha=\alpha_{i,1},\alpha_{i,3}$ to
be the projection  on the  function $({\cal A}_{{\cal S}^{\alpha}})^{-1/2}$ w.r. to the $L^2$-norm 
$(f,g)_{L^2({\cal A}_{{\cal S}^{\alpha}}(p)dp)}:=\int dp\, {\cal A}_{{\cal S}^{\alpha}}(p) f(p)g^*(p)$ in momentum
space,
\BEQ P^{\alpha}(f)(p):=\frac{\int dp'\, f(p') ({\cal A}_{{\cal S}^{\alpha}}(p'))^{1/2}}{\Vol({\cal S}^{\alpha})} ({\cal A}_{{\cal S}^{\alpha}}(p))^{-1/2}, \qquad \supp(f)\subset {\cal S}^{\alpha}
\EEQ
Let $\Vol({\cal S})$ be the volume of ${\cal S}^{\alpha}$ for any sector $\alpha$. Then a short-hand for (\ref{eq:ppppp}) is  simply
\BEQ 
\Big(  \frac{({\cal A}_{{\cal S}^{\alpha_{1}}})^{-1/2}}
{\Vol({\cal S})} \otimes \frac{({\cal A}_{{\cal S}^{\alpha_{n}}})^{-1/2}}
{\Vol({\cal S})} , K^{\alpha_1,\alpha_2}\cdots K^{\alpha_{n-1},\alpha_n} \Big)_{L^2({\cal A}_{{\cal S}^{\alpha_1}}(p)dp)
\otimes L^2({\cal A}_{{\cal S}^{\alpha_n}}(p)dp)}.
\label{eq:KKKKK1}
\EEQ

Define
\BEA &&K^{\alpha_{i,1},\alpha_{i,3}}_{0|0}(p_1,p_3):=(P^{\alpha_{i,1}}\otimes P^{\alpha_{i,3}})(K)(p_1,p_3)\nonumber\\
&&\qquad= \frac{ 
\int_{{\cal S}^{\alpha_{i,1}}} dp'_1\, \int_{{\cal S}^{\alpha_{i,3}}} dp'_3\,  K(p'_1,p'_3) ({\cal A}_{{\cal S}^{\alpha_{i,1}}}(p'_1))^{1/2}
 ({\cal A}_{{\cal S}^{\alpha_{i,3}}}(p'_3))^{1/2} 
}{(\Vol({\cal S}))^2} \ \times \nonumber\\
&&\qquad\qquad \times\   ({\cal A}_{{\cal S}^{\alpha_{i,1}}}(p_1))^{-1/2} ({\cal A}_{{\cal S}^{\alpha_{i,3}}}(p_3))^{-1/2}
 \EEA
 $(p_1\in{\cal S}^{\alpha_{i,1}},p_3\in{\cal S}^{\alpha_{i,3}})$;
\BEA &&K^{\alpha_{i,1},\alpha_{i,3}}_{1|0}(p_1,p_3):=\Big((1-P^{\alpha_{i,1}}\otimes P^{\alpha_{i,3}}\Big)(K)(p_1,p_3) \nonumber\\
&&\ \  \frac{ \int_{{\cal S}^{\alpha_{i,1}}} dp'_1\, \int_{{\cal S}^{\alpha_{i,3}}} dp'_3\,  
\Big\{K(p_1,p'_3) ({\cal A}_{{\cal S}^{\alpha_{i,1}}}(p'_1))^{-1/2}-K(p'_1,p'_3) ({\cal A}_{{\cal S}^{\alpha_{i,1}}}(p_1))^{-1/2}\Big\}
}{(\Vol({\cal S}))^2} \ \times \nonumber\\
&&\qquad\qquad ({\cal A}_{{\cal S}^{\alpha_{i,3}}}(p_3))^{-1/2}
\EEA 
and a similar formula by symmetry
for $K^{\alpha_{i,1},\alpha_{i,3}}_{0|1}(p_1,p_3):=\Big(P^{\alpha_{i,1}}\otimes (1-P^{\alpha_{i,3}})\Big)(K)(p_1,p_3)$; and finally,
\BEA  &&K^{\alpha_{i,1},\alpha_{i,3}}_{1|1}(p_1,p_3):=\Big((1-P^{\alpha_{i,1}})\otimes (1-P^{\alpha_{i,3}})\Big)(K)(p_1,p_3)
\nonumber\\
&&\qquad  = \frac{1}{(\Vol({\cal S})^2}  
\int_{{\cal S}^{\alpha_{i,1}}} dp'_1\, \int_{{\cal S}^{\alpha_{i,3}}} dp'_3\nonumber\\
&&\qquad   \Big\{
\Big( (K(p_1,p_3) ({\cal A}_{{\cal S}^{\alpha_{i,1}}}(p'_1))^{-1/2}  -K(p'_1,p_3) ({\cal A}_{{\cal S}^{\alpha_{i,1}}}(p_1))^{-1/2}  
 \Big) \, ({\cal A}_{{\cal S}^{\alpha_{i,3}}}(p'_3))^{-1/2}
 \nonumber\\
&&\qquad - \Big( (K(p_1,p'_3) ({\cal A}_{{\cal S}^{\alpha_{i,1}}}(p'_1))^{-1/2}  -K(p'_1,p'_3) ({\cal A}_{{\cal S}^{\alpha_{i,1}}}(p_1))^{-1/2}  
 \Big) \, ({\cal A}_{{\cal S}^{\alpha_{i,3}}}(p_3))^{-1/2}
 \Big\}. \nonumber\\
\EEA

\Medskip By $L^2$-orthogonality, (\ref{eq:KKKKK1}) reduces
to a sum over chains $0=\eps_1\to\eps_2\to\eps_{n-1}\to \eps_n=0$ of intermediate $0-1$ states,
\BEA && \sum_{\eps_2,\ldots,\eps_{n-1}=0,1} \Big(  \frac{({\cal A}_{{\cal S}^{\alpha_{1}}})^{-1/2}}
{\Vol({\cal S})} \otimes \frac{({\cal A}_{{\cal S}^{\alpha_{n}}})^{-1/2}}
{\Vol({\cal S})} ,  \nonumber\\
&&\ 
K_{0|\eps_2}^{\alpha_1,\alpha_2}
K_{\eps_2|\eps_3}^{\alpha_2,\alpha_3}\cdots 
K_{\eps_{n-2}|\eps_{n-1}}^{\alpha_{n-2},\alpha_{n-1}} K_{\eps_{n-1}|0}^{\alpha_{n-1},\alpha_n} \Big)_{L^2({\cal A}_{{\cal S}^{\alpha_1}}(p)dp)
\otimes L^2({\cal A}_{{\cal S}^{\alpha_n}}(p)dp)}.
\label{eq:KKKKK2}
\EEA
Also -- and this is the crucial point --, the covariance kernels are nearly constant,
\BEQ \frac{1}{\II p^0-e^*(p)-{\mathbb{\Gamma}}}=-\frac{1}{{{\mathbb{\Gamma}}}}(1+O(2^{-(j'_{\phi}-j_{\phi})})), \EEQ
hence all kernels except
$K_{0|0}$ have a small prefactor, namely, $O(2^{-(j'_{\phi}-j_{\phi})})$ for
$K_{1|0},K_{0|1}$, and $O(2^{-2(j'_{\phi}-j_{\phi})})$ for
$K_{1|1}$. 
We may therefore   {\em resum all contributions} corresponding
to chains $0=\eps_1\to 1\to 1\cdots\to 1\to \eps_n=0$ of
length $n\ge 3$, i.e. including at least one intermediate
state $1$, and reconsider them as {\em scale-neutral four-point
polymers with a small factor} 
\BEQ g':=2^{-2(j'_{\phi}-j_{\phi})}. \label{eq:g'}
\EEQ
Thus we are left (up to a redefinition of $K_{0|0}^{\alpha_1,\alpha_2}$ including these chains, for which we choose not to introduce
a new notation) with purely $0\to 0\cdots\to 0$ chains: {\em there is no norm dependence any more}.  

\item[(iii)] {\em (angle averaging procedure)} One must
still a priori sum over intermediate sector variables
$\alpha_2,\ldots,\alpha_{n-1}$. {\em Assume first that 
$q=0$.} Then $K^{\alpha_i,\alpha_{i+1}}_{0|0}=K^{\beta_i}_{0|0}$, $\beta_i:=\alpha_i-\alpha_{i+1}$  
depends only  (by global rotation invariance) on angle differences. Using a discrete Fourier
transform, $f(\beta)=\sum_{k=0}^{2^{j_+}-1} \hat{f}(k) e^{-2\II
k\beta/2^{j_+}}$, so that 
\BEQ \sum_{\alpha_2,\ldots,\alpha_n} K^{\alpha_1,\alpha_2}_{0|0}\ldots K^{\alpha_{n-1},\alpha_n}_{0|0}=\sum_{k_1,\ldots,k_{n-1}} \hat{K}_{0|0}^{k_1}\ldots
\hat{K}_{0|0}^{k_{n-1}},  \label{eq:KK-angle-dependence}
\EEQ
   we immediately see by standard orthogonality properties of Fourier series that only  {\em diagonal terms} $k=k_1=\ldots=k_{n-1}$ contribute. The terms
$k=0,1,2,\ldots$ are called resp. $s$-, $p$-, $d$-wave
contributions, following the standard chemists' notation for
angular momentum. Dominant contributions here are due to
bubbles, which are in the $s$-wave because the interaction
$\int (\bar{\psi}_{\uparrow}\psi_{\uparrow})(\bar{\psi}_{\downarrow} \psi_{\downarrow})$ 
itself is. Hence the sum of all contributions other than
that in the $s$-wave has a supplementary factor $O(g)$, and
may be reconsidered as in (ii) as   {\em scale-neutral four-point
polymers with a small factor} $O(g)$. Thus there remains
only the $s$-wave contribution.

\end{itemize}

One then obtains a geometric series 
$ {\cal A}+{\cal A}\bar{\Pi}{\cal A}+{\cal A}\bar{\Pi}{\cal A}\bar{\Pi}{\cal A}+\ldots $ resummed as the
{\em single-scale Goldstone boson propagator.}

\Medskip An easy supplementary step consists in \label{eq:p.124}
\begin{itemize}
\item[(iv)] resumming single-scale Goldstone boson propagators connecting a
polymer to itself. Namely, these tadpole like bosonic insertions are convergent.
\end{itemize}

\begin{itemize}
\item[(v)]

The last step consists in resumming by scale induction  two-point functions
of single-scale Goldstone boson propagators. For instance, 
the following scale $j_{+,3}$ two-point function may be built out of the
bosonic tree of Fig. \ref{subsection:complementary}.6
"glued" with its mirror image,

\begin{tikzpicture}[scale=0.5]
\draw(6.3,0) circle(1);
\draw[dashed](6.3+0.707,-0.707)--(6.3+0.707,-6);
\draw[dashed](6.3,-1)--(6.3,-11);
\draw[dashed](6.3-0.707,-0.707)--(6.3-0.707,-16);
\draw[dotted,decorate,decoration=snake](6.3+0.707,-6)--(1+9.3+0.707,-6);
\draw[dotted,decorate,decoration=snake](6.3,-11)--(2.5+9.3,-11);
\draw[dotted,decorate,decoration=snake](6.3-0.707,-16)--(3.3-0.707,-16);
\draw(17,-6) node {$j_{+,1}$};
\draw[color=blue](0,-14)--(15,-14);
\draw[color=blue](0,-9)--(15,-9);
\draw[color=blue](0,-4)--(15,-4);
\draw(17,-11) node{$j_{+,2}$};
\draw(4+0.8+12.2,-10-6) node{$j_{+,3}$};
\begin{scope}[shift={(18,0)},xscale=-1]
\draw(6.3,0) circle(1);
\draw[dashed](6.3+0.707,-0.707)--(6.3+0.707,-6);
\draw[dashed](6.3,-1)--(6.3,-11);
\draw[dashed](6.3-0.707,-0.707)--(6.3-0.707,-16);
\draw[dotted,decorate,decoration=snake](6.3-0.707,-16)--(3.3-0.707,-16);
\draw[color=blue](0,-14)--(15,-14);
\draw[color=blue](0,-9)--(15,-9);
\draw[color=blue](0,-4)--(15,-4);
\end{scope}
\end{tikzpicture}
\end{itemize}

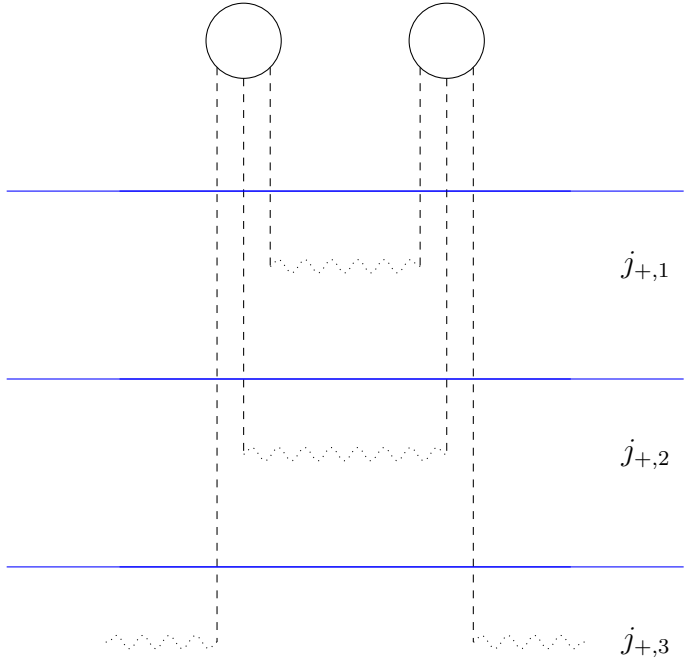
\Medskip {\bf \tiny Fig. \thesubsection.15. Two-point
function of the single-scale Goldstone boson propagator.}

\Bigskip Resumming the geometric series of such diagrams
with lowest transfer momentum scale $<j_{+,3}$ yields the
Goldstone boson propagator, denoted by a wiggling line
\begin{tikzpicture} \draw[decorate,decoration=snake](0,0)--(2,0); 
\end{tikzpicture}

\Bigskip The final outcome of this series of expansion 
is a multi-scale tree $\T$, with {\em scale-neutral polymers}
as vertices, and {\em Goldstone boson propagators} as edges.
Vertices have an arbitrary coordination number $n\ge 3$. 
The case $n=1$ is excluded by the gap equation; the  case $n=2$ is excluded since polymers in sandwich between
two Goldstone boson propagators have already been included
in the geometric series. The only cases when $n=1$ or $2$ are
allowed is when external legs -- either isolated fermions
$\Psi_{ext}=\Psi(\xi_{ext})$ or Cooper pairs $(\bar{\Psi}{\mathbb{\Gamma}}\Psi)_{ext}=(\bar{\Psi}{\mathbb{\Gamma}}\Psi)(\xi_{ext})$ --
are connected to the polymer. See Figure below.

\bigskip

\begin{tikzpicture}[scale=0.7]
\draw[color=blue](0,0.2)--(21,0.2);
\draw[color=blue](0,-0.2)--(21,-0.2);
\draw(4-0.707,2+0.707) node {$\times$}; 
\draw(4-0.707,2+0.707)--(4-2*0.707,2+2*0.707);
\draw(4-2*0.707,2+2*0.707+0.3) node{$\Psi_{ext}$};
\draw(4,2) circle(1); \draw(8,2) circle(1);
\draw(8,3)--(8,4); \draw(8,3) node {$+$};
 \draw(8,4.4) node {$\Psi_{ext}$}; \draw(12,2) 
circle(1); \draw(16,2) circle(1);
\draw[dotted](4+0.707,2-0.707)--(4+0.707,-3+2-0.707);
\draw[dotted](8-0.5,2-0.866)--(8-0.5,-3+2-0.707);
\draw[decorate,decoration=snake](4+0.707,-3+2-0.707)--(8-0.5,-3+2-0.707);
\draw[dotted](8+0.5,2-0.866)--(8+0.5,-6+2-0.866); 
\draw[dotted](12-0.5,2-0.866)--(12-0.5,-6+2-0.866);
\draw[decorate,decoration=snake](8+0.5,-6+2-0.866)--(12-0.5,-6+2-0.866); 
\draw[dotted](8,1)--(8,-9+1);
\draw[dotted](12+0.5,2-0.866)--(12+0.5,-9+1);
\draw[decorate,decoration=snake](8,-9+1)--(12+0.5,-9+1);
\draw[dotted](12+0.25,2-0.9)--(12+0.25,-12+2-0.9);
\draw[dotted](16,1)--(16,-12+2-0.9);
\draw[decorate,decoration=snake](12+0.25,-12+2-0.9)--(16,-12+2-0.9);
\draw[dotted](12-0.25,2-0.9)--(12-0.25,-15+2-0.9);
\draw[dotted](4-0.707,2-0.707)--(4-0.707,-15+2-0.9);
\draw[decorate,decoration=snake](12-0.25,-15+2-0.9)--(4-0.707,-15+2-0.9);
\draw(16-0.707,2+0.707) node {$\times$};
\draw(16+0.707,2+0.707) node {$\times$};
\draw(16-0.707,2+0.707)--(16,3.5);
\draw(16+0.707,2+0.707)--(16,3.5);
\draw(16,3.8) node {$(\bar{\Psi}{\mathbb{\Gamma}}\Psi)_{ext}$};
\end{tikzpicture}

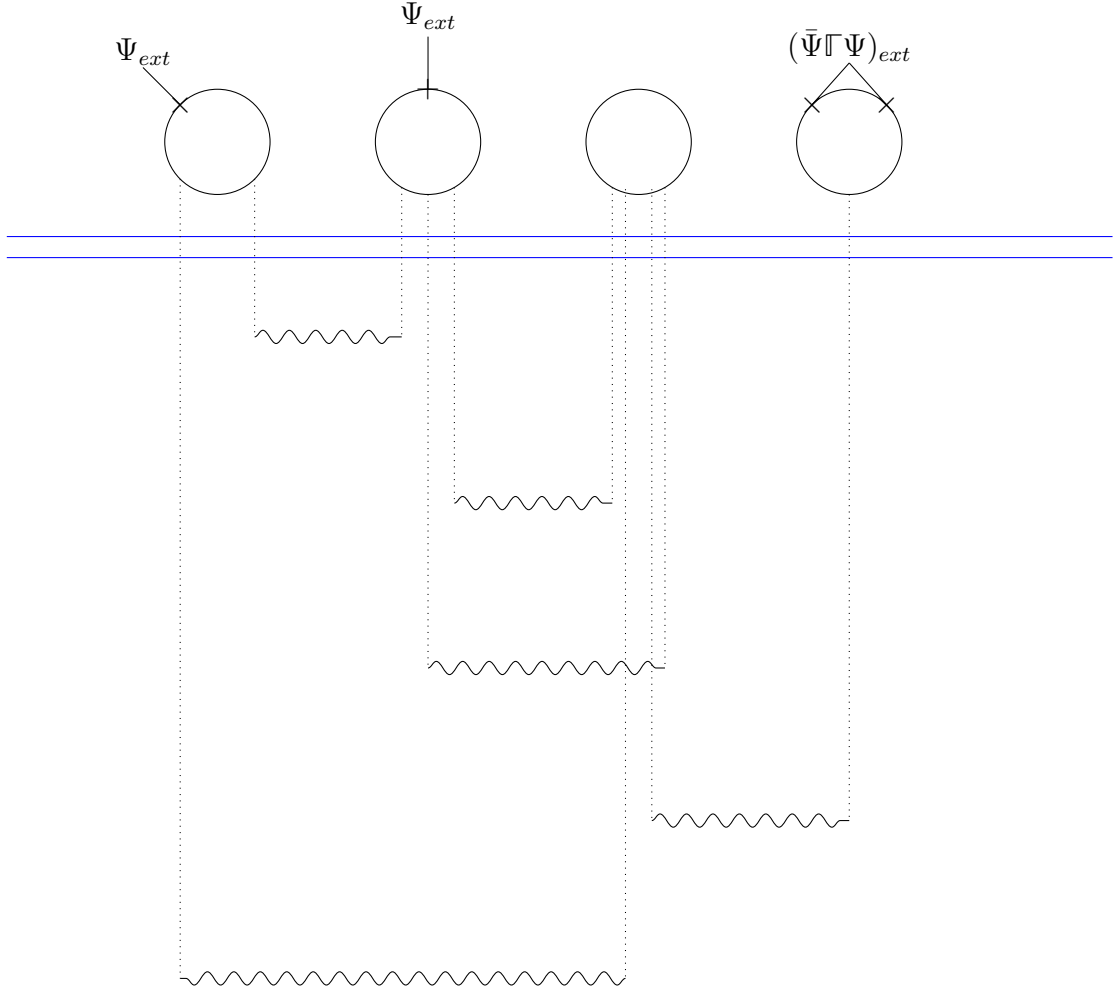
\Medskip {\bf \tiny Fig. \thesubsection.16. Final
bosonic multi-scale tree.}

\vskip 2cm
\noindent
 {\bf E. General principle of the bounds.}

\Medskip The next lines are concerned with the power-counting
of a polymer $\P_i^{j_+\to}$ with Cooper pair external structure 
\BEQ \prod_{i=1}^{N_{ext}} \int d\zeta_i\, \chi_+^{k_{+,i}}(\zeta_i)
\bar{\Psi}^{j'_{\phi},\alpha_{i,1}}(\zeta_i+\xi_i){\mathbb{\Gamma}}_i \Psi^{j'_{\phi},\alpha_{i,2}}(\zeta_i+\xi_i),
\label{eq:Cooper-pair-external-structure}
\EEQ
with  $k_{+,1},\ldots,k_{+,N_{ext}}>j_+$,
  $\alpha_{i,1},\alpha_{i,2}\in\Z/2^{k_{+,i}}\Z$, see
 splitting phase, step \#1 (\ref{eq:zetaxi1xi2}). As in the fermionic scales, see
\S \ref{subsection:cluster}, one is led to separate the
{\em absolute power-counting}, obtained by assuming
that $k_{+,1}=\ldots=k_{+,N_{ext}}=j_++1$, from the
{\em relative power-counting}, i.e. the power of the prefactor
in $2^{-(k_{+,min}-j_+)}$, where $k_{+,min}:=\min(k_{+,i},i=1,\ldots,N_{ext})$ is the highest external scale.

\Medskip It is instructive at this point to redo the power-counting
of \S \ref{subsubsection:single-scale-bounds} for a connected fermion diagram with $N_{ext,+}$ external Cooper pair
legs located at $\xi_1,\ldots,\xi_{N_{ext,+}}$, and with scales $k_{+,1},\ldots,k_{+,N_{ext}}$, i.e. (considering the Fourier transformed
graph) such that transfer momenta $|q_i|_+\approx \mu 2^{-j_{+,i}}$. The worst case is when  Cooper pairs are connected by Goldstone
boson propagators $\Sigma(\xi-\xi')\approx \frac{g_{\phi}/v_{\phi}^2}{|\xi-\xi'|_+}$, scaling like $\big(\Gamma_{\phi} 2^{-j_+/2} \big)^2$.  Hence the power-counting of a Cooper
pair $\int d\zeta \, \chi_+^{k_+}(\zeta) \bar{\Psi}^{j'_{\phi},\alpha_1}(\zeta+\xi){\mathbb{\Gamma}}\Psi^{j'_{\phi},\alpha_2}(\zeta+\xi)= \sum_{\alpha_1,\alpha_2\in \Z/2^{k_+}\Z} \int d\zeta_i\, \chi_+^{k_{+}}(\zeta)
\bar{\Psi}^{j'_{\phi},\alpha_1}(\zeta+\xi){\mathbb{\Gamma}} \Psi^{j'_{\phi},\alpha_2}(\zeta+\xi)$, see (\ref{eq:Cooper-pair-external-structure}),  summed over all possible sectors scales, like $O(\Gamma_{\phi} 2^{-j_+/2})$; note that this
contrasts with the naive scaling (\ref{eq:scaling-psijalpha}) for the product of two 
fermions, $\sum_{\alpha\in\Z/2^{j_+}\Z} \bar{\Psi}^{j_+,\alpha}(\xi)\Psi^{j_+,\alpha}(\xi)\approx 
2^{-3j_+/2}$.  If $j_+\gg j'_{\phi}$, then the $\xi$-integration in a box $\Del\in\D_+^{k_+}$ costs a volume factor $O(2^{3k_+})$.
Thus, the {\em relative power-counting} for a polymer in $\P^{j_+\to}$ with highest external scale $k_+> j_+$ is 
\BEQ O((2^{k_+-j_+})^{\omega(N_{ext,+})}), \qquad 
\omega(N_{ext}):=\half(6-N_{ext,+}) \label{eq:rel-PC},
\EEQ
$\omega(\cdot)$=relative degree of divergence,
 which implies that {\em only diagrams with $\le 6$ external Cooper pair legs are
potentially divergent when $k_+-j_+\to\infty$ with $j_+$
 fixed}. 
 
\Bigskip Let us now bound instead the {\em absolute} power-counting of {\em non-amputated polymers}
 (i.e. including external bosonic legs) for  $k_+-j_+=O(1)$ is fixed; the total power-counting of the full diagram for $k_+>j_+$ arbitrary is by construction the {\em product} of the relative
power-counting (\ref{eq:rel-PC}) by the absolute power-counting
which we shall presently bound. 

\Medskip We restrict here to the case of a {\em fermionic polymer} in ${\cal P}^{j\to}$ with $N_{ext,+}$ external bosonic legs, hence $N_{ext}=2N_{ext,+}$ external fermionic legs. The corresponding amputated power-counting (as shown in
(\ref{eq:GFMSB-P})) is 
$\lesssim (2^{j/2})^{2N_{ext,+}-4}=2^{j(N_{ext,+}-2)}$. 
Polymers in ${\cal P}^{j\to}$, $j<j'_{\phi}-1$ with $N_{ext}=2$
have been resummed into the fermionic two-point function, hence
we may assume that either  $j=j'_{\phi}-1$ or $N_{ext,+}\ge 2$;
thus,  $2^{j(N_{ext,+}-2)}\le 2^{j'_{\phi}(N_{ext,+}-2)}$, 
and the worst bound is obtained when $j_+=k_+=j'_{\phi}$, which
we now assume.
The power-counting of the corresponding {\em full} diagram
is obtained by multiplying by the weight of the  external bosonic
legs, and by the volume of a scale $j'_{\phi}$ box, all together (taking into account that $\Gamma_{\phi}\approx 2^{-j_{\phi}}\mu$) at most:
\BEQ 2^{j'_{\phi}(N_{ext,+}-2)} \times (2^{-j_{\phi}} 2^{-j'_{\phi}/2})^{N_{ext,+}} \times 2^{3j'_{\phi}} 
\le (2^{j'_{\phi}})^{-\frac{N_{ext,+}}{2}+1}. \label{eq:abs-PC}
\EEQ
For $N_{ext,+}\ge 3$ (excluding the case of a Goldstone boson
one-point function, or of a Goldstone boson propagator, for which $N_{ext,+}=2$)
this is small.

\Medskip  {\em Ward identities} -- proved in \S 
\ref{subsection:Ward} -- will  show that Goldstone boson
one-point function vanish, and that 
diagrams with $N_{ext,+}\le 6$ are actually $O(1)$. They
will also allow to compute the absolute power-counting
of non-amputated polymers in ${\cal P}^{j_+\to}$ including
scale $j'_{\phi}$ fermions.


\subsection{Ward identities}  \label{subsection:Ward}


 {\em The general purpose of this subsection is to obtain
supplementary {\em spring factors} $O(2^{-(k_+-j_+)})$  for external Cooper pair
legs of $N_{ext,+}$-point functions or bosonic polymers with 
lowest internal scale $j_+$ and highest external scale $k_+$.
}  As 
is apparent from the previous discussion in \S \ref{subsection:complementary} {\bf E.}, only a small (actually {\em two}, as we shall see), finite  number of spring factors is really required to show that
polymers converge.  In this subsection, we shall be uniquely concerned with {\em showing that the naive relative degree of
divergence $\omega(\P)=\half(6-N_{ext,+}(\P))$, $\P\in{\cal P}^{j_+\to}$ of (\ref{eq:rel-PC}) can be enhanced
to $\tilde{\omega}(\P)=\omega(\P)-\del\omega(\P)$, with 
$\del\omega(\P)$ (called: {\em relative gain}) large enough so that $\tilde{\omega}(\P)<0$.} 

\Bigskip {\bf In guise of introduction: gauge symmetries.}
The Grassmann measure is invariant under $U(1)$ number symmetry,
\BEQ (\bar{\psi},{\psi})(\xi)\longrightarrow 
(\bar{\psi}\,  e^{-\II\alpha},e^{\II\alpha}
\psi)(\xi) \EEQ
or equivalently
\BEQ (\bar{\Psi},{\Psi})(\xi)\longrightarrow 
(\bar{\Psi} e^{-\II \alpha\sigma^3}, e^{\II \alpha\sigma^3}\Psi
)(\xi) \EEQ
if $\alpha\in\R$ is a constant. Furthermore, the interaction $\lambda (\bar{\Psi}\Psi)^2(\xi)$ is invariant
under $U(1)$ gauge transformations 
\BEQ (\bar{\Psi},{\Psi})(\xi)\longrightarrow (
\bar{\Psi} e^{-\II \alpha(\xi)\sigma^3}, e^{\II \alpha(\xi)\sigma^3}\Psi
)(\xi) \EEQ
where now $\alpha=\alpha(\xi)$ is allowed to depend on space-time.

\Medskip 
Multiplying $\alpha$ by $\eps$ and letting $\eps\to 0$, we obtain
an infinitesimal transformation $\del_{\alpha}$ such that 
\BEQ \del_{\alpha}  (\bar{\Psi}\Psi)^2(\xi)=0 \EEQ
\BEA &&  \del_{\alpha} \Big(\int dp\, \bar{\Psi}(-p) \Big(\II p^0- e^*(\vec{p})\sigma^3 - {\mathbb{\Gamma}}(\theta) \Big)\Psi(p)\Big) \nonumber\\
&& \qquad\qquad=\int d\xi\, \alpha(\xi)\ \bar{\Psi}(\xi)  \Big\{ 2\alpha(\xi){\mathbb{\Gamma}}^{\perp}(\theta)-\Big(\partial_{\tau}\alpha(\xi)\, \sigma^3+\frac{1}{m^*}(
\vec{\nabla}\alpha
(\xi)\cdot\vec{\nabla})\,   {\mathbb{1}} \Big)
\Big\} 
\Psi(\xi) \nonumber\\  \label{eq:3.36}
\EEA
where ${\vec{\Gamma}}^{\perp}(\theta):=\left(\begin{array}{c} -\Gamma_2(\theta) \\
\Gamma_1(\theta) \end{array}\right)$ is the image of the vector $\vec{\Gamma}(\theta)\equiv \vec{\Gamma}^{//}(\theta)$
by a rotation of angle $\pi/2$, and ${\mathbb{\Gamma}}^{\perp}(\theta)$ the
associated off-diagonal Hermitian matrix. The inserted kernel
$\II p^0- e^*(\vec{p})\sigma^3 - {\mathbb{\Gamma}}(\theta)$ is
essentially the inverse of the fermion covariance kernel. This is true up to the ultra-violet cut-off  $\chi^{\to j_D}(|p|/\mu)$, however. {\em If $\alpha$ is constant}, then
the multiplication by $e^{\pm \II\alpha(\xi)\sigma^3}$
 commutes with Fourier cut-offs; however, this is not
 true for non-constant gauge transformations. This generates
 {\em error terms} denoted by "err." in the expressions below,
which will be shown  in \S \ref{subsection:error} to be of the same order or smaller
than the terms in (\ref{eq:3.36}).    Exponentiating the infinitesimal transformation
${\mathbb{\Gamma}}(\theta)\mapsto {\mathbb{\Gamma}}(\theta)+2\eps\alpha(\xi){\mathbb{\Gamma}}^{\perp}(\theta)$, one obtains a {\em rotated  order parameter}, ${\mathbb{\Gamma}}(\theta+2\alpha(\xi))$; equivalently (see (\ref{eq:equivariant})),  gauge
transformations locally rotate $\Gamma$-vectors, $\Gamma(\theta)\mapsto \Gamma(\theta+\alpha(\xi))$.

\Medskip  Applying 
gauge transformations specifically to products of Cooper pair fields\\
 $\prod_{i=1}^n \bar{\Psi}(\xi_i) {\mathbb{\Gamma}}(\theta_i)
{\Psi}(\xi_i)$ yields up to  error terms, denoted by "err.",  

\BEA &&  \Big\langle \del_{\alpha}\Big( \prod_{i=1}^n (\bar{\Psi}
{\mathbb{\Gamma}}(\theta_i)
\Psi)(\xi_i) \Big) \Big\rangle^{connected}_{\theta;\lambda} \nonumber\\
&&\ \ = 2\sum_{i=1}^n \alpha(\xi_i)\Big(\prod_{i'\not=i}  (\bar{\Psi}{\mathbb{\Gamma}}(\theta_{i'})\Psi)(\xi_{i'})\Big) (\bar{\Psi}{\mathbb{\Gamma}}(\theta_i)\Psi)(\xi_i)  \nonumber\\
&&\ \  = \int d\xi\  \Big\langle \Big(  \prod_{i=1}^n
(\bar{\Psi}   {\mathbb{\Gamma}}(\theta_i)
\Psi)(\xi_i) \Big) \ \cdot\ \nonumber\\
&&\qquad\cdot\   \bar{\Psi}(\xi) \,   \Big\{ -2\alpha(\xi) {\mathbb{\Gamma}}^{\perp}(\theta)+ \Big(\partial_{\tau}\alpha(\xi)\sigma^3+\frac{1}{m^*}(\vec{\nabla}\alpha(\xi)\cdot 
\vec{\nabla}){\mathbb{1}}\Big) \Big\}
\Psi(\xi) \Big\rangle^{connected}_{\theta;\lambda} \ + {\mathrm{err.}},  \nonumber\\  \label{eq:3.27}
\EEA
where $\del_{\alpha}$ acts as $\II\eps\alpha(\xi_i)\sigma^3$ on 
$\Psi(\xi_i)$, and as $-\II\eps\alpha(\bar{\xi}_i)\sigma^3$ on $\bar{\Psi}(\bar{\xi}_i)$. 

\Bigskip We are however interested in {\em infra-red cut-off}
quantities
obtained  by integrating Goldstone boson
propagators $\Sigma^{k_+}$ with scales $k_+\le j_+$, and
leaving out $\Sigma^{k^+}$-kernels with scales $k_+>j_+$. 
This introduces further error terms due to the infra-red cut-off on the $\Sigma$-kernel this time. Otherwise (\ref{eq:3.27})
still holds true with $\langle \ \cdot\ \rangle^{connected}_{\theta;\lambda}$ replaced by $\langle \ \cdot\ \rangle^{connected}_{\theta;\lambda,j_+\to}$.

\Medskip We choose to apply Ward identities to {\em  amputated
bosonic $n$-point functions}\\
 $f_n^{j_+\to}(\theta_1,\ldots,\theta_n;\xi_1,\ldots,\xi_n)$. By  definition, compare with
(\ref{eq:dphidphilog}) and (\ref{eq:zetaxi1xi2}),
\BEA && f_n^{j_+\to}(\theta_1,\ldots,\theta_n;\xi_1,\ldots,\xi_n)
:=\nonumber\\
&&\ \Big(\prod_{i=1}^n \int d\bar{\xi}_i\,  
\int d\zeta_i \, \chi_+^{j_+\to}(\zeta_i) \Big)\  
\Big(\prod_{i=1}^n\frac{\del}{\del\Phi(\zeta_i+\xi_i)}
\sigma(\theta_i) \frac{\del}{\del\bar{\Phi}(\zeta_i+\bar{\xi}_i)}  \Big)\Big|_{\Phi,\bar{\Phi}=0}
\log({\cal Z}_{\lambda}^{j_+\to}(\Phi,\bar{\Phi})), \nonumber\\ \label{eq:fnj+to}
\EEA
where $\log {\cal Z}_{\lambda}^{j_+\to}(\Phi,\bar{\Phi})$ is the generating function
of connected diagrams excluding four-point diagrams with  transfer momentum scale $>j_+$, and
additive sources, $\Psi\longrightarrow\Psi+\Phi,
\bar{\Psi}\longrightarrow\bar{\Psi}+\bar{\Phi}$.
The multi-scale cluster expansions of \S \ref{subsection:complementary} downto scale $j_+$ typically produce -- after displacing the $\bar{\Psi}$-legs
of each external Cooper pair  -- polymers with low-momentum
external $\Sigma_{\perp,\perp}$-kernels, summed as
\BEQ 
F_n^{j_+\to}(\theta+\frac{\pi}{2},\ldots,\theta+\frac{\pi}{2};\xi_1,\ldots,\xi_n):= \Big(\prod_{i=1}^n \Sigma_{\perp,\perp}^{k_{+,i}} (\xi_{ext,i},\xi_i)
\Big)  f_n^{j_+\to}(\theta+\frac{\pi}{2},\ldots,\theta+\frac{\pi}{2};\xi_1,\ldots,\xi_n)
  \label{eq:F}
\EEQ
 However, there is the possibility that some of the
Cooper pairs are in the $(//,//)$-channel, or that the two
 fermions composing the Cooper pair connect to two different
 polymers. These "massive" cases are treated in {\bf C.} below.
Using linear combinations, one may restrict to $\theta_i\in\theta+\frac{\pi}{2}\Z$ or even to $\theta_i\in\{\theta,\theta+\frac{\pi}{2}\}$.


\Bigskip {\bf A. One-point functions and gap equation.} Let us for a start
consider  the amputated
one-point function $f_1^{j_+\to}(\theta')$, $\theta'=\theta$ or $\theta+\frac{\pi}{2}$.
The latter function is conveniently rewritten as

\Bigskip

\begin{tikzpicture}[scale=0.5]
\draw(-2,0) arc(180:90:0.5 and 2);
\draw(-2,0) arc(180:270:0.5 and 2);
\draw(0,0) node {$-$};  
\draw(2,0)--(7,0);
\draw(4,2) node {$\times$};
\draw[dashed](4,2)--(4,-2);
\draw(3,-2.5)--(4,-2);
\draw(3,-1.5)--(4,-2);
 \draw(3,-2) node {\tiny $\sigma(\theta')$};
\draw(8.5,0) node {$+$};
\draw(10,0)--(16.5,0);
\draw[dashed](12,2)--(12,-2);
\draw(11,-2.5)--(12,-2); \draw(11,-1.5)--(12,-2); 
\draw(11,-2) node {\tiny $\sigma(\theta')$}; \draw(12,-2) node {\textbullet};
\draw[fill=gray](12,2) arc(180:-180:1.5 and 1);
\draw(18.5,0) arc(0:90:0.5 and 2);
\draw(18.5,0) arc(0:-90:0.5 and 2);
\draw(20.5,0) node {$+$};
\begin{scope}[shift={(4,-6)}]
\draw(-2,0) arc(180:90:0.5 and 2);
\draw(-2,0) arc(180:270:0.5 and 2);
\draw(0,0) node {$-$};  
\draw(2,0)--(7,0);
\draw[fill=gray](4,2) arc(180:-180:1.5 and 1);
\draw(6.5,2.5)--(7.5,2); \draw(6.5,1.5)--(7.5,2);
\draw(7.5,2) node {$\times$};
\draw[dashed](4,2)--(4,-2);
\draw(3,-2.5)--(4,-2);
\draw(3,-1.5)--(4,-2);
 \draw(3,-2) node {\tiny $\sigma(\theta')$};
\draw(8.5,0) node {$+$};
\draw(10,0)--(20.5,0);
\draw[dashed](12,2)--(12,-2);
\draw(11,-2.5)--(12,-2); \draw(11,-1.5)--(12,-2); 
\draw(11,-2) node {\tiny $\sigma(\theta')$}; \draw(12,-2) node {\textbullet};
\draw[fill=gray](12,2) arc(180:-180:1.5 and 1);
\draw(14.5,2.5)--(16.5,2.5); \draw(14.5,1.5)--(16.5,1.5);
\draw[fill=gray](15.5,2) arc(180:-180:1.5 and 1);
\draw(22.5,0) arc(0:90:0.5 and 2);
\draw(22.5,0) arc(0:-90:0.5 and 2);
\draw(24.5,0) node {$+\cdots$};
\end{scope}
\end{tikzpicture}

{\bf\tiny Fig. \thesubsection.1. Amputated one-point function.}
\Bigskip

where gray blobs \begin{tikzpicture}[scale=0.25] \draw[fill=gray](12,2) arc(180:-180:1.5 and 1);
\end{tikzpicture} denote strongly connected (see \ref{subsection:complementary}),  bosonic  one-point functions, excluding four-point 
subdiagrams with transfer momentum scale $>j_+$; and \begin{tikzpicture} \draw(0,0) node {$\times$}; \end{tikzpicture}
are two-point vertex insertions coming from off-diagonal propagators computed in the $\sigma(\theta')$-channel and with the
$\Gamma$-coefficient computed at scale $j_+$, therefore evaluated as $\Gamma^{j_+\to}$ when $\theta'=\theta$, resp. $0$ when $\theta'=\theta+\frac{\pi}{2}$.

\Medskip {\em Consider to begin with the one-point function in the parallel $(//)$ direction
$\theta'=\theta$.} Setting to zero the first line of Figure  \thesubsection.1

\Bigskip

\begin{tikzpicture}[scale=0.5]
\draw(-2,0) arc(180:90:0.5 and 2);
\draw(-2,0) arc(180:270:0.5 and 2);
\draw(0,0) node {$-\Gamma^{j_+\to}$};  
\draw(8.5,0) node {$+$};
\draw(10,0)--(16.5,0);
\draw[dashed](12,2)--(12,-2);
\draw(11,-2.5)--(12,-2); \draw(11,-1.5)--(12,-2); 
\draw(11,-2) node {\tiny $\sigma(\theta)$}; \draw(12,-2) node {\textbullet};
\draw[fill=gray](12,2) arc(180:-180:1.5 and 1);
\draw(18.5,0) arc(0:90:0.5 and 2);
\draw(18.5,0) arc(0:-90:0.5 and 2);
\draw(20.5,0) node {$=0$};
\end{tikzpicture}

{\bf\tiny Fig. \thesubsection.2 Gap equation, one-point version.}

\Bigskip

yields an implicit equation for $\Gamma^{j_+\to}$,

\begin{Definition}[gap equation, one-point version] \label{def:gap-one-fct}
\BEQ -\Gamma^{j_+\to}+I^{j_+\to}_{s.c.,//}=0, \EEQ
\end{Definition}
$I^{j_+\to}_{s.c.,//}$ being the sum of strongly connected (s.c.), bosonic one-point function diagrams
(drawn above as a gray blob),
 which is -- as we shall presently prove,
using a Ward identity --  exactly equivalent in the infra-red limit to its
{\em two-point} version  (a refined version of  Definition \ref{def:gap}) stating that the Goldstone
boson propagator has a pole in the transverse $(\perp)$ direction. Let us note
that, provided the above gap equation is verified, {\em all lines
} in Fig \thesubsection.1 vanish {\em up to error terms due to external leg
displacements} which are at most of order $2^{-(j_+-j'_{\phi})}$, hence vanish in the infra-red limit ($j_+\to+\infty$), e.g.

\Bigskip

\begin{tikzpicture}[scale=0.5]
\draw(-2,0) arc(180:90:0.5 and 2);
\draw(-2,0) arc(180:270:0.5 and 2);
\draw(0,0) node {$-$};  
\draw(2,0)--(7,0);
\draw[fill=gray](4,2) arc(180:-180:1.5 and 1);
\draw(6.5,2.5)--(7.5,2); \draw(6.5,1.5)--(7.5,2);
\draw(7.5,2) node {$\times$};
\draw[dashed](4,2)--(4,-2);
\draw(3,-2.5)--(4,-2);
\draw(3,-1.5)--(4,-2);
 \draw(3,-2) node {\tiny $\sigma(\theta')$};
\draw(8.5,0) node {$+$};
\draw(10,0)--(20.5,0);
\draw[dashed](12,2)--(12,-2);
\draw(11,-2.5)--(12,-2); \draw(11,-1.5)--(12,-2); 
\draw(11,-2) node {\tiny $\sigma(\theta')$}; \draw(12,-2) node {\textbullet};
\draw[fill=gray](12,2) arc(180:-180:1.5 and 1);
\draw(14.5,2.5)--(16,2); \draw(14.5,1.5)--(16,2);
\draw[dashed](16,2)--(16,4);
\draw[fill=gray](16,4) arc(180:-180:1.5 and 1);
\draw(22.5,0) arc(0:90:0.5 and 2);
\draw(22.5,0) arc(0:-90:0.5 and 2);
\draw(24.5,0) node {$=0.$};
\end{tikzpicture}

\Bigskip {\em Next,} we apply to the one-point function
in the parallel direction, see Fig. \thesubsection.2, 
the global (constant) Ward identity associated
to $\alpha\equiv 1$. We immediately obtain
\begin{tikzpicture}[scale=0.5]
\draw(10,0)--(16.5,0);
\draw[dashed](12,2)--(12,-2);
\draw(11,-2.5)--(12,-2); \draw(11,-1.5)--(12,-2); 
\draw(11-0.7,-2) node {\tiny $\sigma(\theta+\frac{\pi}{2})$}; \draw(12,-2) node {\textbullet};
\draw[fill=gray](12,2) arc(180:-180:1.5 and 1);
\draw(20.5,0) node {$=0$,};
\end{tikzpicture}
in other words, 
\BEQ \boxed{f_1^{j_+\to}(\theta+\frac{\pi}{2})=0:}
\label{eq:one-pt-fct-perp} \EEQ
{\em the one-point function in the perpendicular $(\perp)$ direction vanishes.} 

\Bigskip Finally, {\em we consider the sum $I^{j_+\to}_{s.c.,\perp}$ of strongly connected (s.c.) contributions to the  one-point function $f_1^{j_+\to}(\theta+\frac{\pi}{2})$ in the perpendicular 
direction $(\perp)$.} Applying to it likewise the global (constant) Ward identity associated
to $\alpha\equiv 1$, we get

\bigskip

\begin{tikzpicture}[scale=0.5]
\draw(-2,0) arc(180:90:0.5 and 2);
\draw(-2,0) arc(180:270:0.5 and 2);
\draw(0,0) node {$-$};  
\begin{scope}[shift={(-1,0)}]
\draw(2,0)--(7,0);
\draw[fill=gray](4,2) arc(180:-180:1.5 and 1);
\draw[dashed](4,2)--(4,-2);
\draw(3,-2.5)--(4,-2);
\draw(3,-1.5)--(4,-2);
 \draw(3,-2) node {\tiny $\sigma(\theta)$};
 \end{scope}
\draw(8.5,0) node {\small $+\int d\xi$};
\draw(10,0)--(20.5,0);
\draw[dashed](12,2)--(12,-2);
\draw(11,-2.5)--(12,-2); \draw(11,-1.5)--(12,-2); 
\draw(10.2,-2) node {\tiny $\sigma(\theta+\frac{\pi}{2})$};  \draw(12,-2) node {\textbullet};
\draw[fill=gray](12,2) arc(180:-180:1.5 and 1);
\draw(14.7,2.5)--(16.5,2); \draw(14.7,1.5)--(16.5,2);
\draw(14.7,2.5) node {\small\textbullet}; \draw(14.7,1.5) node {\small\textbullet};
\draw(14.7,3) node {\small $\xi_3$}; \draw(14.7,1) node {\small $\xi_4$};
\draw(15.7,2) node {\tiny  ${\mathbb{\Gamma}}^{\perp}$};
\draw(16.5,2) node{\small\textbullet}; \draw(17,2) node {\small $\xi$}; 
\draw(22.5,0) arc(0:90:0.5 and 2);
\draw(22.5,0) arc(0:-90:0.5 and 2);
\draw(24.5,0) node {$=0$};
\end{tikzpicture}
  
\Medskip {\bf \tiny Fig. \thesubsection.3. Ward identity connecting one- and two-point functions of the
Goldstone boson.}

\Bigskip The left term is  immediately seen to be $I^{j_+\to}_{s.c.,//}$, which is
equal to $\Gamma^{j_+\to}$ by Definition \ref{def:gap-one-fct}. On the other hand,
the gray blob in the right term is now a strongly connected {\em four-point} diagram. Displacing its external leg $\xi_4$ to $\xi_3$ following the displacement procedure of
the previous subsection yields a product of a $\bar{\Pi}$-kernel by a bubble in the
$(\perp,\perp)$-channel. The error term due to the displacement procedure has  
by symmetry an extra spring factor $2^{-2(j_+-j'_{\phi})}$.  One has found:
\BEQ \bar{\Pi}^{j_+\to}_{\perp,\perp}(0) {\cal A}(\Gamma^{j_+\to},\Upsilon_3)_{\perp,\perp}(0)=O(g^2 2^{-2(j_+-j'_{\phi})}).
\EEQ
As asserted above, this is -- up to error terms in $O(q^2)$ -- equivalent to the
gap equation of \S \ref{subsection:Sigma}. For convenience, {\em we shall define $\Gamma^{j_+\to}$ in 
 \S \ref{subsection:bounds} -- see Definition \ref{def:gap-j} -- in such a way that 
 the infra-red cut-off Goldstone boson propagator in the $(\perp,\perp)$-channel, $\Sigma^{j_+\to}(q)$, has a pole precisely at $q=0$.}   This implies in turn that the 
 one-point function version of the gap equation, see Definition \ref{def:gap-one-fct},
 is satisfied up to error terms,
\BEQ -\Gamma^{j_+\to}+I^{j_+\to}_{s.c.,//}=O(g^2 2^{-2(j_+-j'_{\phi})}).
\EEQ


\Bigskip {\bf B. Local Ward identities.} Let $n\ge 2$. We shall now apply to
\BEQ I^{j_+\to}_{(0,n)}(\xi_1,\ldots,\xi_n):=f^{j_+\to}(\theta+\frac{\pi}{2},\ldots,
\theta+\frac{\pi}{2};\xi_1,\ldots,\xi_n) \label{eq:I0n}
\EEQ
 a general gauge
transformation along $\alpha$. Note that we do {\em not} restrict to strongly connected
diagrams any more. At this point it is useful to introduce the following general notation,
\BEA && I_n^{j_+\to}\left(\begin{array}{ccc} \eps_1 &  \cdots &  \eps_{n} \\
\xi_1 &  \cdots &  \xi_{n} \end{array}\right):=
\Big(\prod_{i=1}^n \int d\bar{\xi}_i\,  
\int d\zeta_i \, \chi_+^{j_+\to}(\zeta_i) \Big)\  
\Big(\prod_{i=1}^n\frac{\del}{\del\Phi(\zeta_i+\xi_i)}
\eps_i \frac{\del}{\del\bar{\Phi}(\zeta_i+\bar{\xi}_i)}  \Big)
\Big|_{\Phi,\bar{\Phi}=0}\nonumber\\
&&\qquad
\log({\cal Z}_{\lambda}^{j_+\to}(\Phi,\bar{\Phi}))  \label{eq:Inj+to}
\EEA
generalizing (\ref{eq:fnj+to}), where $\eps_i$ are general Hermitian (not even necessarily off-diagonal) two-by-two matrices. Gauge transformations rotate the matrix $\eps_i$ to
$\dot{\eps}_i:=\begin{cases} {\mathbb{\Gamma}}^{\perp}, \qquad \eps_i={\mathbb{\Gamma}}^{//}
\\ -{\mathbb{\Gamma}}^{//}, \qquad \eps_i={\mathbb{\Gamma}}^{\perp} \\
0, \qquad \eps_i={\mathbb{1}},\sigma^3\end{cases}.$
The lower pair of indices $(0,n)$ points to the fact that
all angles $\theta_i$, $i=1,\ldots,n$ are in the perpendicular
$(\perp)$ direction.

\Medskip  Then -- generalizing  Fig. \thesubsection.3 -- we
obtain the following order 1 {\bf local Ward identities},
\BEA &&0= \del_{\alpha} I^{j_+\to}_{(0,n-1)}(\xi_1,\ldots,\xi_{n-1})= 2\sum_{i=1}^{n-1}  
\alpha(\xi_i) I_{n-1}^{j_+\to}\left(\begin{array}{ccccccc} 
{\mathbb{\Gamma}}^{\perp} &  \cdots & {\mathbb{\Gamma}}^{\perp}
& -{\mathbb{\Gamma}}^{//} &  {\mathbb{\Gamma}}^{\perp} & \cdots & {\mathbb{\Gamma}}^{\perp} \\
\xi_1 &  \cdots  & \xi_{i-1} & \xi_i &
\xi_{i+1} &  \cdots &   \xi_{n-1} \end{array}\right)
\nonumber\\
&&\qquad +2 \int d\xi_{n} \, \alpha(\xi_{n})  I_{(0,n)}^{j_+\to}(\xi_1,\ldots,\xi_{n})  \nonumber\\
&&\qquad - \int d\xi_{n} \,  \partial_{\tau}\alpha(\xi_{n})  I_{n}^{j_+\to}\left(\begin{array}{cccc} {\mathbb{\Gamma}}^{\perp} &  \cdots & {\mathbb{\Gamma}}^{\perp} &  \sigma^3 \\
\xi_1 &  \cdots & \xi_{n-1} & \xi_{n} \end{array}\right) \nonumber\\
&&\qquad - \int d\xi_{n} \,  \Big\{\frac{1}{m^*}
\vec{\nabla}\alpha(\xi_{n})\cdot\vec{\nabla}_{\xi_{n}}\Big\}\  I^{j_+\to}_{n}\left(\begin{array}{cccc} {\mathbb{\Gamma}}^{\perp} &  \cdots & {\mathbb{\Gamma}}^{\perp} &  {\mathbb{1}} \\
\xi_1 &  \cdots & \xi_{n-1} & \xi_{n} \end{array}\right)
 \ + \ {\mathrm{err.}} \nonumber\\ \label{eq:3.32}
\EEA

\Medskip Take $\alpha$ such that $\alpha(\xi_i)=0$, $i=1,\ldots,n-1$. Then the first term on the r.-h.s. of (\ref{eq:3.32}) vanishes, so that
\BEA && \int d\xi_{n}\, \alpha(\xi_{n}) I_{(0,n)}^{j_+\to}
(\xi_1,\ldots,\xi_{n}) \nonumber\\
&&\qquad =\half \int d\xi_{n} \,  \partial_{\tau}\alpha(\xi_{n})  I_{n}^{j_+\to}\left(\begin{array}{cccc} {\mathbb{\Gamma}}^{\perp} &  \cdots & {\mathbb{\Gamma}}^{\perp} &  \sigma^3 \\
\xi_1 &  \cdots & \xi_{n-1} & \xi_{n} \end{array}\right) \nonumber\\
&&\qquad + \int d\xi_{n} \,  \Big\{\frac{1}{2m^*}
\vec{\nabla}\alpha(\xi_{n})\cdot\vec{\nabla}_{\xi_{n}}\Big\}\  I^{j_+\to}_{n}\left(\begin{array}{cccc} {\mathbb{\Gamma}}^{\perp} &  \cdots & {\mathbb{\Gamma}}^{\perp} &  {\mathbb{1}} \\
\xi_1 &  \cdots & \xi_{n-1} & \xi_{n} \end{array}\right) \ 
+\  {\mathrm{err.}} \nonumber\\ \label{eq:3.53}
\EEA

\Medskip

\Medskip Let us now take the functional derivative
$\frac{\del}{\del \alpha(\xi_n)}(\cdot)$
of (\ref{eq:3.53}) w.r. to $\alpha$:
\BEA && I_{(0,n)}^{j_+\to}
(\xi_1,\ldots,\xi_{n}) = \half 
\partial_{\tau_n}  I_{n}^{j_+\to}\left(\begin{array}{ccccc} 
{\mathbb{\Gamma}}^{\perp} &  \cdots & {\mathbb{\Gamma}}^{\perp} &  \sigma^3  \\
\xi_1 &  \cdots  & \xi_{n-1} & \xi_{n} \end{array}\right) 
\nonumber\\
&&\qquad + \frac{1}{2m^*} \vec{\nabla}^2_{\xi_n} 
 I_{n}^{j_+\to}\left(\begin{array}{ccccc} 
{\mathbb{\Gamma}}^{\perp} &  \cdots & {\mathbb{\Gamma}}^{\perp} &  {\mathbb{1}} \\
\xi_1 &  \cdots  & \xi_{n-1} & \xi_{n} \end{array}\right) 
 \ 
+\  {\mathrm{err.}} \nonumber\\ 
 \label{eq:3.37}
\EEA
   Finally,
integrating w.r. to the external legs except $\xi_1$, one finds in particular by
integrations by parts,
\BEA && F_n^{j_+\to}(\theta+\frac{\pi}{2},\ldots,\theta+
\frac{\pi}{2};\xi_1)=\frac{1}{2}
\Big(\prod_{i=2}^{n-1}\int d\xi_i\, \Sigma^{k_{+,i}}(\xi_{ext,i},\xi_i)\Big) \ \int d\xi_{n}  \nonumber\\
&&\qquad 
\Big\{\ 
I_{n}^{j_+\to}\left(\begin{array}{cccc} {\mathbb{\Gamma}}^{\perp} &  \cdots  &  {\mathbb{\Gamma}}^{\perp} &  \Gamma_{\phi}\sigma^3 \\
\xi_1 &  \cdots & \xi_{n-1} & \xi_{n} \end{array}\right) \Big(\frac{1}{\Gamma_{\phi}}\frac{\partial}{\partial \tau_{n}}\Big)\Sigma^{k_{+,n}}(\xi_{ext,n},\xi_{n})
 \nonumber\\
&&\qquad  + \ I_{n}^{j_+\to}\left(\begin{array}{cccc} {\mathbb{\Gamma}}^{\perp} &  \cdots  &  {\mathbb{\Gamma}}^{\perp} &  \Gamma_{\phi} {\mathbb{1}} \\
\xi_1 &  \cdots & \xi_{n-1} & \xi_{n} \end{array}\right)
\Big(\frac{1}{2m^*\Gamma_{\phi}}\vec{\nabla}^2_{\xi_n}\Big)\Sigma^{k_{+,n}}(\xi_{ext,n},\xi_{n}) \Big\}
\nonumber\\ & \qquad  + {\mathrm{err.}} 
\label{eq:Fn-Ward}
\EEA

\Medskip The second-order operator $\frac{1}{2m^*\Gamma_{\phi}} \vec{\nabla}^2_{\xi_n}$ contributes
a factor of order $2^{j_+}2^{-2k_+})\le 2^{-2(k_+-j_+)}$, a squared spring factor; on the other hand, the first-order operator 
$\frac{1}{\Gamma_{\phi}} \frac{\partial}{\partial\tau_n}$ contributes a single
spring factor $O(2^{-(k_+-j_+)})$, which is not enough for our purposes. Fortunately, the whole procedure may be iterated
twice, using successively two derivatives $\del_{\alpha},\del_{\beta}$, with $\alpha(\xi_i)=0,
i\not=n-1$ and $\beta(\xi_i)=0,\not=n$, and taking the functional derivative $\frac{\del^2}{\del\alpha(\xi_{n-1})
\del\beta(\xi_n)}$. Considering only time-derivatives, one has schematically
\BEQ 0=\del_{\alpha}I^{j_+\to}_{(0,n-1)}\   \longrightarrow \  I^{j_+\to}_{(0,n)} \equiv \partial_{\tau} I^{j_+\to}_n
\left(\begin{array}{cccc} {\mathbb{\Gamma}}^{\perp} &  \cdots & {\mathbb{\Gamma}}^{\perp} &  \sigma^3 \end{array}\right)
\EEQ

\BEA &&   0=\del_{\beta} I^{j_+\to}_{n-1}
\left(\begin{array}{cccc} {\mathbb{\Gamma}}^{\perp} &  \cdots & {\mathbb{\Gamma}}^{\perp} &  \sigma^3 \end{array}\right)
\nonumber\\
&&\qquad 
  \longrightarrow \   I^{j_+\to}_n
\left(\begin{array}{cccc} {\mathbb{\Gamma}}^{\perp} &  \cdots & {\mathbb{\Gamma}}^{\perp} &  \sigma^3 \end{array}\right) \equiv \partial_{\tau} I^{j_+\to}_n
\left(\begin{array}{ccccc} {\mathbb{\Gamma}}^{\perp} &  \cdots & {\mathbb{\Gamma}}^{\perp} & \sigma^3 &   \sigma^3 \end{array}\right)  \nonumber\\
\EEA
Thus  $F_{n}^{j_+\to}(\theta+\frac{\pi}{2},\ldots,\theta+
\frac{\pi}{2};\xi_1)$,  $n\ge 3$   has been rewritten as a sum of four terms 
involving $ I^{j_+\to}_n
\left(\begin{array}{ccccc} {\mathbb{\Gamma}}^{\perp} &  \cdots & {\mathbb{\Gamma}}^{\perp} & * &   * \end{array}\right) $   
with $*={\mathbb{1}},\sigma^3$, and product differential operators of the type
$\big(\frac{1}{\Gamma_{\phi}}\frac{\partial}{\partial\tau_{n-1}},\frac{1}{2m^*\Gamma_{\phi}}\vec{\nabla}_{\xi_{n-1}}^2\big)
\times \big (\frac{1}{\Gamma_{\phi}}\frac{\partial}{\partial\tau_{n-1}},\frac{1}{2m^*\Gamma_{\phi}}\vec{\nabla}_{\xi_{n-1}}^2\big)$, therefore of order $\ge 2$.   Thus
 one has 
 decreased the relative degree of divergence $\omega_n=\half(6-n)$  of the original
 $n$-point function
 by at least two units.  This is sufficient to show the
 convergence of  amputated  $n$-point functions $I^{j_+\to}_{(0,n)}$ provided $n\ge 3$, 
  since  then the corrected degree of divergence is 
  $\tilde{\omega}_n:=\half(6-n)-2<0$.

\Bigskip {\bf C. Massive configurations.}
We have not considered in previous computations the following two cases:

\begin{itemize}

\item[(i)] $n$-point functions of the type
\BEQ I^{j_+\to}_{(k_{//},k_{\perp})}(\xi_1,\ldots,\xi_n):=
I_n^{j_+\to}\left(\begin{array}{cccccc} {\mathbb{\Gamma}}^{//} &  \cdots & {\mathbb{\Gamma}}^{//} & {\mathbb{\Gamma}}^{\perp} &  
\cdots & {\mathbb{\Gamma}}^{\perp}\\
\xi_1 &  \cdots & \xi_{k_{//}} & \xi_{k_{//}+1} & \cdots & \xi_{n} \end{array}\right)
\label{eq:Ikk} 
\EEQ
with $k_{\perp}=n-k_{//}$,  $k_{//}\ge 1$;

\item[(ii)] $n$-point functions $I^{j_+\to}_{(0,n)}(\cdots)$
such that at least one of the external Cooper pairs connects
to two different polymers, as in Fig. \ref{subsection:complementary}.7.

\end{itemize}

Both cases are {\em massive}, featuring at least one external fermionic propagator with quasi-exponential decay at distances
$\gg \Gamma_{\phi}^{-1}$,  implying that the corresponding
polymers are convergent, and furthermore, their contribution decreases
exponentially as $j_+\to\infty$. More precisely:

 \begin{itemize}
\item[(i)] external Cooper pairs along the parallel ($//$) direction form a $\Sigma^{\to(j_+ +1)}_{//,//}$-kernel which is
bounded as in (\ref{eq:decay-PreSigma//}).

\item[(ii)] Referring to the notations on Fig. \ref{subsection:Ward}.7, we have the following scaling factors for 
the compound polymer connected by two fermions (assuming fermion
angular sectors are fixed):

-- $2^{-3j'_{\phi}}$ for the integral over $p$;

-- $2^{-3j_+}$ for the integral over $p':=-p+q$ constrained by the fact that $p+p'=q$ ;

-- $(2^{j'_{\phi}})^2$ for the two fermionic propagators;

-- a scaled quasi-exponential decay factor  of scale $j'_{\phi}$ for the propagator with momentum $p$, resp. $j_+$ for the propagator
with momentum $-p+q$;

-- and finally the squared  spring factor  $(2^{-(j_+-j'_{\phi})})^2$ coming from the Ward identities. The
total factor is $2^{-4j'_{\phi}}2^{-5(j_+-j'_{\phi})}$, multiplied by the two quasi-exponential decay factors. The
first factor $2^{-4j'_{\phi}}=(2^{-j'_{\phi}})^4$ is the expected weight for four fermions  $(\psi^{j,\alpha})^4$. The
bad scale $j_+$ decay factor for the second propagator yields an extra volume factor $2^{3(j_+-j'_{\phi})}$ compared to
the  expected volume implied by a scale $j'_{\phi}$ decay factor. Decomposing $2^{-5(j_+-j'_{\phi})}$ into $2^{-3(j_+-j'_{\phi})}\times 2^{-2(j_+-j'_{\phi})}$, one sees that the extra volume factor is compensated, and that one has a further spring factor 
$2^{-2(j_+-j'_{\phi})}$ which makes it possible to sum over the scale $j_+$.

\end{itemize}


\subsection{Bosonic fixed-point procedure and low-energy multi-scale bounds}
\label{subsection:fixed-point}


Consider a bosonic multi-scale tree as in Fig.
\ref{subsection:complementary}.15. Bounds for {\em fermionic}
polymers with low-momentum Cooper pair legs in \ref{subsection:complementary} {\bf E.} was
based on the naive power-counting for a Goldstone boson propagator $\Sigma_{\perp,\perp}(\xi-\xi')$ of scale $k_+$,
\BEQ O((\Gamma_{\phi} 2^{-k_+/2})^2) =
  O((2^{-j_{\phi}}2^{-k_+/2}\mu)^2),
\EEQ
therefore, $O(2^{-j_{\phi}}2^{-k_+/2}\mu)$ per Cooper pair.  Ward identities (see \ref{subsection:Ward}), as proved
just above (see below (\ref{eq:Fn-Ward})),  imply however that 
 one can get one or two supplementary spring 
 factors $O(2^{-(k_+-j_{\phi})})$ per polymer, thus modifying the  scaling of the corresponding Cooper pairs, 
 \BEQ  2^{-j'_{\phi}}2^{-k_+/2}\mu \longrightarrow
 2^{-(k_+-j_{\phi})} \ \cdot\ 2^{-j'_{\phi}}2^{-k_+/2}\mu =   2^{-3k_+/2}\mu.
 \EEQ 
 Thus the sum over bosonic multi-scale trees may be bounded
 as follows. We first need some notations. Let $\P^{(j'_{\phi}-1)\to}_1,\ldots,\P^{(j'_{\phi}-1)\to}_{n_{j'_{\phi}-1}}$ be the scale-neutral polymers. By assumption, we {\em fix} the location of {\em one} of the
 vertices of $(\P^{(j'_{\phi}-1)\to}_i)_{i\le n_{j'_{\phi}-1}}$, so that their
 power counting is $O(g')$, where $g'=2^{-2(j'_{\phi}-j_{\phi})}$. The scale $j'_{\phi}$ 
Goldstone boson propagators merge these polymers into 
 {\bf bosonic polymers} $\P^{j'_{\phi}\to}_1,\ldots, \P^{j'_{\phi}\to}_{n_{j'_{\phi}}}$ with lowest scale $\le j'_{\phi}$. The power-counting of each of these is computed with the
 same convention, namely, by fixing the location of one of
 the vertices. Repeating this procedure, we get a sequence
 of polymers $(\P^{j_+\to}_i)_{i\le n_{j_+}}$, $j_+\ge j'_{\phi}$ which becomes stationary when all Goldstone boson propagators
 have been integrated into polymers.
 
\Medskip We proceed as in \S \ref{subsection:multi-scale}
(compare with (\ref{eq:GFMSB-P}), (\ref{eq:GFMSB-n}),
(\ref{eq:GFMSB})).

\Medskip {\bf  General bosonic multi-scale bound.} The actual energy gaps $\Gamma_{\phi}$, resp. non-linear sigma model coupling constant $g_{\phi}$, resp. velocity $v_{\phi}$ 
are obtained inductively as the limits of the converging series $\Gamma^{j'_{\phi}\to}+\sum_{j_+=j'_{\phi}+1}^{+\infty} (\Gamma^{j_+\to}-\Gamma^{(j_+-1)\to})$, resp. $g_{\phi}:=g_{\phi}^{j'_{\phi}\to}+
\sum_{j_+=j'_{\phi}+1}^{+\infty} (g_{\phi}^{j_+\to}-g^{(j_+-1)\to})$, resp.
resp. $v_{\phi}:=v_{\phi}^{j'_{\phi}\to}+
\sum_{j_+=j'_{\phi}+1}^{+\infty} (v_{\phi}^{j_+\to}-v_{\phi}^{(j_+-1)\to})$.

\Medskip We start in {\bf A.} by stating the {\em difference estimates}. Such estimates  make it possible to solve inductively a sequence of approximate gap equations (see Definition \ref{def:gap-j}) yielding a Cauchy sequence
$(\Gamma^{j_+\to})_{j_+\ge j'_{\phi}}$.  As an intermediate
step, we first prove in {\bf B.} {\em polymer bounds} by scale induction. Difference
estimates are finally proved in {\bf C.}.

\Bigskip {\bf A. Statement of difference estimates and gap equations.} 
 For each $j_+\ge j'_{\phi}$, we search for the solution $\Gamma:=\Gamma^{j_+\to}$  of largest module of  the 
\begin{Definition}[scale $j_+$  gap equation] \label{def:gap-j}
\BEQ  \boxed{\Big(-{\mathbb{1}}+\bar{\Pi}^{j_+\to}_0(\Gamma^{j_+\to}) ({\cal A}_0(\Gamma^{j_+\to},\Upsilon_3)\Big) 
\left(\begin{array}{c} (\Gamma^{j_+\to})^{\perp}\\
\big((\Gamma^{j_+\to})^{\perp}\big)^* \end{array}\right) =0,}  \label{eq:gap-j} \EEQ
\end{Definition}
 compare with Definition \ref{def:gap} and (\ref{eq:gap-0}).
The parameters $g_{\phi}^{j_+\to}, v_{\phi}^{j_+\to} $ of
the Goldstone boson propagator $\Sigma_{\perp,\perp}^{j_+\to}$ are
defined as in Lemma \ref{lemma:bubble} and  Definition \ref{def:gphivphi} by

\begin{Definition}[scale $j_+$ Goldstone boson propagator
parameters]  \label{def:scalej+param}  Let
\BEQ  \boxed{\ \ \partial_{q^0}^2  \Big\{\bar{\Pi}_q^{j_+\to}(\Gamma^{j_+\to}) {\cal A}_q(
\Gamma^{j_+\to},\Upsilon_{3})\Big\}_{diag}\Big|_{q=0}=:-\frac{1}{g^{j_+\to}_{\phi,diag}}}
\label{eq:delgphidiag}
\EEQ
\BEQ  \boxed{\partial_{q^0}^2 \Big\{
\bar{\Pi}^{j_+\to}_q(\Gamma^{j_+\to}) {\cal A}_q(\Gamma^{j_+\to},\Upsilon_{3})
\Big\}_{off}\Big|_{q=0}=:-\frac{1}{g^{j_+\to}_{\phi,off}} \ \ }  \label{eq:delgphioff}
\EEQ

\BEQ \boxed{\ \  \vec{\nabla}^2  \Big\{\bar{\Pi}_q^{j_+\to}(\Gamma^{j_+\to}){\cal A}_q(\Gamma^{j_+\to},\Upsilon_{3})\Big\}_{diag}\Big|_{q=0}=:
 -\frac{(v^{j_+\to}_{\phi,diag})^2}{g^{j_+\to}_{\phi,diag}} \, \Id}, \label{eq:delvjdiag}
 \EEQ
\BEQ \boxed{ \vec{\nabla}^2 \Big\{\bar{\Pi}_q^{j_+\to}(\Gamma^{j_+\to}){\cal A}_q(\Gamma^{j_+\to},\Upsilon_{3})\Big\}_{off}\Big|_{q=0}=:
 -\frac{(v^{j_+\to}_{\phi,off})^2}{g^{j_+\to}_{\phi,off}} \, \Id\ \ }  \label{eq:delvjoff}
\EEQ

and 
\BEQ \boxed{g^{j_+\to}_{\phi}:=\frac{g^{j_+\to}_{\phi,diag}g^{j_+\to}_{\phi,off}}{g^{j_+\to}_{\phi,diag}+g^{j_+\to}_{\phi,off}},
\qquad v^{j_+\to}_{\phi}:=\sqrt{ \frac{g^{j_+\to}_{\phi,off}(v^{j_+\to}_{\phi,diag})^2+g^{j_+\to}_{\phi,diag}(v^{j_+\to}_{\phi,off})^2}{g^{j_+\to}_{\phi,diag}+g^{j_+\to}_{\phi,off}}}.}
\EEQ 

\end{Definition}

\Bigskip The scale $j_+$ gap equations are solved by induction on $j_+$ using  {\em difference bounds}  of two types:

\begin{itemize}
\item[(i)] {\em (scale differences)}  let $\Gamma$ such that
$|\Gamma|\approx \hbar\omega_D e^{-\pi/g}$, then

\BEQ \Big| \bar{\Pi}_0^{j_+\to}(\Gamma)-\bar{\Pi}_0^{(j_+-1)\to}(\Gamma)\Big|\lesssim 2^{-2(j_+-j_{\phi})} g  \label{eq:delPij} \EEQ

\BEQ |g_{\phi}^{j_+\to}-g_{\phi}^{(j_+-1)\to}|\lesssim 2^{-2(j_+-j_{\phi})} g_{\phi},
\qquad |v_{\phi}^{j_+\to}-v_{\phi}^{(j_+-1)\to}|\lesssim 2^{-2(j_+-j_{\phi})} v_{\phi} \label{eq:delgphidelvphi}
\EEQ
with $g_{\phi}\approx \frac{\Gamma_{\phi}^2}{m^*}, v_{\phi}\approx v_F$;

\item[(ii)] {\em ($\Gamma$ dependence)}  let $\Gamma,\Gamma'$ such that
$|\Gamma|,|\Gamma'|\approx \hbar\omega_D e^{-\pi/g}$, and
$\del\Gamma:=|\Gamma-\Gamma'|$,  then

\BEQ \Big| \bar{\Pi}_q^{j_+\to}(\Gamma)-\bar{\Pi}_q^{j_+\to}(\Gamma')\Big|\lesssim \frac{\del\Gamma}{\Gamma_{\phi}} g   \label{eq:delPiGamma} \EEQ
\BEQ \lambda\Big|{\cal A}_q(\Gamma)-{\cal A}_q(\Gamma')\Big|\lesssim \frac{\del\Gamma}{\Gamma_{\phi}}.  \label{eq:delAGamma} \EEQ

\end{itemize}

\noindent If difference bounds hold, then subtracting  
(\ref{eq:gap-j})$_{(j_+-1)}$ from (\ref{eq:gap-j})$_{j_+}$ and
 applying 
the implicit function theorem yields after an elementary
computations
\BEQ |\Gamma^{j_+\to}-\Gamma^{(j_+-1)\to}|\lesssim 2^{-2(j_+-j_{\phi})}
\Gamma_{\phi} 
\EEQ
where $\Gamma_{\phi}\approx 2^{-j_{\phi}}\mu$. 
In particular, $\Gamma^{j_+\to} \to_{j_+\to +\infty} e^{\II\theta}\Gamma_{\phi}$.

\Medskip Also, by elementary computation,  $v_{\phi}^{j_+\to}\to_{j_+\to\infty} v_{\phi}\approx v_F$, $g_{\phi}^{j_+\to}\to_{j_+\to\infty} g_{\phi}\approx \frac{\Gamma_{\phi}^2}{m^*}$, and
\BEQ \Sigma^{j_+\to}_{\perp,\perp}(q)\to_{j_+\to\infty} \Sigma(q)=
 \frac{g_{\phi}}{(q^0)^2+v_{\phi}^2 |\vec{q}|^2+O(|q|_+^4/\Gamma_{\phi}^2)}, \EEQ
\BEQ \Sigma^{j_+\to}_{\perp,\perp}(\xi-\xi')\to_{j_+\to\infty}\Sigma_{\perp,\perp}(\xi-\xi'), \EEQ
\BEQ   \Sigma_{\perp,\perp}(\xi-\xi')\sim_{|\xi-\xi'|\to\infty} 
\frac{g_{\phi}/4\pi v_{\phi}}{\sqrt{|\vec{x}-\vec{x}'|^2+v_{\phi}^2(\tau-\tau')^2}} 
\Big(1+O(\frac{1}{\Gamma_{\phi}(|\tau-\tau'|+|\vec{x}-\vec{x}'|/v_{\phi})})\Big), \label{eq:Sigmaj+Sigma}
\EEQ
as stated in Theorem 2.

\Bigskip {\bf B. Proof of polymer bounds.}

\Bigskip Let $\P^{j_+\to}\in{\cal P}^{j_+\to}\not=\emptyset$, then  we want to prove, in analogy with  the bounds (\ref{eq:GFMSB-P}),(\ref{eq:GFMSB-P}),(\ref{eq:GFMSB}) of 
section \ref{section:fermion} for fermionic polymers 
in ${\cal P}^{j\to}$, $j\le j'_{\phi}-1$, but now with 
$j\longrightarrow j_+$, $j_+\ge j'_{\phi}$, and {\em bosonic polymers}
obtained by merging polymers also including scale $j'_{\phi}$ fermions,
following the procedure of \S \ref{subsection:complementary}:

\Medskip {\bf  General bosonic multi-scale bound.}
 
 \Medskip {\em For every $\P^{j_+\to}$ belonging to the set  ${\cal P}^{j_+\to}_{N_{ext,+}}$ of polymers with lowest scale $j_+$ and  $N_{ext,+}$ external legs,}
\BEQ  \boxed{ |{\cal A}(\P^{j_+\to};\xi_{ext,1})|\lesssim g^{n(\P^{j_+\to})/ 4} \  (2^{-j_{\phi}}/4)^{N_{ext,+}(\P^{j_+\to})-4} \ 
 (2^{j_+})^{\frac{1}{2}(N_{ext,+}(\P^{j_+\to})-6)} \ (p^*_F)^{-2N_{ext,+}}  }
\label{eq:GBMSB-P}
\EEQ

\Medskip
The exponent of $2^{j_+}$ has been chosen in such a way that 
the factor  $(2^{j_+})^{\frac{1}{2}(N_{ext,+}-6)}$, times
the {\em weight} $(2^{-j_{\phi}}2^{-j_+/2})^{N_{ext,+}} 2^{3j_+}$ 
of the external structure of a polymer with $N_{ext,+}$ external Cooper pair legs (see below), is scale independent. 
Then, if $\P^{j_+\to}$ ranges in the set ${\cal P}_{N_{ext,+}}^{j_+\to}(n)$  of  polymers
in ${\cal P}_{N_{ext,+}}^{j_+\to}$ with   $n\ge 1$ vertices, 
\BEQ  \boxed{\sum_{\P^{j_+\to}\in {\cal P}_{N_{ext,+}}^{j_+\to}(n)} |{\cal A}(\P^{j_+\to};\xi_{ext,1})|\lesssim g^{n/4}   \ (2^{-j_{\phi}/4})^{N_{ext,+}(\P^{j_+\to})-4} \  (2^{j_+})^{\frac{1}{2}(N_{ext,+}(\P^{j_+\to})-6)}  (p^*_F)^{-2N_{ext,+}} }
\label{eq:GBMSB-n}
\EEQ
from which, finally, summing over all polymers in ${\cal P}_{N_{ext,+}}^{j_+\to}$
 with one external leg located at $\xi_{ext,1}$,
 \BEQ  \sum_{\P^{j_+\to}\in{\cal P}_{N_{ext,+}}^{j_+\to}} |{\cal A}(\P^{j_+\to};\xi_{ext,1})|\lesssim g^{N_{ext,+}/4} \ (2^{-j_{\phi}/4})^{N_{ext,+}(\P^{j_+\to})-4} \  (2^{j_+})^{\frac{1}{2}(N_{ext,+}-6)}\ 
  (p^*_F)^{-2N_{ext,+}}
\label{eq:GBMSB}
\EEQ

\Medskip The proof  of {\bf bosonic polymer bounds} (\ref{eq:GBMSB-P}),(\ref{eq:GBMSB-n}),(\ref{eq:GBMSB}) is by induction on $j_+$. Note first that 
(\ref{eq:GBMSB-P}),(\ref{eq:GBMSB-n}),(\ref{eq:GBMSB})
follow from the general {\em fermionic} multi-scale bound  (\ref{eq:GFMSB-P}),(\ref{eq:GFMSB-P}),(\ref{eq:GFMSB}) {\em if one sets $j_+= j_{\phi}$}, which means that {\em all fermionic polymers} (i.e. all polymers having all internal propagators
of scales $\le j'_{\phi}-1$) {\em are assimilated to polymers
with lowest scale $j_{\phi}$}; this is a convenient choice for
the induction, since  field scaling factors (see below)
have a scale-independent prefactor proportional to $\Gamma_{\phi}\approx 2^{-j_{\phi}}\mu$. 
 Namely, identifying
the $N_{ext,+}$ {\em bosonic} legs of a {\em fermionic}
 polymer $\P^{j'_{\phi}\to}\in{\cal P}^{j'_{\phi}\to}$ with
  \BEQ N_{ext}:=2N_{ext,+} \label{eq:NextNext+}
 \EEQ
  {\em fermionic} legs, we remark that
 \BEQ  (2^{-j_{\phi}/4})^{N_{ext,+}-4} (2^{j_+})^{\frac{1}{2}(N_{ext,+}-6)}=
 (2^{j_{\phi}})^{\half(N_{ext}-4)}
 \EEQ
 which shows that (\ref{eq:GBMSB-P}) holds with prefactor
 $g^{n/2}$ instead of $g^{n/4}$ for a fermionic polymer in ${\cal P}^{j_{\phi}\to}$. Since $N_{ext}\ge 4$, the fermionic
 bound  $j\mapsto (2^j)^{\half(N_{ext}-4)}$ increases with
 $j$, hence (\ref{eq:GBMSB-P}) is all the more true for 
 a fermionic polymer in ${\cal P}^{j\to}$ if $j\le j_{\phi}$.
 Finally, for $j_{\phi}\le j\le j'_{\phi}$, a correcting factor
 $\le  (2^{j'_{\phi}-j_{\phi}})^{\half(N_{ext}-4)}=  (2^{j'_{\phi}-j_{\phi}})^{N_{ext,+}-2}$ is required to
 obtain the fermionic bounds $(2^j)^{\half(N_{ext}-4)}$, which
 is compensated by $g^{-n/4}$ since $n\ge N_{ext,+}$. 
 
\Bigskip
We now suppose as in \S \ref{subsection:bounds} that
$\P^{j_+\to}$ is obtained by merging at scale $j_+$ polymers
$\P_1,\ldots,\P_n$ with lowest bosonic scales $k_{+,1},\ldots,
k_{+,n}<j_+$ and $n_i:=N_{ext,+}(\P_i)$ external Cooper pair
legs. 

\Medskip {\bf Ward spring factor.} As discussed in the introduction to this subsection, we have an extra spring
factor $O(2^{-2(j_+-k_+)})$ for each polymer 
$\P_i$. 

\Medskip{\bf Volume factors.}
External legs of $\P^{j_+\to}$ which are contracted at scale $j_+$ are
integrated in a box $\Del^{j_+}$ of scale $j_+$. This yields a volume factor $O(2^{3j_+})$. 

\Medskip {\bf Field scaling factors.} 
Each external Goldstone boson half-propagator of $\P^{j_+\to}$ which is contracted at scale $j_+$ comes with a supplementary
 prefactor $2^{-j_{\phi}}2^{-j_+/2}$.

\Medskip The product of the volume factors and of the field
scaling factors gives the above mentioned {\em weight} $(2^{-j_{\phi}}2^{-j_+/2})^{N_{ext,+}} 2^{3j_+}$. Taking these into account 
for the  polymers $\P^{k_i}, i=1,\ldots,n$ and (in the denominator, in order to avoid double-counting) for $\P^{j_+\to}$,  yields the general multi-scale bound (dismissing dimensional
factors)
\BEA &&  |{\cal A}(\P^{j_+\to};\xi_{ext,1})| \lesssim g^{n(\P^{j_+\to})/2}
(2^{j_{\phi}}2^{j_+/2})^{N_{ext,+}(\P^{j_+\to})} 2^{-3j_+}
\nonumber\\
&&\qquad
\prod_{i=1}^n  \Big\{a(\P_i;\cdot)\ 
2^{-2(j_+-k_{+,i})} (2^{-j_{\phi}}2^{-j_{+}/2})^{n_i} 2^{3j_+}\Big\}.  \label{eq:malditesta}
\EEA
where 
\BEQ a(\P_i):=(2^{-j_{\phi}/4})^{n_i-4} \  (2^{k_{+,i}})^{\frac{1}{2}(n_i-6)} \EEQ
is the scale-dependent part of (\ref{eq:GBMSB-P}). 
 We 
show that this is $\lesssim (2^{-j_{\phi}/4})^{N_{ext,+}-4} (2^{j_+})^{\frac{1}{2}(N_{ext,+}-6)}  \ \cdot\ \prod_{i=1}^n 
2^{-\half(j_+-k_{+,i})}$, where $N_{ext}:=N_{ext}(\P^{j_+\to})$. For this we consider separately {\em scale-independent factors,
scale $j_+$ factors, and rescaling factors:}

\begin{itemize}
\item[(i)] {\em (scale-independent factors)} 
 Each polymer
is connected at least to one other polymer, hence at least one of the external legs of each polymer $\P_i$ does not belong to the $N_{ext}$ external legs of $\P_{j_+\to}$. Thus   $\sum_{i=1}^n n_i\ge N_{ext,+}+n$.  The  total exponent of the scale-independent
factor $2^{-j_{\phi}}$  in (\ref{eq:malditesta})  is
\BEQ -N_{ext,+}+\sum_{i=1}^n (\frac{5}{4}n_i-1)\ge \frac{1}{4}(N_{ext,+}-4).  \label{eq:3.78} \EEQ

\item[(ii)] {\em (rescaling factors)} rewrite $ 2^{-2(j_+-k_{+,i})} (2^{k_{+,i}})^{\half(n_i-6)}$  as $(2^{j_+})^{\half(n_i-6)}$ times\\  $(2^{-(j_+-k_{+,i})})^{2+\half(n_i-6)}$. Since by hypothesis $n_i\ge 3$, 
the exponent $2+\half(n_i-6)$ is $>\half$. The remaining
spring factor
$\le \prod_{i=1}^n (2^{-(j_+-k_{+,i})})^{1/2}$ makes it possible
to sum over all scales $k_{+,1},\ldots,k_{+,n}<j$.

\item[(iii)] {\em (scale $j_+$ factors)} For each $i=1,\ldots,n$, the left-over factor
$(2^{j_+})^{\half(n_i-6)}$ in (ii) cancels with the
scale-dependent part  $(2^{-j_+/2})^{n_{i}} 2^{3j_+}$  of the weight of $\P_i$. Therefore, there remains only
the required overall prefactor $(2^{j_+})^{\half(N_{ext,+}-6)}$.
\end{itemize}

\Medskip {\bf C.   Proof of difference estimates.}  
 
\Medskip 1. {\em Difference estimates for $\Gamma$.} 
The difference
$\bar{\Pi}^{j_+\to}(\Gamma)-\bar{\Pi}^{(j_+-1)\to}(\Gamma)$,
see (\ref{eq:delPij}),
involves terms of three types which we present by {\em decreasing}
order: 

\begin{itemize}
\item[(i)] pure scale $j_+$ irreducible four-point polymers, like the first term of Fig. \S \ref{subsection:complementary}.3.
These contain no inner Goldstone boson, hence enjoy by
Ward identities a spring factor $O(2^{-2(j_+-j_{\phi})})$, multiplied by the scaling (\ref{eq:GFMSB-P}) of an amputated,
four-point, fermionic polymer of lowest scale $j\le j'_{\phi}$, $O((2^j)^{\half(N_{ext}-4)})=O(1)$; all together, one gets
the expected factor $2^{-2(j_+-j_{\phi})}$.

\Medskip The two remaining types are {\em multi-scale} bosonic
trees obtained in the last two resummation steps, 
Step (iv), resp. Step (v) of p. \pageref{eq:p.124}, involving
at least one Goldstone boson of scale $j_+-1$:

\item[(ii)] (internal Goldstone boson propagator)

\bigskip
{\centerline{
\begin{tikzpicture}
\draw[decorate,decoration=snake](-0.707,0.707)--(-0.5-0.707,0.5+0.707);
\draw(0,0) circle(1);  \draw[decorate,decoration=snake](-0.707,-0.707)--(-0.5-0.707,-0.5-0.707);
\draw(0,0) node {$\P_1^{(j_+-1)\to}$};
\draw[decorate,decoration=snake](0.707,0.707) arc(135:-135:1 and 1);
\draw(3,0) node {\tiny $j_+\!-\!1$};
\end{tikzpicture}
}}

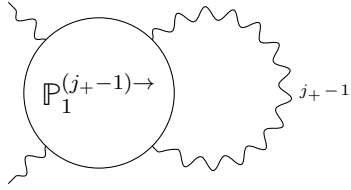
\Medskip{\bf\tiny Fig. \thesubsection.1. Case of a scale 
$(j_+\!-\!1)$ internal Goldstone boson propagator.}

\bigskip involving one internal Goldstone boson propagator 
connecting a polymer $\P_1^{(j_+-1)\to}$ to itself. Leading terms
are as in the above Figure, leading to a contribution larger
than in case (i),
\BEQ \Big( 2^{-j_+}\Big) \times
\Big(2^{-2j_{\phi}}2^{-j_+}\Big)=2^{-2j_{\phi}}2^{-2j_+}, \EEQ
that is, \big(Amplitude of a polymer with 4 external Cooper pair legs $\P_1^{(j_+-1)\to}\big)\times\big($Scale $j_+$ 
Goldstone boson propagator\big). See (\ref{eq:GBMSB-P}).

\item[(iii)] (Goldstone boson connecting two different polymers)

\bigskip

{\centerline{
\begin{tikzpicture}
\draw(0,0) circle(1); \draw(4,0) circle(1);
\draw(0,0) node {$\P_1^{(j_+-1)\to}$}; \draw(4,0) node {$\P_2^{(j_+-1)\to}$};
\draw[decorate,decoration=snake](0.707,0.707)--(4-0.707,0.707);
\draw(2,1) node {\tiny $j_+\!-\!1$};
\draw[decorate,dotted,decoration=snake](0.707,-0.707)--(4-0.707,-0.707); 
\draw[decorate,decoration=snake](-1,0)--(-1.8,0);
\draw[decorate,decoration=snake](5,0)--(5.8,0);
\draw(2,-1.3) node {$j'_+$};
\draw(4-0.707,0.707) node {\textbullet};\draw(4-0.707,1.2) node {$\xi_2$};
\draw(0.707,0.707) node {\textbullet};\draw(0.707,1.2) node {$\xi_1$};
\end{tikzpicture} 
}}
 
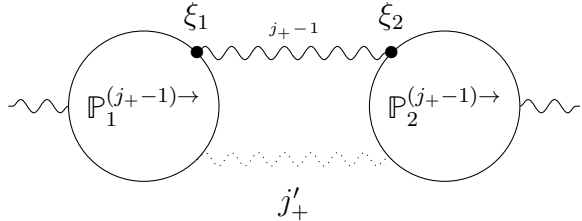
\Medskip{\bf\tiny Fig. \thesubsection.2. Case of a scale 
$j_+$ Goldstone boson connecting two different polymers.}

\bigskip involving {\em two} Goldstone boson propagators
 connecting the {\em same two} polymers $\P_1^{(j_+-1)\to},\P_2^{(j_+-1)\to}\in {\cal P}^{j_+\to}$,
one of scale  $j_+-1$, the other of scale $j'_+\le j_+-1$. 
Spring factors make it possible to sum over $j'_+\le j_+-1$ for
$j_+-1$ fixed, hence we can assume that $j'_+=j_+-1$. 

\Medskip Other more complicated diagrams $\P^{j_+\to}$ with $N_{ext,+}(\P^{j_+\to})=2$  as 
in {\bf B.}, connecting $\P_1^{(j_+-1)\to},
\ldots,\P_n^{(j_+-1)\to}$, with $n\ge 2$ and $\ell\ge 2$ internal Goldstone
boson propagators, are possible.  The power-counting of such diagrams is as in 
the General bosonic multi-scale bound, eq. (\ref{eq:GBMSB-P}).
Eq. (\ref{eq:3.78}) and point (i) around it show that the
largest power-counting is obtained for $n=\ell=2$, that is,
for the situation depicted in the above Figure, for which
one obtains a contribution to $\bar{\Pi}^{j_+\to}(\Gamma)-\bar{\Pi}^{(j_+-1)\to}(\Gamma)$ of order at most
\BEQ 2^{3j_+} \Big(2^{j_{\phi}/4} 2^{-\frac{3}{2}j_+}\Big)^2 \times
\Big(2^{-2j_{\phi}}2^{-j_+}\Big)^2=2^{-\frac{7}{2}j_{\phi}} 2^{-2j_+}, \EEQ
that is, \big(Integration volume of $\xi_2$ w.r. to fixed $\xi_1$\big)$\times\big($Product of the amplitudes of the polymers with 3 external Cooper pair legs $\P_1^{(j_+-1)\to},\P_2^{(j_+-1)\to}\big)\times\big($Squared scale $j_+$ 
Goldstone boson propagator\big).

\end{itemize}

\vskip 1cm

\noindent 2. {\em Difference estimates for $g_{\phi},v_{\phi}$.} 
These are obtained in a straightforward way using (\ref{eq:delgphidiag}),
(\ref{eq:delgphioff}), (\ref{eq:delvjdiag}), (\ref{eq:delvjoff}) from 
$\nabla_q^2\Big\{\big(\bar{\Pi}_q^{j_+\to}(\Gamma)-\bar{\Pi}_q^{(j_+-1)\to}(\Gamma)\big) {\cal A}_q(
\Gamma^{j_+\to},\Upsilon_{3})\Big\}\Big|_{q=0}$. Because the bubble diagram ${\cal A}_q(\Gamma^{j_+\to},\Upsilon_3)$ is made up of fermion propagators with a scale $\le j_{\phi}$
effective infra-red cut-off due to the $|\Gamma|^2$-term in the denominator of
(\ref{eq:C*}), operators $\nabla_q$
change its scaling by a factor $\lesssim 2^{j_{\phi}}$. So let us study instead the effect
of the action of $\nabla_q$ on polymers of the type $ \bar{\Pi}_q^{j_+\to}(\Gamma)-\bar{\Pi}_q^{(j_+-1)\to}(\Gamma)$ studied in 1.
As a general rule, gradients w.r. to the transfer momentum
$q$ of a diagram with two external Cooper pairs can be evaluated by choosing a line of propagators connecting the
two external vertices $\xi_1,\xi_2$,

\Bigskip

{\centerline{
\begin{tikzpicture}[scale=0.5]
\draw[->,decorate,decoration=snake](2,10)--(3,9);
\draw[->,decorate,decoration=snake](3,9)--(4,8);
\draw(5,9) node {$q+\del q$};
\draw[->,decorate,decoration=snake](8,0)--(9,-1);
\draw[,decorate,decoration=snake](9,-1)--(10,-2);
\draw(11,-1)node {$q+\del q$};
\draw(0,0) node {\textbullet};
\draw(4,0) node {\textbullet};
\draw(8,0) node {\textbullet};
\draw(0,0)--(8,0);
\draw(4,4) node {\textbullet};
\draw(8,4) node {\textbullet};
\draw(4,8) node {\textbullet};
\draw(4,8) node {\textbullet};
\draw(4,8)--(4,0);\draw(4,4)--(8,4);
\draw(0,0)--(4,4);
\draw(0,0)--(4,8);
\draw(8,0) arc(60:120:4.3 and 3);
\draw[->](8,4)--(8,2); \draw(8,2)--(8,0);
\draw(10,2) node {$p_2+\del q$};
\draw[->](4,8)--(6,6); \draw(6,6)--(8,4);
\draw(8,6) node {$p_1+\del q$};
\draw(4,4)--(0,0);
\end{tikzpicture}}}

\Medskip {\bf \tiny Fig. \thesubsection.3. Changing the transfer momentum flowing inside
a diagram.}

\Bigskip see Fig. \ref{subsection:cluster}.3. Translating
$q$ by $\del q$ inside the diagram is equivalent to 
substituting $C^*_{\theta}(p_1)\cdots C^*_{\theta}(p_n)\longrightarrow C^*_{\theta}(p_1+\del q)\cdots C^*_{\theta}(p_n+\del q).$ Taking two successive derivatives w.r. to $q$ yields
sums of terms featuring the product
$\big\{ \prod_{k\not=i,i'} C^*_{\theta}(p_k)\big\} \ \nabla C^*_{\theta}(p_i)\, \nabla C^*_{\theta}(p_{i'})$. In principle, the inverse Fourier transform of a gradient fermion propagator
$\nabla C^{j,\alpha}_{\theta}(p)$, $j\le j'_{\phi}$, is
$\II(x-y)C^{j,\alpha}_{\theta}(x-y)$, which is bounded as
 in (\ref{eq:I-II}), but with a further prefactor scaling 
 at most like $2^{j_{\phi}}$, induced by the scale $j$ 
 quasi-exponential decay. This implies directly for 
scale $j_+$ irreducible diagrams $\P_{(i)}$ of type (i)
as above a relative scaling of $\nabla_q^2 {\cal A}(\P_{(i)};\xi_{ext,1})$
w.r. to ${\cal A}(\Pi_{(i)};\xi_{ext,1})$ of order 
$2^{2j_{\phi}}$ at most. Consider now a diagram $\P$
or type (ii) or (iii), involving a scale $(j_+-1)$ Goldstone
boson propagator, assume that the line contains a Goldstone boson
propagator of scale $k_+\le j_+-1$. Differentiating
$\Sigma_{\perp,\perp}^{k_+}(q+\del q)$ w.r. to $\del q$ yields (by the 
resolvent identity) 
\BEQ \Sigma_{\perp,\perp}^{k_+}(q+\del q)-\Sigma^{k_+}(q)\sim_{\del q\to 0} \Sigma_{\perp,\perp}^{k_+}(q) \ \nabla_q (\Pi^{(j_+-1)\to} 
(\Gamma){\cal A}(\Gamma,\Upsilon_3)) \ \Sigma_{\perp,\perp}^{k_+}(q)\  \del q.
\EEQ 
Proceeding by induction,
we see now that $\nabla_q (\Pi^{(j_+-1)\to}(\Gamma))(q)$ involves
a supplementary scaling factor $2^{j_{\phi}}$, exactly as
$\nabla_q {\cal A}_q(\Gamma,\Upsilon_3))$.

\Medskip Concluding, and letting $\del g_{\phi}:=g_{\phi}^{j_+\to}-g_{\phi}^{(j_+-1)\to}$, and similarly $\del v_{\phi}:=v_{\phi}^{j_+\to}-v_{\phi}^{(j_+-1)\to}$:
$\frac{\del g_{\phi}}{g_{\phi}^2}, \frac{v_{\phi}}{g_{\phi}}\del v_{\phi}$ (see (\ref{eq:delgphidiag}),
(\ref{eq:delgphioff}), (\ref{eq:delvjdiag}), (\ref{eq:delvjoff}))
 scale at 
most like 
$(2^{j_{\phi}})^2 \times  2^{-2(j_+-j_{\phi})}$, namely, (scaling factor due to $\nabla_q$)$^2 \times ($scaling factor of $\bar{\Pi}^{j_+\to}(\Gamma)-\bar{\Pi}^{(j_+-1)\to}(\Gamma))$, 
whence the estimates (\ref{eq:delgphidelvphi}).

\Bigskip 3. {\em $\Gamma$-dependence.}  Exactly as for $\nabla_q$, and for the same
reasons (see arguments in 2.), the $\Gamma$-derivative
operator $\frac{\partial}{\partial\Gamma}$ costs a scaling factor $2^{j_{\phi}}$ or 
(reintroducting dimensional constants)
$\frac{1}{\Gamma_{\phi}}$. Thus 
\BEQ \Big| \big(\bar{\Pi}_q^{j'_+\to}(\Gamma)-\bar{\Pi}_q^{(j'_+-1)\to}(\Gamma)\big)
-\big(\bar{\Pi}_q^{j'_+\to}(\Gamma')-\bar{\Pi}_q^{(j'_+-1)\to}(\Gamma')\big)\Big|\lesssim 2^{-2(j_+-j_{\phi})}\frac{\del\Gamma}{\Gamma_{\phi}} g.\EEQ   
Summing over scales $j'_+=j'_{\phi},\ldots, j_+$ yields (\ref{eq:delPiGamma}). 
The other estimate (\ref{eq:delAGamma}) follows as in 2. from the scale $j_{\phi}$ 
effective cut-off scale $|\Gamma|^2$ in the denominator of (\ref{eq:C*}).

\Medskip
 Error terms $\big|\PreSigma(\Gamma^{j_+\to};\xi-\xi')-
\PreSigma(\Gamma^{(j_+-1)\to};\xi-\xi')\big|$, $j_+\ge j'_{\phi}$ (see \S \ref{subsubsection:bubble-sum}), modifying the value of the pre-Goldstone boson
propagators inserted inside fermionic polymers, are
similarly bounded.

\vskip 1cm

{\centerline{***********************************************}}

\vskip 1cm

Next subsections are dedicated to the proof of Theorems 2--5 giving estimates
for various $n$-point functions. The general scheme of proof is as follows. Define {\em infra-red cut-off $n$-point 
functions}
\BEQ f^{j_+\to}(\vec{\xi},\vec{\bar{\xi}}):=\Big\langle \prod_{i=1}^n \Psi(\xi_i) \prod_{i=1}^{\bar{N}}
\bar{\Psi}(\bar{\xi}_i) \Big\rangle_{\theta;\lambda,j_+\to}
\EEQ
for $j_+\ge j'_{\phi}$. When $j_+=j'_{\phi}$, $f^{j_+\to}(\cdot)$
is the sum of entirely fermionic diagrams obtained by
the fermionic expansion of $\Big\langle \prod_{i=1}^n \Psi(\xi_i) \prod_{i=1}^{\bar{N}}
\bar{\Psi}(\bar{\xi}_i) \Big\rangle_{\theta;\lambda}$, i.e. of 
all fermionic polymer contributions including no Goldstone
boson propagator. When $j_+>j'_{\phi}$, the sum is rather
obtained from the fermionic polymer expansion of 
$\Big\langle \prod_{i=1}^n \Psi(\xi_i) \prod_{i=1}^{\bar{N}}
\bar{\Psi}(\bar{\xi}_i) \Big\rangle_{\theta;\lambda}$
by substituting $\Sigma^{j_+\to}$ to all Goldstone boson
propagators $\Sigma$. 

\Medskip To go from the purely fermionic contribution $f^{j'_{\phi}\to}(\vec{\xi},\vec{\bar{\xi}})$ to the 
$n$-point function $\Big\langle \prod_{i=1}^n \Psi(\xi_i) \prod_{i=1}^{\bar{N}}
\bar{\Psi}(\bar{\xi}_i) \Big\rangle_{\theta;\lambda}=\lim_{j_+\to +\infty} f^{j_+\to}(\vec{\xi},\vec{\bar{\xi}})$, one
proceeds by induction, extending downwards the 
polymers to the bosonic scales. Namely, let $\P_1,\ldots,
\P_n$ be the fermionic polymers of a given contribution to
the $n$-point function, and decide that $\P_i$ and $\P_{i'}$
are connected at scale $j_+$ (which we denote by $\P_i\sim^{j_+}\P_{i'}$) if there exists a scale $j_+$ propagator $\Sigma^{j_+}$ connecting $\P_i$ to $\P_{i'}$. Then we define 
\BEQ \P^{j'_{\phi}\to}_i=\P_i, \qquad i=1,\ldots,n;\EEQ
\BEQ \P^{j_+\to}_i=\uplus\{ \P_{i'}\ |\ \exists k_+\le j_+,
\P_i\sim^{k_+}\P_{i'}\}, \qquad j_+>j'_{\phi},i=1,\ldots,n.\EEQ

\Medskip External legs of  a polymer $\P^{j_+\to}_i$ are then
made up of $N_{ext,+}\equiv N_{ext,+}(\P_i^{j_+\to})$ external $\Sigma$-propagators with scales $>j_+$. 
It is important to realize that no supplementary
cluster expansion is needed here: going down the scales inductively, polymers $\P_i^{j_+\to}$ simply
coalesce, until there is only one polymer left at some scale
$j_{+,max}$. 


\subsection{Proof of Theorem 2: Cooper pair $n$-point functions} \label{subsection:proof-Theorem2}


In this section we prove Theorem 2. Fix $n\ge 1$ and $\xi_1,\ldots,\xi_{2n}\in\R\times\R^2$.
The main contribution to 
\BEQ F_{2n}(\vec{\xi})\equiv F_{2n}(\xi_1,\ldots,\xi_{2n}):=  \Big\langle \prod_{i=1}^{2n}\, :\, \Big(
\bar{\Psi}
{\mathbb{\Gamma}}^{\perp} \Psi\Big)(\xi_i)\, :\ 
\Big\rangle_{\theta;\lambda}
\EEQ
is as in Fig. \ref{subsection:heuristic}.2, except that $\PreSigma$-kernels are
replaced by $\Sigma$-kernels,

\Bigskip

{\centerline{
 \begin{tikzpicture}[scale=1.5]
 \draw(-1.2,0) arc(150:30:0.7 and 0.35); \draw(-1.2,0) arc(-150:-30:0.7 and 0.35);
  \draw(-1.2,-4) arc(150:30:0.7 and 0.35); \draw(-1.2,-4) arc(-150:-30:0.7 and 0.35);
   \draw(-2.2,-2) arc(150:30:0.7 and 0.35); \draw(-2.2,-2) arc(-150:-30:0.7 and 0.35);
    \draw(2,0) arc(150:30:0.7 and 0.35); \draw(2,0) arc(-150:-30:0.7 and 0.35);
  \draw(2,-4) arc(150:30:0.7 and 0.35); \draw(2,-4) arc(-150:-30:0.7 and 0.35);
   \draw(3,-2) arc(150:30:0.7 and 0.35); \draw(3,-2) arc(-150:-30:0.7 and 0.35);
 \draw(0,0) node {\textbullet};  \draw(2,0) node {\textbullet};
 \draw(-1,-2) node {\textbullet}; \draw(3,-2) node {\textbullet};
 \draw(0,-4) node {\textbullet}; \draw(2,-4) node {\textbullet};
 \draw(-1.2,0) node {\textbullet}; \draw(3.2,0) node {\textbullet};
 \draw(-2.2,-2) node {\textbullet}; \draw(4.2,-2) node {\textbullet};
 \draw(-1.2,-4) node  {\textbullet}; \draw(3.2,-4) node {\textbullet};
 \draw(-1.2-0.5,0) node {$\xi_1$}; \draw(1.2+2.5,0) node {$\xi_4$};
 \draw(-1.2-1.5,-2) node {$\xi_2$}; \draw(1.2+3.5,-2) node {$\xi_5$};
 \draw(-1.2-0.5,-4) node {$\xi_3$}; \draw(1.2+2.5,-4) node {$\xi_6$};
 \draw[decorate,decoration=snake] (0,0)--(2,0);
 \draw[decorate,decoration=snake] (-1,-2)--(3,-2);
 \draw[decorate,decoration=snake] (0,-4)--(2,-4);
  \draw(-0.9,0) node {\tiny ${\mathbb{\Gamma}}^{\perp}$};
  \draw(3,0) node {\tiny ${\mathbb{\Gamma}}^{\perp}$};
\draw(-0.9-1,0-2) node {\tiny ${\mathbb{\Gamma}}^{\perp}$};
  \draw(3+1,0-2) node {\tiny ${\mathbb{\Gamma}}^{\perp}$};
  \draw(-0.9,-4) node {\tiny ${\mathbb{\Gamma}}^{\perp}$};
  \draw(3,0-4) node {\tiny ${\mathbb{\Gamma}}^{\perp}$};
 \draw(7,-2) node {+ perm.};
 \end{tikzpicture}
 }}
 
\Medskip {\bf \tiny Fig. \thesubsection.2. $6$-point function
of Cooper pairs.}

\Bigskip where "perm." indicates the sum over all possible pairings. Each bubble
diagram is equal to $(1+O(g'))\frac{1}{\lambda}$ by the gap equation. This yields
the leading term in Theorem 2. Corrections are due to more complicated fermionic diagrams
than bubble diagrams, possibly containing other Goldstone bosons. They all contain at least
one more vertex, implying a further prefactor in $O(g)$.


\subsection{Proof of Theorem 3: fermion quasi-exponential decay}


Main diagrams  contributing to (\ref{eq:0.39}), resp. 
(\ref{eq:0.40}) in Theorem 3 are 

\Bigskip

{\centerline{
\begin{tikzpicture}
\draw(0,0) node {\textbullet};
\draw(0,0)--(2,0); \draw(2,0) node {\textbullet};
\draw(0,0.4) node {$\xi$}; \draw(2,0.4) node {$\xi'$};
\end{tikzpicture}
}}

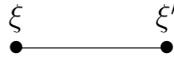
\Medskip {\bf\tiny Fig. \thesubsection.1. Two-point functions for isolated fermions.}

\Bigskip

{\centerline{
\begin{tikzpicture}[scale=1]
\draw(0,0) node {\textbullet};
\draw(0,0)--(2,0); \draw(2,0) node {\textbullet};
\draw(0,0.4) node {$\xi_1$}; \draw(2,0.4) node {$\xi_2$};
\draw(0,-1) node {\textbullet};
\draw(0,-1)--(2,-1); \draw(2,-1) node {\textbullet};
\draw(0,-1.4) node {$\xi_3$}; \draw(2,-1.4) node {$\xi_4$};
\end{tikzpicture}
}}

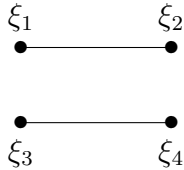
\Medskip {\bf\tiny Fig. \thesubsection.2. Four-point functions for fermions either in non-Cooper pairing, or in parallel $(//)$ 
Cooper pairing.}

\Bigskip
in the limit when $\xi_3=\xi_1,\xi_4=\xi_2$ and $|\xi-\xi'|$, 
resp. $|\xi_1-\xi_2|\to\infty$. 

\Bigskip This,  and the quasi-exponential decay of 
the $C^j$-kernels (see Proposition \ref{prop:C}) and of the $\Sigma_{//,//}$-kernel
 (see  (\ref{eq:decay-PreSigma//})
 and discussion in \S \ref{subsection:Ward} {\bf C.}), accounts for the
quasi-exponential decay of such correlation functions.


\subsection{Proof of Theorem 4: phase transition}  \label{subsection:phase-transition}


In this section we prove Theorem 4. By global rotation
invariance, we may assume that $\theta=0$. We want
to prove 
\BEQ F_1(\xi):=\langle (\bar{\Psi}\sigma^1\Psi)(\xi)\rangle_{0;\lambda}=2(1+O(g'))\frac{\Gamma_{\phi}}{\lambda}. \EEQ
 On the other hand,  $\langle (\bar{\Psi}\sigma^2\Psi)(\xi)\rangle_{0;\lambda}=0$, as follows from the vanishing of
the bosonic one-point function in the perpendicular direction
(see
 (\ref{eq:one-pt-fct-perp})), proved using a Ward identity. 
Note that, for any cut-off scale $j_+$,  the gray blob in  the figure above (\ref{eq:one-pt-fct-perp}) cannot be connected to the external Cooper pair
$\bar{\Psi}\sigma^2\Psi$  by a Goldstone boson by momentum
conservation, hence the limit $j_+\to +\infty$ can be taken; also, error terms in $O(2^{-(j_+-j'_{\phi})}$  vanish in that limit.  All together, this implies Theorem 4.
Clearly (by translation invariance) $F_1(\xi)=F_1$ is a constant. 

\Medskip Main contribution to $\langle (\bar{\Psi}\sigma^1\Psi)(\xi)\rangle_{0;\lambda}$ is  the bubble diagram

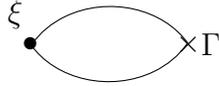
\Bigskip

{\centerline{
\begin{tikzpicture}
\draw(14,2) node {\textbullet}; \draw(13.8,2.4) node {$\xi$};
\draw(16.1,2) arc(30:150:1.2 and 1); \draw(14,2) arc(210:330:1.2 and 1);
\draw(16.1,2) node {$\times$};
 \draw(16.4,2) node {$\Gamma$};
\end{tikzpicture}}}

\noindent which is indeed equal to $\Tr((\sigma^1)^2)(1+O(g'))
\Gamma_{\phi}{\cal A}_0(\Upsilon_{3,diag})=2(1+O(g'))\frac{\Gamma_{\phi}}{\lambda}$ by the gap equation.
By number conservation (see e.g. discussion before Definition \ref{def:averaging}), all diagrams contributing to $F_1$
contain at least one off-diagonal fermionic propagator, hence
they all have at least one $\Gamma_{\phi}$ in factor.
Since the one-point function of the Goldstone boson vanishes
by construction, the Cooper pair $(\bar{\Psi}\sigma^1\Psi)(\xi)$
may not be directly connected to a Goldstone boson propagator.
On the other hand, more complicated fermionic diagrams than the 
bubble diagram, potentially containing Goldstone bosons are possible. However, they all contain at least one more vertex,
implying a further prefactor in $O(g)$.


\subsection{Proof of Theorem 5: local transverse behavior of the Goldstone boson}  \label{subsection:transverse-mass}


We are interested in this subsection in large deviations for the random variable
\BEQ X_{\Omega}:=\Re \int_{\Omega} d\xi\, (\bar{\Psi}\sigma(\theta)\Psi)(\xi)\EEQ
-- interpreted as the integral over $\Omega$ of the projection along the
direction $\theta$ of the Goldstone
boson --,
where $\sigma(\theta)=\cos(\theta)\, \sigma^1 + \sin(\theta)\, \sigma^2$, and
$\Omega\subset V$ is some volume comparable to a box of  transition scale $j_{\phi}$,
say, $\Omega$ is a roughly cubic connected union of boxes $\Del\in\D^{j_{\phi}}$ with dimensions $\approx\frac{L}{v_{\phi}}\times L\times L$, $L\approx \frac{v_{\phi}}{\Gamma_{\phi}}$.  By rotation invariance,
 we may and shall assume that $\theta=0$.

\Medskip   We let $\ell$ be the ratio
$\frac{L}{v_{\phi}/\Gamma_{\phi}}$. The  validity of the 
argument below extends in part to  $\ell\gg 1$, so we simply
assume that $\ell\gtrsim 1$. Note that 
$\Vol(\Omega)\approx \ell^3
\frac{v_{\phi}^2}{\Gamma_{\phi}^3}$.

\Bigskip {\em Standard large-deviation theory }(see e.g. Cramer's theorem in \cite{DemZei},
\S 2.2) {\em implies the following large deviation bound.}  Let $\bar{X}_{\Omega}:=
\langle X_{\Omega}\rangle_{0;\lambda}$ be the average of $X_{\Omega}$ in
the symmetry-broken measure along $\theta=0$; as
proved in \S \ref{subsection:phase-transition}, $ \bar{X}_{\Omega}=\Vol(\Omega) 
  \langle (\bar{\Psi}\sigma^1\Psi)(0)\rangle_{0;\lambda}\sim \Vol(\Omega)
 \frac{\Gamma_{\phi}}{\lambda}$. Let 
 \BEQ \Lambda(A):=  \log \, \langle e^{A X_{\Omega}} \rangle_{0;\lambda} \EEQ
be the {\em log-cumulant} of $X_{\Omega}$, and 
\BEQ \tilde{\Lambda}(x):=\sup_{A\in\R} \Big(Ax-\Lambda(A) \Big) \EEQ
its {\em Legendre transform. Then, for $\eta>0$,  
\BEQ \proba[X_{\Omega}>\bar{X}_{\Omega}+\eta \frac{\Gamma_{\phi}}{\lambda}\,  \Vol(\Omega)]
\le e^{-\tilde{\Lambda}(\bar{X}_{\Omega}+\eta \frac{\Gamma_{\phi}}{\lambda}\,  \Vol(\Omega))}.
\label{eq:LDP}
\EEQ}
Similarly,
\BEQ \proba[X_{\Omega}<\bar{X}_{\Omega}-\eta \frac{\Gamma_{\phi}}{\lambda}\,  \Vol(\Omega)]
\le e^{-\tilde{\Lambda}(\bar{X}_{\Omega}-\eta \frac{\Gamma_{\phi}}{\lambda}\,  \Vol(\Omega))}.
\EEQ
We concentrate on the consequences of (\ref{eq:LDP}) in the sequel; similar bounds, obtained
by substituting $\eta\longrightarrow-\eta$,  hold
for $\proba[X_{\Omega}<\bar{X}_{\Omega}-\eta \frac{\Gamma_{\phi}}{\lambda}\,  \Vol(\Omega)]$.

\Medskip 
Note that $\Lambda(0)=0$ and (by Jensen's inequality) $\Lambda(A)\ge A\bar{X}_{\Omega}$ for all $A$. As a consequence,  $\min_{x\in\R} \tilde{\Lambda}(x)=\tilde{\Lambda} (\bar{X}_{\Omega})=0$, so that $\tilde{\Lambda}'(\bar{X}_{\Omega})=0$,
and $\tilde{\Lambda}''(x)=1/\Lambda''(A)$, where 
\BEQ A\equiv A(x)={\mathrm{argmax}}\Big(A\mapsto Ax-\Lambda(A) \Big)
=(\Lambda')^{-1}(x)  \label{eq:A(x)}
\EEQ
 is the unique $A\in\R$
realizing the maximum of the function $A\mapsto  Ax-\Lambda(A) $; in particular, $A(\bar{X}_{\Omega})=0$.  Thus 
\BEA  && -\log\Big( \proba[X_{\Omega}>\bar{X}_{\Omega}+\eta\frac{\Gamma_{\phi}}{\lambda}\,  \Vol(\Omega)] \Big)\ge 
\half \eta^2 (\frac{\Gamma_{\phi}}{\lambda})^2\,  \Vol(\Omega)^2\, 
\Big(\max_{0\le A\le A(\bar{X}_{\Omega}+\eta\frac{\Gamma_{\phi}}{\lambda}\,  \Vol(\Omega))} \Lambda''(A) \Big)^{-1}
\nonumber\\
&& = \half \eta^2 (\frac{\Gamma_{\phi}}{\lambda})^2\,  \Vol(\Omega)^2\,  \Big( \max_{0\le A\le A(\bar{X}_{\Omega}+\eta \frac{\Gamma_{\phi}}{\lambda}\, \Vol(\Omega))}  \langle 
X_{\Omega} X_{\Omega}\rangle^c_{\lambda,\Omega,A}\Big)^{-1},  \label{eq:log-P}
\EEA
where $\langle \ \cdot\ \rangle^c_{\lambda,\Omega,A}$ is the connected expectation w.r. to
the perturbed measure $d\mu_{\lambda,\Omega,A}\propto e^{AX_{\Omega}} d\mu_{0;\lambda}$.

\Medskip Now, define
\BEQ A_{\eta}:=A(\bar{X}_{\Omega}+\eta \frac{\Gamma_{\phi}}{\lambda}\, \Vol(\Omega));
\EEQ
By (\ref{eq:A(x)}), 
\BEQ \langle X_{\Omega}\rangle_{\lambda,\Omega,A_{\eta}} =\Lambda'(A_{\eta})\equiv \bar{X}_{\Omega}+\eta \frac{\Gamma_{\phi}}{\lambda}\, \Vol(\Omega); \label{eq:extremum-condition} 
\EEQ
equivalently, $A_{\eta}=(\Lambda')^{-1}(\bar{X}_{\Omega}+\eta \frac{\Gamma_{\phi}}{\lambda}\, \Vol(\Omega))$.

\Bigskip   By construction, $\langle X_{\Omega}\rangle_{\lambda,\Omega,A_{\eta}}$ is
equal to the sum of truncated expectation values $\sum_{n\ge 0} A_{\eta}^n \langle X^{n+1}_{\Omega}\rangle^c_{0;\lambda}$.
{\em If $n\ge 2$}, the main contributions to $A_{\eta}^n \langle X_{\Omega}^{n+1}\rangle^c_{0;\lambda}$ are in the form of $(n+1)$ bubble
diagrams connected through scale $\approx j_{\phi}$ Goldstone bosons to a fermionic polymer of scale $\approx j_{\phi}$, represented as the central
gray blob in the Figure below.  

\Bigskip

{\centerline{
\begin{tikzpicture}[scale=0.5]
\draw[fill,color=gray](0,0) circle(2);
\draw[decorate,decoration=snake](2,0)--(5,0);
\draw(5,0) arc(150:30:1 and 0.5);
\draw(5,0) arc(-150:-30:1 and 0.5);
\draw(6.6,0) node {\textbullet}; 
\draw(7.4,0) node {$\sigma^1$};
\begin{scope}[rotate=72]
\draw[decorate,decoration=snake](2,0)--(5,0);
\draw(5,0) arc(150:30:1 and 0.5);
\draw(5,0) arc(-150:-30:1 and 0.5);
\draw(6.6,0) node {\textbullet}; 
\draw(7.4,0) node {$A_{\eta}\sigma^1$};
\end{scope}
\begin{scope}[rotate=144]
\draw[decorate,decoration=snake](2,0)--(5,0);
\draw(5,0) arc(150:30:1 and 0.5);
\draw(5,0) arc(-150:-30:1 and 0.5);
\draw(6.6,0) node {\textbullet}; \draw(7.4,0) node {$A_{\eta}\sigma^1$};
\end{scope}
\begin{scope}[rotate=216]
\draw[decorate,decoration=snake](2,0)--(5,0);
\draw(5,0) arc(150:30:1 and 0.5);
\draw(5,0) arc(-150:-30:1 and 0.5);
\draw(6.6,0) node {\textbullet}; 
\draw(7.4,0) node {$A_{\eta}\sigma^1$};
\end{scope}
\begin{scope}[rotate=288]
\draw[decorate,decoration=snake](2,0)--(5,0);
\draw(5,0) arc(150:30:1 and 0.5);
\draw(5,0) arc(-150:-30:1 and 0.5);
\draw(6.6,0) node {\textbullet}; 
\draw(7.4,0) node {$A_{\eta}\sigma^1$};
\end{scope}
\end{tikzpicture}
}}

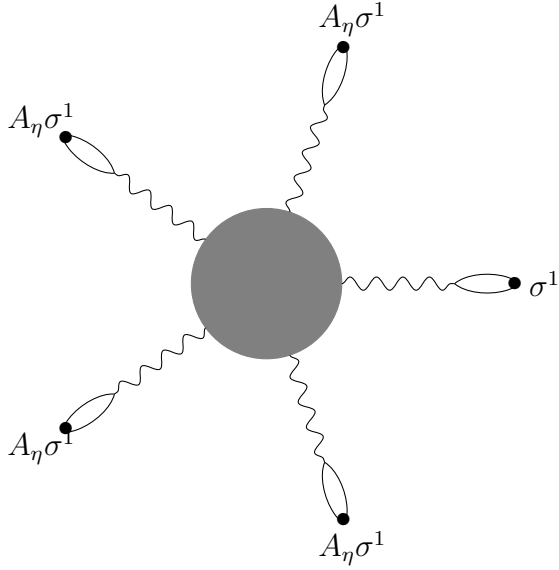
{\bf \tiny Fig. \thesubsection.1. Main contribution to
$A_{\eta}^n \langle X_{\Omega}^{n+1}\rangle^c_{0;\lambda}$
for $n=4$.}

\Bigskip Similarly, the truncated expectation $\langle X_{\Omega}X_{\Omega}\rangle^c_{\lambda,\Omega,A_{\eta}}$ 
may be expanded into $\sum_{n\ge 0} A_{\eta}^n \langle X^{n+2}_{\Omega}\rangle^c_{0;\lambda}$.

\Medskip Let
\BEA  V_{\Omega} &:=& \int\int_{\Omega\times\Omega} d\xi\, d\xi'\ \frac{1}{\lambda^2} \Sigma(\xi-\xi') \nonumber\\
&\approx& \frac{1}{\lambda^2} \Vol(\Omega) \int_0^L \frac{g_{\phi}/4\pi v^2_{\phi}}{r}\, 4\pi r^2\, dr \approx \Vol(\Omega)\frac{m^* \Gamma_{\phi}^2 L^2}{g^2 v_{\phi}^2}  \approx \Vol(\Omega)\ell^2
\frac{m^*}{g^2}
\EEA
be the  Cooper pair two-point function (see (\ref{eq:expected}) 
for the intuition, or Theorem 2 for a more precise statement)
integrated over the volume $\Omega$.  Since $\Vol(\Omega)\approx \ell^3\frac{v_{\phi}^2}{\Gamma_{\phi}^3}$, 
$V_{\Omega}\approx \ell^5 \frac{m^* v_{\phi}^2}{g^2 \Gamma_{\phi}^3}$.

\Medskip A diagram as in the above Figure contributes
at most $A_{\eta}^n$ times  : (i) $\frac{1}{\Vol(\Omega)} V_{\Omega}\approx\ell^2 \frac{m^*}{g^2}$ per Goldstone boson propagator with one fixed extremity, terminated in the other
direction by a bubble,
all together:  $O((\ell^2\frac{m^*}{g^2})^{n+1})$; (ii) $ g^{n+1} (2^{j_{\phi}/2})^{2(n+1)-4} (p^*_F)^{-2(n+1)}=\frac{g\Gamma_{\phi}}{(m^*)^3 
v_{\phi}^4} \times \left(\frac{g}{m^*\Gamma_{\phi}}\right)^n$ (power-counting (\ref{eq:GFMSB-P}) for an amputated fermionic diagram with lowest scale $j_{\phi}$ and $2(n+1)$ external
legs), times a non-dimensional localization factor breaking
translation invariance, $\approx (\ell 2^{j_{\phi}})^3 \approx \ell^3\frac{\mu^3}{\Gamma_{\phi}^3}\approx  \ell^3
\frac{(m^*)^3 v_{\phi}^6}{\Gamma^3_{\phi}}$, all together :
 $O(\ell^3 \frac{g v_{\phi}^2}{ \Gamma_{\phi}^2}(\frac{g}{m^*\Gamma_{\phi}})^n)$. 
Hence, assuming $\frac{A_{\eta}}{g\Gamma_{\phi}}\ll 1$
(which will be checked later on by self-consistency  as soon
as $\eta\ll 1$)
\BEQ  \sum_{n\ge 2}  \frac{A_{\eta}^n}{n!} \langle X_{\Omega}^{n+1}\rangle^c_{0;\lambda}
\lesssim \ell^5 \frac{m^*v_{\phi}^2}{g\Gamma_{\phi}^2} \exp_2(\ell^2 A_{\eta}/g\Gamma_{\phi}) \lesssim \ell^9 \frac{m^* v_{\phi}^2} {g^3\Gamma_{\phi}^2} 
(\frac{A_{\eta}}{\Gamma_{\phi}})^2
\EEQ
where $\exp_2(t):=e^t-1-t\approx t^2 e^t$ if $|t|\ll 1$. 

\Medskip On the other hand, the terms of order $n=0,1$ are
resp. $\langle X_{\Omega}\rangle_{0;\lambda}=\bar{X}_{\Omega}$ and $A_{\eta} \langle X_{\Omega}X_{\Omega}\rangle^c_{0;\lambda}\approx A_{\eta} V_{\Omega}\approx A_{\eta} \Vol(\Omega)
\ell^2 \frac{m^*}{g^2}$. Concluding:

\BEQ \langle X_{\Omega}\rangle_{\lambda,\Omega,A_{\eta}} -\bar{X}_{\Omega} \approx  A_{\eta}V_{\Omega} + O\Big(\ell^9 \frac{m^* v_{\phi}^2} {g^3\Gamma_{\phi}^2} 
(\frac{A_{\eta}}{\Gamma_{\phi}})^2\Big). \label{eq:XOmega}
\EEQ
Comparing with
 (\ref{eq:extremum-condition}), we get, neglecting the error
 term in (\ref{eq:XOmega}): 
 \BEQ A_{\eta}\approx \ell^{-2} g \Gamma_{\phi} \eta. \label{eq:Aeta}
 \EEQ
 
   Taking this as the value for $A_{\eta}$, we check
 that the error term in (\ref{eq:XOmega}) is $\lesssim \ell^5 m^*  \left(\frac{v_{\phi}}{\Gamma_{\phi}}\right)^2 \frac{1}{g} \eta^2$, which is $\ll \Big|\langle X_{\Omega} \rangle_{\lambda,\Omega,A_{\eta}}-\bar{X}_{\Omega} \Big|=\eta\frac{\Gamma_{\phi}}{\lambda}\Vol(\Omega)\approx 
 \eta\frac{\Gamma_{\phi}}{\lambda}  \ell^3 \frac{v_{\phi}^2}{\Gamma_{\phi}^3} \approx \ell^3
 m^* \left(\frac{v_{\phi}}{\Gamma_{\phi}}\right)^2 \frac{1}{g} \eta $   {\em provided $\ell^2\eta\ll 1$}. 

\Bigskip 
Now the covariance $\max_{0\le A\le A(\bar{X}_{\Omega}+\eta\frac{\Gamma_{\phi}}{\lambda} \Vol(\Omega))}  \langle 
X_{\Omega} X_{\Omega}\rangle^c_{\lambda,\Omega,A}$ in (\ref{eq:log-P}) is similarly computed  as
\BEQ \langle X_{\Omega} X_{\Omega}\rangle^c_{\lambda}+ O(\ell^5\frac{m^* v_{\phi}^2}{g\Gamma^2_{\phi}}\times \frac{\ell^2}{g\Gamma_{\phi}}) \exp_1(\ell^2 A_{\eta}/g\Gamma_{\phi}) \approx V_{\Omega}  +  O(\ell^7 \frac{m^* v_{\phi}^2}{g^2\Gamma^3_{\phi}}
) \exp_1(\ell^2 A_{\eta}/g\Gamma_{\phi})\EEQ
where $\exp_1(t):=e^t-1\approx te^t$.  The error term,
computed using (\ref{eq:Aeta}),  is
now $\lesssim  \ell^7\frac{m^* v_{\phi}^2}{g^2\Gamma_{\phi}^3} \eta \ll V_{\Omega}\approx \ell^5 \frac{m^* v_{\phi}^2}{\Gamma_{\phi}^3}
 \frac{1}{g^2}$, provided $\ell^2 \eta\ll 1$ once again.

\Bigskip  Concluding:  assume  $\ell^2\eta\ll 1$, 
then

\BEA   -\log\Big( \proba[X_{\Omega}>\bar{X}_{\Omega}+\eta\frac{\Gamma_{\phi}}{\lambda}\,  \Vol(\Omega)] \Big) &\gtrsim& 
\eta^2 (\frac{\Gamma_{\phi}}{\lambda})^2\,  (\frac{ \ell^3 v_{\phi}^2}{\Gamma^3_{\phi}})^2 \, \frac{ \Gamma_{\phi}^3 g^2}{\ell^5 m^* v_{\phi}^2} \nonumber\\
& \approx& \ell  mv^2_{\phi} \frac{1}{\Gamma_{\phi}} \eta^2 \approx \ell \frac{\mu}{\Gamma_{\phi}}\eta^2\approx \ell 2^{j_{\phi}}\eta^2.  \label{eq:large-deviation-rate}
\EEA
For $\ell\approx 1$ one has proved Theorem 5. For $\ell$ 
larger one observes that the large deviation rate (\ref{eq:large-deviation-rate}) increases like $\ell\propto 
\Vol(\Omega)^{1/3}$, and not $\Vol(\Omega)$ as expected from
the discussion in the Introduction based on the effective
potential approach. Note however that this volume cubic root
prefactor is not reliable, since it is valid only for small deviations
$\eta\ll \ell^{-2}$. Another method would be needed to study
large  transverse deviations of the Goldstone boson  over a
large volume.


\section{Generalizations and perspectives} \label{section:conclusion}


This work is meant to be the first of a series of articles investigating
the superconducting phase transition and quantum fluctuations of 
superconducting materials in various situations. The spectacular 
experimental and theoretical advances of the last 30 years or so,
see e.g.  
\cite{Fra,SanSer}, is a
strong incentive to draw detailed predictions from our microscopic approach in a variety of situations.

\Bigskip{\bf A first generalization.} As mentioned in the Introduction,
our approach extends in a  straightforward way to more general two-body
potentials $U$ as in (\ref{eq:intro-U}). Substituting the kernel 
$\frac{\omega^2((\vec{p}_1-\vec{p}_3)/\hbar)}{(p_1^0-p_3^0)^2 + \omega^2((\vec{p}_1-\vec{p}_3)/\hbar)}$ to the delta-interaction
only changes numerical constants, since the phonon interaction is a {\em short-range} one. The same conclusion holds if one adds
a small  two-body potential
$\hat{v}$ with high enough infra-red cut-off.

\Bigskip{\bf Three-dimensional theory.}  The phase space analysis of the 3D model is more complicated because the sector counting of Proposition
\ref{prop:sector-counting}
is not available, due to the fact that fixing two sectors does {\em not} fix the remaining two sectors of a
 vertex any more. Thus a precise sector-counting argument is needed already
 in  Regime I. However, it can be proved  perturbatively that there is just {\em at most one sum over sectors per vertex} in any dimension, as proved here in \S \ref{subsubsection:single-scale-bounds} by following a loop; we plan to extend this argument non-perturbatively (see \cite{MR-single} for a first step in that direction). After resumming explicitly the leading contribution in the form of chains of Cooper pair
 bubbles as in dimension 2, remaining terms should be small by virtue of
 a  1/N argument. Consequently, we expect the present proof to extend more or less straightforwardly to the 3d case. 

\Bigskip{\bf BCS theory at positive temperature.} The finite-temperature
formalism using Matsubara frequences, as presented e.g. in Chapter 7
of the book by Fetter and Walecka, makes it possible to compute
Green's function of the system in thermal equilibrium in terms of
a modified functional integral with discretized energies $p^0$  now ranging in the set 
$\{\hbar\omega_n,n\in\Z\}, \omega_n:=\frac{(2n+1)\pi }{\hbar\beta}$, where $\beta$ is the inverse
temperature. At first sight, the same scheme should produce similar
conclusions when the lowest Matsubara frequencies are smaller than
the energy gap $\Gamma_{\phi}$, i.e. when $k_B T\ll \Gamma_{\phi}$. 
In the inverse regime ($k_B T\gg \Gamma_{\phi}$), Cooper pairing
should be shown to be statistically unfavorable. Interesting things should
occur in 3D around a critical temperature $T_c\approx \frac{\Gamma_{\phi}}{k_B}$, where one should be able to investigate more precisely
the quantum second-order phase transition, in connection to ideas due to
M. Salmhofer \cite{Sal,Sal2}.

\Bigskip
{\bf  Real time.}  The Hamitonian dynamics may be discussed in a $(1+d)$-dimensional quantum-field
theoretic framework with  real-time variable $t$  using advanced
and retarded propagators at temperature zero, or more generally the Keldysh
formalism \cite{Ram,LeBel2} at $T>0$. The major difficulty is now that the infra-red singularity is carried by a {\em paraboloid} in energy-momentum variables, $p_0-\frac{1}{2m}|\vec{p}|^2=\mu$, a two-dimensional manifold instead of a circle.


\section{Appendix}


We collect in this section some technical formulas and bounds required for the proof. 
An essential bound is Proposition \ref{prop:Iago} below, ensuring the convergence
of the series of perturbations.


\subsection{Grassmann integrals}  \label{subsection:Grassmann}


We briefly recall here usual conventions regarding
Grassmann integrals, see e.g. \cite{PesSch}, \S 9.5. 
By definition, if $\psi_1,\ldots,\psi_n$ are Grassmann
anticommuting
variables, and $F=F(\psi_1,\ldots,\psi_n)$ is a
polynomial function, then $\int d\psi_n \cdots d\psi_1\
 F$ is the coefficient of $\psi_1\cdots\psi_n$ in $F$.
In particular, if $\psi,\bar{\psi}$ is a couple of conjugate Grassmann
variables, i.e. $\psi=\frac{\psi_1+\II\psi_2}{\sqrt{2}}, \bar{\psi}=\frac{\psi_1+\II\psi_2}{\sqrt{2}}$, then
\BEQ \int d\bar{\psi} \, d\psi \ \psi\bar{\psi}=1,
\EEQ
and
\BEQ \int d\bar{\psi} \, d\psi \ e^{-\bar{\psi} b\psi}=b,
\qquad  \int d\bar{\psi} \, d\psi \  \psi\bar{\psi} \, e^{-\bar{\psi} b\psi}=1=\frac{1}{b}\cdot b.
\EEQ 
In case there are several couples $(\psi_i,\bar{\psi}_i)_{i=1,2,\ldots}$, the above rules generalized easily using
some linear algebra, 
\BEQ \Big(\prod_i \int d\bar{\psi}_i \, d\psi_i\Big)\, e^{-\sum_{i,j} \bar{\psi}_i B_{ij}\psi_j} = \det(B),\ \ 
\Big(\prod_i \int d\bar{\psi}_i \, d\psi_i\Big)\,  \psi_k \bar{\psi}_{\ell}\, e^{-\sum_{i,j} \bar{\psi}_i B_{ij}\psi_j} = \det(B)\ (B^{-1})_{k,\ell}.
\EEQ
The measure $\frac{1}{\det(B)} \Big(\prod_i \int d\bar{\psi}_i \, d\psi_i\Big)\, e^{-\sum_{i,j} \bar{\psi}_i B_{ij}\psi_j}$ is called the (normalized)
{\em Grassmann Gaussian measure with covariance kernel
$B^{-1}$}, where $B=B(\psi,\bar{\psi}):=\sum_{i,j}
\bar{\psi}_i B_{ij}\psi_j$.


\subsection{Integration by parts formulas}  \label{subsection:integration-by-parts}


\begin{Proposition} \label{prop:IPP}
\begin{enumerate}
\item Let $\phi$ be a (fermionic or bosonic) Gaussian field with
translation-invariant
covariance kernel $C_{\phi}$ and functional measure $d\mu(\phi)$. Then, letting ${\cal F}\equiv {\cal F}[\phi]$ be a functional of $\phi$,
\BEQ \int d\mu(\phi)\ \phi(\xi){\cal F}[\phi] = \int d\xi'\, 
C_{\phi}(\xi-\xi')\ \cdot \  \int d\mu(\phi) \frac{\del {\cal F}[\phi]}{\del \phi(\xi')}.
\EEQ

\item Consider a smooth variation of the covariance kernel,
$\eps\mapsto C_{\phi}(\eps)$ and the associated functional
measures $d\mu_{\eps}(\phi)$, with $C_{\phi}(0)\equiv C_{\phi}$. Then
\BEQ \frac{d}{d\eps} \Big(\int d\mu_{\eps}(\phi)\ {\cal F}[\phi] 
\Big) \Big|_{\eps=0}= \int d\xi \, \int d\xi'\, \int d\mu(\phi)\  \Big[
\frac{\partial C_{\phi}}{\partial\eps}(\eps=0;\xi-\xi') \frac{\del^2}{\del \phi(\xi)
\del\phi(\xi')}\Big] {\cal F}[\phi] . 
\EEQ
\end{enumerate}
\end{Proposition}

The proof of these elementary identities in a quantum field theoretic context can be found e.g. in \cite{Unt-rev}. 
They imply in particular the following elegant formula
for {\em horizontal cluster expansions}, called 
Brydges-Kennedy-Abdesselam-Rivasseau (BKAR) formula.  For that we first
need some definitions.
Introduce a partition of $\R\times\R^2$ into a disjoint union
of "objects" $o^j$, which are clusters, i.e. connected unions of 
scale $j$ boxes; let  ${\cal O}^j$ denote the set of objects.
(The simplest example is when objects are elementary scale $j$ boxes, and  ${\cal O}^j=\D^j$). A {\em weakening function}
$s^j:{\cal O}^j\times{\cal O}^j\to[0,1]$ is a function assigning
a weight to each link $\ell=(o_{\ell},o'_{\ell})$ between two
different objects; by hypothesis, $s^j_{o,o'}=s^j_{o',o}$
$(o\not=o')$ and $s^j_{o,o}=1$. Alternatively,
$s^j$ may be considered as a weakening function $s^j:\D^j\times\D^j\to[0,1]$ which is a constant on a box:  letting $o_{\Del}$ being the cluster containing $\Del$,   $s^j_{\Del,\Del'}:=s^j_{o_{\Del},o_{\Del'}}$.  Then $C^j_{\theta}(s^j):(\R\times
\R^2)\times(\R\times\R^2)\to\C$ is the Hermitian kernel defined by
\BEQ C^j_{\theta}(s^j;\xi,\xi'):=\sum_{\Del,\Del'\in\D^j} 
\chi_{\Del}(\xi)\chi_{\Del'}(\xi') s^j_{\Del,\Del'}C^j_{\theta}(\xi,\xi').
\EEQ
A weakening function $s^j$ may be chosen independently for
each scale $j=j_D,\ldots,j'_{\phi}-1$. 
The associated Grassmann Gaussian measure is denoted by
$d\mu^*_{\theta}(\vec{s})$, where $\vec{s}:=(s^{j_D},\ldots, 
s^{j'_{\phi}-1})$. We now  present the outcome of a single
 application of the  BKAR formula at scale $j$, so that
 $s^{j'}\equiv 1$ if $j'\not=j$. The result is in terms of a sum over 
 scale $j$ forests, whose set is denoted by ${\cal F}^j$. 
 Forests are disjoint unions of trees.  We write $o\sim_{\F^j}o'$ $(\F^j\in{\cal F}^j,o,o'\in {\cal O}^j)$ if $o$ and $o'$
 are in the same connected component of $\F^j$.

\begin{Proposition}[BKAR formula](see \cite{MagUnt2}, Proposition 2.6)  \label{prop:BKAR}
Let $F^j\equiv F^j(\Psi^j,\bar{\Psi}^j)$ be a functional of the scale $j$
fermion fields. Then
\BEA && \langle F^{j} \rangle_{\theta;\lambda}=\sum_{\F^{j}\in {\cal F}^j} \left[ \prod_{\ell\in L(\F^j)} \int_0^1 dw_{\ell} \int_{o_{\ell}} d\xi_{\ell} \int_{o'_{\ell}}d\xi'_{\ell}\,  \right] \nonumber\\
&& \qquad \qquad \frac{1}{{\cal Z}^*_{\lambda}} \int d\mu^*_{\theta}(s^j(\vec{w}))(\Psi,\bar{\Psi})\,  {\mathrm{Hor}}^j \left( F^j(\Psi^j,\bar{\Psi}^j) e^{ -{\cal L}_{\theta}(\Psi,\bar{\Psi})} \right),\EEA
where 
\BEA && {\mathrm{Hor}}^j:=\prod_{\ell\in L(\F^j)}  D^j_{\ell},
\label{eq:Horj}\\
&&\qquad \qquad D^j_{\ell}:=
\sum_{\alpha_{\ell}\in\Z/2^j\Z}  C^{j,\alpha_{\ell}}_{\theta}(s^j(\vec{w});\xi_{\ell},\xi'_{\ell}) \frac{\del}{\del\Psi^{j,\alpha_{\ell}}(\xi_{\ell})} \frac{\del}{\del\Psi^{j,\alpha_{\ell}}(\xi'_{\ell})} , \label{eq:Djl}
\EEA
 and $s(\vec{w})=(s_{o,o'}(\vec{w}))_{o,o'\in{\cal O}^j}$, 
$s_{o,o'}(\vec{w}),o\not=o'$ being the infimum of the $w_{\ell}$ for $\ell$ running over the unique path from $o$ to $o'$ in $\F^j$ if $o \sim_{\F^j} o'$, and $0$ else.
\end{Proposition}


\subsection{Local bounds for determinants}  \label{subsection:local-bounds-det}


We prove here the statement (\ref{eq:Iago-factor}) made
in \S \ref{subsubsection:single-scale-bounds} concerning bounds for diagrams with a large number of vertices per box. Before we do 
so however, let us discuss this problem briefly in general
terms,  and cite
a result (Proposition \ref{prop:Iago} below) proved in a different context in \cite{IagMag}. 
The proposition is not directly applicable in our context, so
we shall give an independent proof of  (\ref{eq:Iago-factor}), which however relies on the same arguments.

\Medskip
The expansion produces multilinear expressions
of the type 
\BEQ G(\vec{\xi},\vec{\bar{\xi}}):=\int d\mu^*_{\theta}(\Psi^j,
\bar{\Psi}^j)\  \Big(\prod_{i=1}^n {\Psi}^{j,\alpha_i}_{\sigma_i}(\xi_i) \Big)
\Big(\prod_{\bar{i}=1}^n \bar{\Psi}^{j,\bar{\alpha}_{\bar{i}}}_{\bar{\sigma}_{\bar{i}}}(\bar{\xi}_{\bar{i}}) \Big)
, \EEQ
$\sigma_i,\bar{\sigma}_{\bar{i}}=\uparrow,\downarrow$.  Applying Wick's formula and taking signs into account,
one gets 
\BEQ G({\vec{\xi}},\bar{\vec{\xi}})=
\det (A) \EEQ
where
\BEQ A=(a_{i,\bar{i}})_{1\le i,\bar{i}\le n}, a_{i,\bar{i}}=\del_{\alpha_i,\bar{\alpha}_{\bar{i}}} C^{j,\alpha_i}_{\sigma_i,\bar{\sigma}_{\bar{i}}}(\xi_i,\bar{\xi}_{\bar{i}}).  \label{eq:aii'} \EEQ

Expanding naively the determinant, one gets  a sum over $n!$ terms indexed by permutations $\pi\in{\cal P}_n$ of the set of $\bar{i}$-indices,
making the sum over $n$ naively divergent.
Looking at this factorial into details,  however, one sees that this apparent 
divergence comes from accumulations of large numbers of fields  $\Psi^j(\xi_i),\bar{\Psi}^j(\xi_i)$ inside boxes. Namely, assuming e.g. that the $\xi_i$'s and $\bar{\xi}_{\bar{i}}$'s are
all located in different boxes, Proposition \ref{prop:C} implies
\BEQ \sum_{\pi\in {\cal P}_n} \prod_{i=1}^n |C^{j,\alpha_i}_{\sigma_i,\bar{\sigma}_{\pi(\vec{i})}}(\xi_i,\bar{\xi}_{\pi(\bar{i})})| \le (2^{-j}p^*_F)^{2n}  \Big\{ \sup_{\Del\in\D^j}\sum_{\Del'\in\D^j}  C_N
\Big(1+ d^j(\Del,\Del')\Big)^{-N} \Big\}^n =O(C^n (2^{-j}p^*_F)^{2n})
\EEQ
for some constant $C$, since the sum between $\Big\{\ \Big\}$ converges
for $N>3$.  There remains to discard the possibility of {\em  local factorials}
in the numerator, i.e. combinatorial factors of the form $\prod_{\Del} (n_{\Del} !)$ to a certain power, where $n_{\Del}$ is the number of fields contained in a box $\Del\in\D^j$. 
This does happen in the case of bosonic theories, which requires more elaborated, truncated expansions of the exponentiated action, in order to
produce converging series. However, here, due to the {\em fermionic} character
of the theory, Pauli's principle {\em stronly suppresses} local accumulations
of fields. As a matter of fact, one has the following

\begin{Proposition}[Local bounds for determinants]
\cite{IagMag}   \label{prop:Iago}
Consider an $n\times n$ complex-valued matrix $A=(a_{i,\bar{i}})_{1\le i,\bar{i}\le n}$, $a_{i,\bar{i}}=\del_{\alpha_i,\bar{\alpha}_{\bar{i}}} \tilde{C}^{j,\alpha_i}_{\sigma_i,
\bar{\sigma}_{\bar{i}}}(\xi_i,\bar{\xi}_{\bar{i}})$ of the same form as  (\ref{eq:aii'}). Assume that the kernel $\tilde{C}^{j,\alpha}_{\sigma,\bar{\sigma}}$ is smooth, and that it has a scale $L$ quasi-exponential decay , i.e. that, for every integer $N\ge 0$ and 
multi-index $\vec{\kappa}$, 
\BEQ |\nabla_{\xi}^{\vec{\kappa}} \tilde{C}^{j,\alpha}_{\sigma,\bar{\sigma}}(\xi,\bar{\xi})|\le C_{N,|\vec{\kappa}|}
J^{2} (L/v^*_F)^{-\kappa_0} L^{-(\kappa_1+\kappa_2)} \Big( 1+\frac{|\xi-\bar{\xi}|}{L/v^*_F}\Big)^{-N}
\EEQ
for some $J>0$ (compare with Proposition \ref{prop:C}).
Pave $\R\times\R^2$ by cubes $\Del$ of side $L$ and define $n_{\Del}$,
resp. $\bar{n}_{\Del}$ to be the number of $\xi_i$'s, resp. 
$\bar{\xi}_{\bar{i}}$'s inside $\Del$. Then, for every couple
of  integers
$N,r\ge 0$,  there exists a constant $C_{N,r}$, such that 
\BEQ |\det (A)|\lesssim (C_{N,r})^n J^{2n} \prod_{\Del} 
\prod_{\alpha\in\Z/2^j\Z} \frac{1}{(n_{\Del,\alpha}!)^r (\bar{n}_{\Del,\alpha}!)^r} \sup_{\pi\in {\cal P}_n} \prod_{i=1}^n  \Big( 1+\frac{|\xi_i-\bar{\xi}_{\pi(\bar{i})}|}{L/v^*_F}\Big)^{-N}.  \label{eq:Iago}
\EEQ
\end{Proposition}
 Proposition \ref{prop:Iago} applies in our context to the evaluation of determinants,  with $J=2^{-j}p_F^*$, $L=2^j (p^*_F)^{-1}$ and $\tilde{C}^{j,\alpha}_{\sigma,\bar{\sigma}}(\xi,\bar{\xi})=\int d\mu^*_{\theta}(\Psi^j,\bar{\Psi}^j) \  e^{-\II (p^{\alpha},\xi)} {\Psi}^{j,\alpha}_{\sigma}(\xi) \! \cdot\! \bar{\Psi}^{j,\alpha}_{\bar{\sigma}}(\bar{\xi}) e^{\II (p^{\alpha},\bar{\xi})}$ is
 the covariance kernel of the fields $\Psi^{j,\alpha}$,
 $\bar{\Psi}^{j,\alpha}$ accompanied by their sector oscillation, so that $|\nabla_{\xi} \tilde{C}^{j,\alpha}_{\sigma,\bar{\sigma}}(\xi,\bar{\xi})|=|(\nabla_{\xi}-\II p^{\alpha}) C^{j,\alpha}_{\sigma,\bar{\sigma}}(\xi,\bar{\xi})|
 $. The above Proposition has been proven only for fields
 with a single component (in our context, lying in a single angular sector), however the extension to fields with arbitrary angular sectors is straightforward (as can be checked from
 the arguments below).

\Bigskip As mentioned before, we cannot use this result directly in the context of \S \ref{subsubsection:single-scale-bounds}, because we want to bound sums of diagrams with a given
loop structure, sector assignment and choice of scale $j$ boxes
for vertex locations. So let us rewrite briefly the
main arguments of the proof of Proposition \ref{prop:Iago} and
prove that they do apply to such sums of diagrams. Fix 
some integer $k\ge 0$ (later on to be identified with
some multiple of $r$ in (\ref{eq:Iago})). 
The idea is, for each sector $\alpha\in\Z/2^j\Z$,  to split  each scale $j$ box $\Del$ into 
$\approx n_{\Del,\alpha}/3^k$ equal sub-boxes $\del$ of 
side length scaling like $(n_{\Del,\alpha}/3^k)^{-1/3} \, \times
\, 2^j$, centered in $\xi_{\del}$, containing each
$n_{\del,\alpha}$ fields, with $\sum_{\del\subset\Del} n_{\del,\alpha}=n_{\Del,\alpha}$. Then each $\alpha$-sector propagator line $C_{\theta}^{j,\alpha}(\xi,\cdot)$ connecting a $\Psi^{j,\alpha}$-field
located in $\xi\in\del\subset\Del$ is rewritten using Taylor
expansion as
\BEA && C^{j,\alpha}_{\theta}(\xi,\cdot)=e^{\II (p^{j,\alpha},\xi)}
\ \Big(\Big\{ \sum_{|\kappa|<k} \frac{(\xi-\xi_{\del})^{\kappa}}{\kappa!} \nabla^{\kappa} \Big\} \big(e^{-\II (p^{j,\alpha},\xi_{\del})}
C^{j,\alpha}_{\theta}(\xi_{\del},\cdot)\big) \nonumber\\
&&\qquad + \Big\{\sum_{|\kappa|=k} (\xi-\xi_{\del})^{\kappa}
\int_0^1 dt\, \frac{(1-t)^{k-1}}{(k-1)!} \nabla^{\kappa} \Big\} \big(e^{-\II (p^{j,\alpha},(1-t)\xi_\del+t\xi)}
C^{j,\alpha}_{\theta}((1-t)\xi_{\del}+t\xi,\cdot)\big) \, \Big\} \nonumber\\ \label{eq:Iago-Cnabla}
\EEA

This leads to a rewriting of an individual Feynman diagram as a sum of Feynman diagrams with displaced vertices and differentiated $\Psi^{j,\alpha}$-fields.
Because of Pauli's constraint,  $\Big(\nabla^{\kappa} (e^{-\II(p^{j,\alpha},\xi_{\del})}\Psi_{\sigma}^{j,\alpha}(\xi_{\del}))
\Big)^2=0$. This means that Feynman diagrams with
$n_{\del,\alpha}\gg 3^k$ ($3^k$ being roughly the number of
terms on the first line of  (\ref{eq:Iago-Cnabla})) involve
a large number of $k$-order derivative fields $\Psi^{j,\alpha}$
inside $\del$, enjoying by (\ref{eq:I-II})  a supplementary prefactor $O((2^{-j})^k)$, multiplied by $(\xi-\xi_{\del})^{\kappa}=O((2^j (n_{\Del,\alpha}/3^k)^{-1/3})^k)$,
all together a small factor $O((n_{\Del,\alpha}/3^k)^{-k/3})$. 
Multiplying all these factors yields an overall factor
\BEQ \lesssim C^n \prod_{\Del}\prod_{\alpha} \prod_{\del\subset\Del} 
(n_{\Del,\alpha}^{-1/3})^{\frac{k}{3}(n_{\del,\alpha}-O(3^k))} \EEQ
where $n=\sum_{\Del}\sum_{\alpha}n_{\Del,\alpha}$ is the total number of $\Psi$-fields, and $C\lesssim 3^{k^2/3}$. Summing
all exponents corresponding to sub-boxes $\del$ included in
a given box $\Del$ yields $ n_{\Del,\alpha}^{-ckn_{\Del,\alpha}}$ for some
constant $c>0$, or equivalently, inverse factorials
$(n_{\Del,\alpha}!)^{-1}$ to a power $r\approx k$.  


\subsection{Cooper pair bubble and $\Sigma$-kernel estimates}  \label{subsection:anisotropic}


{\em Proof of (\ref{eq:Aq3}).} We need to show that, assuming
$|q|_+\gtrsim 2^{-j}\mu$,
\BEQ |{\cal A}_q^{j\to}(\Upsilon_{3,diag})- {\cal A}_0^{k\to}(\Upsilon_{3,diag})
|\lesssim m^*, \qquad k:=\lfloor \log(\mu/|q|_+)\rfloor
\EEQ

Assume $q=(0,\vec{q})$ (the easy generalization to $q^0\not=0$ is left to the reader). Copying the derivation of (\ref{eq:sinh-1}), we have
${\cal A}_q^{j\to}(\Upsilon_{3,diag})=\frac{1}{(2\pi)^2} \int d\vec{p}\, \chi^{j\to}(\vec{p}) f_{\vec{q}}(\vec{p})+O(m^*)$, with
\BEQ f_{\vec{q}}(\vec{p}):=\frac{1}{e^*_{|\Gamma|}(\vec{p}+\vec{q})+e^*_{|\Gamma|}(\vec{p})}  \ 
\Big(\sgn(e^*_{|\Gamma|}(\vec{p}+\vec{q}))+\sgn(e^*_{|\Gamma|}(\vec{p}))\Big).
\EEQ
If both signs are equal, then $f_{\vec{q}}(\vec{p})=
\Big|\frac{1}{e^*_{|\Gamma|}(\vec{p}+\vec{q})+e^*_{|\Gamma|}(\vec{p})} \Big|$. Split $\int d\vec{p} (\cdots)$ into $\int_{|p_{\perp}|\gg |\vec{q}|} (\cdots) +
\int_{|p_{\perp}|\lesssim |\vec{q}|}$. In the first regime,
both signs are always equal, and $f_{\vec{q}}(\vec{p})\simeq
f_0(\vec{p})$; more precisely, letting $\theta$ be the
angle $(\vec{p},\vec{q})$, and $q_{//}=\vec{q}\cdot \frac{\vec{p}}{|\vec{p}|}$, 
\BEA &&   \Big|\int_{|p_{\perp}|\gg |\vec{q}|} d\vec{p}\  \chi^{j\to}(\vec{p}) \, \Big(f_{\vec{q}}(\vec{p})-
f_0(\vec{p}) \Big)\Big| \lesssim \int_{|p_{\perp}|\gg |\vec{q}|} d\vec{p}\   \frac{|e^*_{|\Gamma|}(\vec{p}+
\vec{q}) - e^*_{|\Gamma|}(\vec{p})|}{(\frac{p^*_F}{m^*} p_{\perp})^2}
\nonumber\\
&&
\approx m^*\int_{|p_{\perp}|\gg |\vec{q}|} dp_{\perp} \int d\theta\, 
\frac{|q_{//}|}{p_{\perp}^2} \lesssim m^* |\vec{q}| 
\int_{|p_{\perp}|\gg |\vec{q}|}  \frac{dp_{\perp}}{p_{\perp}^2} = O(m^*). 
\EEA
In the opposite regime ($p_{\perp}|\lesssim |\vec{q}|$),
$\sgn(e^*_{|\Gamma|}(\vec{p}+\vec{q}))=\sgn(e^*_{|\Gamma|}(\vec{p}))$ implies either (i) $|\cos(\theta)|\lesssim
\frac{|p_{\perp}|}{|\vec{q}|}$, which amounts to $\vec{p}\in\Omega_{\vec{q}}$, $\Omega_{\vec{q}}:=\{\vec{p}\ |\ |\frac{\pi}{2}\pm\theta|\lesssim \frac{|p_{\perp}|}{|\vec{q}|}\}$;
(ii)
or $\cos\theta\gtrsim \frac{|p_{\perp}|}{|\vec{q}|}$, $e^*_{|\Gamma|}(\vec{p}+\vec{q})\gtrsim \frac{1}{m^*} \Big(2p^*_F |\vec{q}|\cos(\theta)+ |\vec{q}|^2\Big)\gtrsim  e^*_{|\Gamma|}(\vec{p})$ (briefly said,
this is a sub-regime where $\vec{q}$ dominates w.r. to $\vec{p}$).

The small integration volume  in $\theta$ in regime (i) makes the integral
 restricted to $\Omega_{\vec{q}}$ convergent,
\BEA && \int_{(|p_{\perp}|\lesssim |\vec{q}|)\cap \Omega_{\vec{q}}} d\vec{p}\,  \chi^{j\to}(\vec{p})
|f_{\vec{q}}(\vec{p})| \lesssim p^*_F \int _{|p_{\perp}|\lesssim |\vec{q}|} \frac{dp_{\perp}}{|e^*_{|\Gamma|}(\vec{p}+\vec{q})|+|e^*_{|\Gamma|}(\vec{p})|}\ 
O(  \frac{|p_{\perp}|}{|\vec{q}|}) \nonumber\\
&&\qquad  \lesssim \frac{m^*}{|\vec{q}|} \int _{|p_{\perp}|\lesssim |\vec{q}|} dp_{\perp}=O(m^*),
\EEA

In sub-regime (ii), letting $\bar{\theta}:=\frac{\pi}{2}-\theta$, 
\BEA &&  \int_{(|p_{\perp}|\lesssim |\vec{q}|)\cap \Omega_q^c}
d\vec{p}\, \chi^{j\to}(\vec{p})\,  |f_{\vec{q}}(\vec{p})|\lesssim 
\frac{m^* p^*_F}{|\vec{q}|} \int dp_{\perp}\,  
\int_{\frac{|p_{\perp}|}{|\vec{q}|}}^{\frac{\pi}{2}} \frac{d\bar{\theta}}{p^*_F\bar{\theta}+|\vec{q}|} \nonumber\\
&&\qquad= \frac{m^* p^*_F}{|\vec{q}|}
\int_0^{\frac{\pi}{2}} \frac{d\bar{\theta}}{p^*_F\bar{\theta}+|\vec{q}|}\ |\vec{q}| \bar{\theta} =O(m^*) \nonumber\\
\EEA
too. \hfill\eop

\bigskip\noindent {\em Proof of Lemma \ref{lemma:bubble}.}

\Medskip We must evaluate the second derivatives at zero momentum of the
 Cooper pair bubble diagram of \S \ref{subsection:bubble}, $\partial_{q^0}^2 {\cal A}_q^{j\to}(\Upsilon_3)\Big|_{q=0}$
and $\vec{\nabla}^2 {\cal A}_q^{j\to}(\Upsilon_3)\Big|_{q=0}$.

This is a tedious task in Fourier coordinates. It is easier to use an argument
in direct space, where the  positivity argument (\ref{eq:positivity-argument}) can be used.
First
\BEQ -\frac{d^2}{d(q^0)^2}\Big|_{q=0} \int d\xi\, |C^{j\to}(\xi)|^2 \, e^{-\II q^0\tau}= 
-\int d\xi\, \tau^2 \, |C^{j\to}(\xi)|^2,
\EEQ
\BEQ - \sum_{i=1}^2 \frac{d^2}{d(q^i)^2}\Big|_{q=0} \int d\xi\, |C^{j\to}(\xi)|^2 \, e^{-\II \vec{q}\cdot \vec{x}}= 
-\int d\xi\, |\vec{x}|^2 \, |C^{j\to}(\xi)|^2.
\EEQ

We now need an estimate of $C^{j\to}(\xi)=\sum_{k=j_D}^j C^k(\xi)$. The covariance function $C^k$ is decomposed in \cite{FMRT-infinite} into $2^{k/2}$ anisotropic angular
sectors of size $\sim 2^{-k/2}p_F^*$ along the Fermi circle. This makes it easy
to understand where the major contribution to $C^{j\to}(\xi)$ comes. Decompose
$\chi^k$ into $\sum_{\tilde{\alpha}\in\Z/2^{j/2}\Z} \tilde{\chi}^{k,\tilde{\alpha}}$, where
the $(\tilde{\chi}^{k,\tilde{\alpha}})_{\tilde{\alpha}\in\Z/2^{j/2}\Z}$ are anisotropic as indicated, but otherwise similar
to the $(\chi^{k,\alpha})_{\alpha\in\Z/2^j\Z}$ of section \ref{section:fermion}.
Take $\xi=(\tau,\vec{x})\equiv (\tau, x\vec{e}_1)$, where $(\vec{e}_1,\vec{e}_2)$ is an orthonormal basis, with $\vec{e}_1$  following the
direction of some sector $\tilde{\alpha}$, and $x>0$. Let $\tilde{\beta}$ be the index of some
angular sector, and $\vec{p}^{k,\tilde{\beta}}$ the projection onto the Fermi
sphere of some momentum contained in the given angular sector. Then (as proven in \cite{FMRT-infinite}), for every $\kappa\ge 0$, there exists $C_{\kappa}>0$ s.t.  \BEQ |C^{k,\tilde{\beta}}(\xi)|\le C_{\kappa} (p^*_F)^2\,  2^{-3k/2}  (1+2^{-k}\mu |\tau|)^{-\kappa} (1+2^{-k}p_F |x_{//}|)^{-\kappa} (1+2^{-k/2}p_F |x_{\perp}|)^{-\kappa},
\label{eq:5.12}
 \EEQ
(compare with Proposition \ref{prop:C})  where $x_{//}$ is the projection of $\vec{x}$ along $\vec{p}^{k,\tilde{\beta}}$, and $x_{\perp}$ its
projection along the orthogonal direction. So $|C^{k,\tilde{\beta}}(\xi)|$ 
decreases quasi-exponentially outside a box of dimensions $\approx 2^k \mu^{-1}\times 2^k p_F^{-1}\times 2^{k/2} p_F^{-1}$ and volume $\approx  \frac{1}{m^* \mu^2}\, 2^{5k/2} $. For $|\xi|\approx 2^j$, this means
that the main contribution comes from sectors $(k,\tilde{\beta})$ with $k\simeq j$ and $\tilde{\beta}\simeq \tilde{\alpha}$. Then
\BEA  C^{k,\tilde{\beta}}_{1,1}(\xi) &=& (2\pi)^{-3} \int dp\, e^{\II p\cdot \xi} \chi^{k,\tilde{\beta}}(p) \frac{
-\II p_0-e^*(\vec{p})}{(p^0)^2+(e^*_{|\Gamma|}(\vec{p}))^2} \nonumber\\
& \sim&  - (2\pi)^{-2}\, 2^{-k/2} \, e^{-\II \vec{p}^{k,\tilde{\beta}}\cdot\vec{x}} \int dp^0 \, \int p_F^*\, dp_{\perp} \  \chi^{k}(p_{\perp}) \, 
\frac{p^*_F}{m^*}\frac{p_{\perp}}{((p^0)^2+(e^*_{|\Gamma|}(\vec{p}))^2}  e^{\II \del p\cdot\xi} \nonumber\\   \label{eq:5.13}
\EEA
where $\del p:=(p^0,\vec{p}-\vec{p}^{k,\tilde{\beta}})$ and (to leading order)
$|\vec{p}|=p_F^*+p_{\perp}$ has been replaced with $p_F^*$. If $|\xi|\ll 2^j \mu^{-1}$,
then the first-order Taylor expansion $e^{\II \del p\cdot\xi}\sim 1+\II \del p\cdot\xi$ is a good approximation. 
By symmetry, the term of order $0$ vanishes, so the maximum order of 
magnitude of $|C_{1,1}^{k,\tilde{\beta}}(0,\vec{x})|$ is obtained not in a neighborhood of 
$0$, but for $\vec{x}\approx 2^j p_F^{-1}\,  \vec{e}_1$. Replacing $e^{-\II \del \vec{p}\cdot\vec{x}}-1$ by $-\II \del\vec{p}\cdot\vec{x}$, one obtains for such $\vec{x}$
\BEQ  |C^{k,\tilde{\beta}}_{1,1}(0,\vec{x})|  \approx 2^{-k/2}  |\vec{x}|\,  m^* \int dp^0\,
dp_{\perp} \approx  2^{-3j/2}\,  (p_F^*)^2   \label{eq:5.14}
\EEQ
as expected from (\ref{eq:5.12}), whereas
\BEQ |C^{k,\tilde{\beta}}_{1,1}(\xi)| \approx 2^{-5j/2} |\xi| \, (p^*_F)^3, \qquad |\xi|\ll 2^j p_F^{-1}.  \label{eq:5.15}
\EEQ
 Hence (integrating over a box and summing over the $\approx 2^{j/2}$
angular sectors $\tilde{\beta}$) one may conjecture that 
\BEQ \int  d\xi\, |\vec{x}|^2 \, |C^{j\to}_{1,1}(\xi)|^2 \approx  2^{j/2}\ \cdot\  \frac{1}{m^* \mu^2}\, 2^{5j/2} \ \cdot\  \Big(2^{j} (p^*_F)^{-1} \times 2^{-3j/2}(p^*_F)^2 \Big)^2=2^{2j} \mu^{-1}. \EEQ 
Letting $2^{-j}\mu \approx \Gamma_{\phi}$ be near the transition scale, this yields 
$-\vec{\nabla}^2 {\cal A}_q(\Upsilon_3)\Big|_{q=0}\approx  (\frac{p^*_F}{m^*})^2 \frac{m^*}{\Gamma_{\phi}^{2}} \, \Id$, which
is what one wanted to prove. Considering instead $\partial^2_{q^0}{\cal A}_q(\Upsilon_3)\big|_{q=0}$, one expands similarly $e^{\II p^0 \tau}=1+\II p^0\tau+\cdots$, replaces $e^{\II p^0\tau}-1$ by $\II p^0\tau$ and obtains a bound
of the same magnitude but without the dimensionful prefactor $v^2_{\phi}\approx
(\frac{p^*_F}{m^*})^2$.
 
\Bigskip  This is however not a rigorous proof, since there is some overlap
between sectors, namely, e.g.   $\int d\xi\, |\vec{x}|^2\, C^{j,\tilde{\alpha}}(\xi) (
C^{k,\tilde{\beta}}(\xi))^*\not=0$  in general for neighboring sectors $(j,\tilde{\alpha})$,
$(k,\tilde{\beta})$ with $|j-k|,|\tilde{\alpha}-\tilde{\beta}|=O(1)$. It is
simpler -- though less instructive -- to integrate over the angular coordinate
along the Fermi sphere; making computations for an arbitrary value of $\Gamma$ and
$j$ will allow us to produce at the same time the asymptotics (\ref{eq:1pt-density-kernel0}) for the one-point density kernel, and the desired order of
magnitude for the second derivatives of ${\cal A}(\Upsilon_3)$. Let $\rho:=|\vec{p}|$. We use the
standard  formula
\BEQ \frac{1}{2\pi}\, \int d\theta\, e^{\II \rho |\vec{x}|\cos(\theta)}=J_0(\rho|\vec{x}|)\sim_{|x|\to\infty} \sqrt{\frac{2}{\pi\rho|\vec{x}|}}\ \cos(\rho|\vec{x}|-\frac{\pi}{4}) + O (\frac{1}{\rho |\vec{x}|})
\EEQ   
in terms of Bessel functions of the first kind, and neglect the
ultra-violet cut-off at scale $j_D$. Then the theorem of residues yields
\BEA && C^{j\to}_{1,1}(\xi)= (2\pi)^{-2} \int d\vec{p}\,  \sgn(p_{\perp})f(p_{\perp})\chi^{j\to}(p)\,
e^{-|e^*_{|\Gamma|}(\vec{p})\tau|+\II \vec{p}\cdot\vec{x}} \nonumber\\
&&\ \ = (2\pi)^{-1}  \int dp_{\perp}\,  \chi^{j\to}(p)\,\sgn(p_{\perp})
f(p_{\perp})\, 
e^{-|e^*_{|\Gamma|}(p_{\perp})\tau|}\ \cdot\  (p_F^*+p_{\perp}) J_0((p^*_F+p_{\perp})|\vec{x}|) \nonumber\\ \label{eq:resinter} \\
&&\ \ \sim_{|\vec{x}|\to\infty} (2\pi)^{-1} \sqrt{\frac{2}{\pi|\vec{x}|}}  \int dp_{\perp}\, \chi^{j\to}(p)\, \sgn(p_{\perp})f(p_{\perp})\,  e^{-|e^*_{|\Gamma|}(p_{\perp})\tau|}\ \cdot\ \nonumber\\
&&\qquad\qquad\qquad\qquad\qquad \ \cdot\    (p_F^*+p_{\perp})^{1/2} \, \cos((p_F^*+p_{\perp})|\vec{x}|).
\nonumber\\   \label{eq:5.18}
\EEA
where $f(p_{\perp}):=\begin{cases} \frac{(e^*-e^*_{|\Gamma|})(p_{\perp})}{2e^*_{|\Gamma|}(p_{\perp})} \qquad (p_{\perp}\tau>0) \\ \frac{(e^*_{|\Gamma|}+e^*)(p_{\perp})}{2e^*_{|\Gamma|}(p_{\perp})} \qquad (p_{\perp}\tau<0)
\end{cases}$. 
Replacing $\chi^{j\to}(p)$ by $\chi(p)\equiv \lim_{j\to +\infty}\chi^{j\to}(p)$,
and letting (as a vestige of the smoothing due to ultra-violet cut-off) $C^{j\to}_{1,1}(0,\vec{x}):=\half\lim_{\tau\to 0^+} \Big(C_{1,1}^{j\to}(\tau,\vec{x})+C_{1,1}^{j\to}(-\tau,\vec{x})\Big)$, so that the coefficient
$f(p_{\perp})$ is replaced by the even function $\tilde{f}(p_{\perp}):=\half(f(p_{\perp})+f(-p_{\perp}))=\half \frac{e^*(p_{\perp})}{e^*_{|\Gamma|}(p_{\perp})}$ in the above integrals,
one obtains (\ref{eq:1pt-density-kernel0}) in the limit $\Gamma\to 0$ by remarking that
\BEQ \int_0^{+\infty} dp_{\perp}\, \chi(p) \, (p_F^*+p_{\perp})^{1/2} \, e^{\pm\II
(p^*_F+p_{\perp}) |\vec{x}|} \sim_{|\vec{x}|\to\infty} -\frac{e^{\pm \II p^*_F
|\vec{x}|}\, (p^*_F)^{1/2}}{|\vec{x}|}.   \label{eq:5.19}
\EEQ  
Assume on the other hand  that  $|\vec{x}|\approx 2^{j_{\phi}}(p^*_F)^{-1}$. Hence one may (up to an exponentially small error) replace
$\chi(p) (p^*_F+p_{\perp})^{1/2}$ in the above expressions by $(p^*_F)^{1/2}$, and
$\sgn(p_{\perp})\tilde{f}(p_{\perp})$ by $ \frac{p_{\perp}}{\sqrt{p^2_{\perp}+p^2_{\phi}}}$, with $p_{\phi}:=\frac{m^*}{p_F^*}\Gamma_{\phi}$. Hence  
\BEA  && C_{1,1}(0,\vec{x})\sim \frac{C}{\sqrt{|\vec{x}|}}\, \Re\, \Big\{e^{\II p^*_F |\vec{x}|}\,  (p^*_F)^{1/2} \, \int dp_{\perp}\,  \frac{p_{\perp}}{\sqrt{p^2_{\perp}+p^2_{\phi}}} \, e^{\II p_{\perp}|\vec{x}|}   \Big\} \nonumber\\
&&\sim  \frac{C}{\sqrt{|\vec{x}|}}\,\Re\, \Big\{  e^{\II p^*_F |\vec{x}|}\,  (p^*_F)^{1/2} \,
\frac{-\II p_{\phi}^2}{|\vec{x}|} \int \frac{ e^{\II p_{\perp} |\vec{x}|} dp_{\perp}}{(p^2_{\perp}+p^2_{\phi})^{3/2}} \Big\} \,  \nonumber\\
&&\sim \frac{C}{\sqrt{|\vec{x}|}}\, \sin(p^*_F |\vec{x}|)\,   (p^*_F)^{1/2} \, p_{\phi}\,  K_1(p_{\phi}|\vec{x}|)
\EEA
with $C=2(2\pi)^{-1} \sqrt{\frac{2}{\pi}}$, where $K_1(p_{\phi}|\vec{x}|)$ is a modified Bessel
function (exponentially decreasing at infinity for $|\vec{x}|\gg p_{\phi}^{-1}$), see \cite{Bat} (7) p. 11. All together, one has found that
$C_{1,1}(0,\vec{x})$ is equal for $|\vec{x}|\approx p_{\phi}^{-1}$  to $\approx 
(p^*_F)^{1/2} p_{\phi}^{3/2}\approx 2^{-3j_{\phi}/2} (p^*_F)^2$, in conformity
with  (\ref{eq:5.14}), times a fast oscillating phase function.

\Medskip The above argument is easily adapted to prove that $|C_{1,1}(\tau,\vec{x})|\approx 2^{-3j_{\phi}/2} (p^*_F)^2$ for $|\tau|\approx \Gamma_{\phi}$.
Namely, the change of variable $p_{\perp}\mapsto P:=\sqrt{p^2_{\perp}+p_{\phi}^2}$
($p_{\perp}>0$)
reduces the problem to the evaluation of the Laplace transform of the
function $P\mapsto \frac{P}{\sqrt{P^2-p_{\phi}^2}} \, {\bf 1}_{P>p_{\phi}}$,
which is also equal to the function $p_{\phi}K_1(p_{\phi}|\vec{x}|)$, see
\cite{Bat} (29) p. 136. 
 
\hfill\eop


\vskip 2cm
\noindent {\em We further prove eq. (\ref{eq:Sigma-error-terms}) and (\ref{eq:bound-nabla-kappa-Sigma}).}
 The infra-red cut-off bound (\ref{eq:bound-nabla-kappa-Sigma}) actually plays a minor
 r\^ole for the proof of (\ref{eq:Sigma-error-terms}), so let us prove it first. We concentrate on spatial gradients
 $(\frac{\mu}{p^*_F}\nabla_{\vec{q}})^{\kappa}f(q)$, with $f={\cal A}^{j\to}(\Upsilon_3)$, Pre$\Sigma^{j\to}$ or $\Sigma^{j\to}$, but the bounds we write down also hold
 for homogeneized gradients $\nabla^{\kappa}_{q}:=\nabla_{q^0}^{\kappa^0} (\frac{\mu}{p_F^*}\nabla_{q^1})^{\kappa^1} (\frac{\mu}{p_F^*}
 \nabla_{q^2})^{\kappa^2}$. 
 Let $|q|_+\gg \Gamma_{\phi}$. Then 
$(\frac{\mu}{p^*_F}\nabla_{\vec{q}})^{\kappa} {\cal A}_q^{j\to}(\Upsilon_3) \lesssim \frac{m^*}{|q|_+^{\kappa}} $  as shown by following the arguments in
the proof of (\ref{eq:Aq3}); $1-\lambda{\cal A}_q^{j\to}(\Upsilon_{3,diag})
\approx \lambda m ^*\log(|q|_+/\Gamma_{\phi})$; and
$\frac{\mu}{p^*_F}\nabla_{\vec{q}} {\mathrm{Pre}}\Sigma^{j\to}(q)=\frac{\mu}{p^*_F}\nabla_{\vec{q}} \frac{\lambda}{1-\lambda{\cal A}_q^{j\to}(\Upsilon_3)} = \frac{\lambda}{(1-\lambda{\cal A}_q(\Upsilon_3))^2} \lambda \frac{\mu}{p^*_F} \nabla_{\vec{q}} {\cal A}_q^{j\to}(\Upsilon_3)
=O(\frac{1}{m|q|_+ \log^2(|q|_+/\Gamma_{\phi})})$, from
which by an easy induction (\ref{eq:bound-nabla-kappa-Sigma})
holds.

\Medskip We now return to corrections to leading-order behavior of $\Sigma(q)$,

\BEQ \Sigma(\xi)\sim_{|\xi|_+\to\infty} \frac{1}{4\pi}  \frac{g_{\phi}/v_{\phi}^2}{|\xi|_+} (1+O(\frac{1}{\Gamma|\xi|_+})). 
\EEQ
 Error terms involve both the correction to leading-order term
in (\ref{eq:1.59}) for $|q|_+\lesssim \Gamma_{\phi}$, and a subleading
contribution obtained by integrating over large transfer momenta $q$ such
that $\Gamma_{\phi}\ll |q|_+\lesssim \hbar \omega_D$.  In either case, the idea is
to apply repeated integrations by parts on $f(q)=\Sigma(q)$ or $f(q)=\Sigma(q)-\frac{g_{\phi}}{q_0^2+v_{\phi}^2 |\vec{q}|^2}$,
$$\Big|\int dq \, e^{\II (q,\xi)}
f(q)\Big| \lesssim \frac{1}{((\xi^1)^2+(\xi^2)^2))^{|\kappa|/2}} \int dq \, | \nabla_q^{\kappa}f(q)|,$$
 where $\kappa\in\N$ is 
arbitrarily large and $\nabla_q^{\kappa}=\partial_{q^0}^{\kappa_0}
\prod_{i=1}^2 (\frac{1}{v_{\phi}}\partial_{q^i})^{\kappa_i}$. The contribution of large
transfer momenta is easily
bounded using (\ref{eq:bound-nabla-kappa-Sigma}):
if $|\kappa|> 3$, 
\BEQ \int_{|q|_+\gg\Gamma_{\phi}} dq\, |\nabla_{\vec{q}}^{\kappa}\Sigma(q)|
=O(\frac{1}{m^* v_{\phi}^2 \Gamma_{\phi}^{\kappa-3}}). \EEQ
So assume that $|q|_+\lesssim\Gamma_{\phi}$;  we leave it to the reader to check that
\BEQ \nabla_{\vec{q}}^{\kappa} \Big(
\Sigma(q)-\frac{g_{\phi}}{q_0^2+v_{\phi}^2 |\vec{q}|^2}\Big)=
\frac{O(g_{\phi})}{\Gamma_{\phi}^2 |q|_+^{|\kappa|}}
\EEQ
 for $|q|_+\lesssim \Gamma_{\phi}$. Hence
\BEQ \int_{|q|_+\lesssim\Gamma_{\phi}} dq\, |\nabla_{\vec{q}}^2\Sigma(q)|
\lesssim (\frac{p^*_F}{m^*})^{-2} \frac{g_{\phi}}{\Gamma_{\phi}}
\approx \frac{g_{\phi}}{v_{\phi}^2\Gamma_{\phi}},
\EEQ
 from which, finally, (\ref{eq:Sigma-error-terms}) holds.
\hfill \eop


\subsection{Error terms for  Ward identities}
\label{subsection:error}


We bound here the error terms "err." due to the
propagator cut-offs neglected in \S \ref{subsection:Ward}.
We consider to begin with the variation of the fermionic covariance
kernel $C_{\theta}^{\to j_D}(p)\equiv\chi^{\to j_D}(|p|/\mu) C_{\theta}(p)$, $C_{\theta}(p):=\frac{1}{\II p^0-e^*(\vec{p})
\sigma^3-{\mathbb{\Gamma}}(\theta)}$ under an infinitesimal gauge transformation $\alpha$. The commutator 
$[\alpha(q),C_{\theta}^{\to j_D} (p)]$ is a sum of two terms,
$[\alpha(q),C_{\theta}(p)] \chi^{\to j_D}(|p|/\mu) +
[\alpha(q),\chi^{\to j_D}(|p|/\mu)] C_{\theta}(p)$. Since
$\alpha(\xi)$ acts multiplicatively on fields, its Fourier transform
$\alpha(q)$ acts by convolution,
\BEA  [\alpha\ast,\chi^{\to j_D}(\cdot)]f(p) &=& \int dq\ \Big\{ \alpha(q) \chi^{\to j_D}(|p-q)|/\mu-
\chi^{\to j_D}(|p|/\mu)\alpha(q) \Big\}\ 
f(p-q) \nonumber\\
&=&\int dq\  \Big\{ \chi^{\to j_D}(|p-q|/\mu)-\chi^{\to j_D}(|p|/\mu) \Big\}\ \alpha(q)f(p-q).
\EEA
Write $(\del_{\alpha})_{cut}I_{n-1}(\xi_1,\ldots,
\xi_{n-1})$ the contribution of this term to (\ref{eq:3.32}),
 and proceed as done in the proof of Ward identities, see
 \S \ref{subsection:Ward} {\bf B.}, namely, take  the functional derivative
 $\frac{\del}{\del\alpha(\xi_n)}$, multiply by
 the low-momentum kernel $\Sigma^{k_{+,n}}(\xi_{ext,n}-\xi_n)$,
 and integrate w.r. to $\xi_n$. After a Fourier transform, this
 is equivalent to the expression
\BEQ \int dq\, \Sigma^{k_{+,n}}(q)\  \Big\{\chi^{\to j_D}(|p-q|/\mu)-\chi^{\to j_D}(|p|/\mu)\Big\}\,  f(p-q). \label{eq:5.40}
\EEQ
The above integral is restricted to $|q|_+\approx 2^{-k_+}\mu$. 
 By symmetry, the difference in (\ref{eq:5.40}) vanishes
 to order 1 when $q\to 0$.   Hence the kernel  $\chi^{\to j_D}(|p-q|/\mu)-\chi^{\to j_D}(|p|/\mu)$  in (\ref{eq:5.40}) may be replaced by a second-order derivative bounded like $|q|^2_+\ |\nabla^2_p\chi^{\to j_D}(|p|/\mu)|=O(2^{-2k_+}\mu)$.
{\em Thus the error term exhibits a squared  spring term $2^{-2(k_+-j_+)}$ as in \S \ref{subsection:Ward}}, and a further
absolute small factor $O(2^{-2j_+})$.

\Bigskip Let us illustrate this on an example.
\Bigskip

{\centerline{
\begin{tikzpicture}[scale=1.7]
\draw(-3.7,0) arc(195:345:0.8 and 0.4);
\draw(-2.15,0) arc(15:165:0.8 and 0.4);
\draw[dashed](-3.7,0)--(-3.7,-1);
\draw[dashed](-2.15,0)--(-2.15,-1.4);
\draw[dashed](-2.8,0.3)--(-2.8,-2.1);
\draw[dashed](-3.1,-0.3)--(-3.1,-2.8);
\draw(-3.7,0) node {\textbullet}; 
\draw(-2.15,0) node {\textbullet}; 
\draw(-2.8,0.3) node {\textbullet}; 
\draw(-3.1,-0.3) node {\textbullet}; 
\draw[decorate,decoration=snake](-3.7,-1)--(-4.3,-1);
\draw[decorate,decoration=snake](-2.15,-1.4)--(-2.15+0.6,-1.4);
\draw[decorate,decoration=snake](-2.8,-2.1)--(-2.8+0.6,-2.1);
\draw[decorate,decoration=snake](-3.1,-2.8)--(-3.1-0.6,-2.8);
\draw(-4,-0.85) node {\tiny $q_1$};
\draw(-2.15+0.3,-1.45+0.2) node {\tiny $q_4$};
\draw(-2.8+0.3,-2.15+0.2) node {\tiny $q_3$};
\draw(-3.4,-2.85+0.2) node {\tiny $q_2$};
\draw(-3.4,0.4) node {\tiny $p_{13}$};
\draw(-4.4,-0.7)--(-2.15+0.6,-0.7);
\draw(-4.4,-1.7)--(-2.15+0.6,-1.7);
\draw(-4.3,-0.5) node {\tiny $j_+$};
\draw(-4.3,-1.5) node {\tiny $k_+$};
\end{tikzpicture}
}}

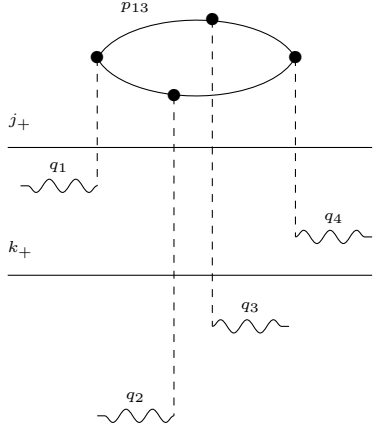
\Medskip {\bf\tiny Fig. \ref{subsection:error}.1. An
example of multi-scale diagram.}

\Bigskip Internal momenta are $p_{ij}$, $(i,j)=(1,2),(1,3),(2,4),(3,4)$ connecting $\xi_{ext,i}$ with external transfer
momentum $q_i$ to $\xi_{ext,j}$ with external transfer
momentum $q_j$. Let $p=p_{12}$. The amplitude of the
amputated diagram $\Upsilon$ is given by an integral in $p$,
\BEQ {\cal A}(\Upsilon;\xi_{ext,1})=\int dp\ 
\prod_{(i,j)} (\chi^{\to j_D} C_{\theta})(p_{ij}).
\EEQ
Let $f_{ij}(p):=\prod_{(i',j')\not=(i,j)} \chi(p_{i'j'})
\ \cdot\ \prod_{(i',j')} C_{\theta}(p_{i',j'}).$ Then
\BEQ (\del_{\alpha})_{cut} {\cal A}(\Upsilon;\xi_{ext,1}) = \sum_{(i,j)} \int dp\, \int dq\ 
\ \Big\{\chi^{\to j_D}(|p-q|/\mu)-\chi^{\to j_D}(|p|/\mu)\Big\}\ \alpha(q)  f_{ij}(p-q).  \label{eq:5.42}
\EEQ

\vskip 1cm

\Medskip A similar analysis may be done for the variation under
$\alpha$ of the {\em Goldstone boson} propagator 
$\Sigma^{j_+\to}(p)=\chi_+^{j_+\to}(|p|_+/\mu)\Sigma(p)$.  This time,  $\chi^{j_+\to}(|p-q|_+/\mu)-\chi^{j_+\to}(|p|_+/\mu)$
in (\ref{eq:5.42}) may be replaced by a second-order derivative bounded like $|q|^2_+\ |\nabla^2_p\chi^{j_+\to}(|p|_+/\mu)|= O(2^{-2(k_+-j_+)})$, yielding precisely the desired squared
spring factor.


\subsection{Mayer expansion}


We introduce here a variant of the BKAR formula already described in \S \ref{subsection:integration-by-parts}, to be
used for the Mayer expansion (see \S \ref{subsubsection:Mayer}). The idea is to test overlaps between polymers
in ${\cal P}^{j\to}$ with a {\em small} number of external legs,
 so as to extract leading-order translation-invariant quantities.

\Medskip Let ${\cal O}$ be the
set of  {\em polymers} in $\P^{j\to}$. Among these polymers, there are polymers with 
$<N_0=8$  external legs, making up a subset ${\cal O}_1\subset{\cal O}$. The complementary set ${\cal O}_2$ is made up of polymers
with $\ge N_0$ external legs, which need no particular further
treatment. 
The following variant of BKAR's formula, found originally in \cite{AbdRiv1}, is
 stated in the present form in \cite{MagUnt2}. We now denote by $\{\P_{\ell},\P'_{\ell}\}$ a pair of polymers connected by a link $\ell\in L({\cal O})$.

 \begin{Proposition}[restricted 2-type cluster or BKAR2 formula] \label{prop:BK2}

Assume ${\cal O}={\cal O}_1\amalg {\cal O}_2$. Choose as initial object an object
$o_1\in {\cal O}_1$ of type 1, and stop the Brydges-Kennedy-Abdesselam-Rivasseau expansion as soon as a
link to an object of type 2 has appeared. Then choose a new object of type $1$, and so
 on. This  leads to a  restricted expansion, for which {\em only} the  link variables
  $z_{\ell}$, with $\ell\not\in {\cal O}_2\times {\cal O}_2$, have been weakened. The
   following closed formula holds. Let $\vec{S}:L({\cal O})\to [0,1]$ be a link
   weakening of ${\cal O}$, and $F=F((\vec{S}_{\ell})_{\ell\in L({\cal O})})$ 
 a smooth function. Let ${\cal F}_{res}({\cal O})$ be the set of  forests
    $\G$ on $\cal O$, each component of which is (i) either a tree of
objects of type $1$, called {\em unrooted tree}; (ii)or a {\em rooted tree} such that only the root is of type $2$. Then
\BEQ F(1,\ldots,1)=\sum_{\G\in {\cal F}_{res}({\cal O})}
\left( \prod_{\ell \in L(\G)}
\int_0^1 dW_{\ell} \right) \left( \left(\prod_{\ell\in L(\G)}  
\frac{\partial}{\partial
S_{\ell} } \right) F(S_{\ell}(\vec{W})) \right),  \label{eq:Mayer-BKAR}
 \EEQ

where $S_{\ell}(\vec{W})$ is either $0$ or the minimum of the $w$-variables running along the
unique path in $\bar{\G}$ from $\P_{\ell}$ to $\P'_{\ell}$, and $\bar{\G}$ is the
forest obtained from $\G$ by merging all roots of $\G$ into a single vertex.

\end{Proposition}

\medskip\noindent The above Proposition is applied to
the non-overlap function $F$=NonOverlap in \S \ref{subsubsection:Mayer}.



\section*{Index of notations} \label{section:index}


\begin{tabular}{cc}
${\cal A}(\Upsilon)$, diagram amplitude \pageref{subsection:bubble} \\ ${\cal A}(\P)$, polymer
amplitude \pageref{eq:A(P)} & 
$C^*_{\theta}$, $C^{j(,\alpha)}_{\theta}$, symmetry-broken  covariance 
\pageref{eq:C*}, \pageref{eq:Cj} \\
$\D^j$, set of scale $j$ boxes \pageref{def:boxes} &
$d\mu^*_{\theta}$,  symmetry-broken Gaussian measure \pageref{def:Grassmann-measure} \\ 
$d\mu_{\theta;\lambda}$, symmetry-broken measure  \pageref{def:Grassmann-measure} &
$e(p)$, dispersion relation \pageref{eq:cut-off3} \\
  $e^*(p)$,
renormalized dispersion relation \pageref{eq:e*(p)} 
& $e^*_{|\Gamma|}(p)$, \pageref{eq:e*Gamma} \\
$g$, non-dimensional coupling constant \pageref{eq:g} &
$G$, Green function \pageref{eq:G}  \\
$g^0_{\phi}$, pre-Goldstone boson coupling constant, \pageref{eq:gphi} & $g_{\phi}$, Goldstone boson coupling constant, \pageref{eq:gphi} \\
Hor$^j$, horizontal expansion operator, \pageref{eq:jDBKAR},
\pageref{prop:BKAR} &  $I^{j_+\to}_{(0,n)}, I^{j_+\to}_{(k_{//},k_{\perp})}$, $n$-point functions \pageref{eq:I0n},\pageref{eq:Ikk} \\
$I^{j_+\to}_{s.c.,//}$, $I^{j_+\to}_{s.c.,\perp}$,  one-point function \pageref{def:gap-one-fct} \\
$j_D$, highest momentum scale (Debye cut-off) \pageref{eq:jD}
& $j_+$, bosonic scale \pageref{eq:xi+}\\
$j_{\phi}$, transition scale,   \pageref{eq:jphi}  & $j'_{\phi}$, modified transition scale \pageref{eq:j'phi},\pageref{def:j'phi} \\
${\cal L}_{\theta}$, action \pageref{def:Grassmann-measure} &
${\cal L}_{\theta}^{\to }(\vec{t})$, dressed action 
\pageref{def:dressing} \\
LN, linking number \pageref{eq:LN} \\
$m$, mass  & $m^*$, renormalized mass \pageref{eq:jphi} \\
$\mu$, chemical potential & $\mu^*$, renormalized chemical
potential \pageref{eq:jphi} \\
$N_j$ number of scale $j$ angular sectors 
\pageref{eq:intro-N},\pageref{def:angular},\pageref{eq:viivii'} \\
$N_{ext}$, \pageref{eq:Next} & $N_{ext,+}$, \pageref{eq:NextNext+} \\
$\P$, polymer \pageref{subsection:cluster} &
$\P^{j\to},{\cal P}^{j\to}$, (set of) lowest scale $j$ polymers
\pageref{Pjto} \\
$p_{\perp}$, transverse momentum coordinate \pageref{subsection:multi-scale} &    $p_c$, \pageref{eq:g'phi} \\
$p_F$, Fermi momentum \pageref{eq:A0} & $p^*_F$, renormalized
Fermi momentum \pageref{eq:p*F} \\
$p^{j,\alpha}$, momentum in ${\cal S}^{j,\alpha}$ \pageref{def:Cjm} & $s^j$, scale $j$ cluster parameters 
\pageref{subsubsection:hor-cluster} \\ 
${\cal S}^{j(,\alpha,k)}$ angular sectors \pageref{def:angular},
\pageref{eq:further-angular} &
$S$, Mayer expansion parameters \pageref{eq:Mayer2}, \pageref{subsubsection:Mayer} \\
 $t^j$, vertical expansion
parameters \pageref{subsubsection:vert-cluster} \\
$v_F$, Fermi velocity \pageref{eq:vF} \\
$v^0_{\phi}$, Goldstone pre-boson velocity \pageref{eq:vphi} & $v_{\phi}$, Goldstone boson velocity \pageref{eq:vphi}  \\
Vert$^j(\P)$, vertical expansion parameter \pageref{subsubsection:vert-cluster} & 
$Z^j$, wave-function renormalization \pageref{def:counterterms} \end{tabular}

\Medskip
\begin{tabular}{cc}
$\alpha$, angular sector \pageref{def:angular} &
${\mathbb{\Gamma}},{\mathbb{\Gamma}}(\theta)$, $\Gamma$-matrix
\pageref{Theorem2} \\ 
$\Gamma$, complex \pageref{eq:equivariant} & 
$\vec{\Gamma}$, vector \pageref{eq:equivariant} \\
${\mathbb{\Gamma}}^{//,\perp}$ or ${\mathbb{\Gamma}}^{//,\perp}(\theta)$, \pageref{eq:Gamma//Gammaperp} &   $\Gamma_{\phi}$, energy gap \pageref{eq:Gammaphi} \\
$\Del$, box \pageref{def:boxes} & $\chi_{\Del}$, box
cut-off function \pageref{def:boxes}\\
$\lambda$,  coupling constan  \pageref{eq:lambda} \\
$\bar{\Pi}$, Bethe-Salpeter kernel \pageref{eq:Pibar(q)} &
$\tilde{\Pi}$, \pageref{eq:Pitilde} \\ 
$\tilde{\Pi}^{disp}$, \pageref{eq:Sigmadispkernel} & 
$\tilde{\Pi}^{proj}$, \pageref{eq:PitildePi} \\
$\PreSigma$, pre-Goldstone boson propagator \pageref{eq:PreSigma} \\
    $\Sigma$, Goldstone boson propagator
\pageref{eq:PreSigma},\pageref{eq:Sigmaj+Sigma}  
& $\tilde{\Sigma}^{disp}$, \pageref{eq:Sigmadispkernel}\\ 
$\Sigma_F$, Fermi circle \pageref{eq:SigmaF} & 
$\Sigma^*_F$, renormalized Fermi circle \pageref{eq:SigmaF*}\\
$\Upsilon_{3,diag}$, Cooper pair bubble diagram \pageref{Cooper-pair-bubble} &  $\Upsilon_{3,off}$, off-diagonal bubble
diagram \pageref{eq:A0Upsilon3off} \\
$\Psi,\bar{\Psi}$, Nambu spinors  \pageref{subsection:Nambu}
& $\Psi^{j,(\alpha)},\bar{\Psi}^{j,(\alpha)}$ \pageref{eq:Cj}
\\ 
$\omega(N_{ext})$, degree of divergence \pageref{eq:rel-PC} &
$\omega_D$, Debye energy \pageref{eq:cut-off3} \\
$|\, \cdot\, |$, fermionic norm \pageref{eq:equivariant} &
$|\, \cdot\, |_+$, bosonic norm \pageref{eq:equivariant}
\end{tabular}



\end{document}